\documentclass[12pt]{article}
\usepackage{jheppub}
\pdfoutput=1
\usepackage{epstopdf}

\usepackage{mathrsfs}
\usepackage{psfrag}
\usepackage{color}
\usepackage{hyperref}

\usepackage{slashed}
\usepackage{feynmp-auto}
\usepackage{simplewick}
\usepackage{cancel}

\usepackage[table]{xcolor}

\usepackage{amsmath,bbm,array,amsfonts,graphicx,wrapfig,arydshln,lscape,float,multirow,longtable,rotating,makecell,todonotes}
\usepackage{url}

\newcommand{\la}{\langle}
\newcommand{\ra}{\rangle}

\newcommand{\ab}[1]{\langle #1 \rangle}

\def\dlog{d\log}

\def\N{\mathcal{N}}

\newcommand{\NeqFour}{\mathcal{N}=4\text{ sYM}}

\def\MHVbar{\overline{\text{MHV}}}

\definecolor{lightyellow}{rgb}{1,1,0.85}
\definecolor{lightblue}{rgb}{0.87, 0.94, 1}
\definecolor{lightred}{rgb}{1, 0.85, .85}
\definecolor{lightorange}{rgb}{1, 0.9, .8}
\definecolor{lightgreen}{rgb}{0.88, 1, .88}
\definecolor{lightpurple}{rgb}{0.94, .88, .94}
\definecolor{mygrey2}{rgb}{0.9, 0.9, .9}
\definecolor{lightbrown}{rgb}{0.94, 0.91, .88}
\definecolor{lightmagenta}{rgb}{1, 0.9, 1}
\definecolor{mygreen}{rgb}{0.6,0.8,0.3}

\definecolor{airforceblue}{rgb}{0.36, 0.54, 0.66}
\definecolor{bananayellow}{rgb}{1.0, 0.88, 0.21}
\definecolor{bittersweet}{rgb}{1.0, 0.44, 0.37}
\definecolor{blue(ncs)}{rgb}{0.0, 0.53, 0.74}
\definecolor{bole}{rgb}{0.47, 0.27, 0.23}
\definecolor{brass}{rgb}{0.71, 0.65, 0.26}
\definecolor{bronze}{rgb}{0.8, 0.5, 0.2}
\definecolor{brgreen}{rgb}{0.0, 0.26, 0.15}
\definecolor{burgundy}{rgb}{0.5, 0.0, 0.13}
\definecolor{cherry}{rgb}{1.0, 0.72, 0.77}
\definecolor{cocao}{rgb}{0.82, 0.41, 0.12}
\definecolor{citrine}{rgb}{0.99, 0.82, 0.07}
\definecolor {celadon} {rgb} {0.67, 0.88, 0.69}
\definecolor {chromeyellow} {rgb} {1.0, 0.65, 0.0}
\definecolor{lavenderblue}{rgb}{0.8, 0.8, 1.0}

\newcommand{\tred}[1]{\textcolor{red}{#1}}
\newcommand{\tblue}[1]{\textcolor{blue}{#1}}

\newcommand{\twhite}[1]{\textcolor{white}{#1}}

\newcommand{\tbrgreen}[1]{\textcolor{brgreen}{#1}}

\title{Positive geometry, local triangulations, and the dual of the Amplituhedron}

\author{Enrico Herrmann,$^1$}
\author{Cameron Langer,$^{2,3}$}
\author{Jaroslav Trnka,$^{2}$}
\author{Minshan Zheng,$^{2}$}

\affiliation{$^1$ SLAC National Accelerator Laboratory, Stanford University, Stanford, CA 94039, USA}
\affiliation{$^2$ Center for Quantum Mathematics and Physics (QMAP),\\ 
Department of Physics, University of California, Davis, CA 95616, USA}
\affiliation{$^3$ Institute for Gravitation and the Cosmos, Department of Physics,\\Pennsylvania State University, University Park, PA 16892, USA}

\emailAdd{eh10@stanford.edu, 
cklanger@ucdavis.edu, 
trnka@ucdavis.edu,mszheng@ucdavis.edu}

\preprint{}
\abstract{We initiate the systematic study of \emph{local positive spaces} which arise in the context of the Amplituhedron construction for scattering amplitudes in planar maximally supersymmetric Yang-Mills theory. We show that all local positive spaces relevant for one-loop MHV amplitudes are characterized by certain sign-flip conditions and are associated with surprisingly simple logarithmic forms. In the maximal sign-flip case they are finite one-loop octagons. Particular combinations of sign-flip spaces can be glued into new local positive geometries. These correspond to local pentagon integrands that appear in the local expansion of the MHV one-loop amplitude. We show that, geometrically, these pentagons do \emph{not} triangulate the original Amplituhedron space but rather its twin ``Amplituhedron-Prime." This new geometry has the same boundary structure as the Amplituhedron (and therefore the same logarithmic form) but differs in the bulk as a geometric space. On certain two-dimensional boundaries, where the Amplituhedron geometry reduces to a polygon, we check that both spaces map to the same dual polygon. Interestingly, we find that the pentagons internally triangulate that dual space. This gives a direct evidence that the chiral pentagons are natural building blocks for a yet-to-be discovered dual Amplituhedron.}

\begin{document}
\maketitle

\section{Introduction}
\label{sec:introduction}

The Amplituhedron \cite{Arkani-Hamed:2013jha} is a geometric object encapsulating the tree-level amplitudes and all-loop integrands of planar maximally supersymmetric Yang-Mills theory ($\N{=}4$ sYM) \cite{sYM1,sYM2}. It is a particular example of a \emph{positive geometry} \cite{Arkani-Hamed:2017tmz} and is defined as a certain geometric region in the space of positive external data. Another example of a positive geometry is the Associahedron \cite{Arkani-Hamed:2017mur}, which is relevant for $\phi^3$ amplitudes and is also connected to cluster polytopes and string amplitudes \cite{Arkani-Hamed:2019plo,Arkani-Hamed:2019mrd}. For a review of these geometric ideas, see \cite{Ferro:2020ygk}, and for attempts to include other interactions and matter, c.f.~\cite{Banerjee:2018tun,Herderschee:2019wtl,Herderschee:2020lgb,Jagadale:2020qfa,Aneesh:2019cvt}. Recent advances have also uncovered positive geometries in conformal field theory correlation functions \cite{Arkani-Hamed:2018ign} and in cosmology \cite{Arkani-Hamed:2017fdk,Arkani-Hamed:2018bjr,Benincasa:2020uph}. 

The original definition of the Amplituhedron \cite{Arkani-Hamed:2013jha} involved an auxiliary kinematic space. In contrast, the more recent reformulation of the Amplituhedron defines the geometry directly in momentum-twistor space \cite{Hodges:2009hk} using certain sign-flip conditions \cite{Arkani-Hamed:2017vfh} on sequences of twistor invariants. In this geometric setup, amplitudes correspond to differential forms defined by the property of having logarithmic singularities on all boundaries of the Amplituhedron. All physical properties of scattering amplitudes, such as unitarity and locality, are consequences of the Amplituhedron geometry \cite{Arkani-Hamed:2013kca,YelleshpurSrikant:2019meu}. 

While there is no direct derivation of this geometric construction from first principles of quantum field theory, by now we have ample evidence for the relation between the Amplituhedron geometry and scattering amplitudes. In particular, one can explicitly compute the differential forms by triangulating the Amplituhedron and comparing them to existing tree-level amplitudes and loop integrands available in the literature \cite{Arkani-Hamed:2017vfh,Kojima:2018qzz,Kojima:2020tjf,Kojima:2020gxs}. The geometric picture has also been used to obtain a large amount of all-loop data \cite{Arkani-Hamed:2013kca,Arkani-Hamed:2018rsk,Langer:2019iuo} not currently accessible by any conventional amplitude methods. In parallel, there has also been significant progress trying to understand more formal aspects of the Amplituhedron, including its boundary and combinatorial structure \cite{Franco:2014csa,Karp:2016uax,Karp:2017ouj,Lukowski:2019kqi,Lukowski:2020bya}, the connection to symbol alphabets, properties of final amplitudes (rather than integrands) \cite{Prlina:2017tvx,Prlina:2017azl,Dennen:2016mdk,Dixon:2016apl}, and many others (see e.g., \cite{Salvatori:2018fjp,Bai:2015qoa,Bai:2014cna,Lam:2014jda,Eden:2017fow}). Despite this progress, there are many open questions, including the determination of the all-loop four-point integrand (see recent progress \cite{Arkani-Hamed:2013kca,An:2017tbf,Rao:2018uta}). 

The Amplituhedron arises as a generalization of the positive Grassmannian \cite{ArkaniHamed:2009dn,Mason:2009qx,ArkaniHamed:2009vw,ArkaniHamed:2009dg,ArkaniHamed:2009sx,Arkani-Hamed:2016byb} which appears in the context of on-shell diagrams and all-loop recursion relations \cite{ArkaniHamed:2010kv} in planar $\N=4$ sYM. Despite their origin in maximally supersymmetric Yang-Mills theory, on-shell diagrams and the associated construction in the Grassmannian can be defined in various other theories \cite{Arkani-Hamed:2016byb,Benincasa:2015zna,Benincasa:2016awv,Herrmann:2016qea,Heslop:2016plj,Frassek:2015rka,Kim:2014hva,Huang:2014xza}. This naturally leads to the question whether or not Amplituhedron-like structures also exist more generally. Even within $\N=4$ sYM theory, the striking similarity of the analytic structure of planar and non-planar integrands \cite{Arkani-Hamed:2014via,Bern:2014kca,Bern:2015ple}, and the natural non-planar extension of on-shell diagrams \cite{Arkani-Hamed:2014bca,Franco:2015rma,Bourjaily:2016mnp} suggests that an Amplituhedron-like object should exist for $\N=4$ sYM amplitudes beyond the planar limit. A crucial first step in this direction has been achieved at tree-level by formulating the geometry of the Amplituhedron directly in momentum space \cite{Damgaard:2019ztj,Ferro:2020lgp}. The momentum space formulation seems crucial beyond the planar limit, since we no longer have access to momentum-twistor variables \cite{Hodges:2009hk} which are at the heart of the kinematic space underlying the original Amplituhedron. At loop-level, the main obstruction is the non-uniqueness of the non-planar loop momentum, although recent works \cite{Tourkine:2019ukp,Ben-Israel:2018ckc} suggest possible paths to remedy this situation.

In this paper, we revisit one major open question central to Amplituhedron research. In particular, we investigate the r\^{o}le of ``local triangulations" of the Amplituhedron and their relation to a putative dual Amplituhedron space. The original idea of interpreting scattering amplitudes as volumes appeared in the work of Andrew Hodges \cite{Hodges:2009hk} where the original momentum twistor space formulation for planar $\N=4$ sYM amplitudes was defined. In \cite{Hodges:2009hk}, the six-point NMHV tree-level amplitude is given by the volume of a certain polyhedron in \emph{dual} momentum-twistor space and the BCFW expansion of the amplitude \cite{Britto:2004ap,Britto:2005fq} was interpreted as a particular triangulation of this volume. This picture was later extended to all NMHV tree-level amplitudes \cite{ArkaniHamed:2010gg}. In contrast, higher $k$ tree-level amplitudes and loop integrands have \emph{not} as of yet been directly identified as volumes. Instead, the Amplituhedron is defined in momentum twistor space, where tree-level amplitudes and all-loop integrands are differential forms rather than volumes. The existence of a \emph{dual Amplituhedron} where amplitudes are interpreted as literal volumes is a very natural and important open problem. Nontrivial evidence for the existence of such a dual picture was given in \cite{Arkani-Hamed:2014dca,Ferro:2015grk,Arkani-Hamed:2017tmz}. In particular, it was shown for many examples that both tree-level amplitudes and loop integrands are \emph{positive} when evaluated inside the Amplituhedron, which is reminiscent of the volume positivity. Interestingly, this property also seems to hold for some IR safe quantities post integration \cite{Dixon:2016apl}. Here, we make a further major step in this direction. In particular, we show that the local expansion of the one-loop MHV amplitudes in terms of chiral pentagon integrals \cite{ArkaniHamed:2010gh} can be naturally interpreted as the internal triangulation of the putative dual Amplituhedron. We make this statement precise on two-dimensional boundaries of the full geometry, where the space reduces to polygons and the dualization procedure is well-defined. Our main tool is the reformulation of the Amplituhedron using sign flips \cite{Arkani-Hamed:2017vfh}. In the process of interpreting the chiral pentagons geometrically, we introduce more general positive spaces defined by sign-flip conditions. Surprisingly, the logarithmic forms of the maximal sign-flip spaces are chiral octagons, originally introduced in \cite{ArkaniHamed:2010gh} as local building blocks for one-loop integrands.  \\

Our paper is organized as follows: In section \ref{sec:Amplituhedron_geometry}, we introduce the Amplituhedron geometry and then focus on the MHV one-loop case, where we discuss the projective geometry of lines in $\mathbb{P}^3$ and the positions of physical and unphysical singularities. We also pose the question of the r\^{o}le of the chiral pentagon expansion in the context of the dual Amplituhedron. In sections \ref{sec:geometry_dlog_forms} and \ref{sec:sign_flip_regions}, we discuss how to associate a geometric space to a $d\log$ form and classify all such one-loop geometries. We show that the set of these one-loop spaces is bounded by the number of sign flips in the propagator space, and we determine the logarithmic forms for all of these spaces. In section \ref{sec:geometry_chiral_pentagons}, we discuss the geometry of chiral pentagons in detail and show that they naturally triangulate a different positive space which we call the Amplituhedron-Prime. This space has an identical boundary structure as the original Amplituhedron and therefore corresponds to the same logarithmic form. In section \ref{sec:triangulation_dual_Amplituhedron}, we show that on all two-dimensional boundaries both spaces map under dualization to the same geometry, and the chiral pentagons internally triangulate this dual Amplituhedron space. We finally conclude with some future directions in section \ref{sec:conclusion}. Various technical details are covered in four appendices.

\section{Amplituhedron geometry}
\label{sec:Amplituhedron_geometry}

\subsection*{Momentum twistors and the Amplituhedron's kinematic space}

Scattering amplitudes in planar $\NeqFour$ are naturally described in momentum twistor space \cite{Hodges:2009hk}. Since the same kinematic space also plays a major r\^{o}le in the definition of the Amplituhedron geometry \cite{Arkani-Hamed:2013jha}, we briefly review some of the salient features of projective geometry and momentum twistor space that are relevant to our discussions. These concepts are of course well known (see e.g.~\cite{Hodges:2009hk,ArkaniHamed:2010gh,Arkani-Hamed:2013kca}), but for convenience we recall them telegraphically. 

A key feature of the twistor correspondence \cite{Penrose:1967wn} is the relation between points $x$ in Minkowski spacetime and lines $\mathcal{L}$ in twistor space. For planar scattering amplitudes, the relevant ``spacetime" is dual momentum space where region momenta $x_a$ are related to cyclically ordered external momenta $p_a$ via $p_a = x_a - x_{a-1}$. Dual momentum space trivializes momentum conservation by the identification $x_{n+1} = x_1$.

The kinematic data for $n$-point massless planar scattering amplitudes can be efficiently encoded in $n$ momentum twistors \cite{Hodges:2009hk}, denoted $Z_a^I$, where $a=1,\ldots,n$ labels the particles and $I=1,\ldots,4$ is an $SL(4)$ index on which dual conformal symmetry acts linearly. The natural $SL(4)$ invariant is the antisymmetric contraction of four momentum twistors with the Levi-Civita tensor, which we denote as
\begin{equation}
\ab{abcd}:=\epsilon_{IJKL}Z_a^IZ_b^JZ_c^KZ_d^L \,.
\end{equation}
Scattering amplitudes in planar $\mathcal{N}=4$ sYM are invariant under both $SL(4)$ transformations and the action of the little group i.e., the rescaling $Z_a\mapsto t_aZ_a$. This rescaling invariance implies that the external data lives in the projective space $\mathbb{P}^3$. Moreover, in terms of the $Z_a$ the amplitudes are invariant under a (twisted) cyclic symmetry: for the $\text{N}^k$MHV helicity sector the amplitude is invariant under the transformation $Z_1\mapsto Z_2,Z_2\mapsto Z_3,\ldots,Z_n\mapsto \hat{Z}_1:=(-1)^{k-1}Z_1$ of the external data. 

Incorporating (dual) loop momenta $y$ into the picture is achieved by associating a line $\mathcal{L}$ to $y$. We will often make use of the following notation: since lines in twistor space are characterized by the linear span of two representative points, we denote the line associated to $x_a \leftrightarrow \mathcal{L}_a := (Z_a,Z_{a+1}):=(a\,a{+}1)$ and $y \leftrightarrow \mathcal{L}_y := (AB)$. For each loop, we can represent the corresponding line by two arbitrary momentum twistors $Z_A$ and $Z_B$, defined modulo $GL(2)$ transformations that leave the line invariant. We can furthermore expand $Z_A$ and $Z_B$ in an arbitrary basis of four linearly independent twistors $Z_i,\, Z_j,\, Z_k$, and $Z_{\ell}$ and fix the $GL(2)$ redundancy,\footnote{Note that this choice of coordinates fixes the bracket $\ab{ABk\ell}=\ab{ijk\ell}$ to be a loop-momentum independent function of the external kinematics.}
\begin{align}
\label{eq:loop_line_par_gen}
   Z_A = Z_i + \alpha_1 Z_k + \alpha_2 Z_{\ell}\,, 
   \qquad
   Z_B = Z_j + \alpha_3 Z_k + \alpha_4 Z_{\ell}\,.
\end{align}
In this work, we make regular use of such expansions, and depending on the purpose, we chose convenient expansion twistors. Quite nicely, the four unconstrained parameters $\alpha_i$ in the twistor expansion of eq.~(\ref{eq:loop_line_par_gen}) match the four degrees of freedom of an off-shell Feynman loop momentum. 

If we consider the analog of a generic Lorentz-invariant in dual momentum space $x^2_{ab} = (p_a{+}p_{a+1} + \cdots +p_{b-1})^2$ in momentum twistor space, we realize that this quantity breaks conformal invariance,
\begin{align}
 x^2_{ab} = \frac{\ab{a a{+}1 b{+}1}}{\ab{a a+1}\ab{b b+1}},
\end{align}
as signaled by the appearance of the two-brackets in the denominator. In dual-conformally-invariant quantities, all two-brackets drop out and we therefore do not discuss them any further. Crucially, such non-light-like Lorentz invariants are important to characterize loop-propagators, where any local pole of the integrand is given by
\begin{align}
   \text{local pole} \leftrightarrow \ab{AB ii{+}1}\,.
\end{align}
In the following it will be important that any other four-bracket involving the loop-line $(AB)$ together with an arbitrary ``external" line $X$ constitutes a \emph{spurious pole} if it appears in the denominator of some integrand
\begin{align}
   \text{spurious pole} \leftrightarrow \ab{ABX}, \quad \text{if}\quad X\neq(ii{+}1)\,.
\end{align}
Below, we will see many examples of spurious poles that occur in various expansions of a scattering amplitude. 

One additional feature of the twistor correspondence is the following fact: whenever two spacetime points $x_a$ and $x_b$ are null separated, the associated lines in twistor space $\mathcal{L}_a$ and $\mathcal{L}_b$ intersect. This leads to special configurations of lines which can be concisely summarized in the following image \cite{ArkaniHamed:2010gh}
\begin{equation}
\includegraphics[scale=.3]{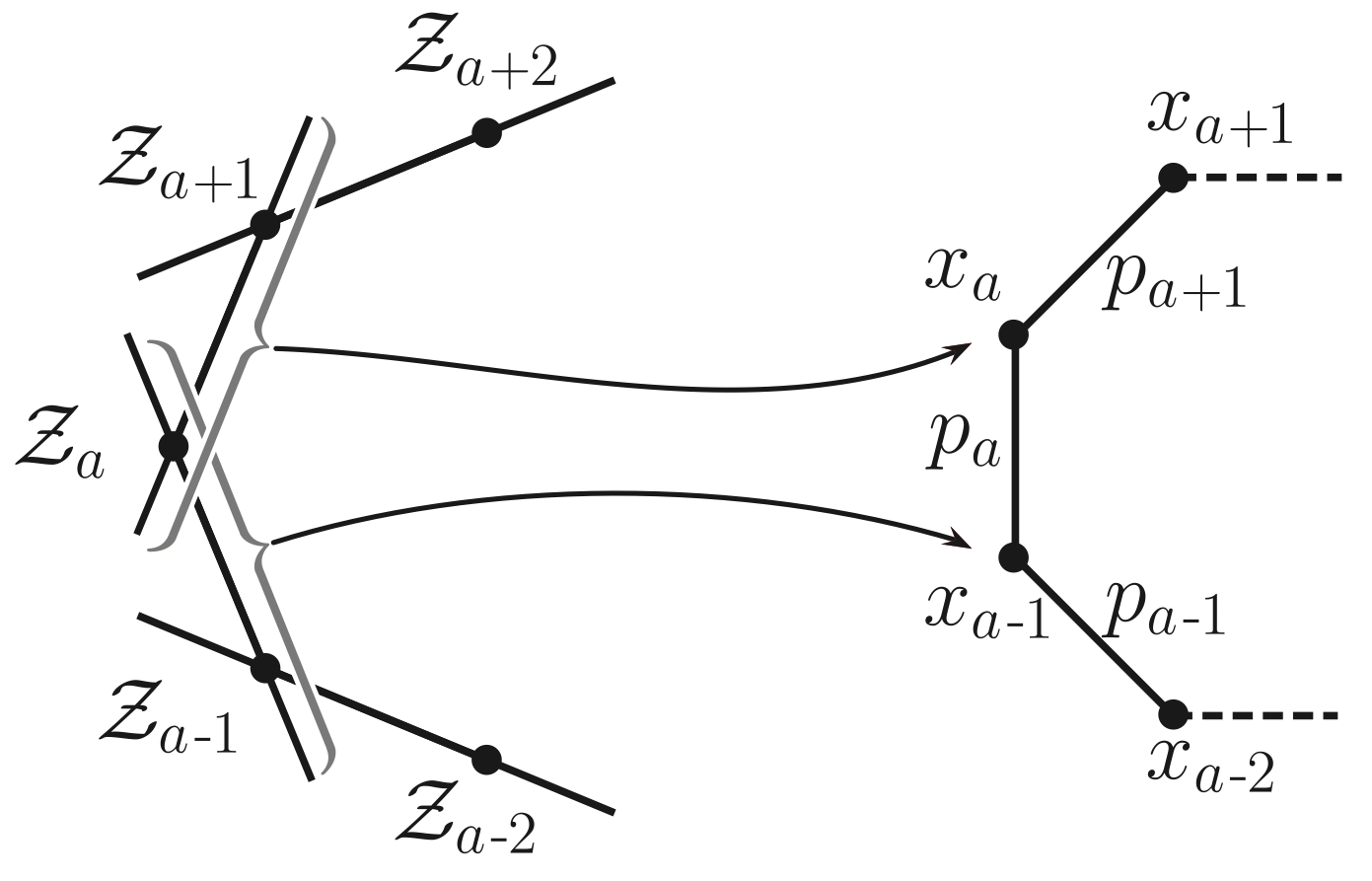}
\end{equation}
For a more detailed introduction to the twistor correspondence, we refer directly to the above references \cite{Hodges:2009hk,ArkaniHamed:2010gh,Arkani-Hamed:2013kca}. 

Finally, let us elaborate on the connection between twistor geometry and the analytic structure of scattering amplitudes that will play an important r\^{o}le in this work. There is an intimate relation between configurations of (loop) lines in momentum twistor space and certain restricted kinematic configurations of loop momenta on unitarity cuts of loop integrands or local integrals. At one loop, we can depict the off-shell configuration of lines in twistor space corresponding to a generic loop integrand (either of the amplitude or of an integral) by a set of lines corresponding to external dual momenta, together with a line $(AB)$ in a generic configuration (parameterized via eq.~(\ref{eq:loop_line_par_gen})),
\begin{align}
    \raisebox{-52pt}{\includegraphics[scale=.5]{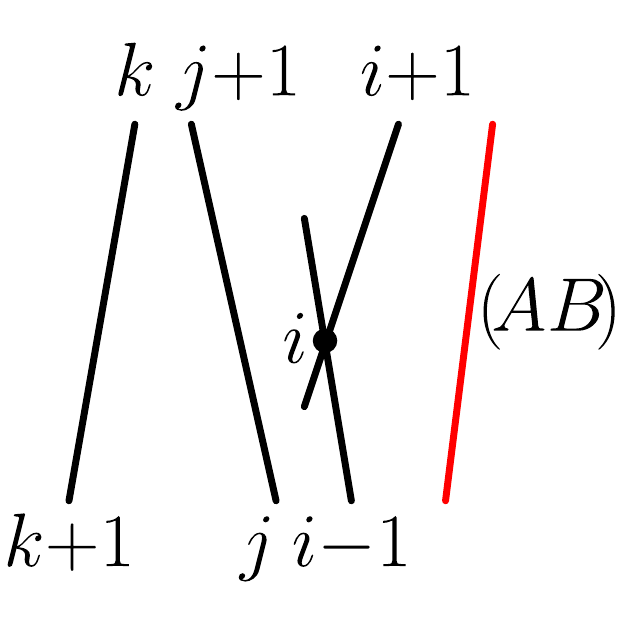}}
    \qquad
    \leftrightarrow
    \qquad
    \raisebox{-50pt}{\includegraphics[scale=.5]{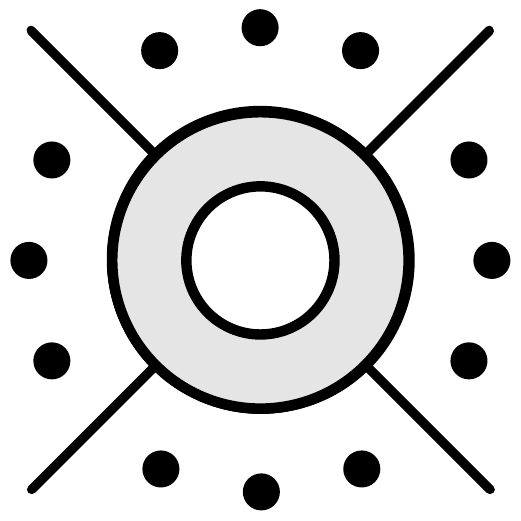}}
\end{align}
In this setup, the loop-line $(AB)$ does not intersect any of the lines associated to external kinematic points. In the next step, one could go to codimension one configurations by imposing one condition, e.g.~$\ab{ABii{+}1} = 0$, which geometrically means that the lines $(AB)$ and $(ii{+}1)$ intersect. 
\begin{align}
\label{fig:codim_1_cut_and_line_config}
    \raisebox{-52pt}{\includegraphics[scale=.5]{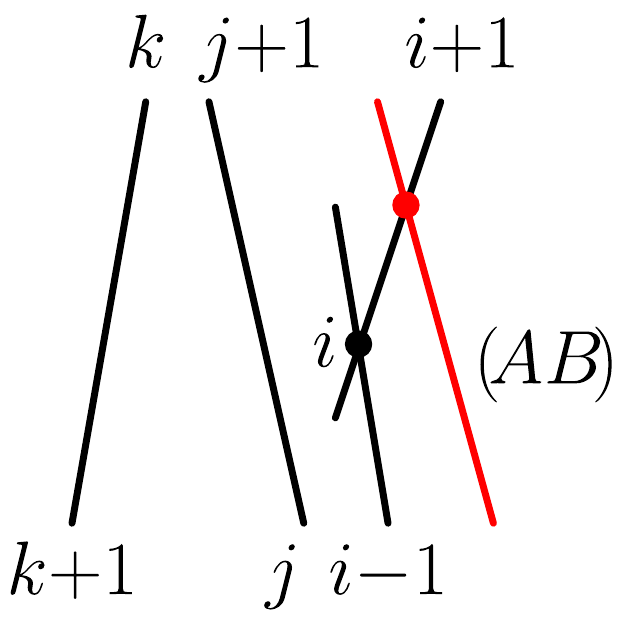}}
    \qquad
    \leftrightarrow
    \qquad
    \raisebox{-50pt}{\includegraphics[scale=.5]{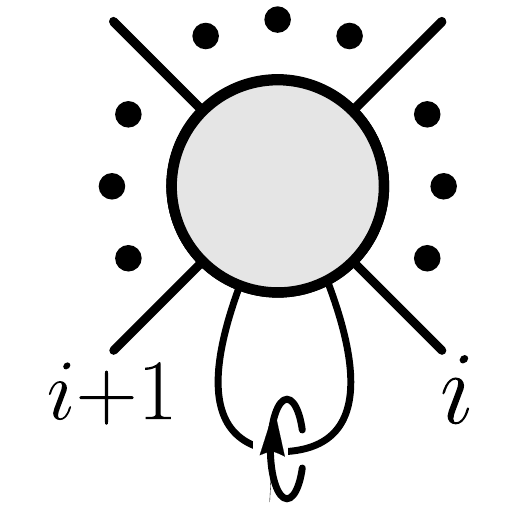}}
\end{align}
At the level of cuts, this corresponds to setting a single propagator $\ab{ABii{+}1} = 0$ to zero. This codimension-one configuration for the line $(AB)$ can be parameterized by three degrees of freedom. The intersection implies that one of the defining points of the $(AB)$-loop lies on the line $(i i{+1})$. Taking into account the projectivity of the external data, one possible parametrization is 
\begin{align}
    Z_A = Z_i + \alpha_1 Z_{i+1}\,, \quad
    Z_B = Z_j + \alpha_2 Z_{k} + \alpha_3 Z_{\ell}\,.
\end{align}
In subsequent steps, one can impose additional constraints to end up on codimension-two, three, or four configurations of the line $(AB)$. For completeness, we have included a list of higher codimension boundaries and their associated cut configurations in appendix \ref{app:line_configs}.

\subsection{Topological sign-flip definition of the Amplituhedron}
\label{subsec:Amplituhedron_def}
%
The Amplituhedron is defined by three types of positivity conditions \cite{Arkani-Hamed:2017vfh}. First, we have inequalities which carve out the positive part of the external kinematic space; these conditions depend only on the helicity configuration of interest and are loop-momentum independent. For $\text{N}^k$MHV amplitudes the conditions read 
\begin{equation}
\begin{split}
\label{eq:zIneq}
\text{N$^k$MHV ext. positivity:} \quad
 &  \langle ii{+}1jj{+}1\rangle>0\,, \\
 &  \text{sequence } \{\langle abb{+}1i\rangle\}_{i\neq a,b,b{+}1} \text{ has $k$ sign flips},
\end{split}
\end{equation}
where the twisted cyclic symmetry implies that $(-1)^{k-1}\ab{n1ii{+}1}>0$. For the simplest case of MHV ($k=0$) amplitudes the positivity of the external data as defined in eq.~(\ref{eq:zIneq}) is equivalent to all ordered brackets being positive, 
\begin{align}
\label{eq:mhv_ext_positivity}
 \text{MHV ext. positivity:}\quad   \ab{ijkl}>0\,, \quad \text{for }  i<j<k<l\,. 
\end{align}
Second, there are inequalities between the loop variables $(AB)$ and the external data, again naturally expressed in terms of sign flips,\footnote{As we will see in subsection \ref{subsec:1_loop_MHV_space}, for MHV and $\MHVbar$ amplitudes, there is an equivalent formulation of the loop-positivity given in eqs.~(\ref{eq:mhv_alt}) and (\ref{eq:mhvBar_alt}), respectively.}
\begin{equation}
\begin{split}
\label{eq:abIneq}
\text{N$^k$MHV loop positivity:}\quad 
    & \langle ABii{+}1\rangle>0\,, \\
    & \text{sequence } \{\ab{ABij}\}_{j\neq i} \text{ has $k{+}2$ sign flips}.
\end{split}
\end{equation}
Finally, for multi-loop calculations, the third type of positivity conditions demand that all loops are mutually positive, 
\begin{equation}
\label{eq:mutualIneq}
\text{multi-loop positivity: }\quad 
 \ab{(AB)_i(AB)_j}>0,\text{ for $i\neq j=1,\ldots,L$}. 
\end{equation}
The set of conditions in eqs.~(\ref{eq:zIneq}), (\ref{eq:abIneq}), and (\ref{eq:mutualIneq}) define the general loop-level Amplituhedron $\mathcal{A}^{(n,k,L)}$ relevant for $n$-particle N$^k$MHV $L$-loop integrands in the space of $L$ lines $(AB)_1,\ldots,(AB)_L$ and $n$ momentum twistors $Z_1,\ldots,Z_n$. As such, the space defined by eqs.~(\ref{eq:zIneq}), (\ref{eq:abIneq}), and (\ref{eq:mutualIneq}) constitutes a highly nontrivial example of a \emph{positive geometry} \cite{Arkani-Hamed:2017tmz}. In this setup, the $n$-point $\text{N}^k$MHV $L$-loop integrand is conjectured to be the unique degree $4(k{+}L)$ differential form on this space defined by having logarithmic singularities on all its boundaries. We denote this form (that is, the \emph{integrand}) by $\Omega^{(n,k,L)}$,\footnote{We commonly use the notation that all capital forms $\Omega$ include the loop measure, whereas for lower-case forms $\omega$ the measure has been stripped off.} 
\begin{equation} 
\label{eq:integrand_form_measure_stripping_def}
\Omega^{(n,k,L)}=\prod_{i=1}^L \langle (AB)_id^2A_i\rangle\langle (AB)_id^2 B_i\rangle\times \omega^{(n,k,L)},
\end{equation}
where $\omega^{(n,k,L)}$ is a $4k$-form in external momentum twistors $Z_a$, but is simply a rational function in the loop variables $(AB)_1,\ldots,(AB)_L$.  The loop integrand can be obtained by replacing the differentials $dZ_a$ with the fermionic Grassmann variables $\eta_a$ to work directly in on-shell superspace. For MHV amplitudes, $\omega^{(n,0,L)}$ is just a rational function of $Z_a$ and $(AB)_i$ and directly constitutes the loop integrand. 

\subsection*{Rudiments of positive geometry}
\label{subsec:pos_geometry}
In the course of exploring the geometric properties of the local representation of loop integrands, we shall naturally encounter a variety of ``spaces.'' Loosely speaking, we define a \emph{geometric space} in this work to be a collection of inequalities imposed on four-brackets involving the loop line $(AB)$. By parametrizing $(AB)$ as above in terms of four (real) degrees of freedom $x_1,\ldots,x_4$, any four-bracket between $(AB)$ and a bitwistor $X_i$ becomes a polynomial $\mathfrak{p}_i(x_1,\ldots,x_4):=\ab{ABX_i}$. A more precise definition of a geometric space $S$ is then a semialgebraic set defined by some number say, $d$, of inequalities
\begin{equation}
\label{geometric_space}
S=\{(x_1,\ldots,x_4)\vert \mathfrak{p}_1>0,\ldots,\mathfrak{p}_d>0\}.
\end{equation}
Note that this definition does not require the existence of a canonical form, and thus may or may not be a positive geometry satisfying the recursive definition of \cite{Arkani-Hamed:2017tmz}. It may happen that no line in $\mathbb{P}^3$ satisfies the conditions we impose, which means eq.~(\ref{geometric_space}) would be equivalent to the empty set; we shall refer to such spaces as \emph{empty}.

The existence of a canonical form with logarithmic singularities on all boundaries puts further constraints on the general geometric space $S$. In particular, since any logarithmic form in the one-loop space is of degree four and can be written as the $d\log$ of four (projectively invariant) ratios,\footnote{Since we can always rescale the entry of any $d\log$ form by a four-bracket involving only the external data, it is always trivial to make any entry projective in all $Z_i$, without modifying the canonical form (or, in the case of MHV data, affecting the overall sign of the entry). In this work we shall often neglect to write such factors, meaning our $d\log$ entries will be manifestly projective only in $(AB)$.} any one-loop \emph{positive geometry} must be defined by at least five inequalities. Any space defined by four or fewer inequalities must have only the trivial form $\Omega=0$; throughout this work we refer to such geometries as \emph{zero form spaces}.

\subsection{One-loop MHV and $\MHVbar$ spaces}
\label{subsec:1_loop_MHV_space}

Following the general Amplituhedron definitions above, let us now specialize to the geometry relevant to MHV one-loop amplitudes. As stated previously, for MHV integrands (i.e.,~$k=0$) the positivity conditions on the external data eq.~(\ref{eq:zIneq}) simplify significantly to eq.~(\ref{eq:mhv_ext_positivity}). Besides the MHV integrand, there is another one-loop integrand which naturally lives in the same kinematic space: namely, its parity conjugate. In twistor space, spacetime parity is implemented by the duality under the exchange of points and planes, $Z_a \leftrightarrow W_a:=(a{-}1aa{+}1)$ (often, we will use the alternative notation $\overline{a}:=(a{-}1aa{+}1)$). Thus, we will refer to this ``dual" space, as well as its corresponding canonical form, with the moniker ``$\MHVbar$,'' despite the fact that the actual $\MHVbar$ Amplituhedron corresponds to setting $k=n{-}2$, which yields a slightly different set of positivity constraints on the external data (see eq.~(\ref{eq:zIneq})). Despite the fact that these two spaces are not equivalent (because the definition of each involves different positivity conditions on the external data!), their canonical forms are trivially related by stripping the overall tree-level amplitude prefactor. 

Both the MHV and $\MHVbar$ integrands have only \emph{local} poles; equivalently, the only codimension-one boundaries of both geometries are the loci where $\ab{ABii{+}1}=0$. We can think of the MHV and $\MHVbar$ spaces as being cut out from a ``larger'' space defined only by $\ab{ijkl}>0$, for $i<j<k<l$, and $\ab{ABii{+}1}>0$, \emph{without any additional constraints}. For reasons which will become more clear below, we call this space an ``achiral'' one-loop space. By construction, its codimension-one boundaries are the same as those of the MHV and $\MHVbar$ subspaces. From this perspective, it is quite nontrivial that there even exist two distinct ways of slicing the achiral space without introducing additional codimension-one (spurious) boundaries. Since the notion of chiralization plays a pivotal r\^{o}le in our analysis of local triangulations throughout this work, in the following subsection we review some salient features of the MHV and $\MHVbar$ one-loop geometries.

\subsection{Boundaries of MHV amplitudes}
\label{subsec:MHV_boundaries_kermit}

In this subsection, we briefly discuss the geometric boundaries of the ``Kermit'' expansion of one-loop MHV amplitudes \cite{ArkaniHamed:2010kv}. In particular, we will show how to detect whether or not certain boundaries are present in the geometry and if the corresponding poles appear in the logarithmic form. This discussion allows us to introduce the necessary notation and concepts used in our later analysis of chiral pentagons.

The inequalities defining the one-loop MHV Amplituhedron are a special case of eq.~(\ref{eq:abIneq}), and involve the usual positivity of adjacent brackets as well as a sign-flip condition:
\begin{align}
\begin{split}
\text{MHV loop positivity:} \quad
                &  \ab{ABii{+}1}>0 \,, \\
                &\{\ab{AB12},\ldots,\ab{AB1n}\} 
                      \text{ has two sign flips},
\label{mhv_loop_amphedron_flip_cond}
\end{split}
\end{align}
where we have chosen the sequence $\{\ab{AB1i}\}_{i=2,\ldots,n}$ only for specificity and to easily handle the twisted cyclic symmetry in the last bracket of the sequence. There are a number of sign choices for the brackets appearing in eq.~(\ref{mhv_loop_amphedron_flip_cond}) consistent with two sign flips, which can be labelled by two indices $i,j$ indicating the brackets $\ab{AB1i}$ and $\ab{AB1j}$ where the sign flip occurs. The geometry of these individual sign-flip patterns is completely well understood: the pattern where the sign flips occur at indices $i$ and $j$ is directly associated to a ``Kermit'' which corresponds to a particular cell in the positive Grassmannian. In the following, we make repeated use of these sign-flip patterns and often denote brackets by their respective sign, e.g., for the $(i,j)$ flip term, 
\begin{align}
\label{kermit_ij_example}
\begin{split}
    \{\tblue{\ab{AB12},\ldots,\ab{AB1i}}&,
      \tred{\ab{AB1i{+}1},\ldots \ab{AB1j}},
      \tblue{\ab{AB1j{+}1},\ldots,\ab{AB1n}}\}  
      \\ 
     & \hspace{-1.5cm} \mapsto \{\tblue{+,\ldots,+}, \tred{-,\ldots-},\tblue{+,\ldots,+}\} .
\end{split}
\end{align}
For the $(i,j)$ sign-flip region, the line $(AB)$ can be conveniently parametrized as
\begin{equation}
\label{kermit_param}
A=Z_1+\alpha_i Z_i+\alpha_{i+1}Z_{i+1},\quad B=-Z_1+\alpha_j Z_j+\alpha_{j+1}Z_{j+1}\,.
\end{equation}
In this coordinate chart, the inequalities in eq.~(\ref{kermit_ij_example}) are equivalent to $\alpha_i,\alpha_{i{+}1},\alpha_j,\alpha_{j{+}1}{>}0$ and therefore the canonical form is simply the $d\log$ form in all variables.\footnote{For detailed discussions and numerous examples of going from inequalities to forms, see e.g.~\cite{Arkani-Hamed:2017vfh,Kojima:2020tjf}.}
Written projectively, this yields the expression for the canonical form (stripping off the measure $\ab{ABd^2A}\ab{ABd^2B}$)
\begin{equation}
\label{eq:Kermit}
\hspace{-.5cm}
\omega^{(n,0,1)}\!=\!
\sum_{i<j} \omega^{(i,j)}_n \!\equiv\!
\sum_{i<j}\!\frac{\langle AB(1ii{+}1){\cap}(1jj{+}1)\rangle^2}{\langle AB1i\rangle\langle AB1i{+}1\rangle\langle ABii{+}1\rangle\langle AB1j\rangle\langle AB1j{+}1\rangle\langle ABjj{+}1\rangle},
\hspace{-.3cm}
\end{equation}
where $(1ii{+}1){\cap}(1jj{+}1)$ is the line in which the planes $(1ii{+}1)$ and $(1jj{+}1)$ intersect.\footnote{A review of the projective geometry relevant to this work is given in section~\ref{sec:Amplituhedron_geometry} and appendix~\ref{app:line_configs}.} By construction, each term in this expansion is associated with a non-overlapping piece of the amplitude; in other words, this collection of cells provides an honest triangulation by introducing term-wise spurious poles of the form $\ab{AB1i}$ in the denominator. The fact that these individual sign-flip spaces are non-overlapping follows directly from the observation that different Kermit cells differ by the signs of at least one of the inequalities defining their regions.

Let us provide a simple concrete five-point example where there are three different sign patterns with two sign-flips. Since the $\ab{AB12}$ and $\ab{AB15}$ brackets are forced to be positive by the loop conditions in eq.~(\ref{mhv_loop_amphedron_flip_cond}), the three sign patterns are labelled by the signs of $\ab{AB13}$ and $\ab{AB14}$, i.e.,
\begin{equation}
\begin{tabular}{|c|c|c|c|c|}
\hline $(i,j)$&$\ab{AB12}$ & $\ab{AB13}$ & $\ab{AB14}$ & $\ab{AB15}$\\ \hline
(3,4)&+ & + & $-$ & +\\
(2,3)&+ & $-$ & + & +\\
(2,4)&+ & $-$ & $-$ & + \\ \hline
\end{tabular} .\label{sign5}
\end{equation}
In the first column we indicate the locations $(i,j)$ at which the sign flips occur, which directly match the labels in eq.~(\ref{eq:Kermit}). Note that although all spaces are na\"{i}vely defined by the same number of inequalities, in this table, the spaces where $j=i{+}1$ are geometrically simpler than the generic case. To illustrate this point and make our notion of ``geometric boundaries'' clear, we now consider the $(3,4)$ cell in more detail. The analysis of the properties of this space is particularly amenable to the following parametrization:
\begin{equation}
A=Z_2+w Z_1+xZ_5, \quad B=Z_3+y Z_1+zZ_5.
\label{eq:param_kermit_5pt}
\end{equation}
This choice of coordinates gauge-fixes $\ab{AB15}=\ab{1235}>0$. Imposing $\ab{ABii{+}1}>0$, $\ab{AB13}>0$ and $\ab{AB14}<0$ defines the $(3,4)$ Kermit space:
\begin{align}
\label{eq:kermit_example_param}
\begin{split}
&\ab{AB12}=-x\ab{1235}>0,\\
&\ab{AB23}=(wz-xy)\ab{1235}>0,\\
&\ab{AB34}=-y\ab{1234}+(wz-xy)\ab{1345}+z\ab{2345}>0,\\
&\ab{AB45}=-y\ab{1245}+w\ab{1345}+\ab{2345}>0,\\
&\ab{AB15}=\ab{1235}>0,\\
&\ab{AB13}=-z\ab{1235}>0,\\
&\ab{AB14}=\ab{1234}-z\ab{1245}+x\ab{1345}<0.
\end{split}
\end{align}
The codimension-one boundaries of this space correspond to localizing one (and only one) of the variables $w,x,y,z$ by setting one of the brackets in eq.~(\ref{eq:kermit_example_param}) to zero. However, not all of the inequalities in eq.~(\ref{eq:kermit_example_param}) are boundaries. In particular, we will now demonstrate that setting either $\ab{AB12}\rightarrow0$, $\ab{AB23}\rightarrow0$, or $\ab{AB14}\rightarrow0$ leads to an inconsistent set of inequalities in the three remaining variables, thus proving that none of these brackets are geometric boundaries of the $(3,4)$ Kermit. First, consider $\ab{AB12}\rightarrow0$ which, in our parametrization, corresponds to sending $x\rightarrow0$. From eq.~(\ref{eq:kermit_example_param}), we have
\begin{equation}
\ab{AB14}\underset{x\rightarrow0}{\longrightarrow}\ab{1234}-z\ab{1245}<0,
\end{equation}
which implies that $z>\ab{1234}/\ab{1245}>0$. This is inconsistent with $\ab{AB13}=-z\ab{1235}>0$, which demands that $z<0$. Thus, $\ab{AB12}\rightarrow0$ is not a codimension-one boundary of the space. Next, consider $\ab{AB23}\rightarrow0$, which can be parametrized as $w\rightarrow xy/z$. In this case, the space defined by eq.~(\ref{eq:kermit_example_param}) is equivalent to $x,z<0$ and, after some algebraic manipulation,
\begin{align}
\label{eq:kermit_example_param2}
\begin{split}
 \ab{AB34} \underset{w\rightarrow xy/z}{\longrightarrow} 
                    &-y\ab{1234}+z\ab{2345}>0,\\
  \ab{AB45} \underset{w\rightarrow xy/z}{\longrightarrow}                       &y(x\ab{1345}-z\ab{1245})+z\ab{2345}<0,\\
 \ab{AB14} \underset{w\rightarrow xy/z}{\longrightarrow}   
   &\ab{1234}-z\ab{1245}+x\ab{1345}<0.
\end{split}
\end{align}
The first inequality in eq.~(\ref{eq:kermit_example_param2}) together with $z<0$ implies that $y<0$. Multiplying the third inequality in eq.~(\ref{eq:kermit_example_param2}) by $y$ and adding $z\ab{2345}$ to both sides, we find
\begin{equation}
y(x\ab{1345}-z\ab{1245})+z\ab{2345}>-y\ab{1234}+z\ab{2345}>0,
\end{equation}
which is incompatible with the second inequality of eq.~(\ref{eq:kermit_example_param2}). Finally, consider setting $\ab{AB14}\rightarrow0$, which corresponds to $x\rightarrow(z\ab{1245}-\ab{1234})/\ab{1345}$. In this case, eq.~(\ref{eq:kermit_example_param}) yields
\begin{align}
\label{eq:kermit_example_param3}
\begin{split}
   &\ab{AB34}\rightarrow z\left(-y\ab{1245}+w\ab{1345}+\ab{2345}\right)>0,\\
    &\ab{AB45}\rightarrow-y\ab{1245}+w\ab{1345}+\ab{2345}>0,
\end{split}
\end{align}
which, together with $z<0$, are incompatible.

In contrast to the three examples just considered, it is straightforward to verify that no such inconsistency is found upon setting either $\ab{AB34}\rightarrow0$, $\ab{AB45}\rightarrow0$ or $\ab{AB13}\rightarrow0$. Thus, these three brackets are geometric boundaries of the space defined in eq.~(\ref{eq:kermit_example_param}). In fact, $\ab{AB15}\rightarrow0$ is also a geometric boundary of the full projective space; while this fact is obscured in our choice of gauge-fixing, it can be proven with a similar argument as above by working in an alternative parametrization (which fixes the sign of some other four-bracket).

Thus, the $(3,4)$ cell has four codimension-one boundaries $\ab{AB34}$, $\ab{AB45}$, $\ab{AB15}$ and $\ab{AB13}$ which also correspond to the allowed poles in the associated canonical form. An analogous argument holds for the $(2,3)$ cell, while the $(2,4)$ cell is a generic Kermit with six accessible codimension-one boundaries. A straightforward computation yields the following expressions for the canonical forms:
\begin{align}
&
\omega_5^{(3,4)} = \frac{\ab{1345}^2}{\ab{AB13}\ab{AB34}\ab{AB45}\ab{AB15}},
\quad 
\omega_5^{(2,3)} = \frac{\ab{1234}^2}{\ab{AB12}\ab{AB23}\ab{AB34}\ab{AB14}},
\nonumber\\
&\hspace{3cm}
\omega_5^{(2,4)} = \frac{\ab{AB(123){\cap}(145)}^2}
{\ab{AB12}\ab{AB23}\ab{AB13}\ab{AB14}\ab{AB15}\ab{AB45}} .
\label{forms5}
\end{align}
The spurious boundary $\ab{AB13}=0$ cancels geometrically between the first and third sign-flip terms in eq.~(\ref{sign5}). This can be seen in the parametrization of eq.~(\ref{eq:param_kermit_5pt}) by sending $z\rightarrow0$ in both spaces, and observing that the contributions completely overlap and therefore cancel. A similar analysis shows that the spurious boundary $\ab{AB14}=0$ cancels between the second and third space in eq.~(\ref{sign5}).  Algebraically, the cancellation of spurious poles can be observed between the corresponding forms in eq.~(\ref{forms5}). We will see later that the algebraic cancellation of a spurious pole in the form is not, in general, sufficient to imply a geometric cancellation of the corresponding spurious boundary.

As indicated earlier, the one-loop MHV Amplituhedron space is a slice of a larger ``achiral'' positive space given only by the $\ab{ABii{+}1}{>}0$ conditions. In the five-point example, the achiral space allows for arbitrary signs of the ``spurious-pole" brackets $\ab{AB13}$, $\ab{AB14}$. This means that in addition to the three sign-flip patterns of eq.~(\ref{sign5}), we have one additional sign-flip-zero space where both $\ab{AB13}$ and $\ab{AB14}$ are positive,
\begin{equation}
\begin{tabular}{|c|c|c|c|}
\hline $\ab{AB12}$ & $\ab{AB13}$ & $\ab{AB14}$ & $\ab{AB15}$\\ \hline
+ & + & + & + \\ \hline
\end{tabular} \label{sign5a}
\end{equation}
Having zero sign flips in the sequence $\{\ab{AB1i}\}$ actually implies that all $\ab{ABij}>0$ are positive for any $i<j$. The corresponding logarithmic form reproduces the five-point $\MHVbar$ amplitude, and the $\{++\}$ sign pattern is an example of the $\MHVbar$ space discussed above and is directly related to the MHV space by parity conjugation. This has a direct generalization to higher points: the MHV space has two sign flips in the $\{\ab{AB1i}\}$ sequence, while the $\MHVbar$ parity conjugate has zero sign flips. At higher points, na\"{i}vely, there are additional subregions of the larger achiral space with more than two sign-flips. For example, at six points we can have 
\begin{equation}
\begin{tabular}{|c|c|c|c|c|}
\hline $\ab{AB12}$ & $\ab{AB13}$ & $\ab{AB14}$ & $\ab{AB15}$& $\la AB16\ra$\\ \hline
+ & $-$ & + & $-$ & +\\ \hline
\end{tabular} \label{sign6}
\end{equation}
However, this combination of inequalities is an example of an \emph{empty space} discussed in subsection \ref{subsec:pos_geometry}. In fact, it is known \cite{Arkani-Hamed:2017vfh} that, for MHV external data, the same is true for all spaces with more than two sign flips! Thus, remarkably the achiral space can be cut into two (and only two) chiral subspaces, MHV and $\MHVbar$, which have only local boundaries of the form $\ab{ABii{+}1}{=}0$. 

Note that the sign-flip-zero ($\MHVbar$) space in the five-point example of eq.~(\ref{sign5a}) was not triangulated further into simpler subspaces (with spurious poles). Therefore, the generalization of this space at $n$-points has $n$ boundaries, while any individual sign-flip-two region only has (up to) six codimension-one boundaries, as can be seen from the Kermit canonical form in eq.~(\ref{eq:Kermit}). It is therefore natural to ask whether or not we can analogously slice the sign-flip-zero region into simpler spaces with fewer boundaries, just as we did with the Kermits for the sign-flip-two MHV pattern. This slicing can be achieved by realizing that both MHV and $\MHVbar$ spaces are related by parity, $Z_a\leftrightarrow W_a$. In particular, the $n$-point $\MHVbar$ space has an equivalent definition as the sign-flip-two space in the $W_a$ coordinates, i.e., two sign flips in the sequence
\begin{align}
\begin{split}
``\MHVbar\text{'' positivity}: \quad
&  \ab{ijkl}>0\,, \quad \text{for }i<j<k<l, \\
&  \ab{AB i i{+}1}>0,                      \\
&  \{\ab{AB\overline{12}},\ab{AB\overline{13}},
            \ldots,\ab{AB\overline{1n}}\} 
\quad 
\text{has two sign flips}, 
\label{conj}
\end{split}
\end{align}
where the $\ab{AB1i}$ term in the MHV sequence eq.~(\ref{mhv_loop_amphedron_flip_cond}) has been replaced by $\ab{AB\overline{1i}}$, and $(\overline{1i})$ denotes the intersection of the planes $\bar{1}=-(n12)$ and $\bar{i}=(i{-}1ii{+}1)$. By expanding the intersections in the first and last brackets in this sequence, we see they are invariant under conjugation (up to a factor that only depends on the external data) and are therefore both positive as a consequence of $\ab{AB12},\ab{AB1n}>0$. In this definition, the $\MHVbar$ space is triangulated by the collection of sign-flip-two regions of eq.~(\ref{conj}), analogous to the Kermit triangulation of the MHV space. Using the sequence of brackets in eq.~(\ref{conj}), we can conversely define the MHV space as the single sign-flip-zero region with all $\ab{AB\overline{ij}}>0$, for $i<j$ (always accounting for the twisted cyclic symmetry when $i=1$ or $j=n$).

While breaking up the MHV and $\MHVbar$ spaces into smaller regions using sign flips is a useful triangulation strategy, we can also characterize them in a uniform way as the sign-flip-zero regions in eq.~(\ref{mhv_loop_amphedron_flip_cond}) and eq.~(\ref{conj}) respectively. In addition to $\ab{ABii{+}1}>0$ and the $k{=}0$ conditions on the external data eq.~(\ref{eq:mhv_ext_positivity}), we get:
\begin{align}
\label{eq:mhv_alt}
\text{alternative MHV}:& 
\quad \ab{AB\overline{ij}}>0,
\quad \text{for all $i<j$},\\
\label{eq:mhvBar_alt}
\text{alternative $\MHVbar$}:&
\quad \ab{ABij}>0,\quad \text{for all $i<j$}.
\end{align}
As we have seen, the Kermit expansion \cite{ArkaniHamed:2010kv} of the MHV one-loop integrand in eq.~(\ref{eq:Kermit}) is in one-to-one correspondence with the sign-flip representation of the Amplituhedron, and it provides a very natural triangulation. We call this triangulation {\it internal}, emphasizing that it cuts the Amplituhedron into smaller pieces by introducing internal (spurious) boundaries. Each geometric space associated to a Kermit lies inside the Amplituhedron, but in addition to \emph{physical} (Amplituhedron) boundaries it also has a number of spurious boundaries of the form $\ab{AB1i}=0$, which cancel geometrically when taking the collection of all Kermits. The analogues of Kermit expansions for higher $k$ and $L$ have been found in \cite{Kojima:2020tjf}, and they involve more complicated spurious poles whose cancellations are nontrivial to demonstrate analytically.
%
\subsubsection*{MHV and $\MHVbar$ geometry and allowed singularity structure}
%
As we have laid out in section \ref{sec:Amplituhedron_geometry} and appendix \ref{app:line_configs}, taking residues of one-loop integrands in loop-momentum space is equivalent to localizing the line $(AB)$ to special configurations with respect to the external $(ii{+}1)$ lines. On the other hand, imposing a set of inequalities $\{\ab{AB X_1}>0,\ldots,\ab{AB X_r}>0\}$, where the $X_i$ are some lines (involving external data) in $\mathbb{P}^3$, effectively slices the full configuration space of $(AB)$ into smaller subspaces. A given subspace will generically contain a rather small subset of the possible special configurations associated to cuts of the integrand. The constraint that a cut-configuration of the loop line $(AB)$ be compatible with the inequalities defining the positive geometry of interest becomes rather severe once we go deep into the cut structure to, say, codimension-four boundaries. Checking the compatibility of certain cut configurations and the geometric inequalities will then tell us which singularities are physical and which ones are spurious.

Let us illustrate this point explicitly in the case of the MHV one-loop geometry. As we discussed in eq.~(\ref{mhv_loop_amphedron_flip_cond}) and eq.~(\ref{eq:mhv_alt}), this space can be characterized by a simple set of inequalities, $\ab{ABii{+}1}>0$ and $\ab{AB\overline{ij}}>0$ (together with the conditions on the external data in eq.~(\ref{eq:zIneq})). It is easy to verify that none of the codimension-one (eq.~(\ref{fig:codim_1_cut_and_line_config})) or two configurations (listed in eq.~(\ref{eq:codim_2_line_configs}) of appendix \ref{app:line_configs}) violate any of these inequalities, so all of these singularities are physical. However, there are two spurious codimension-three configurations which violate the MHV inequalities: the three-mass triple cut where $(AB)$ intersects three non-adjacent lines,
\begin{align}
\label{eq:spurious_three_mass}
   &\text{not allowed in MHV:}\quad
   \raisebox{-42pt}{\includegraphics[scale=.5]{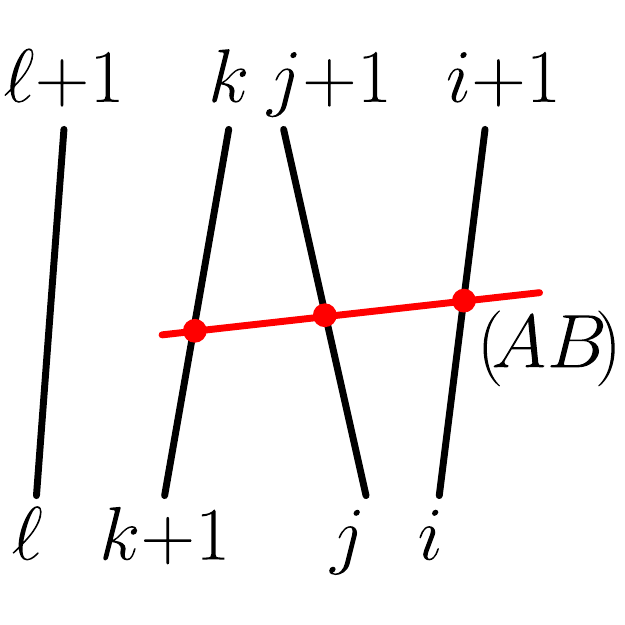}}
   \leftrightarrow
    \raisebox{-42pt}{\includegraphics[scale=.5]{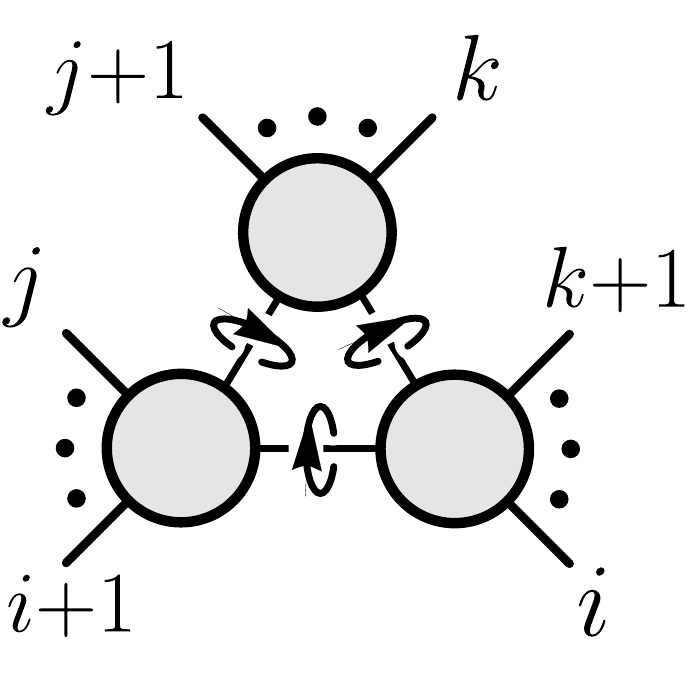}},
\end{align}
as well as the configuration where $(AB)$ is in the plane $(\overline{i})$ and intersects a non-adjacent line $(kk{+}1)$,
\begin{align}
      &\text{not allowed in MHV:}\quad 
      \raisebox{-42pt}{\includegraphics[scale=.5]{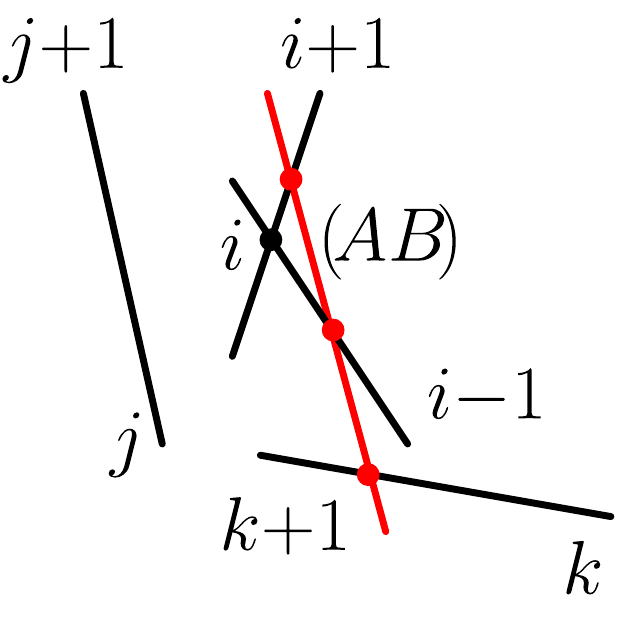}}
   \leftrightarrow
    \raisebox{-42pt}{\includegraphics[scale=.5]{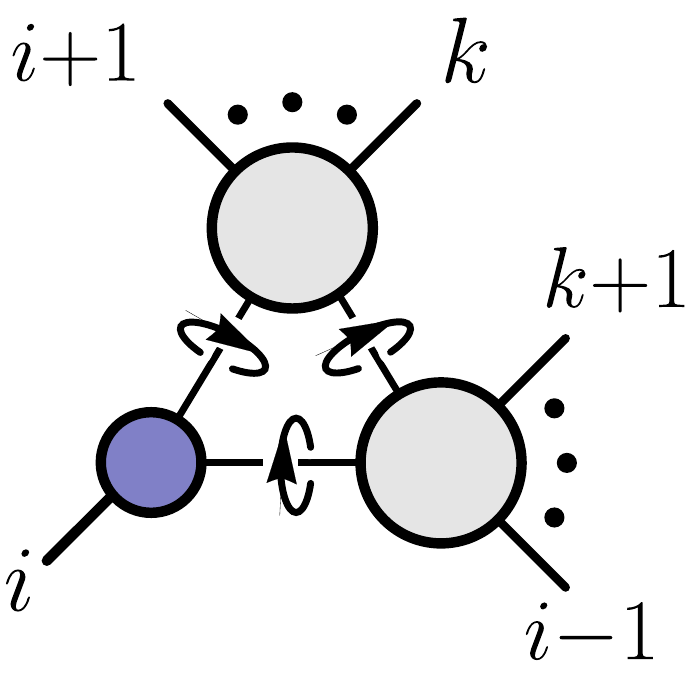}}.
\end{align}
To prove that the three-mass triple cut of eq.~(\ref{eq:spurious_three_mass}) lies outside of the MHV Amplituhedron, we first consider the simplest six-point case where this cut first arises, $\ab{AB12}=\ab{AB34}=\ab{AB56}=0$. This cut can be parametrized by setting 
\begin{align}
\label{eq:triple_cut_param}
\begin{split}
(AB)&=((Z_1+xZ_2)34){\cap}((Z_1+xZ_2)56)\\ &=(134){\cap}(156)+x(134){\cap}(256)+x(234){\cap}(156)+x^2(234){\cap}(256), 
\end{split}
\end{align}
where $x$ parametrizes the intersection point of the lines $(AB)$ and $(12)$. Because this cut is inconsistent with both the MHV and $\MHVbar$ geometries, the proof that this cut configuration is spurious must follow directly from the inequalities which are valid in both spaces. For this example, these are the three inverse propagators \begin{align}
\label{eq:triple_cut_ineq_eg1}
\begin{split}
\ab{AB23}&=\ab{1234}(\ab{1356}+x\ab{2356})>0,\\
\ab{AB45}&=-(\ab{1345}+x\ab{2345})(\ab{1456}+x\ab{2456})>0,\\
\ab{AB16}&=\ab{1256}x(\ab{1346}+x\ab{2346})>0.
\end{split}
\end{align}
This set of inequalities is equivalent to
\begin{align}
-\frac{\ab{1345}}{\ab{2345}}<x<-\frac{\ab{1346}}{\ab{2346}}\quad\text{and}\quad x>-\frac{\ab{1356}}{\ab{2356}}.
\end{align}
Non-trivially, the upper bound on $x$ in the first inequality is inconsistent with the second lower bound given. Namely, by combining $x<-\ab{1346}/\ab{2346}$ and $x>-\ab{1356}/\ab{2356}$ we find, using the Schouten identity,
\begin{equation}
0<\frac{\ab{1356}\ab{2346}-\ab{1346}\ab{2356}}{\ab{2346}\ab{2356}}=-\frac{\ab{1236}\ab{3456}}{\ab{2346}\ab{2356}}<0,    
\end{equation}
which is a contradiction. This proof crucially depends on the simple external data positivity conditions relevant for the MHV and $\MHVbar$ spaces. The general three-mass triple cut $\ab{ABii{+}1}=\ab{ABjj{+}1}=\ab{ABkk{+}1}=0$ can be shown to be spurious by the obvious generalization of the parametrization of eq.~(\ref{eq:triple_cut_param}), and the proof depends only on the inequalities adjacent to the cut propagators. 

In contrast, the codimension-four leading singularities are much easier to analyze, as any inequality evaluated on such a configuration just reduces to a condition on the external data. In fact, simply demanding compatibility of the inequalities defining the MHV space eq.~(\ref{eq:mhv_alt}) evaluated on the leading singularity configurations listed in appendix \ref{app:line_configs} with the positivity constraints on the external data eq.~(\ref{eq:zIneq}) is enough to fix the positions of all MHV leading singularities to $(AB)=(ij)$. Said differently, any leading singularity \emph{not} of this form will explicitly violate at least one of the inequalities defining the MHV space. For example, at five points, the line $(13)$ is an allowed leading singularity, but its parity conjugate $(\overline{13})=-(512){\cap}(234)$ is not. 
\begin{align}
    \text{allowed in MHV:}
    \raisebox{-42pt}{\includegraphics[scale=.6]{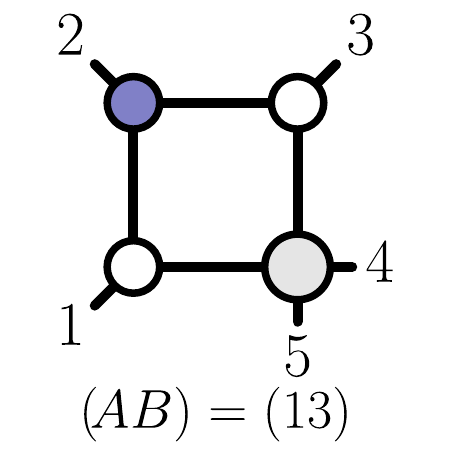}}
    \qquad
    \text{not allowed in MHV:}
    \raisebox{-42pt}{\includegraphics[scale=.6]{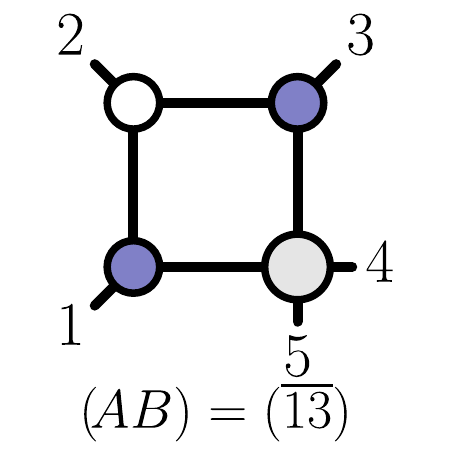}}
\end{align}
From the inequalities point of view, this can be clearly seen by evaluating $\ab{AB\overline{24}} = \ab{AB(123){\cap}(345)}$ on the forbidden leading singularity $(AB)=(\overline{13})$, which gives 
\begin{equation}
\ab{AB\overline{24}}\big\vert_{(\overline{13})} 
{=}
{-}\ab{(512){\cap}(234)(123){\cap}(345)}
{=}
{-}\ab{1235}\ab{2345}\ab{1234}{<}0,
\end{equation}
and is incompatible with being inside the MHV Amplituhedron. This argument goes through at $n$-points. The leading singularities of the form $(ij)$, for $i<j$, are allowed precisely because of the nontrivial positivity conditions \cite{Arkani-Hamed:2014dca} 
\begin{equation}
\ab{AB\overline{k\ell}}\big\vert_{(ij)}
{=}
\ab{ij\overline{k\ell}}{>}0, 
\quad \text{for } i<j,
\end{equation}
while the parity conjugate lines $(\overline{ij})$, as well as all other codimension-four configurations listed in appendix~\ref{app:line_configs}, are always inconsistent with at least one of the positivity constraints. Note that related analyses were performed in \cite{Dennen:2016mdk} in order to check admissible Landau singularities in planar $\NeqFour$.

\subsection{Positivity and the dual Amplituhedron}
\label{subsec:positivity_dual_Amplituhedron}

The original idea of linking scattering amplitudes to projective geometry appeared in the seminal work by Andrew Hodges \cite{Hodges:2009hk} (which pre-dates the Amplituhedron) who showed that the six-particle NMHV tree-level amplitude can be interpreted as the \emph{volume} of a polytope in \emph{dual} momentum twistor space. Later it was shown that the same amplitude can also be associated with a \emph{logarithmic differential form}, $\Omega$, directly in momentum twistor space. It is this picture of amplitudes as differential forms with logarithmic singularities that was later generalized to the Amplituhedron for all $n$, $k$, and $L$. For many reasons, the original volume interpretation seems more fundamental than thinking about amplitudes as differential forms. This led to the conjecture of the existence of a \emph{dual Amplituhedron} \cite{Arkani-Hamed:2014dca} whose volume calculates all scattering amplitudes in planar ${\cal N}=4$ sYM theory.

The only case for which we fully understand the volume interpretation, however, remains the NMHV tree-level amplitude. There, both the Amplituhedron and the dual Amplituhedron are certain polytopes in projective space $\mathbb{P}^4$ which are related by the standard dualization procedure: vertices in the Amplituhedron space are mapped into faces of the dual Amplituhedron, and similarly for other boundaries (codimension $r$ subspaces of the Amplituhedron map into codimension $5-r+1$ subspaces of the dual Amplituhedron). It is not clear how to repeat the same procedure beyond NMHV ($k>1$) and/or at higher loops ($L>0$) where Amplituhedra are no longer polytopes but rather their generalizations to Grassmannians and beyond \cite{Arkani-Hamed:2013jha}. 

Even without an explicit definition of the dual Amplituhedron, there are two important properties which follow from its presumed existence. First, volumes are naturally \emph{positive} and therefore, we would expect that amplitudes exhibit a similar positivity property. While the definition of the Amplituhedron is based on a set of positivity conditions eq.~(\ref{eq:zIneq}) and eq.~(\ref{eq:abIneq}), nothing a priori predicts any positivity properties of the differential form $\Omega$. However, if the amplitude also has a volume interpretation, then we expect that $\Omega$ is positive in some suitable sense. Indeed, it was shown in \cite{Arkani-Hamed:2014dca} for many nontrivial examples that if we strip off the measure from $\Omega$ i.e.,
\begin{equation}
\Omega^{(n,k,L)} = \mathrm{d}\mu\,\omega^{(n,k,L)}\,,
\end{equation}
then the integrand $\omega^{(n,k,L)}$ (which is just the scattering amplitude) is in fact positive if evaluated inside the positive region for both $Z_a$ and ${\cal L}_i=(AB)_i$. The positivity of $\omega^{(n,k,L)}$ therefore serves as indirect evidence for the existence of a dual Amplituhedron and the volume interpretation of scattering amplitudes,
\begin{equation}
    \omega^{(n,k,L)}=\int_{\widetilde{\mathcal{A}}} dV.
\end{equation}
where $\widetilde{\mathcal{A}}$ is the dual Amplituhedron space and $dV$ is the appropriate volume form. An important clue on how to proceed in the search for the dual Amplituhedron is to investigate whether or not positivity is respected in the context of different triangulations of the Amplituhedron. Despite the fact that the full integrand $\omega^{(n,k,L)}$ is positive inside the positive region, individual terms in e.g., the BCFW triangulation of ${\cal A}_{n,k,L}$ do not have definite signs inside the positive space. While individual BCFW terms \emph{internally} triangulate the Amplituhedron, in the dual picture, they get mapped to spaces that are partially outside of the dual Amplituhedron and therefore do not have a uniform sign. This is easiest to understand with a simple toy example of a quadrilateral in the projective plane:
\vspace{-5pt}
\begin{align}
 \label{toy_dualization}
    \raisebox{-45pt}{\includegraphics[scale=.55]{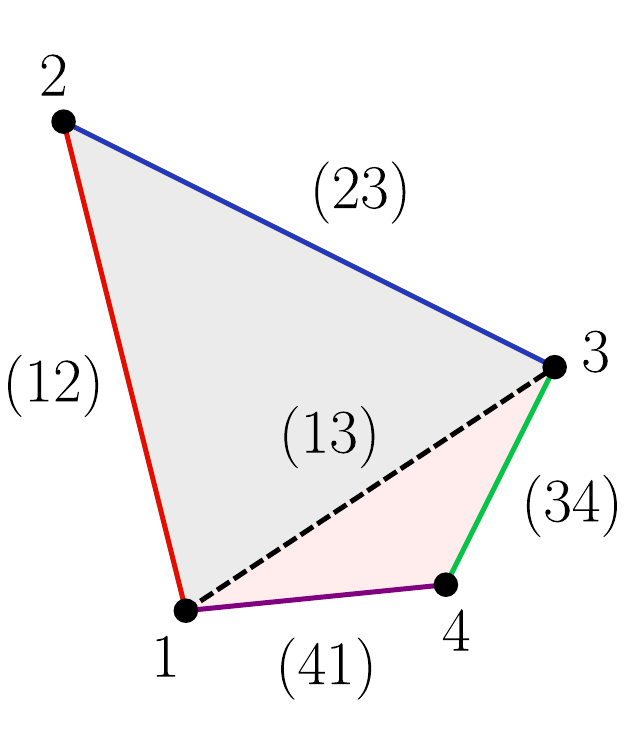}}
    \quad
    &\underset{\text{dual to}}{\longleftrightarrow}
    \quad
     \raisebox{-52pt}{\includegraphics[scale=.6]{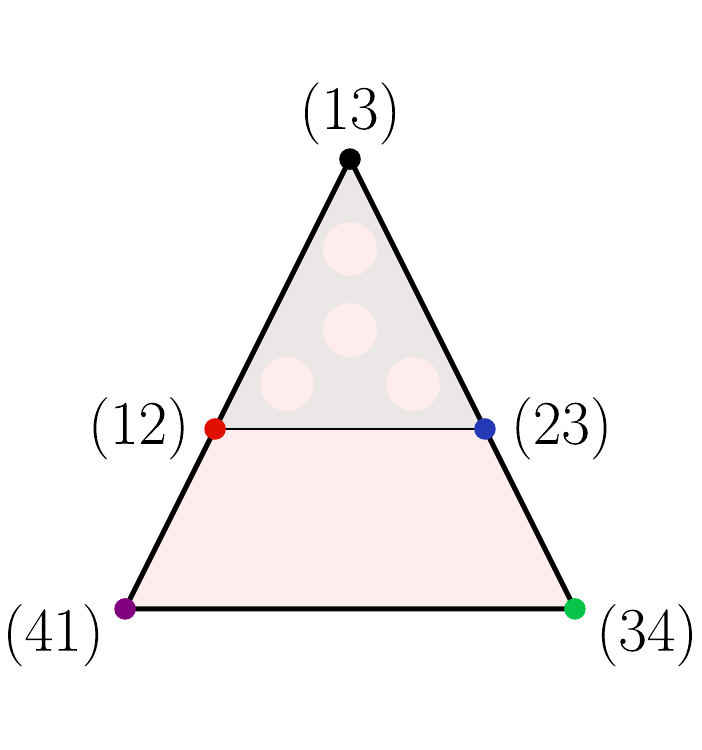}}
    \\[-18pt]
    \hspace{-1cm}
    \raisebox{-32pt}{\includegraphics[scale=.4]{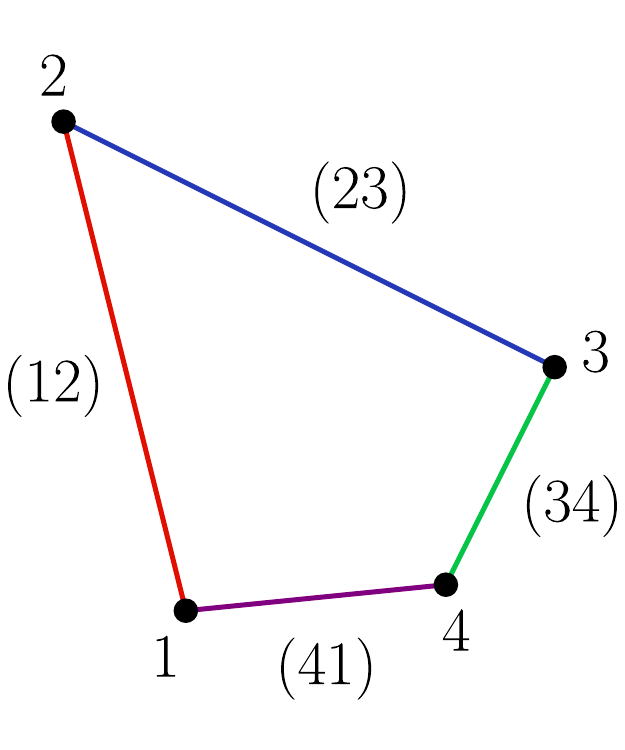}}
    =
     \raisebox{-32pt}{\includegraphics[scale=.4]{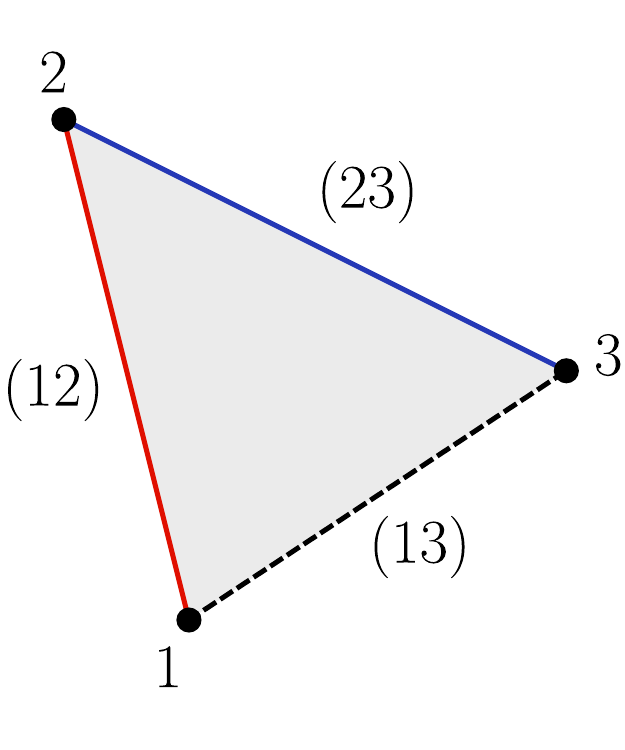}}
     \hspace{-1cm}
     \raisebox{-22pt}{+}
     \hspace{-1.5cm}
     \raisebox{-70pt}{\includegraphics[scale=.4,angle=0]{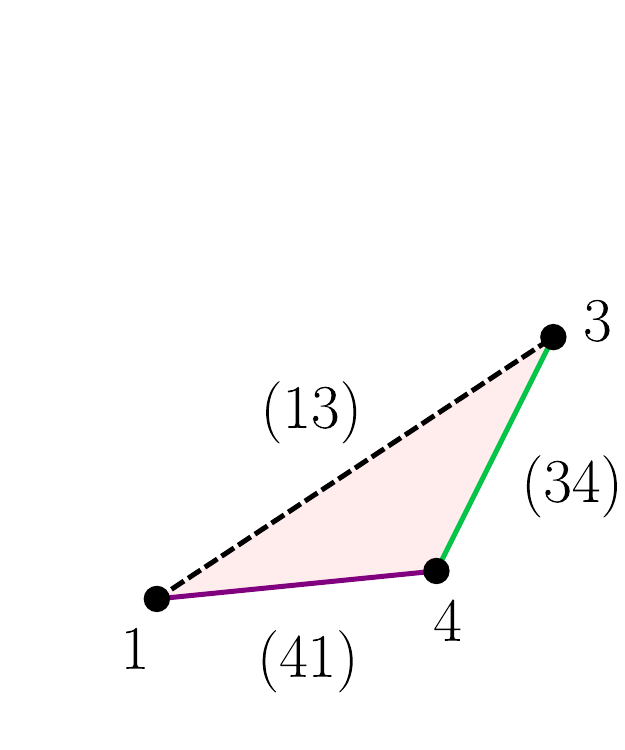}}
    &\underset{\text{dual to}}{\longleftrightarrow}
    \raisebox{-32pt}{\includegraphics[scale=.4]{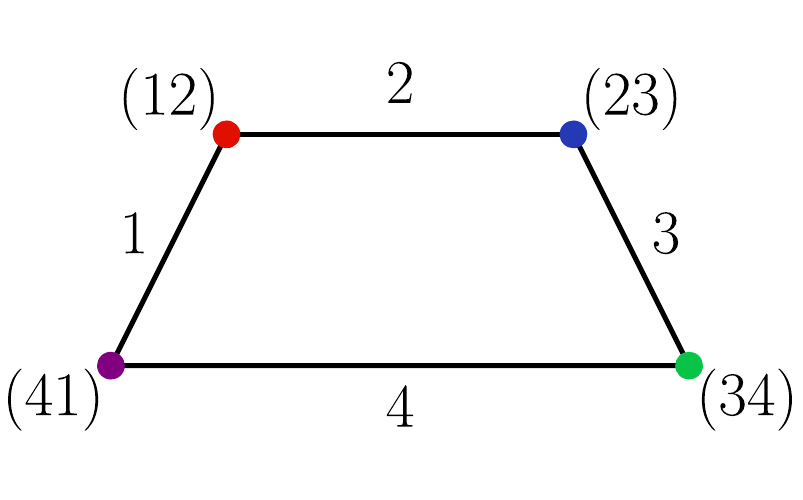}}
    =
    \hspace{-.4cm}
    \raisebox{-45pt}{\includegraphics[scale=.4]{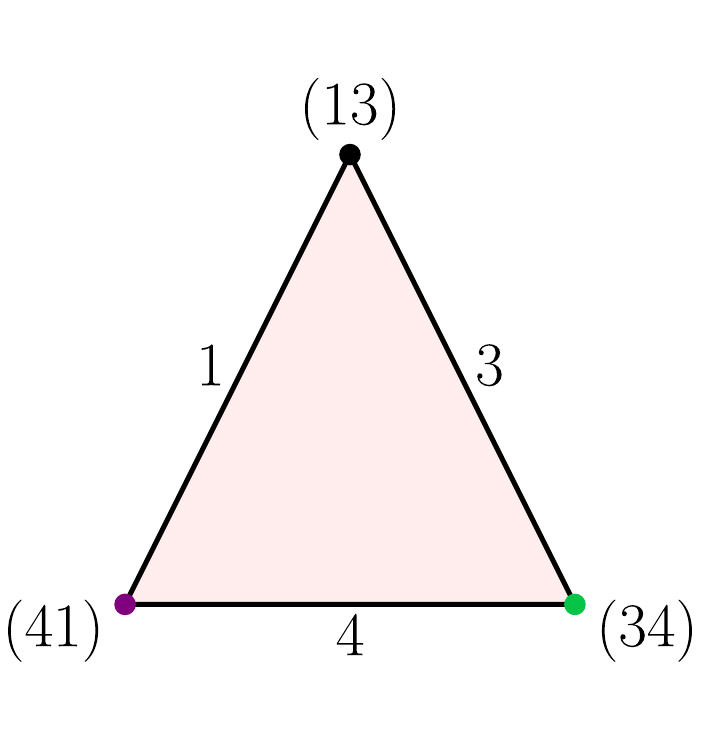}}
    \hspace{-.6cm}
    -
    \hspace{0cm}
    \raisebox{-40pt}{\includegraphics[scale=.4]{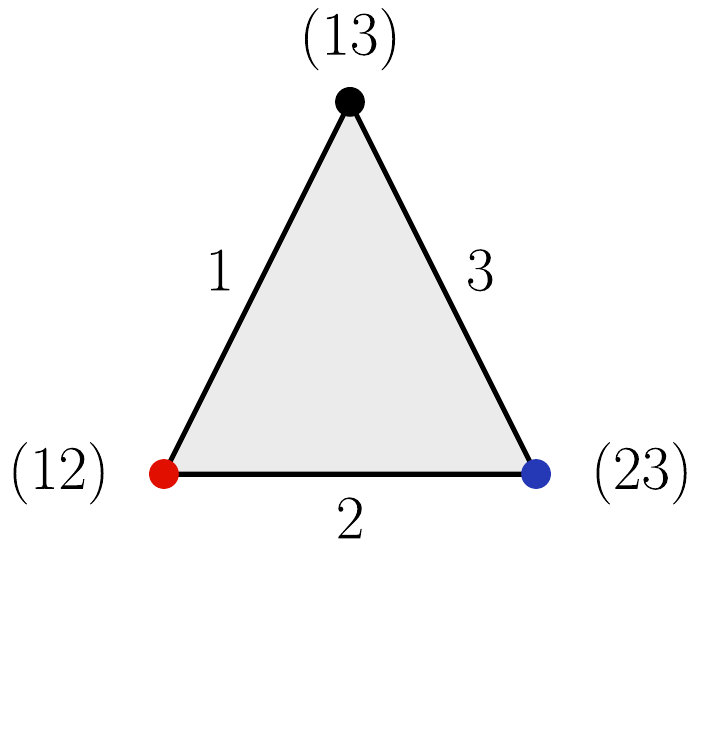}}\nonumber
\end{align}
The two triangles with vertices (123) and (134) triangulate the quadrilateral, but in the dual picture they correspond to triangles which lie partially outside of the dual quadrilateral. Therefore, while the triangles on the left-hand-side of (\ref{toy_dualization}) give an \emph{internal} triangulation of the original space via the line $(13)$, in the dual picture, this corresponds to an \emph{external} triangulation with an external triangulation point $(13)$ that is outside the dual space. In the same sense, internal triangulations of the Amplituhedron are expected to externally triangulate the dual Amplituhedron. 

In particular, the Kermit expansion of MHV amplitudes internally triangulates the one-loop Amplituhedron \cite{Arkani-Hamed:2013jha} and---following our discussion above---is expected to externally triangulate the putative dual Amplituhedron. This external triangulation of the dual space therefore suggests that individual Kermits will not be positive term-wise (after stripping the measure) despite the positivity of the full one-loop integrand. We can see this non-positivity directly by looking at the denominators of the kermit forms: each Kermit contains denominator factors like $\ab{AB1i}$ which do not have a fixed sign inside the Amplituhedron. Consequently, for fixed positive external data and an arbitrary point $(AB)$ inside the MHV one-loop Amplituhedron, the stripped forms $\omega^{(i,j)}_n$ in eq.~(\ref{eq:Kermit}) can have arbitrary signs. 

It is very natural to ask about a term-wise positive expansion of the amplitude, which geometrically would provide an internal triangulation of the putative dual Amplituhedron. While we will not definitively establish the existence of a dual Amplituhedron in this work, it is easy to at least find a term-wise positive expansion of the amplitude. A priori, any such candidate expansion must have only \emph{local} poles, $\ab{ABii{+}1}$ as only these terms have fixed signs, and in addition the numerator factors must be uniformly positive inside the Amplituhedron. As it turns out, the right candidate is the chiral pentagon expansion \cite{ArkaniHamed:2010gg,ArkaniHamed:2010gh,ArkaniHamed:2010kv}\footnote{In the representation here, we have singled out a particular propagator $\ab{AB1n}$ that appears in all diagrams and will also take a special r\^{o}le in our geometric considerations below. Furthermore, we point out that the chiral pentagon expansion contains various ``boundary" terms that correspond to one-mass, and two-mass hard box integrands. This should be contrasted to the older representations of one-loop $n$-point MHV amplitudes \cite{Bern:1994zx} in terms of two-mass easy boxes that only match the parity-even sector at the integrand level but are of course equivalent upon loop integration.}
\begin{align}
\begin{split}
\label{pent}
\omega^{(n,0,1)}=
&
\sum_{i<j} \omega^{(n)}_{ij} =
\sum_{i<j}\ \  \raisebox{-40pt}{\includegraphics[scale=.6]{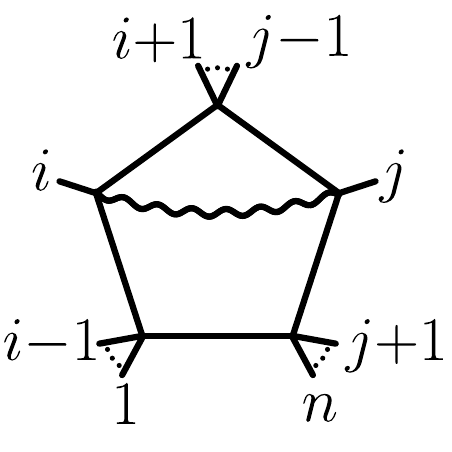}}  
\\
=&\sum_{i<j}
\frac{\ab{1ijn}\ab{AB\overline{ij}}}
{\ab{ABi{-}1i}\ab{ABii{+}1}\ab{ABj{-}1j}\ab{ABjj{+}1}\ab{AB1n}}.
\end{split}
\end{align}
As a consequence of the MHV loop-positivity eq.~(\ref{eq:mhv_alt}), together with the positivity of external data $\ab{ijkl}>0$ (for $i<j<k<l$) in eq.~(\ref{eq:mhv_ext_positivity}), it can be seen that all four-brackets in eq.~(\ref{pent}) are manifestly positive, including the loop-momentum dependent part of the numerator. Based on this positivity property of individual terms, the natural conjecture is that these chiral pentagons internally triangulate the dual Amplituhedron (but externally triangulate the original Amplituhedron). In our quadrilateral analogy, this would correspond to introducing a spurious triangulation point $(12){\cap}(34)$ \emph{outside} the quadrilateral,
\begin{align}
\begin{split}
& \\[-25pt]
\label{toy_dual_internal_triangulation}
& \hspace{-1cm}
    \raisebox{-32pt}{\includegraphics[scale=.4]{./figures/2d_toy_amplituhedron_quadrilateral.pdf}}
    =
    \hspace{-1cm}
    \raisebox{-32pt}{\includegraphics[scale=.33]{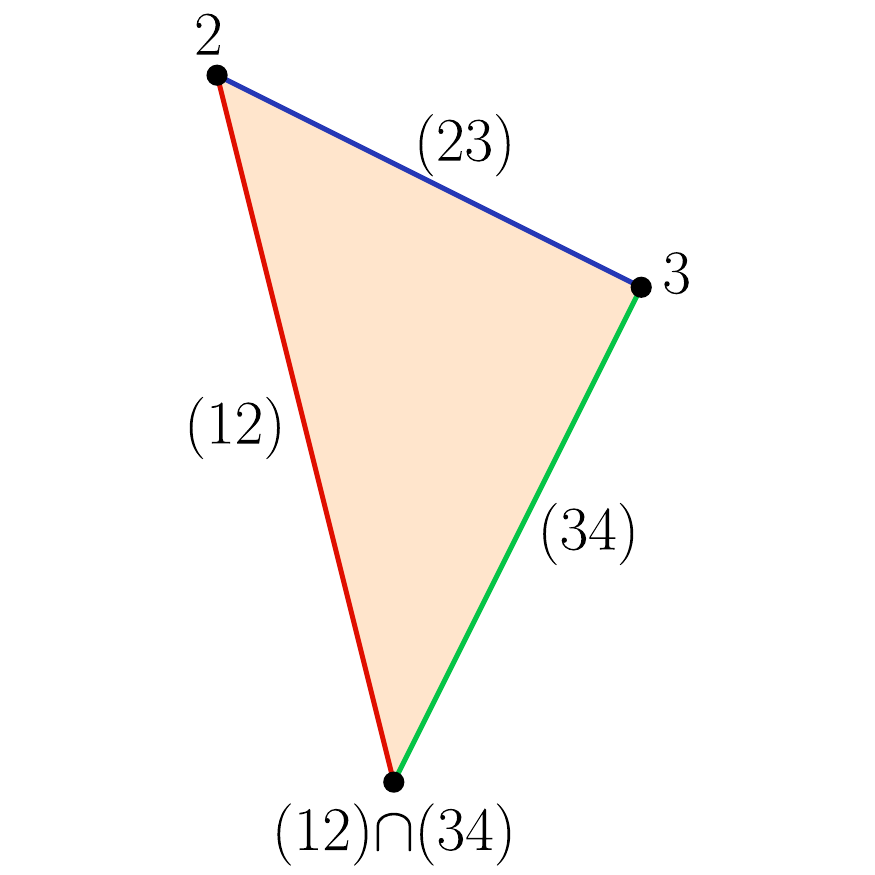}}
    \hspace{-.8cm}
     -
    \hspace{-1.2cm}
    \raisebox{-52pt}{\includegraphics[scale=.4,angle=0]{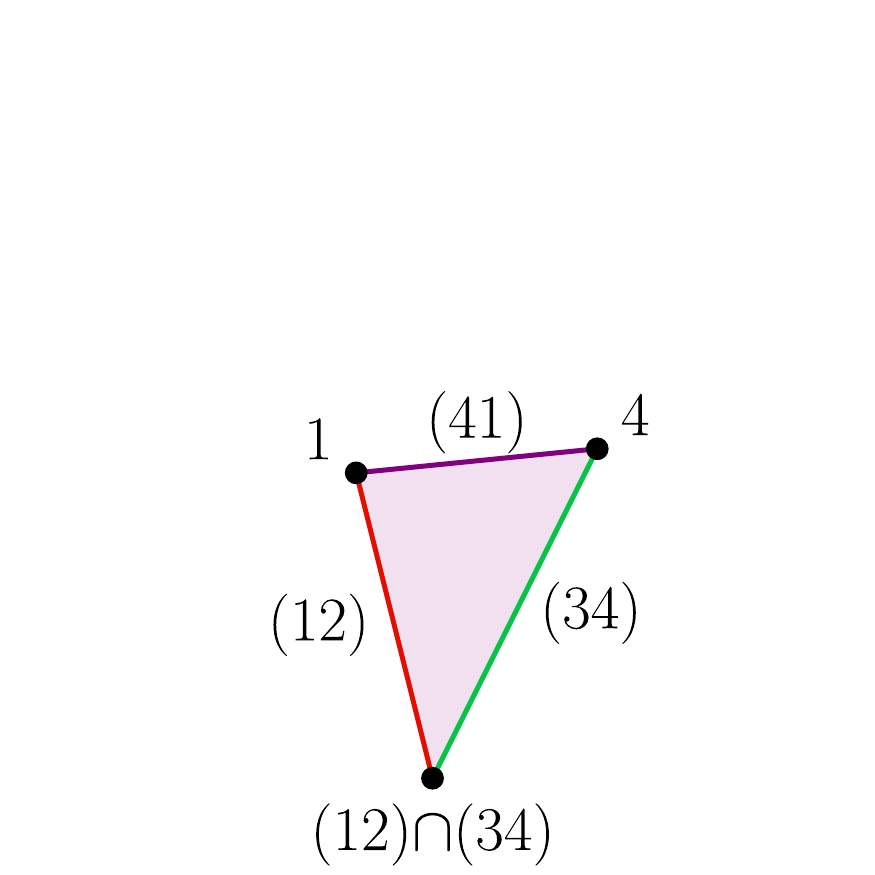}}
    \hspace{-.7cm}
   \underset{\text{dual to}}{\longleftrightarrow}
    \raisebox{-32pt}{\includegraphics[scale=.4]{./figures/2d_toy_dual_quadrilateral.pdf}}
    \hspace{-.2cm}
    =
    \hspace{-.2cm}
    \raisebox{-32pt}{\includegraphics[scale=.4]{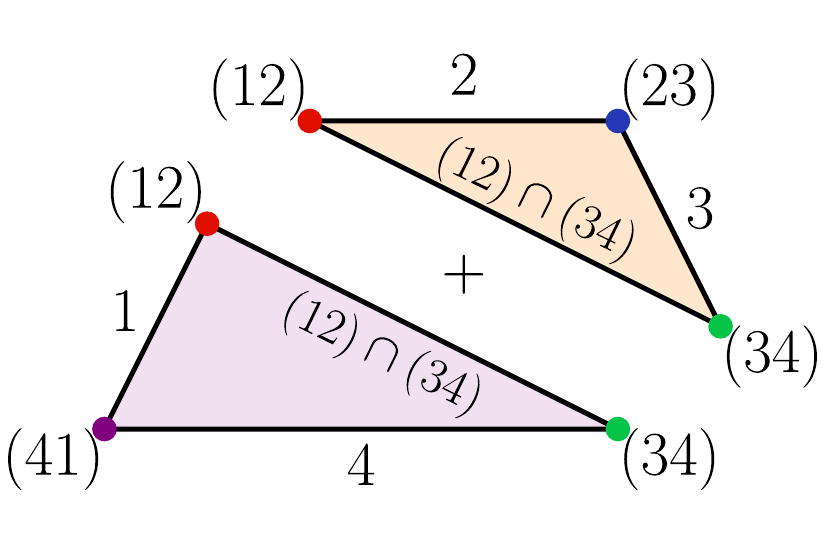}}
\hspace{-.8cm}  
\end{split}
\end{align}
While the chiral pentagons have only physical codimension-one poles $\ab{AB i i{+}1}$ (we also refer to them as local poles), their higher codimension singularities are not all physical. For example, the generic chiral pentagon eq.~(\ref{pent}) has a non-zero residue at the leading singularity location $(AB)=(ijj{+}1){\cap}(in1)$, which is not a physical leading singularity of MHV amplitudes and corresponds to a cut solution labeled by the following on-shell function\footnote{Here, we make use of on-shell functions to label solutions to on-shell conditions and not the value of field theory cuts. This dual meaning of on-shell functions is common in the literature, see e.g.~\cite{Cachazo:2008vp}.}
\begin{align}
    \raisebox{-45pt}{
    \includegraphics[scale=.7]{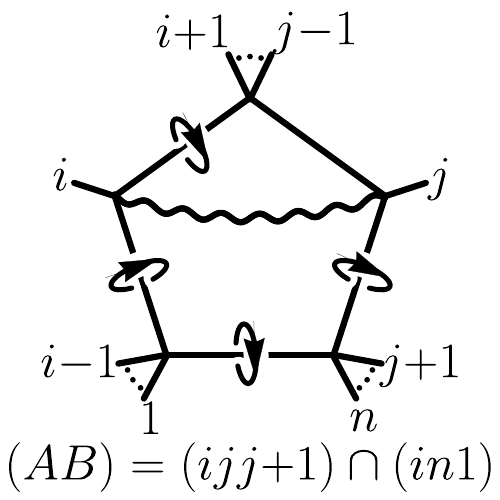}}
    \quad
    \leftrightarrow
    \quad
    \raisebox{-45pt}{
    \includegraphics[scale=.7]{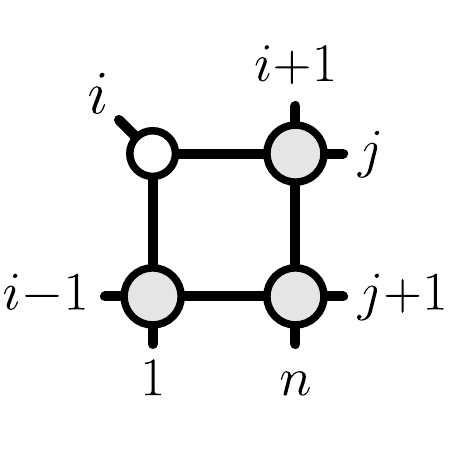}}
\end{align}
As we will show, in the original Amplituhedron geometry this means that all codimension-one boundaries are given by physical $\ab{ABii+1}=0$ singularities analogous to the lines $(i\,i{+}1)$ in eq.~(\ref{toy_dual_internal_triangulation}), whereas higher codimension boundaries can be spurious (such as the spurious triangulation vertex $(12){\cap}(34)$). In the dual picture, some of the first boundaries of chiral pentagon spaces are spurious, analogous to the spurious triangulation line of our quadrilateral example on the right-hand-side of eq.~(\ref{toy_dual_internal_triangulation}).

\section{Geometry of $d\log$ forms}
\label{sec:geometry_dlog_forms}

In this section we lay the groundwork for further study of the relation between positive geometry and local triangulations of the Amplituhedron. As a first step, we need to associate a geometric region to a particular local loop-integrand form. The most natural starting point for such an endeavour is the $d\log$ representation of local integrands.

\subsection{$d\log$ forms for pentagon integrands}

It is our goal to associate a ``local'' positive geometry to each chiral pentagon and explore how they glue together into a larger geometric object. In a subsequent step, we define these new objects and find their connection to the Amplituhedron in section~\ref{sec:geometry_chiral_pentagons}. Starting from the local representation of the one-loop MHV amplitude eq.~(\ref{pent}), our only input information is the rational pentagon integrand. Due to the natural connection between $d\log$ forms and positive geometry, we therefore rewrite the integrands appearing in eq.~(\ref{pent}) as $d\log$ forms \cite{Arkani-Hamed:2016byb},
\begin{equation}
\label{eq:dlog_pent}
\Omega^{(n)}_{ij} = \ab{AB\,d^2A}\ab{AB\,d^2B}\, \omega^{(n)}_{ij} 
       =  d\log f_1\, d\log f_2\, d\log f_3\,d\log f_4\,,
\end{equation}
where we suppress the wedge notation for differential forms and $f_j$ are ratios of four-brackets (to be specified in eq.~(\ref{chiral})). We would like to interpret this $d\log$ form as the form with logarithmic singularities on a region which has fixed signs of all $f_j$. In particular, each $f_j$ can be either positive or negative which leads to $2^4=16$ different regions that each have the same $d\log$ form. Furthermore, the change of variables eq.~(\ref{eq:dlog_pent}) is not unique and there are many different looking $d\log$ forms that all have the same rational form appearing in eq.~(\ref{pent}).

For the massless scalar box integral, which can be viewed as a particular $n=4$ degeneration of the chiral pentagon, there are at least two different $d\log$ forms,
\begin{align}
\label{box4}
   \Omega_{23}^{(4)} & = \raisebox{-20pt}{\includegraphics[scale=.5]{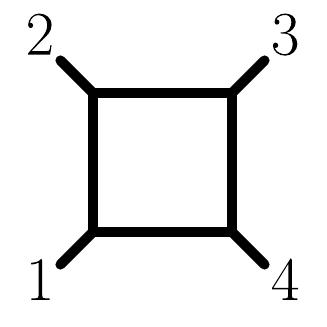}} 
   = \frac{\la ABd^2A\ra\la ABd^2B\ra\la1234\ra^2}{\ab{AB12}\ab{AB23}\ab{AB34}\ab{AB14}}\\ 
   &= d\log\frac{\ab{AB12}}{\la ABX\ra} d\log\frac{\ab{AB23}}{\la ABX\ra} d\log\frac{\ab{AB34}}{\la ABX\ra}d\log\frac{\ab{AB14}}{\la ABX\ra}\nonumber
\end{align}
where we can choose between the two solutions of the quadruple cut, $X=(13)$ or $X=(24)$. For $X=(13)$ the $d\log$ form eq.~(\ref{box4}) explicitly depends on $\ab{AB13}$. Therefore, one might worry that this spurious pole could show up in the rational form.\footnote{The presence of spurious poles in the arguments of $d\log$'s is a very general feature, and in fact is relevant for understanding simplified differential equations based on $d\log$ integrands \cite{Herrmann:2019upk}.} However, by construction, the $d\log$ form cannot have $\ab{AB13}$ or $\ab{AB24}$ as an actual singularity. Algebraically, the absence of the spurious poles from the rational integrand derived from the $\dlog$ form eq.~(\ref{box4}) follows from non-trivial kinematic identities. 

For more complicated integrands, various alternative $d\log$ forms for the same rational integrand might look remarkably different. This is also the case for the generic chiral pentagon we discuss next. One particular $d\log$ form for this integrand has been written down in ref.~\cite{Arkani-Hamed:2016byb} which manifestly breaks the $ij$ flip symmetry of the diagram,
\begin{align}
\label{chiral}
\Omega^{(n)}_{ij} & = 
\raisebox{-30pt}{\includegraphics[scale=.6]{./figures/chiral_pent.pdf}} 
= \frac{\ab{ABd^2A}\ab{ABd^2B}\ab{1ijn} \ab{AB\overline{ij}}}
        {\ab{ABi{-}1i}\ab{ABii{+}1}\ab{ABj{-}1j}\ab{ABjj{+}1}\ab{AB1n}}
\\
& 
\hspace{-1cm} 
= d\log\frac{\ab{AB i{-}1i}}{\ab{ABn1}} 
  d\log\frac{\ab{ABii{+}1}}{\ab{ ABn1}}
  d\log\frac{\ab{ABj{-}1j}}{\ab{AB(in1){\cap}\overline{j}}}
  d\log\frac{\ab{ABjj{+}1}}{\ab{AB(in1){\cap}\overline{j}}}\,,\nonumber
\end{align}
with an analogous expression for $i\leftrightarrow j$. Here, we present a \emph{new} $d\log$ form for $\Omega_{ij}^{(n)}$:
\begin{equation}
\label{eq:dlog_chiral_symmetric}
\Omega_{ij}^{(n)}{=}
d\log\frac{\ab{ABi{-}1i}}{\ab{ABii{+}1}}
d\log\frac{\ab{ABj{-}1j}}{\ab{ABjj{+}1}}
d\log\frac{\ab{AB(n1i){\cap}(ij\star)}}{\ab{AB1n}}
d\log\frac{\ab{AB(n1j){\cap}(ij\star)}}{\ab{AB1n}}\,,
\end{equation}
where $\star$ corresponds to an \emph{arbitrary} point in momentum twistor space. It is a highly non-trivial statement that the integrand form does not depend on the choice of $\star$. Let us point out that our new $d\log$ representation in eq.~(\ref{eq:dlog_chiral_symmetric}) has two nice features: $(i)$ it is manifestly $ij$ symmetric, and $(ii)$ it makes the fact that the integrand vanishes for $(AB)\in(i{-1}ii{+}1)$ or $(AB)\in(j{-}1jj{+}1)$ obvious since e.g. only one $d\log$ blows up on this cut -- $\ab{ABi{-}1i}$ and $\ab{ABii{+}1}$ appear together in one ratio -- and this is not enough to produce a non-zero residue. However, the other solution of the double cut $\ab{ABi{-}1i}=\ab{ABii{+}1}=0$, which is $A=i$, does produce a non-zero residue because $\ab{AB(n1i){\cap}(ij\star)}$ also vanishes. In the simplest five-point case, the $d\log$ form reads
\begin{align}
\Omega_{24}^{(5)}& = 
\raisebox{-25pt}{\includegraphics[scale=.4]{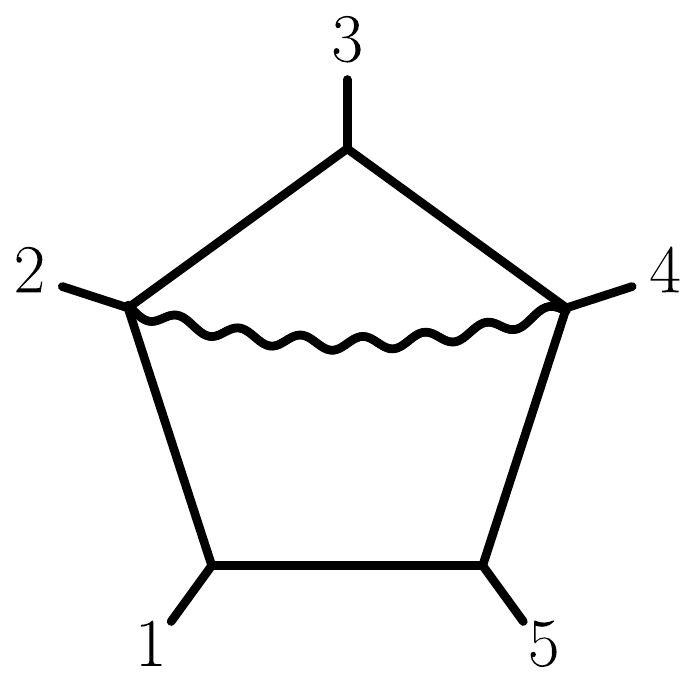}} 
= \frac{\ab{AB\,d^2A}\ab{AB\,d^2B}\ab{1245}\ab{AB\overline{24}}}
       {\ab{AB12}\ab{AB23}\ab{AB34}\ab{AB45}\ab{AB15}}\\
&= d\log \frac{\ab{AB12}}{\ab{AB23}}\,
   d\log\frac{\ab{AB34}}{\ab{AB45}}
   d\log\frac{\la AB(512){\cap}(243)\ra}{\ab{AB15}}
   d\log\frac{\la AB(514){\cap}(243)\ra}{\ab{AB15}}\,,\nonumber
\end{align}
where we have chosen the special point $\star = 3$.

Similar to the chiral pentagon expansion of the MHV amplitude in eq.~(\ref{pent}), we can write an analogous formula for the $\MHVbar$ amplitude where the pentagon with a wavy-line numerator is replaced by a dashed-line numerator of opposite chirality,
\begin{align}
\begin{split}
\label{mhvBar_pent}
\omega^{(n,n{-}2,1)}=
&
\sum_{i<j} \overline{\omega}^{(n)}_{ij} =
\sum_{i<j} \raisebox{-30pt}{\includegraphics[scale=.5]{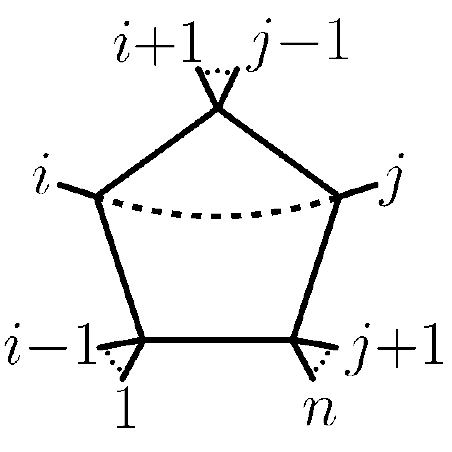}}
\\
=&\sum_{i<j}
\frac{\ab{1\overline{ij}n}\, \ab{ABij}}
{\ab{ABi{-}1i}\ab{ABii{+}1}\ab{ABj{-}1j}\ab{ABjj{+}1}\ab{AB1n}}.
\end{split}
\end{align}
and all statements about the ``MHV" pentagons can be readily transferred to the ``$\MHVbar$" pentagons as well. For completeness, we write a \emph{novel} symmetric $d\log$ form for $\overline{\Omega}^{(n)}_{ij}$ obtained by dualizing eq.~(\ref{eq:dlog_chiral_symmetric}),
\begin{align}
\label{eq:dlog_chiral_dash_symmetric}
    &\overline{\Omega}^{(n)}_{ij} = 
    \raisebox{-30pt}{\includegraphics[scale=.5]{figures/chiral_pent_dashed.pdf}} \\
    &{=} 
    d\log \frac{\ab{ABi{-}1 i}}{\ab{ABii{+}1}}
    d\log \frac{\ab{ABj{-}1 j}}{\ab{ABjj{+}1}}
    d\log \frac{\ab{AB((n1){\cap}\overline{i})(\overline{\star}{\cap}\overline{j}{\cap}\overline{i})}}{\ab{AB1n}}
    d\log \frac{\ab{AB((n1){\cap}\overline{j})(\overline{\star}{\cap}\overline{i}{\cap}\overline{j})}}{\ab{AB1n}},
    \nonumber
\end{align}
where $\overline{\star}=(x_1x_2x_3)$ is an arbitrary plane. 

Besides the two chiral pentagons appearing in eq.~(\ref{chiral}) and eq.~(\ref{mhvBar_pent}), there is one additional pentagon that will play an important r\^{o}le, namely the \emph{parity-odd} pentagon. In momentum space its numerator is proportional to the Levi-Cevita tensor and evaluates to zero upon integration over real Minkowski space. In momentum twistor space, the parity-odd numerator is the difference of two chiral numerators 
\begin{equation}
\label{eq:odd_pent_num}
N_{ij}^{\text{odd}} = \ab{AB\overline{ij}}\ab{1ijn}{-} 
                      \ab{AB\,ij} \ab{1\overline{ij}n}.
\end{equation}
The relative minus sign between the the wavy and dashed numerators is necessary for the integrand to be unit on all codimension-four residues; a relative plus sign would give some leading singularities equal to $\pm2$, rather than $\pm1$. The denominator of the parity-odd pentagon is the same as in eq.~(\ref{chiral}). The $d\log$ form for this integral is very simple and involves only physical propagators:
\begin{equation}
\label{eq:dlog_odd_pentagon}
\Omega_{ij}^{(n),\text{odd}}=
d\log\frac{\langle ABi{-}1i\rangle}{\langle AB1n\rangle}
d\log\frac{\langle ABii{+}1\rangle}{\langle AB1n\rangle}
d\log\frac{\langle ABj{-}1j\rangle}{\langle AB1n\rangle}
d\log\frac{\langle ABjj{+}1\rangle}{\langle AB1n\rangle}\,,
\end{equation}
where we could also reshuffle the propagators that appear in the denominator. 
The fact that only propagators appear as $d\log$ arguments also leads to a simple understanding of why this integral evaluates to zero from a differential equation point of view \cite{Herrmann:2019upk}.

The difference of MHV and $\MHVbar$ amplitudes is the \emph{parity-odd} amplitude, which integrates to zero on the parity invariant contour. However, it will play a very important r\^{o}le in our further discussion. Combining the expansions eq.~(\ref{pent}) and eq.~(\ref{mhvBar_pent}), we see that the chiral numerators combine precisely into the numerator $N^{\text{odd}}_{ij}$ of eq.~(\ref{eq:odd_pent_num}) and we get a sum of parity-odd pentagons
\begin{align}
\begin{split}
\label{odd_pent}
\omega^{\text{odd}}=
&
\sum_{i<j} \left(\omega^{(n)}_{ij} {-}\overline{\omega}^{(n)}_{ij}\right) =
\sum_{i<j}\left( 
\raisebox{-30pt}{\includegraphics[scale=.5]{figures/chiral_pent.pdf}}
{-}
\raisebox{-30pt}{\includegraphics[scale=.5]{figures/chiral_pent_dashed.pdf}}
\right)
\\
=&\sum_{i<j}\frac{N^{\text{odd}}_{ij}}
{\ab{ABi{-}1i}\ab{ABii{+}1}\ab{ABj{-}1j}\ab{ABjj{+}1}\ab{AB1n}}\,.
\end{split}
\end{align}
Note that all ``boundary terms" (i.e.~$j=i{+}1$) cancel between the MHV and $\MHVbar$ sectors in this expansion so that there are no parity-even box integrals remaining. Therefore, the sum is zero at four points, gives one parity-odd pentagon at five points, and so on.

\subsection{From $d\log$'s to geometry}
\label{subsec:dlog_form_to_geometry}

In the usual Amplituhedron setup, we \emph{start} with a positive geometry, and from it calculate the canonical form with logarithmic singularities on its boundaries. However, in our case the situation is reversed: we have the $d\log$ form for e.g.~the pentagon integrand in eq.~(\ref{chiral}) or eq.~(\ref{eq:dlog_chiral_symmetric}), but we do not a priori know the correct geometric space associated to that differential form, let alone whether or not there is even a unique answer to that question. By construction, any space defined by demanding definite signs for the \emph{ratios} of the arguments of each $d\log$ in e.g., eq.~(\ref{eq:dlog_chiral_symmetric}) gives \emph{some} geometry with the appropriate canonical form. More generally, if we start with a $d\log$ form
\begin{equation}
    \Omega = d\log f_1\, d\log f_2\, d\log f_3\, d\log f_4
\end{equation}
there are $2^4=16$ geometric spaces associated to it via the inequalities 
\begin{equation}
\label{eq:dlog_sign_choices_gen}
    f_1\lessgtr 0,\, f_2 \lessgtr 0,\, f_3 \lessgtr 0,\, f_4 \lessgtr0\,.
\end{equation}
Each $d\log f_i$ factor is the correct form for both inequalities $f_i>0$ and $f_i<0$, up to a sign. (If we did not impose any restrictions on $f_i$, the form would vanish as there are no boundaries.)

\subsubsection*{Faithful geometries and $d\log$ forms}
%
As we have alluded to above, starting from a rational form of the integrand that only has logarithmic singularities there are often numerous ways to change variables to bring the integrand into a $d\log$ form. For the simple four-point box integrand, we found at least two solutions specified by the choice of $\ab{ABX}$ in eq.~(\ref{box4}). Likewise, for the chiral pentagons, we also have at least two alternative $d\log$ forms in eqs.~(\ref{chiral}) and (\ref{eq:dlog_chiral_symmetric}). Combining this multitude of $d\log$ forms with the 16-fold multiplicity of geometric spaces associated to a given one-loop form encoded in the choice of inequalities in eq.~(\ref{eq:dlog_sign_choices_gen}), it is not hard to imagine that the number of possible geometric spaces associated to a given integrand quickly grows. It is therefore natural to ask, whether or not there exist any special subsets of geometries that have certain desirable properties. Here, and in the following, we are going to argue for \emph{faithful geometries}. 

What we mean by \emph{faithful geometries} is the following: for our purposes, simply getting the correct canonical form is insufficient---we also require the correct boundary structure of the geometric space itself. Concretely, we demand that \emph{all} boundaries of the geometry show up as poles in the form, and furthermore that these are the \emph{only} boundaries. (Note that looking at the entries of the $d\log$ forms can be misleading as certain entries of the $d\log$s are in fact not poles of the form. If one were to compute the residue on such a pole, one in fact finds zero. The simplest example of this is $\ab{ABX}$ in eq.~(\ref{box4}). This bracket is manifestly absent in the rational representation of the form so that it is clear that there is no singularity at this location.) Checking the ``faithfulness'' of a given geometry is more intricate and requires a detailed analysis similar in spirit to the discussion of the geometric boundaries of the kermit expansion in section \ref{subsec:MHV_boundaries_kermit}. As we will explain shortly, there are ``rare'' cases of $d\log$ forms that give rise to faithful geometries, i.e., there is at least one set of inequalities akin to (\ref{eq:dlog_sign_choices_gen}) for which the resulting geometry \emph{only} has the geometric boundaries appearing as singularities of the form and no others. These are the spaces of our interest in this paper. 

The presence of a boundary in the geometry which does not appear as a corresponding pole in the integrand form might seem strange, but this is in fact one of the defining features of Grassmannian geometry. As a simple example, consider the four-point box integral in eq.~(\ref{box4}). As shown earlier, one version of the $d\log$ form contains the $\ab{ABX}=\ab{AB13}$ bracket while the rational integrand form does not. This $d\log$ form naively implies that there are $2^4=16$ different geometries, reflecting the various sign choices for the four ratios of brackets. Geometrically, each choice would give rise to the same $d\log$ form. However, these 16 geometries are all different, and some of them actually have $\ab{AB13}=0$ as a boundary despite its absence in the differential form. One concrete example of this occurs for the following sign choice\footnote{Let us note that for the `local spaces' we define in this work, we do not insist on $\ab{ABii{+}1}>0$ which was a crucial part of the Amplituhedron definition (\ref{eq:abIneq}). We comment on this in section \ref{subsec:no-go-local-triangulation}.}
\begin{align}
\label{eq:geometry_counterexample}
    \frac{\ab{AB12}}{\ab{AB13}}>0, \quad
    \frac{\ab{AB23}}{\ab{AB13}}>0, \quad
    \frac{\ab{AB34}}{\ab{AB13}}>0, \quad
    \frac{\ab{AB14}}{\ab{AB13}}<0.
\end{align}
We can always fix one of the brackets to have a definite sign, as the only relevant information is encoded in the ratios. Fixing the sign of $\ab{AB12}>0$ then implies that all four-brackets appearing in eq.~(\ref{eq:geometry_counterexample}) are positive, except for $\ab{AB14}$, which is negative. The geometric region associated to the sign choice eq.~(\ref{eq:geometry_counterexample}) still has eq.~(\ref{box4}) as the logarithmic form, but geometrically we can now access the $\ab{AB13}=0$ boundary without violating any of the inequalities. We can see this fact explicitly in coordinates: expanding $Z_A = Z_3 + x Z_1 + y Z_2$, and $Z_B=Z_4 + zZ_1 + wZ_2$ the inequalities eq.~(\ref{eq:geometry_counterexample}) are equivalent to
\begin{equation}
x>0,\,\,y<0,\,\, w>0,\,\, (xw-yz)>0.
\end{equation}
In this parametrization we have $\ab{AB13}=-y\ab{1234}$, so accessing the boundary corresponds to setting $y\rightarrow0$. In this case, the only remaining inequality sensitive to this choice is $xw-yz\rightarrow xw>0$, which is clearly consistent with $x,w>0$. Therefore, $\ab{AB13}=0$ is an accessible boundary of the geometry. Since this singularity is absent in the differential form, we conclude that this geometry is not faithful according to our definition above.

However, if we instead choose the geometric space where all ratios appearing in eq.~(\ref{eq:geometry_counterexample}) are positive, in the parametrization above the space becomes
\begin{align}
\label{parametric_inequality_eg}
    x>0,\,\,y<0,\,\, w<0,\,\, (xw-yz)>0\,,
\end{align}
i.e., we now have the opposite sign for $w<0$. This time, sending $y\rightarrow0$ yields the three inequalities $x>0,w<0$ and $xw>0$, which are clearly incompatible, thus demonstrating that $y=0$ (and therefore $\ab{AB13}=0$) is not a geometric boundary of the space. According to our definition above, we would call the positive geometry associated to this choice of inequalities \emph{faithful}. 

In summary, we see that each $d\log$ form gives rise to a large set of geometric spaces (one for each sign choice of the ratios of four-brackets appearing in the $d\log$ form) which all have the same integrand form. However, only a subset of these spaces are free of unphysical boundaries. For admissible (faithful) positive geometries, we demand that all boundaries of the space are directly reflected in the pole structure of the integrand form. This is true for the Amplituhedron, and we want to preserve this property here. 

\subsubsection*{Faithful geometries from the chiral pentagon $d\log$ form}
%
For the generic chiral pentagon of eq.~(\ref{chiral}), we found two possible $d\log$ forms in eqs.~(\ref{chiral}) and (\ref{eq:dlog_chiral_symmetric}). Starting with the original $d\log$ form in eq.~(\ref{chiral}) and checking the boundary structure of the $2^4$ geometries arising from the respective sign choices for the entries of the $d\log$s, we find that \emph{none} of these spaces gives rise to a \emph{faithful geometry}, i.e., these spaces always have certain additional geometric boundaries that do not appear as singularities of the form and are therefore unacceptable to us. 

This encourages us to consider our novel $d\log$ form presented in eq.~(\ref{eq:dlog_chiral_symmetric}). While the rational integrand in eq.~(\ref{chiral}) does not have the spurious poles $\ab{AB(n1i){\cap}(ij\star)}$ or $\ab{AB(n1j){\cap}(ij\star)}$, certain sign choices for the arguments of the $d\log$ form in eq.~(\ref{eq:dlog_chiral_symmetric}) do lead to geometric spaces with boundaries when $(AB)$ intersects the lines $(n1i){\cap}(ij\star)$ or $(n1j){\cap}(ij\star)$. Only the special sign combinations where the spurious boundaries are geometrically absent are of our interest. In this case, there are exactly two choices of signs for the ratios of four-brackets in the $d\log$ form eq.~(\ref{eq:dlog_chiral_symmetric}) which do have the correct boundary structure to represent a faithful geometry. The two possibilities correspond to choosing the reference point $Z_\star$ to be either in the set $Z_\star {\in} \{i{+}1,{\ldots},j{-}1\}$, or $Z_\star {\in} \{1,\ldots,i{-}1\}{\cup}\{j{+}1,\ldots,n\}$.\footnote{Here, we restrict ourselves to simple momentum twistor choices for $Z_\star$ that are already part of the diagram. We do not exclude the possibility that there may exist more complicated choices that yield additional choices satisfying our boundary structure criterion. Furthermore, it is also possible that there are other representations of the chiral pentagon $d\log$ form that opens up yet more options. It would be interesting to prove exhaustively what the set of the most general geometries that can consistently be assigned to the pentagon integrand are.} 
In the first case, $Z_\star\in\{i{+}1,\ldots,j{-}1\}$, we find a consistent geometry defined by the following set of inequalities
\begin{equation}
\hspace{-.5cm}
\label{ratio1}
\left\{
\frac{\ab{ABi{-}1i}}{\ab{ABii{+}1}}{<}0,\,
\frac{\ab{ABj{-}1j}}{\ab{ABjj{+}1}}{<}0,\,
\frac{\ab{AB(n1i){\cap}(ij\star)}}{\ab{AB1n}}{<}0,\,
\frac{\ab{AB(n1j){\cap}(ij\star)}}{\ab{AB1n}}{>}0
\right\}\,.
\hspace{-.5cm}
\end{equation}
We fix the sign of $\ab{AB1n}>0$ which in turn fixes the signs for the brackets with intersections, but leaves four options for the signs of the four individual propagator-type brackets, $\{\ab{ABi{-}1i},\,\ab{ABii{+}1},\ab{ABj{-}1j},\,\ab{ABjj{+}1}\}$. As a result, the first consistent chiral pentagon space is a union of four sign patterns:
\begin{equation}
P^{(1)}_{ij} \leftrightarrow
\begin{tabular}{|c|c|c|c|c|c|c|}
\hline \small   $\!\!\!\ab{ABi{-}1i}\!\!\!$ & $\!\!\!\ab{ABii{+}1}\!\!\!$ & 
                $\!\!\!\ab{ABj{-}1j}\!\!\!$ & $\!\!\!\ab{ABjj{+}1}\!\!\!$ & 
                $\!\!\!\ab{AB1n}\!\!\!$     & $\!\!\!\ab{ABX_i}\!\!\!$    & 
                $\!\!\!\ab{ABX_j}\!\!\!$\\ \hline
$-$ & + & $-$ & + & + & $-$ & +\\ 
+ & $-$ & $-$ & + & + & $-$ & +\\ 
$-$ & + & + & $-$ & + & $-$ & +\\ 
+ & $-$ & + & $-$ & + & $-$ & +\\ \hline
\end{tabular} \label{pentsign1}
\end{equation}
where we denoted $X_i=(n1i){\cap}(ij\star)$ and $X_j=(n1j){\cap}(ij\star)$.

The second consistent option we found is to pick $\star\in\{1,\dots,i{-}1\}$,\footnote{In fact, we can also choose $\star\in\{(j{+}1,\dots,n\}$, for which the inequalities of the last two ratios in eq.~(\ref{ratios2}) flip sign. For this option the geometric space is identical to the one defined by eq.~(\ref{ratios2}). As such, it is just a different representation of the same geometry.} together with the following signs for the ratios of four-brackets in the $d\log$ form eq.~(\ref{eq:dlog_chiral_symmetric})
\begin{equation}
\label{ratios2}
\hspace{-.5cm}
\left\{
\frac{\ab{ABi{-}1i}}{\ab{ABii{+}1}}{<}0,\,
\frac{\ab{ABj{-}1j}}{\ab{ABjj{+}1}}{<}0,\,
\frac{\ab{AB(n1i){\cap}(ij\star)}}{\ab{AB1n}}{>}0,\,
\frac{\ab{AB(n1j){\cap}(ij\star)}}{\ab{AB1n}}{>}0
\right\}\,.
\hspace{-.5cm}
\end{equation}
One can check that out of the four possibilities for the signs of individual four-brackets consistent with the ratios eq.~(\ref{ratios2}), only one region is actually non-empty, and we get
\begin{equation}
\label{pentsign2}
P^{(2)}_{ij} \leftrightarrow
\begin{tabular}{|c|c|c|c|c|c|c|}
\hline \small $\!\!\!\ab{ABi{-}1i}$\!\!\! & $\!\!\!\ab{ABii{+}1}\!\!\!$ & 
              $\!\!\!\ab{ABj{-}1j}\!\!\!$ & $\!\!\!\ab{ABjj{+}1}\!\!\!$ & 
              $\!\!\!\ab{AB1n}\!\!\!$     & $\!\!\!\ab{ABX_i}\!\!\!$    & $\!\!\!\ab{ABX_j}\!\!\!$\\ 
\hline
$-$ & + & + & $-$ & + & + & +\\ 
\hline
\end{tabular}
\end{equation}
When some legs of the pentagon become massless, there are degenerate configurations which allow more sign choices in the $d\log$ form eq.~(\ref{eq:dlog_chiral_symmetric}) than in the generic case. However, upon gluing different integrands, none of these choices lead to a global geometry which is free of spurious boundaries. We return to this point in greater detail in section \ref{sec:geometry_chiral_pentagons} as well as appendix \ref{app:2d_gluing_details}.

\subsection*{Faithful geometries from the box $d\log$ form}

A particular degeneration of the chiral pentagon in eq.~(\ref{pent}) leads to the two-mass-hard integral which arises as a special case, $j=i{+}1$. The $d\log$ form for the general box integral is similar to eq.~(\ref{box4}). In the context of the two-mass-hard box, it reads
\begin{align}
\label{2mh_boxform}
   \Omega_{ii{+}1}^{(n)} & = 
   \raisebox{-35pt}{\includegraphics[scale=.6]{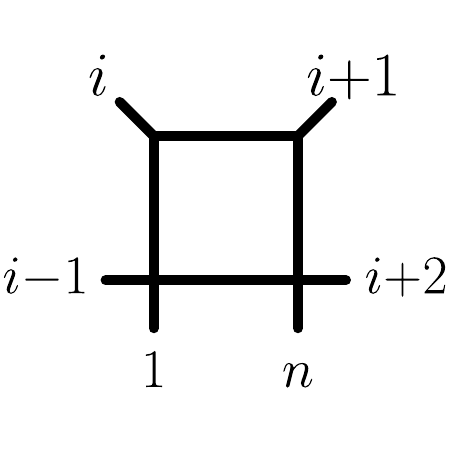}}
   = \frac{\ab{ABd^2A}\ab{ABd^2B}\ab{i{-}1ii{+}1i{+}2}\ab{1ii{+}1n}}
          {\ab{ABi{-}1i}\ab{ABii{+}1}\ab{ABi{+}1i{+}2}\ab{AB1n}}
          \\ 
   &=   d\log\frac{\ab{ABi{-}1i}}{\ab{ABX}}
        d\log\frac{\ab{ABii{+}1}}{\ab{ABX}}
        d\log\frac{\ab{ABi{+}1i{+}2}}{\ab{ABX}}
        d\log\frac{\ab{AB1n}}{\ab{ABX}}\,,
        \nonumber
\end{align}
where $X$ is one of the two solutions of the quadruple cut of the box,
\begin{align}
\hspace{-1cm}
  \raisebox{-32pt}{\includegraphics[scale=.5]{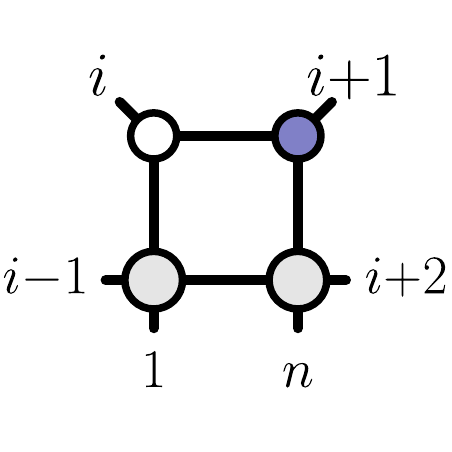}}
  \hspace{-.5cm}
  \leftrightarrow
  X_i=(ii{+}1i{+}2){\cap}(in1), 
 \quad \text{or} 
  \raisebox{-32pt}{\includegraphics[scale=.5]{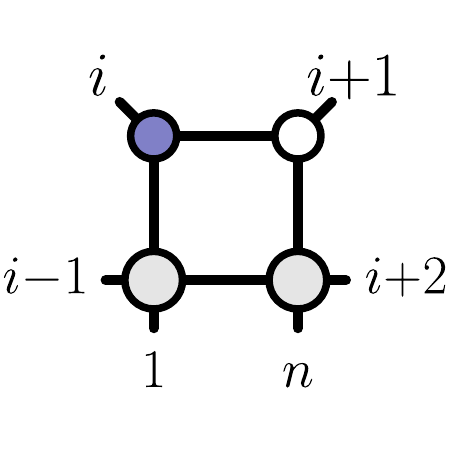}}
  \hspace{-.5cm}
  \leftrightarrow
 X_{i{+}1}=(i{-}1ii{+}1){\cap}(i{+}1n1)\,.
 \hspace{-.4cm}
\end{align}
We now repeat the same exercise as for the chiral pentagon and determine the consistent positive geometries associated to this $d\log$ form. As it turns out, fixing one of the signs of the brackets involving either $X_i$ or $X_{i+1}$ is sufficient to specify the space, as one inequality automatically enforces the other. Fixing $\ab{AB1n}>0$ then forces $\ab{ABX}$ to have a definite sign depending on the sign of the ratio in the last $d\log$. We find two different consistent sign choices for the ratios of the $d\log$s in eq.~(\ref{2mh_boxform}), both of which have fixed signs for the $\ab{ABii{+}1}$ brackets of the diagram. The first region is given by:
\begin{equation}
B^{(1)}_{1i{-}1,i{+}2\,n} \leftrightarrow
\begin{tabular}{|c|c|c|c|c|c|}
\hline \small 
      $\!\!\!\ab{ABi{-}1i}\!\!\!$       & $\!\!\!\ab{ABii{+}1}\!\!\!$ 
    & $\!\!\!\ab{ABi{+}1i{+}2}\!\!\!$   & $\!\!\!\ab{AB1n}\!\!\!$
    & $\!\!\!\ab{ABX_i}\!\!\!$          & $\!\!\!\ab{ABX_{i{+}1}}\!\!\!$ 
            \\ \hline
            $-$ & + & $-$ & + & + & $-$
            \\ \hline
\end{tabular} 
\label{2mh_boxsign1}
\end{equation}
Consistent with the statements above, let us reiterate that fixing the sign of only one of the brackets $\ab{ABX_i}$ or $\ab{ABX_{i{+}1}}$ together with the signs of the propagator brackets is sufficient to fix the region; the sign of the other bracket is implied by the rest. 

The second allowed region has the same signs as in eq.~(\ref{2mh_boxsign1}) for the first four brackets, but the signs for $\ab{ABX_i}<0$, $\ab{ABX_{i{+}1}}>0$ are flipped.  
\begin{equation}
B^{(2)}_{1i{-}1,i{+}2\,n} \leftrightarrow
\begin{tabular}{|c|c|c|c|c|c|}
\hline \small $\!\!\!\ab{ABi{-}1i}\!\!\!$       & $\!\!\!\ab{ABii{+}1}\!\!\!$ 
            & $\!\!\!\ab{ABi{+}1i{+}2}\!\!\!$   & $\!\!\!\ab{AB1n}\!\!\!$
            & $\!\!\!\ab{ABX_i}\!\!\!$          & $\!\!\!\ab{ABX_{i{+}1}}\!\!\!$ 
            \\ \hline
            $-$ & + & $-$ & + & $-$ & +
            \\ \hline
\end{tabular} 
\label{2mh_boxsign2}
\end{equation}
Note that the union of the two regions eq.~(\ref{2mh_boxsign1}) and eq.~(\ref{2mh_boxsign2}) constitutes a larger achiral geometric region defined only by the first four inequalities with no constraints on $\ab{ABX_i}$ and $\ab{ABX_{i{+}1}}$. The canonical form for this achiral region trivially vanishes because four inequalities are insufficient to produce a non-trivial $d\log$ form (simply because we cannot form four independent ratios). Therefore, this achiral space is a particular example of a zero-form space defined in subsection \ref{subsec:pos_geometry}. The larger space can be sliced into two subspaces by fixing the sign of $\ab{ABX_i}$, corresponding to exactly the two chiral subspaces we found in  eq.~(\ref{2mh_boxsign1}) and eq.~(\ref{2mh_boxsign2}). Because the form for the achiral space vanishes, both subspaces have the same form (up to a sign). We can once again illustrate this feature of zero-form spaces with a simple two-dimensional example in the $xy$-plane. The space defined by $-\infty<x<\infty,0<y<b$ has zero canonical form, so cutting the space into two pieces with $x\lessgtr a$, the respective canonical forms $\Omega_<,\Omega_>$ differ only by a sign:
\begin{align}
    \twhite{.}& \nonumber \\[-30pt]
    \hspace{-.8cm}
    \raisebox{-44pt}{\includegraphics[scale=.7]{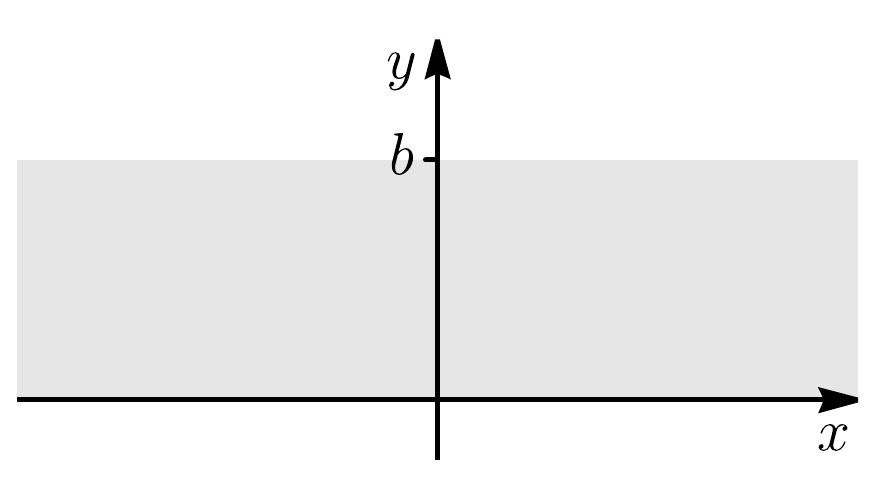}} 
    \ \ 
    &
    \underset{\text{slicing}}{\longrightarrow}
    \ \ 
    \raisebox{-44pt}{\includegraphics[scale=.7]{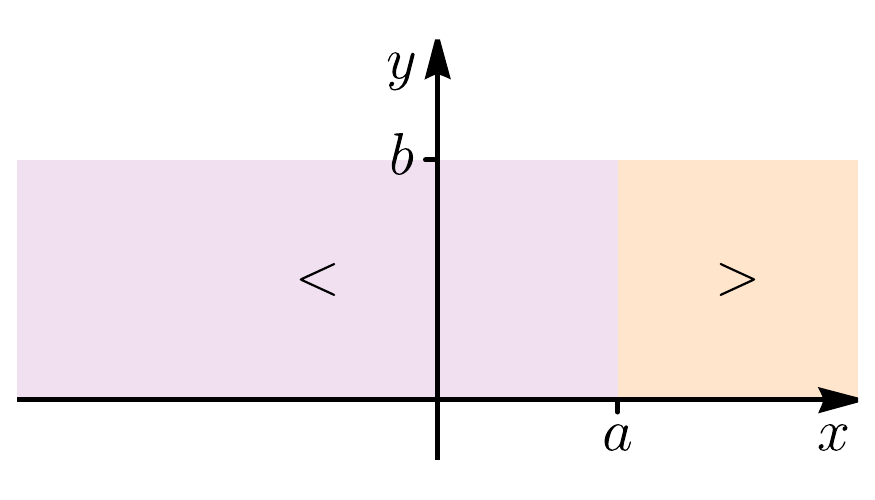}} \\[-10pt]
\Omega =0 
 \twhite{dddddddddddddi}& 
 \longrightarrow
\ 
\Omega_{<} + \Omega_{>} 
= \frac{dx \, dy}{(x{-}a) y (y{-}b)}{-}\frac{dx \, dy}{(x{-}a) y (y{-}b)} \nonumber
\end{align}
We can also use an alternative $d\log$ form representation for the box integrand derived directly from the degeneration of the chiral pentagon eq.~(\ref{eq:dlog_chiral_symmetric}) by setting $j=i{+}1$. For this leg configuration, the integrand form eq.~(\ref{chiral}) na\"{i}vely generates a double pole $\ab{ABii{+}1}$ in the denominator. However, the chiral numerator exactly cancels one power of this pole so that we end up with a two-mass-hard box. The resulting $d\log$ form obtained by this procedure has different arguments than eq.~(\ref{2mh_boxform}), but reproduces the same rational integrand. This is in complete agreement with our earlier statements about the non-uniqueness of $d\log$ representations. 

If we repeat the exercise of subsection~\ref{subsec:dlog_form_to_geometry} to associate consistent geometric regions with the appropriate boundary structure to the $d\log$ form, we find exactly the same geometric regions as in eq.~(\ref{2mh_boxsign1}) and eq.~(\ref{2mh_boxsign2}), albeit described by different inequalities. The brackets involving the lines $X_i=(n1i){\cap}(ii{+}1\star)$ and $X_{i{+}1} = (n1i{+}1){\cap}(ii{+}1\star)$ inherited from the pentagon simply provide an equivalent description of the same spaces. Once again, this simply reflects the non-uniqueness of the $d\log$ form of an integrand, and emphasizes that demanding we get the \emph{geometry} correct (rather than just the form) is a very strong constraint. More generally, we expect (but have no proof) that alternative $d\log$ forms for the local integrals discussed in this section will not yield any new consistent geometries.

In addition to the two-mass hard box discussed above, the only other box topology relevant for the local expansion eq.~(\ref{pent}) are ``one-mass'' boxes. The first of these arises as a boundary term when $i=n{-}2,j=n{-}1$. The second one-mass box is a degeneration of the pentagon when $i=2,j=3$, and can be obtained by trivial relabelling.
\begin{align}
    B_{1\,n{-}3} = 
    \raisebox{-42pt}{\includegraphics[scale=.6]{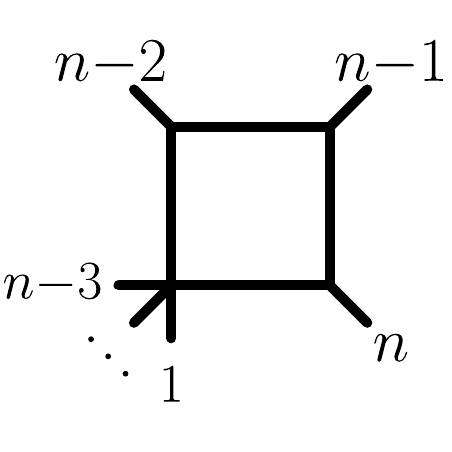}}\,,
      \label{one_mass_left}
    \qquad   
    B_{4n} =  
    \raisebox{-38pt}{\includegraphics[scale=.55]{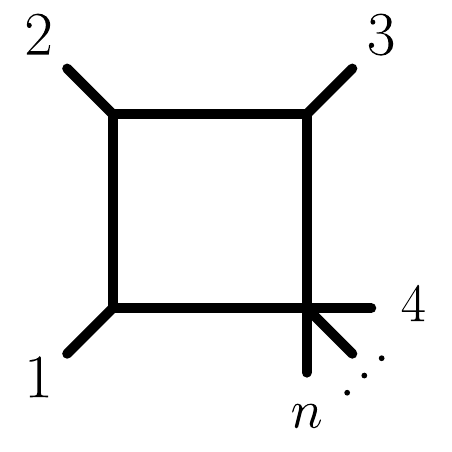}}. 
\end{align}
The one-mass geometries with the correct boundary structure can be specified by imposing conditions on the propagators appearing in the diagram, as well as one bracket involving (one of) the solutions to the quadruple cut. Thus, for e.g.,~$B_{1\,n{-}3}$ in eq.~(\ref{one_mass_left}) we can define the space by correctly choosing the signs of
\begin{equation}
\label{eq:signList}
    \{
    \langle ABn{-}3n{-}2\rangle,
    \langle ABn{-}2n{-}1\rangle,
    \langle ABn{-}1n\rangle,
    \langle AB1n\rangle\} 
    \quad\text{and}\quad 
    \langle ABn{-}2n\rangle.
\end{equation}
In the two-mass hard case there was only one choice of signs for the list of propagators i.e., eq.~(\ref{sign6}) and eq.~(\ref{2mh_boxsign2}) differed only in the signs of the brackets $\langle ABX_i\rangle$ and $\langle ABX_{i+1}\rangle$. In the one-mass case, both sign choices for the bracket $\langle ABn{-}2n\rangle$ are also allowed. In addition, there are two allowed choices for the signs of the sequence of brackets in eq.~(\ref{eq:signList}) which are geometrically free of spurious boundaries---giving a total of \emph{four} consistent spaces. The first two can be written compactly as
\begin{equation}
B^{(1,2)}_{1n{-}3} \leftrightarrow
\begin{tabular}{|c|c|c|c|c|}
\hline      $\!\!\!\ab{ABn{-}3n{-}2}\!\!\!$   & $\!\!\!\ab{ABn{-}2n{-}1}\!\!\!$ 
            & $\!\!\!\ab{ABn{-}1n}\!\!\!$       & $\!\!\!\ab{AB1n}\!\!\!$
            & $\!\!\!\ab{ABn{-}2n}\!\!\!$ 
            \\ \hline
            $+$ & $-$ & $-$ & $+$ & $\pm$
            \\ \hline
\end{tabular} \,,
\label{1m_boxsign12}
\end{equation}
while the second pair of solutions is 
\begin{equation}
B^{(3,4)}_{4\,n} \leftrightarrow
\begin{tabular}{|c|c|c|c|c|}
\hline      $\!\!\!\ab{ABn{-}3n{-}2}\!\!\!$   & $\!\!\!\ab{ABn{-}2n{-}1}\!\!\!$ 
            & $\!\!\!\ab{ABn{-}1n}\!\!\!$       & $\!\!\!\ab{AB1n}\!\!\!$
            & $\!\!\!\ab{ABn{-}2n}\!\!\!$ 
            \\ \hline
            $-$ & $+$ & $-$ & $+$ & $\pm$
            \\ \hline
\end{tabular} \,.
\label{1m_boxsign34}
\end{equation}
In subsections~\ref{subsec:gluing_regions} and \ref{subsec:Amplituhedron_prime} we consider the problem of \emph{combining} the geometries associated to individual terms in the chiral pentagon expansion of the MHV one-loop integrand eq.~(\ref{pent}). A priori, we have no reason to prefer any one of the individually well-defined spaces we have just constructed. Remarkably, however, as we shall demonstrate in section \ref{sec:geometry_chiral_pentagons}, it turns out that demanding a sensible global geometry whose boundary structure is identical to that of the original Amplituhedron is restrictive enough to give a \emph{unique} choice for the one-mass, the two-mass hard, as well as the chiral pentagon spaces subject to the assumption that we treat all graph isomorphic topologies in a uniform way.

\subsection*{Faithful geometries from the chiral hexagon $d\log$ form}
%
In this subsection, we consider the problem of assigning faithful geometries to logarithmic forms more generally. The first example which goes beyond the results of the previous two subsections is the chiral hexagon,
\begin{equation}
\label{wavyHex}
\Omega_{\text{hex}}=
 \raisebox{-52pt}{\includegraphics[scale=.45]{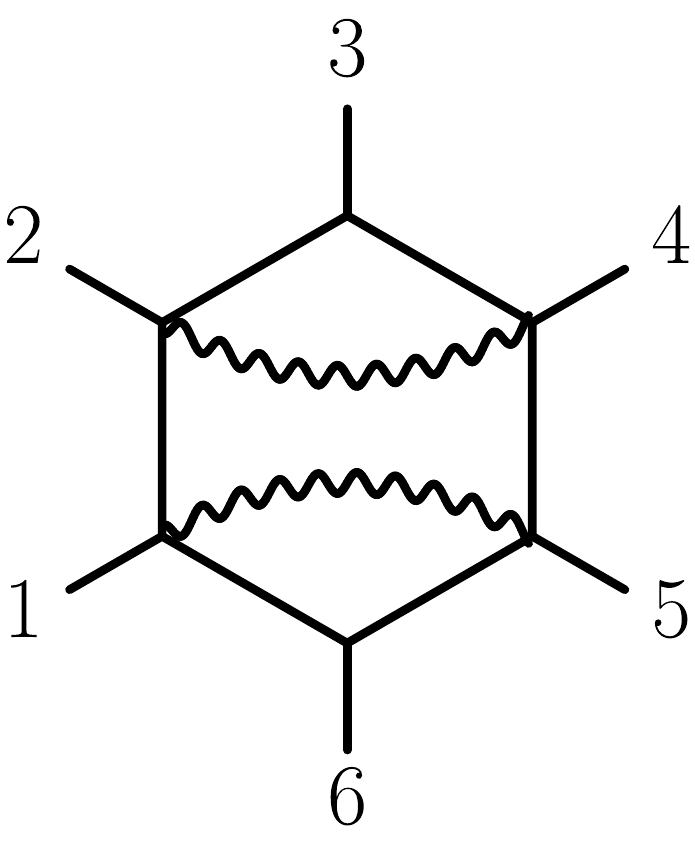}}
   =\frac{\ab{ABd^2A}\ab{ABd^2B}\,\ab{AB\overline{24}}\ab{AB\overline{51}}}{\ab{AB12}\ab{AB23}\ab{AB34}\ab{AB45}\ab{AB56}\ab{AB16}}.
\end{equation}
To identify some candidate geometries the most natural starting point is, as we have seen, a rewriting of this integrand as a single $d\log$ form. For integrals with more than five local poles, any such form must involve at least one ratio in which both the numerator and denominator depend quadratically on $(AB)$. For the hexagon of eq.~(\ref{wavyHex}) our starting point is a novel expression which is a single $d\log$ form,
\begin{equation}
\label{wavyHexDLog}
\Omega_{\text{hex}}
   =d\log\frac{\ab{AB13}}{\ab{AB23}}
   d\log\frac{\ab{AB34}}{\ab{AB45}}
   d\log\frac{\ab{AB46}}{\ab{AB56}}
   d\log\frac{\ab{AB23}\ab{AB16}}{\ab{AB12}\ab{AB36}}.
\end{equation}
Demanding that the geometry defined by sign conditions on the four ratios in this expression be faithful requires that the three codimension-one loci $\ab{AB13}{=}0$, $\ab{AB46}{=}0$ and $\ab{AB36}{=}0$ do not appear as boundaries of the space. There are two sign choices which accomplish this:
\begin{align}
H^{(1)}=&\left\{\frac{\ab{AB13}}{\ab{AB23}}<0,\frac{\ab{AB34}}{\ab{AB45}}<0,\frac{\ab{AB46}}{\ab{AB56}}<0,\frac{\ab{AB23}\ab{AB16}}{\ab{AB12}\ab{AB36}}>0\right\},\\
H^{(2)}=&\left\{\frac{\ab{AB13}}{\ab{AB23}}<0,\frac{\ab{AB34}}{\ab{AB45}}<0,\frac{\ab{AB46}}{\ab{AB56}}>0,\frac{\ab{AB23}\ab{AB16}}{\ab{AB12}\ab{AB36}}>0\right\}.
\end{align}
Decomposing $H^{(i)}$ into subspaces where all brackets appearing in the ratios have fixed sign, we na\"{i}vely generate a considerable number of non-overlapping regions. Although this is indeed the case for $H^{(2)}$, the first solution is surprisingly equivalent to a single region described by significantly simpler inequalities
\begin{align}
\begin{split}
H^{(1)}=\Big\{
&\ab{AB12}<0,\ab{AB23}>0,\ab{AB34}>0,\ab{AB45}<0, \\
& \ab{AB56}>0,\ab{AB16}>0,\ab{AB14}<0
\Big\},
\end{split}
\end{align}
which will arise in a different context in section~\ref{sec:sign_flip_regions} below. 

\subsection{No-go theorem for external triangulation}
\label{subsec:no-go-local-triangulation}

In all previous cases, assigning a sensible (faithful) geometric space to a given $d\log$ form led us to consider situations where at least a subset of signs of $\ab{ABii{+}1}$ were negative. In particular, for all faithful geometries of the local integrands, we did not encounter any region for which \emph{all} such brackets were positive. On the other hand, the Amplituhedron itself is defined with all $\ab{ABii{+}1}>0$, and is cut further either by the sign-flip conditions in eq.~(\ref{eq:zIneq}) or by the positivity of $\ab{AB\overline{ij}}>0$ in eq.~(\ref{eq:mhv_alt}). This means that all geometric regions we discussed in the context of the chiral pentagon expansion are \emph{outside} the Amplituhedron. 

It is actually very easy to see that even if there was some other $d\log$ representation for the pentagon or boxes, with correspondingly different geometric regions, they can not possibly be even partially inside the Amplituhedron space for the following reason: In order to fix, e.g., the MHV (or $\MHVbar$) Amplituhedron space, we need to specify $n-3$ conditions in addition to the positivity of all $\ab{ABii+1}>0$. While there are many equivalent ways how to express these conditions, e.g., via sign flips in the sequence $\{\ab{AB1i}\}_{i={2,\ldots,n}}$, or via sign flips in the sequence $\{\ab{AB2i}\}_{i={3,\ldots,n,1}}$, or via the positivity of $\ab{AB\overline{ij}}>0$, we always need at least $n-3$ of them. This number is irreducible and cannot be decreased. If we attempted to specify fewer conditions, the resulting space would lie inside the Amplituhedron, alas it would contain spurious boundaries. One example of this scenario is given by the individual sign-flip spaces associated to the BCFW Kermits eq.~(\ref{eq:Kermit}). The unique space with only physical boundaries inside the Amplituhedron is the whole Amplituhedron itself, and can not be cut into smaller spaces with purely physical boundaries. 

A natural follow-up question is whether or not the Amplituhedron can be contained inside a space whose logarithmic form is given by the chiral pentagon or the box. While the logarithmic form for the Amplituhedron is generally very complicated, combining the Amplituhedron space with many other spaces outside can lead to a simpler space whose form can be as simple as that of the chiral pentagon or the box. In our two-dimensional geometry toy example in subsection \ref{subsec:positivity_dual_Amplituhedron}, this is precisely what happened in the external triangulation of the quadrilateral on the left-hand-side of eq.~(\ref{toy_dual_internal_triangulation}). There, adding a triangle outside the space gave rise to a larger triangle, for which the logarithmic form would be simpler than the one for the quadrilateral itself. So the question is whether or not we can replicate something similar for the Amplituhedron.

For the simplest five-point case, the answer is yes. For $n>5$, however, this is no longer true and there is \emph{no} consistent local space based on the pentagon $d\log$ form which \emph{contains} the Amplituhedron space. This can be understood heuristically by noting that in the $n$-point case, to define pentagon geometries that contain the amplitude we would need $n-3$ extra conditions, analogous to the intersections appearing in eq.~(\ref{eq:mhv_alt}). However, the $d\log$ form of the pentagon contains only four ratios of brackets, indicating that the set of conditions defining the pentagon should not grow with $n$. Thus, without resorting to increasingly complicated ratios of brackets at higher multiplicities a simple pentagon space cannot possibly contain the complicated Amplituhedron. The $n{=}5$ case is exceptional, as the two conditions required to cut out the Amplituhedron matches the number of conditions required to define the pentagon.

This argument suggests the chiral pentagon expansion cannot represent a geometric triangulation (internal or external) of the Amplituhedron. As we will see later in section \ref{sec:geometry_chiral_pentagons}, the chiral pentagons triangulate a different region which has only physical boundaries and has (almost) all the same properties as the Amplituhedron. However, our suspicion, for which we give evidence in section~\ref{sec:triangulation_dual_Amplituhedron}, is that the primary purpose of the pentagons is to internally triangulate the dual Amplituhedron.  

\section{Sign-flip regions}
\label{sec:sign_flip_regions}

In this section we take a step back from the chiral pentagon expansion, and look more generally at the positive geometries which arose in the study of the local integrands, such as eqs.~(\ref{pentsign1}), (\ref{pentsign2}), (\ref{2mh_boxsign1}), (\ref{2mh_boxsign2}), (\ref{1m_boxsign12}) and (\ref{1m_boxsign34}). In particular, we constructed spaces with only physical boundaries that were defined by both positive and negative signs of various $\ab{ABii{+}1}$ brackets. In \cite{Herrmann:2020oud} we provided an intriguing classification of these geometries using certain sign-flip conditions which are reminiscent of, but distinct from, the sign-flip characterization of the Amplituhedron \cite{Arkani-Hamed:2017vfh}. In this section we will review these results and provide further discussions and elaborate on the properties of the various sign-flip spaces.

\subsection*{Classification of sign-flip regions}

In the first step we discuss achiral positive spaces that are defined by imposing fixed signs for $\ab{ABii{+}1}$ brackets only, without further constraints on any other four-bracket. In other words, we study geometric regions defined by the following set of $n$ signs for the sequence of brackets
\begin{equation}
\label{seq}
S{=}\big\{\ab{ ABii{+}1} \big\}_{i\in\{1,\ldots,n\}} 
 {:=}  
 \big\{\ab{AB12},\ab{AB23},\ldots,\ab{AB1n}\big\}.
\end{equation}
From the study of the Amplituhedron, we already know of one example of such a region: the union of the MHV and $\MHVbar$ one-loop spaces.
The corresponding logarithmic form is the parity-odd amplitude, which is given by the sum of MHV and $\MHVbar$ one-loop amplitudes in eq.~(\ref{odd_pent}) (defined with appropriate sign). This function integrates to zero on the parity-invariant Feynman contour. However, from a geometric point of view, it is the most natural space we can consider where all signs in $S$ are positive,
\begin{equation}
    S^{(0)} = \{+,+,\dots,+\}.
\end{equation}
For obvious reasons, we also call $S^{(0)}$ a \emph{sign-flip-zero region} because the sequence $S$ has no sign flips.  By drawing the $n$ points on a circle, we can represent the sign-flip-zero space as
\begin{equation}
 S^{(0)} = \raisebox{-45pt}{\includegraphics[scale=.4]{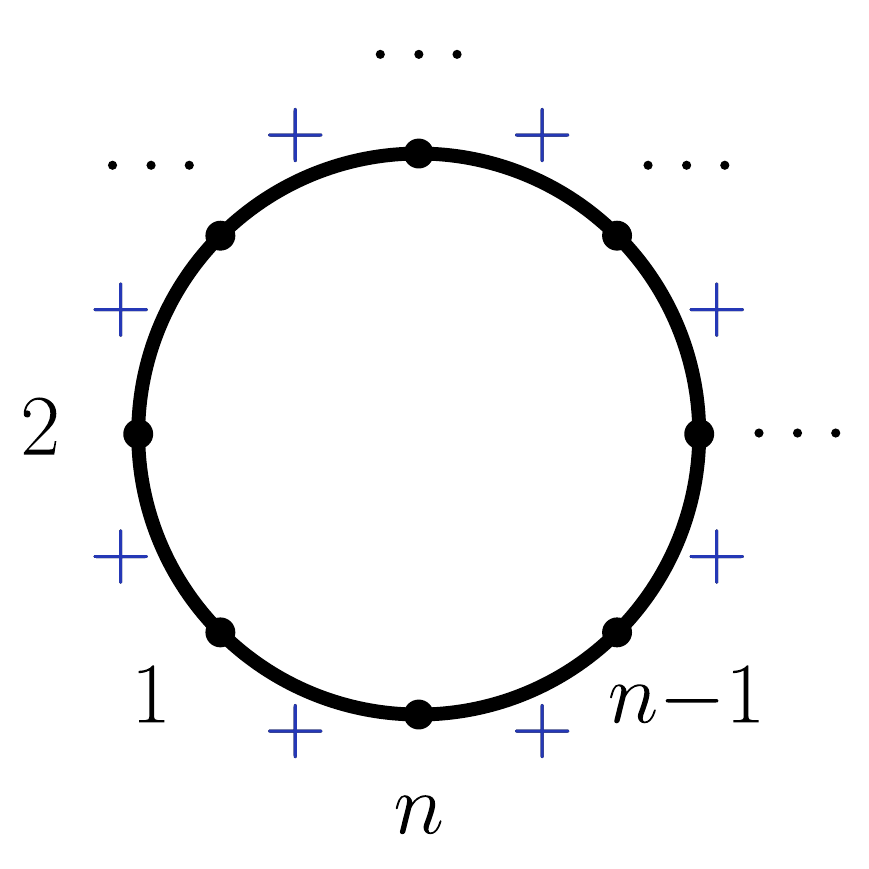}}.
\end{equation}
where the $+$ sign between points 1 and 2 denotes $\ab{AB12}>0$, and similarly for the other brackets. The only subtlety arises when we reach the arc from $n$ to 1, where we draw a $+$ sign to denote the positivity of $\ab{ABn\hat{1}}=\ab{AB1n}$ in line with the twisted cyclic symmetry. 

Following the same logic, we define a \emph{sign-flip-two region} as sequence of signs
\begin{equation}
\label{eq:sf2}
    S^{(2)}_{ij} 
    =\,\, 
\raisebox{-45pt}{\includegraphics[scale=.4]{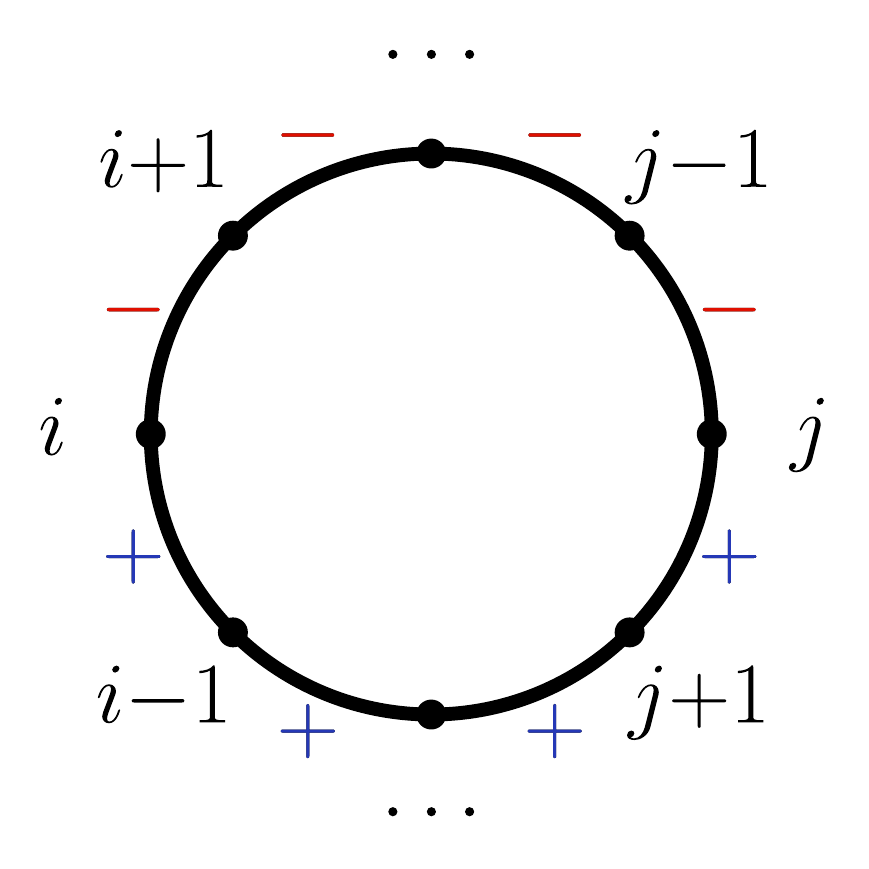}},
\end{equation}
where the labels $i$ and $j$ denote the two positions where the first and second sign flip occurs, respectively. 
In particular, this implies that 
\begin{align}
\label{eq:sf2_bracket_signs}
\begin{split}
    &\{\ab{ABi\,i{+}1},\ab{ABi{+}1\,i{+}2},\ldots, \ab{ABj{-}1j} < 0\}, \\
    &\{\ab{ABjj{+}1},\ab{ABj{+}1j{+}2},\ldots, \ab{ABi{-}1i} > 0\},
\end{split} 
\end{align}
where the labels 1 and $n$ can be in either the positive or negative regions\footnote{As written, eq.~(\ref{eq:sf2}) and eq.~(\ref{eq:sf2_bracket_signs}) suggest that $i<j$ and therefore $\ab{AB1n}>0$ is in the positive region. In general, spaces $\ab{AB1n}<0$ presents no additional complications whatsoever, although we will have no use for them in this paper.} (appropriately taking into account the twisted cyclic symmetry as commented on above).

Next, we define a \emph{sign-flip-four region} with $i,k,\ell,j$ labelling the four positions where the signs in the sequence eq.~(\ref{seq}) flip, 
\begin{align}
\label{eq:sf4}
    S^{(4)}_{ik \ell j} 
    &=\,\,\,\,\,\, \raisebox{-52pt}{\includegraphics[scale=.5]{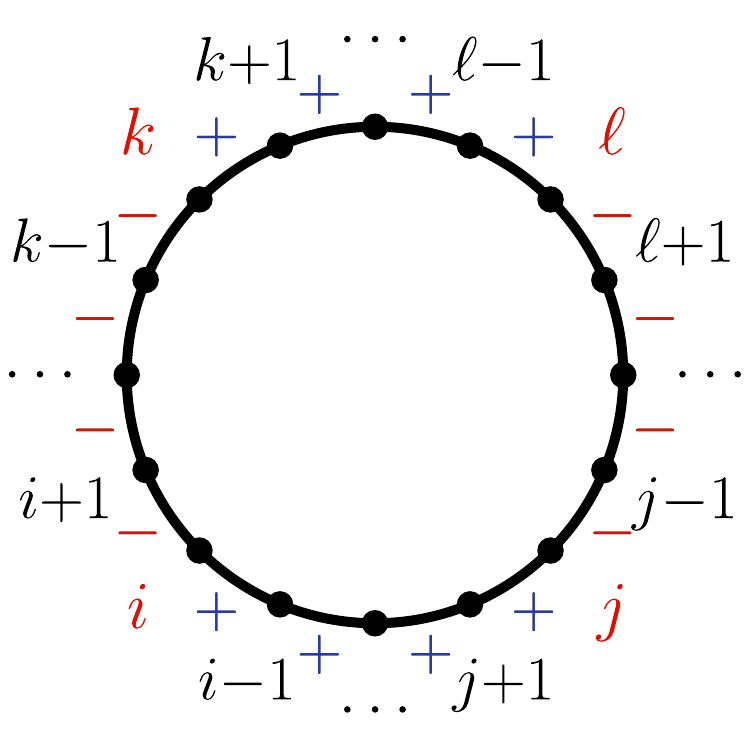}}.
\end{align}
Na\"{i}vely, we can continue to consider sequences of brackets eq.~(\ref{seq}) with ever more sign flips. Remarkably, all higher sign-flip patterns correspond to \emph{empty} geometric regions. 

\subsection*{Sign-flip-zero regions}
%
All the sign-flip-zero, two, and four spaces are positive geometries; as such, all of these spaces have associated canonical forms with only logarithmic singularities on all boundaries. By definition, all codimension-one boundaries of the achiral spaces defined in the previous subsection have to be of the form $\ab{ABii{+}1}=0$. Na\"{i}vely, one may expect that, at $n$ points, all such inverse propagators $\ab{AB12},\ldots,\ab{AB1n}$ will indeed be boundaries of the spaces leading to very complicated canonical forms. For the sign-flip-zero space, this is true: the associated form has $n$ poles 
\begin{equation}
\label{eq:sf0_achiral_form_abstract}
S^{(0)} = 
\raisebox{-45pt}{\includegraphics[scale=.4]{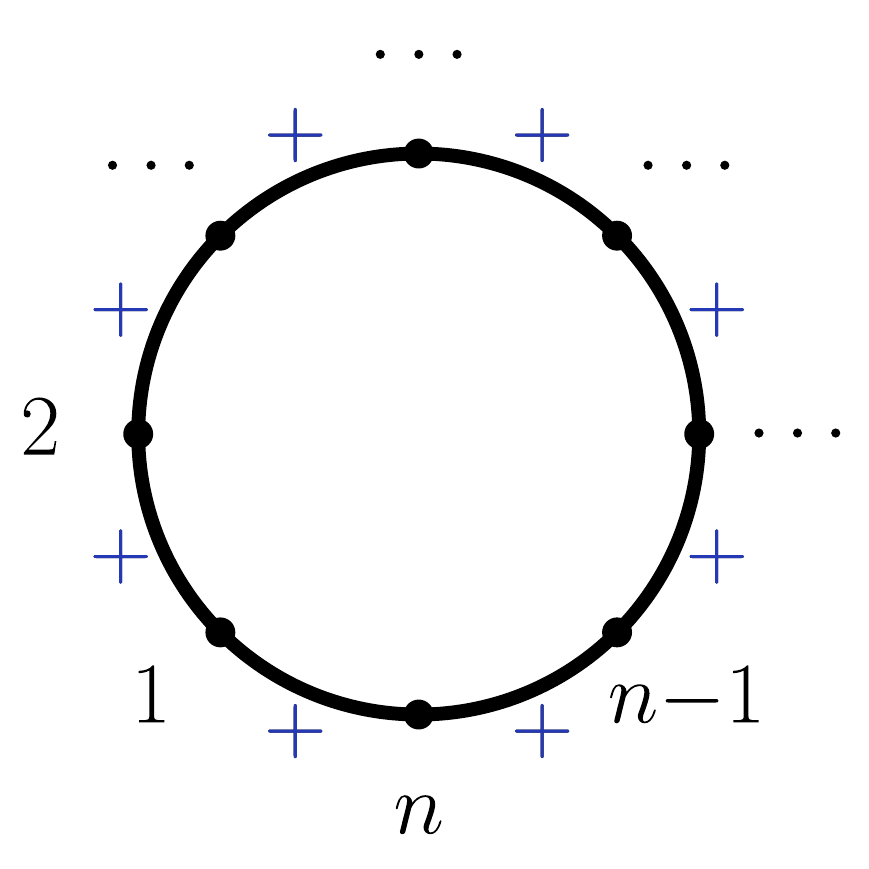}}
\leftrightarrow 
\frac{N^{(0)}}{\ab{AB12}\ab{AB23}\cdots\ab{ABn{-}1n}\ab{AB1n}}.
\end{equation}
This means the complexity of the numerator of the canonical form grows with $n$, just as the complexity of MHV and $\MHVbar$ amplitudes does. This should not be surprising, because both of these chiral amplitude spaces in fact live inside the larger achiral space $S^{(0)}$! The MHV amplitude is the subspace of the achiral space defined by the additional conditions $(i)$ $\ab{ AB\overline{ij}}>0$, eq.~(\ref{eq:mhv_alt}), while the parity conjugate $\MHVbar$ space uses $(ii)$ $\ab{ABij}>0$, eq.~(\ref{eq:mhvBar_alt}). It is quite nontrivial that imposing either set of these conditions does not introduce new codimension-one boundaries. In fact, it is straightforward to verify there is no way to impose a mixed set of conditions of type $(i)$ and $(ii)$, without at least one of these conditions becoming a boundary. Thus, we may think of the achiral space as having two components, neither of which has any spurious boundaries; for obvious reasons, we refer to these as \emph{chiral} components.  

In other words, we can cut the sign-flip-zero space defined by $\ab{ABii{+}1}>0$ into two chiral components which are MHV and $\MHVbar$ Amplituhedra. Both of these spaces have only physical boundaries. As alluded to in subsection~\ref{subsec:no-go-local-triangulation}, we need to impose $n{-}3$ conditions to specify either one of these spaces. 

Interestingly, the achiral sign-flip zero space can be externally triangulated in terms of simpler spaces. In fact, the ``parity-odd" pentagon expansion in eq.~(\ref{odd_pent}) exactly gives this geometric triangulation:
\begin{align}
\label{eq:sf0_odd_pentagon_triangulation_circles_main}
    \raisebox{-45pt}{\includegraphics[scale=.4]{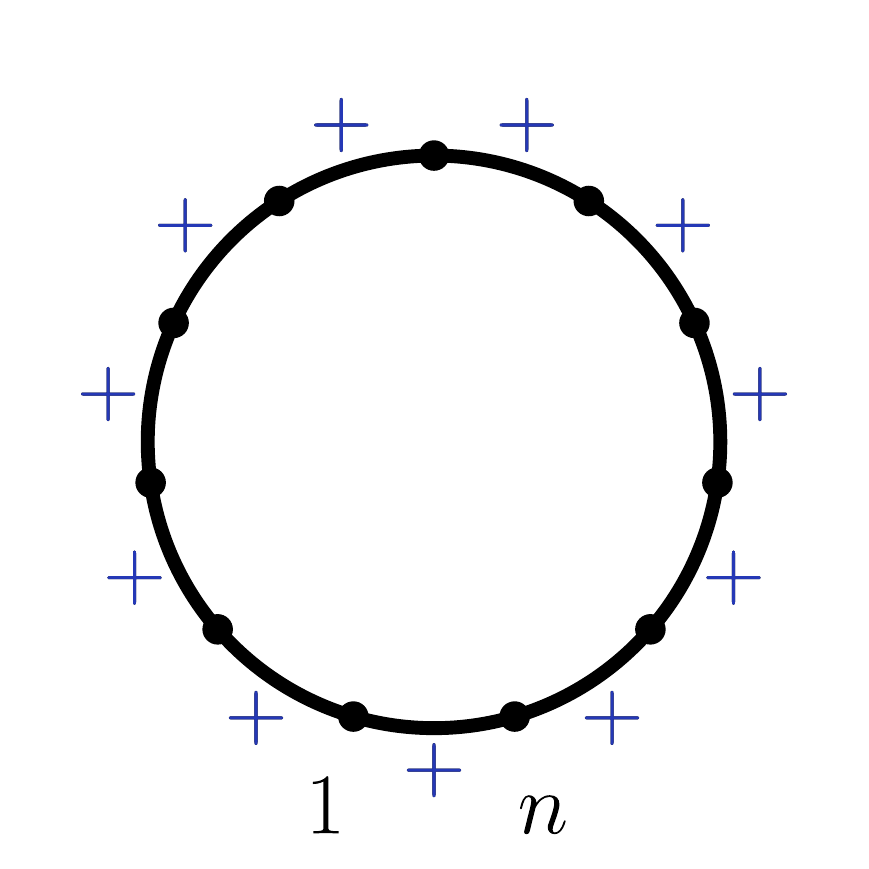}}
    =
    \raisebox{-45pt}{\includegraphics[scale=.4]{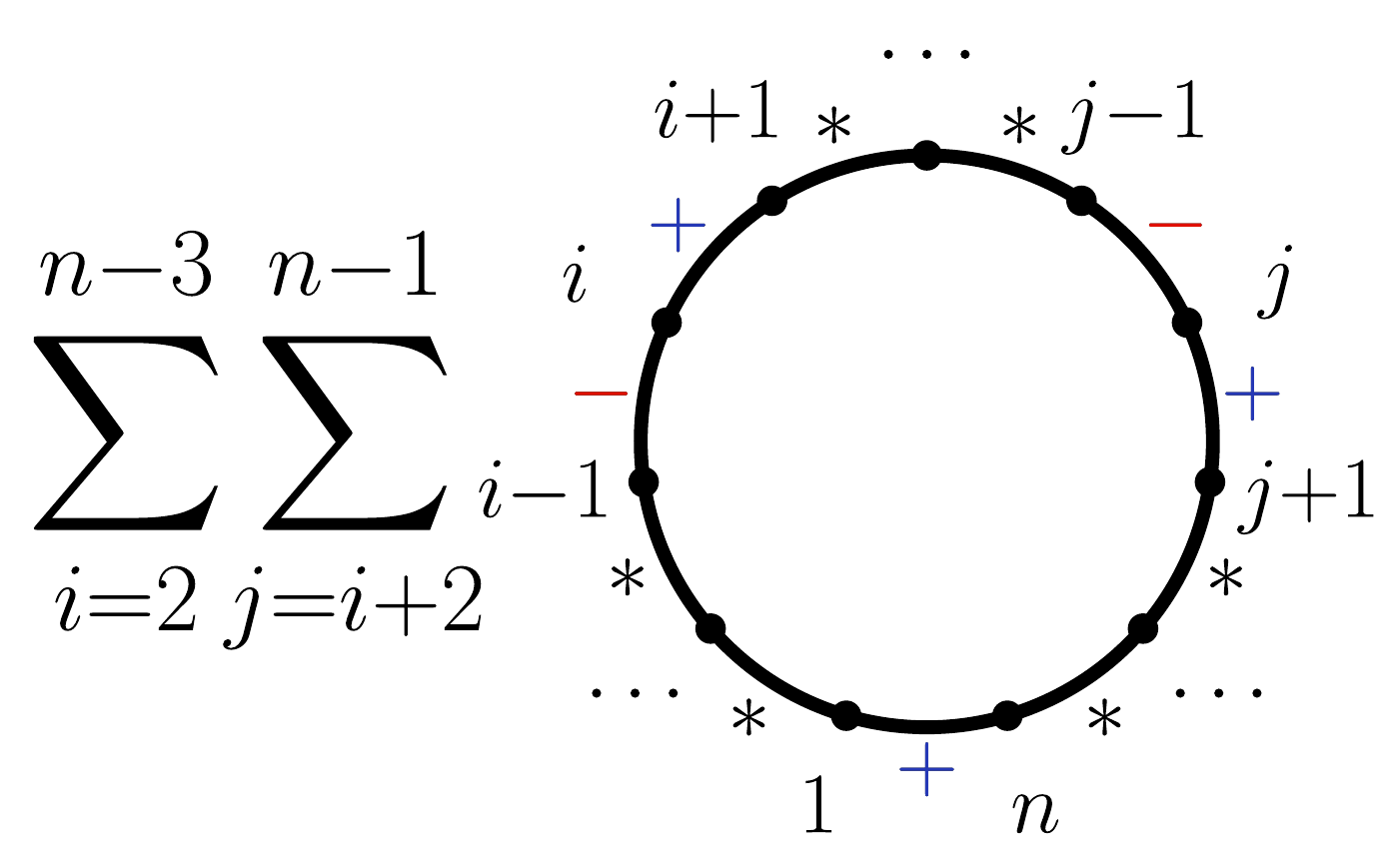}}.
\end{align}
Each parity-odd pentagon corresponds to a positive geometry with only five boundaries. The terms in the set are overlapping and provide an external triangulation of $S^{(0)}$. On the right-hand-side of eq.~(\ref{eq:sf0_odd_pentagon_triangulation_circles_main}), $\ast=+\oplus-$ instructs us to marginalize over both signs of the corresponding bracket. More details, including the derivation of (\ref{eq:sf0_odd_pentagon_triangulation_circles_main}) are given in appendix \ref{sec:external_triangulations}. As argued above, the same procedure does not work for chiral spaces, and the chiral pentagon expansion does not provide an external triangulation of the Amplituhedron. We will get back to the precise r\^{o}le of this expansion in the next section.

\subsection*{Sign-flip-two regions}

Let us continue our discussion with the achiral sign-flip-two regions defined in eq.~(\ref{eq:sf2}) and eq.~(\ref{eq:sf2_bracket_signs}). First, one can check that this space has all $n$ codimension-one boundaries so that the logarithmic form a priori has a similar structure as the one of $S^{(0)}$ in eq.~(\ref{eq:sf0_achiral_form_abstract}), but with a different numerator. 

However, something surprising happens when we slice the achiral sign-flip-two region eq.~(\ref{eq:sf2}) into smaller components. First, similar to the sign-flip-zero case we find that the sign-flip-two region can be again cut into two chiral components without introducing spurious boundaries. Na\"{i}vely, we would expect that in order to specify each subspace we have to impose ${\cal O}(n)$ additional inequalities, but in fact only a single inequality suffices; for the sign-flip-two space eq.~(\ref{eq:sf2}) the chiral components correspond to the two signs of the bracket $\ab{ABij}\gtrless0$, which we represent diagrammatically as
\begin{equation}
\label{sf2_chiral}
S^{(2),\pm}_{ij}=
\raisebox{-45pt}{\includegraphics[scale=.4]{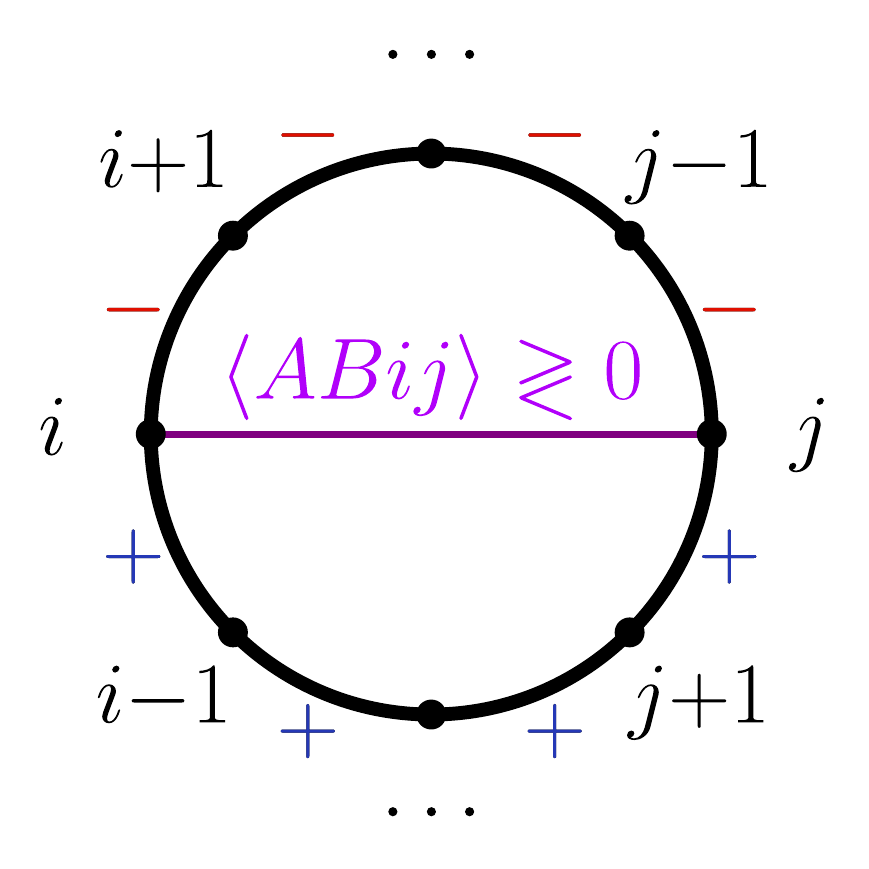}}\,, \qquad \text{for } i<j\,.
\end{equation}
This additional inequality is very natural as $i$ and $j$ are the positions where the two sign-flips occur. 

Looking more closely at one of the subspaces defined by $\ab{ABij}{>}0$, we find two interesting features: First, all brackets inside the ``plus region" have fixed positive sign,
\begin{equation}
\la AB pq\ra >0\,\,\,\mbox{and}\,\,\,
\la AB\overline{pq}\ra >0,
\end{equation}
for $p<q$ and $p,q\in \{j,j{+}1,\dots,i{-}1,i\}$. No similar statement is true for the indices inside the ``minus region." Second, from the collection of terms $\la ABpq\ra$ inside the plus region \emph{only} the boundary terms $\ab{ABjj{+}1}$ and $\ab{ABi{-}1i}$ represent boundaries. In other words, $\la ABpp{+}1\ra$ for $p\in(j{+}1,\dots,i{-}2)$ are not poles of the logarithmic form. Therefore, the canonical form for this chiral region is considerably simpler, 
\begin{equation}
\label{sf2form}
S^{(2),+}_{ij}{=}
\raisebox{-45pt}{\includegraphics[scale=.4]{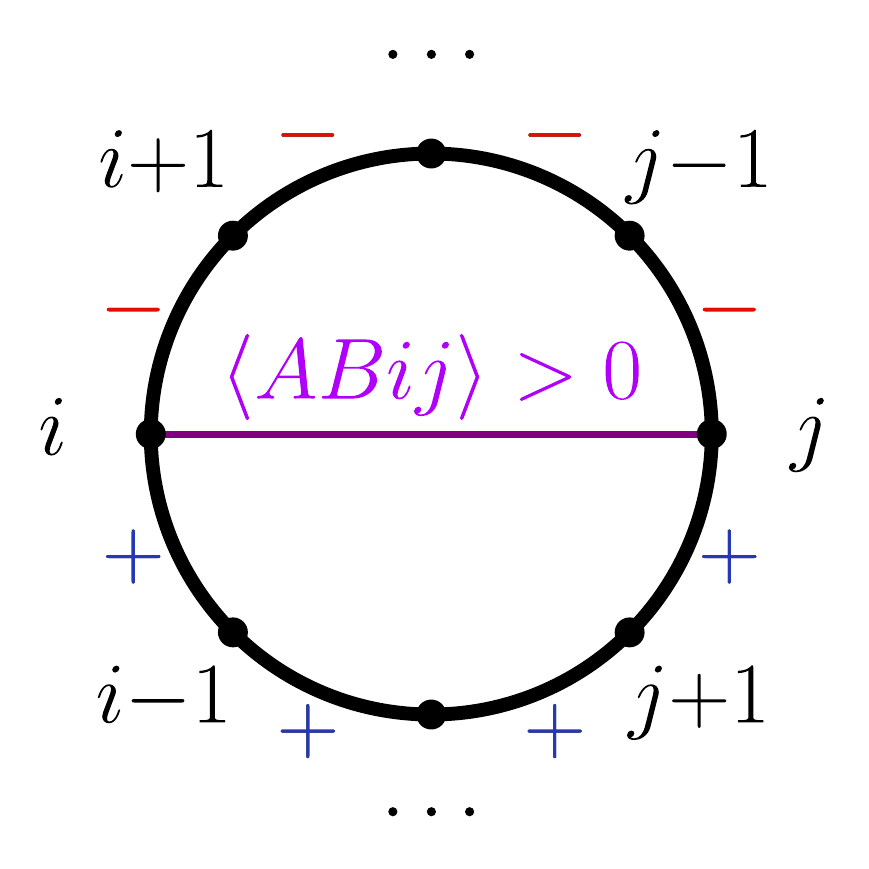}} 
{\leftrightarrow}
\frac{N^{(2),+}_{ij}}{\la ABii{-}1\ra \ab{ABii{+}1} {\cdots} \ab{ABj{-}1j}\ab{ABjj{+}1}}.
\end{equation}
In the general $n$-point case the form eq.~(\ref{sf2form}) is still non-trivial and we will give a precise formula and its derivation in eq.~(\ref{eq:sf2_general_local_int_form_rep}) of appendix \ref{sec:external_triangulations}. For some special cases, the structure of the form simplifies considerably. In particular, if the `$-$' region shrinks, the number of poles decreases, as does the complexity of the form. For the special case where $j=i{+}2$ there are only two negative brackets, $\ab{ABii{+}1},\ab{ABi{+}1i{+}2}<0$, which means that the chiral sign-flip-two space has only four boundaries and the logarithmic form must be a box. 
\begin{align}
\label{eq:sf2_2minus_signs_npt_special_eg}
\begin{split}
 S^{(2),+}_{i\, i{+}2} = 
 \raisebox{-45pt}{\includegraphics[scale=.4]{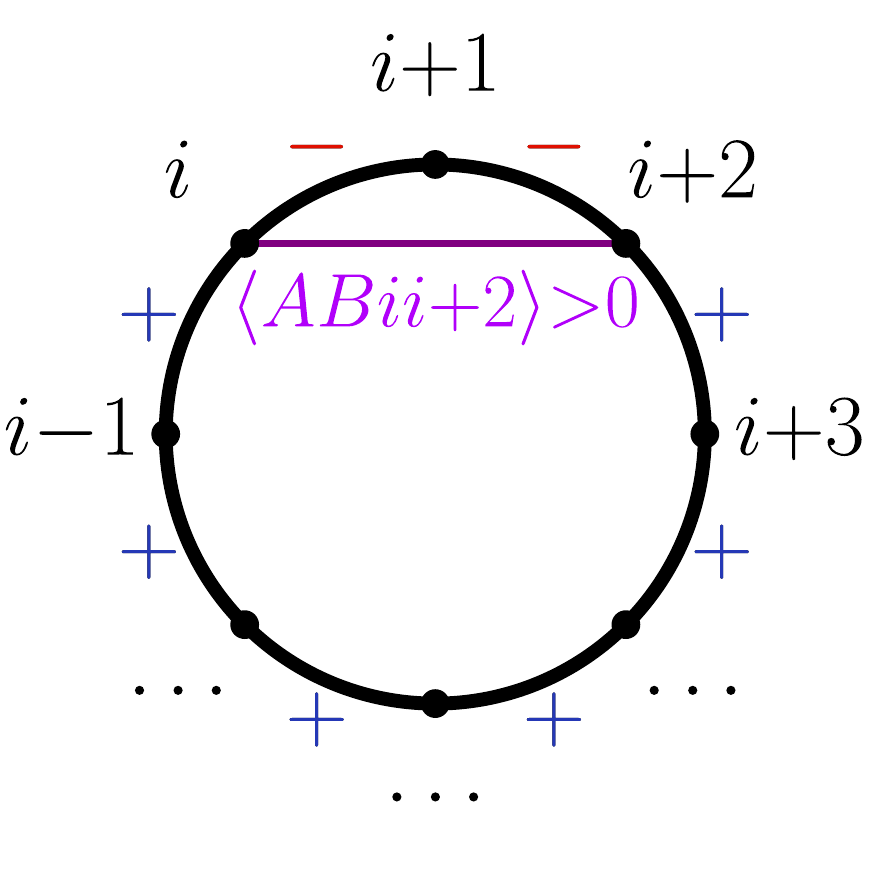}}
 \leftrightarrow & 
 \frac{\ab{i{-}1\,i\,i{+}1\,i{+}2} \ab{i\,i{+}1\,i{+}2\,i{+}3}}
 {\ab{ABi{-}1i}\ab{ABii{+}1}\ab{ABi{+}1 i{+}2} \ab{AB i{+}2 i{+}3}} \\[-35pt]
 & \hspace{1.5cm} = 
 \raisebox{-45pt}{
 \includegraphics[scale=.5]{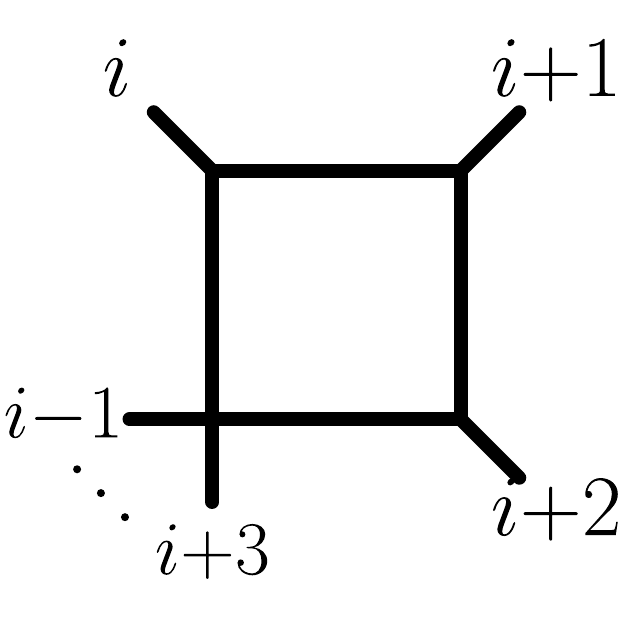}}.
\end{split} 
\end{align}
If we shrink the `$-$' region even further and consider $j{=}i{+}1$, the negative region consists of a single term, $\ab{ABii{+1}}{<}0$, i.e.,
\begin{equation}
\label{eq:sf2_special_chiral_region_single_minus}
\raisebox{-45pt}{\includegraphics[scale=.4]{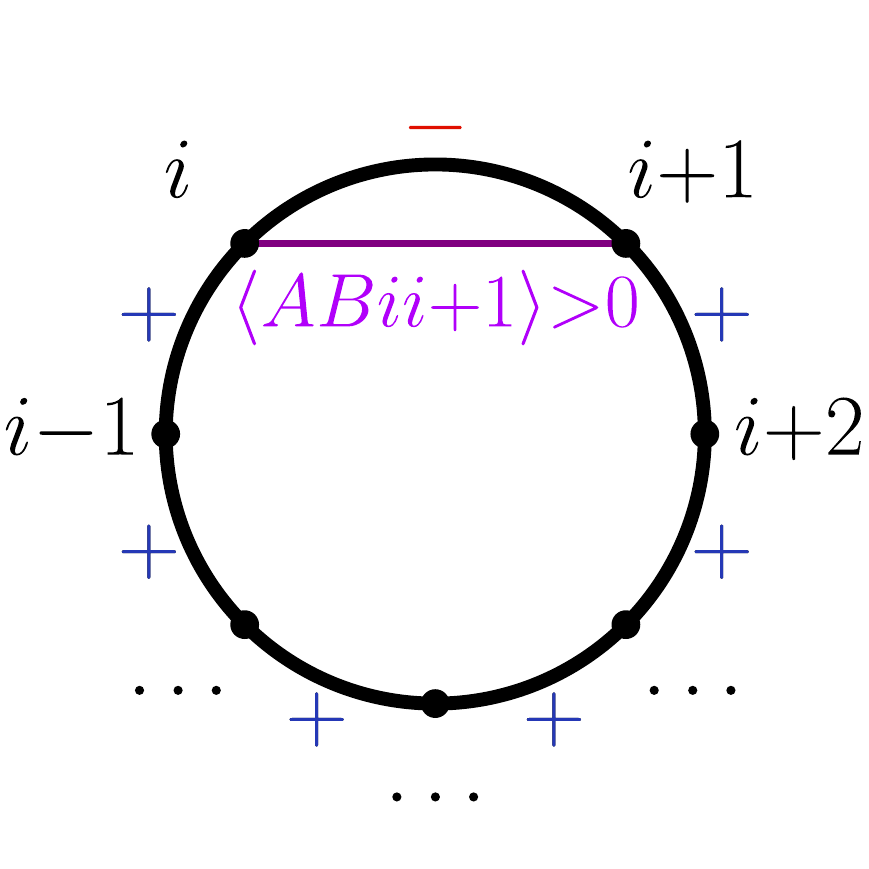}} \qquad \leftrightarrow \qquad 0.
\end{equation}
Note, however, that this bracket $\ab{ABii{+}1}$ is exactly the one used to cut the space into chiral components. Therefore, the $\la ABii+1\ra>0$ subspace is actually empty, i.e., the achiral space is now a single region which cannot be cut further without introducing spurious boundaries.

Above, we have discussed $S^{(2),+}_{ij}$ where $\ab{ABij}>0$, but the same analysis can be done for the opposite chirality where $\la ABij\ra<0$ where the r\^{o}les of $+\leftrightarrow -$ are interchanged. Going back to the achiral space, we divide the region into chiral components, $S^{(2)}_{ij} = S^{(2),+}_{ij} + S^{(2),-}_{ij}$
\begin{align}
\label{eq:sf2_achiral_abstract_def}
\begin{split}
&
\hspace{-.5cm}
\raisebox{-45pt}{\includegraphics[scale=.4]{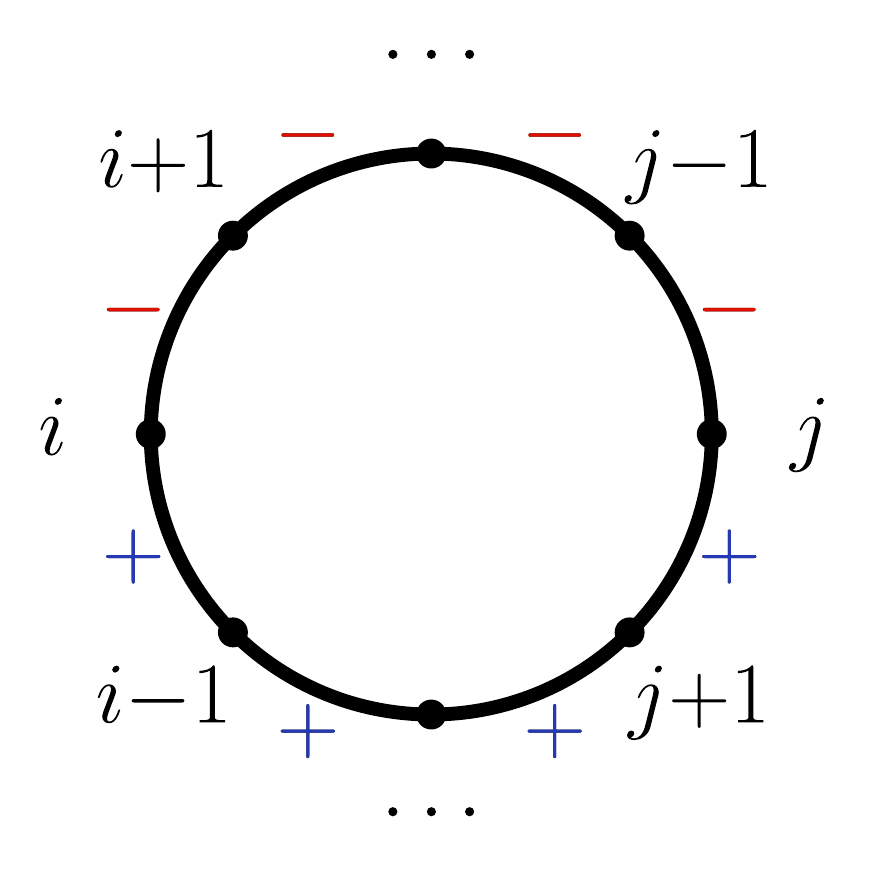}} \quad = \quad 
\raisebox{-45pt}{\includegraphics[scale=.4]{./figures/sf2_plus_components.pdf}} 
\quad +\quad 
\raisebox{-45pt}{\includegraphics[scale=.4]{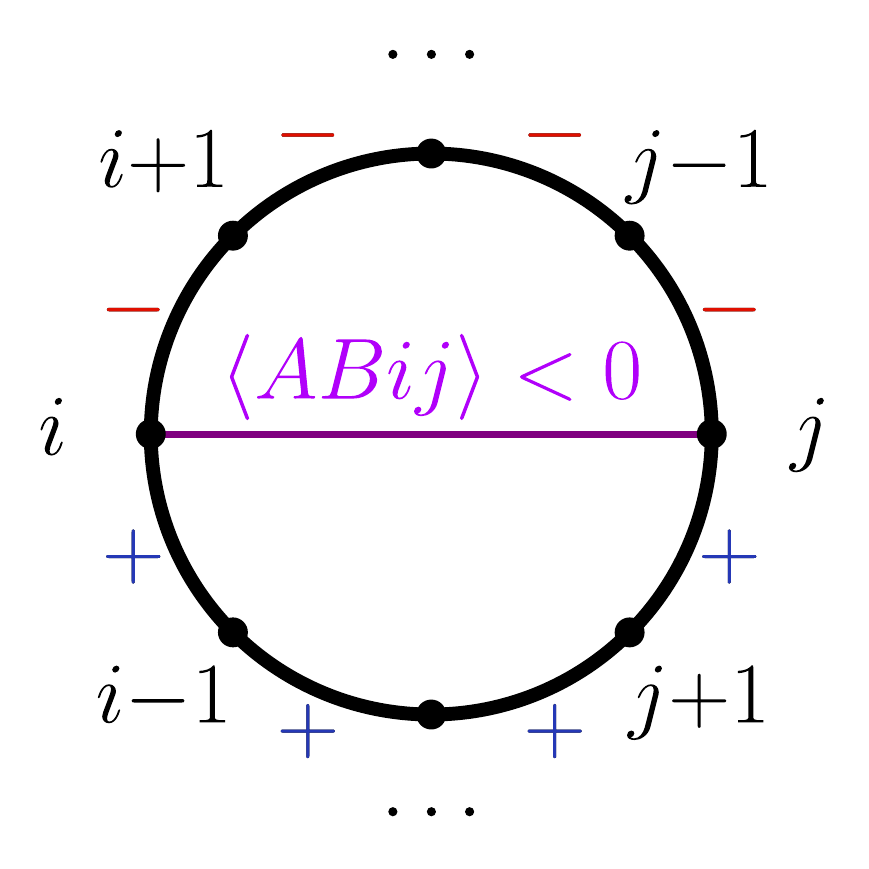}} 
\hspace{-.5cm}
\\[-5pt]
&\leftrightarrow 
\frac{N^{(2),+}_{ij}}{\ab{ABi{-}1i}\ab{ABii{+}1} \dots \ab{ABjj{+}1}} 
+ 
\frac{N^{(2),-}_{ij}}{\ab{ABj{-}1j}\ab{ABjj{+}1} \dots \ab{ABii{+}1}} 
\,. 
\hspace{-.5cm}
\end{split}
\end{align}
While the canonical form for the achiral space eq.~(\ref{eq:sf2}) has all $\ab{ABii{+}1}$ codimension-one boundaries present, it is in fact the sum of two simpler forms with fewer poles coming from two chiral subspaces. This feature makes the boundary structure of $S^{(2)}_{ij}$ simpler than that of the sign-flip-zero space $S^{(0)}$. For example, in $S^{(2)}_{ij}$ there is no codimension-two boundary corresponding to $\la ABpp{+}1\ra=\la ABqq{+}1\ra=0$ where $p\in(i{+}1,\dots j{-}2)$ and $q\in(j{+}1,\dots i{-}2)$. There are several additional interesting aspects of the sign-flip-two spaces which we discuss at greater length in appendix \ref{app:fixed_signs_sf2_sf4_spaces}.

\subsection*{Sign-flip-four regions}

In our discussion of sign-flip-four regions we start with a few simple examples before discussing the general case (which will be surprisingly simple!). Going back to our original description of one of the faithful chiral pentagon geometries in eq.~(\ref{pentsign2}), we see that the five-point pentagon with $i=2$ and $j=4$ can naturally be phrased as a chiral sign-flip-four region, 
\begin{equation}
\label{five_point_pentagon}
\hspace{-1cm}
\raisebox{-55pt}{\includegraphics[scale=.35]{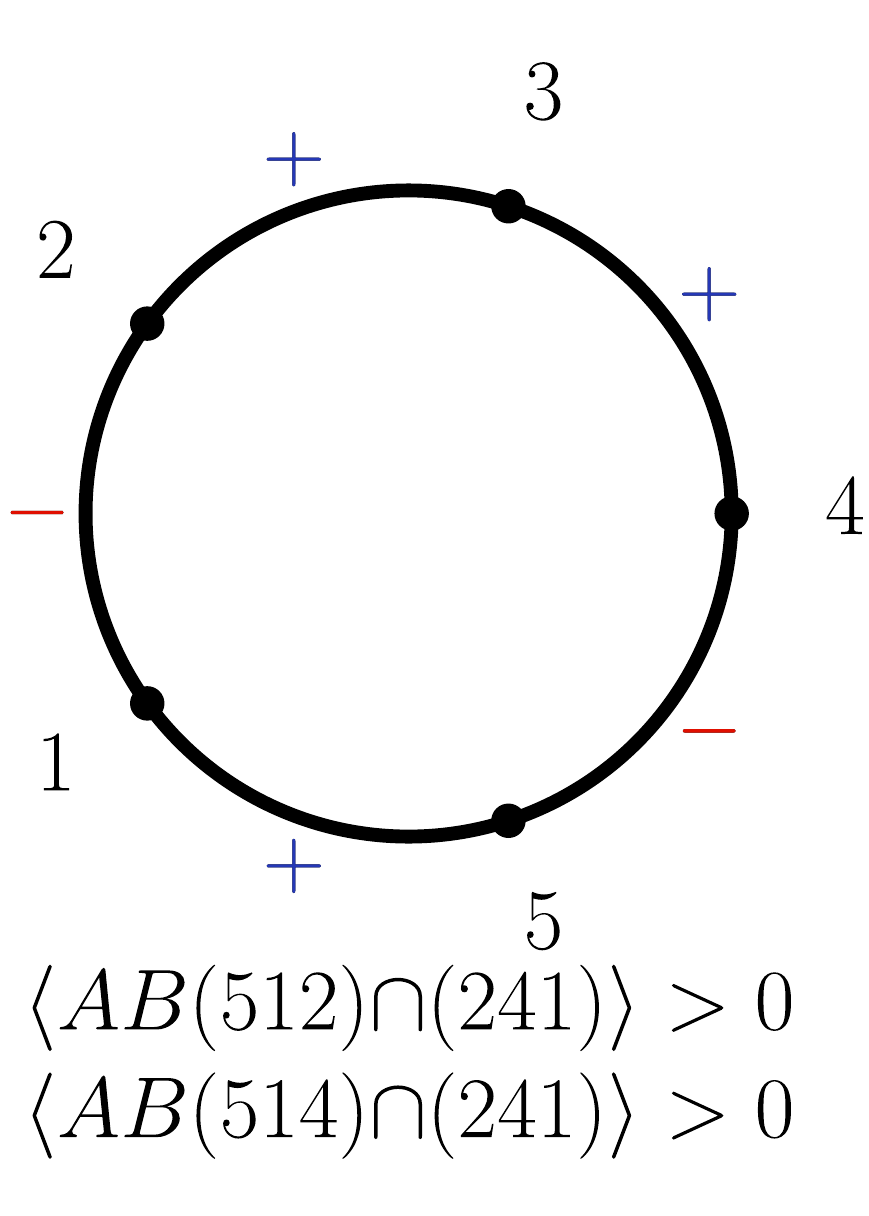}} 
\leftrightarrow 
\frac{\la AB\overline{24}\ra\la 1245\ra}{\ab{AB12}\ab{AB23}\ab{AB34}\ab{AB45}\ab{AB15}}
{=}
\hspace{-.3cm}
\raisebox{-28pt}{\includegraphics[scale=.4]{./figures/chiral_pent_5.pdf}},
\hspace{-1cm}
\end{equation}
where we have also indicated the additional inequalities ($\la AB(512){\cap}(241)\ra>0$ and $\la AB(514)\cap(241)\ra>0$) imposed to define the space. Similar to what we have seen above, this is one chiral subspace of a larger achiral region defined by the signs of $\ab{ABii{+}1}$ only. In fact, the intersections appearing in eq.~(\ref{five_point_pentagon}) can be replaced by a single inequality, either
\begin{equation}
    \la AB25\ra<0 \,\,\mbox{or}\,\,\ab{AB14}<0.
\end{equation}
to define exactly the same space as eq.~(\ref{five_point_pentagon}),
\begin{align}
    \raisebox{-62pt}{\includegraphics[scale=.35]{./figures/sf4_chiral_pentagon_5pts_def1.pdf}} 
    \qquad
  =
    \qquad
    \raisebox{-44pt}{\includegraphics[scale=.4]{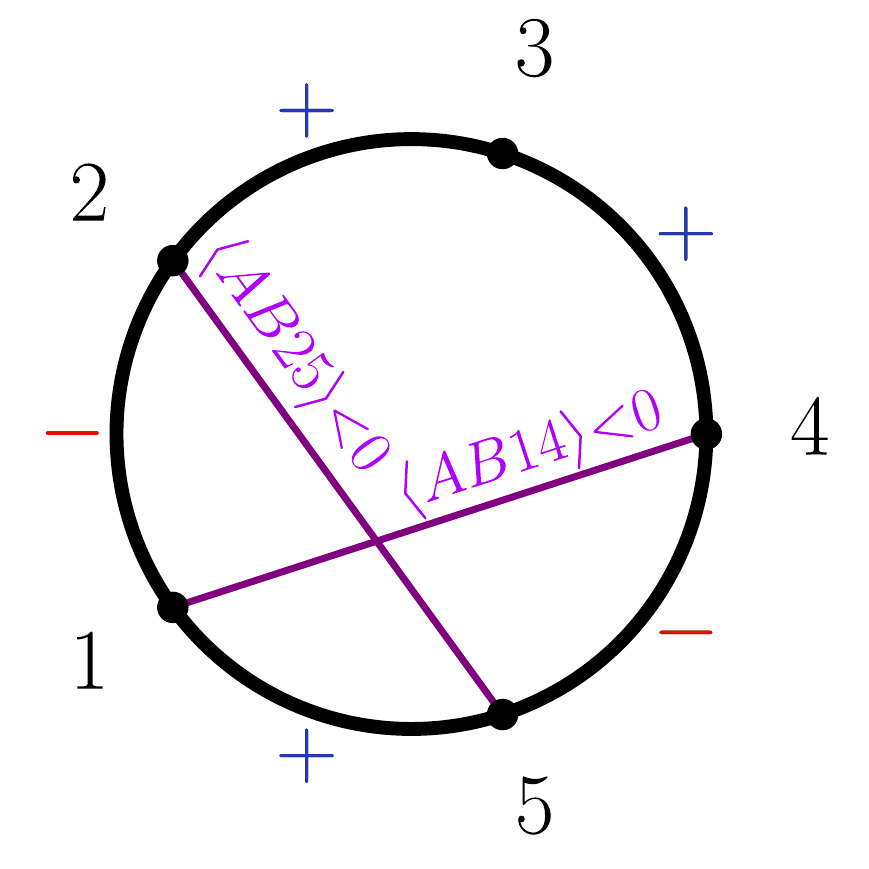}} \,.
\end{align}
Note that the bracket $\ab{AB24}>0$ is positive as a consequence of the $\ab{ABii{+}1}$ signs \emph{only}. From a Schouten identity,
\begin{equation}
 \underset{-}{\ab{AB12}}\underset{-}{\ab{AB45}} + \underset{+}{\ab{AB15}}\underset{+}{\ab{AB24}} = \ab{AB14}\la AB25\ra > 0,
\end{equation}
it then follows that fixing the sign of $\ab{AB14}$ determines the sign of $\ab{AB25}$, and vice versa. Only one of the two four-brackets is necessary to define the chiral sign-flip-four space, and the other is redundant. The five-point pentagon eq.~(\ref{five_point_pentagon}) space has both signs negative, while the second chiral subspace (which corresponds to the opposite chirality pentagon diagrammatically represented by a dashed line) has both signs positive, 
\begin{align}
&\hspace{-1cm}\raisebox{-45pt}{\includegraphics[scale=.4]{./figures/sf4_chiral_pentagon_5pts_minus.pdf}} 
\leftrightarrow 
\frac{\la AB\overline{24}\ra\la 1245\ra}{\ab{AB12}\ab{AB23}\ab{AB34}\ab{AB45}\ab{AB15}}
\equiv 
\hspace{-.3cm}
\raisebox{-30pt}{\includegraphics[scale=.4]{./figures/chiral_pent_5.pdf}},
\hspace{-1cm}
\\
&\hspace{-1cm}\raisebox{-45pt}{\includegraphics[scale=.4]{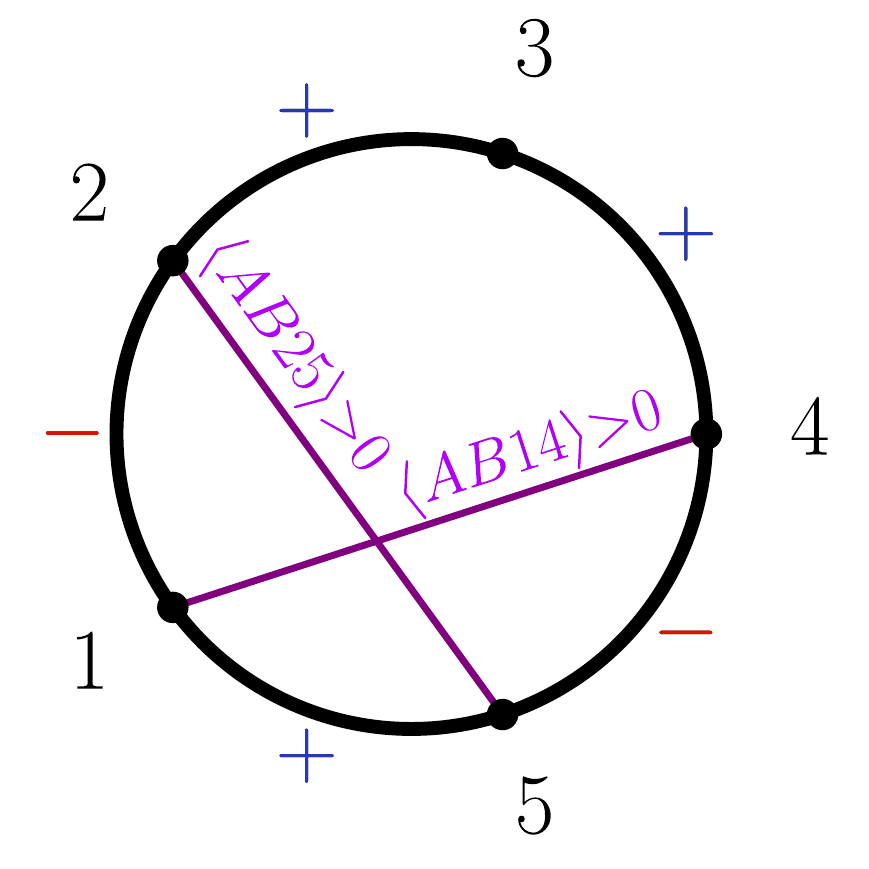}} 
\leftrightarrow 
\frac{\ab{AB24}\la 1235\ra\la 1345\ra}{\ab{AB12}\ab{AB23}\ab{AB34}\ab{AB45}\ab{AB15}}
\equiv 
\hspace{-.3cm}
\raisebox{-30pt}{\includegraphics[scale=.4]{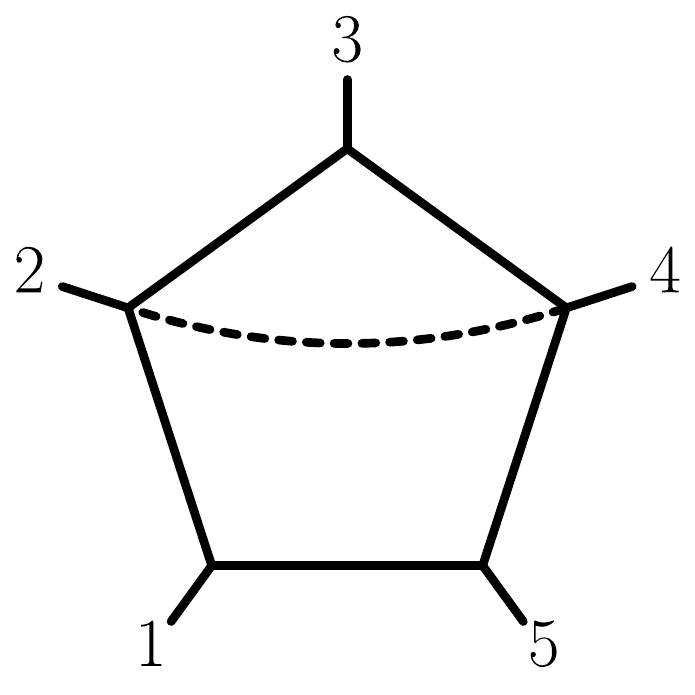}} .
\hspace{-1cm}
\end{align}
The union of these two spaces is a larger achiral space whose form is the difference of the two chiral pentagons, earlier introduced as parity-odd pentagon eq.~(\ref{eq:odd_pent_num}),
\begin{equation}
\hspace{-1cm}
\raisebox{-45pt}{\includegraphics[scale=.4]{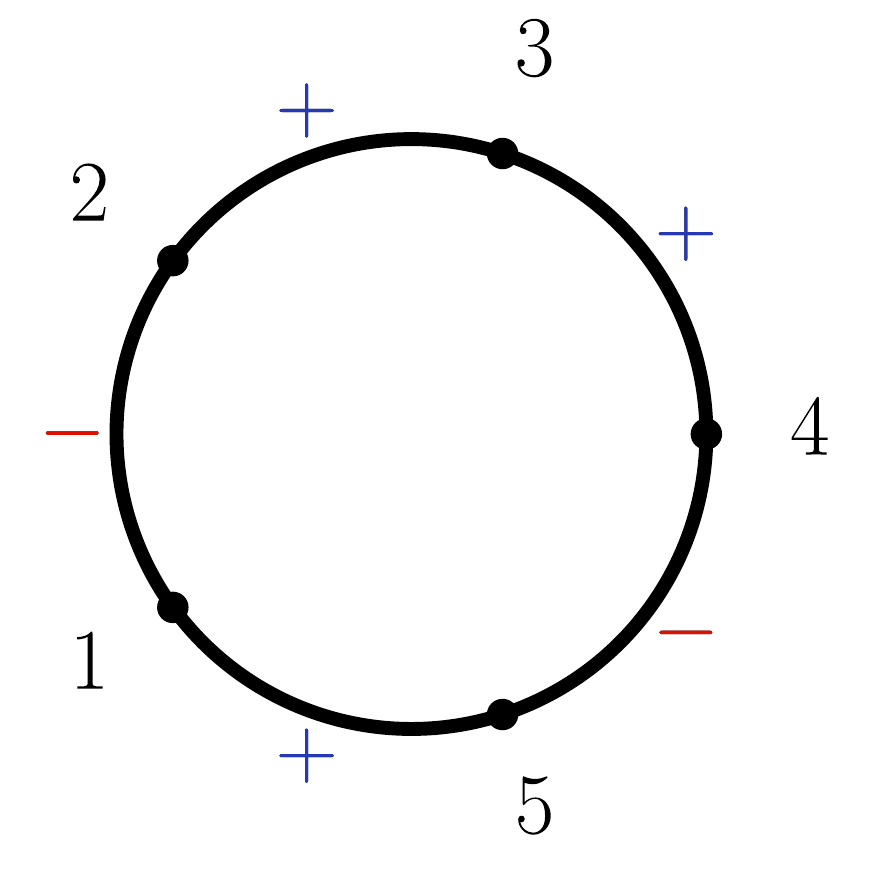}}
\leftrightarrow 
\frac{N_{\text{odd}}}{\ab{AB12}\ab{AB23}\ab{AB34}\ab{AB45}\ab{AB15}}\,,
\hspace{-.3cm}
\end{equation}
where the numerator is $N_{\text{odd}} = \ab{ AB\overline{24}}\ab{1245}- \ab{AB24}\ab{1235}\ab{1345}$.

Our next example is the achiral six-point region which can likewise be cut into two chiral components using a single additional inequality, 
\begin{equation}
\raisebox{-45pt}{\includegraphics[scale=.4]{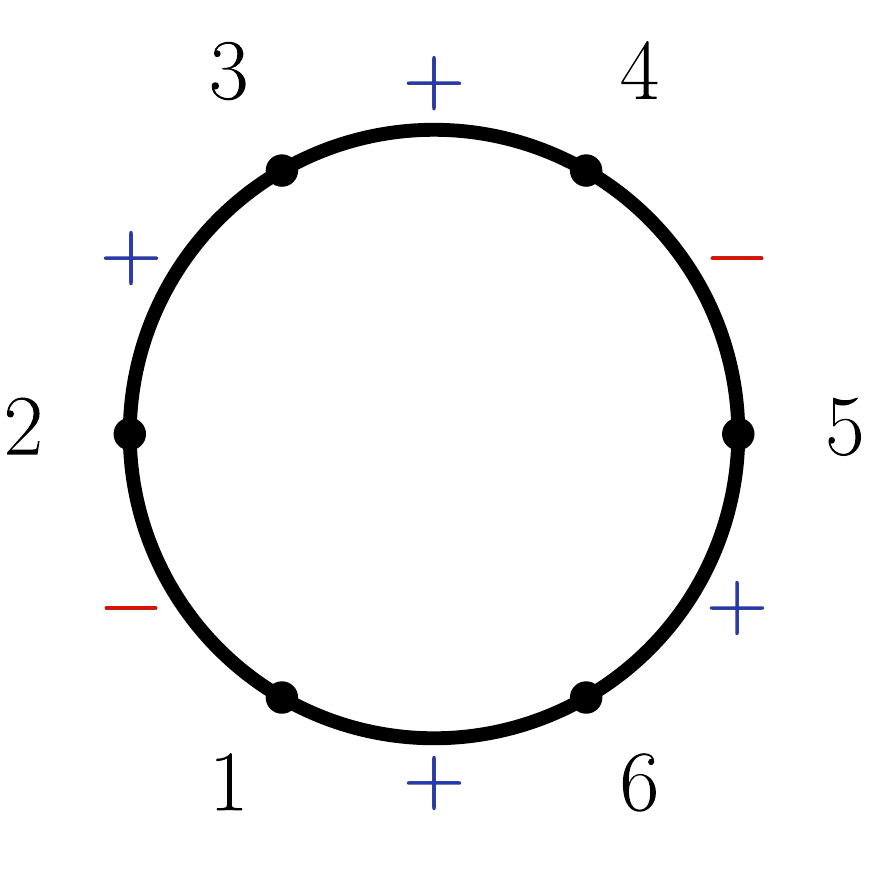}} \,\,=\,\, \raisebox{-45pt}{\includegraphics[scale=.4]{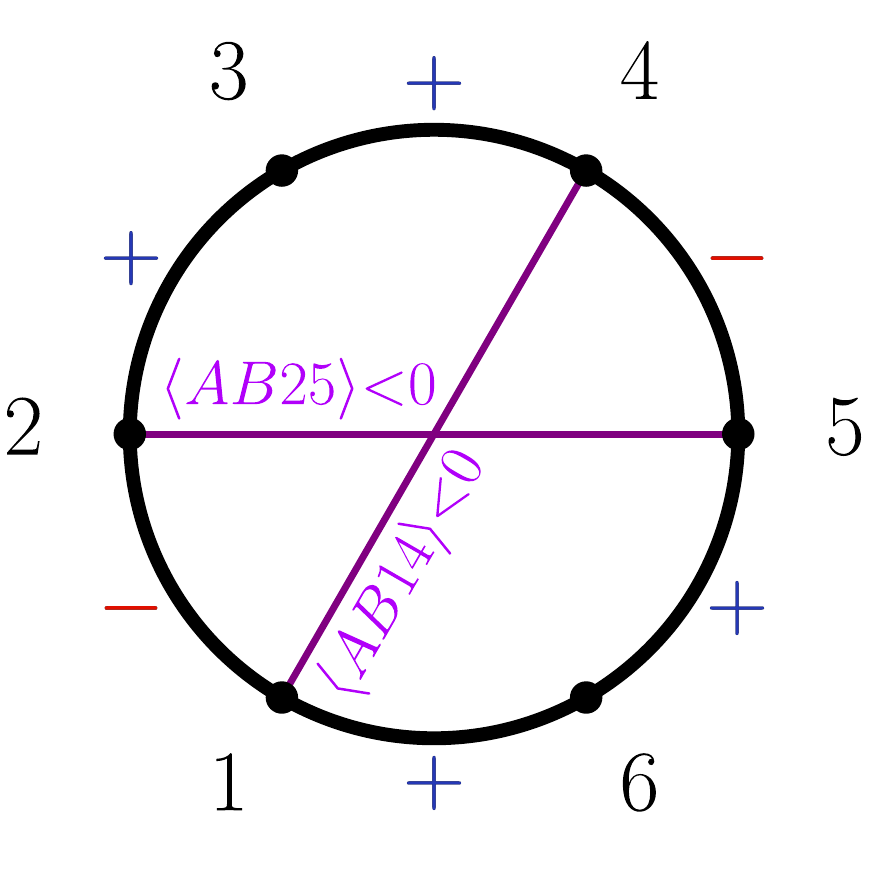}} \,\,+\,\, 
\raisebox{-45pt}{\includegraphics[scale=.4]{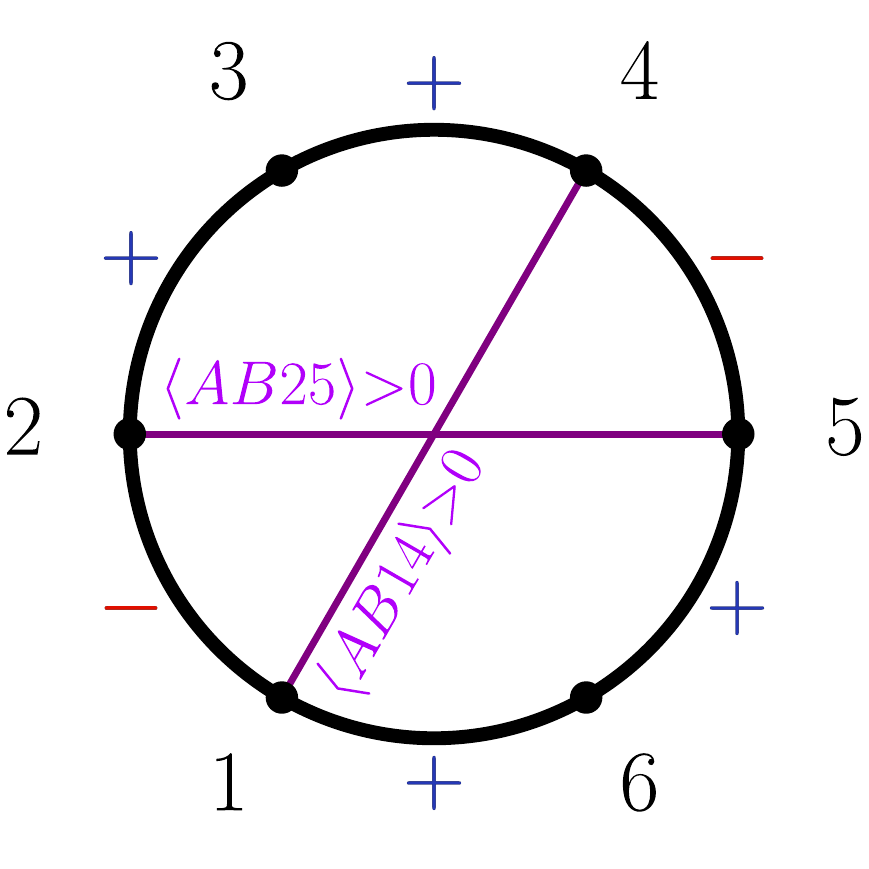}} ,
\end{equation}
where, again, the signs of the ``diagonal brackets" are either both positive or both negative as a consequence of a Schouten identity. Calculating the forms associated to these spaces, we find two chiral hexagons which were introduced in \cite{ArkaniHamed:2010gh} as examples of IR-finite integrands,
\begin{align}
\hspace{-1cm}
   \raisebox{-45pt}{\includegraphics[scale=.4]{./figures/sf4_chiral_hexagon_6pts_minus.pdf}}
   &\leftrightarrow
   \frac{\ab{AB\overline{24}}\ab{AB\overline{51}}}
        {\ab{AB12}\ab{AB23}\cdots\ab{AB16}}
   \equiv 
   \hspace{-.5cm}
   \raisebox{-52pt}{
   \includegraphics[scale=.45]{figures/chiral_hexagon_6pt_wavy.pdf}} ,
   \\[-9pt]
   \raisebox{-45pt}{\includegraphics[scale=.4]{./figures/sf4_chiral_hexagon_6pts_plus.pdf}}
   &\leftrightarrow
   \frac{\ab{AB24}\ab{AB51}\ab{3456}\ab{6123}}{\ab{AB12}\ab{AB23}\cdots\ab{AB16}}
   \equiv 
   \hspace{-.5cm}
   \raisebox{-52pt}{
   \includegraphics[scale=.45]{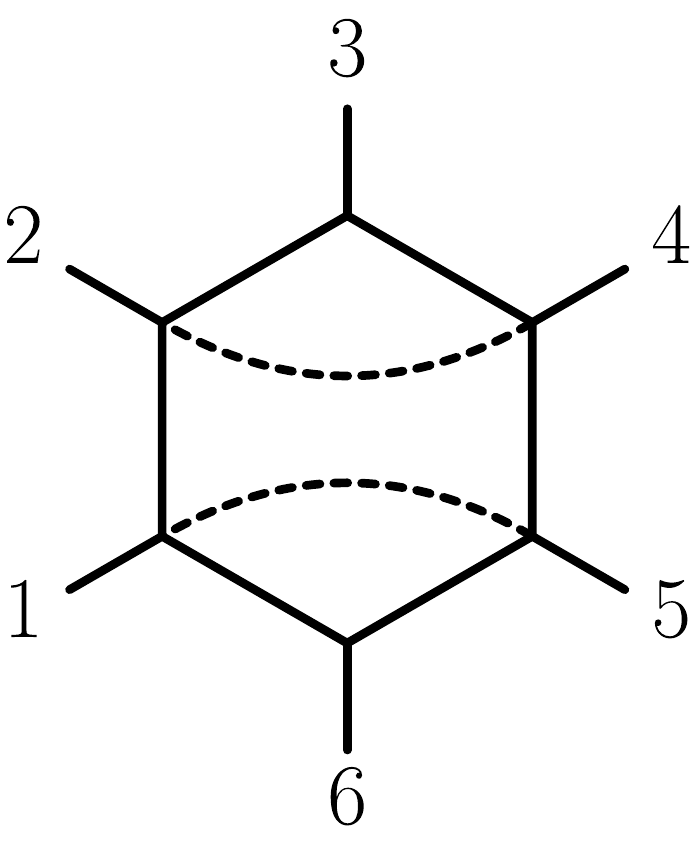}} .
\end{align}
Finally, we consider an additional six-point region corresponding to yet another chiral pentagon integrand:
\begin{equation}
\label{eq:sf4_6pt_chiral_pent}
\hspace{-1cm}
\raisebox{-45pt}{\includegraphics[scale=.4]{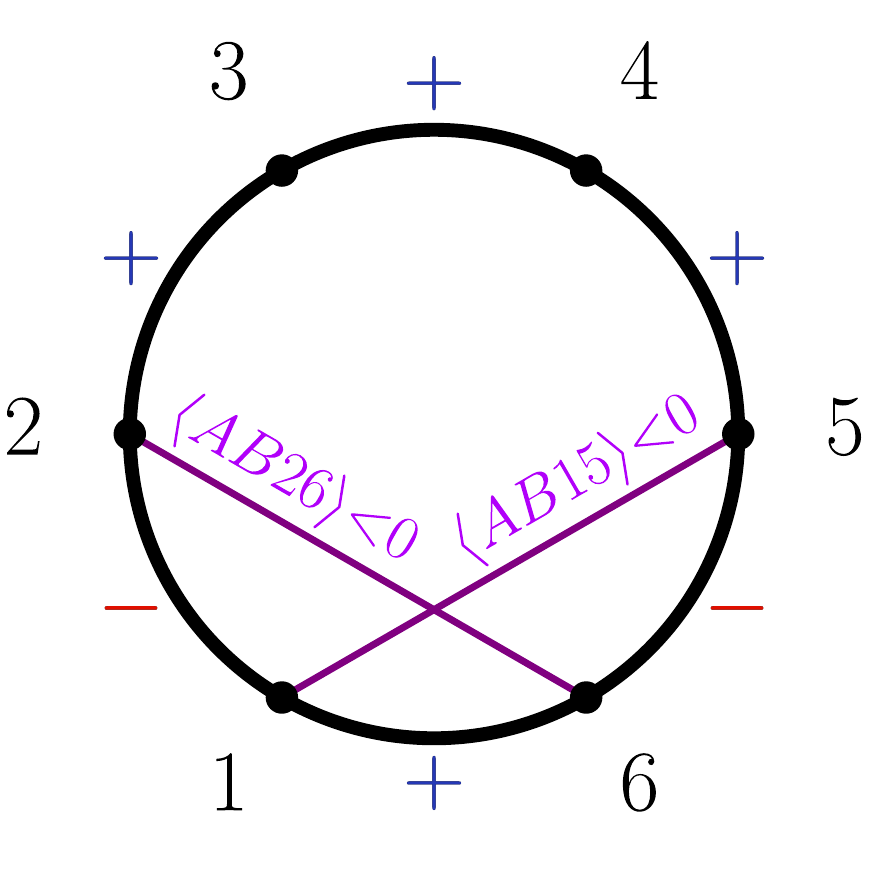}} 
\leftrightarrow 
\frac{\ab{AB\overline{25}}\ab{1256}}
    {\ab{AB12}\ab{AB23}\ab{AB45}\ab{AB56}\ab{AB16}}
\equiv
\raisebox{-35pt}{\includegraphics[scale=.5]{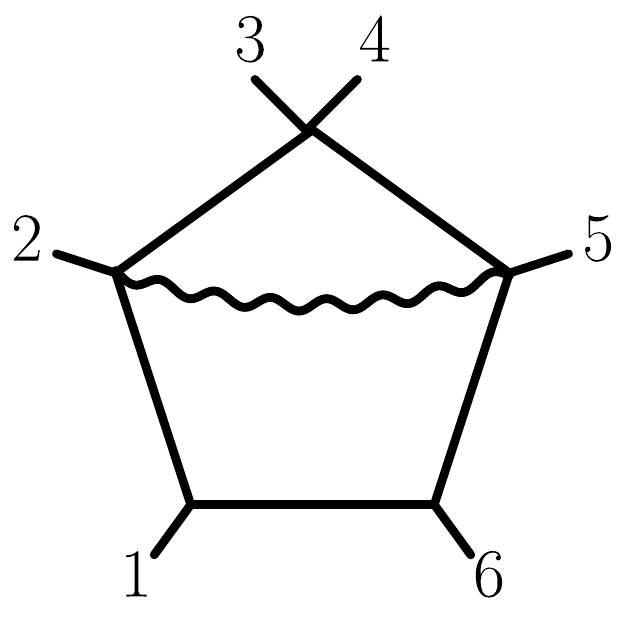}} \,.
\\
\end{equation}
We see the same pattern as in the sign-flip-two spaces: inside the `$+$' region the ``inner boundaries'' are absent; e.g., in eq.~(\ref{eq:sf4_6pt_chiral_pent}), $\ab{AB34}$ is not a pole of the form nor a boundary of the geometric space. 

Having discussed several illuminating examples at low multiplicity, we are now ready to present the general sign-flip-four case:
\begin{equation}
\label{eq:sf4_chiral_comps_fully_dressed}
S^{(4),\pm}_{ik\ell j}
=\raisebox{-68pt}{\includegraphics[scale=.55]{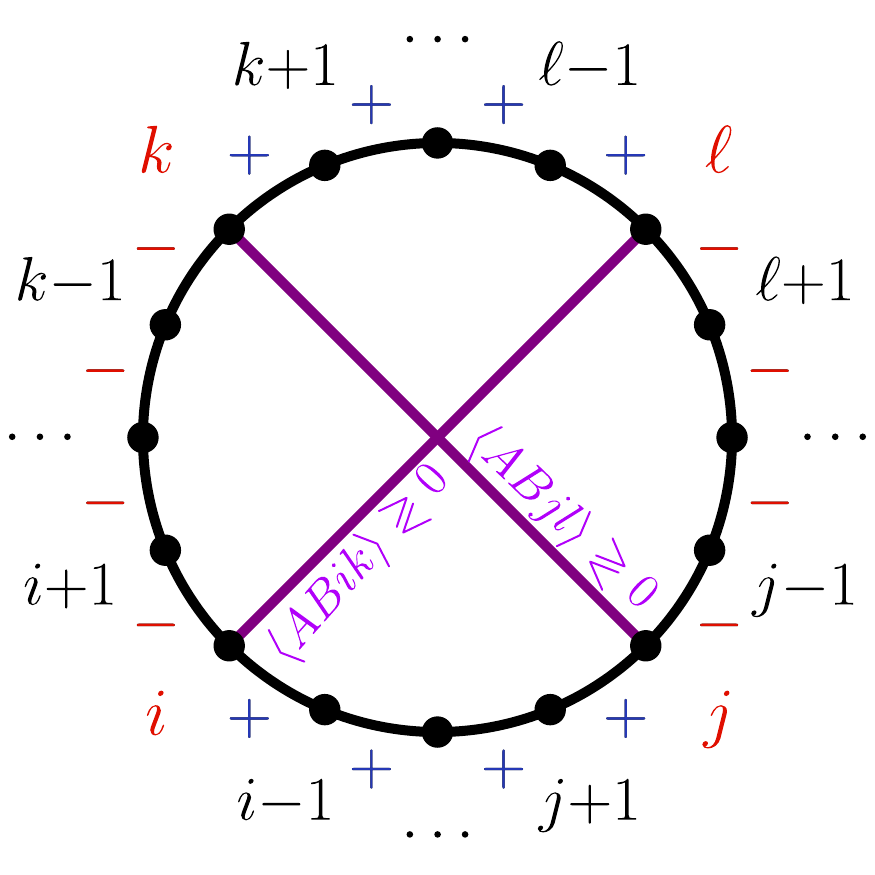}}. 
\end{equation}
where the large achiral region $S^{(4)}_{ik\ell j}$ of eq.~(\ref{eq:sf4}) is cut into two chiral subspaces by specifying the signs of $\ab{ABi\ell}$ \emph{or} $\ab{ABkj}$. In one subspace they are both positive while in the other they are both negative. In addition, all signs are uniformly fixed inside the four sign sectors. For example, all brackets of the form
\begin{equation}
    \la ABpq\ra,\,\la AB\overline{pq}\ra>0
    \quad 
    \mbox{for $p<q\in \{j,j{+}1,\ldots i\}$,}
\end{equation}
are positive, whereas the analogous brackets in the ``minus region'' are negative. We give further details on the fixed sign structure in appendix \ref{app:fixed_signs_sf2_sf4_spaces}. Practically, this means that the only boundaries from each of these sectors are the ones adjacent to the sign-flip positions, so the logarithmic form has exactly eight boundaries in the general case. In fact, each chiral component can be identified with a \emph{chiral octagon} integrand
\begin{align}
\label{eq:sf4_chiral_octagon_association}
&\raisebox{-74pt}{\includegraphics[scale=.5]{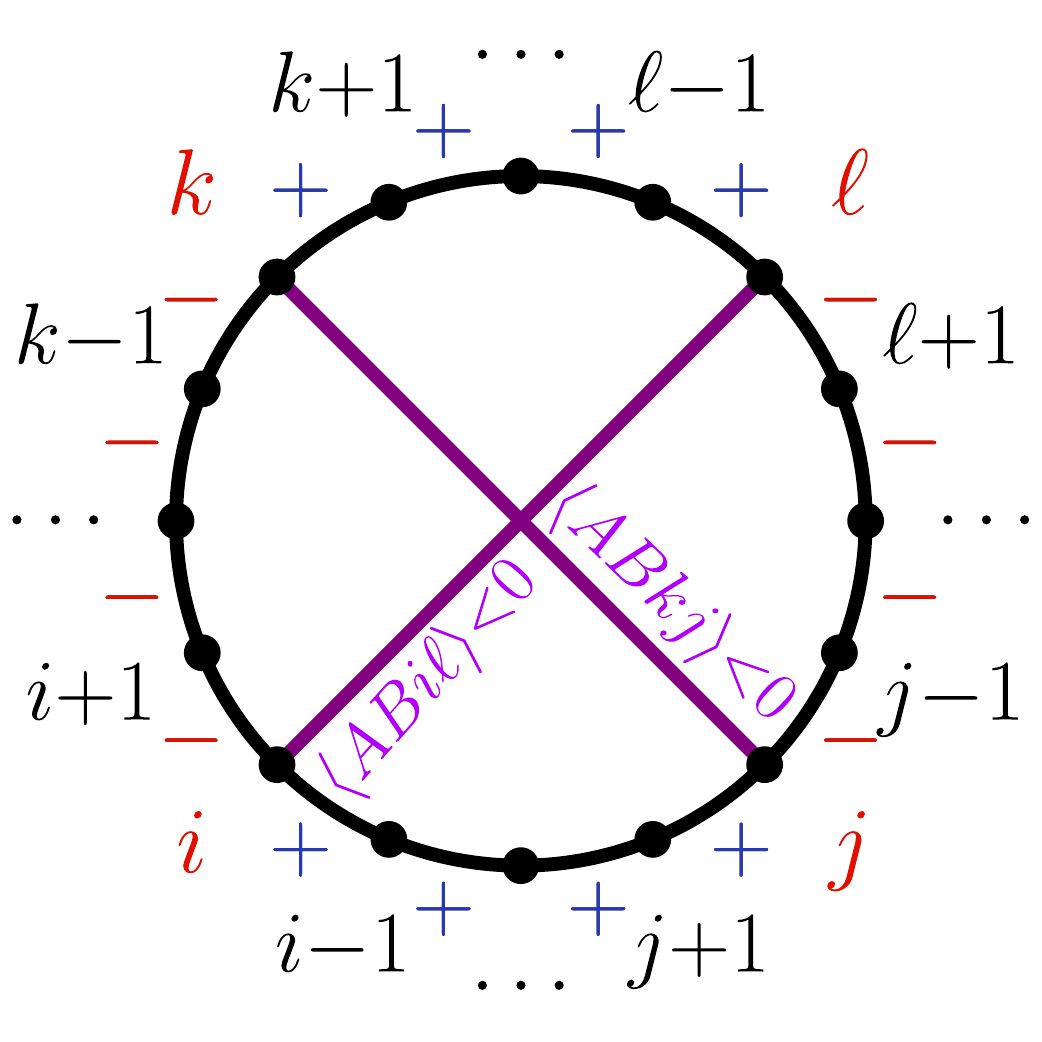}}
\quad \longleftrightarrow \quad 
\raisebox{-80pt}{\includegraphics[scale=.5]{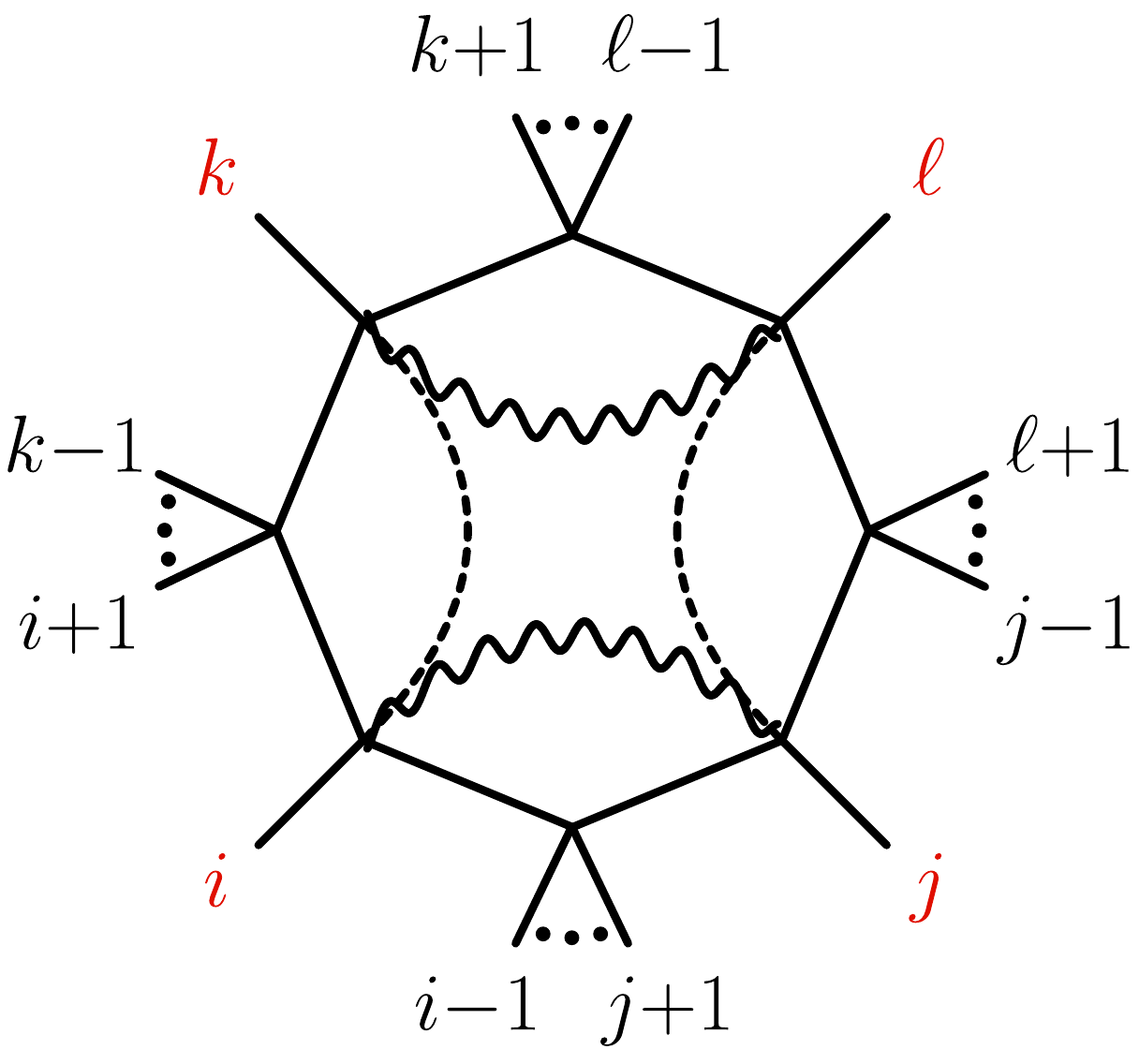}},
\end{align}
where the wavy and dashed lines indicate the respective numerators \cite{ArkaniHamed:2010gh},
\begin{align}
\omega^{(4),-}_{ik\ell j}
{=}\frac{\ab{AB\overline{ij}}\ab{ABik}
               \ab{AB\overline{k\ell}}\ab{AB\ell j}}
               {\begin{array}{c}\ab{ABi{-}1i}\ab{ABii{+}1}
               \ab{ABk{-}1k}\ab{ABkk{+}1}\\
               \times
               \ab{AB\ell{-}1\ell}\ab{ AB\ell\ell{+}1}
               \ab{ABj{-}1j}\ab{ABjj{+}1} 
               \end{array}}.
\end{align}
The form for the other chiral space, $\omega^{(4),+}_{ik\ell j}$,  for which $\ab{ABi\ell},\ab{ABkj}>0$, is obtained by flipping the wavy and dashed lines. The chiral octagons were introduced in \cite{ArkaniHamed:2010gh} as the one-loop integrand basis elements which split the basis into parity-odd (which integrate to zero), IR-finite and IR-divergent integrands. While the expression for the integrand may look complicated, because of the special form of the numerator in the generic case the integrand is IR finite and evaluates to a simple combination of dilogarithms. It is very surprising that we see the same objects here in a very different setup as the integrand forms for maximal sign-flip regions.

The chiral octagons naturally degenerate to simpler spaces when the labels $i,k,\ell,j$ become adjacent. Exactly the same happens with our regions as well, and we can indeed identify the pentagon and hexagon examples discussed above as boundary cases of the generic octagon.

\section{Local geometries and the Amplituhedron-Prime}
\label{sec:geometry_chiral_pentagons}
Having discussed various aspects of more general sign-flip spaces, let us come back to the local integrands of section \ref{sec:geometry_dlog_forms} that enter the chiral pentagon expansion of the one-loop MHV amplitude, eq.~(\ref{pent}). After a careful analysis of consistent sign patterns for the individual local geometries, we concluded in section \ref{sec:geometry_dlog_forms} that there are only two faithful geometric spaces which can be associated with the general chiral pentagon eqs.~(\ref{pentsign1}) and (\ref{pentsign2}), two choices for the two-mass-hard boxes eqs.~(\ref{2mh_boxsign1}) and (\ref{2mh_boxsign2}), and four choices for the one-mass boxes eqs.~(\ref{1m_boxsign12}) and (\ref{1m_boxsign34}), respectively. 

In the next step we discuss how to glue these geometric regions together. We are going to show that only a single choice for the pentagon and box spaces is globally consistent (at all multiplicities) upon gluing. By consistency, we mean the requirement that there are no unphysical boundaries left in the resulting geometry akin to our discussion of \emph{faithful geometries} in section \ref{subsec:dlog_form_to_geometry}. As a tool, we make use of the classification of all relevant positive geometries in terms of the sign-flip-two and four regions summarized in sections \ref{sec:sign_flip_regions} and appendix \ref{sec:external_triangulations}.  This allows us to write a (conjectured) closed formula for the final geometric space in eq.~(\ref{amp_prime_npt}). 

The primary result of this section is a \emph{new} positive geometry, which we name the \emph{Amplituhedron-Prime}, comprised of a particular collection of chiral sign-flip-two and four regions with only physical boundaries of all codimensions. While this new space has exactly the same singularity structure as the Amplituhedron, their bulk geometries are entirely non-overlapping. This follows directly from the fact that the original Amplituhedron is comprised of a single chiral sign-flip-zero space.

\subsection{Chiral regions for boxes and pentagons}
\label{sec:five_point_choices}
%
We start our discussion with the five-point one-loop MHV amplitude. Specializing eq.~(\ref{pent}) to $n{=}5$, the integrand is a sum of one chiral pentagon and two boxes,
\begin{equation}
\label{eq:1loop_5pt_local_int_exp}
\omega^{(5,0,1)} =
        \raisebox{-32pt}{\includegraphics[scale=.4]{./figures/chiral_pent_5.pdf}} 
        +
        \raisebox{-32pt}{\includegraphics[scale=.5]{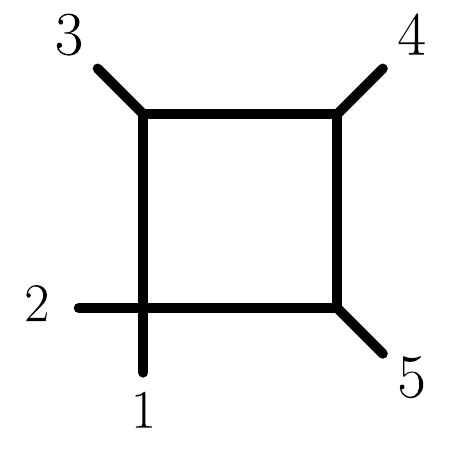}} 
        +
        \raisebox{-32pt}{\includegraphics[scale=.5]{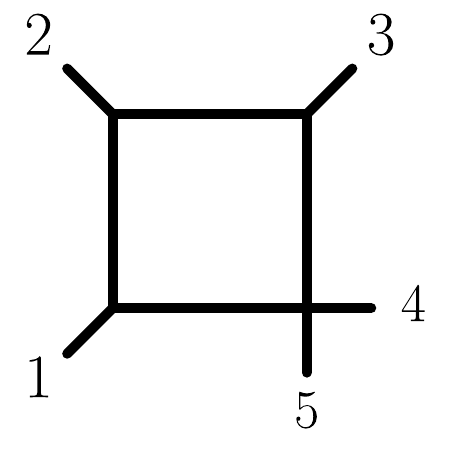}}.
\end{equation}
The first box in this expression, which we label as $B_{12}$, has four individually well-defined faithful geometries as described in eqs.~(\ref{1m_boxsign12})--(\ref{1m_boxsign34}), where the sign of the bracket $\ab{AB12}$ is unfixed. In section~\ref{sec:geometry_dlog_forms} we constructed the candidate spaces for the box (all of which had no spurious boundaries) by imposing conditions on $\ab{ABii{+}1}$ brackets as well as additional conditions involving its leading singularities. By expanding the unfixed sign $\ast=+\oplus-$, it is straightforward to identify all four choices as particular instances of sign-flip-two and four spaces described in section~\ref{sec:sign_flip_regions}. For the box $B_{12}$ we may write the options as
\begin{align}
\label{eq:b12_spaces}
    \begin{split}
    B_{12}^{(1)}&=\raisebox{-45pt}{\includegraphics[scale=.4]{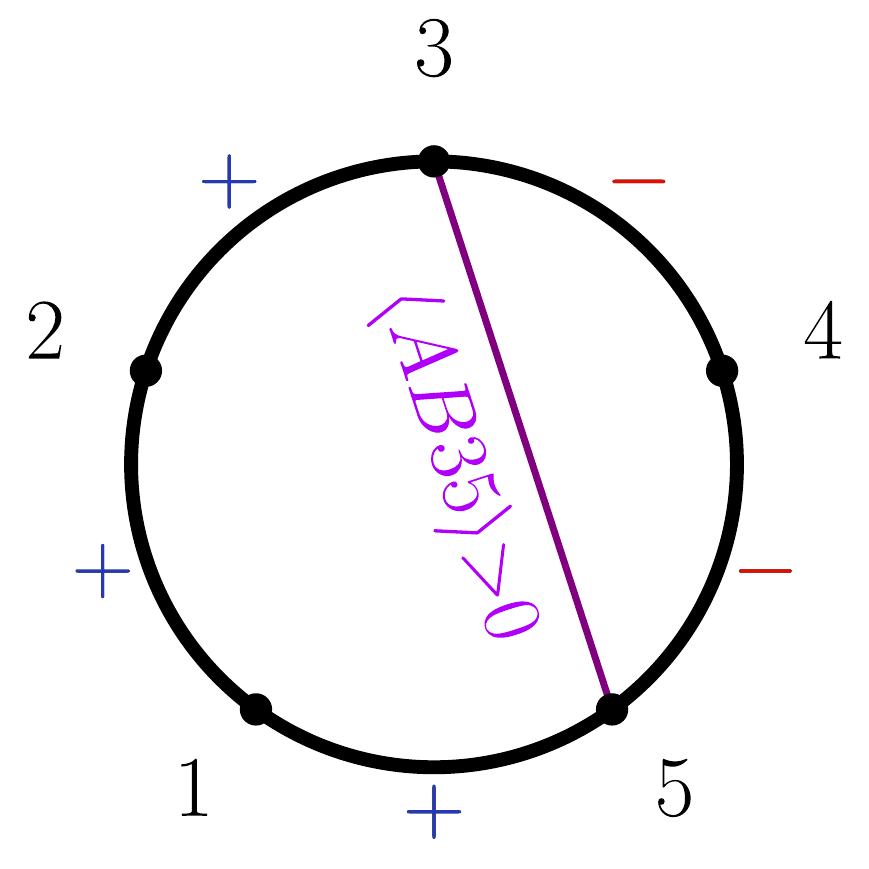}},
    \qquad 
    B_{12}^{(2)}=\raisebox{-45pt}{\includegraphics[scale=.4]{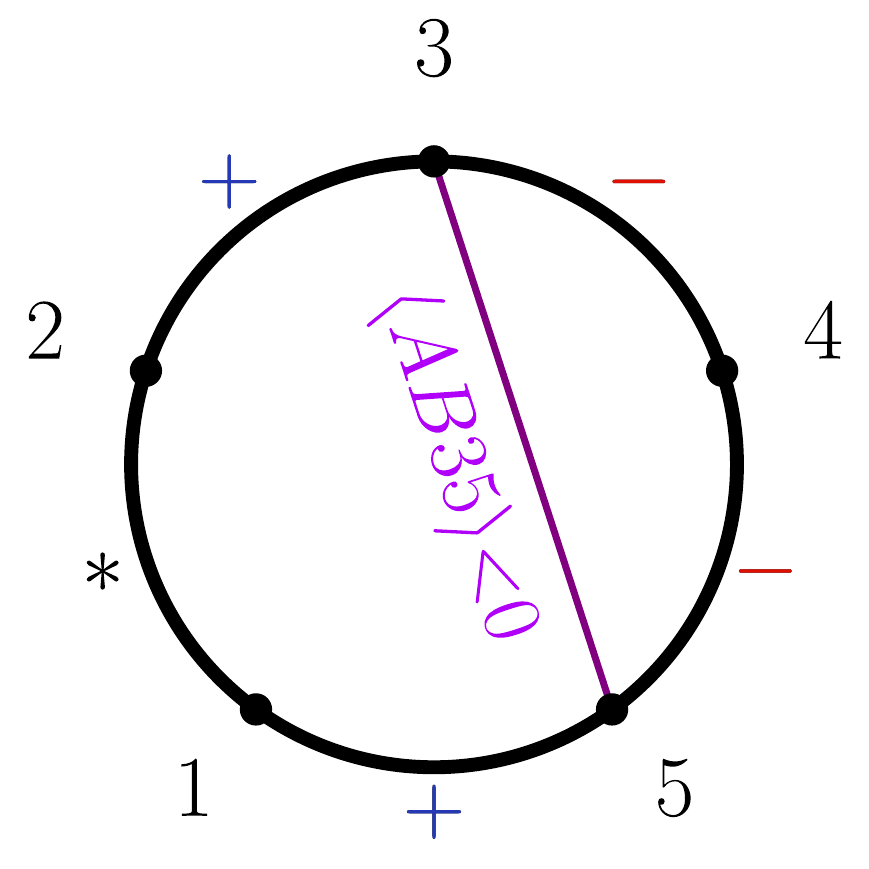}},
    \\ 
    B_{12}^{(3)}&=\raisebox{-45pt}{\includegraphics[scale=.4]{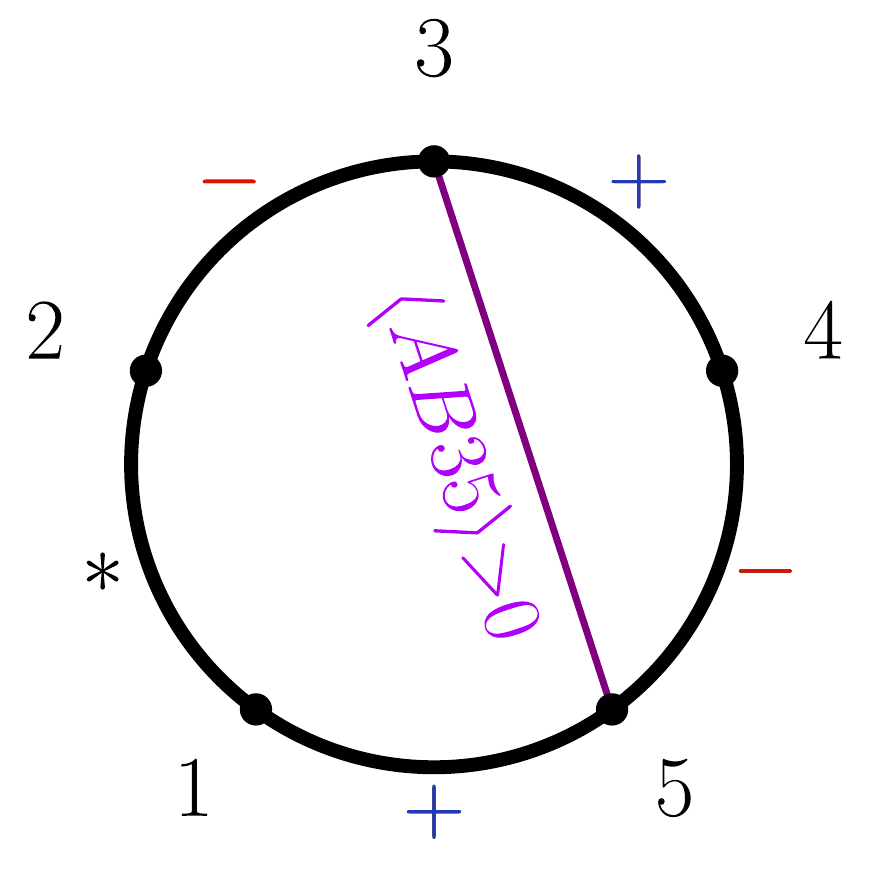}},
    \qquad  
    B_{12}^{(4)}=\raisebox{-45pt}{\includegraphics[scale=.4]{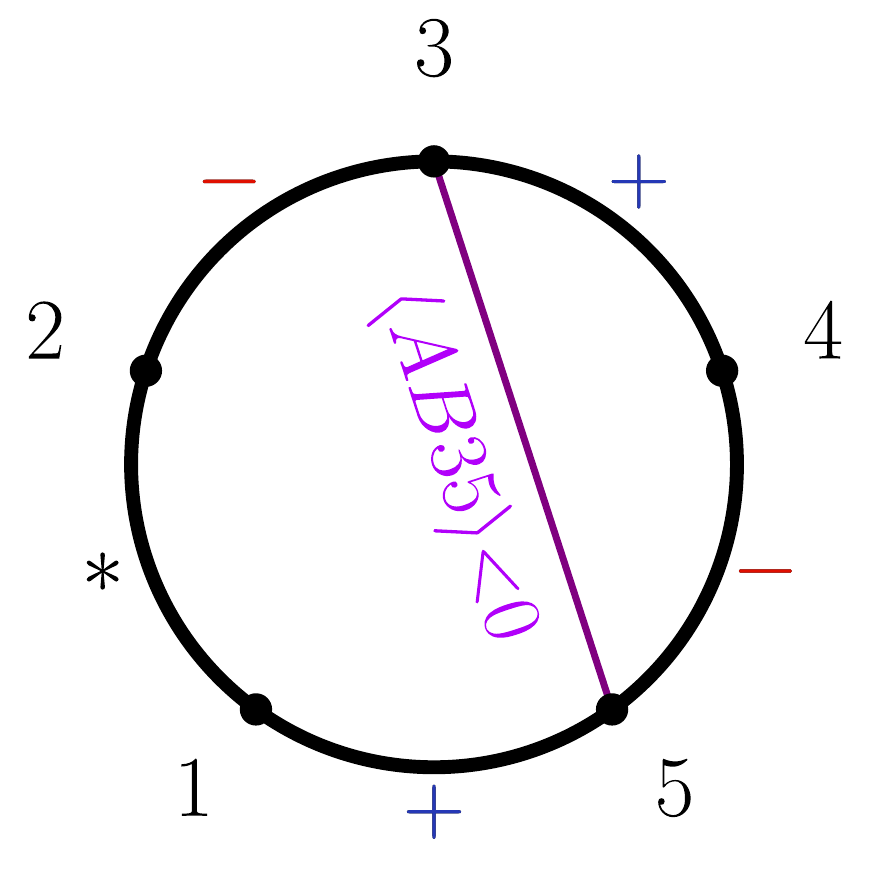}}.
    \end{split}
\end{align}
Note that for the choice labelled $B_{12}^{(1)}$, the missing space where $\ab{AB12}<0$ is empty. The additional inequality $\ab{AB35}\lessgtr0$ is needed to select one of the chiral components of the larger achiral space. As we stressed above, all four spaces $B^{(i)}_{12}$ are \emph{a priori} locally satisfactory as they have the correct boundary structure to represent the associated integral. However, this box has singularities which are absent in the MHV one-loop integrand. Indeed, we consider such singularities spurious from the point of view of the global geometry of the Amplituhedron. Thus, the manner in which $B_{12}$ glues together with the other terms in eq.~(\ref{eq:1loop_5pt_local_int_exp}) must be such that all spurious singularities cancel and we are left with exactly the physical singularity structure of the MHV Amplituhedron. 

The discussion of the other box integrand, $B_{45}$, is analogous and we once again have four different options:
\begin{align}
\label{eq:b45_spaces}
    \begin{split}
    B_{45}^{(1)}&=\raisebox{-45pt}{\includegraphics[scale=.4]{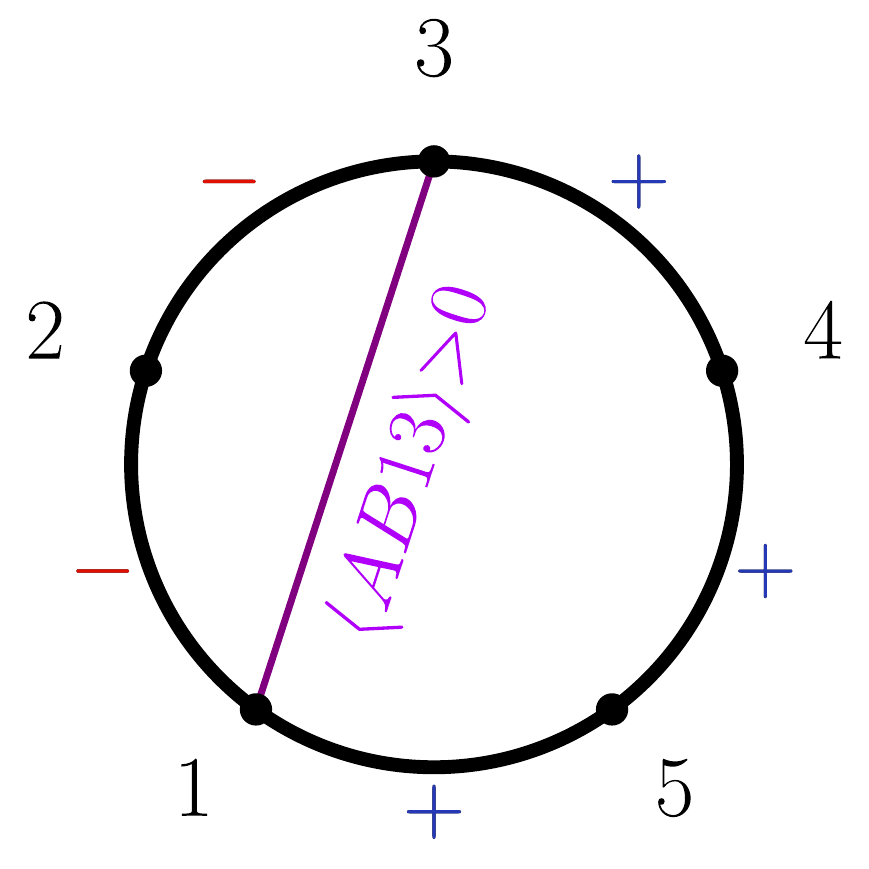}},
    \qquad 
    B_{45}^{(2)}=\raisebox{-45pt}{\includegraphics[scale=.4]{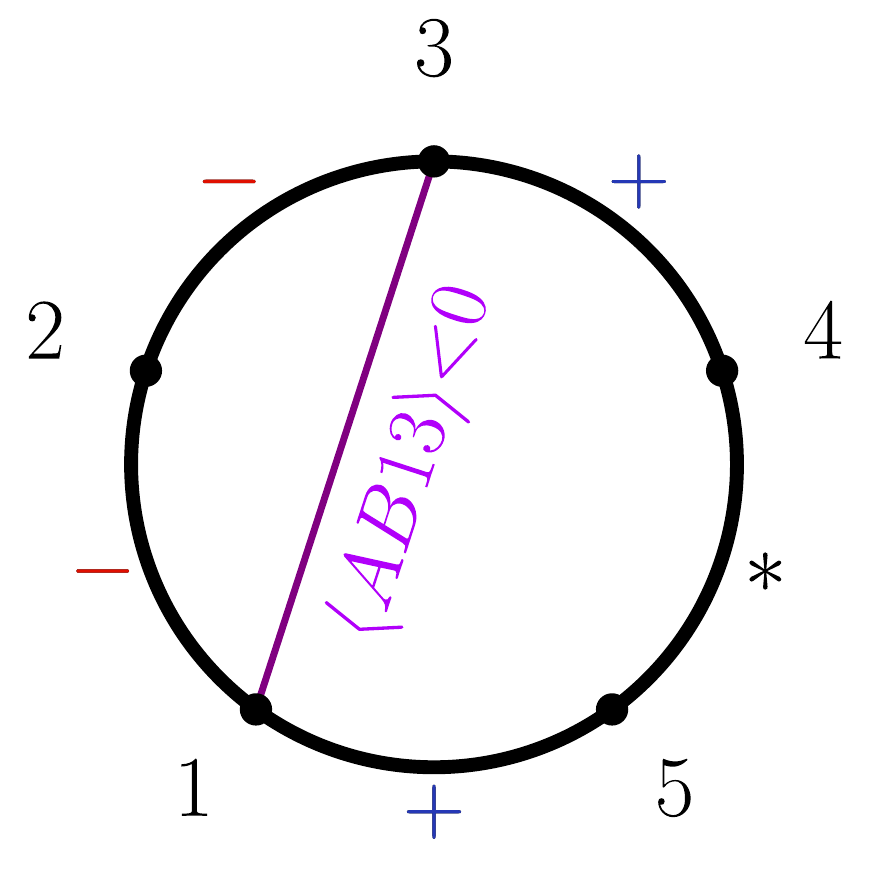}},
    \\ 
    B_{45}^{(3)}&=\raisebox{-45pt}{\includegraphics[scale=.4]{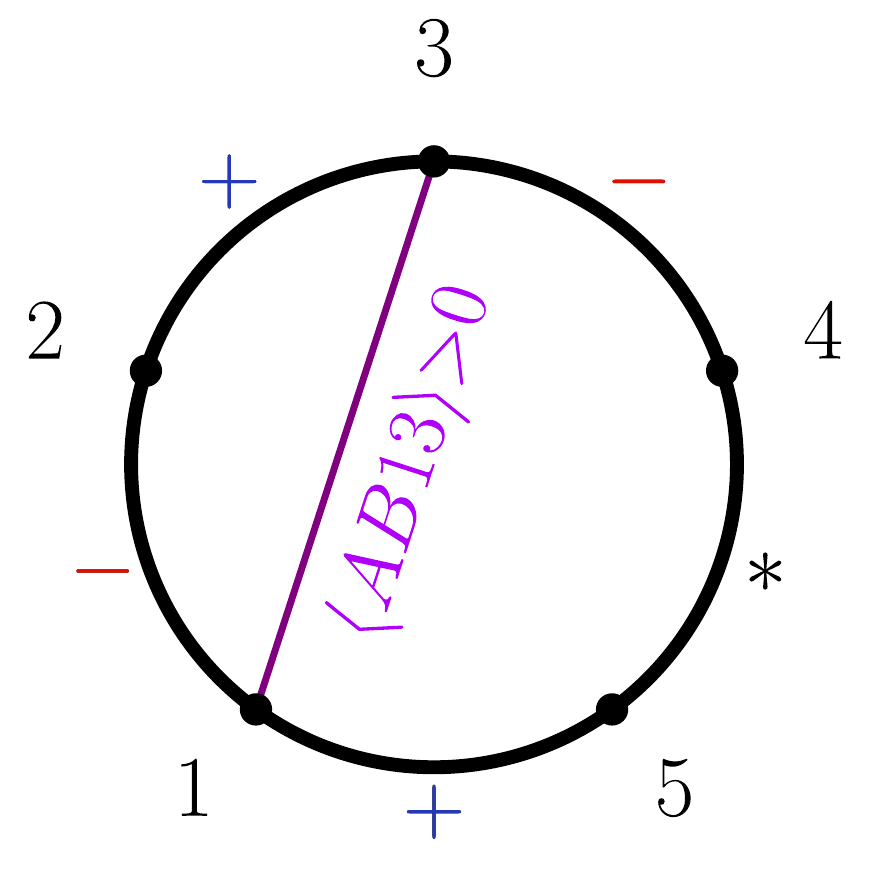}},
    \qquad  
    B_{45}^{(4)}=\raisebox{-45pt}{\includegraphics[scale=.4]{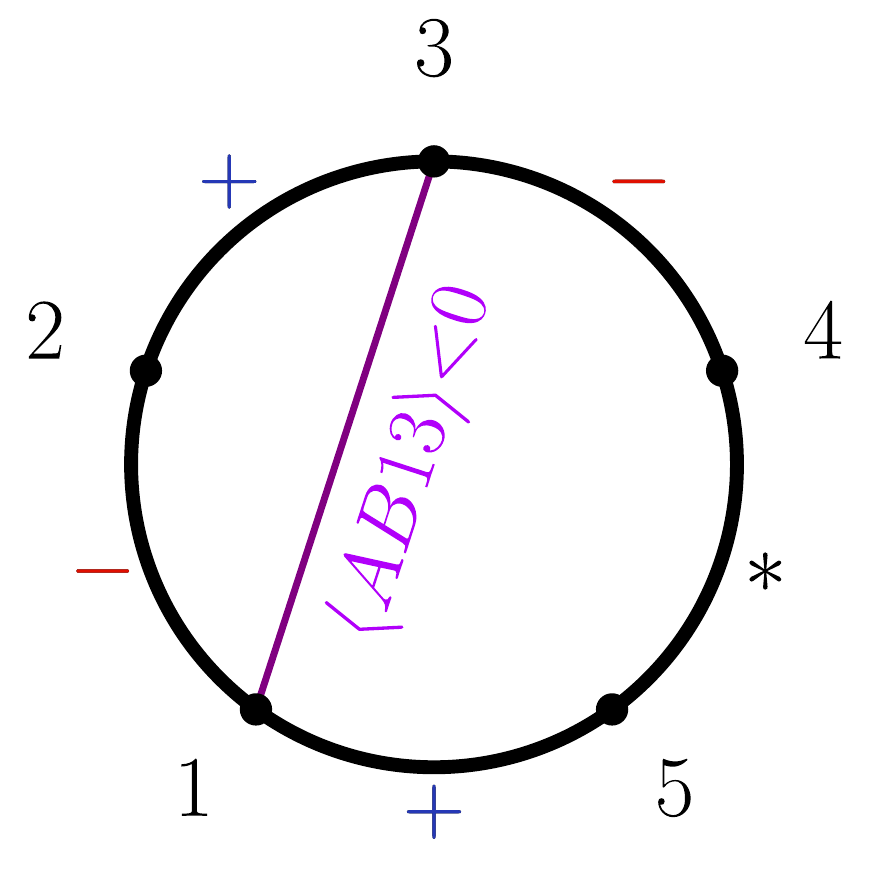}}.
    \end{split}
\end{align}
Although this suggests a total of $4^2$ possible choices for combining the two boxes, in our analysis below we will only consider the four choices $\{B_{12}^{(i)},B_{45}^{(i)}\}$. This corresponds to assigning uniform geometries to all permutations of one-mass boxes. While at five points it is possible to mix-and-match the box spaces and get a consistent global geometry, this is simply due to the degenerate kinematics and does not extend to the all multiplicity case.

For the pentagon space, we have two possibilities which are geometrically consistent, as discussed in subsection~\ref{subsec:dlog_form_to_geometry}. In this discussion we carved out the chiral spaces via inequalities involving $\ab{AB(n1i){\cap}(ij\star)}$ and $\ab{AB(n1j){\cap}(ij\star)}$, where $\star$ was in either of the two sets $\{i{+}1,\ldots,j{-}1\}$ or $\{1,\ldots,i{-}1\}$. However, it follows directly from the complete characterization of \emph{all} sign-flip spaces given in section~\ref{sec:sign_flip_regions} that both solutions for the pentagon (once expanded using $\ast=+\oplus-$) must be a direct sum of either sign-flip-two or four chiral spaces. For the first option eq.~(\ref{pentsign1}) (which corresponds to the choice $\star=3$), the equivalent sign-flip characterization of the space is
\begin{align}
\label{eq:pent_5_space_1}
    \begin{split}
    P^{(1)}_{24}{=}&
    \raisebox{-43pt}{\includegraphics[scale=.38]{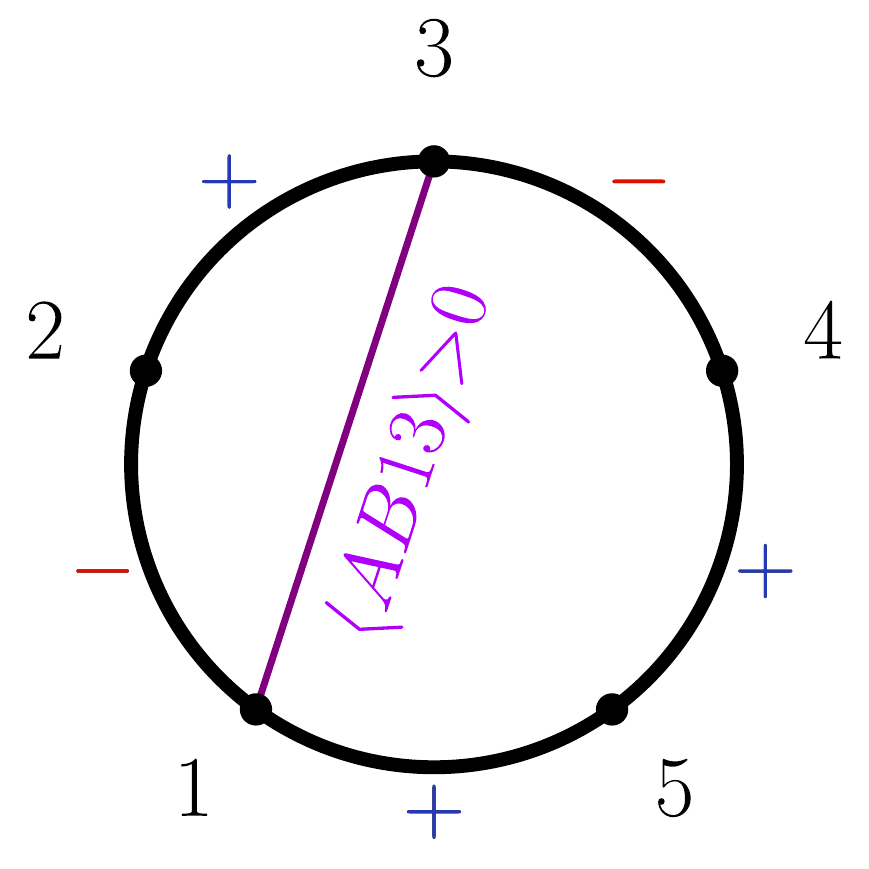}}
    {+}\raisebox{-43pt}{\includegraphics[scale=.38]{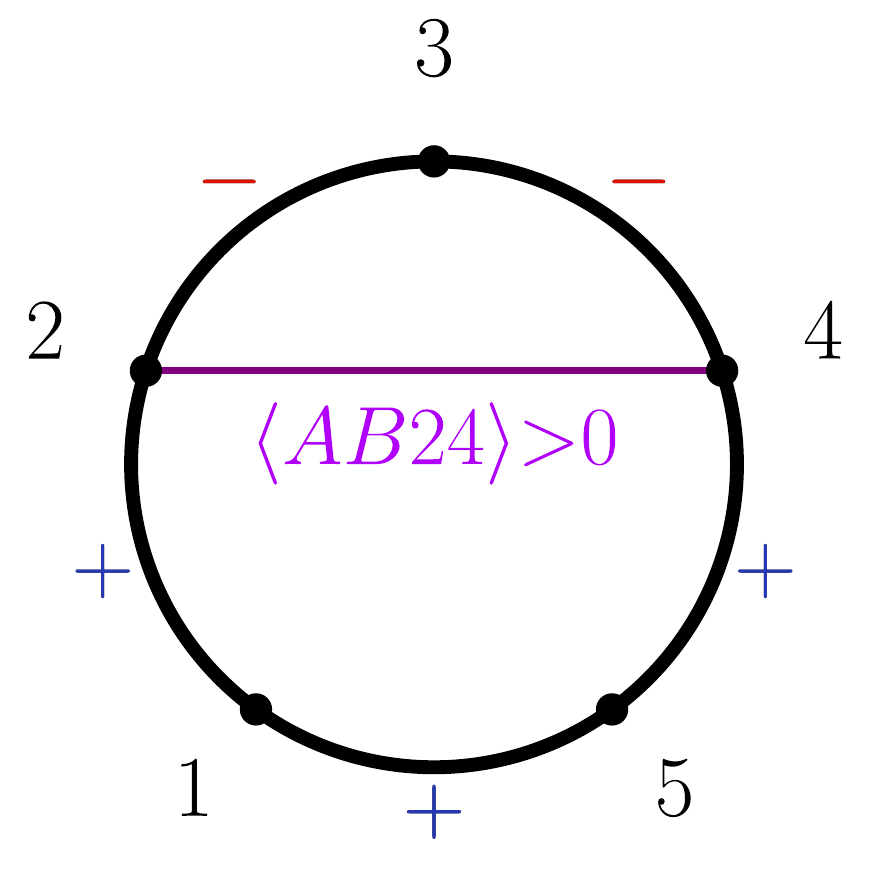}}
    {+}\raisebox{-43pt}{\includegraphics[scale=.38]{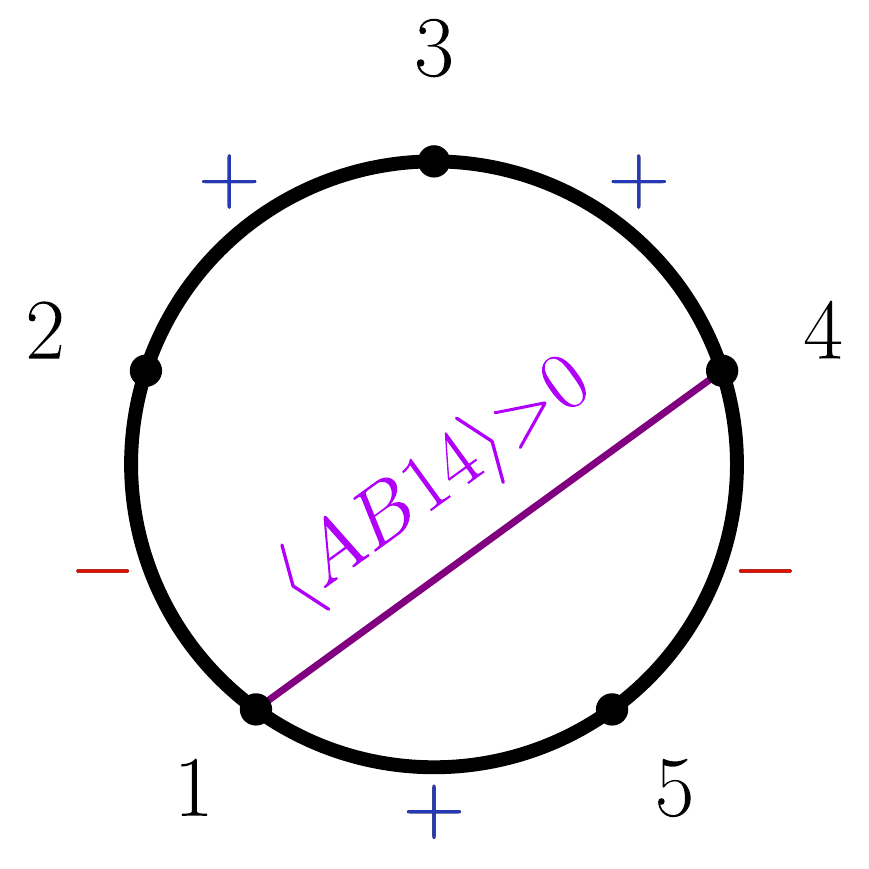}}
    {+}\raisebox{-43pt}{\includegraphics[scale=.38]{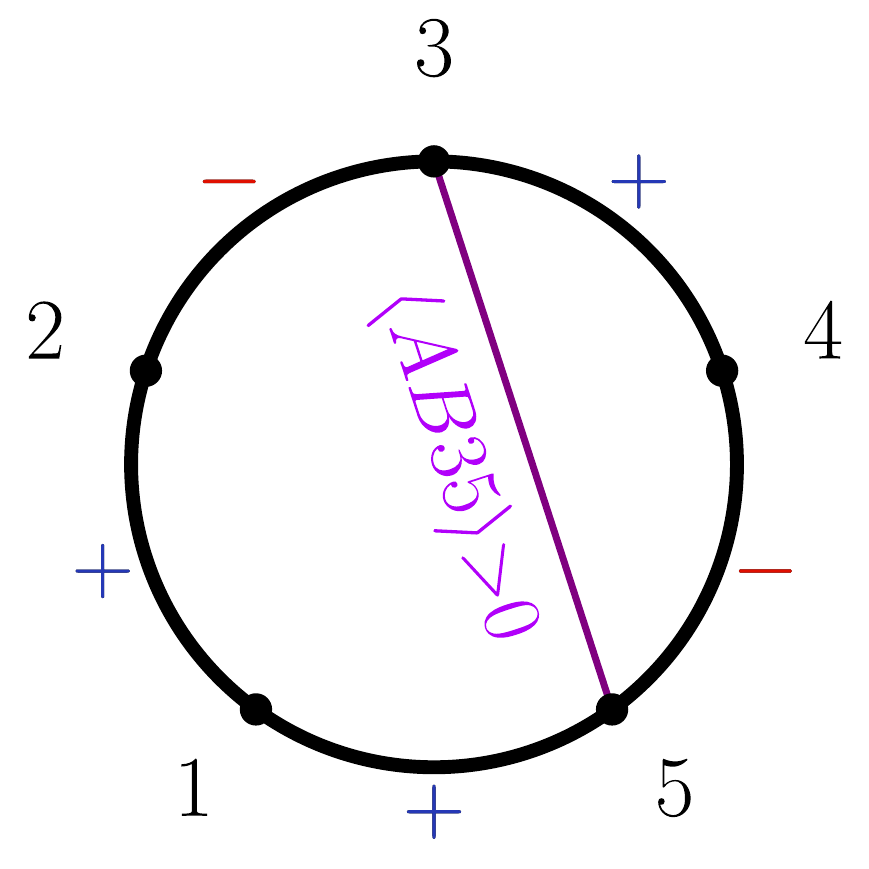}}
    \end{split}
\end{align}
where for brevity we simply label the sign-flip four spaces by a single non-adjacent chord of the circle. As discussed in section \ref{sec:sign_flip_regions} above, a single condition is always sufficient to fix this space. The second option eq.~(\ref{pentsign2}) corresponds to either choice of $\star=5$ or $\star=1$ and is equivalent to a single chiral sign-flip-four space:
\begin{equation}
\label{eq:pent_5_space_2}
    P^{(2)}_{24} = \raisebox{-45pt}{\includegraphics[scale=.4]{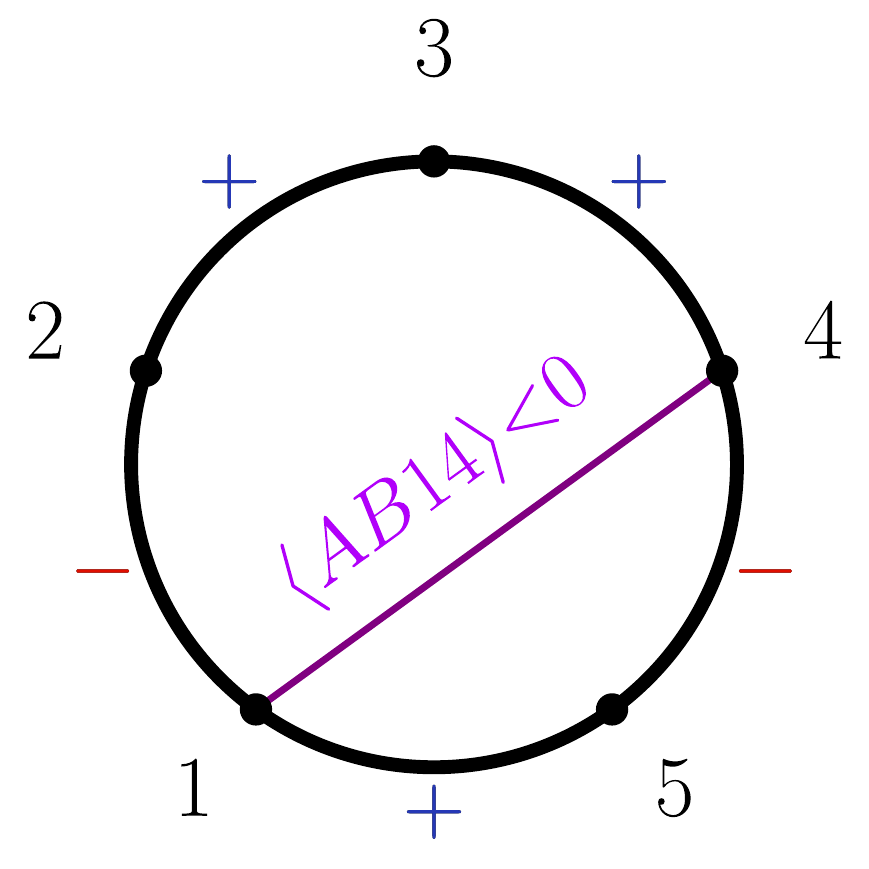}}.
\end{equation}
As a result, considering a uniform choice for the one-mass boxes eqs.~(\ref{eq:b12_spaces})--(\ref{eq:b45_spaces}) together with the two choices for the chiral pentagon eqs.~(\ref{eq:pent_5_space_1})--(\ref{eq:pent_5_space_2}) yields eight possible global geometries as a result of gluing the individual spaces. Our task is to see which (if any) of these are consistent with the \emph{boundary structure} of the original five-point MHV Amplituhedron. This is a nontrivial constraint in spite of the fact that each individual space is free of spurious boundaries from the perspective of the individual local integrals. As described in appendix~\ref{app:line_configs}, each local integral has many unphysical cuts from the perspective of the MHV one-loop geometry; a codimension-four example of this is the point $(AB)=(\overline{13})$, which is an allowed singularity of the one-mass box $B_{45}$ and the pentagon $P_{24}$ but is not an allowed singularity of the MHV Amplituhedron,
\begin{align}
    \begin{split}
       \text{not allowed in MHV:}\qquad \raisebox{-35pt}{\includegraphics[scale=.5]{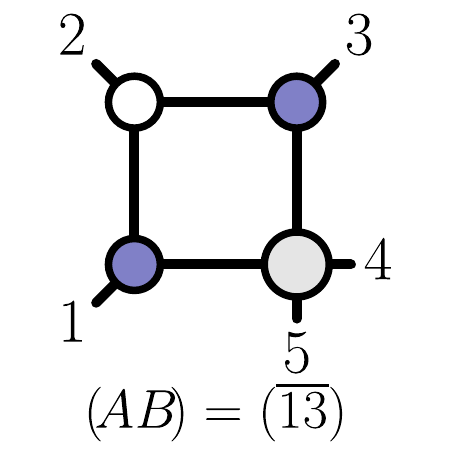}}.
    \end{split}
\end{align}
In fact, while the cancellation of leading singularities is nontrivial, there is an even stronger constraint at the level of triple cuts which we explain in detail shortly. 

\subsection{Two-dimensional projections}
\label{subsec:2d_projections}
%
In deciding which individual geometric spaces can be consistently glued together, it is best to focus on certain two-dimensional boundaries where everything can be visualized geometrically as regions, lines, and points in the projective plane. The codimension-two boundaries of interest corresponds to configurations in momentum twistor space where the line $(AB)$ intersects two adjacent lines $(i{-}1i),(i\,i{+}1)$. Codimension-two boundaries of this type are defined by solutions to $\ab{ABi{-}1i}=\ab{ABii{+}1}=0$. 

As discussed in more detail in appendix~\ref{app:line_configs} in eqs.~(\ref{eq:codim_2_line_configs}) and (\ref{eq:codim_2_line_param}) there are two solutions to these conditions, the first of which has the geometric interpretation that the line $(AB)$ passes through the point $Z_i$. On this boundary, there are four codimension-three boundaries which correspond to setting one of the remaining three $\ab{ABjj{+}1}{=}0$ together with one additional boundary which corresponds to setting $\ab{ABi{-}1i{+}1}{=}0$. Geometrically, this additional cut forces $(AB)$ to also lie in the plane $(i{-}1 i i{+}1)$. Physically, this special configuration is a collinear cut, see eq.~(\ref{eq:codim_3_line_configs}). We can furthermore look at all the higher-codimension residues that are accessible for MHV and $\MHVbar$ amplitudes. This can be summarized in the following picture of on-shell functions that label the respective cut solutions \cite{Arkani-Hamed:2016byb}.
\begin{align}
\label{fig:on-shell-diags-boundary-structure}
\hspace{-1cm}
    \raisebox{-100pt}{\includegraphics[scale=.23]{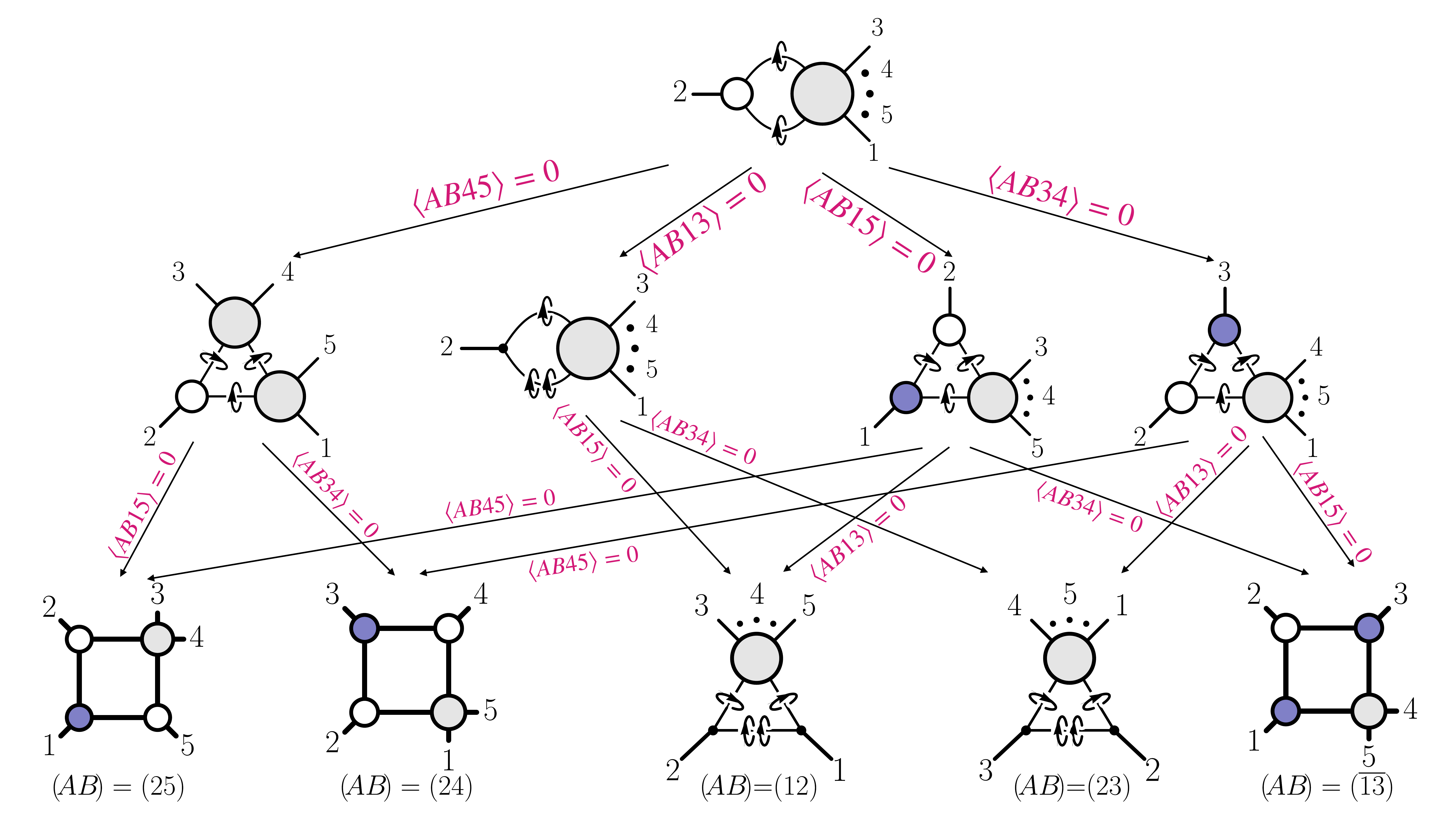}}
\hspace{-1cm}
\end{align}
Besides the physical singularities of either MHV or $\MHVbar$ amplitudes depicted in eq.~(\ref{fig:on-shell-diags-boundary-structure}), there is one additional unphysical leading singularity that could in principle appear in individual integrals,
\begin{align}
\label{eq:5pt_unphys_LS}
\text{unphysical leading singularity: }\qquad 
    \raisebox{-35pt}{\includegraphics[scale=.5]{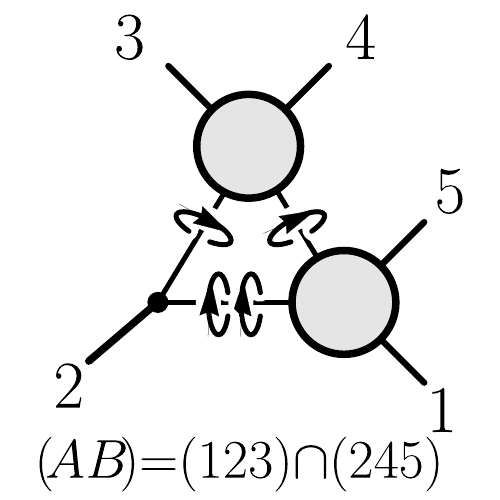}}.
\end{align}
However, for the local representation eq.~(\ref{eq:1loop_5pt_local_int_exp}), this spurious singularity is absent term-by-term. Geometrically, the content of eq.~(\ref{fig:on-shell-diags-boundary-structure}) can be encoded in the two-dimensional configuration space for the line $(AB)$ that is passing through $Z_2$,
\begin{align}
\begin{split}
    \raisebox{-32pt}{\includegraphics[scale=.6]{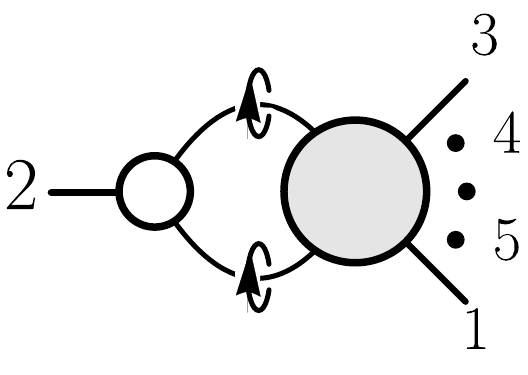}}
    \qquad
    \longleftrightarrow
    \qquad
    \raisebox{-90pt}{
    \includegraphics[scale=0.7]{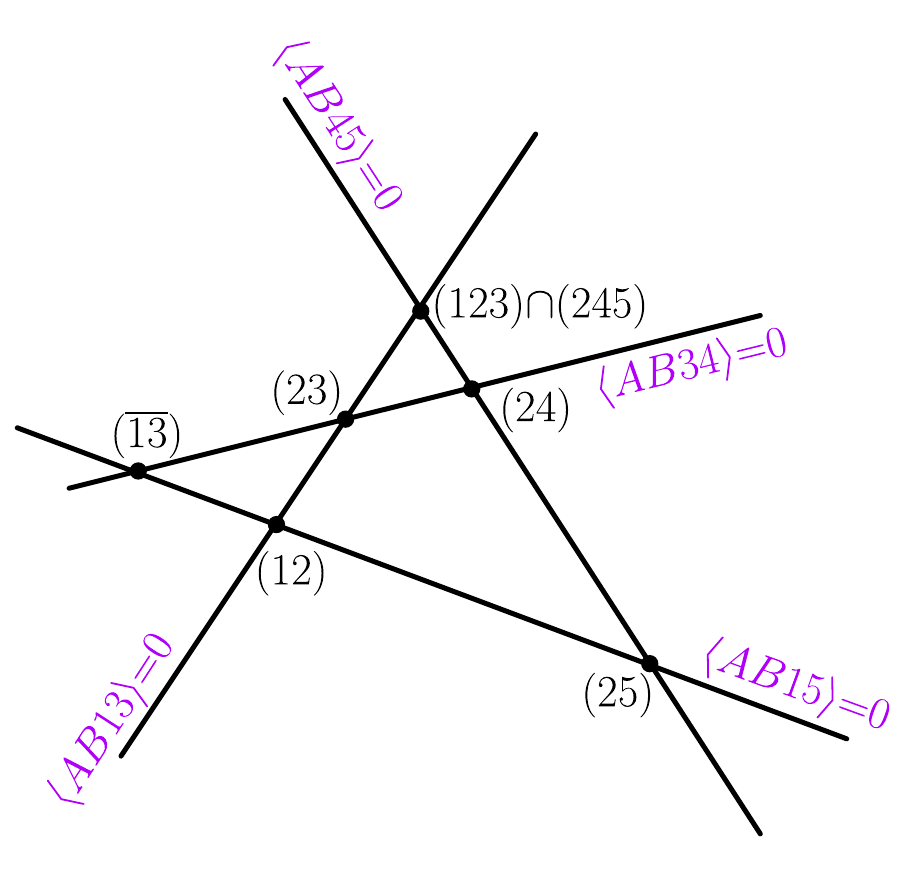}}
\end{split}    
    \label{2_proj_no_labels}
\end{align}
In this picture, lines correspond to codimension-three boundaries, i.e., configurations where $(AB)$ intersects one more line in addition to passing through point $Z_2$. The codimension-four boundaries are the points in this picture where two lines intersect, and correspond to completely fixing all four degrees of freedom in $(AB)$. In other words, the line $(AB)$ intersects two additional lines and passes through $Z_2$. These vertices correspond to positions of leading singularities which are accessible from the codimension-two surface where $(AB)=(A2)$. Note first that all triple cuts in this picture are physical, i.e., the MHV amplitude form has a non-zero residue when evaluated on all codimension-three boundaries in this picture. However, only a subset of the vertices represent actual leading singularities of the amplitude, which are what we consider physical. In particular, for the MHV amplitude only the points $(AB)=(ij)$ are physical and all others are spurious.

The positive geometries associated to the local integrals in eqs.~(\ref{eq:b12_spaces})--(\ref{eq:pent_5_space_2}) above correspond to \emph{regions} in the plane in  eq.~(\ref{2_proj_no_labels}), which can be identified by the signs of brackets which are nonvanishing when evaluated on the boundary $(AB){=}(A2)$. For $(AB){=}(A2)$, the non-vanishing brackets of interest are
\begin{equation}
    \{\ab{AB34},\ab{AB45}, \ab{AB15}, \ab{AB13}\}.
    \label{ser}
\end{equation}
Note that because this cut surface is defined by the conditions $\ab{AB12}{=}\ab{AB23}{=}0$, all information about the signs of these two brackets is lost upon accessing the $(AB){=}(A2)$ boundary. While the first three brackets in eq.~(\ref{ser}) are the usual $\ab{ABii{+}1}$ propagator-type boundaries, the bracket $\ab{AB13}$ corresponds to a spurious codimension-one boundary that only becomes physical \emph{when evaluated on the support of this cut}. On the $(AB){=}(A2)$ codimension-two boundary, sign conditions on $\ab{AB13}$ are equivalent to many other expressions, e.g., $\ab{AB\overline{24}}\rightarrow -\ab{AB13}\ab{2345}$. 

It is a relatively simple exercise to deduce the correspondence between regions in eq.~(\ref{2_proj_no_labels}) and sequences of signs for the brackets in eq.~(\ref{ser}) by looking at the relative positions of vertices with respect to certain lines and using the positivity of the external data, eq.~(\ref{eq:mhv_ext_positivity}). For example, the vertex $(AB)=(23)$ is to the left of the line $\ab{AB45}$. Taking into account $\ab{2345}>0$, we therefore conclude that the whole region to the left of the $\ab{AB45}{=}0$ line corresponds to $\ab{AB45}>0$ while the region to the right corresponds to $\ab{AB45}<0$. Again using the vertex $(AB)=(23)$, we get similar information about the regions where $\ab{AB15}\lessgtr 0$. For information on $\ab{AB34}$ we can use the vertex $(AB)=(12)$. Note that $\ab{AB13}$ vanishes for both $(AB)=(12)$ and $(AB)=(23)$ since both points are on this line. Instead, we can use $(AB)=(24)$ which is on the side of $\ab{AB13}<0$. These arguments fix the labeling of all regions
\begin{align}
\begin{split}
    \raisebox{-32pt}{\includegraphics[scale=.6]{./figures/cut_config_geometry_codim_2_point_2_5pt}}
    \qquad
    \longleftrightarrow
    \qquad
    \raisebox{-80pt}{
   \includegraphics[scale=0.75]{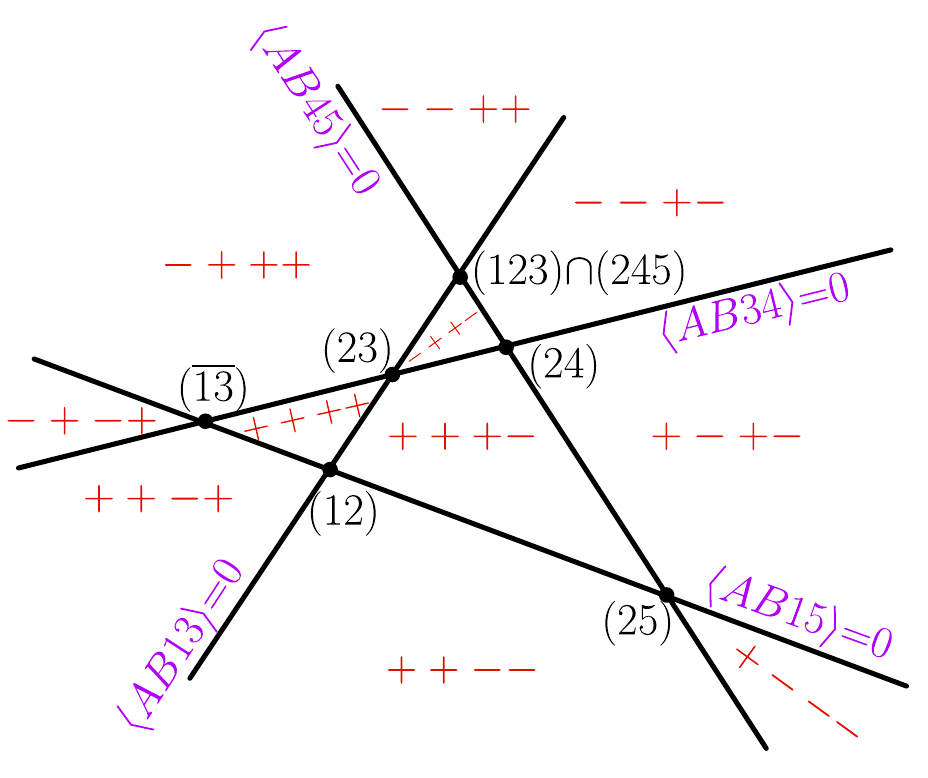}}
\end{split}    
    \label{2_proj_with_labels}
\end{align}
where we simply replace the sequence of brackets eq.~(\ref{ser}) by their respective fixed signs in a given region. Note that since the plane is projective, certain regions ``wrap around'' the point at infinity and come back on the other side of the picture, and naively have all signs opposite; an example of this is region $\{+,-,+,-\}$, which wraps around at infinity to join with $\{-,+,-,+\}$. This is a simple consequence of the fact, already alluded to above in our discussion of $d\log$ forms and geometry, that it is not the inequality $\ab{ABX}>0$ which is projectively meaningful, but rather an inequality involving a \emph{ratio} of two four brackets, $\ab{ABX}/\ab{ABY}>0$. In this context, this means that flipping all signs in the definition of a space gives a completely equivalent description of it. In eq.~(\ref{2_proj_with_labels}) the regions $\{+,-,+,-\}$ and $\{-,+,-,+\}$ are precisely the same space. Therefore, we will use the same signs for all such instances of regions which wrap around at infinity, i.e.,~we identify $\{+,-,+,-\} \sim \{-,+,-,+\}$.

Note that in eq.~(\ref{2_proj_with_labels}) the MHV Amplituhedron corresponds to the region labelled as $\{+,+,+,-\}$. This region is a quadrilateral with the four vertices $(12)$, $(23)$, $(24)$ and $(25)$ that correspond to the four physical leading singularities accessible from the $(AB){=}(A2)$ cut surface. In contrast, the $\MHVbar$ region corresponds to $\{+,+,+,+\}$ which is the triangle with vertices $(12)$, $(23)$, $(\overline{13})$.

The second solution to the cut conditions $\ab{AB12}=\ab{AB23}=0$ has the geometric interpretation that the line $(AB)$ is completely contained in the plane $(123)$,
\begin{align}
\label{eq:os_diag_5pt_AB_in_123}
    \begin{split}
        \raisebox{-32pt}{\includegraphics[scale=.6]{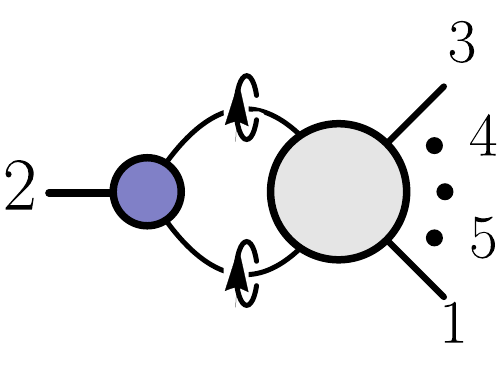}}
        \qquad
        \leftrightarrow
        \qquad
        \raisebox{-42pt}{\includegraphics[scale=.7]{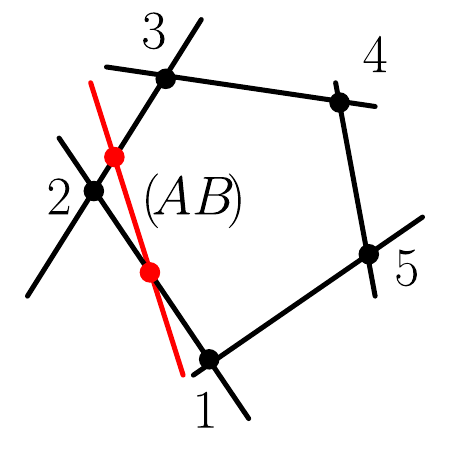}}.
    \end{split}
\end{align}
Starting from the on-shell function for the double-cut, we could again write down all possible higher codimension residues that are accessible from this surface. For the sake of brevity, we refrain from doing so here and proceed directly to the geometric picture for the configuration space of the line $(AB)$ that is contained in the plane $(123)$. Analogous to eq.~(\ref{ser}), the accessible codimension-three boundaries 
\begin{equation}
    \{\ab{AB34},\ab{AB45}, \ab{AB15}, \ab{AB24}\}
    \label{ser2}
\end{equation}
correspond to lines in the two-dimensional pictures. We chose a particular bracket, $\ab{AB24}$, to indicate on which side of the \emph{collinear boundary} the line $(AB)$ is. As mentioned above, on support of the $\ab{AB12}{=}\ab{AB23}{=}0$ cut conditions this can be re-written in various equivalent ways. The corresponding two-dimensional picture for the configuration space of the line $(AB)\subset(123)$ is,

\begin{align}
\begin{split}
\raisebox{-32pt}{\includegraphics[scale=.6]{./figures/cut_config_codim_2_plane_5pt.pdf}}
        \qquad
        \longleftrightarrow
        \qquad
\raisebox{-80pt}{        
\includegraphics[scale=0.75]{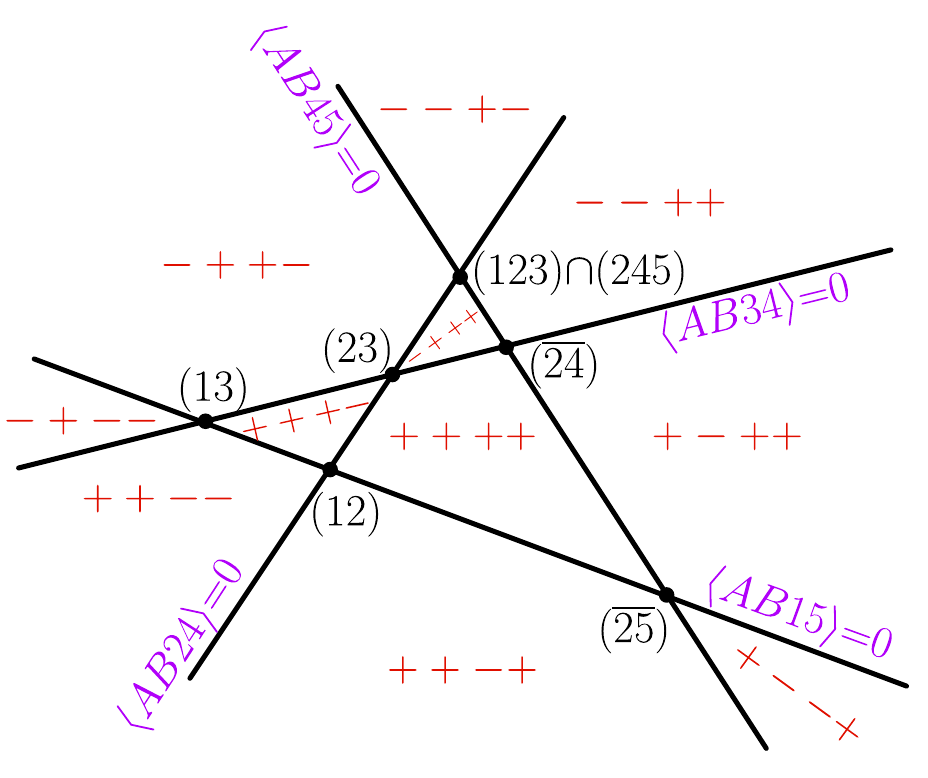}}
    \label{2_proj_with_labels_in_plane_config}
\end{split}    
\end{align}
where we have labeled the regions in terms of the signs of the brackets in the sequence eq.~(\ref{ser2}). The MHV Amplituhedron is the region $\{+,+,+,-\}$ (which on this cut surface has only three vertices) while the $\MHVbar$ region corresponds to $\{+,+,+,+\}$. Note that for the MHV geometry the entire line $\ab{AB45}=0$ is an unphysical codimension-three boundary corresponding to the on-shell function
\begin{align}
\label{eq:unphys_MHV_codim3_bdy_5pt}
\begin{split}
    \includegraphics[scale=.6]{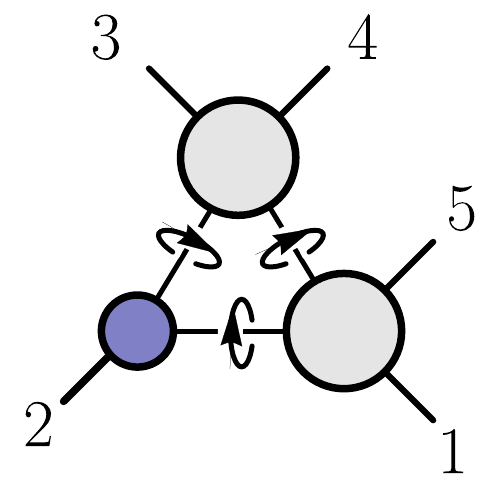}
\end{split}    
\vspace{-.6cm}
\end{align}
Geometrically, none of the physical leading singularities (vertices) lie on $\ab{AB45}{=}0$. In the same spirit, the line $\ab{AB45}=0$ was unphysical for the $\MHVbar$ Amplituhedron in the previous two-dimensional picture eq.~(\ref{2_proj_with_labels}).

All two-dimensional boundaries of the type where $(AB)\subset(i{-}1ii{+}1)$ or $(AB)$ passes through $Z_i$ have the same geometry, and the corresponding projections can be obtained by cyclically relabelling the above examples. In principle, there is one additional class of codimension-two boundaries where we set two non-adjacent $\ab{ABii{+}1}$ brackets to zero, e.g.~$\ab{AB12}{=}\ab{AB34}{=}0$. This boundary, while also two-dimensional, has a significantly more complicated stratification and is not easily visualized in the projective plane. While boundaries of this form can lead to spurious higher codimension cuts which a priori could place additional constraints on gluing together local integral spaces (discussed in the following subsection \ref{subsec:gluing_regions}), we find in practice that matching all simple two-dimensional projections mentioned above is sufficient to fix a unique consistent gluing of spaces into the ``Amplituhedron-Prime''.

\subsection{Gluing regions}
\label{subsec:gluing_regions}

While all codimension-two boundaries of the type $(AB){\subset}(i{-}1ii{+}1)$ or $(AB){=}(Ai)$ have the same geometry, in the context of the particular local expansion eq.~(\ref{pent}) we have to consider each case separately. The reason is that our global choice of $\ab{AB1n}{>}0$ for the local spaces introduced in section \ref{subsec:dlog_form_to_geometry} breaks the cyclic symmetry of the integrand and the individual contributions are different depending on the boundary we consider.

At five points we must consider ten two-dimensional projections of the form described in section~\ref{subsec:2d_projections}. For each projection, we demand that the combination of all local integrals has exactly the same \emph{boundary} (but not necessarily bulk) structure as the original Amplituhedron, i.e., there are no spurious boundaries. In this section, we first give the answer for the correct spaces for the boxes and pentagons at five and six points, and subsequently state the result for the general $n$-point geometries. We only schematically illustrate how all spurious boundaries cancel for the final correct choice of geometries for the boxes and chiral pentagons on some representative two-dimensional planes. The details on how we find a unique solution (under the assumptions discussed in subsection~\ref{sec:five_point_choices}) that holds for an arbitrary number of external points requires a careful analysis of multiple two-dimensional projections which is deferred to appendix~\ref{app:2d_gluing_details}. 

Let us briefly start with the five-point geometries, where further details are relegated to appendix \ref{app:2d_gluing_details}. From eq.~(\ref{sec:five_point_choices}), there are two box integrands, $B_{12}\,,B_{45}$, and one chiral pentagon $P_{24}$, which we can a priori associate with a number of valid geometric spaces, see (\ref{eq:b12_spaces})--(\ref{eq:pent_5_space_2}). Now we would like to combine these individual pieces into a single geometric object which has the same singularity structure as the original Amplituhedron. In order to find the consistent gluings which cancel all spurious singularities for the eight possible combinations of the box and pentagon spaces $\{B^{(i)}_{12},B^{(i)}_{45},P^{(1)}_{24},P^{(2)}_{24}\}$, we consider projections to ten codimension-two boundaries where the loop line $(AB)$ passes either through a point $(AB)=(Ai)$ or is in a plane $(AB)\subset (i{-}1\, i\, i{+}1)$. On these codimension-two boundaries, the configuration spaces for $(AB)$ become the simple two-dimensional geometries of the form of eqs.~(\ref{2_proj_with_labels}) and (\ref{2_proj_with_labels_in_plane_config}). As discussed in section \ref{subsec:2d_projections}, in these projective pictures codimension-three boundaries are lines and codimension-four boundaries (i.e.,~leading singularities) are points. In order to determine the consistent global geometries, we demand that all \emph{spurious boundaries}, i.e. boundaries that are not part of the original Amplituhedron, cancel. The result of this analysis is that, at five points, only two combinations of box and pentagon spaces survive:
\begin{equation}
\label{five_point_solns}
\{B_{12}^{(2)},B_{45}^{(2)},P^{(1)}_{24}\},
\quad\text{and}\quad \{B_{12}^{(3)},B_{45}^{(3)},P^{(1)}_{24}\}.
\end{equation}
For more details on how this was determined we refer the interested reader to appendix \ref{app:2d_gluing_details}. Of these two solutions, it is the second option, $\{B_{12}^{(3)},B_{45}^{(3)},P^{(1)}_{24}\}$, which generalizes to six and higher points, as we show in appendix~\ref{app:2d_gluing_details}. Let us briefly discuss why the second option of eq.~(\ref{five_point_solns}) is consistent on two representative two-dimensional projections. For illustrative purposes, we discuss $(i)$ the projection where $(AB)\subset(234)$, and $(ii)$ the projection where $(AB)=(A2)$. Physically, from the on-shell function point-of-view these correspond to,
\vspace{-10pt}
\begin{align}
\label{eq:os_diag_5pt_234}
    \begin{split}
        \raisebox{-32pt}{\includegraphics[scale=.6]{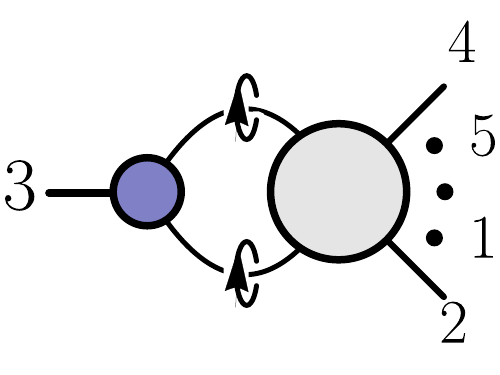}}
        \qquad \text{and}\qquad
       \raisebox{-32pt}{\includegraphics[scale=.6]{./figures/cut_config_geometry_codim_2_point_2_5pt}}
    \end{split}
    \vspace{-10pt}
\end{align}
On these two-dimensional boundaries the local spaces of eqs.~(\ref{eq:b12_spaces})--(\ref{eq:pent_5_space_2}) correspond to certain regions in the projective plane which are labeled by the signs of non-vanishing brackets in the sequences $ \{\ab{AB45}\,, \ab{AB15}\,, \ab{AB12}\,, \ab{AB35}\}$ and eq.~(\ref{ser}), respectively. The (projections of the) second solution of eq.~(\ref{five_point_solns}), which correctly generalizes to the all-multiplicity Amplituhedron-Prime, are
\begin{align}
\begin{split}
    &
    \hspace{-1cm}
    \raisebox{-130pt}{\includegraphics[scale=0.65]{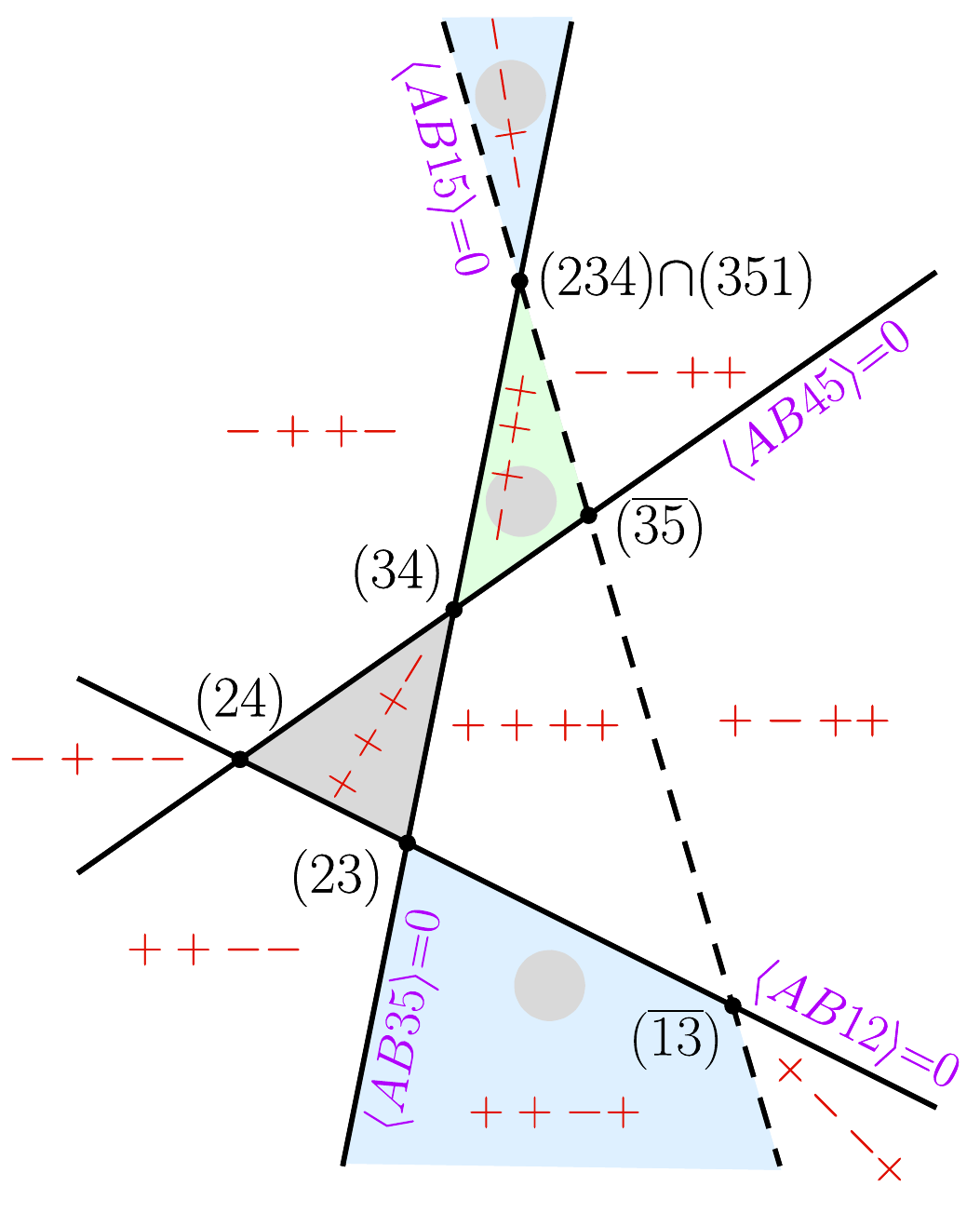}}
    \qquad
    \raisebox{-130pt}{\includegraphics[scale=0.65]{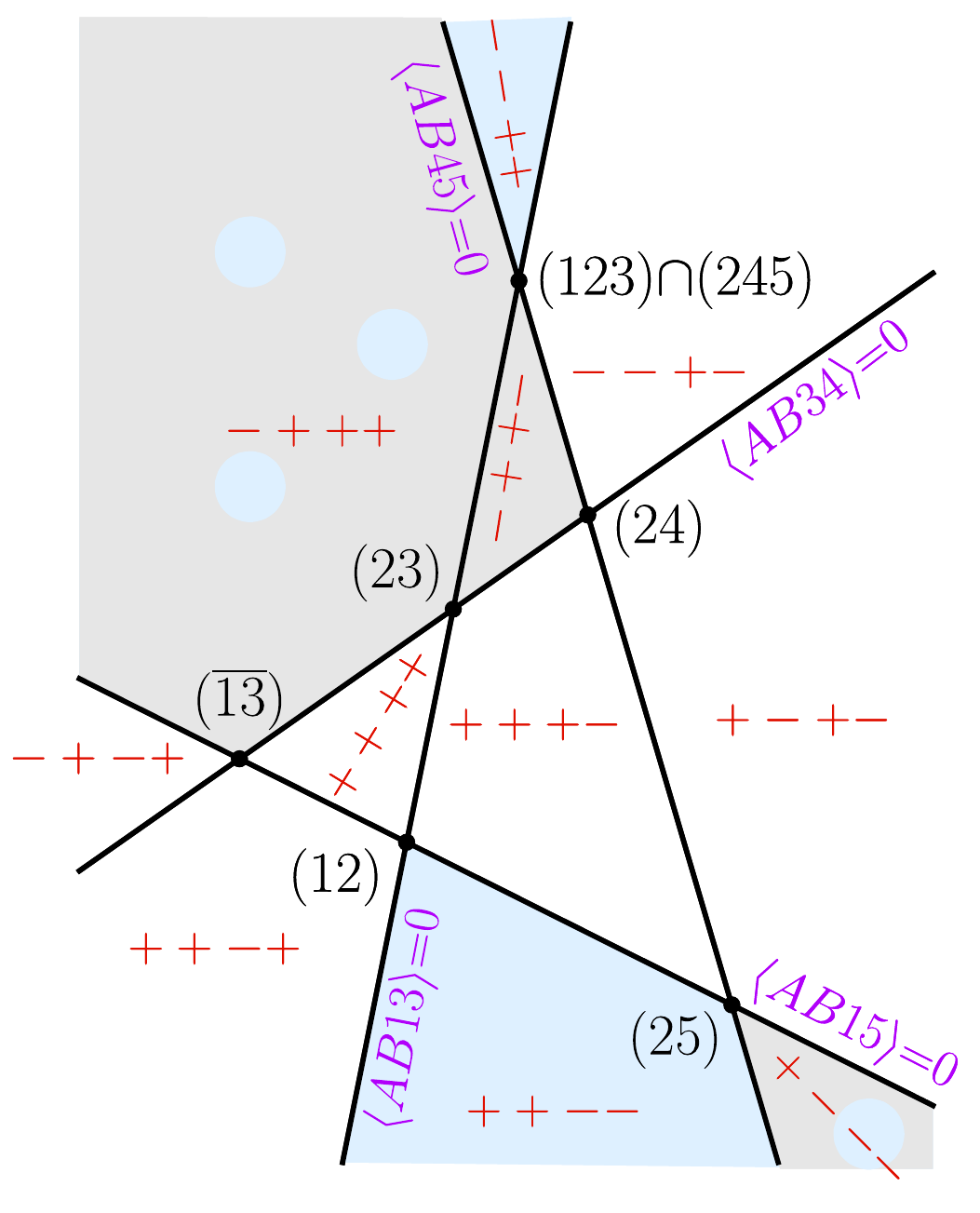}}
    \\
    & \hspace{-1cm}
        \begin{array}{|c|c|c|c|c|}
        \hline
                     & \ab{AB45} & \ab{AB15}  & \ab{AB12} & \ab{AB35} \\
        \hline    
            \cellcolor{lightgreen} B^{(3)}_{12}  &  - & + & + & + \\
       \hline    
            \cellcolor{lightblue} B^{(3)}_{45} & + & + & - & + \\
            \hline
            \cellcolor{mygrey2} P^{(1)}_{24}  &  + & + & - & + \\
            \cellcolor{mygrey2}          & + & + & + & - \\
          \cellcolor{mygrey2}          & - & + & + & + \\
        \hline  
        \end{array}
        \hspace{.8cm}
        \begin{array}{|c|c|c|c|c|}
        \hline
                     & \ab{AB34} & \ab{AB45}  & \ab{AB15} & \ab{AB13} \\
        \hline    
             \cellcolor{lightblue} B^{(3)}_{45}  & - & + & + & +\\
            \cellcolor{lightblue}             & - & - & + & +\\
       \hline
           \cellcolor{mygrey2} P^{(1)}_{24}  &  - & + & + & + \\
            \cellcolor{mygrey2}          &  - & + & + & - \\     \hline  
        \end{array}
\end{split} 
\hspace{-.5cm}
\label{eq:5pt_solns_2d_projection_plane_234}
\end{align}
where we have color-coded the regions of the relevant local spaces 
eqs.~(\ref{eq:b12_spaces})--(\ref{eq:pent_5_space_2}) on both $(AB){\subset}(234)$ (l.h.s.) and  $(AB){=}(A2)$ (r.h.s.) codimension-two boundaries. As already mentioned below eq.~(\ref{2_proj_with_labels}), spaces where we flip all signs lead to equivalent geometries; therefore, we only list one representative in the tables summarizing the contributing regions of each local integrand. This is why, e.g. on the l.h.s. of (\ref{eq:5pt_solns_2d_projection_plane_234}), the pentagon  $P^{(1)}_{24}$ fills in both regions $\{++-+\}\sim\{--+-\}$ in the above left-hand-side picture. Furthermore, not all sign patterns which are present in the full local integral contribute on a given cut surface. Thus, for $P^{(1)}_{24}$ only three of the four sign patterns in eq.~(\ref{eq:pent_5_space_1}) have access to the codimension-two boundary where $(AB)\subset(234)$. 

In order to show the consistency of our claimed solutions eq.~(\ref{five_point_solns}) for the local integrand spaces on this boundary, we have to identify the cancellation of the spurious $\ab{AB15}=0$ line. In the solution $\{B^{(3)}_{12},B^{(3)}_{45},P^{(1)}_{24}\}$ depicted on the left of eq.~(\ref{eq:5pt_solns_2d_projection_plane_234}), we trivially see the cancellation of the entire spurious line, because the regions $\{--+-\}\,, \{-+++\}\,,\{++-+\}$ are double-covered by parts of the pentagon spaces as well as the boxes. For the $(AB)=(A2)$ projection depicted on the right of eq.~(\ref{eq:5pt_solns_2d_projection_plane_234}), we see that after cancelling overlapping regions we are left with the union of three regions, $\{++--\}$, $\{--++\}$ and $\{-++-\}$. While this space has the correct boundary structure, here we can see it is non-overlapping with the original Amplituhedron which is the region $\{+++-\}$ on this projection.

Thus, at five points there is only one consistent space for the pentagon and we may cancel all spurious boundaries using two different (uniform) choices for the boxes. Both choices have exactly the same boundaries as the original Amplituhedron and are completely satisfactory at this multiplicity. However, only one of these solutions generalizes to higher points. This can be seen directly at six points, where an additional constraint arises: our five-point choice must be compatible with (at least) one of the two spaces eqs.~(\ref{2mh_boxsign1}) and (\ref{2mh_boxsign2}) for the two-mass hard box. 

Next, we discuss the local representation of the one-loop six-point MHV integrand
\begin{align}
    \label{eq:1loop_6pt_local_int_exp}
\begin{split}    
    \omega^{(6,0,1)}  = & 
    \raisebox{-30pt}{\includegraphics[scale=.4]{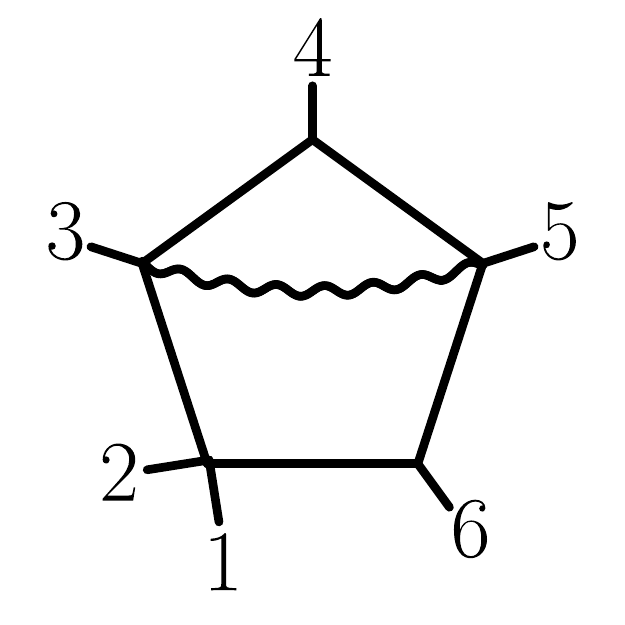}}
    +
    \raisebox{-30pt}{\includegraphics[scale=.4]{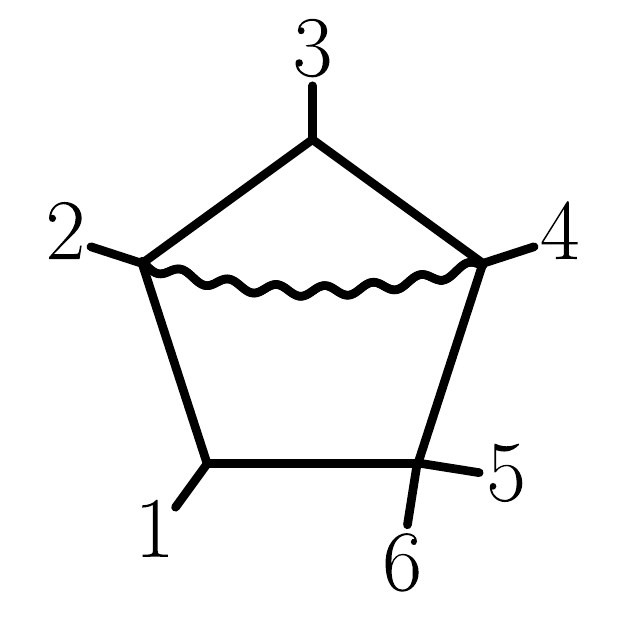}}
    +
    \raisebox{-30pt}{\includegraphics[scale=.4]{./figures/chiral_pent_6pt_34_massive}} \\
    & +
    \raisebox{-32pt}{\includegraphics[scale=.38]{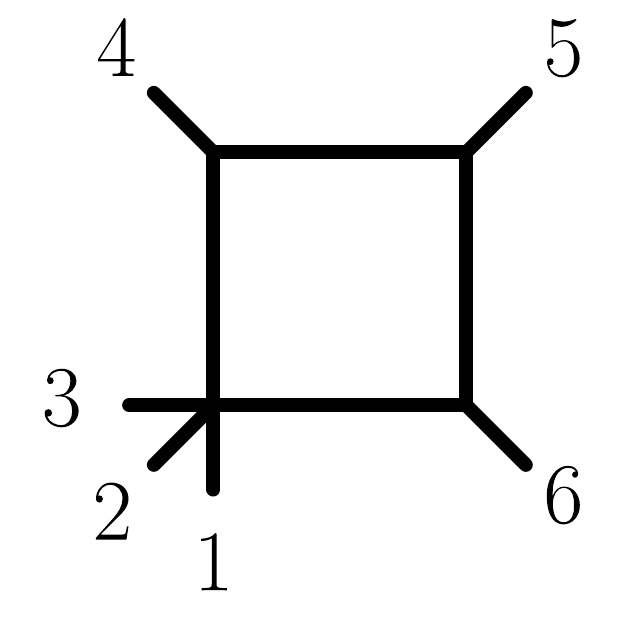}}
    +
    \raisebox{-32pt}{\includegraphics[scale=.38]{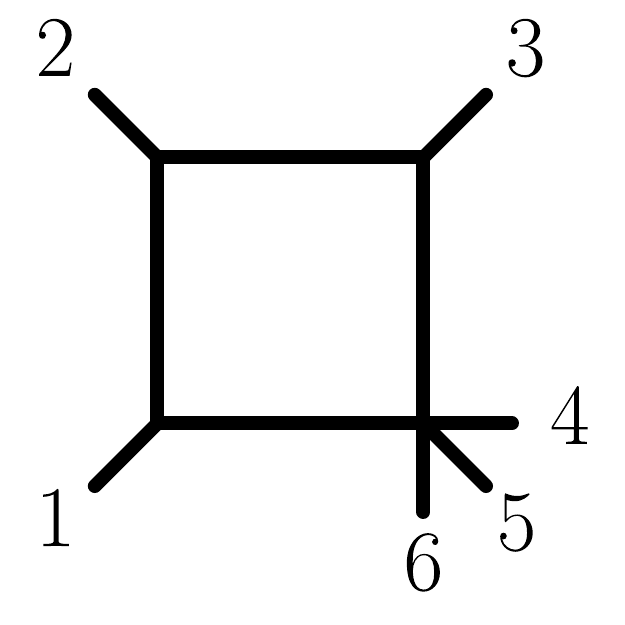}}
    +
    \raisebox{-32pt}{\includegraphics[scale=.38]{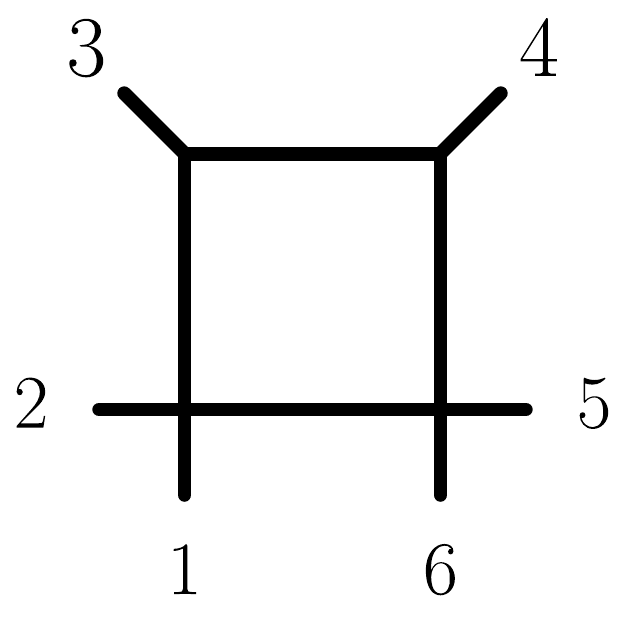}}.
\end{split}    
\end{align}
Let us consider the six-point analogue of eq.~(\ref{2_proj_no_labels}), i.e., the two-dimensional projection where $(AB)\subset(234)$. 
\begin{align}
\begin{split}
\raisebox{-32pt}{\includegraphics[scale=.6]{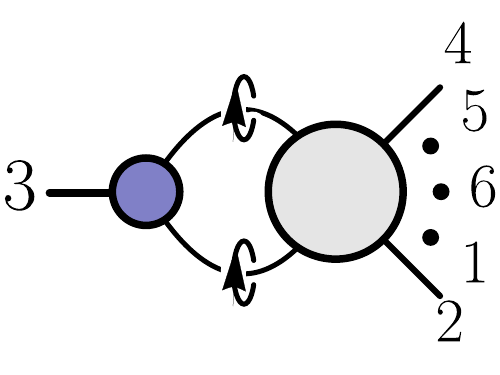}}
\quad
\longleftrightarrow
\quad
\raisebox{-100pt}{\includegraphics[scale=.7]{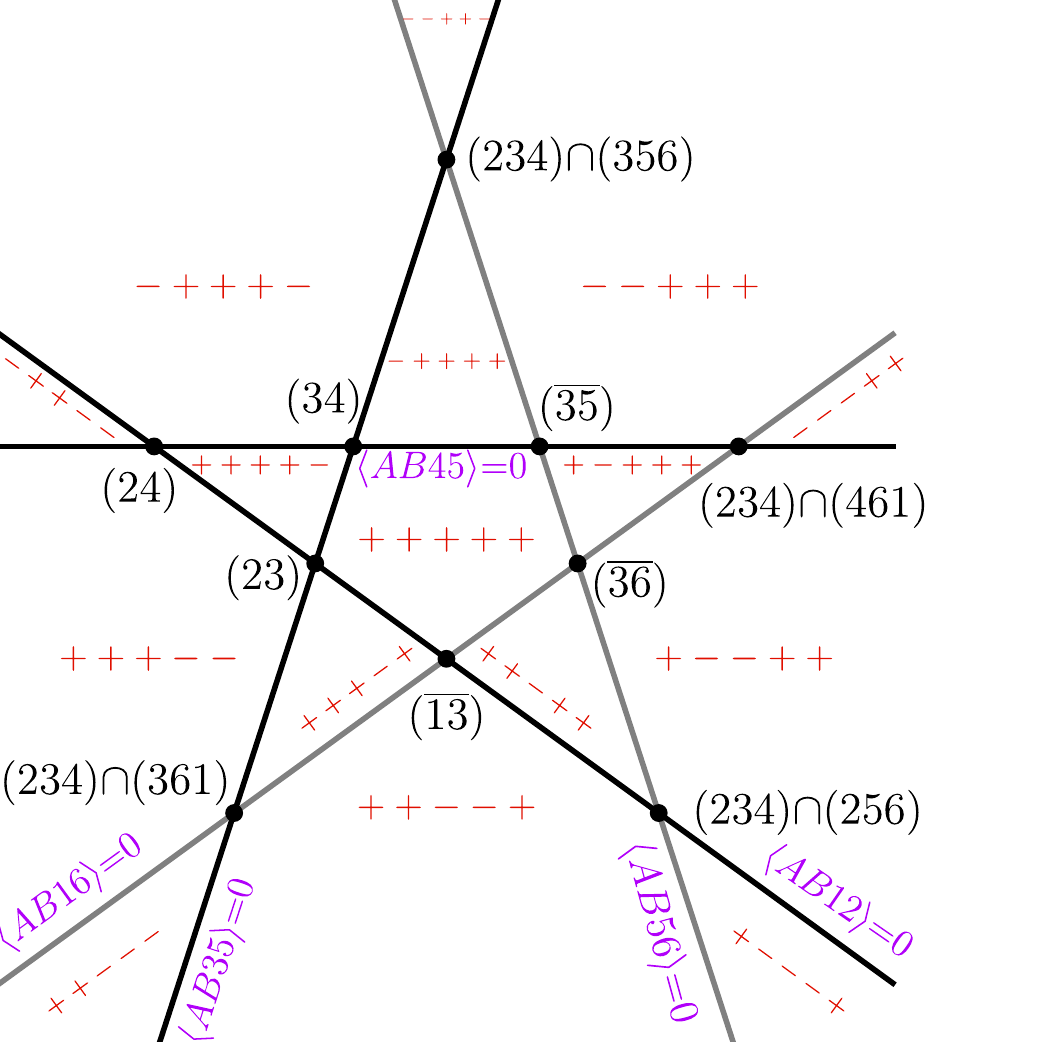}}
\end{split}    
\label{2d_proj_6pt_with_labels}
\end{align}
We have labeled the regions by the signs of the sequence of brackets 
\begin{equation}
\{\ab{AB45},\ab{AB56},\ab{AB16},\ab{AB12},\ab{AB35}\},
\end{equation}
so that the MHV Amplituhedron is the $\{+,+,+,+,-\}$ triangle region with vertices $(23), (24), (34)$. At six points there are now \emph{two} spurious lines, defined by the conditions $\ab{AB56}{=}0$ and $\ab{AB16}{=}0$. These have  simple on-shell function interpretations analogous to eq.~(\ref{eq:unphys_MHV_codim3_bdy_5pt}), but for brevity we do not write them explicitly here. In the chiral pentagon expansion eq.~(\ref{eq:1loop_6pt_local_int_exp}), three terms contribute on this cut configuration:
\begin{equation}
\hspace{-1cm}
\omega^{(6,0,1)}\bigg|_{\text{cut}}  =
\hspace{-.2cm}
\left.
\raisebox{-32pt}{\includegraphics[scale=.4]{./figures/chiral_pent_6pt_56_massive}}
+ 
\raisebox{-32pt}{\includegraphics[scale=.4]{./figures/1mass_box_right_6pt}}
+
 \raisebox{-32pt}{\includegraphics[scale=.4]{./figures/2mass_hard_box_6pt}}
 \right|_{\text{cut}}.
\end{equation}
This is the first example where the two-mass-hard box plays a r\^{o}le. We now take the generalizations of the two consistent spaces for the one-mass box and chiral pentagons from our five-point analysis eq.~(\ref{five_point_solns}) and augment them by two options for the two-mass hard spaces, see eqs.~(\ref{2mh_boxsign1}) and (\ref{2mh_boxsign2}). The details of this analysis can be found in appendix~\ref{app:2d_gluing_details}, but roughly speaking the strategy involves consistently canceling spurious boundaries on all two-dimensional projections.  

Ultimately, we find only the generalization of the second solution in eq.~(\ref{five_point_solns}) together with the first choice $B_{12,56}^{(1)}$ defined in eq.~(\ref{npt_two_mass}) is the \emph{unique} (subject to the assumption that we make uniform choices for all boxes and pentagons, respectively) candidate space which is free of all spurious boundaries on all cut surfaces. For the $(AB)\subset(234)$ projection discussed above, the relevant spaces are defined by the following by-now familiar circle diagrams: 
\begin{align}
\label{6pt_b3_456_space}
    B_{456}^{(3)}=& \raisebox{-45pt}{\includegraphics[scale=.37]{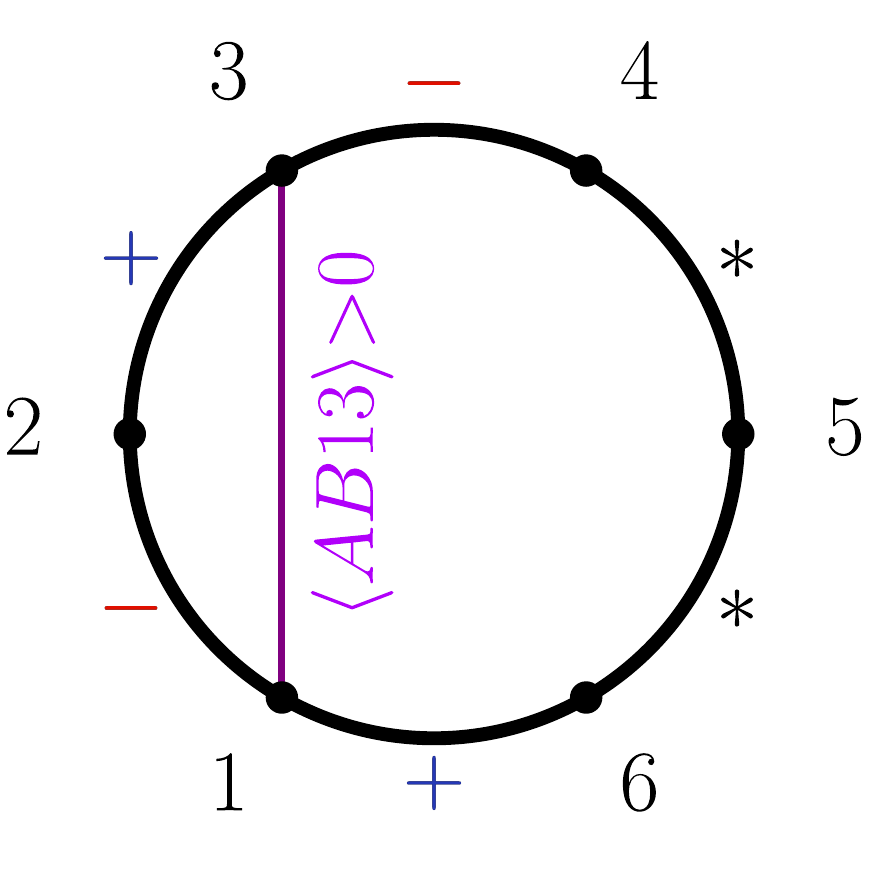}} \,,
   \\
   \label{6pt_p1_24_space}
    P_{24}^{(1)}{=}&
     \raisebox{-45pt}{\includegraphics[scale=.37]{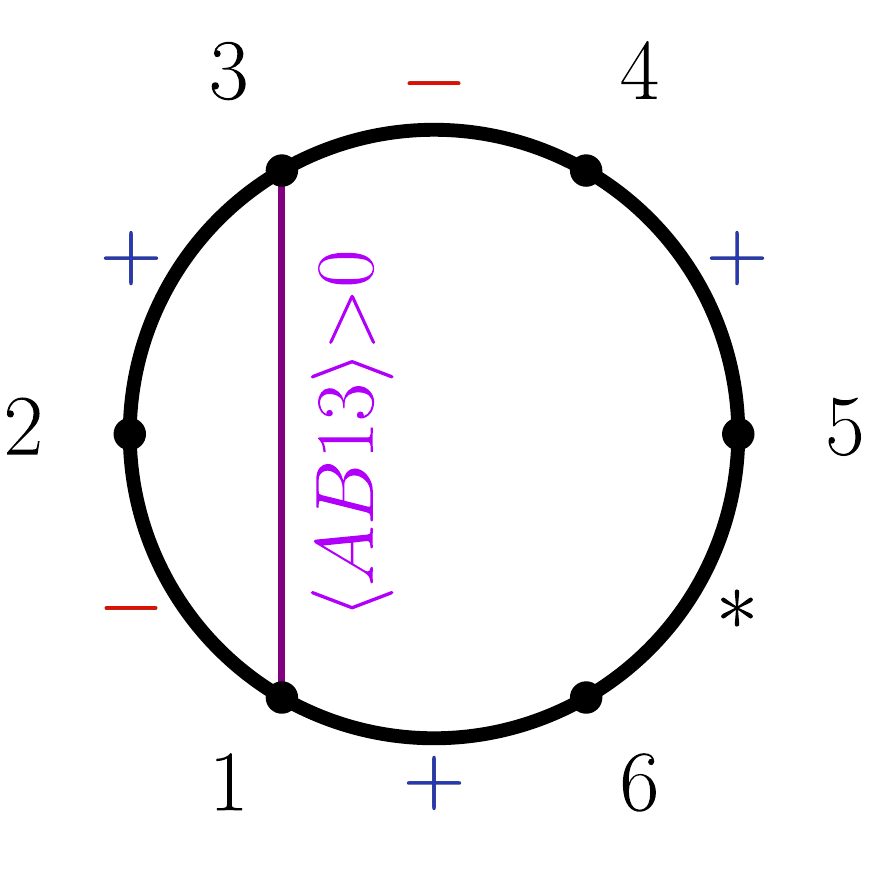}}
   {+}\raisebox{-45pt}{\includegraphics[scale=.37]{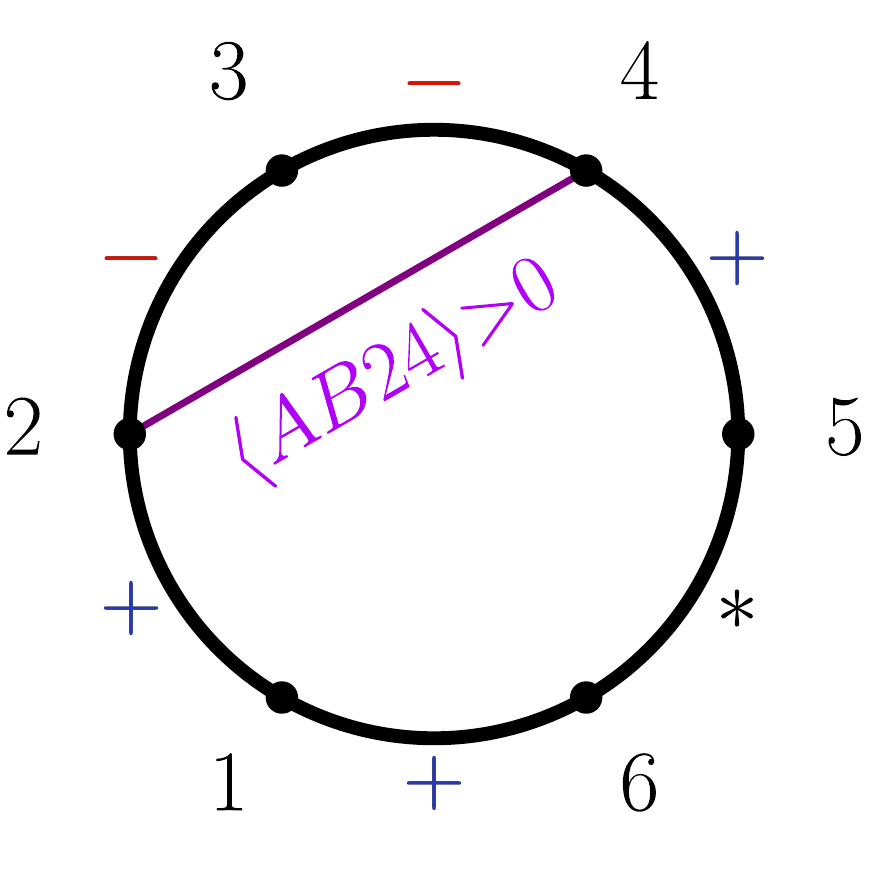}}
    {+}\raisebox{-45pt}{\includegraphics[scale=.37]{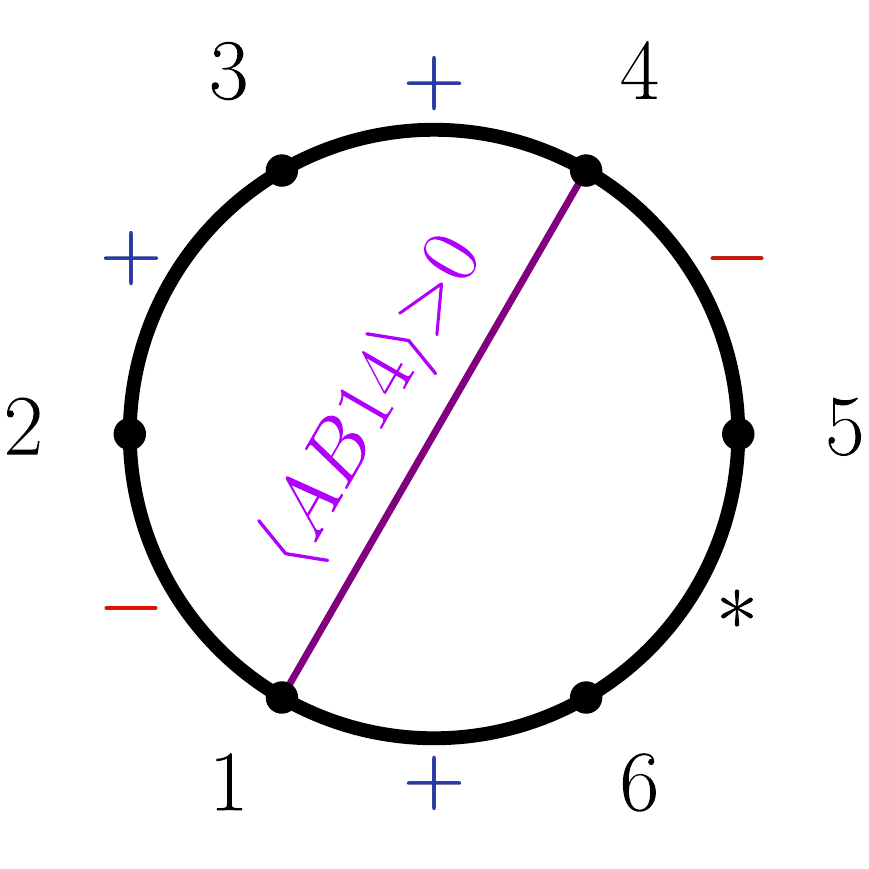}}
    {+}\raisebox{-45pt}{\includegraphics[scale=.37]{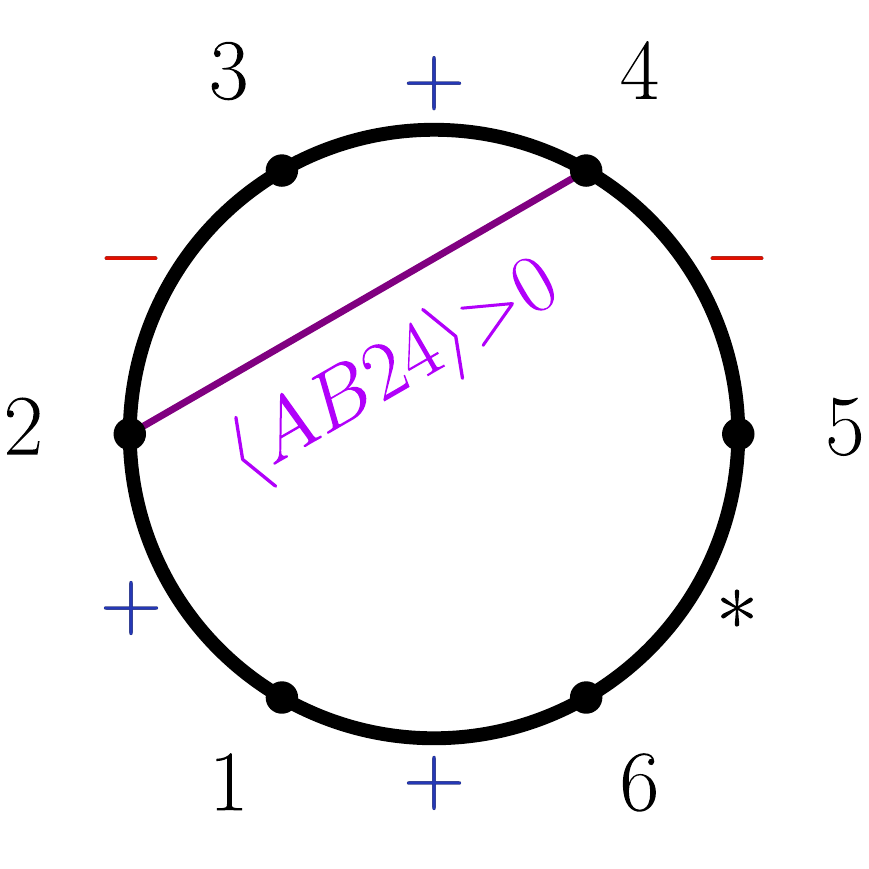}} 
    \\
    \label{6pt_b1_1256_space}
    B_{12,56}^{(1)}=&\raisebox{-45pt}{\includegraphics[scale=.37]{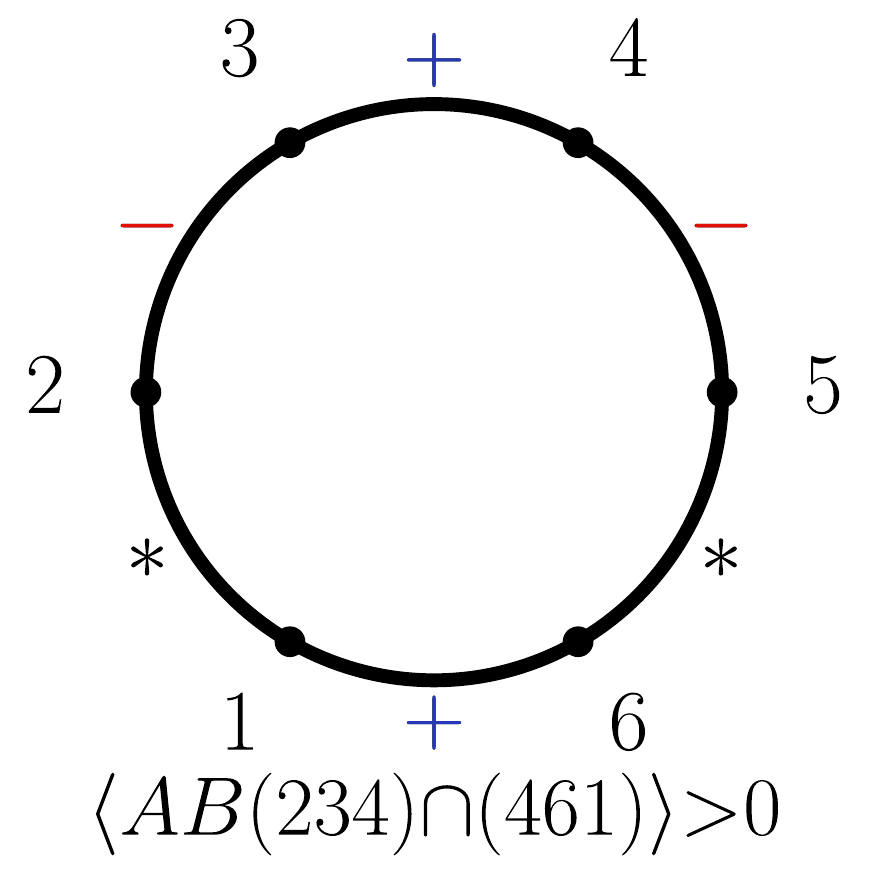}}.
\end{align}
Filling in the regions corresponding to the spaces defined in eqs.~(\ref{6pt_b3_456_space})--(\ref{6pt_b1_1256_space}) in the two-dimensional projection eq.~(\ref{2d_proj_6pt_with_labels}), the result is:

\begin{align}
\label{eq:6pt_solns_2d_projection_plane_234}
\begin{split}
&
\raisebox{-130pt}{\includegraphics[scale=0.6]{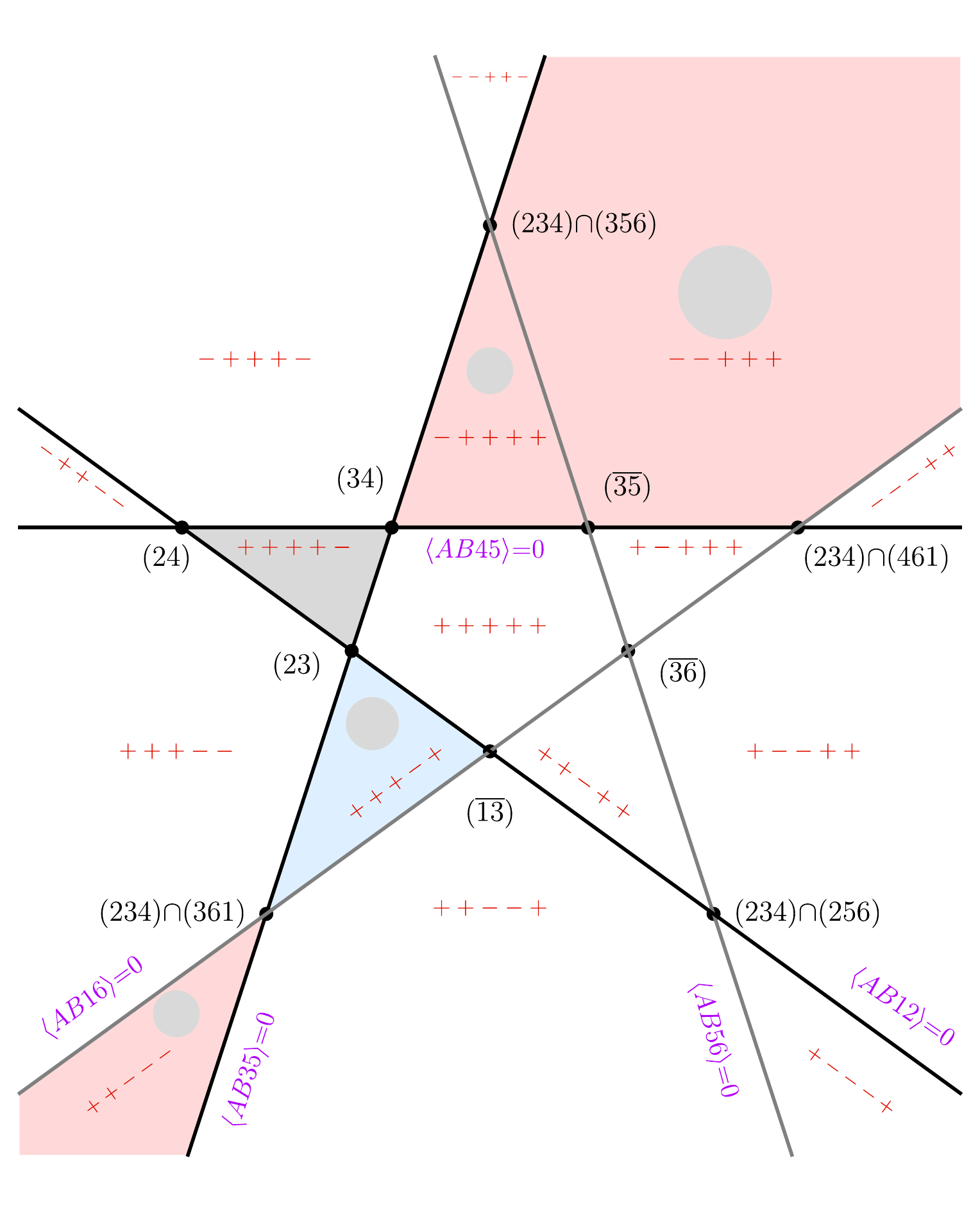}}
    \\[-15pt]
& 
    \begin{array}{|c|c|c|c|c|c|}
        \hline
                     & \ab{AB45} & \ab{AB56}  & \ab{AB16} & \ab{AB12} & \ab{AB35} \\
         \hline    
            \cellcolor{lightblue} B_{456}^{(3)}   & + & + & + & - & +\\
        \hline    
            \cellcolor{mygrey2}    P_{24}^{(1)}   & + & + & + & - & +\\[0pt]
            \cellcolor{mygrey2}                  & + & + & + & + & - \\[0pt]   
            \cellcolor{mygrey2}                  & - & + & + & + & + \\[0pt]   
            \cellcolor{mygrey2}                  & - & - & + & + & + \\             
        \hline    
            \cellcolor{lightred}    B_{12,56}^{(1)}   & - & + & + & + & +\\
            \cellcolor{lightred}                      & - & - & + & + & + \\   
       \hline    
    \end{array}  
\end{split}    
\end{align}
Again, we see that the chiral pentagon overlaps with the one-mass and two-mass-hard box in all regions that have spurious boundaries and therefore cancels those geometrically. On this particular two-dimensional projection, we are left with the triangle with vertices $(23),(24)$ and $(34)$ corresponding to the MHV space. We have verified the cancellation of spurious boundaries on all other two-dimensional projections. \\

\noindent
The five- and six-point examples heretofore considered suggest a unique conjecture for the all-multiplicity Amplituhedron-Prime. Namely, we consider the union of the one-mass box spaces eq.~(\ref{1m_boxsign34}) (choosing the positive sign for the bracket involving the leading singularity), the first pentagon space eq.~(\ref{pentsign1}) and the first two-mass hard box space eq.~(\ref{2mh_boxsign1}). We give explicit formulae for these spaces in terms of the sign-flip spaces of section~\ref{sec:sign_flip_regions} in the following subsection. This encompasses all integrand topologies that enter the $n$-point amplitude in the chiral pentagon expansion eq.~(\ref{pent}).\footnote{Although the fully massive chiral pentagon first appears at eight points, this generates no additional complications in our analysis.} Checking higher-point generalizations of projections such as eqs.~(\ref{eq:5pt_solns_2d_projection_plane_234}) and (\ref{eq:6pt_solns_2d_projection_plane_234}) is a straightforward, if tedious, exercise. At seven points, we have verified that all spurious boundaries accessible from the codimension-two configurations $(AB)\subset(i{-}1ii{+}1)$ cancel geometrically. At eight points, we have also verified by parametrizing $(AB)$ with four real degrees of freedom that the spurious triple cuts which cut three non-adjacent propagators (which are not visible in the two-dimensional projections considered above) cancel as functions of the last remaining degree of freedom in $(AB)$. 

\subsection{Amplituhedron-Prime}
\label{subsec:Amplituhedron_prime}
%
In the previous subsection, we have seen that demanding a consistent gluing of spaces associated to general chiral pentagon and box integrands led to a unique definition for the individual local geometries. In terms of the original integrals appearing in the chiral pentagon expansion, we associate to the generic chiral pentagon the space eq.~(\ref{pentsign1}). From the results of section~\ref{sec:sign_flip_regions} it follows that the additional inequalities involving $\ab{ABX_i}$ and $\ab{ABX_j}$ can always be replaced by one of the conditions defining the sign-flip-two or four spaces. As such, the space in eq.~(\ref{pentsign1}) can be represented in terms of our sign-flip circle-diagrams as a direct sum of four spaces:
\begin{align}
\label{npt_pent}
 P^{(1)}_{ij} \quad  \leftrightarrow    \raisebox{-35pt}{\includegraphics[scale=.6]{figures/chiral_pent.pdf}}  
 \leftrightarrow & \quad 
 \raisebox{-45pt}{\includegraphics[scale=.4]{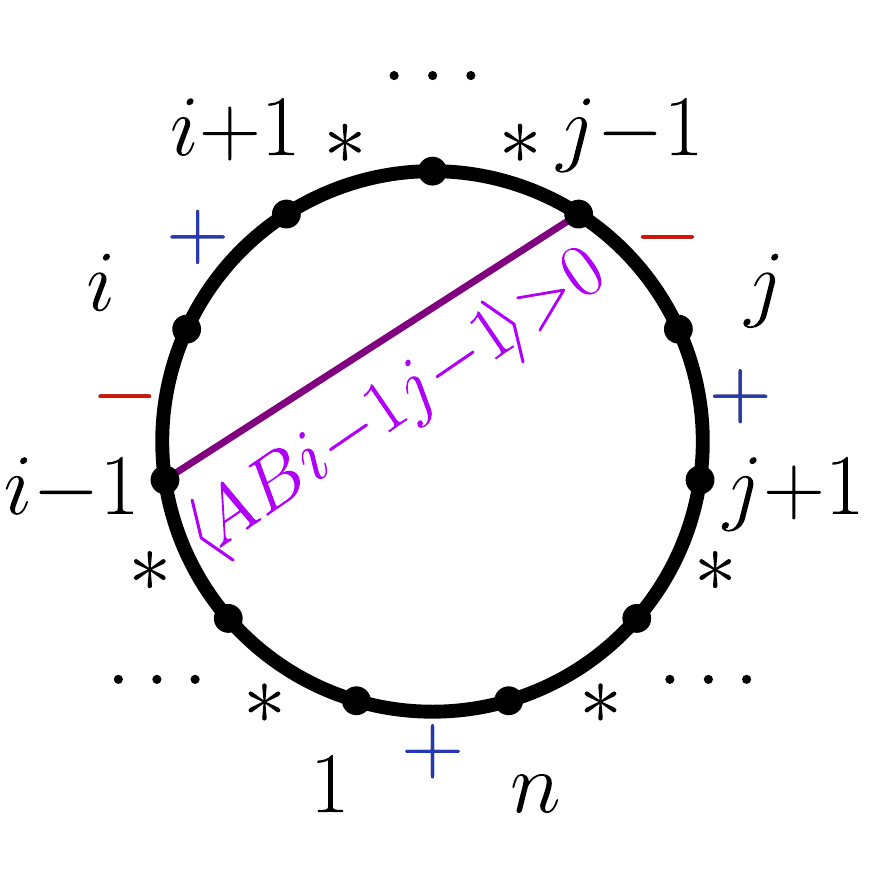}}
 +
  \raisebox{-45pt}{\includegraphics[scale=.4]{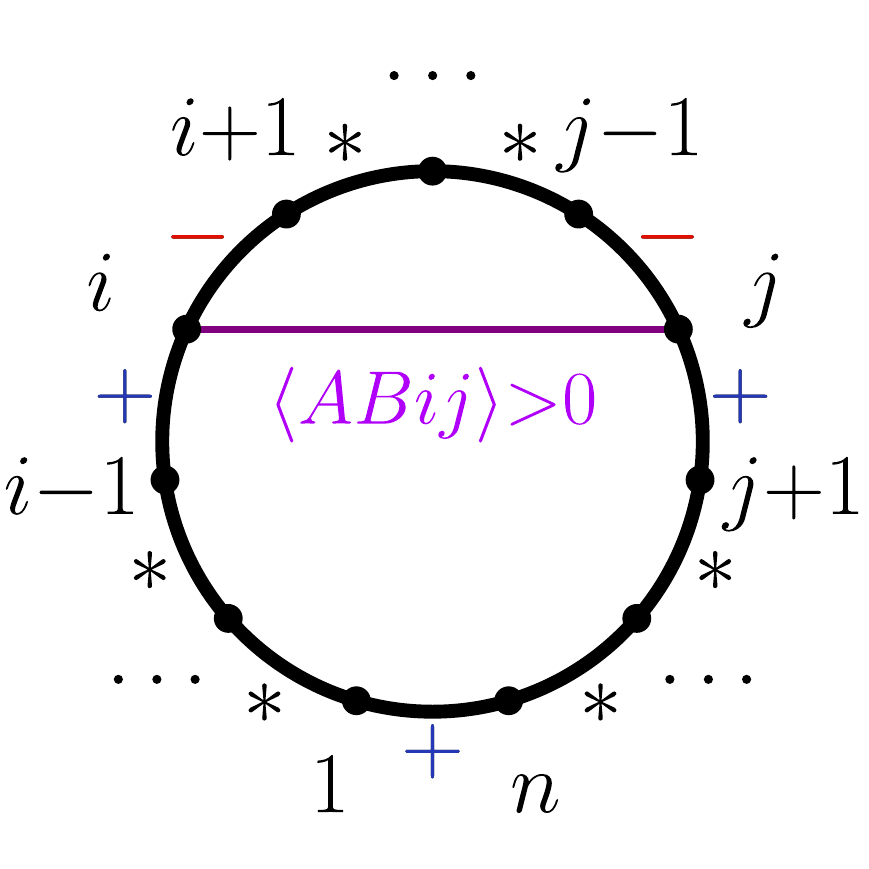}} 
  \\[-10pt] &
  +
  \raisebox{-45pt}{\includegraphics[scale=.4]{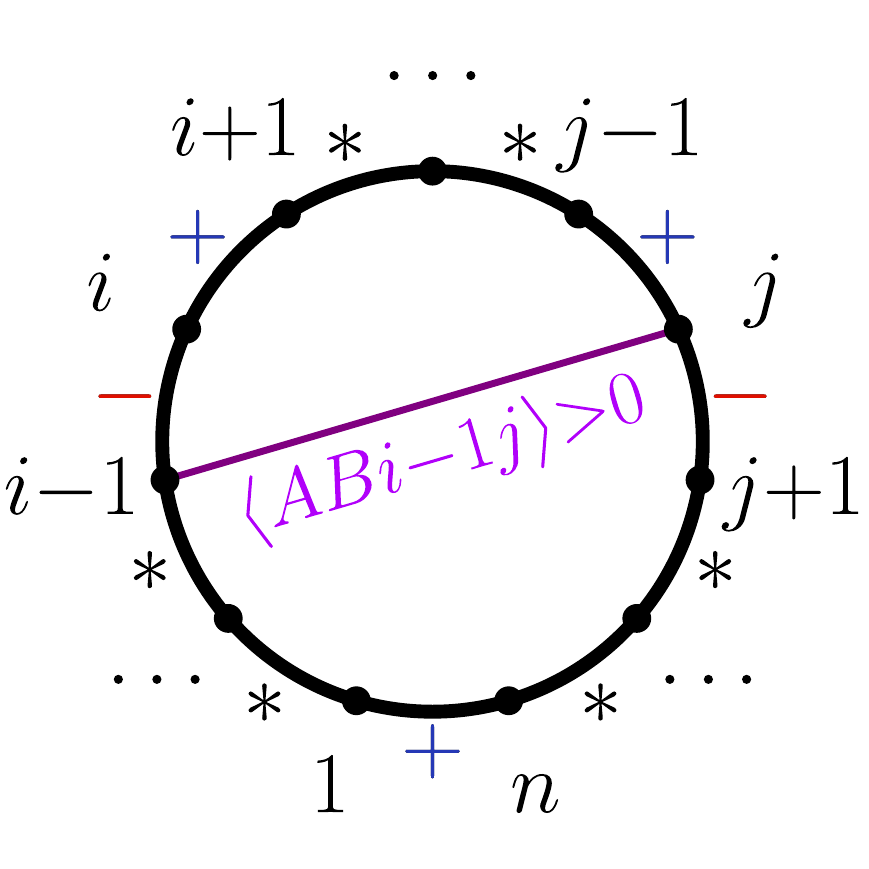}}
   +
  \raisebox{-45pt}{\includegraphics[scale=.4]{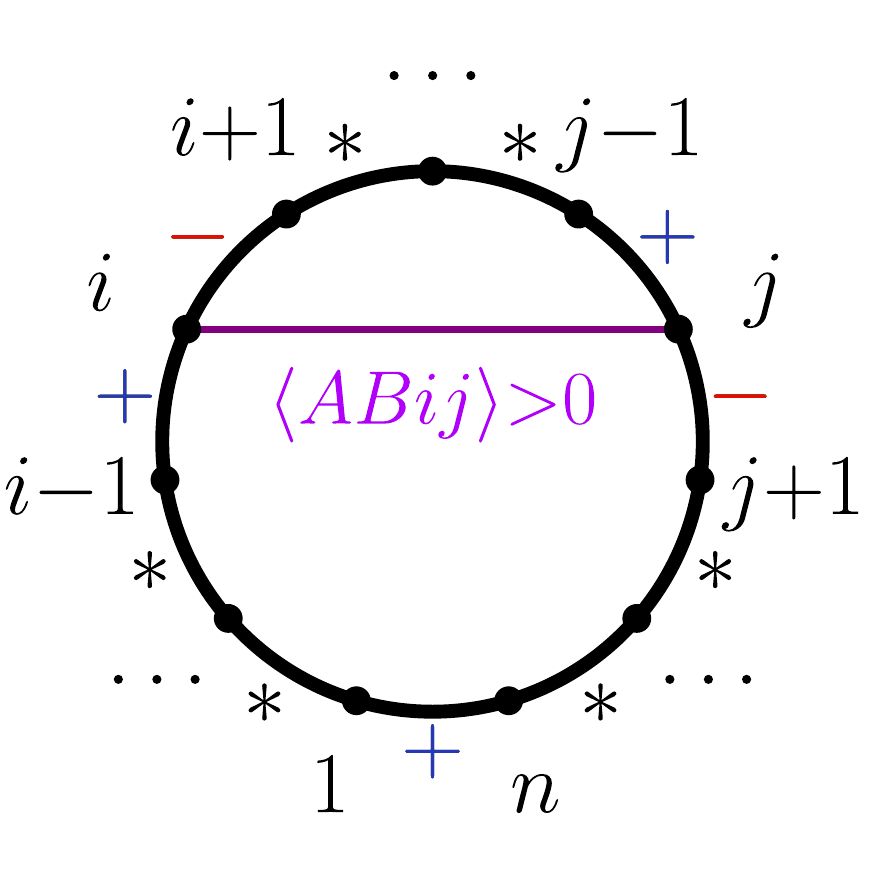}}. \nonumber
\end{align}
The one-mass and two-mass-hard box spaces in eq.~(\ref{1m_boxsign34}) and eq.~(\ref{2mh_boxsign1}), respectively, have the following sign-flip representations:
\begin{align}
\begin{split}
B^{(3)}_{1\,n{-}3} 
\quad  &  \quad \leftrightarrow \quad \raisebox{-50pt}{\includegraphics[scale=.7]{./figures/1mass_box_left.pdf}}
\quad \leftrightarrow \quad 
\raisebox{-48pt}{\includegraphics[scale=.43]{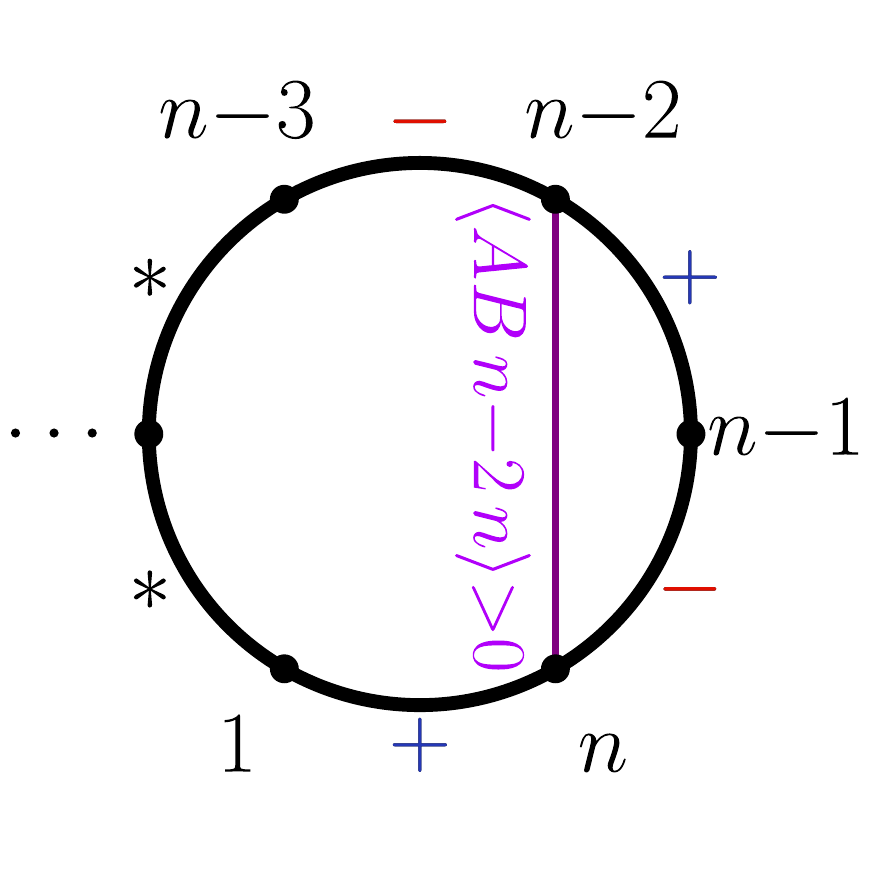}}
\\[-10pt]
B^{(3)}_{4\,n}
\quad  &  \quad \leftrightarrow \quad\quad
\raisebox{-38pt}{\includegraphics[scale=.6]{./figures/1mass_box_right.pdf}}
\quad  \leftrightarrow \quad  \hspace{.1cm}
\raisebox{-45pt}{\includegraphics[scale=.4]{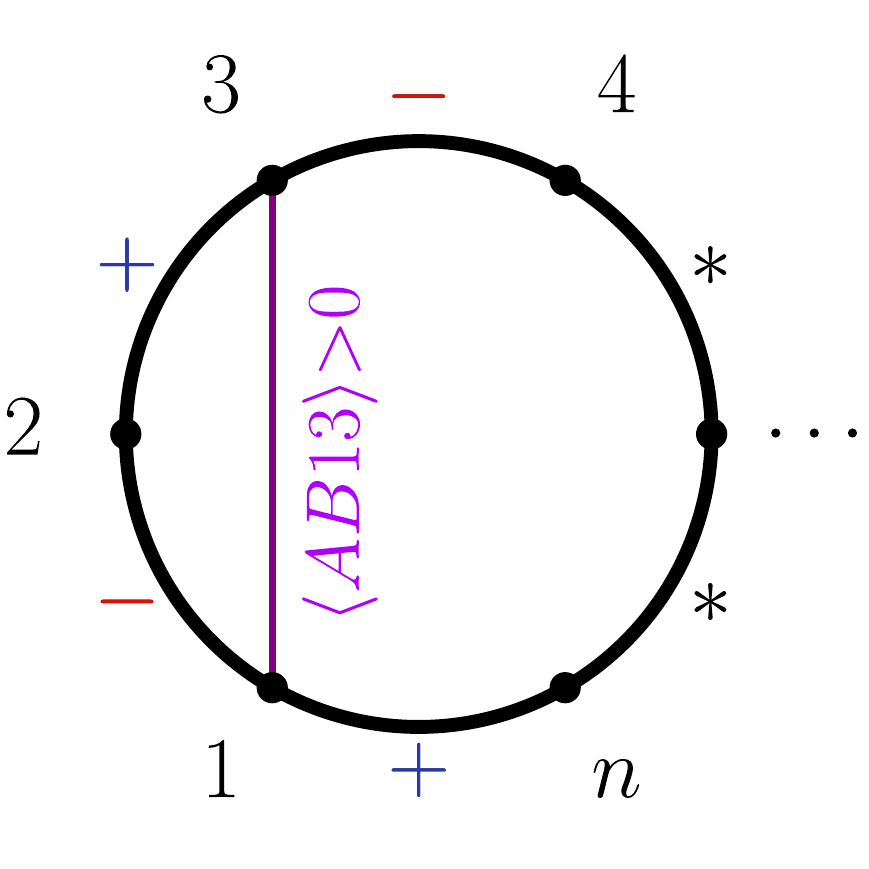}}
    \label{npt_one_mass}
\end{split}    
\end{align}
\begin{equation}
B^{(1)}_{1\,i{-}1,\,i{+}2\,n} \quad \leftrightarrow \quad 
\raisebox{-50pt}{\includegraphics[scale=.7]{./figures/2mass_hard_box.pdf}}
\quad \leftrightarrow \quad 
\raisebox{-55pt}{\includegraphics[scale=.4]{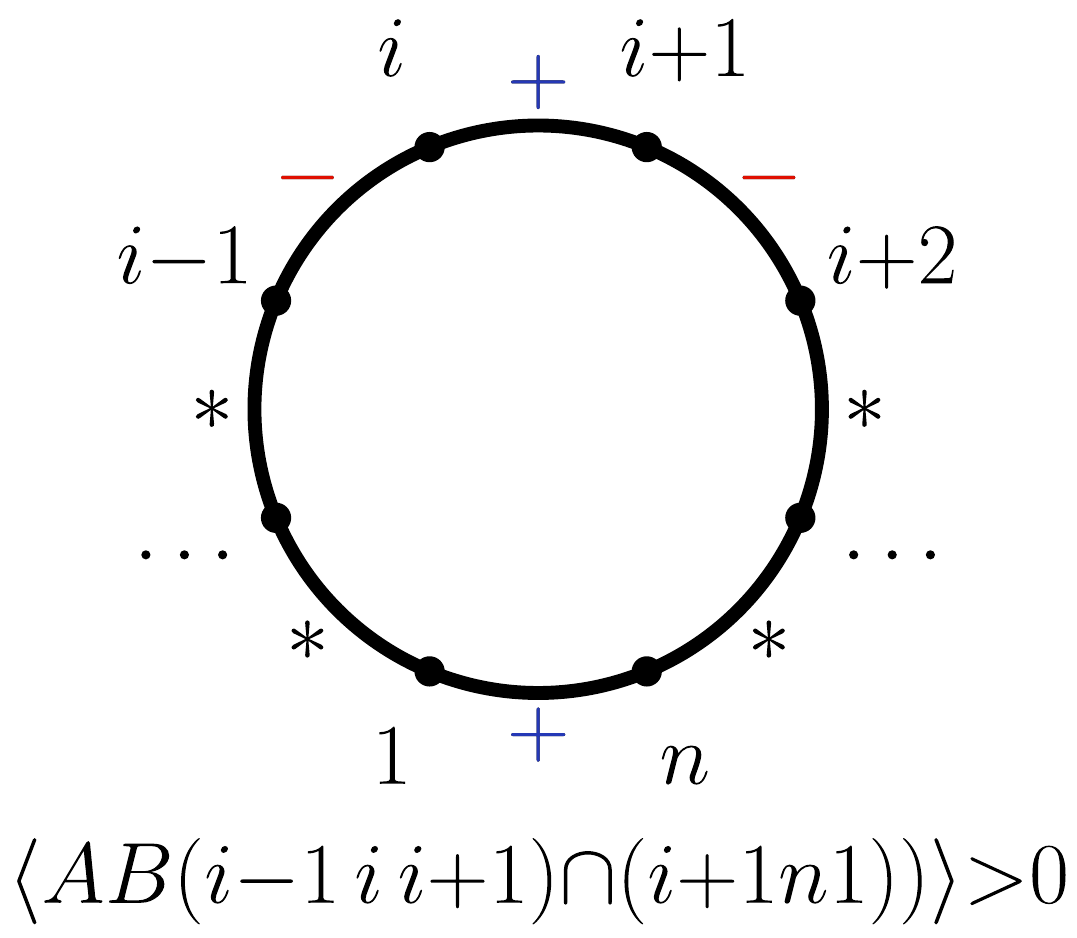}}
     \label{npt_two_mass}
\end{equation}
Let us now take the chiral pentagon expansion for the five- and six-particle amplitudes of eq.~(\ref{eq:1loop_5pt_local_int_exp}) and eq.~(\ref{eq:1loop_6pt_local_int_exp}) and write it in terms of the sign-flip spaces eqs.~(\ref{npt_pent})--(\ref{npt_two_mass}).
As argued in subsection~\ref{subsec:gluing_regions}, the resulting space is free of spurious singularities. Therefore, we call the resulting collection of geometric regions the five- and six-point \emph{Amplituhedron-Prime}: 
\vskip -.7cm
\begin{align}
\mathcal{A}'^{\,(5,0,1)} & =
    \overbrace{
     \raisebox{-33pt}{\includegraphics[scale=.3]{./figures/choice1_p_1.pdf}}
    +\raisebox{-33pt}{\includegraphics[scale=.3]{./figures/choice1_p_2.pdf}}
    +\raisebox{-33pt}{\includegraphics[scale=.3]{./figures/choice1_p_3.pdf}}
    +\raisebox{-33pt}{\includegraphics[scale=.3]{./figures/choice1_p_4.pdf}}}^{P^{(1)}_{24}} \nonumber 
    \\[-5pt]
    & + 
    \underbrace{\raisebox{-33pt}{\includegraphics[scale=.3]{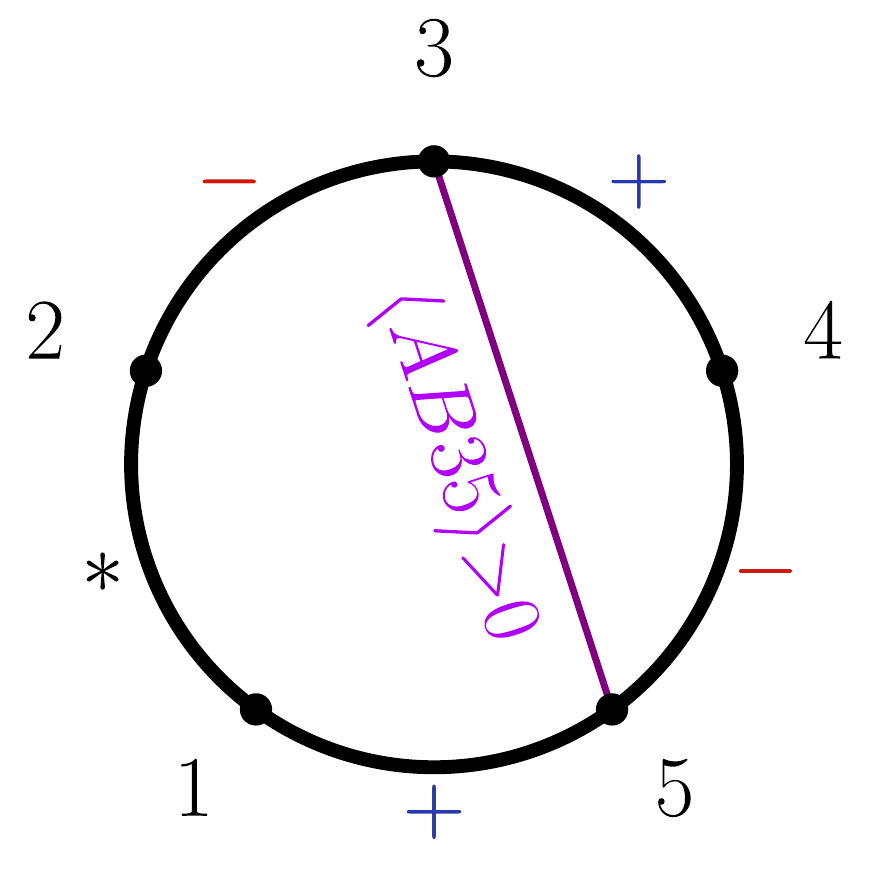}}}_{B^{(3)}_{12}}
     + 
    \underbrace{\raisebox{-33pt}{\includegraphics[scale=.3]{./figures/sf2_b45_3.pdf}}}_{B^{(3)}_{45}}
    \label{5pt_prime}
\end{align}
\vskip -.6cm
\begin{align}
\begin{split}
\mathcal{A}'^{\,(6,0,1)}=&
\overbrace{
    \raisebox{-33pt}{\includegraphics[scale=.3]{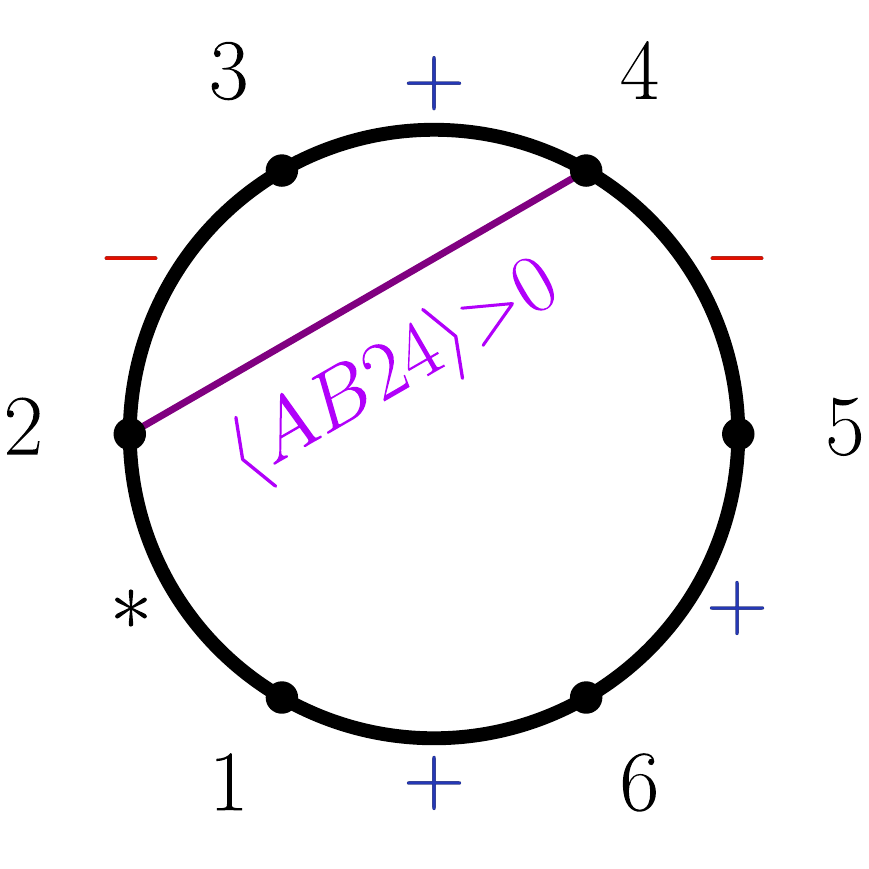}}
    +\raisebox{-33pt}{\includegraphics[scale=.3]{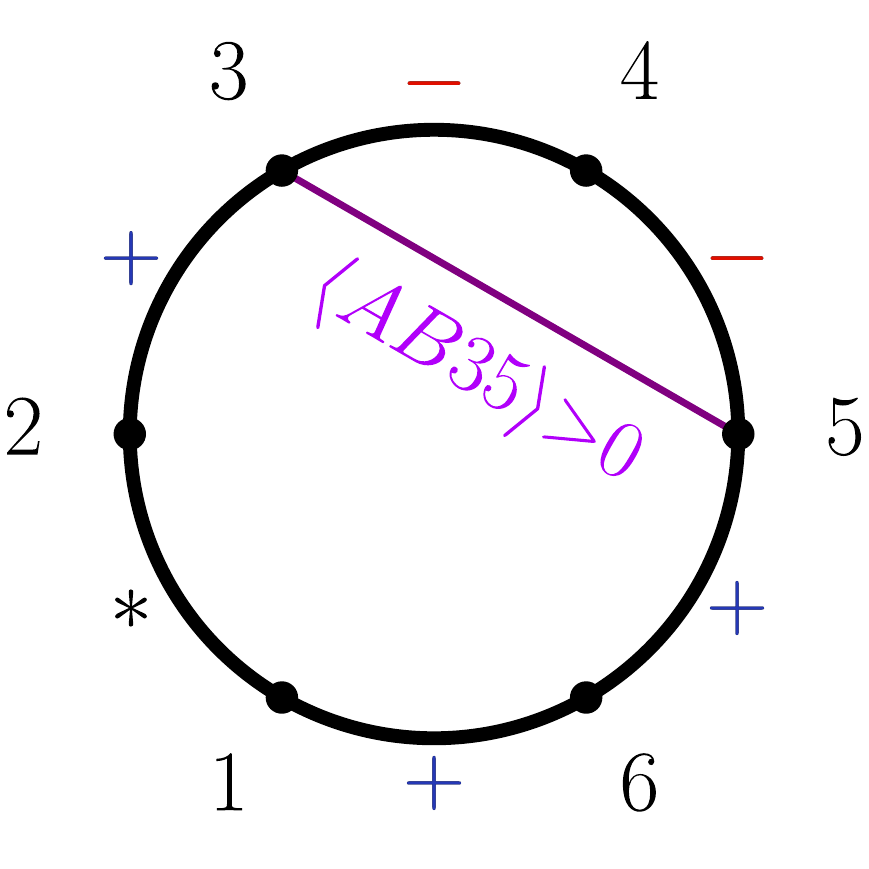}}
    +\raisebox{-33pt}{\includegraphics[scale=.3]{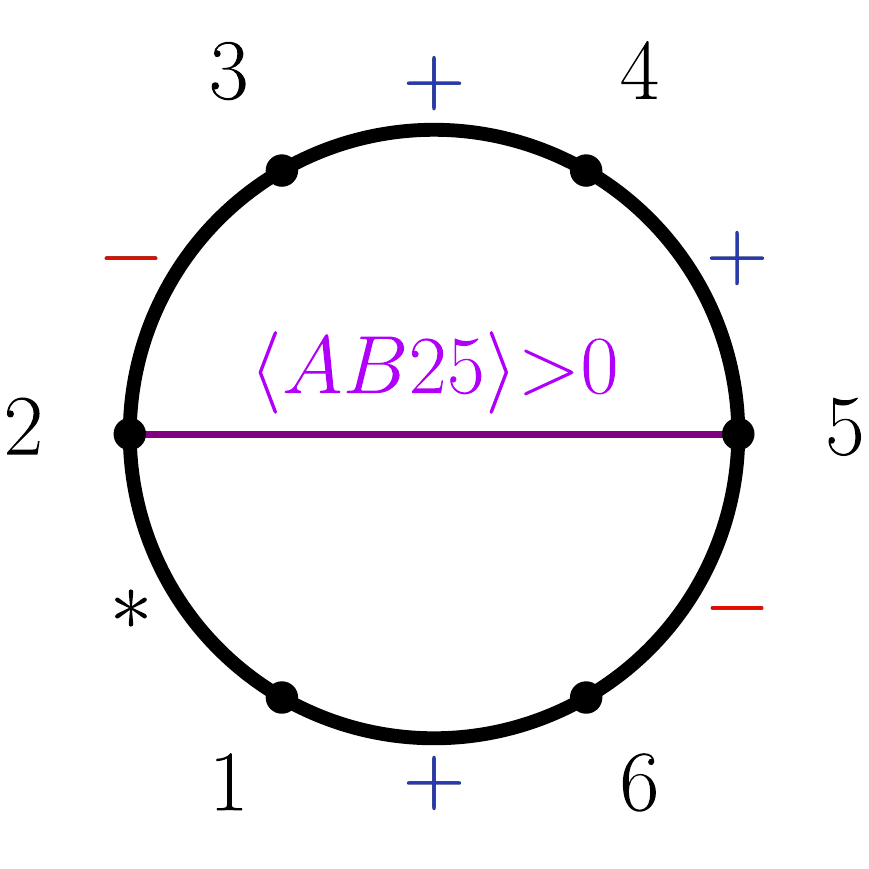}}
    +\raisebox{-33pt}{\includegraphics[scale=.3]{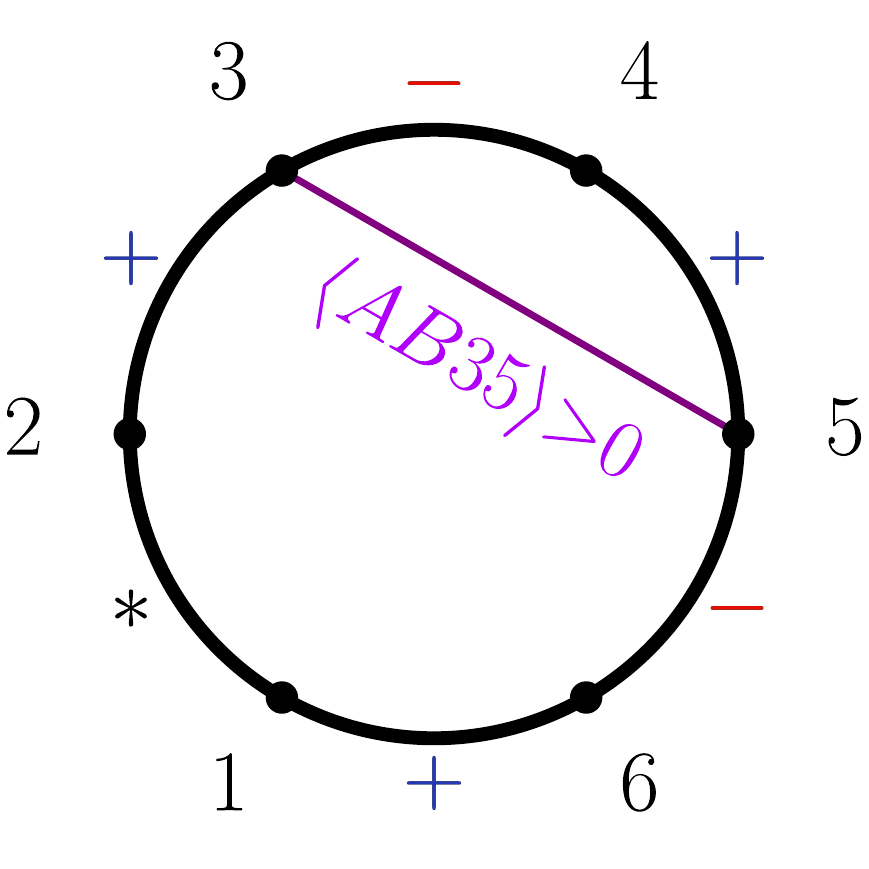}}}^{P^{(1)}_{35}} 
    \\[-10pt]
    + &  
    \overbrace{
    \raisebox{-33pt}{\includegraphics[scale=.3]{./figures/choice1_p24_1.pdf}}
    +\raisebox{-33pt}{\includegraphics[scale=.3]{./figures/choice1_p24_2.pdf}}
    +\raisebox{-33pt}{\includegraphics[scale=.3]{./figures/choice1_p24_3.pdf}}
    +\raisebox{-33pt}{\includegraphics[scale=.3]{./figures/choice1_p24_4.pdf}}}^{P^{(1)}_{24}} 
    \\[-10pt]
    + &
    \overbrace{
    \raisebox{-33pt}{\includegraphics[scale=.3]{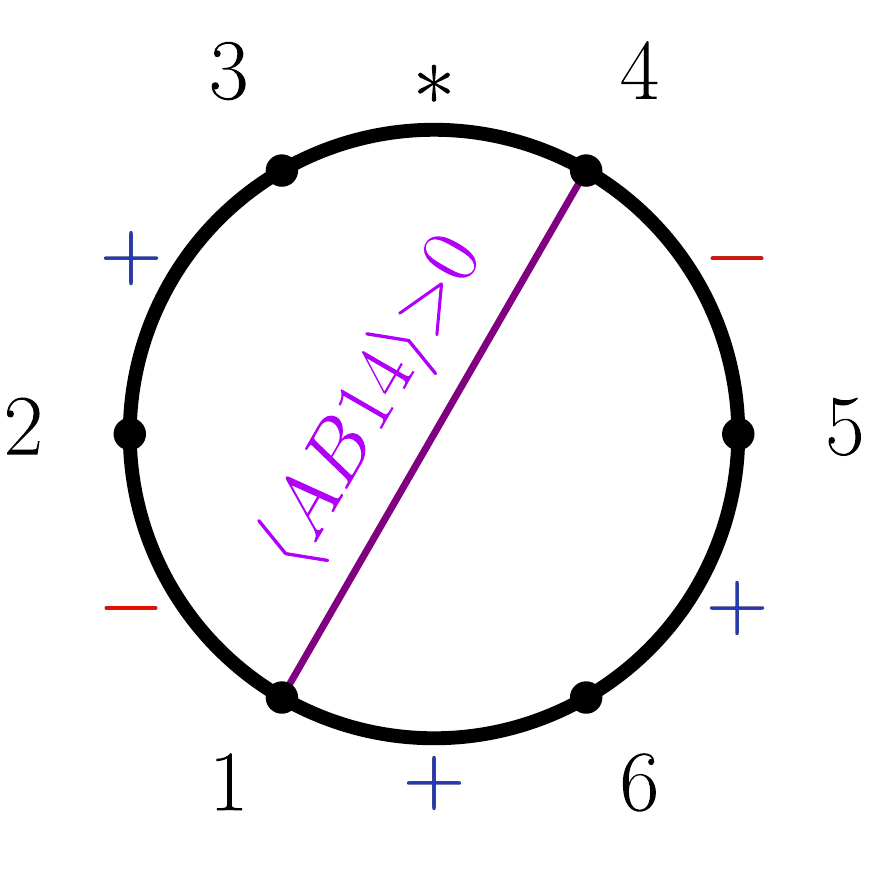}}
    +\raisebox{-33pt}{\includegraphics[scale=.3]{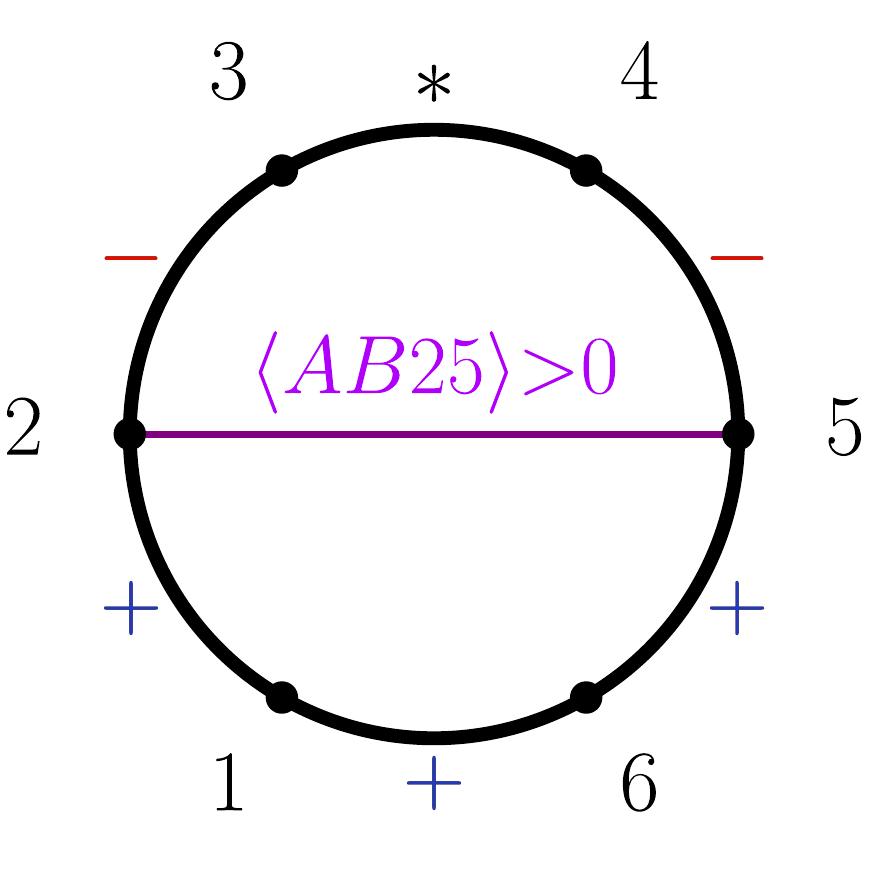}}
    +\raisebox{-33pt}{\includegraphics[scale=.3]{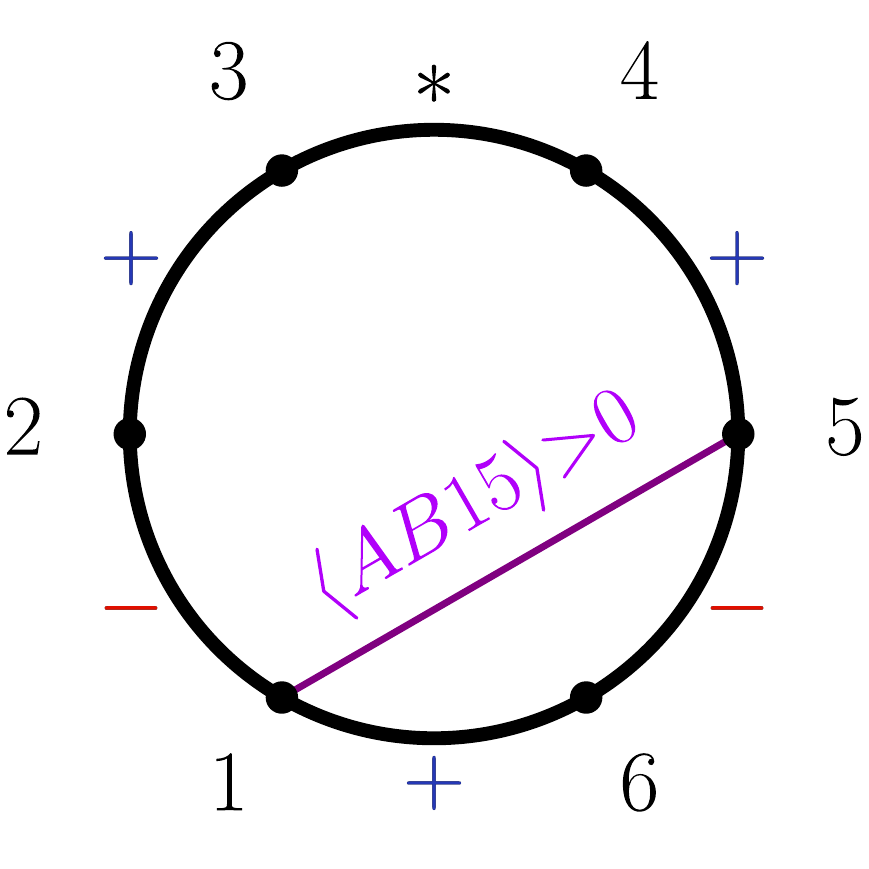}}
    +\raisebox{-33pt}{\includegraphics[scale=.3]{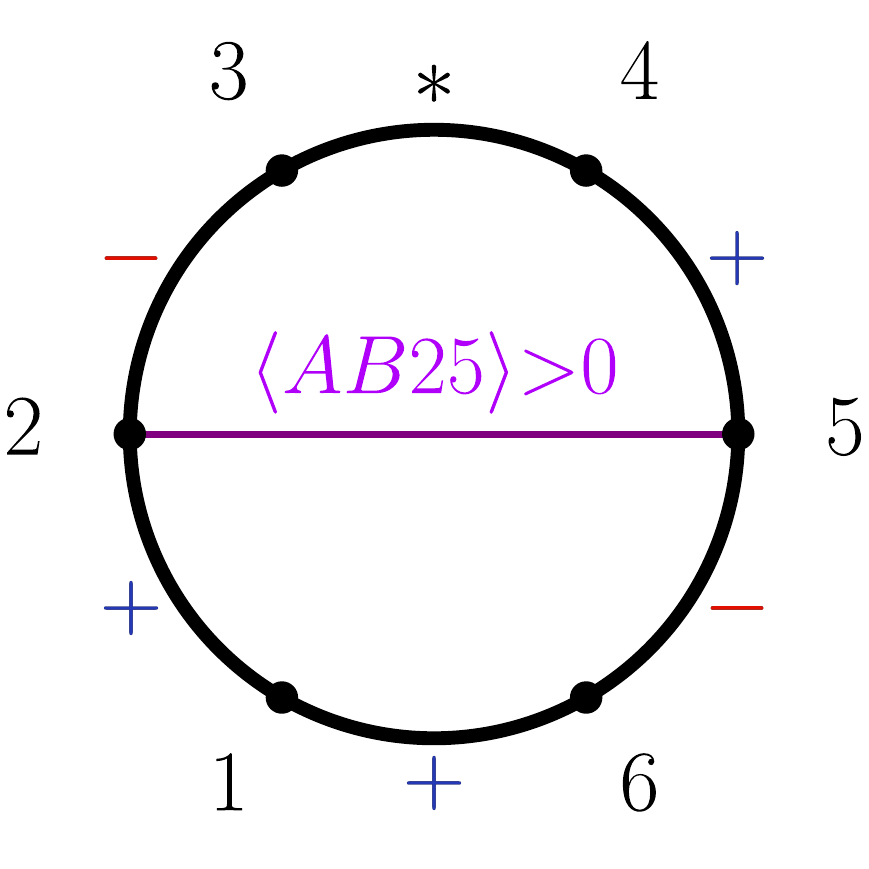}}}^{P^{(1)}_{25}} 
    \\[-5pt]
    + &
    \underbrace{
    \raisebox{-33pt}{\includegraphics[scale=.3]{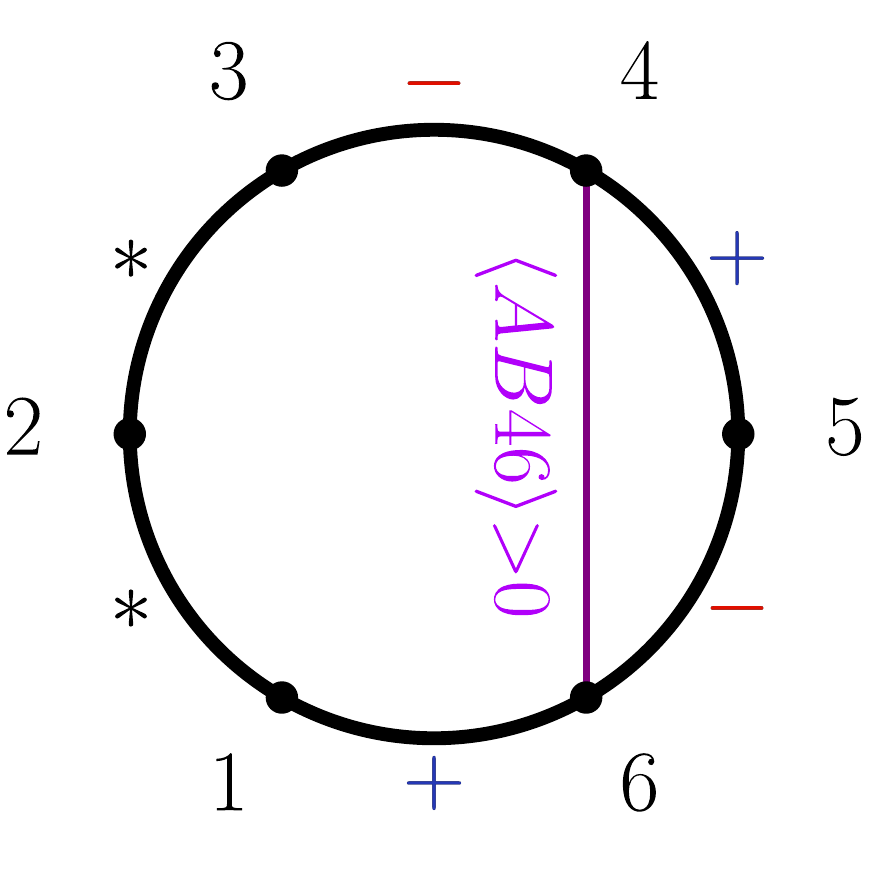}}}_{B^{(3)}_{13}}
    +
    \underbrace{
    \raisebox{-33pt}{\includegraphics[scale=.3]{./figures/choice3_b456.pdf}}}_{B^{(3)}_{46}}
    +
    \underbrace{
    \raisebox{-33pt}{\includegraphics[scale=.3]{./figures/choice1_b12_56.pdf}}}_{B^{(1)}_{12,56}}
\label{6pt_prime}
\end{split}
\end{align}
In subsection~\ref{subsec:gluing_regions}, we have seen on various two-dimensional projections that the representation of the full space in terms of the local building blocks is massively overlapping. From the sign-flip representation in terms of the circle-diagrams in eqs.~(\ref{5pt_prime})--(\ref{6pt_prime}), this overlap is visible as many terms with the same sign patterns appear in different building blocks. In fact, we can exploit the results of section~\ref{sec:sign_flip_regions} and expand all $\ast{=}+{\oplus}-$ present in these spaces\footnote{After expanding all $\ast$ in the two-mass-hard box (\ref{npt_two_mass}) into a collection of spaces with definite signs for all $\ab{ABii{+}1}$, we can relate these spaces to the chiral sign-flip regions in (\ref{sf2_chiral}) and (\ref{eq:sf4_chiral_comps_fully_dressed}).}, throwing out all patterns with more than four sign flips. It is easy to see that many terms appear multiple times throughout the expansion, which geometrically means the same space gets multiply-covered. If we cover a space an even number of times it cancels completely, while for an odd number of covers we are left with a single copy of the space. In the end, there is a surprisingly uniform non-overlapping description of the Amplituhedron-Prime directly in terms of the sign-flip-two and four spaces of section~\ref{sec:sign_flip_regions}, which naturally generalizes to all-multiplicities. Our conjecture for the $n$-point, one-loop MHV Amplituhedron-Prime is\footnote{Note that in the sum on the second line the term where $j=i{+}1$ is an empty space as it simultaneously requires both $\ab{ABij}>0$ and $\ab{ABij}<0$.}
\begin{align}
\hspace{-1cm}
\begin{split}
    \mathcal{A}'^{(n,0,1)} & =
    \hspace{-.5cm}
    \raisebox{-55pt}{\includegraphics[scale=.45]{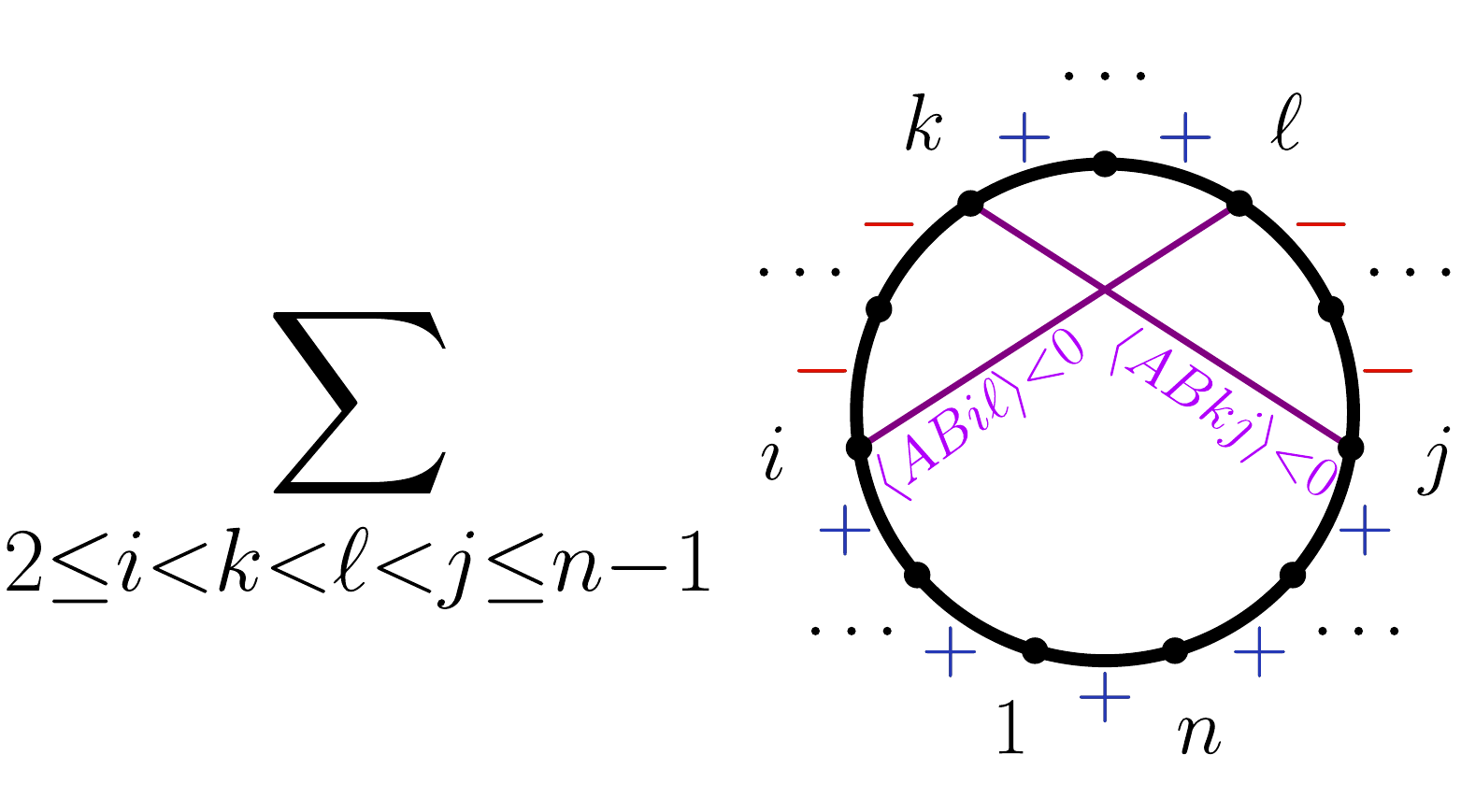}}
    \quad
    {+}
    \raisebox{-55pt}{\includegraphics[scale=.45]{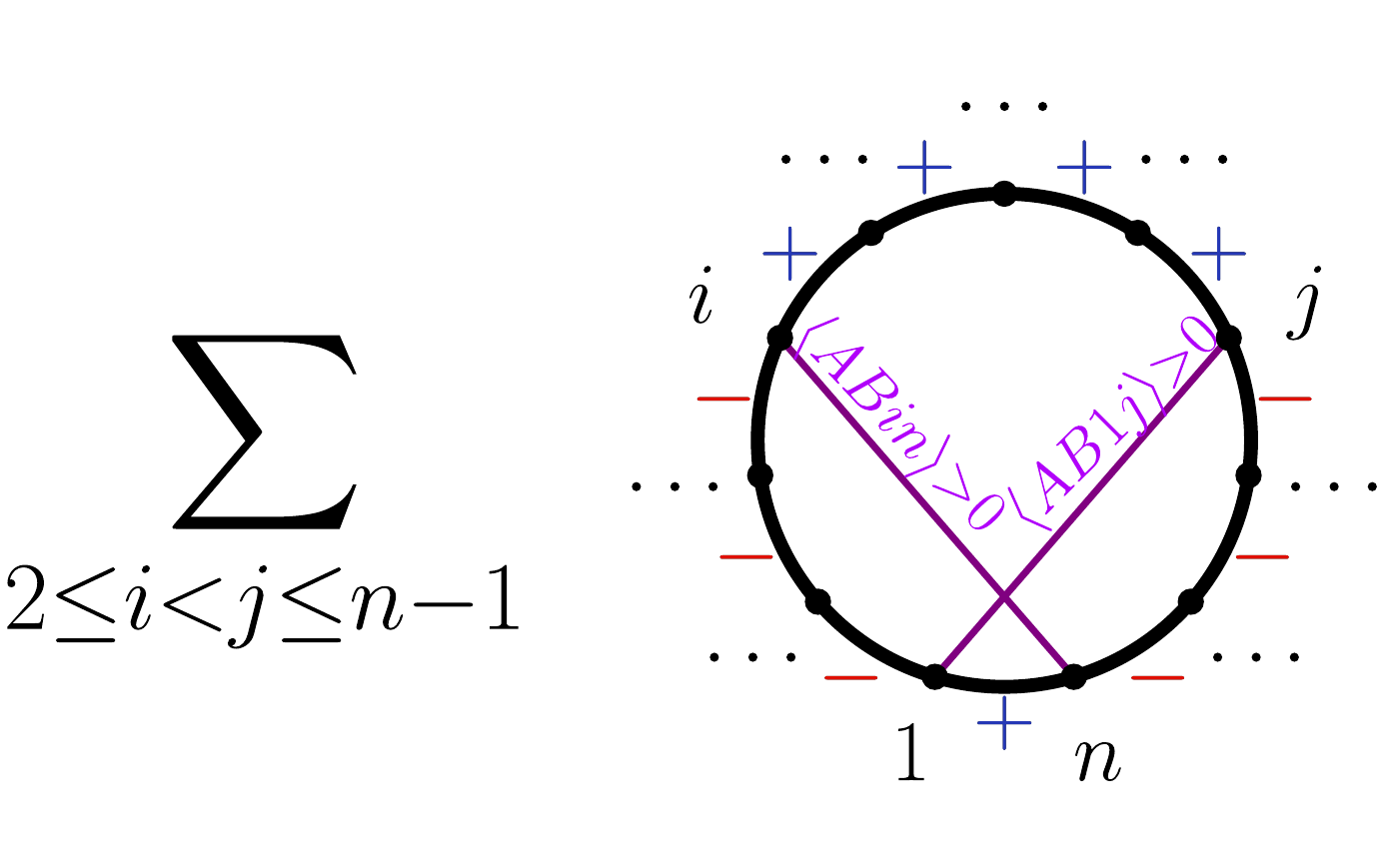}} 
    \\[-20pt]
    & +
    \hspace{.1cm}
    \raisebox{-55pt}{\includegraphics[scale=.45]{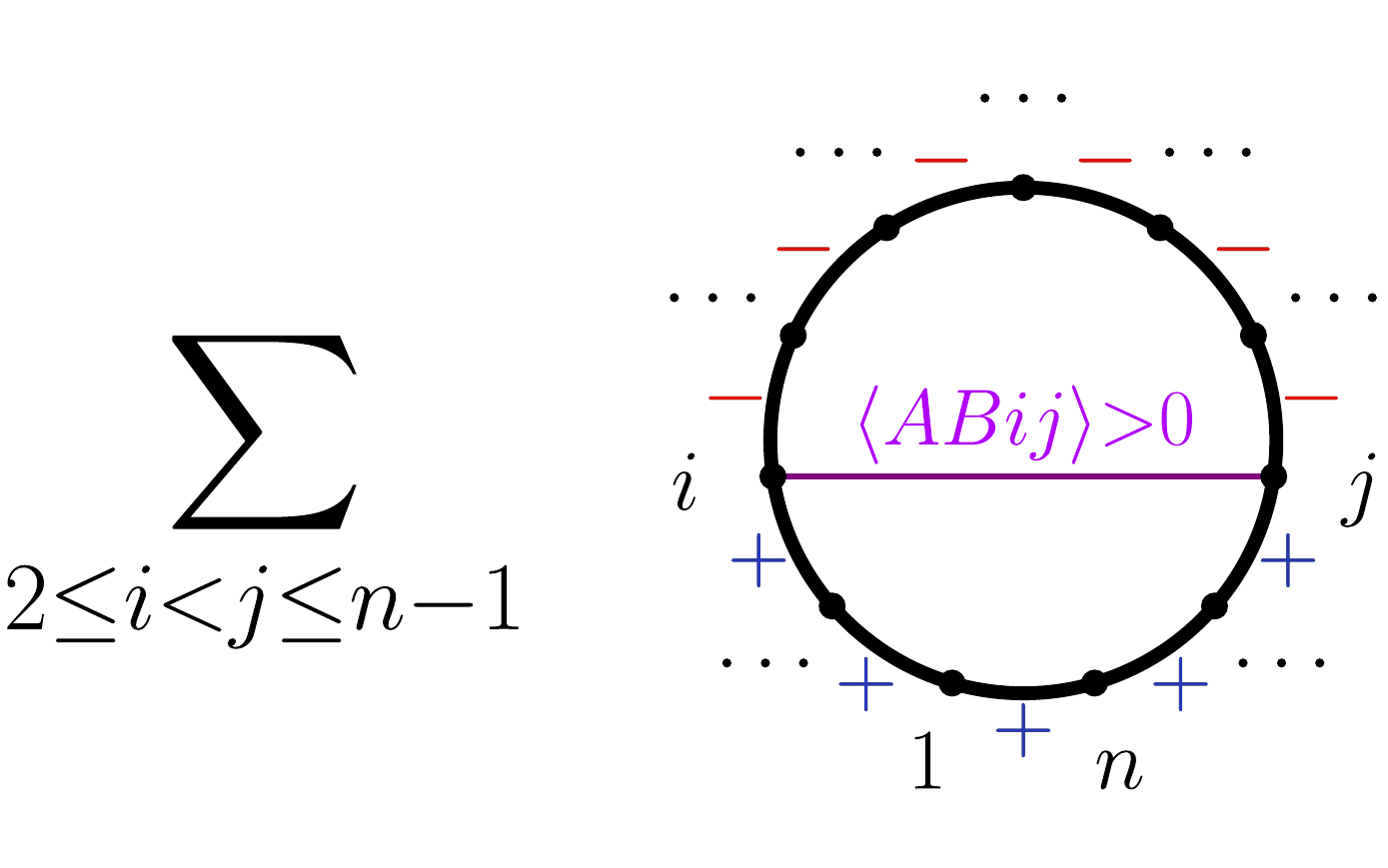}}.
\end{split}  
\hspace{-1cm}
\label{amp_prime_npt}
\end{align}
Extending the five and six-point analysis of section~\ref{sec:geometry_chiral_pentagons}  to test our all-$n$ expression eq.~(\ref{amp_prime_npt}) is a straightforward exercise. We demand that all spurious boundaries present in the individual sign-flip spaces disappear upon gluing. Just as in the five and six-point examples, the spurious codimension-three boundaries are 
\begin{enumerate}
\item[(1)] $(AB)\subset(i{-}1ii{+}1)$ and $(AB)$ cuts $(jj{+}1)$, 
\item[(2)] triple cuts of non-adjacent propagators, $\ab{ABii{+}1}{=}\ab{ABjj{+}1}{=}\ab{ABkk{+}1}{=}0$
\end{enumerate}
We have verified at seven points that all spurious boundaries of type (1) and (2) are absent from the final space. We also performed extensive numerical checks at eight (and higher) points that eq.~(\ref{amp_prime_npt}) satisfies many nontrivial constraints. In principle, we could repeat the exercise of section~\ref{sec:geometry_chiral_pentagons} at higher points and attempt to find all positive geometries which consistently glue together. Our conjecture is that the unique spurious-boundary free combination is equivalent to (after cancelling overlapping regions) the result eq.~(\ref{amp_prime_npt}). Note that it is clear by construction, that our new Amplituhedron-Prime space is externally triangulated by the chiral pentagon expansion. 

Since the spaces constituting the Amplituhedron-Prime are always defined by (at least) one inequality of the form $\ab{ABii{+}1}{<}0$, the bulk of this new geometry is entirely non-overlapping with the original Amplituhedron. At the same time, it has only physical boundaries and exactly the same integrand form as the Amplituhedron. This construction demonstrates there are multiple positive spaces which can be associated with loop integrands in planar ${\cal N}=4$ sYM and the Amplituhedron of \cite{Arkani-Hamed:2013jha} is only a particular example (albeit possibly the most canonical one).

It is also interesting to note that while the chiral pentagon expansion externally triangulates the Amplituhedron-Prime, it also plays an even more natural r\^{o}le in the presumptive dual geometry. We will argue in section \ref{sec:triangulation_dual_Amplituhedron} that the chiral pentagons \emph{internally} triangulate the (yet-to-be discovered) dual Amplituhedron.

As an aside, investigating the structure of eq.~(\ref{amp_prime_npt}) more carefully, we note the absence of certain sign-flip regions, such as 
\begin{align}
    \raisebox{-55pt}{\includegraphics[scale=.45]{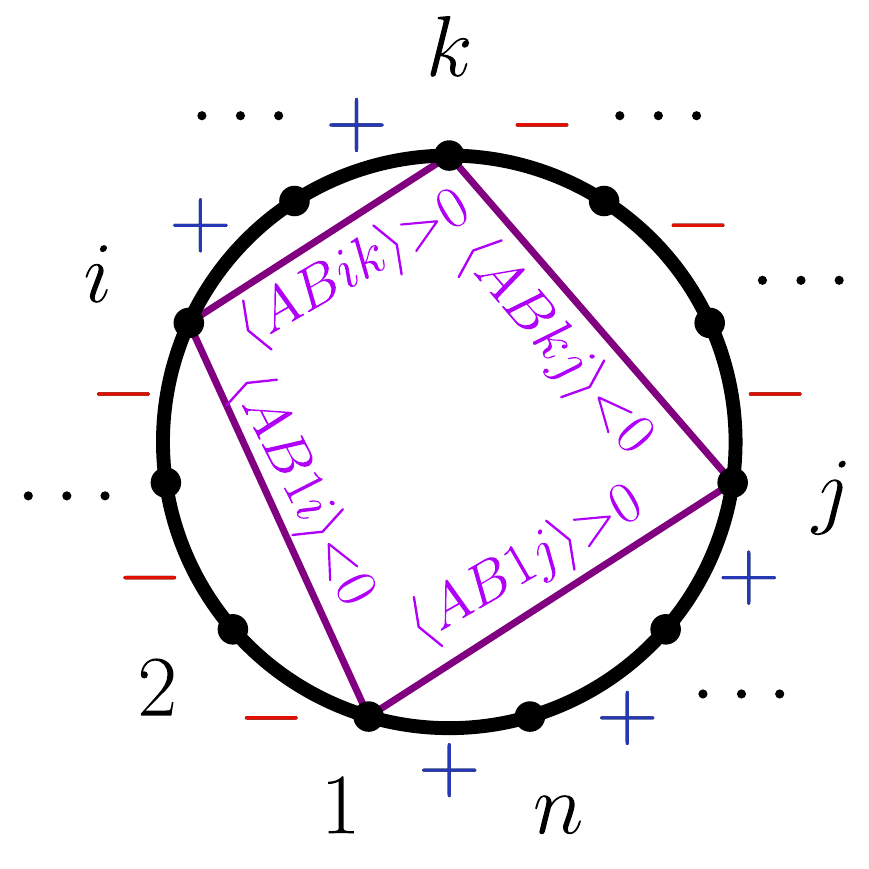}}
    \quad \text{and} \quad
        \raisebox{-55pt}{\includegraphics[scale=.45]{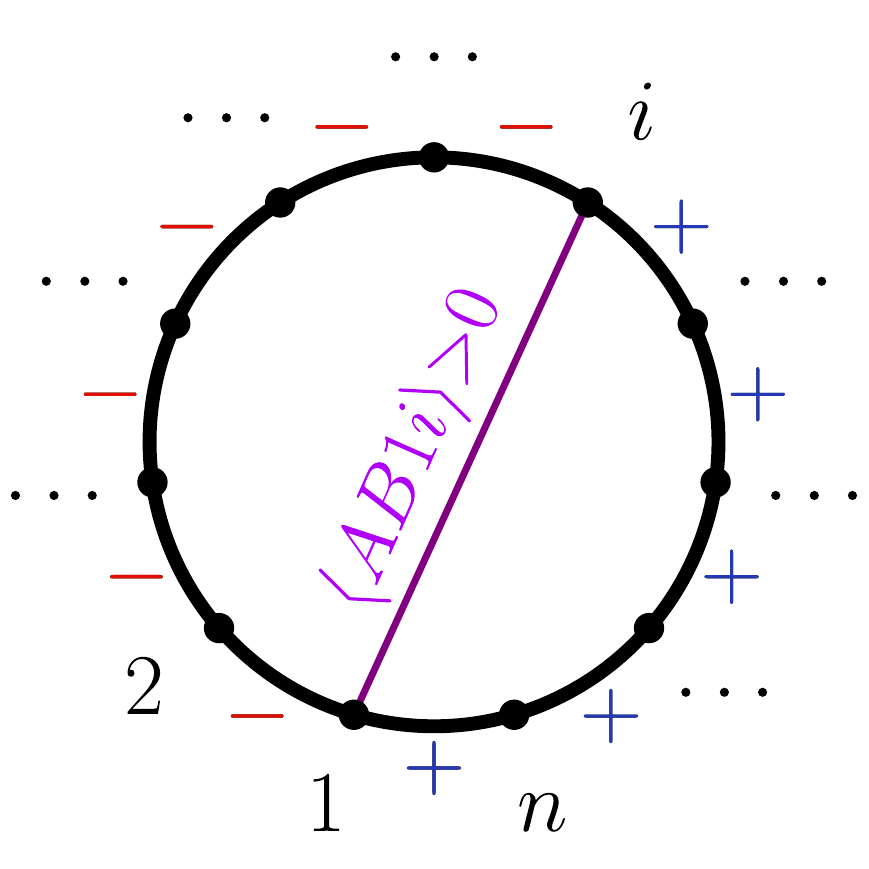}}\,.
        \label{non_appearing_spaces_amp_prime}
\end{align}
While eq.~(\ref{amp_prime_npt}) does provide the complete definition of the Amplituhedron-Prime space, it is an expansion in terms of elementary regions and it would be desirable to find some more uniform definition, much like the definition of the MHV one-loop Amplituhedron in eq.~(\ref{eq:mhv_alt}). One can verify that the following definition is equivalent to eq.~(\ref{amp_prime_npt}): the Amplituhedron-Prime is the space of all lines $(AB)$ which satisfy
\begin{align}
\label{finale}
\begin{split}
& \{\ab{AB12},\ab{AB23},\ldots \ab{ABn{-}1n}\} \,\,\mbox{ has even number of sign flips}\\
& \{ \ab{AB1n}>0, \ab{AB i_1 n}>0,
    \ab{AB i_1 i_{-1}} > 0, \ab{ABi_2i_{-1}}<0\},
\end{split}
\end{align}
where $i_{1,2}$ is the position of the first (second) sign flip and $i_{-1}$ that of the last sign flip.

The final two conditions in the second line of (\ref{finale}) are empty for the third term in eq.~(\ref{amp_prime_npt}) which only has two sign flips. Also, the representative spaces (\ref{non_appearing_spaces_amp_prime}) of terms that do \emph{not} appear in (\ref{amp_prime_npt}) are ruled out by the first condition in (\ref{finale}) which only includes brackets up to $\ab{ABn{-}1n}$ and does not ``wrap around'' to $\ab{AB1n}$. While the definition in eq.~(\ref{finale}) is very simple, we do not quite understand its deeper meaning at the moment and leave a detailed investigation to future work.

Note that the chiral pentagon expansion of eq.~(\ref{npt_pent}) singles out the the line $(n1)$ as special, as does our definition of the Amplituhedron-Prime, where $\ab{AB1n}>0$ is the only uniformly positive quantity throughout the space. While the $d\log$ form for the whole space is cyclic, the space is obviously not, as can be seen in eq.~(\ref{amp_prime_npt}). By using the chiral pentagon expansion with $(ii{+}1)$, rather than $(1n)$, fixed, we can construct $n$ other Amplituhedron-Prime spaces by cyclically shifting eq.~(\ref{amp_prime_npt}).

\subsubsection*{Relation between $\mathcal{A}$ and $\mathcal{A}'$}

The results of this subsection suggests a natural question: how is the Amplituhedron-Prime ${\cal A}'$ related to the original Amplituhedron space ${\cal A}$? They are non-overlapping positive geometries which have only physical boundaries and the same canonical form. Therefore, it must be possible to identify a collection of zero-form spaces (with no spurious boundaries) which can be added to $\mathcal{A}'$ to directly yield $\mathcal{A}$. 

As it turns out, identifying the correct zero-form space which relates $\mathcal{A}$ and $\mathcal{A}'$ is nontrivial, and at the moment we have no closed-form expression for this space. However, the process is relatively straightforward for the simplest case of the four-point one-loop integrand, as we will now demonstrate. While the Amplituhedron is the space given by $\la ABii{+}1\ra>0$ and $\la AB13\ra<0$, the Amplituhedron-Prime is given by a single term from the second sum in eq.~(\ref{amp_prime_npt}),
\begin{align}
\mathcal{A}^{(4,0,1)}&= \raisebox{-45pt}{\includegraphics[scale=.4]{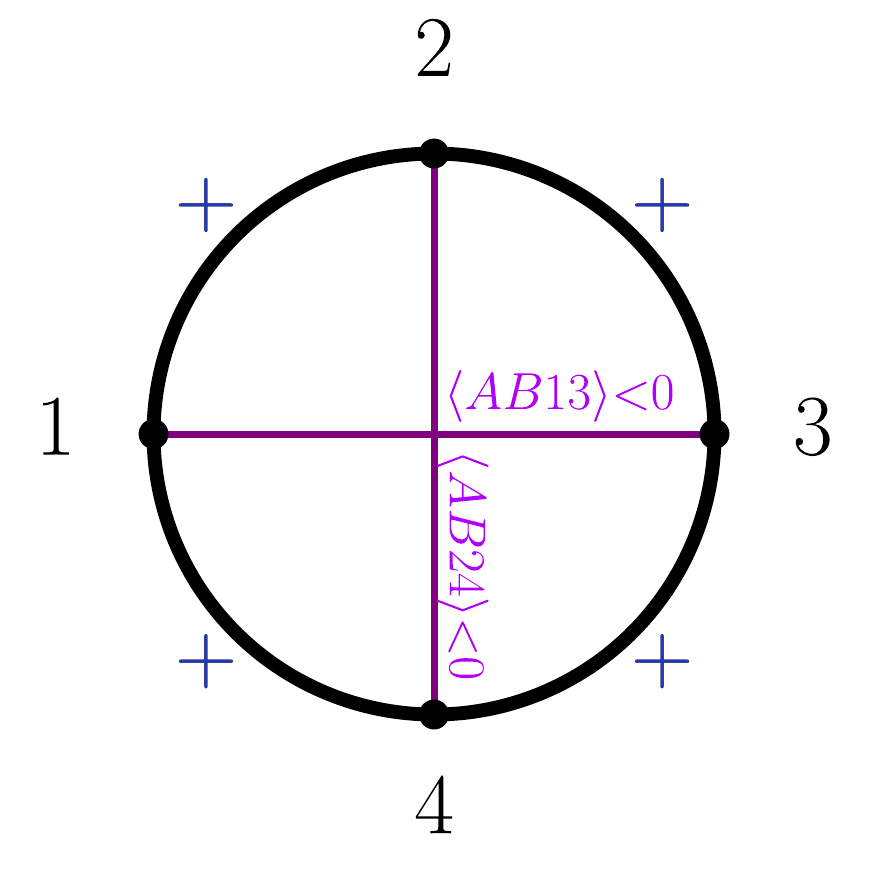}},\quad \mathcal{A}'^{(4,0,1)}= \raisebox{-45pt}{\includegraphics[scale=.4]{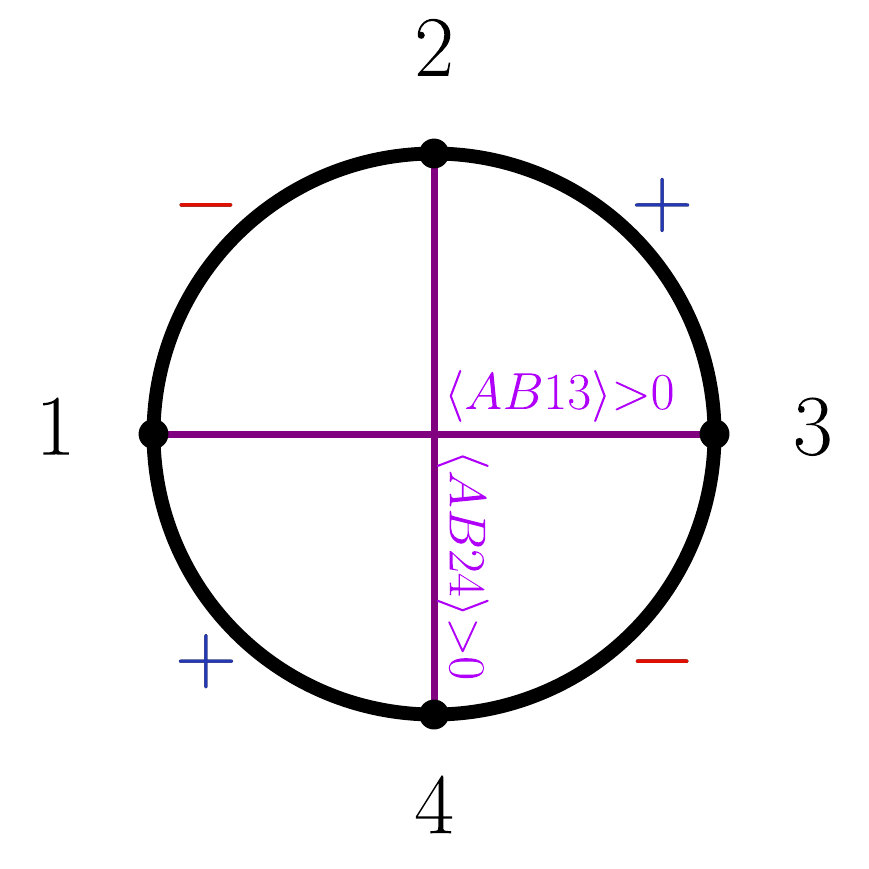}}.    
\end{align}
Starting with ${\cal A}'$, we first add the achiral space 
\begin{equation}
    {\cal B}_1=\raisebox{-45pt}{\includegraphics[scale=.4]{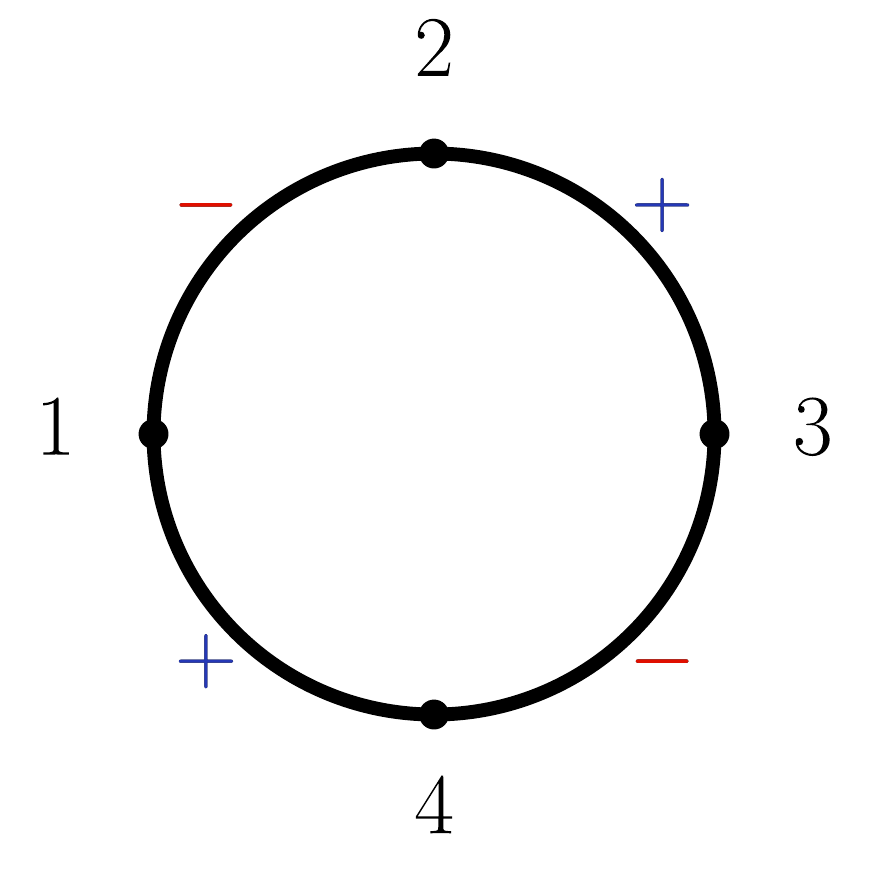}},   
\end{equation} 
which effectively flips the sign of $\la AB13\ra$ in ${\cal A}'$. Next, we add a combination of chiral spaces ${\cal B}_2$ which is defined
\begin{equation}
{\cal B}_2=\left\{\frac{\ab{AB12}}{\ab{AB34}}>0,\ab{AB23}>0,\ab{AB14}>0,\ab{AB13}<0\right\},
\label{eq:b2}
\end{equation}
and obtain ${\cal A}$ as a result. Expanding ${\cal B}_2$ in terms of sign-flip-spaces, we have 
\begin{align}
\mathcal{A}^{(4,0,1)}&=\mathcal{A}'^{(4,0,1)}+\raisebox{-45pt}{\includegraphics[scale=.4]{./figures/5pt_achiral_b1.pdf}}+\raisebox{-45pt}{\includegraphics[scale=.4]{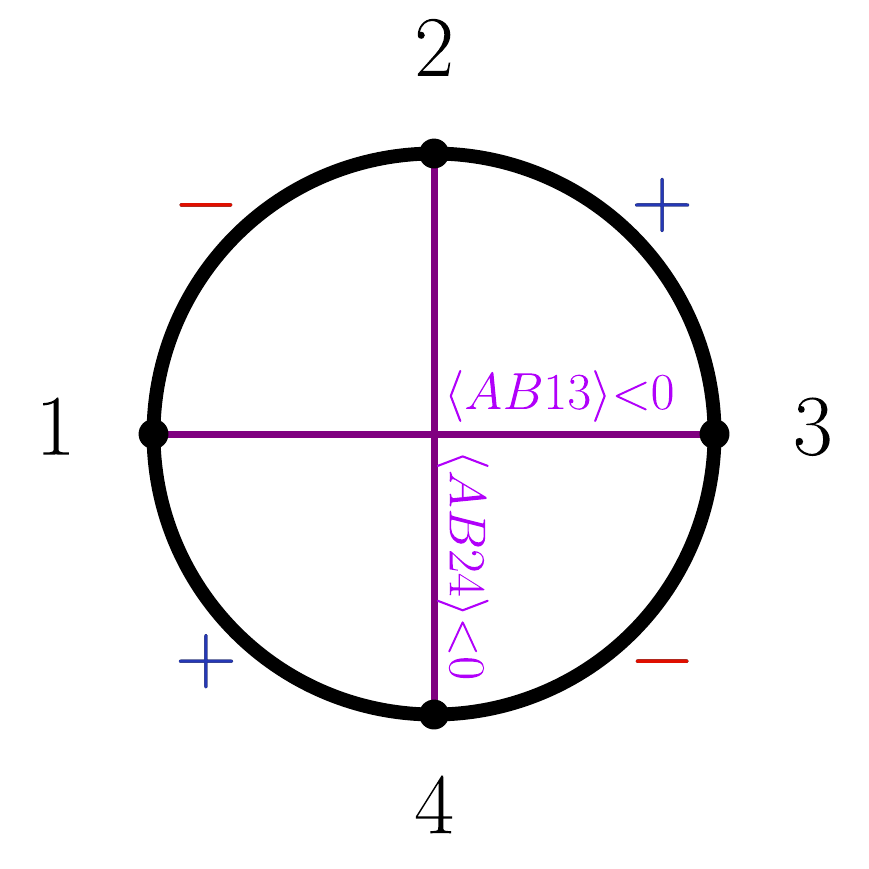}}+\raisebox{-45pt}{\includegraphics[scale=.4]{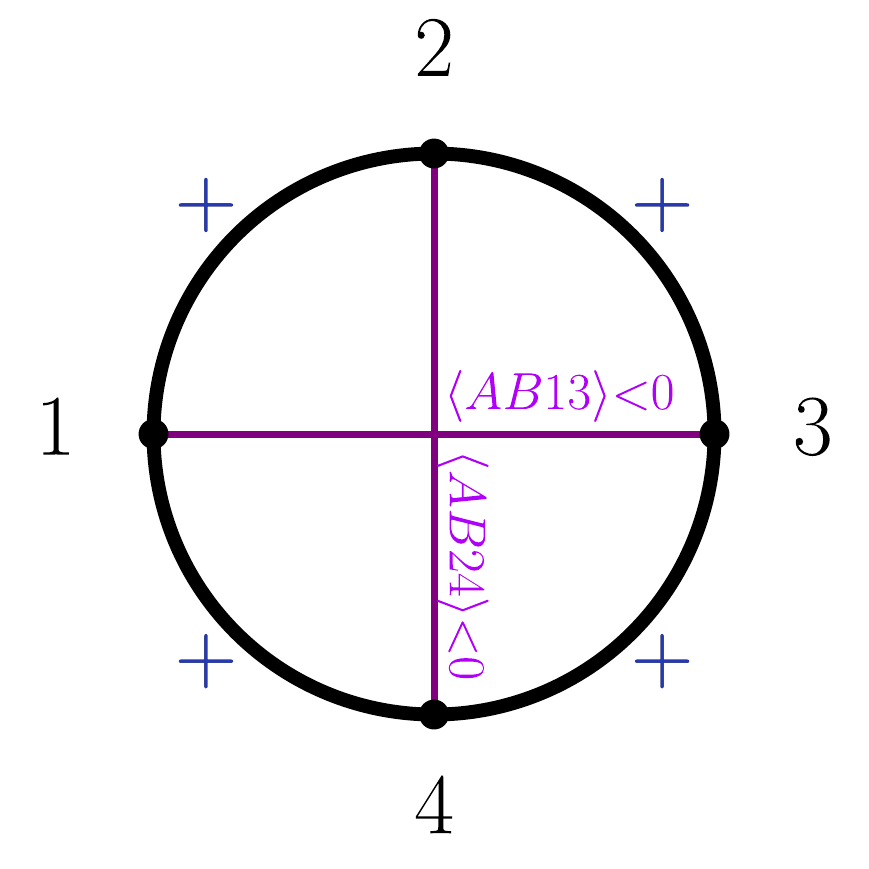}}.
\label{eq:zero_form_space_4pt}
\end{align}
Note that even in this simple example, it is actually quite non-trivial that the spaces with vanishing form we add have only physical boundaries. In particular, if we had flipped the sign of $\la AB23\ra$ in the definition of ${\cal B}_2$ in eq.~(\ref{eq:b2}), the resulting space would still have zero form, but would have the spurious boundary where $\ab{AB13}=0$. 

As our conjecture is that ${\cal A}'$ has only physical boundaries, the geometric difference between ${\cal A}$ and ${\cal A}'$ must be a collection of zero-form spaces with physical boundaries only. Finding the exact combination becomes very laborious at higher points, and we do not have a closed formula for it. However, as discussed in the motivations of section~\ref{sec:introduction} and as we will see in the details of section~\ref{sec:triangulation_dual_Amplituhedron}, the real purpose in life of the chiral pentagons is to triangulate the dual Amplituhedron, where both ${\cal A}$ and ${\cal A}'$ are mapped under dualization.

\newpage
\section{Triangulation of the dual Amplituhedron}
\label{sec:triangulation_dual_Amplituhedron}

In the previous section we have seen that the chiral pentagons externally triangulate the Amplituhedron-Prime, which is free of all spurious boundaries, has the same canonical form as the Amplituhedron, but is geometrically distinct. In fact, both the $\mathcal{A}$ and $\mathcal{A}'$ spaces only intersect on various codimension boundaries. While the Amplituhedron-Prime is certainly an interesting positive geometry in its own right, we believe that the real purpose of the chiral pentagon expansion is more directly associated with the internal triangulation of the \emph{dual Amplituhedron}. 

This belief was first raised in \cite{Arkani-Hamed:2014dca} based on the simple observation that the chiral pentagon forms are positive when evaluated inside the Amplituhedron region, and therefore provide a term-wise positive expansion for the MHV one-loop integrand. In this picture, the positivity of the loop integrand is reminiscent of the volume interpretation of the dual Amplituhedron. The volume is naturally a positive function of (real) geometric data (momentum twistors) and slicing this volume into smaller pieces via internal triangulation preserves a term-wise positivity.

For the simplest $k=1$ tree Amplituhedron the dualization procedure is well understood and involves a simple map between polytopes and their duals. However, for $k>1$ tree-level (and all loop-level amplituhedra) the geometries become non-polytopal, and in these cases the dualization procedure has not been defined as of yet. While we do not give a complete solution of this problem in this work, in this section we provide a more direct link between the chiral pentagon expansion and the yet-to-be-found dual one-loop MHV Amplituhedron. 

In subsection~\ref{subsec:gluing_regions} we considered a significant subset of simple codimension-two faces of the MHV one-loop Amplituhedron. In these pictures, we localize two degrees of freedom of the line $(AB)$, so that the resulting projection can be viewed as a point (and polygons) on the projective plane. Although the correct dual of the fully off-shell line $(AB)$ is not known, we can avoid this problem by working directly on codimension-two surfaces which reduce to the projective plane $\mathbb{P}^2$. By exploiting the elementary geometrical fact that polygons dualize to polygons, we can explicitly construct the associated codimension-two boundaries of the dual Amplituhedron by mapping points$\leftrightarrow$lines, in a precise way that we outline below. On these dual codimension-two boundaries, we show that the chiral pentagon expansion corresponds to an internal triangulation of the dual of the MHV Amplituhedron.

\subsection{Dualizing polygons}
\label{subsec:dualizing_polygons}

As discussed in section~\ref{subsec:gluing_regions} there are two different codimension-two boundaries which reduce to the geometry of a point inside a polygon on $\mathbb{P}^2$: either we localize the line $(AB)$ in a plane $(i{-}1ii{+}1)$ or it passes through the point $Z_i$.

In order to connect the chiral pentagons and the dual Amplituhedron, we dualize the $\mathbb{P}^2$ geometries of subsection~\ref{subsec:gluing_regions}. In fact, these projections are structurally identical to the toy model of appendix~\ref{sec:external_triangulations}; only the labelling of the points and lines is different. Dualizing a polygon is a straightforward procedure and yields another (dual) polygon; this was discussed in the (pre-)Amplituhedron context in \cite{ArkaniHamed:2010gg}. 

We begin with the five-point codimension-two boundary (\ref{2_proj_no_labels}) where $(AB)$ passes through $Z_2$. The dualization procedure maps points$\leftrightarrow$lines. Thus, the vertices (leading singularities) in the original projection become the edges of the dual polygon, while the codimension-one boundaries $\ab{ABii{+}1}=0$ become the vertices in the dual space. The dualization of eq.~(\ref{2_proj_no_labels}) is
\begin{equation}
    \raisebox{-75pt}{\includegraphics[scale=.45]{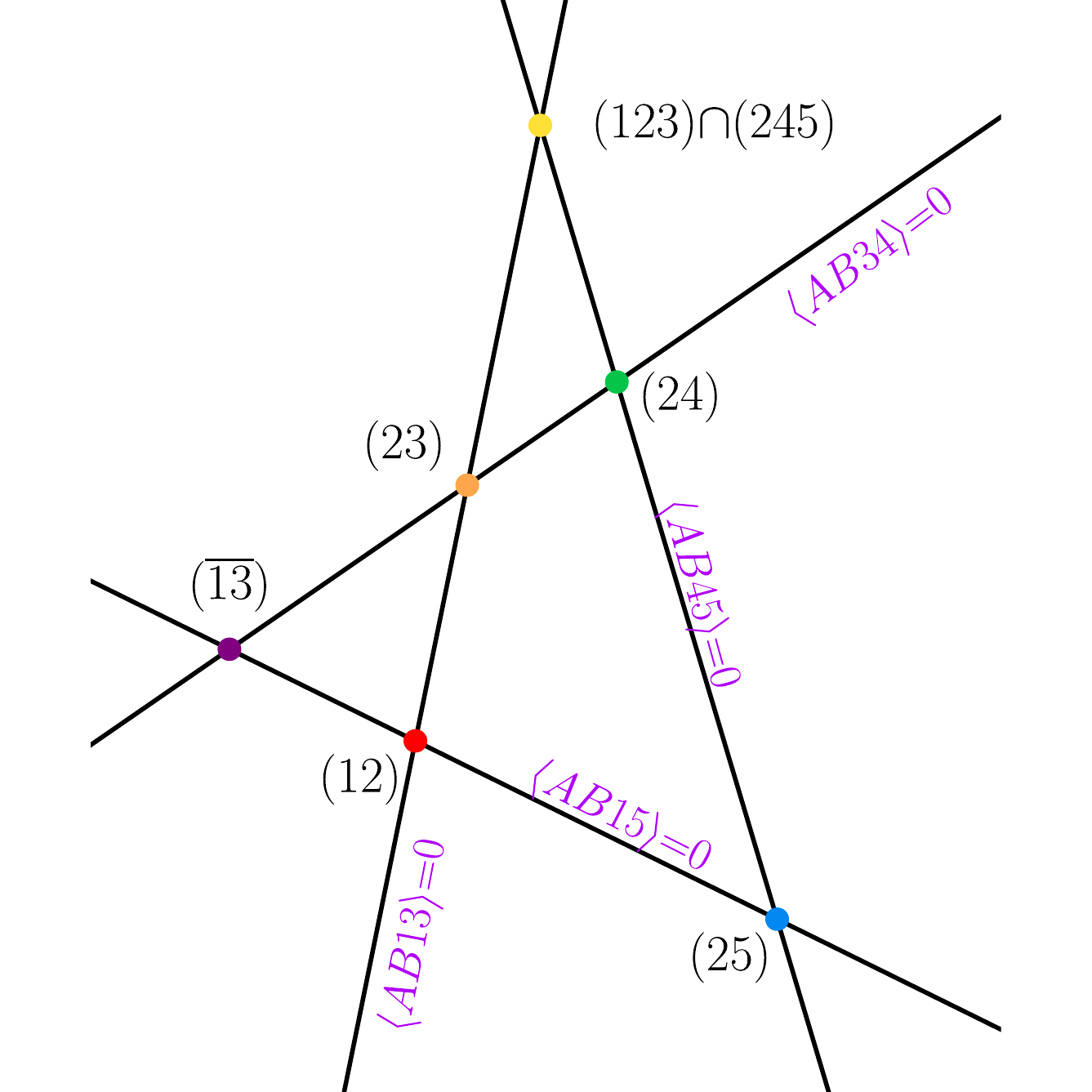}} \underset{\text{dual to}}{\iff}
    \raisebox{-75pt}{\includegraphics[scale=.45]{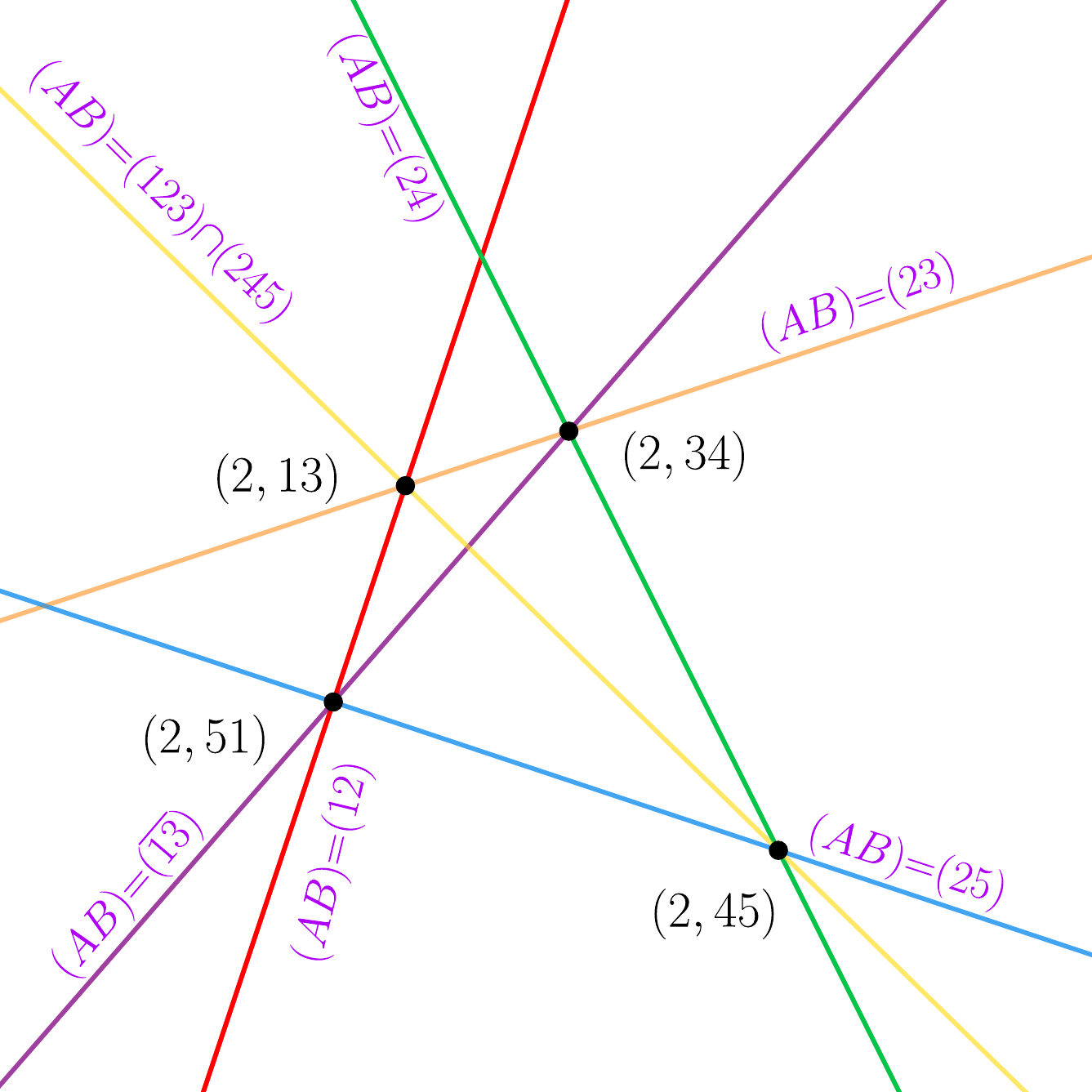}}
\label{2_projection_5pt_dualization}    
\end{equation}
where we use the notation that e.g., $(2,51)$ corresponds to $(AB)$ passing through $Z_2$ and cutting the line $(15)$. We have also color-coded the leading singularities (and their dual lines) to direct the eye of the reader. For example, the red vertex (leading singularity) $AB=(12)$ in the left figure gets mapped to the red line in the right figure. Note that we label the dual picture with conditions imposed on the line $(AB)$, despite the fact that this projection is actually describing the localization of some dual line $\widetilde{(AB)}$ to an associated codimension-two boundary. From this perspective, it is crucial that the space of $(AB)$ lines is four-dimensional, so the dual of a two-dimensional geometry is another two-dimensional geometry! This gives a concrete way of constructing the faces of the dual geometry. Ideally, in the dual picture we would like to dispense with $(AB)$ altogether and identify the regions and boundaries in the projection with the signs of some brackets, \`{a} la $\la \widetilde{AB} \dots \ra \gtrless 0$; however, we do not yet know the constraints which $\widetilde{(AB)}$ should satisfy. Nevertheless, the simple structure of the dualization on these two-dimensional projections allows us to take a region in the original space and find the corresponding region in the dual space just by working in the original $(AB)$ space, using a very simple prescription. The rule is that any line which intersected the region in the original space corresponds to a point in the dual space; moreover, that point must be outside the dual region. Similarly, a line which was outside the region in the original space maps to a point which must be inside the dual region. These simple rules suffice to uniquely determine the image of any region in the dual space. As an example, the dual of the Amplituhedron in the above example where $(AB)$ passes through $Z_2$ is the light-blue shaded region on the right-hand-side of:
\begin{equation}
    \raisebox{-75pt}{\includegraphics[scale=.45]{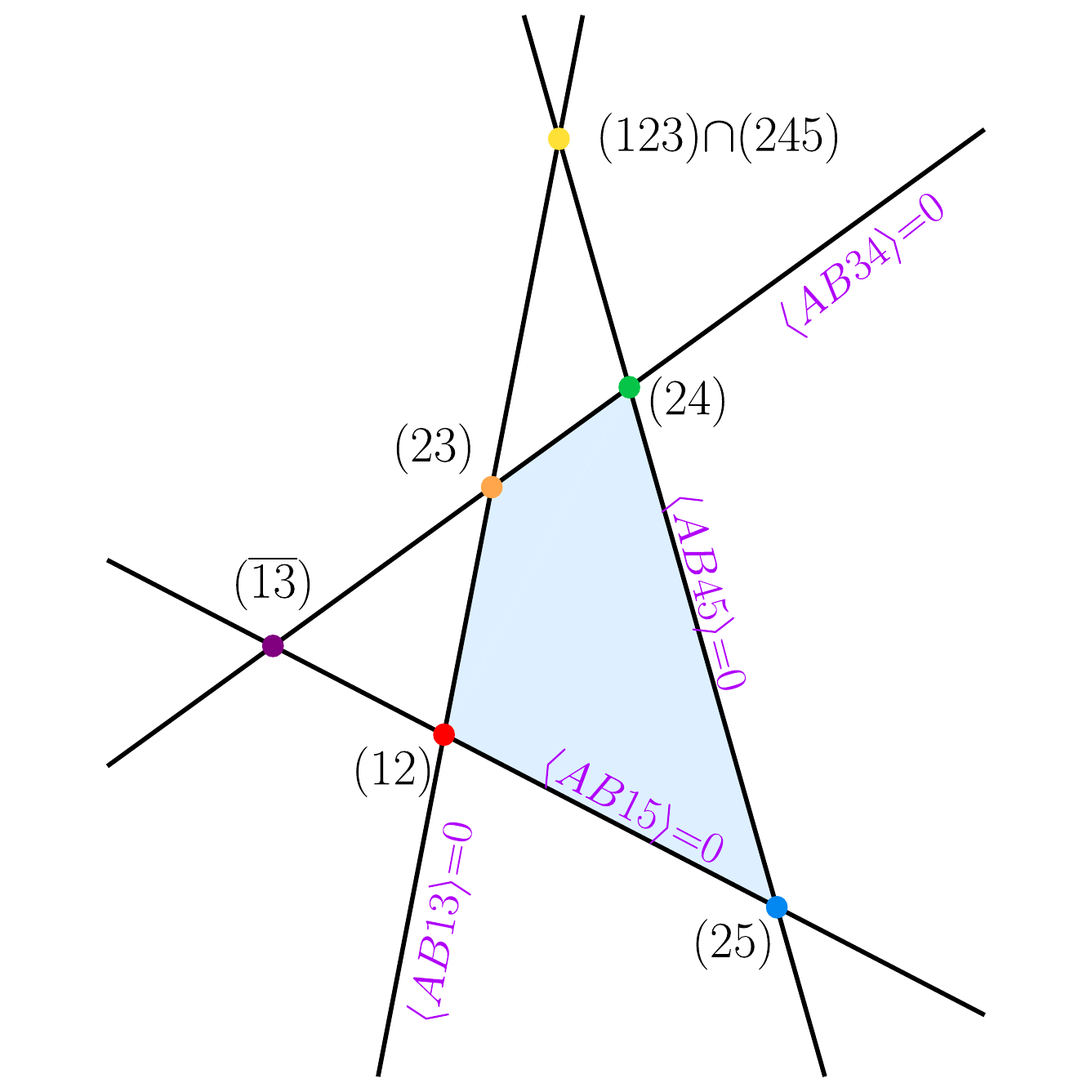}} \underset{\text{dual to}}{\iff}
    \raisebox{-75pt}{\includegraphics[scale=.45]{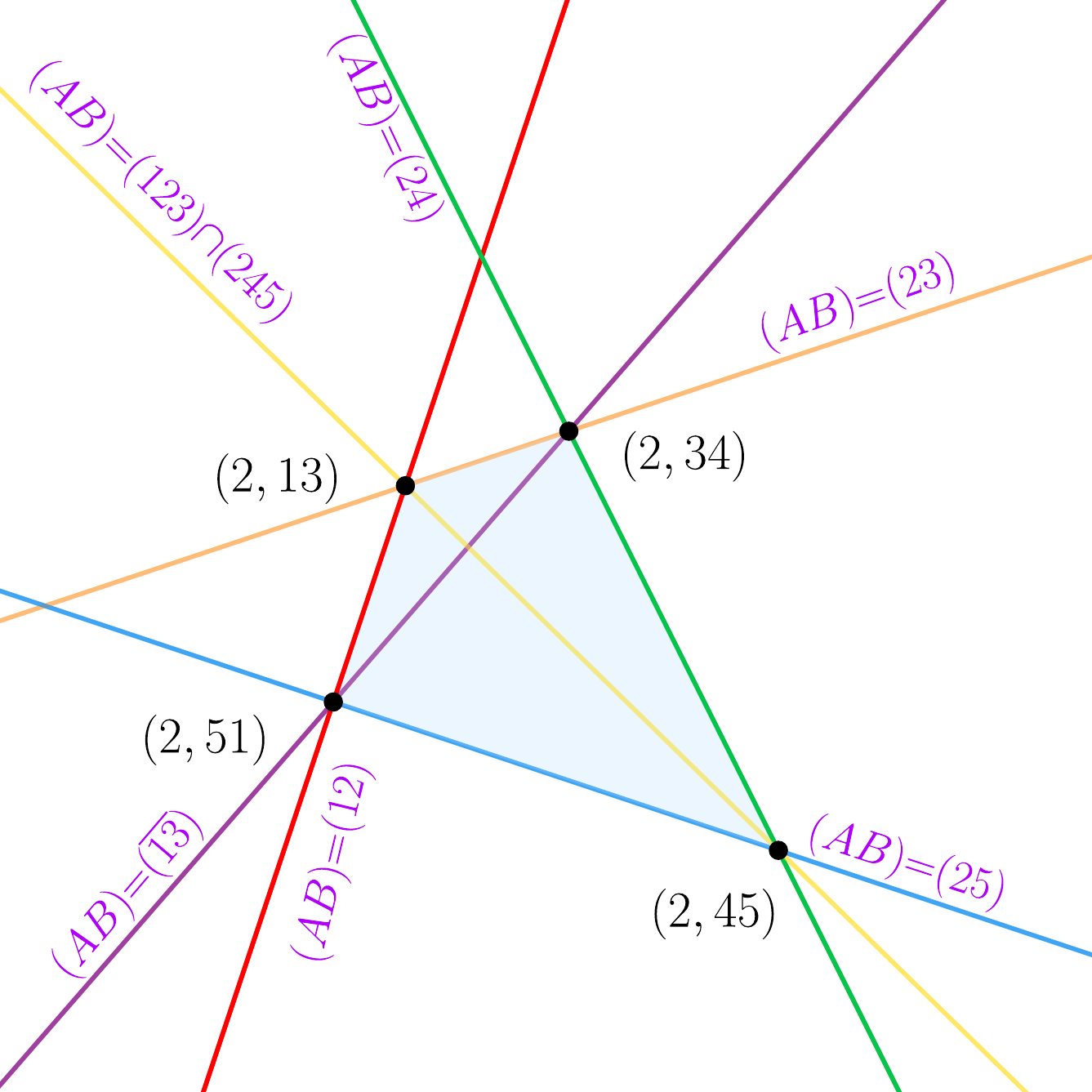}}
    \label{dual_amp_2_projection}
\end{equation}
In particular, the spurious leading singularities $(\overline{13})$ and $(123){\cap}(245)$, which are outside the original Amplituhedron on the left, map to lines in the dual space which pass through the dual Amplituhedron on the right. 

The parity conjugate projection, where $(AB)$ lies in the plane $(123)$, is a slightly less trivial example of the correct dualization procedure
\begin{equation}
    \raisebox{-75pt}{\includegraphics[scale=.45]{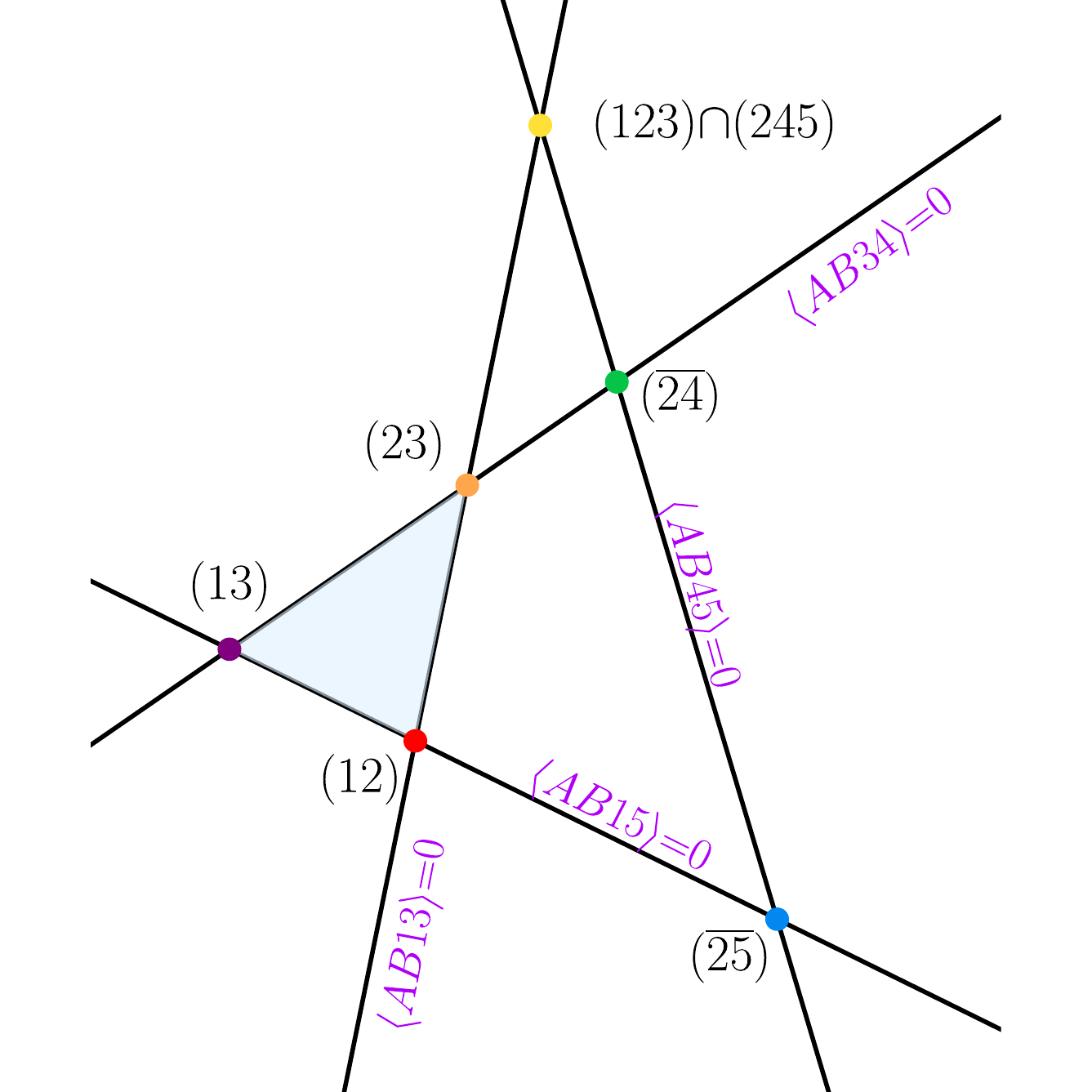}} \underset{\text{dual to}}{\iff}
   \raisebox{-65pt}{\includegraphics[scale=.45]{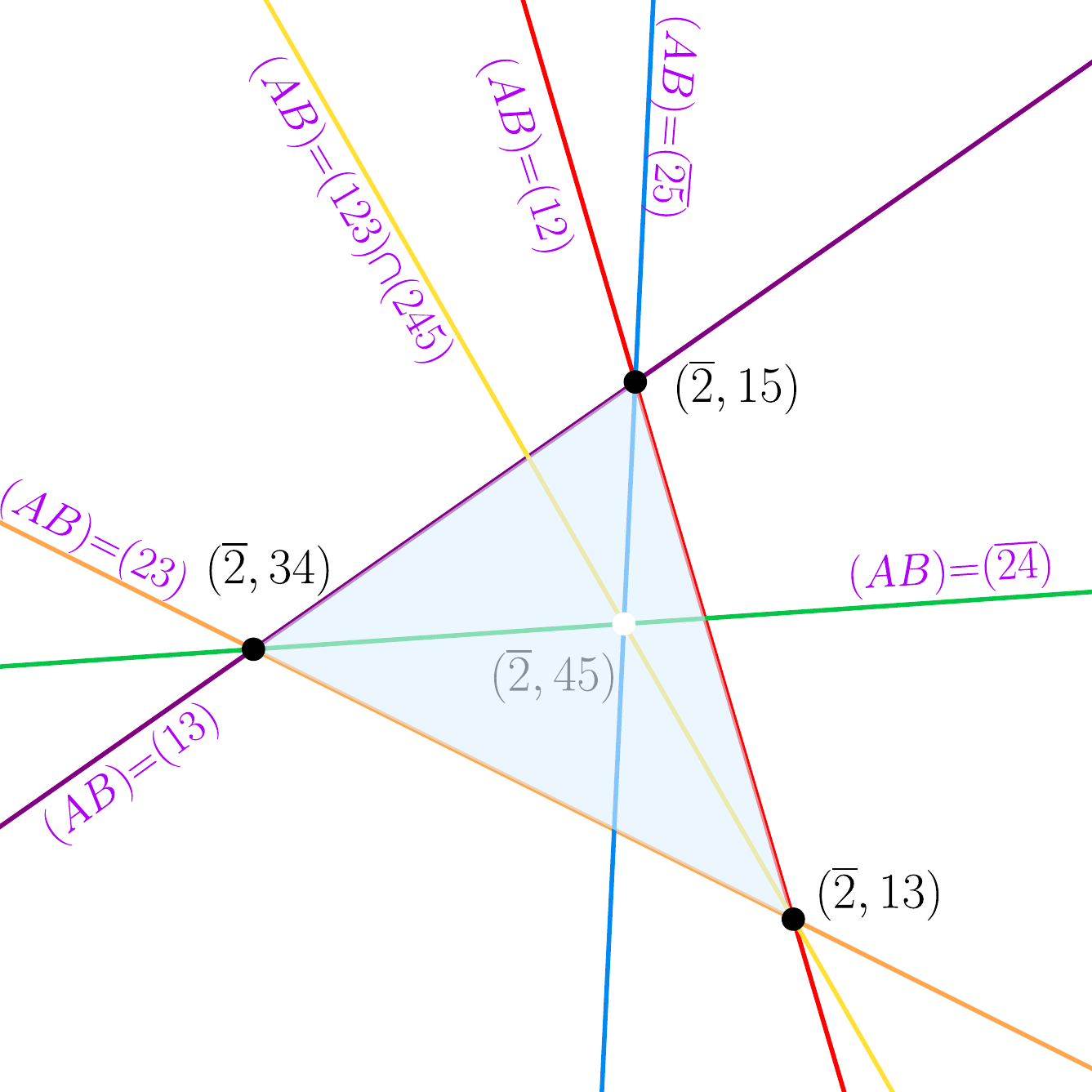}}
   \label{dual_123}
\end{equation}
The relative positions of lines/vertices in the dual picture on the right merit an explanation. Once again, the shaded region is the (dual) Amplituhedron, indicated primarily for illustrative purposes. In this case, the spurious leading singularities $(\overline{24}),(\overline{25})$ and $(123){\cap}(245)$ must all pass through the dual Amplituhedron region. Our convention throughout this work has been to assign the Amplituhedron to a region with finite area. To maintain this convention here requires the vertex $(\overline{2},45)$, which denotes the codimension-three boundary where the line $(AB)$ lies in the plane $(123)$ and cuts $(45)$, to lie \emph{inside} the triangle bounded by the edges corresponding to the accessible MHV leading singularities. On this cut surface, these are $(AB)=\{(12),(13),(23)\}$. Importantly, the point $(\overline{2},45)$ is spurious, so it is not included in the amplitude region; to indicate this, we use an empty (white) vertex. All similar codimension-two boundaries (at five points) are obtained by simply relabeling the above examples. 

\subsection{Dual spaces of chiral pentagons}
%
Having introduced the dual two-dimensional projections, we now turn to identifying the image of the boxes and chiral pentagons under dualization using the prescription discussed above. Let us return to the five-point case, where the amplitude is a sum of two boxes and a single pentagon, c.f. eq.~(\ref{eq:1loop_5pt_local_int_exp}). Using the results of section~\ref{subsec:gluing_regions} and appendix~\ref{app:2d_gluing_details}, we can identify the dual regions corresponding to the projection summarized in eq.~(\ref{2_proj_box45}), where $(AB)=(A2)$. The box space $B_{45}^{(3)}$ relevant for the Amplituhedron-Prime and its na\"{i}ve dualization are
\begin{equation}
    B_{45}^{(3)}\leftrightarrow
    \hspace{-.4cm}
    \raisebox{-75pt}{\includegraphics[scale=.45]{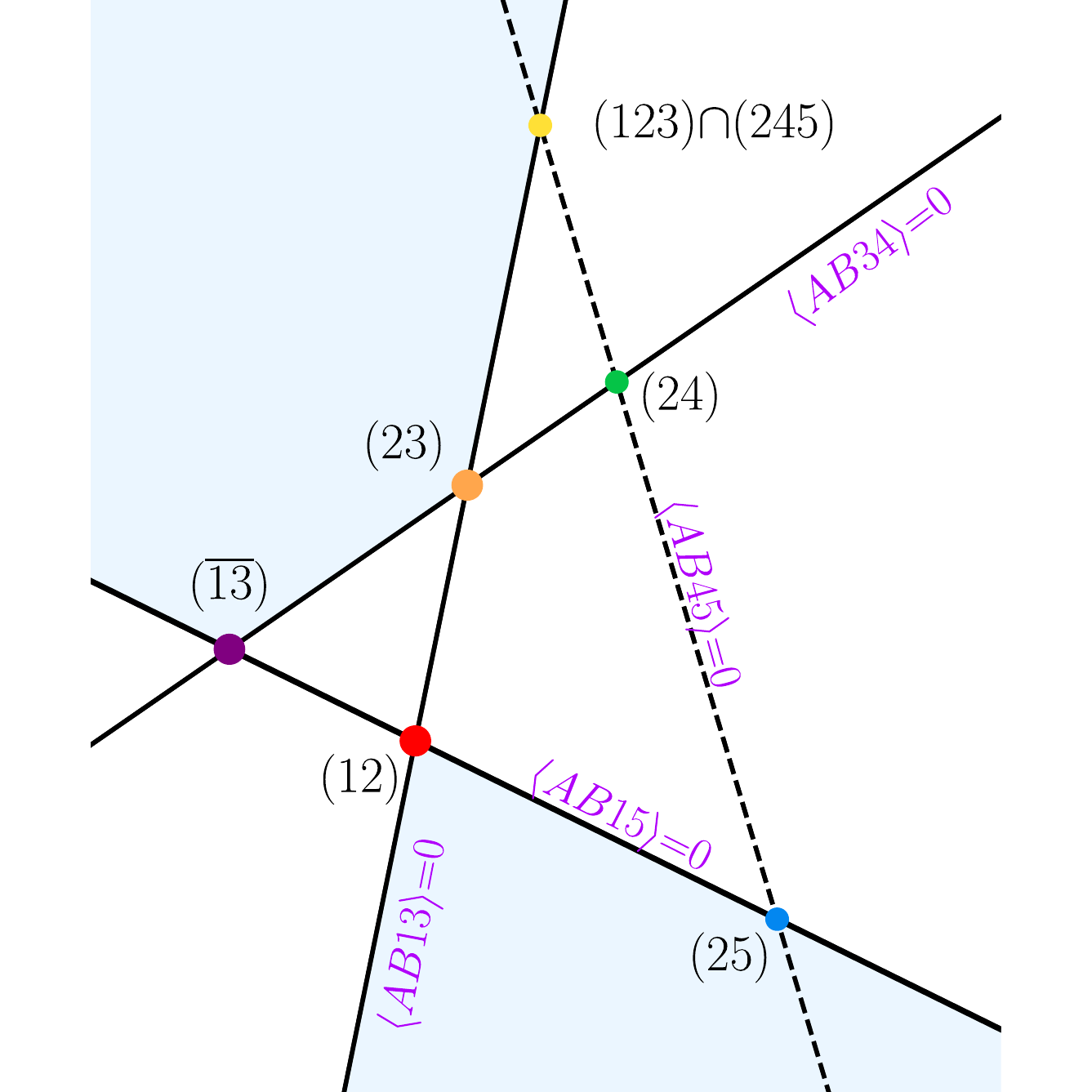}}
    \hspace{-.4cm}
    \underset{\text{dual to}}{\iff}
   \raisebox{-75pt}{\includegraphics[scale=0.45]{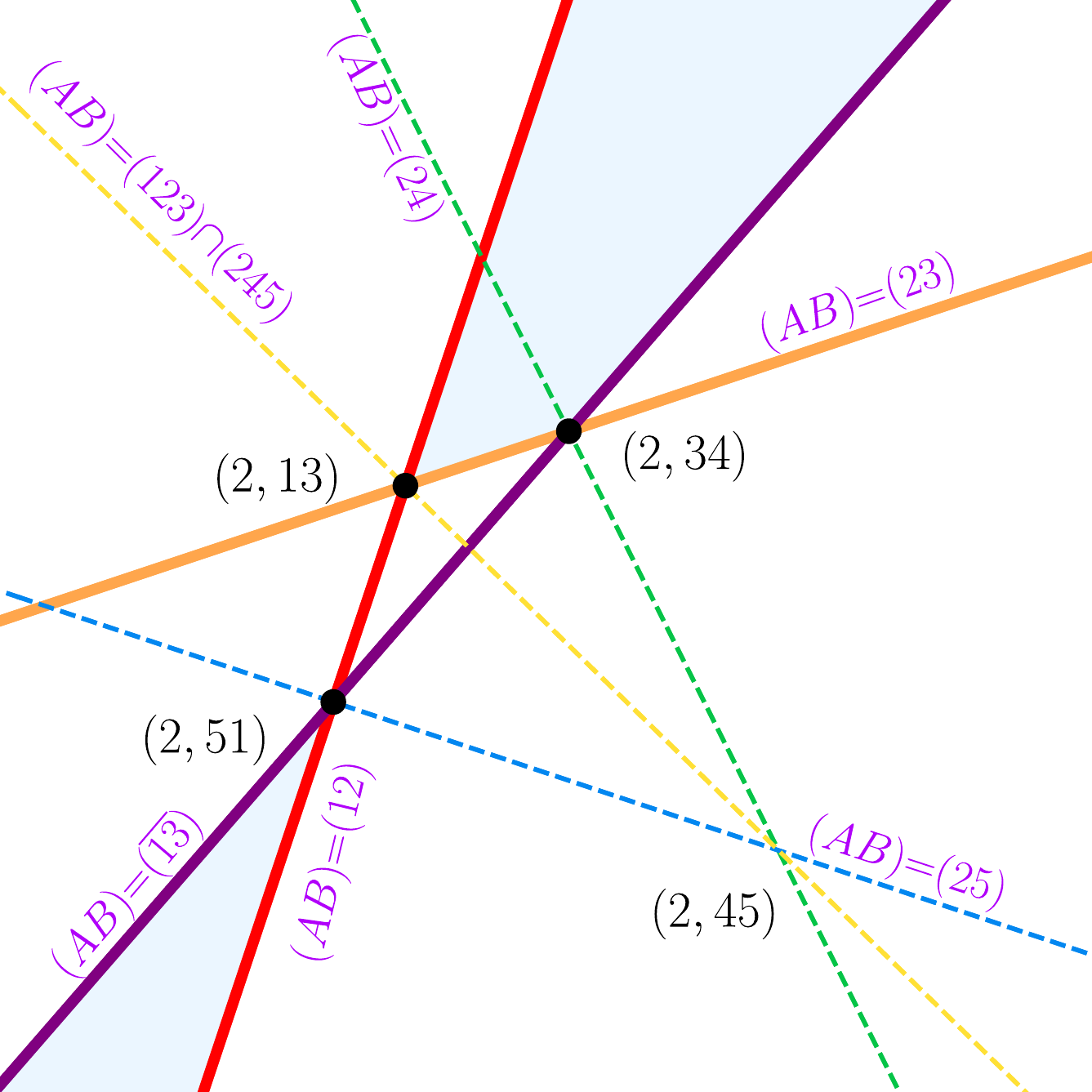}}
    \label{box_dual_2_projection_naive}
\end{equation}
This seems to suggest that the dual of the Amplituhedron-Prime is \emph{not} an internal triangulation of the dual Amplituhedron as the dual of the $B_{45}^{(3)}$ box region naively lies outside of the dual of the Amplituhedron in the right figure of (\ref{box_dual_2_projection_naive}). However, there is a critical feature of the dualization which has been neglected in eq.~(\ref{box_dual_2_projection_naive}): namely, the \emph{lower-dimensional boundaries} in the original projection on the left-hand-side, which map under dualization to infinite wedges in the dual picture on the right. The cavalier treatment of the lines and vertices on the left-hand-sides of eq.~(\ref{box_dual_2_projection_naive}) causes no issue from the perspective of the canonical forms because any less-than-full-rank subspace of $\mathbb{P}^3$ has vanishing form. However, these same boundaries play a pivotal r\^{o}le in the dual picture because they dualize to larger spaces with nonzero (in fact, infinite) volume. Note that while the dual of a line is a point, the dual of a line \emph{segment} is an infinite wedge (with two codimension-one boundaries in the dual) defined by the two leading singularities which bookend the line segment. Thus, the right-hand-side of eq.~(\ref{box_dual_2_projection_naive}) only represents the bulk component of the dualization and is incomplete. 
%
%
%
In fact, there is a simpler way of identifying the correct dual spaces which exploits the fact that zero-form spaces dualize to lower-dimensional boundaries. First, we can identify the region which dualizes to the triangle with vertices $(2,51),(2,13)$ and $(2,34)$, with \emph{all boundaries} included:
\begin{equation}
    \raisebox{-75pt}{\includegraphics[scale=0.45]{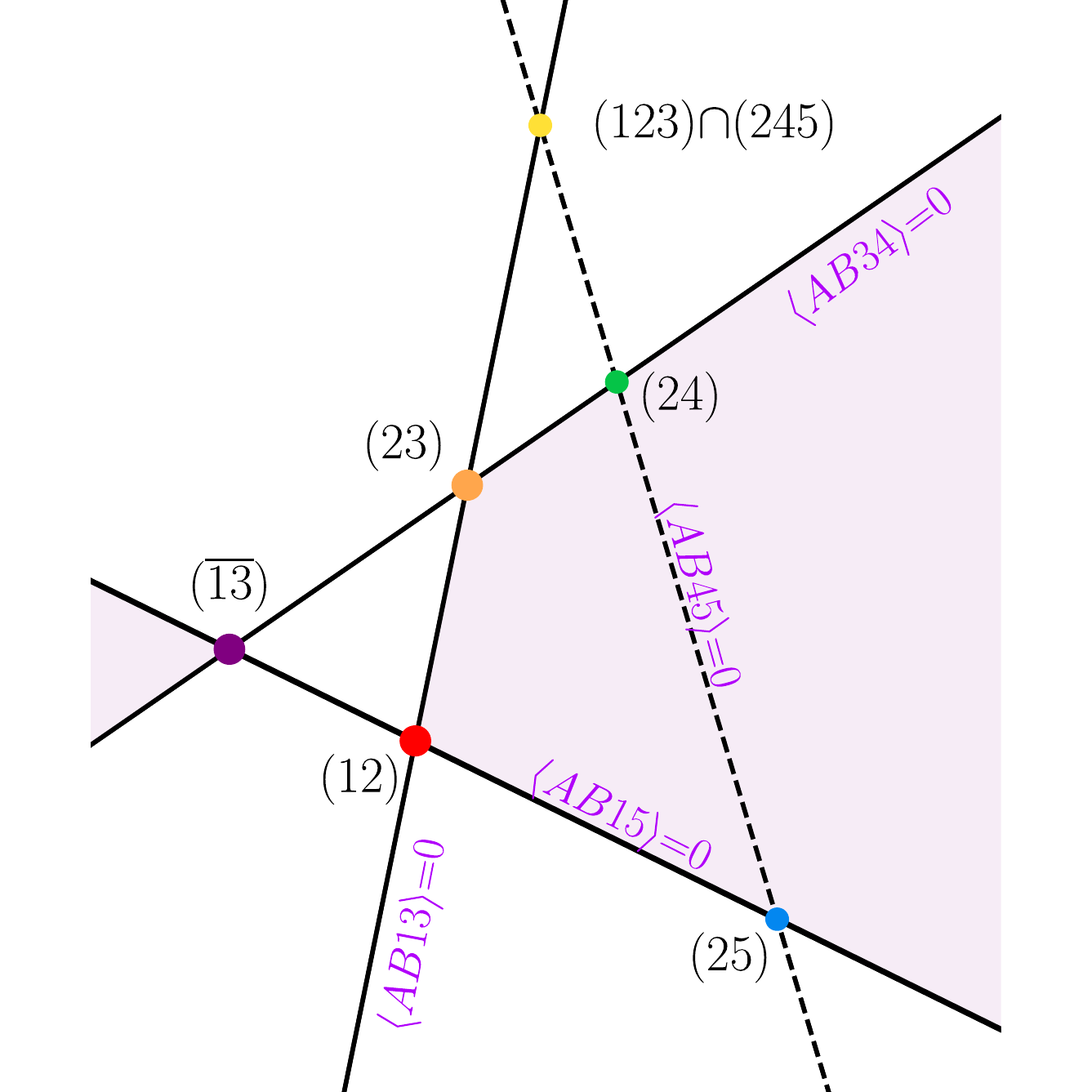}}
    \hspace{-.4cm}
    \underset{\text{dual to}}{\iff}
   \raisebox{-75pt}{\includegraphics[scale=0.45]{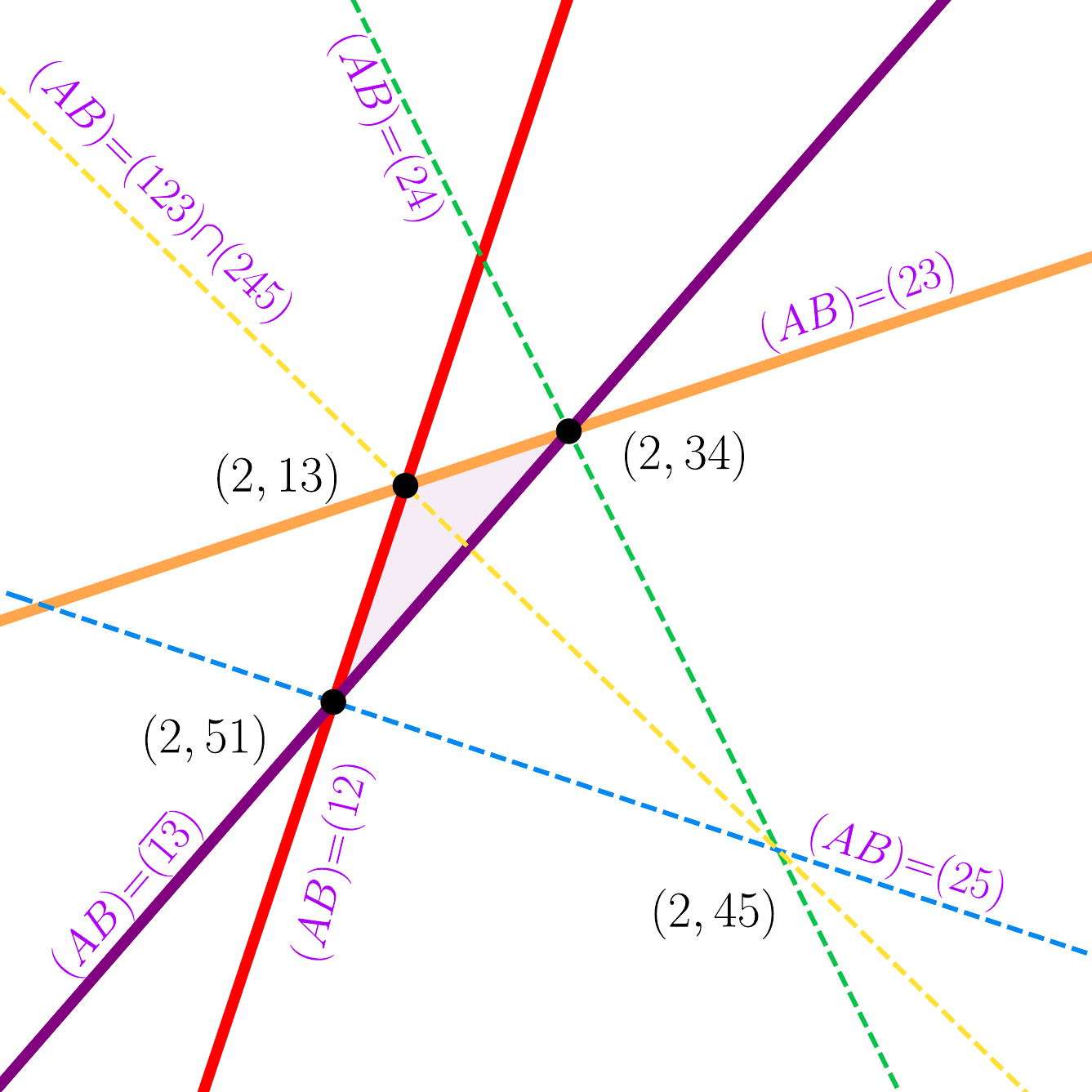}}
   \label{eq:dual_2_b45_ideal}
\end{equation}
In the dual picture, we use filled vertices to indicate these are included as boundaries of the region. Now, the box space on the left-hand-side of eq.~(\ref{box_dual_2_projection_naive}) can be related to the left-hand-side of eq.~(\ref{eq:dual_2_b45_ideal}) by adding a zero-form wedge in the original space, namely
\begin{equation}
     \raisebox{-75pt}{\includegraphics[scale=0.45]{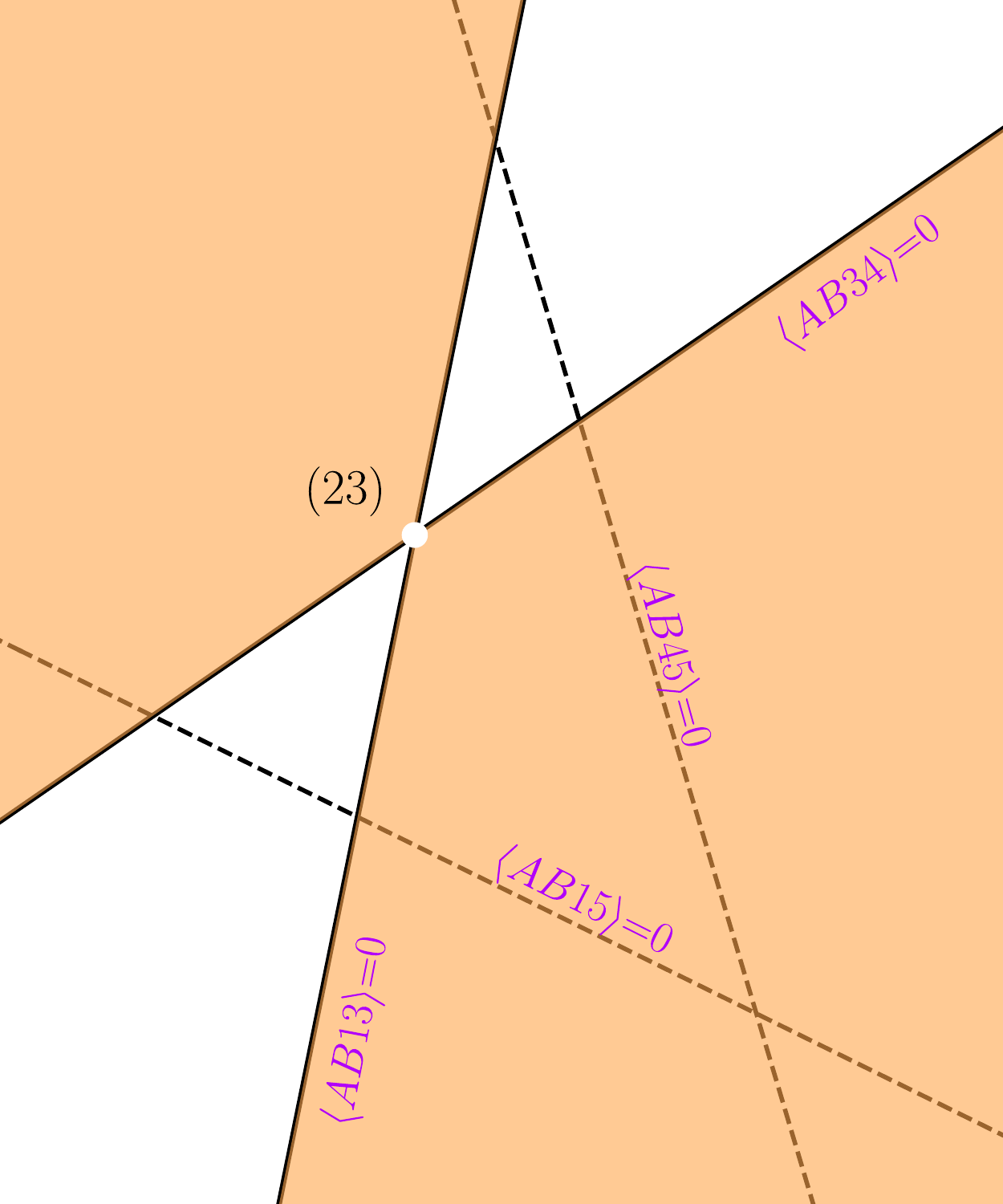}} \underset{\text{dual to}}{\iff}
   \raisebox{-75pt}{\includegraphics[scale=0.45]{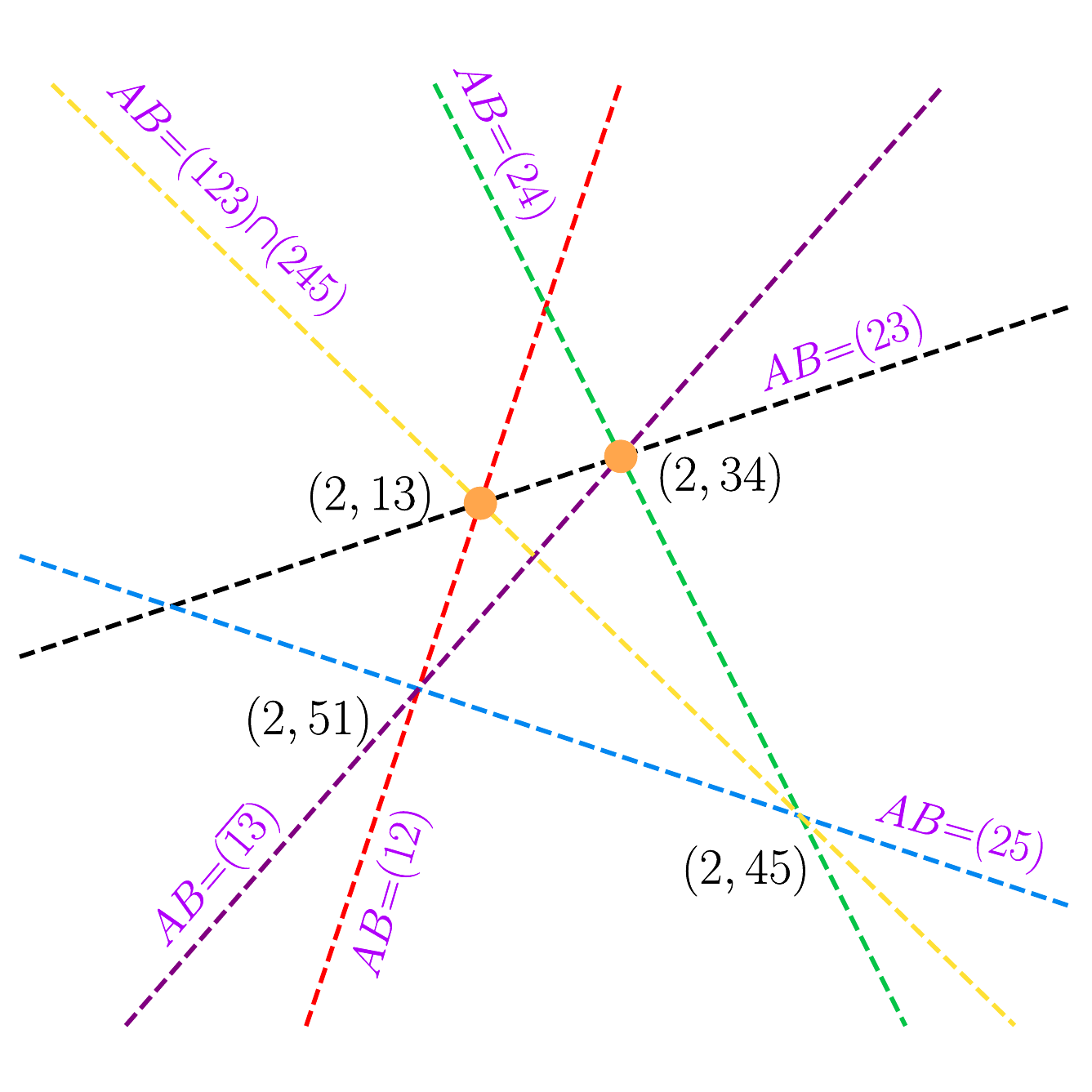}}
   \label{eq:2_projection_zero_form_example_box45}
\end{equation}
We use a white vertex to indicate that this point is \emph{excluded} from boundary of the region. In the dual picture, we draw ``dashed'' lines to indicate that only the two \textcolor{orange}{orange} vertices constitute the dual region. Therefore, the dualization of the box space of the left-hand-side of eq.~(\ref{box_dual_2_projection_naive}) is the internal triangle with precisely this edge absent,
\begin{equation}
    B_{45}^{(3)}\leftrightarrow
    \hspace{-.4cm}
    \raisebox{-75pt}{\includegraphics[scale=0.45]{./figures/B45_projection_A2_original.pdf}}
    \hspace{-.4cm}
    \underset{\text{dual to}}{\iff}
   \raisebox{-75pt}{\includegraphics[scale=0.45]{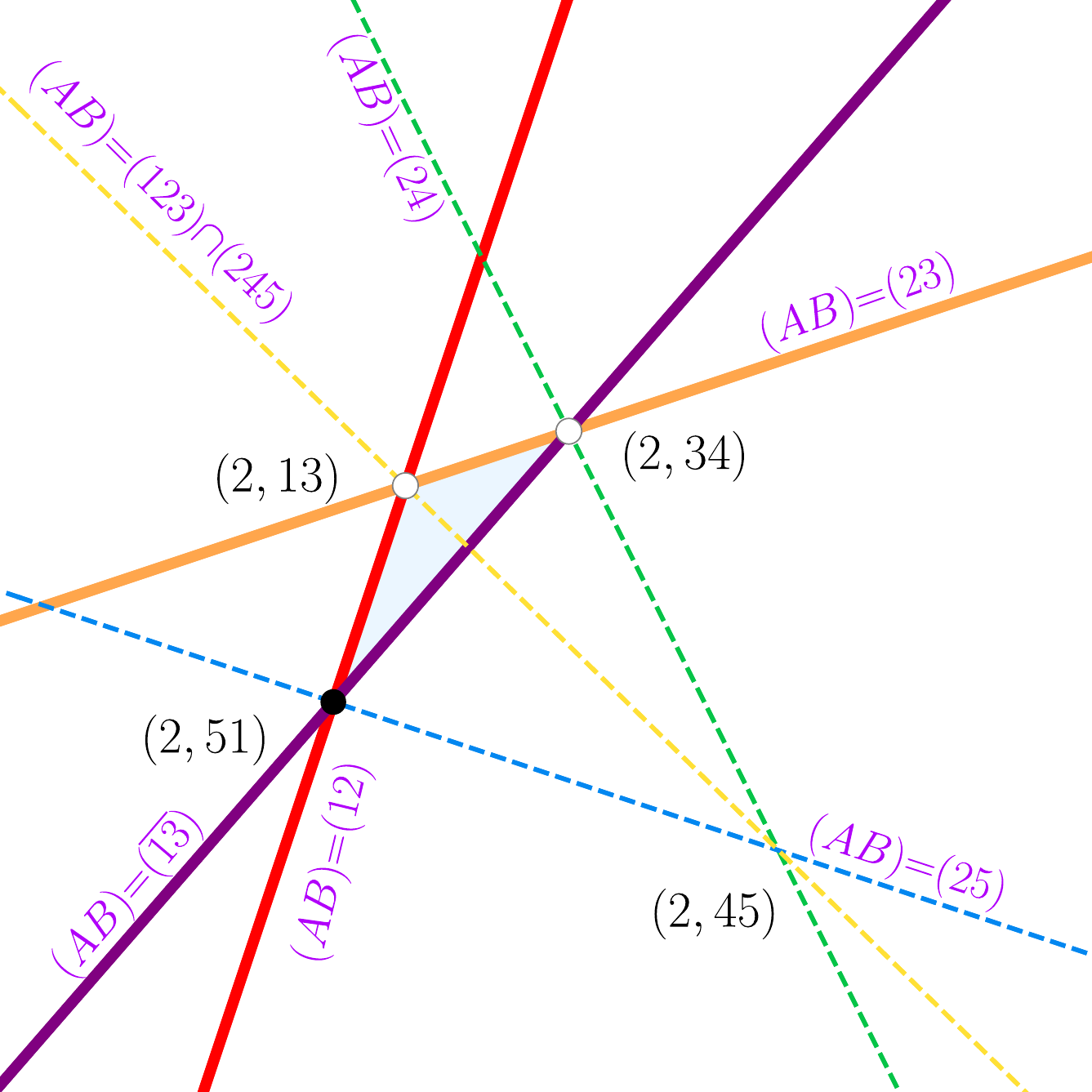}}
   \label{dual_b45}
\end{equation}
The dualization of both pentagon spaces (\ref{2_proj_pent}) on the boundary where $(AB)=(A2)$ can be constructed in a similar fashion. Similarly to (\ref{eq:dual_2_b45_ideal}), we first identify the region in the original space which maps to the second internal triangle of the dual space:
\begin{equation}
    \raisebox{-75pt}{\includegraphics[scale=0.45]{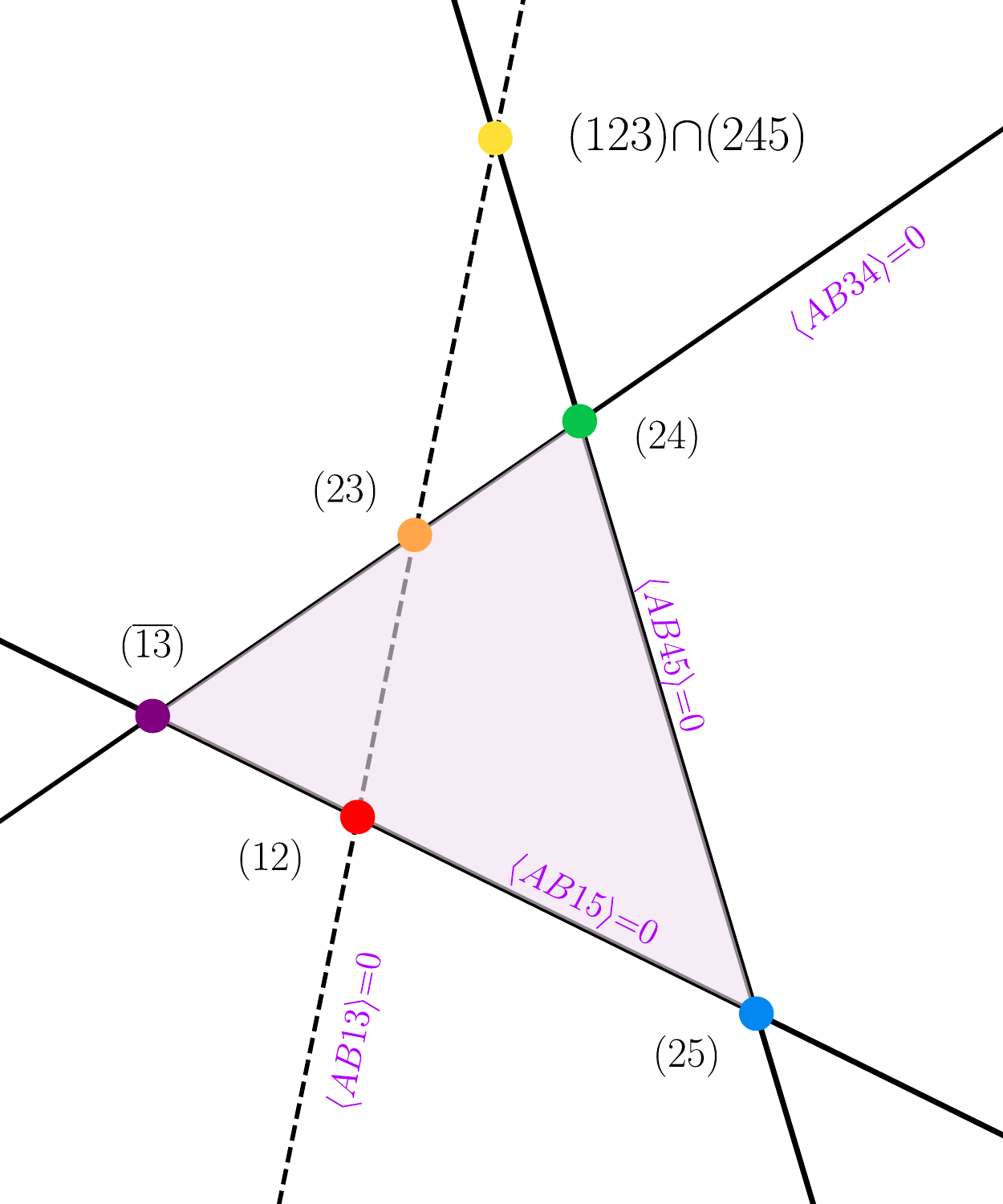}} \underset{\text{dual to}}{\iff}
   \raisebox{-75pt}{\includegraphics[scale=0.45]{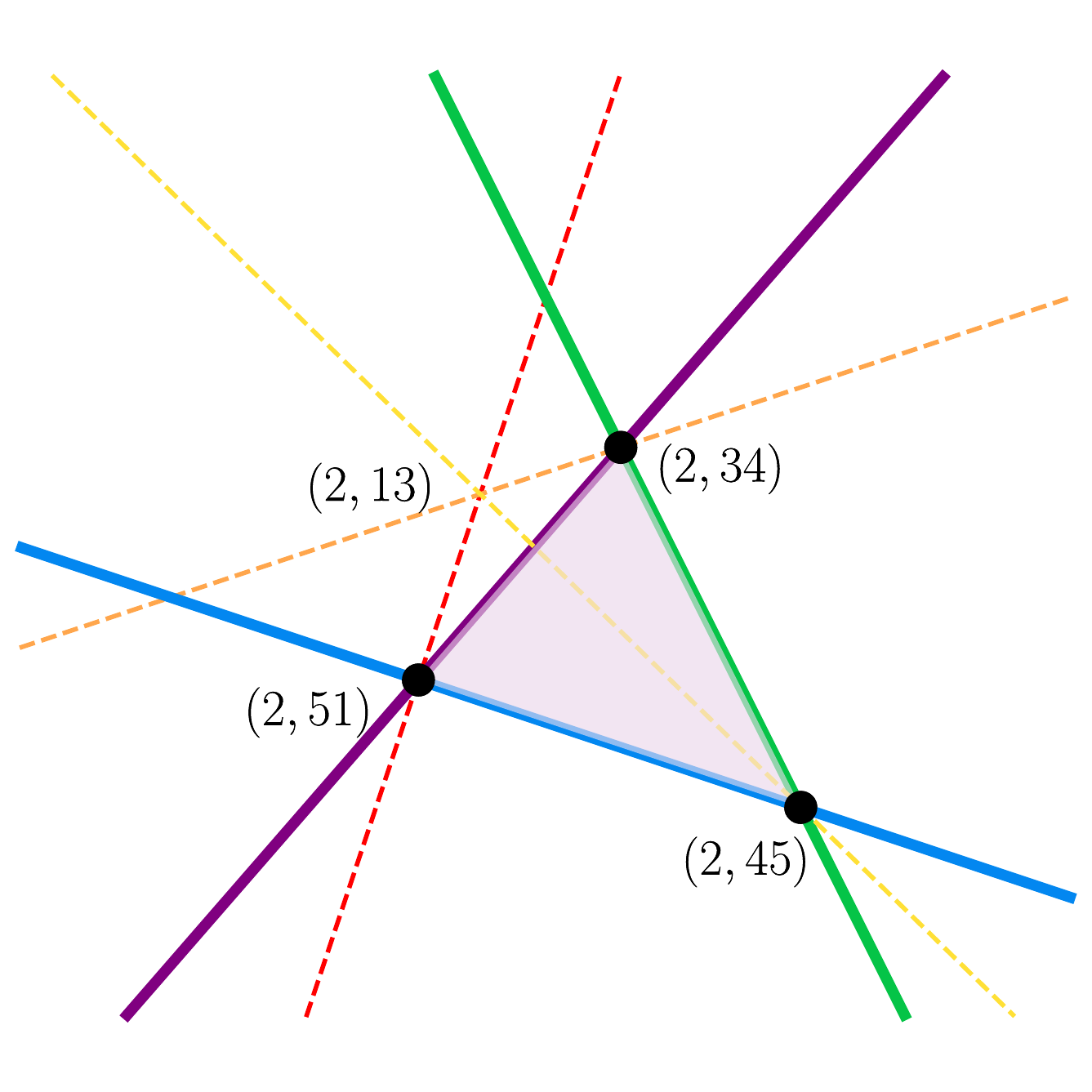}}
   \label{eq:dual_2_p24_ideal}
\end{equation}
The pentagon space $P_{24}^{(1)}$, relevant for the Amplituhedron-Prime, can be related to eq.~(\ref{eq:dual_2_p24_ideal}) by the addition of the zero form region
 \begin{equation}
     \raisebox{-70pt}{\includegraphics[trim={0cm 1cm 0cm 2cm},clip,scale=0.45]{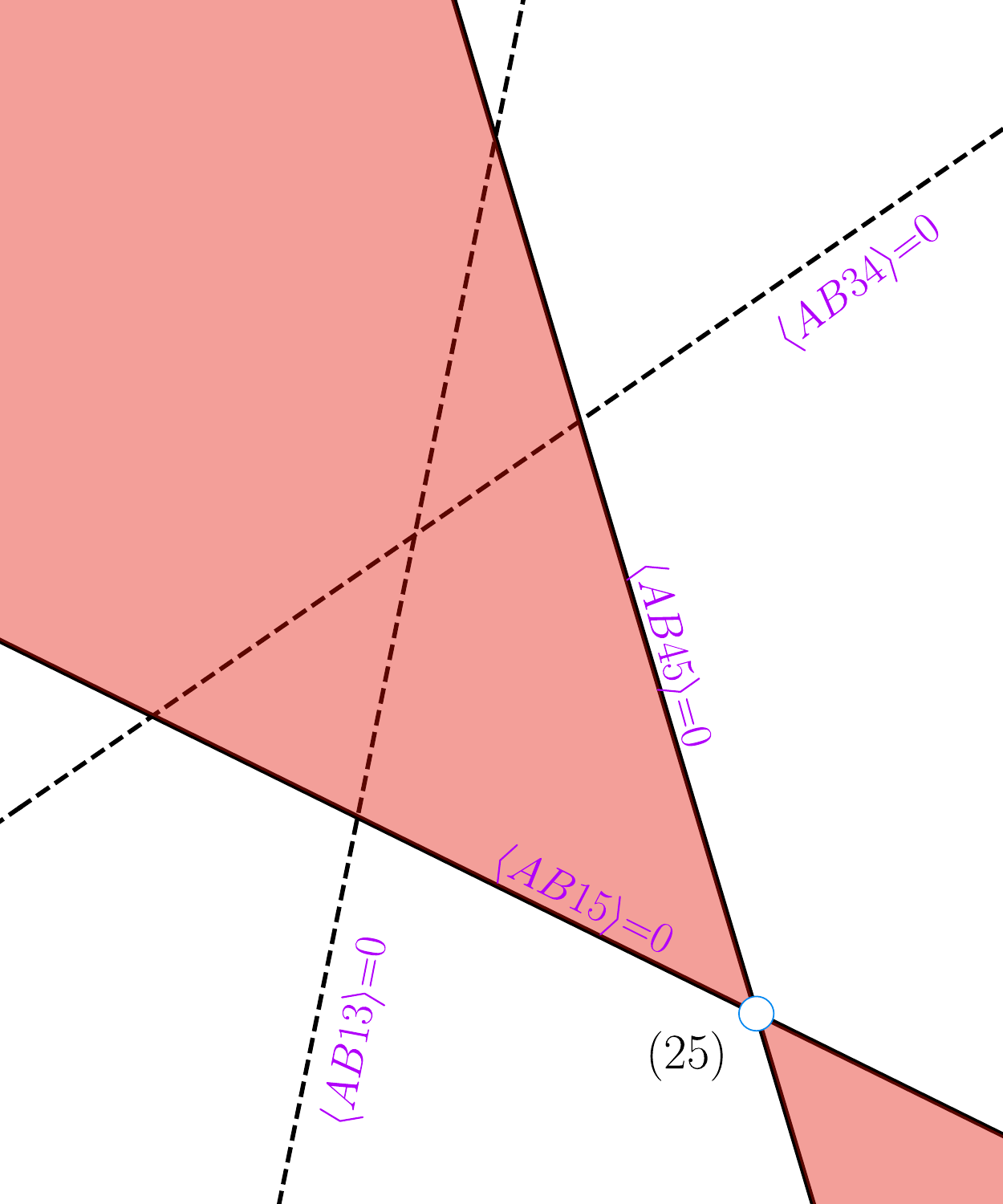}} 
    \hspace{-.1cm} 
    \underset{\text{dual to}}{\iff}
   \raisebox{-75pt}{\includegraphics[scale=0.45]{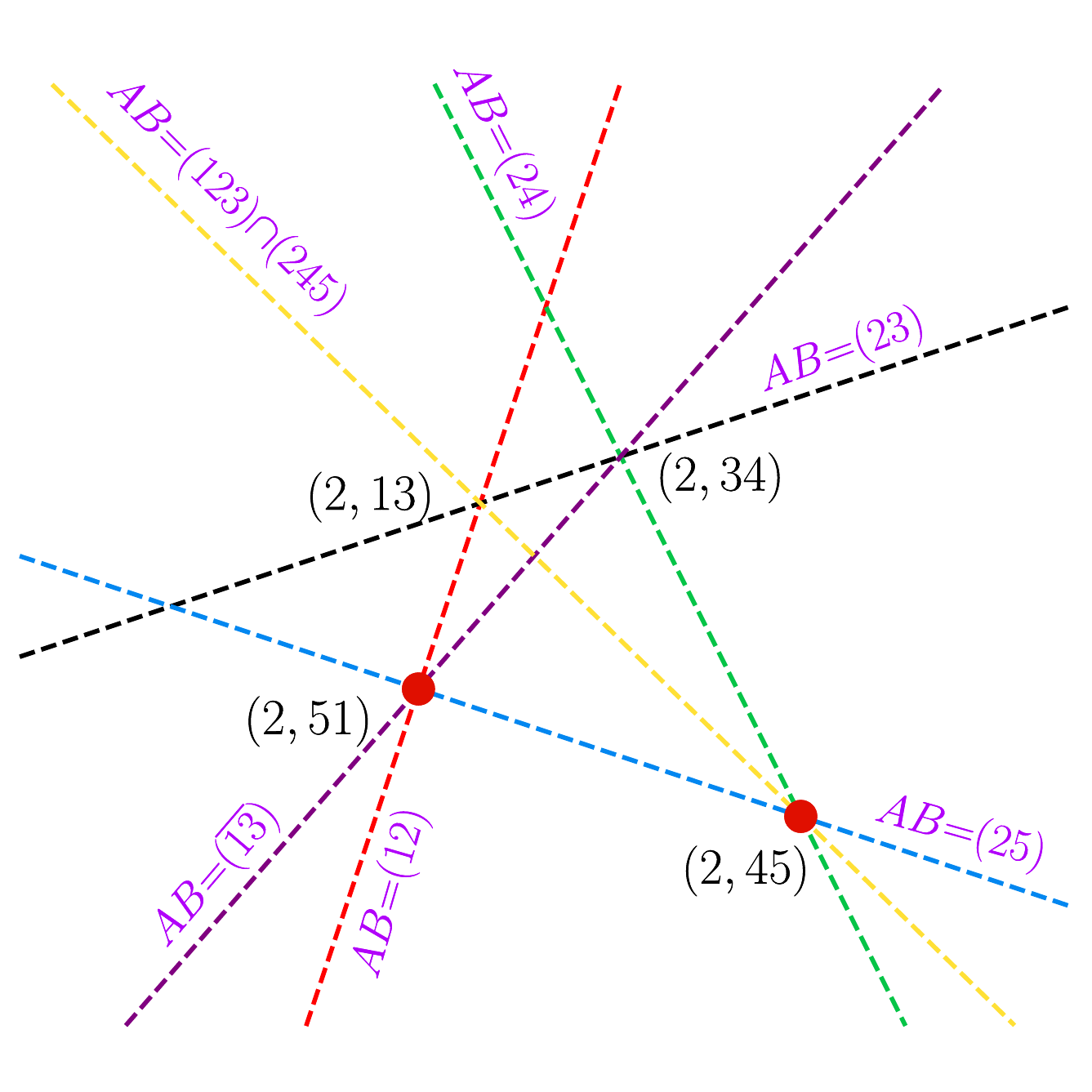}}
   \label{eq:2_projection_zero_form_example_p24}
\end{equation}
so the correct dualization for the pentagon $P_{24}^{(1)}$ is
\begin{equation}
    P_{24}^{(1)} \leftrightarrow
    \hspace{0cm}
    \raisebox{-70pt}{\includegraphics[trim={0cm 1.5cm 0cm 1cm},clip,scale=0.45]{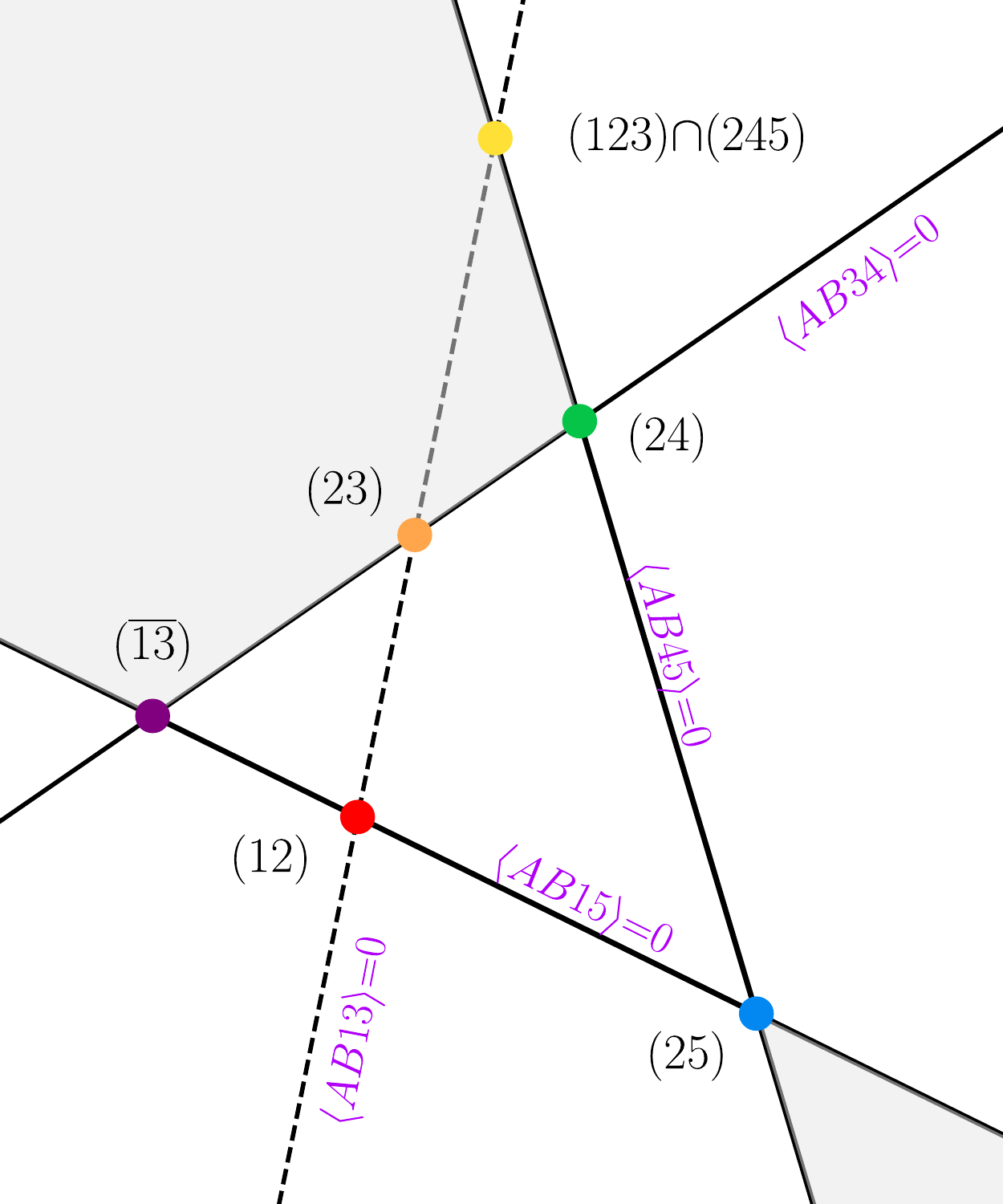}}
    \hspace{-.1cm}
    \underset{\text{dual to}}{\iff}
   \raisebox{-75pt}{\includegraphics[scale=0.45]{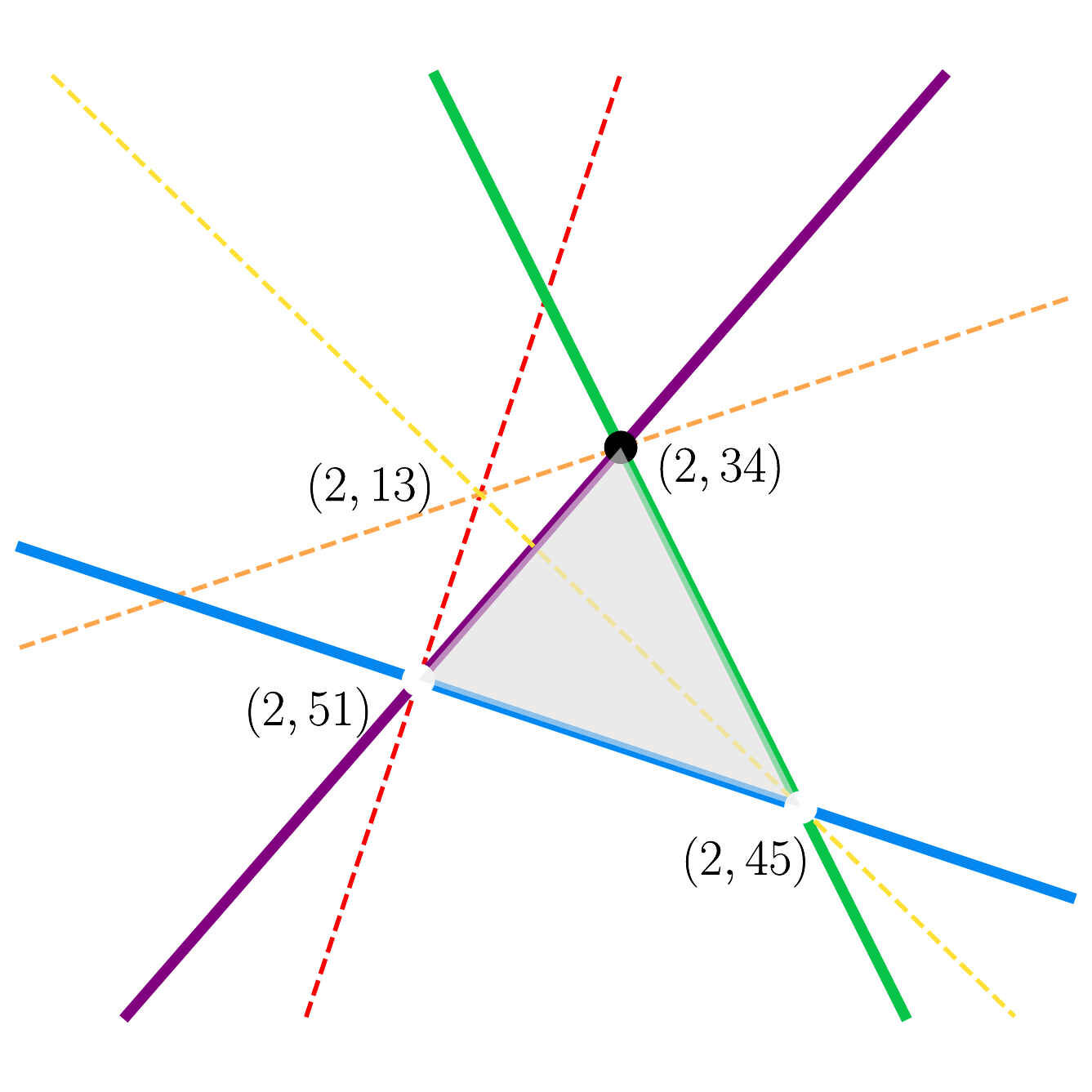}}
   \label{dual_p24}
   \hspace{-.3cm}
\end{equation}
For reference, the dualization for the alternative pentagon space $P_{24}^{(2)}$ is the same dual bulk region, but with the vertices $(2,51)$ and $(2,34)$ missing.

\subsection{Two-dimensional triangulations}
%
The chiral pentagon expansion triangulated the Amplituhedron-Prime space $\mathcal{A}'$. Since the logarithmic forms for $\mathcal{A}$ and $\mathcal{A}'$ are equal, by definition, their dual spaces have the same volume. Thus, a priori in the dual pictures $\mathcal{A}$ and $\mathcal{A}'$ can only differ by zero-volume lower-dimensional boundaries which are dual to zero-form wedges in the original space. In fact, the same argument suggests that \emph{any} choice of individual box and pentagon geometries must match the dual Amplituhedron up to possibly its vertices and edges. To demonstrate this correspondence explicitly, we carefully account for all lower-dimensional boundaries in the dualization procedure. As discussed in the previous subsection, these line segments dualize to infinite wedges which dramatically affect the resulting dual region. In general, it is easiest to understand the dualization by utilizing zero form regions in the original two-dimensional projection.

Let us now compare the behavior of the Kermit eq.~(\ref{eq:Kermit}) and chiral pentagon eq.~(\ref{chiral}) expansions in the original and dual two-dimensional projections. As shown in section~\ref{sec:Amplituhedron_geometry}, the Kermit representation is by construction an internal triangulation of the Amplituhedron. By the schematic arguments of section~\ref{sec:introduction}, the expectation is that internal triangulations map to external triangulations of the dual, and vice versa. We now establish this for the Kermit expansion at five points on the boundary when $(AB)$ passes through $Z_2$. On this cut, the Kermits triangulate the quadrilateral using the line $\ab{AB14}{=}0$. \emph{Excluding} this line from consideration, the na\"{i}ve dualization reads 
\begin{equation}
    \raisebox{-75pt}{\includegraphics[scale=0.45]{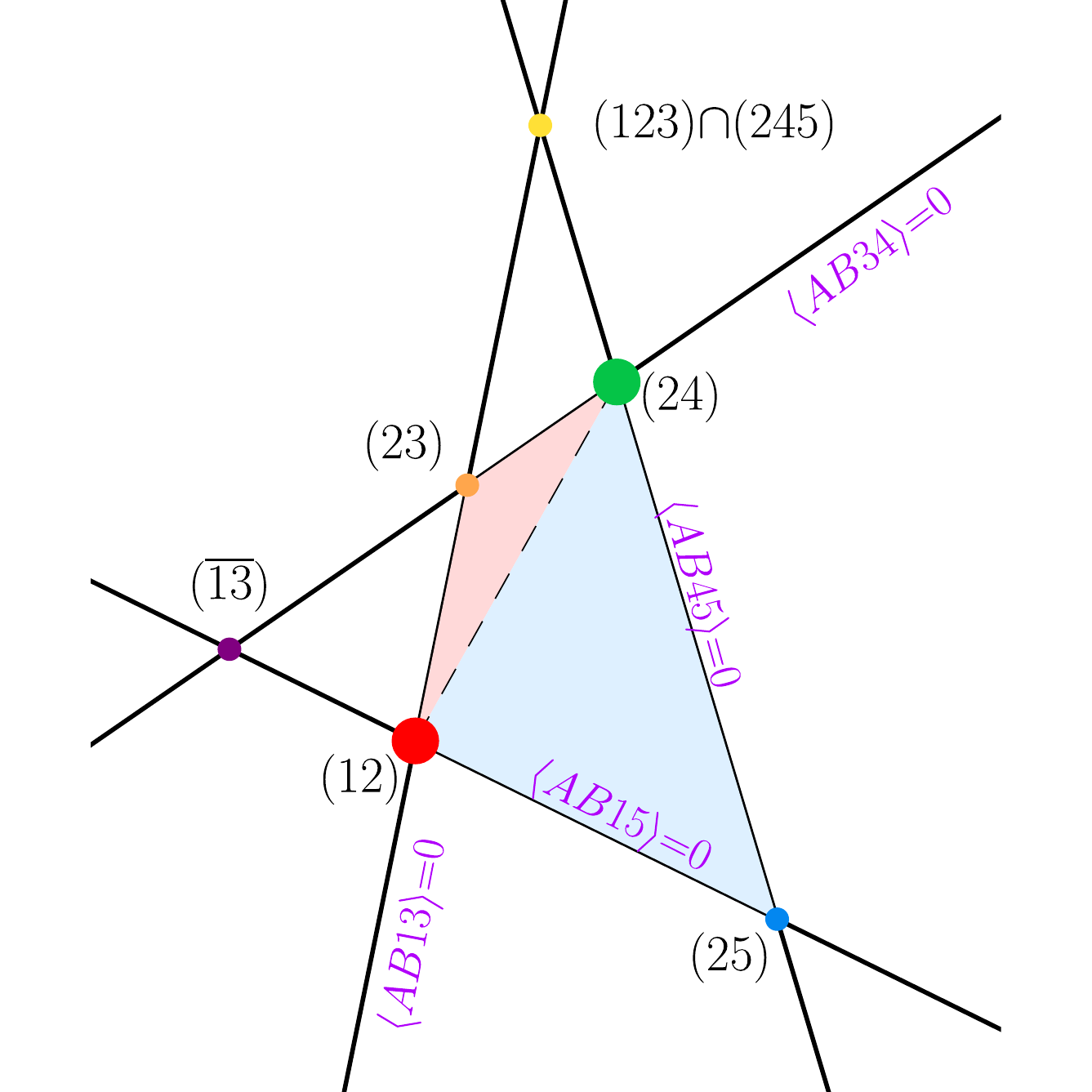}}
    \hspace{-.4cm}
    \underset{\text{dual to}}{\iff}
   \raisebox{-75pt}{\includegraphics[scale=0.45]{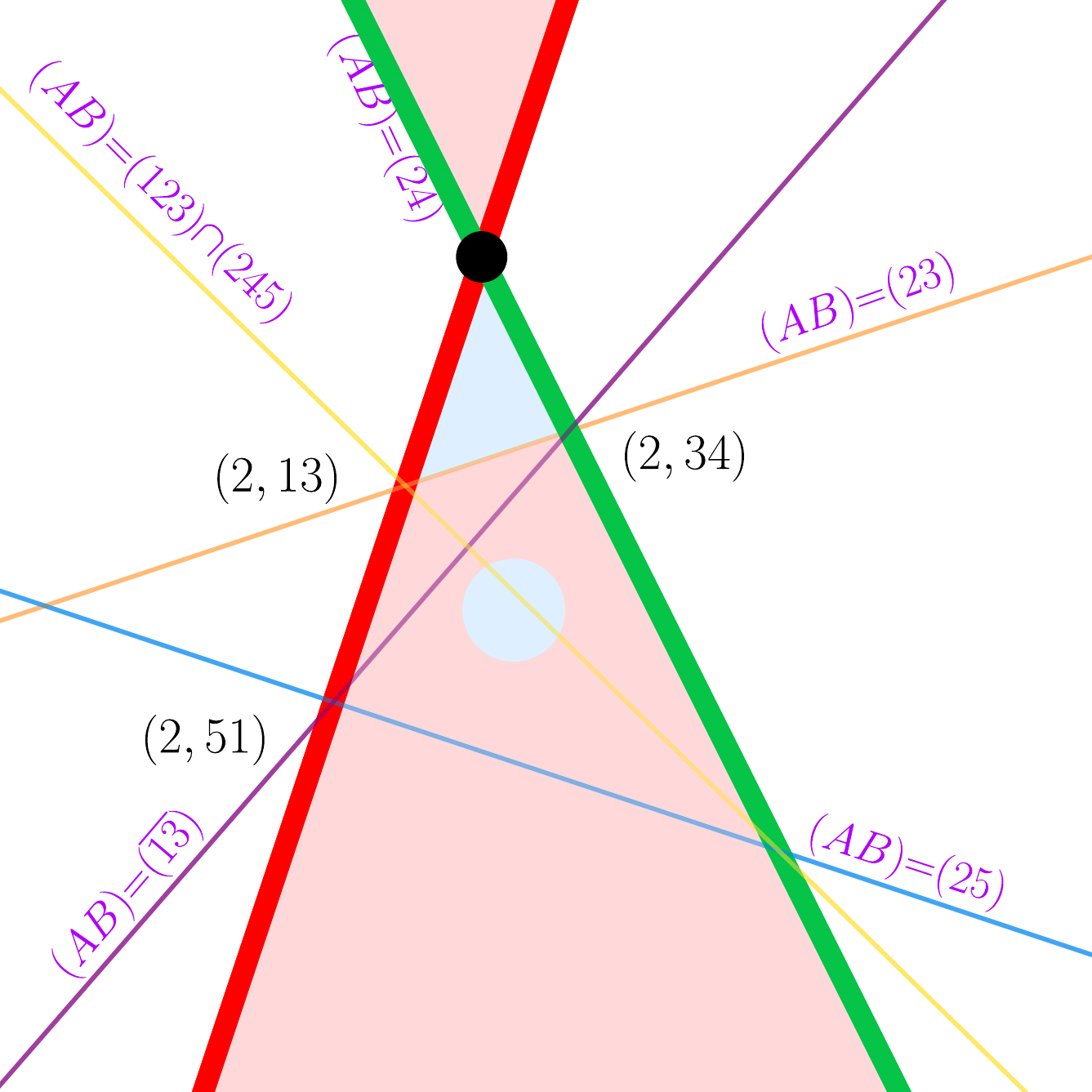}}
    \label{kermit_dual_2_projection}
\end{equation}
We see that just as in eq.~(\ref{box_dual_2_projection_naive}) the space remaining is not the quadrilateral representing the dual Amplituhedron. The issue here is the same as in the na\"{i}ve dualization attempt of the previous subsection: we have been glib about the lower-dimensional boundaries in the original space. Specifically, in the example on the left-hand-side of eq.~(\ref{kermit_dual_2_projection}), the segment of the line $\ab{AB14}=0$ between the points $(12)$ and $(24)$ dualizes to an infinite wedge with exactly these two codimension-one dual boundaries. To account for this in the dual picture requires that we add the dual of this line segment to our na\"{i}ve picture eq.~(\ref{kermit_dual_2_projection}), i.e.,
\begin{equation}
    \raisebox{-75pt}{\includegraphics[scale=0.45]{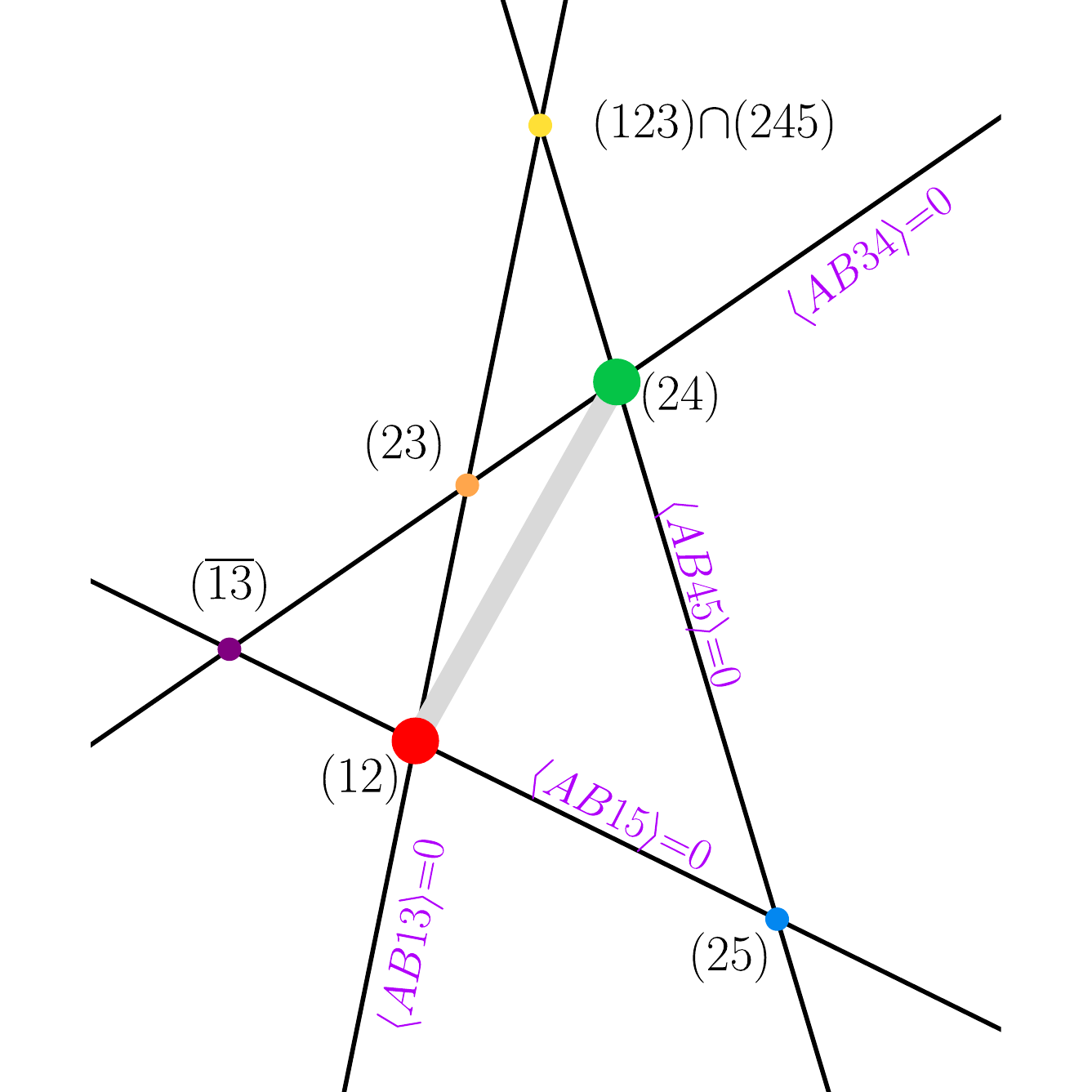}}
    \hspace{-.4cm}
    \underset{\text{dual to}}{\iff}
   \raisebox{-75pt}{\includegraphics[scale=0.45]{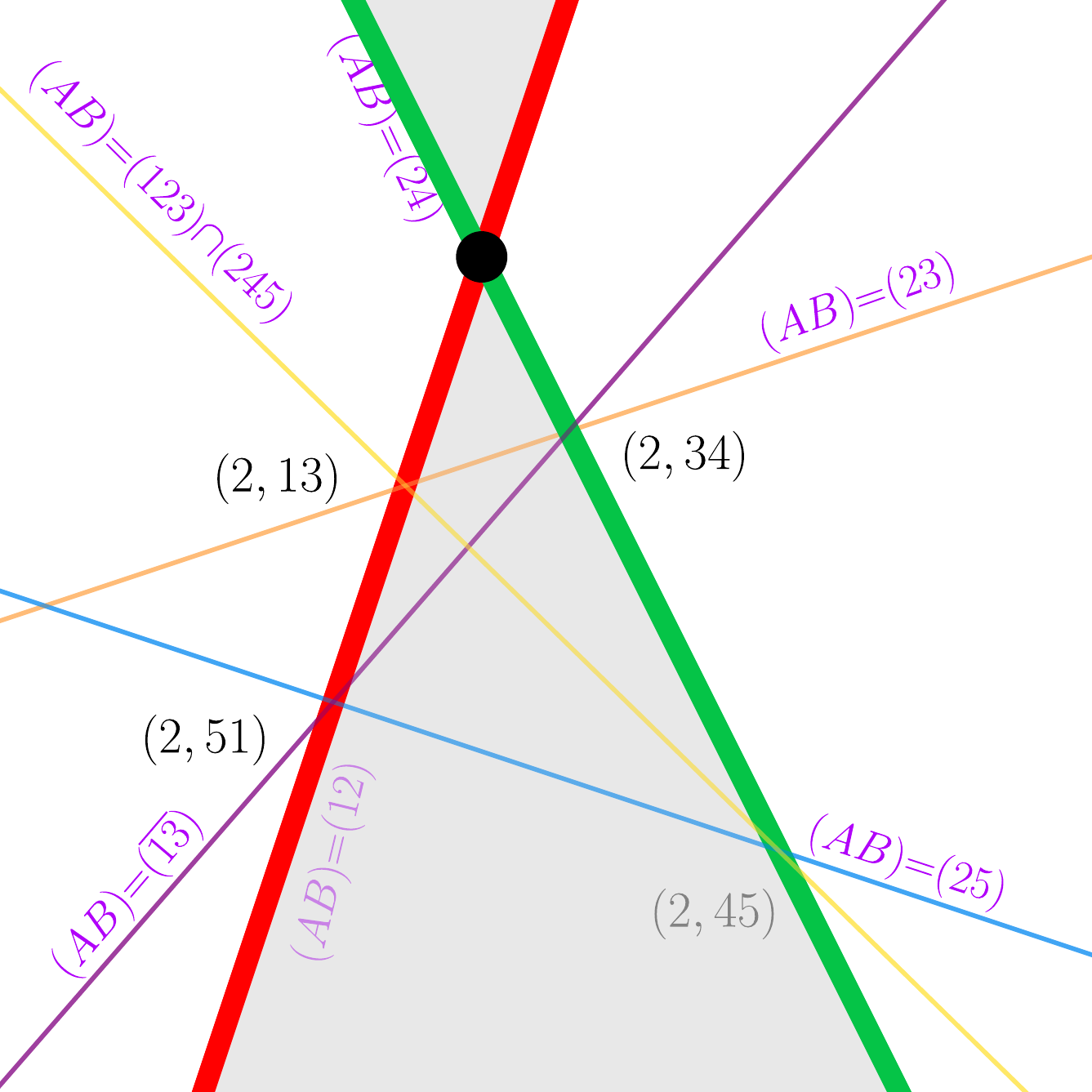}}
   \label{dual_of_line_segment}
\end{equation}
To be clear, this picture includes both the points $(12),(24)$ as well as the line segment itself. Including this boundary with both Kermit regions eq.~(\ref{kermit_dual_2_projection}), we double-cover the points $(12)$ and $(24)$. Thus, to recover the dual space we must add these points back, without the line segment in between them. The dual of this piece is, by completeness, the wedge in eq.~(\ref{dual_of_line_segment}) minus the single point $(2,14)$:
\begin{equation}
    \raisebox{-75pt}{\includegraphics[scale=0.45]{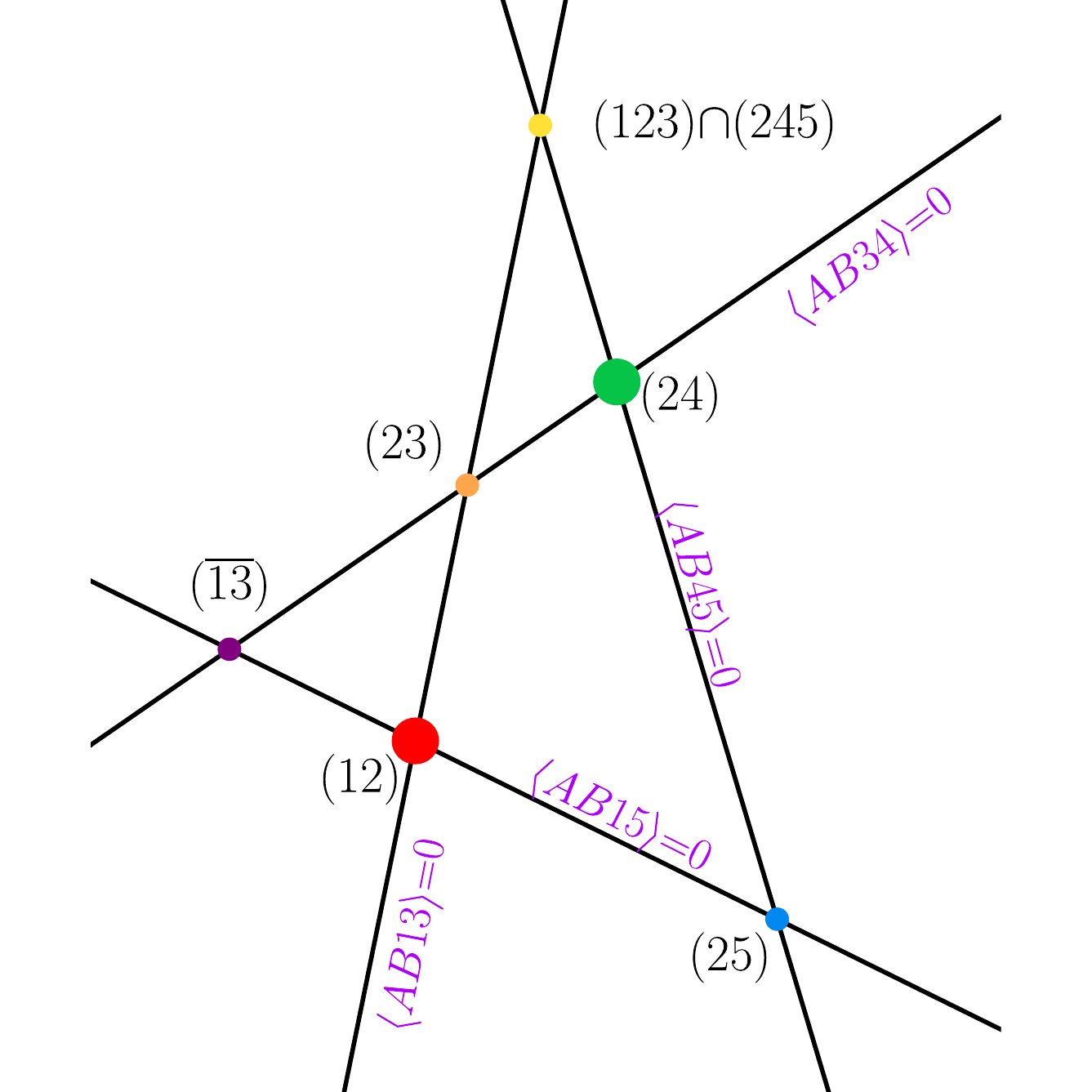}}
    \hspace{-.4cm}
    \underset{\text{dual to}}{\iff}
   \raisebox{-75pt}{\includegraphics[scale=0.45]{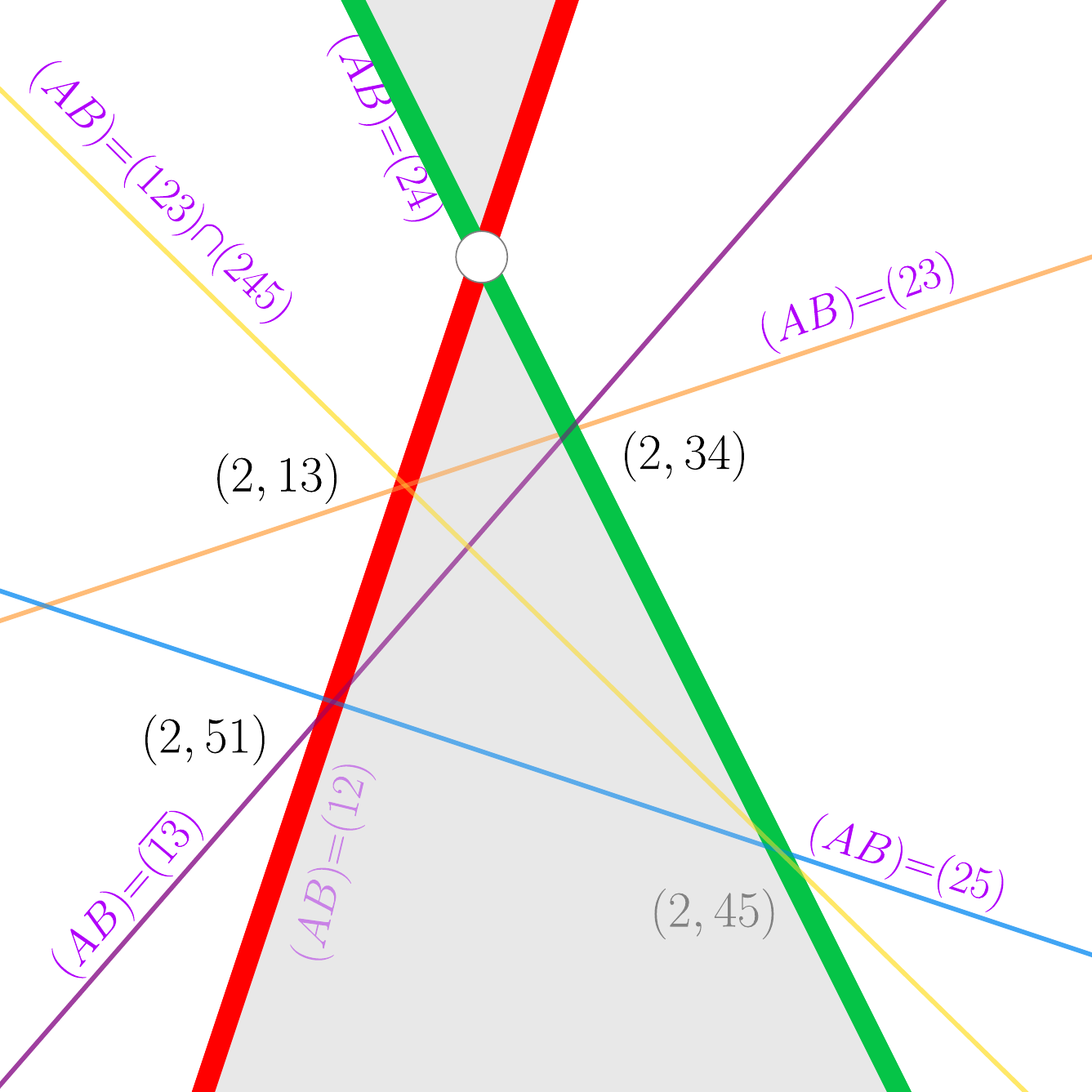}}
    \label{dual_of_line_segment_without_pt}
\end{equation}
The net effect of these subtleties on our na\"{i}ve picture eq.~(\ref{kermit_dual_2_projection}) is the addition of the infinite wedge eq.~(\ref{dual_of_line_segment_without_pt}), which gives exactly the dual Amplituhedron.

(An alternative resolution to this problem is to include the triangulation line with one of the Kermits but not the other. Upon removing the extra point $(14)$ in the dual space, we recover exactly the same (up to relabelling) external triangulation as in the motivational example of eq.~(\ref{toy_dualization}). This is sensible from a purely geometrical perspective. However, from the point of view of canonical forms it seems more natural to include the triangulation line \emph{with} its endpoints in both terms, as both forms do have nonzero residues on this boundary.)

Using the results of eq.~(\ref{dual_b45}) and eq.~(\ref{dual_p24}), we see that the dual of the Amplituhedron-Prime on this cut surface triangulates the dual of the Amplituhedron, \emph{except for the two vertices $(2,13)$ and $(2,45)$}, i.e.,
\begin{equation}
    \raisebox{-85pt}{\includegraphics[scale=0.45]{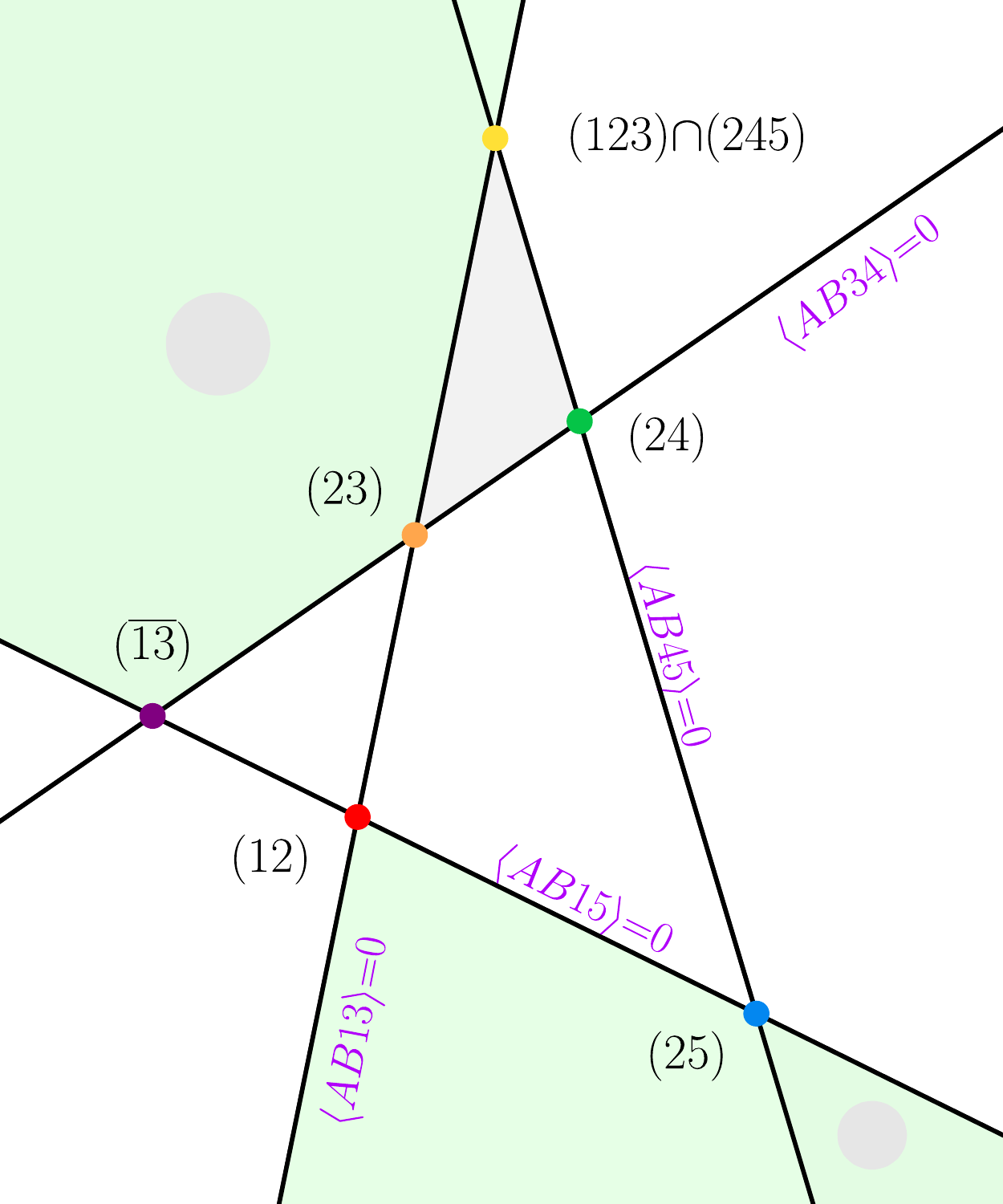}}
    \hspace{-.4cm}
    \underset{\text{dual to}}{\iff}
   \raisebox{-75pt}{\includegraphics[scale=0.45]{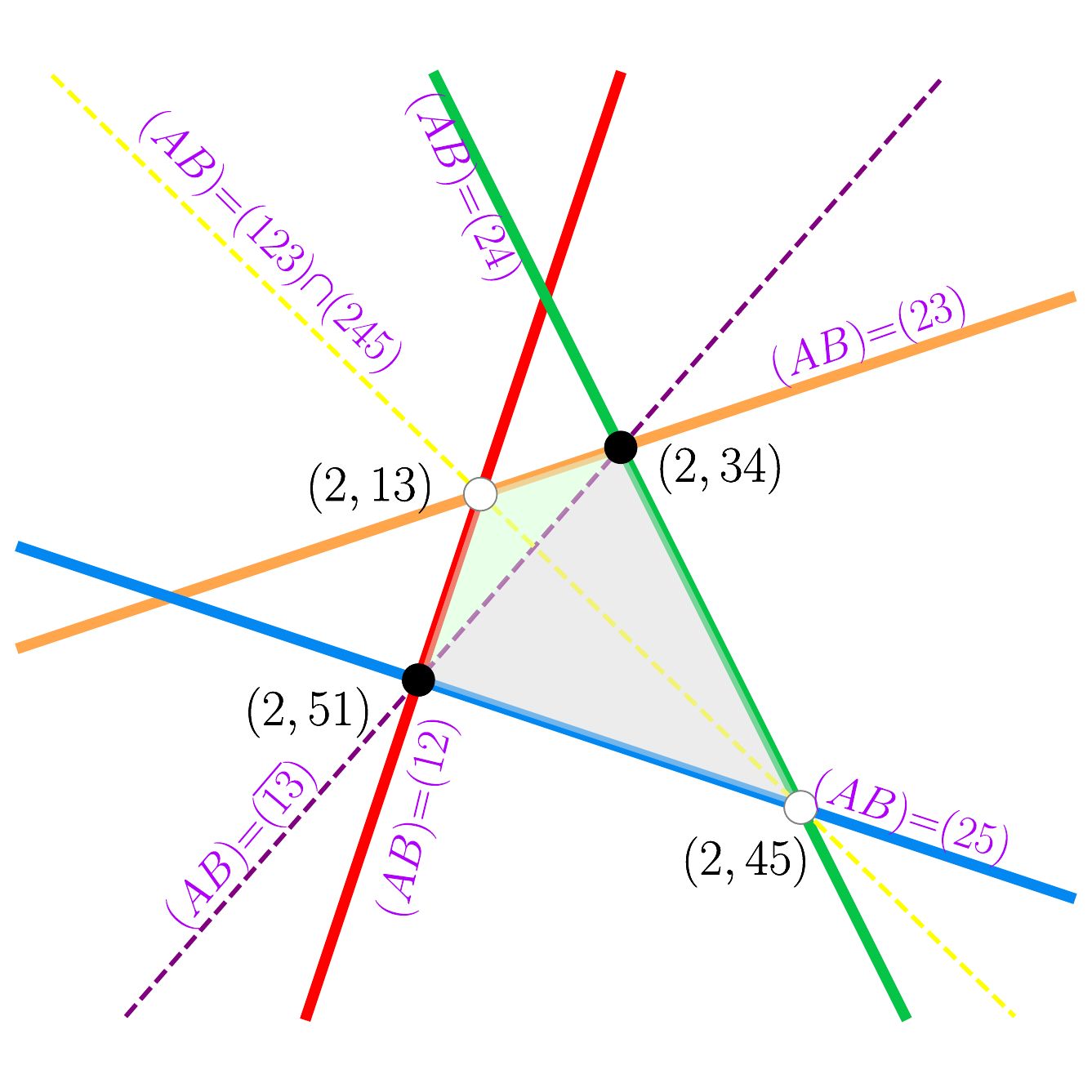}}
   \label{dual_A_prime_A2}
\end{equation}
Note that the original non-overlapping regions with fixed signs of $\ab{ABij}$ brackets are now overlapping in the dual space. Therefore, it is very non-trivial that the regions corresponding to chiral pentagons triangulate internally the dual Amplituhedron without any overlaps. 

The fact that the logarithmic forms for Amplituhedron and Amplituhedron-Prime are identical means that their (conjectured) dual spaces have the same volume and are identical up to spaces which have zero volume. This matches the result of eq.~(\ref{dual_A_prime_A2}), where we can see that $\mathcal{A}$ and $\mathcal{A}'$ differ by lower-dimensional boundaries. This line of reasoning is also suggestive of an ambiguity in the definition of the Amplituhedron-Prime. Namely, we are always free to add any spaces which have zero form (such as the infinite wedge of eq.~(\ref{eq:2_projection_zero_form_example_box45})) because in the dual space they correspond to zero-volume lines or points.

To provide additional evidence that the Amplituhedron-Prime dualizes to an internal triangulation of the (bulk) dual Amplituhedron, we can repeat the above exercise for the cut surface $(AB)\subset(234)$ which was analyzed in eq.~(\ref{eq:5pt_solns_2d_projection_plane_234}). For the Amplituhedron-Prime, all three local integrals contribute:
\begin{equation}
    \hspace{-.4cm}
    B^{(3)}_{45}\leftrightarrow
    \raisebox{-75pt}{\includegraphics[scale=0.45]{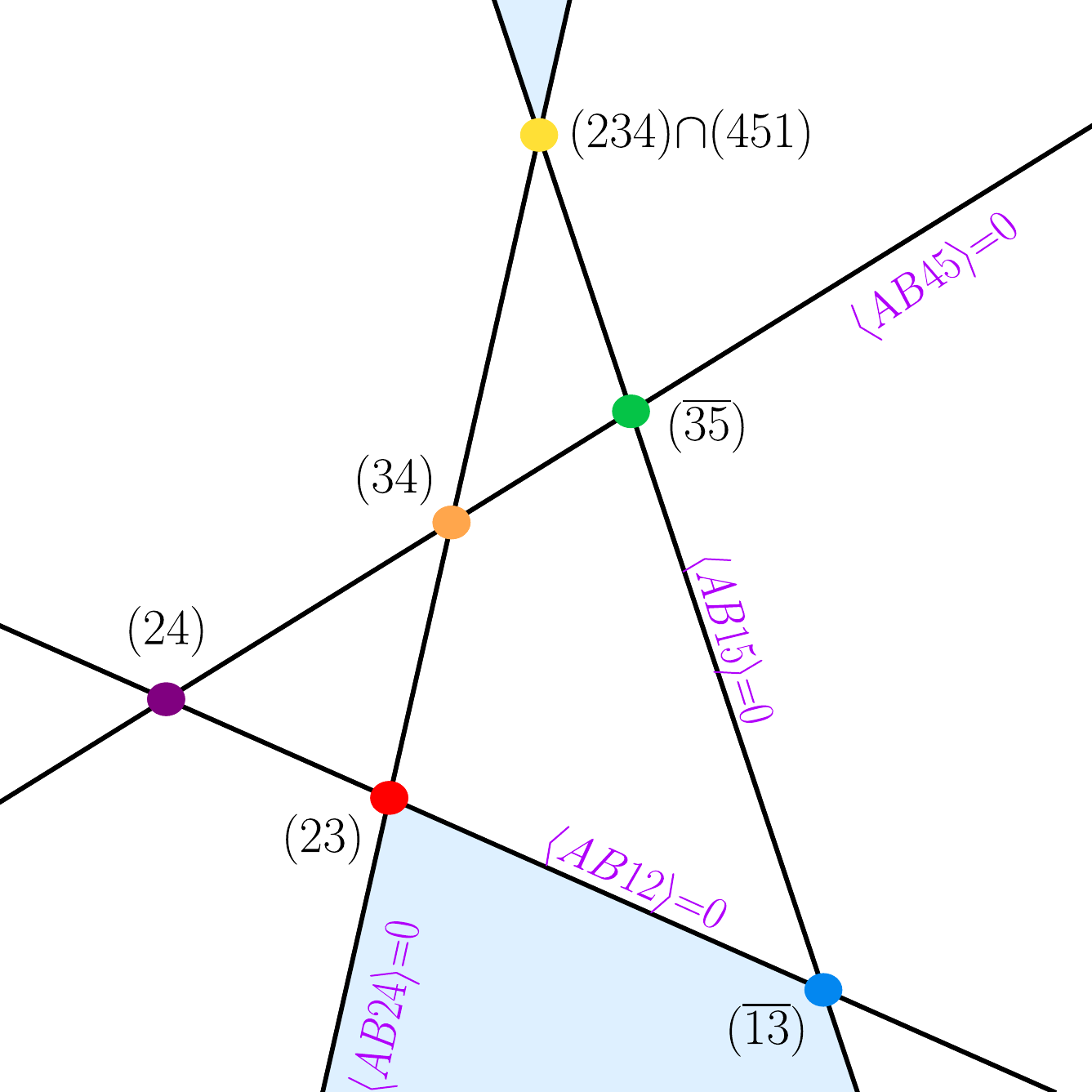}}
    \hspace{-.4cm}
    \underset{\text{dual to }}{\iff}
   \raisebox{-75pt}{\includegraphics[scale=0.45]{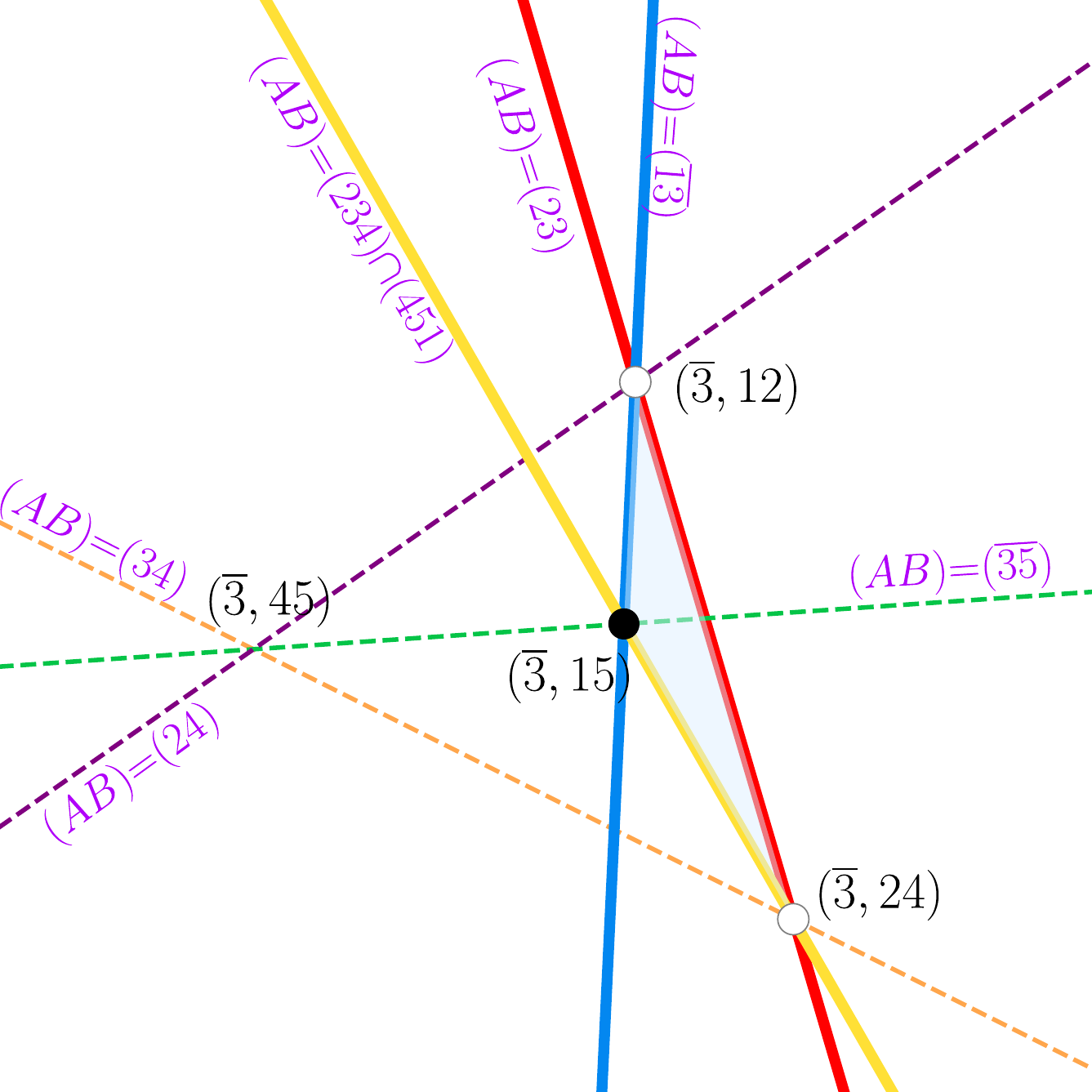}}
   \label{dual_b45_234}
   \hspace{-.2cm}
\end{equation}
\begin{equation}
    \hspace{-.4cm}
    B^{(3)}_{12}\leftrightarrow\raisebox{-75pt}{\includegraphics[scale=0.45]{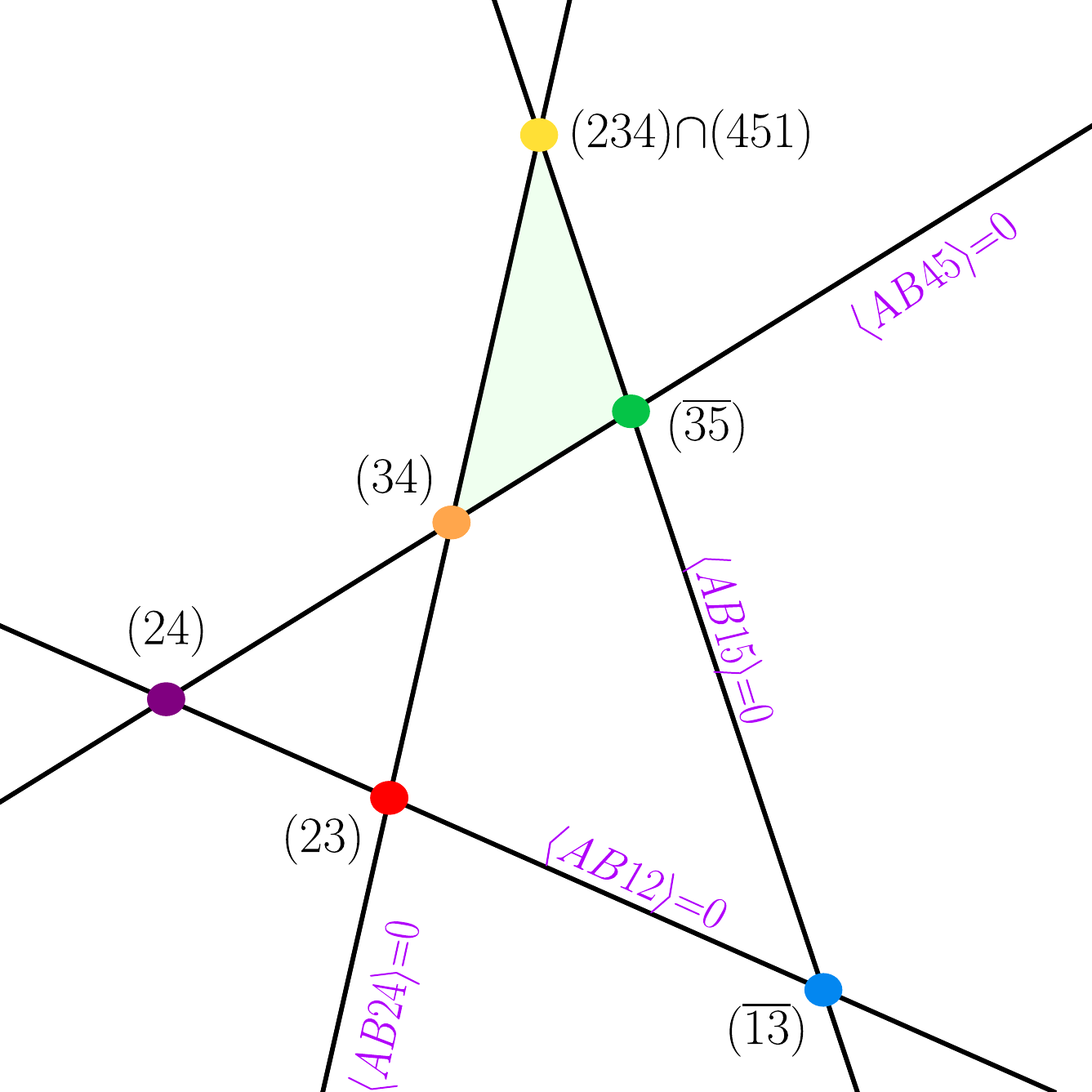}} 
    \hspace{-.4cm}
    \underset{\text{dual to }}{\iff}
   \raisebox{-75pt}{\includegraphics[scale=0.45]{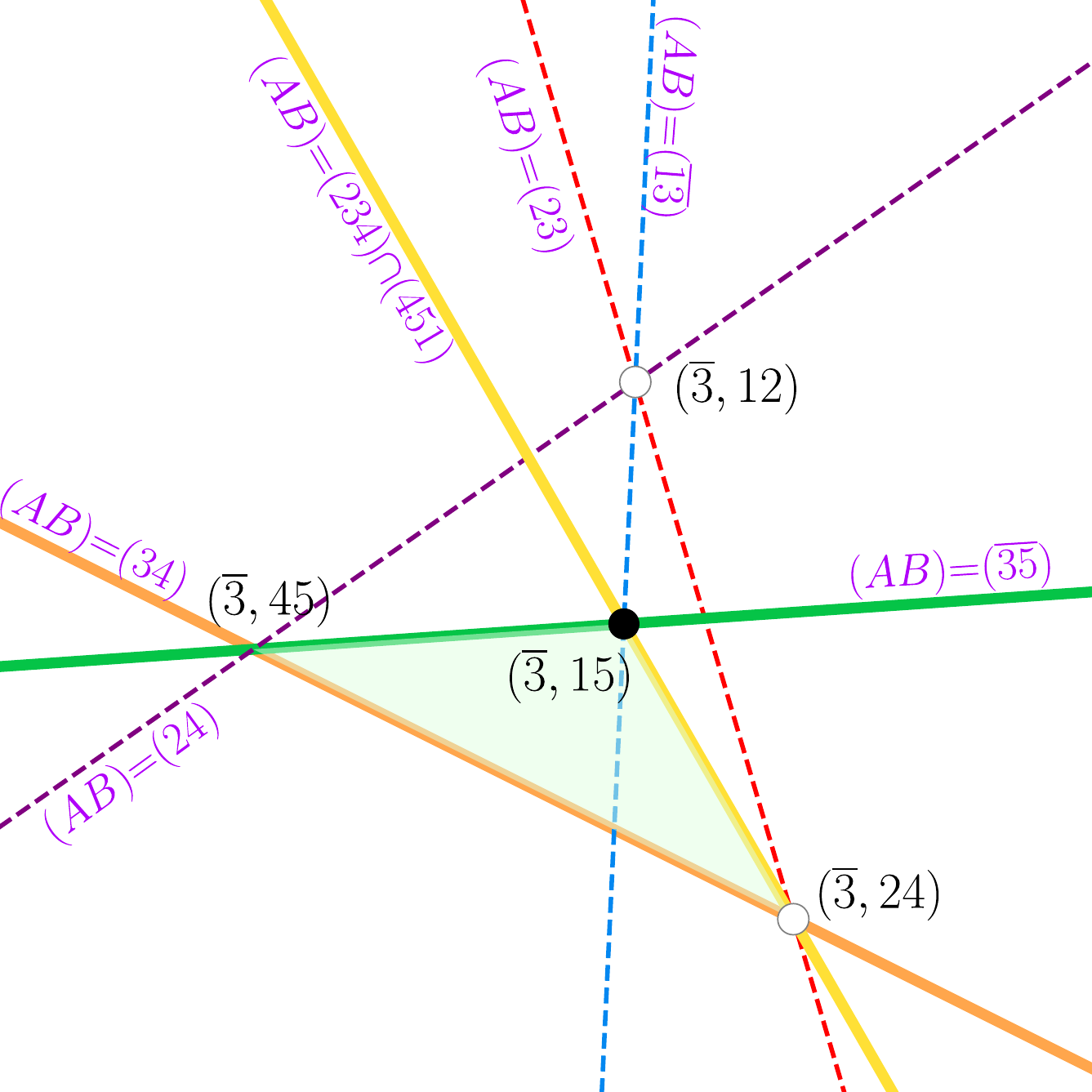}}
   \label{dual_b12_234}
   \hspace{-.2cm}
\end{equation}
\begin{equation}
    \hspace{-.4cm}
    P^{(1)}_{24}\leftrightarrow\raisebox{-75pt}{\includegraphics[scale=0.45]{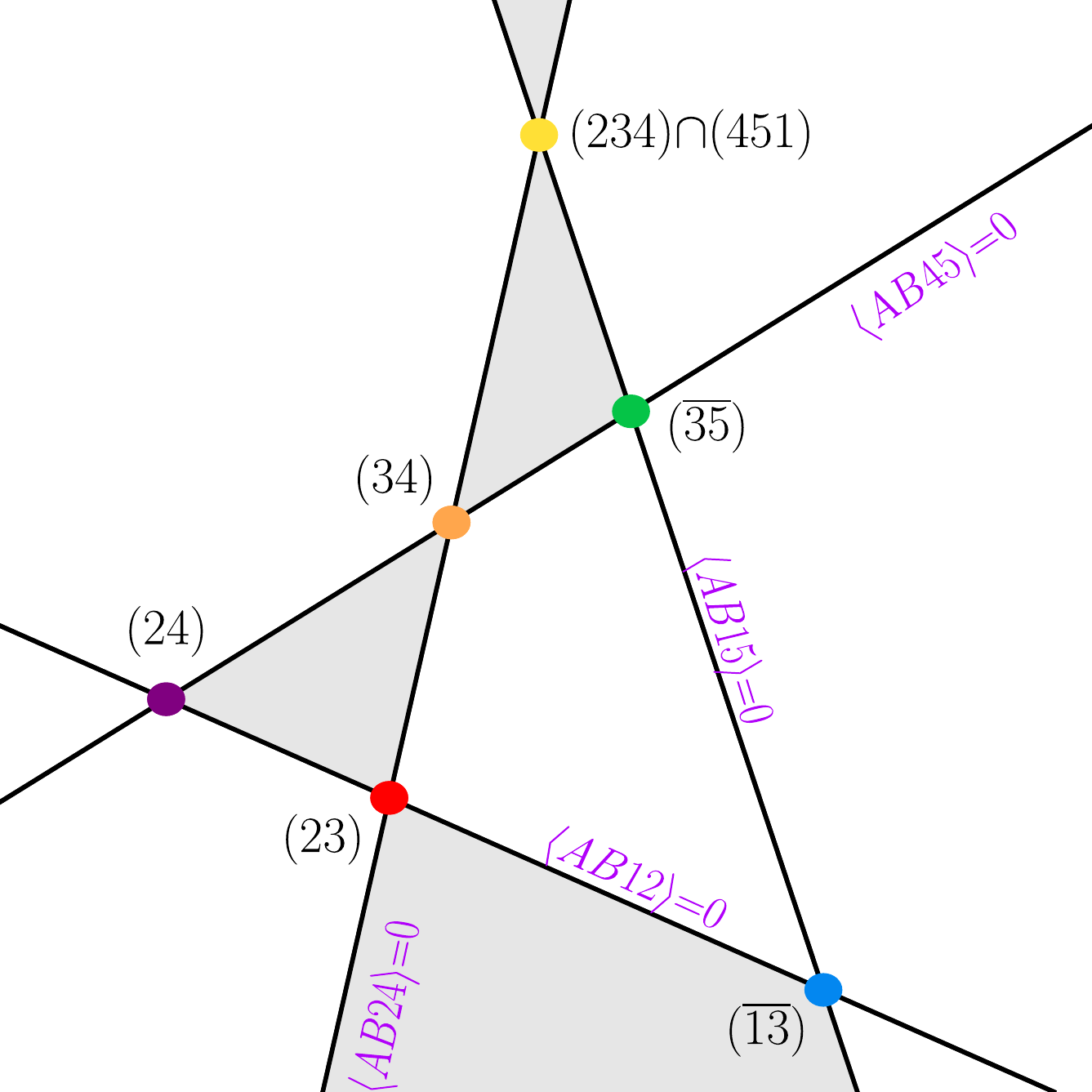}} 
    \hspace{-.4cm}
    \underset{\text{dual to }}{\iff}
   \raisebox{-75pt}{\includegraphics[scale=0.45]{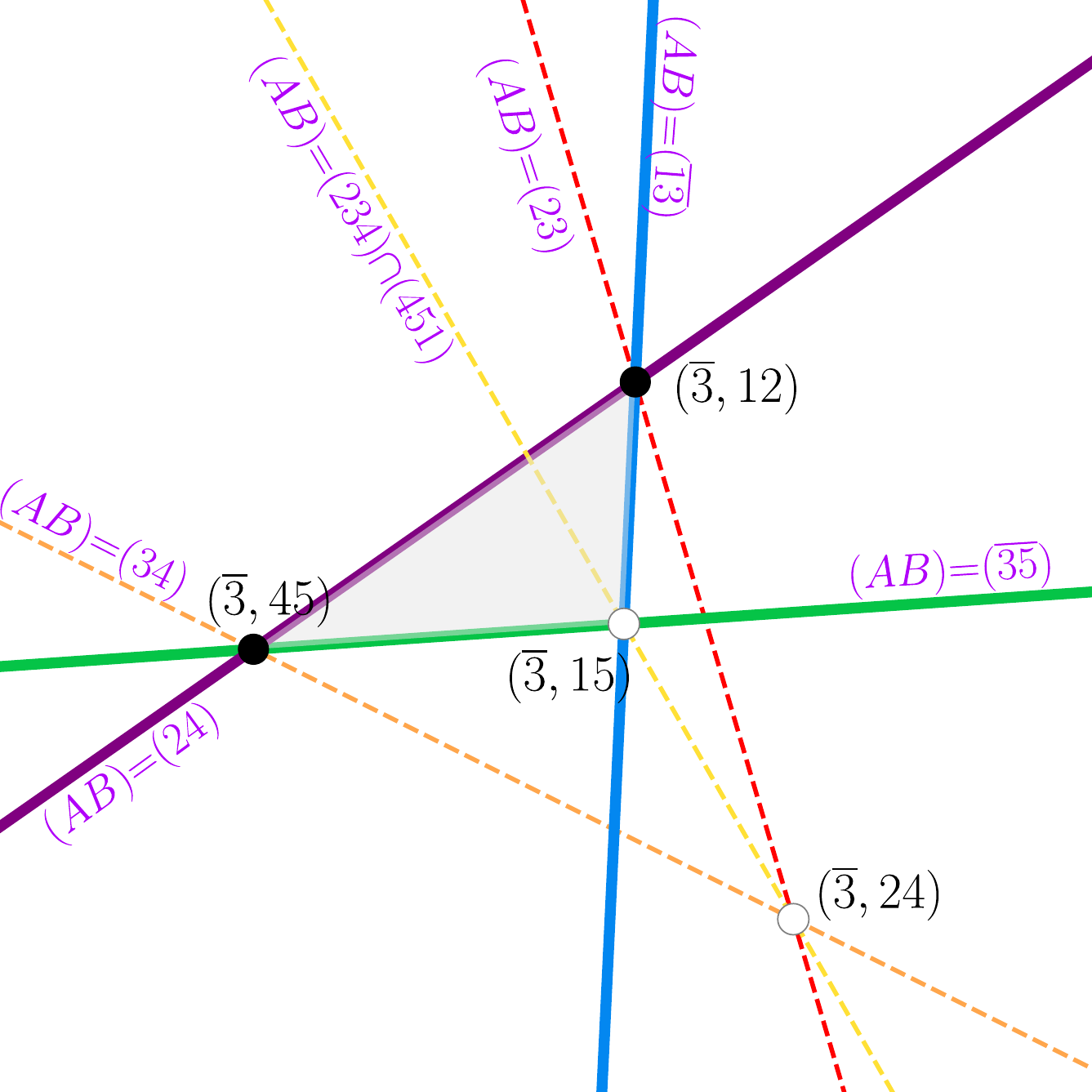}}
   \label{dual_p24_234}
   \hspace{-.2cm}
\end{equation}
Hence the dual of the Amplituhedron-Prime triangulates the dual Amplituhedron, up to a single vertex $(\overline{3},24)$ on the two-dimensional projection where $(AB)\subset(234)$,
\begin{equation}
   \raisebox{-65pt}{\includegraphics[scale=0.45]{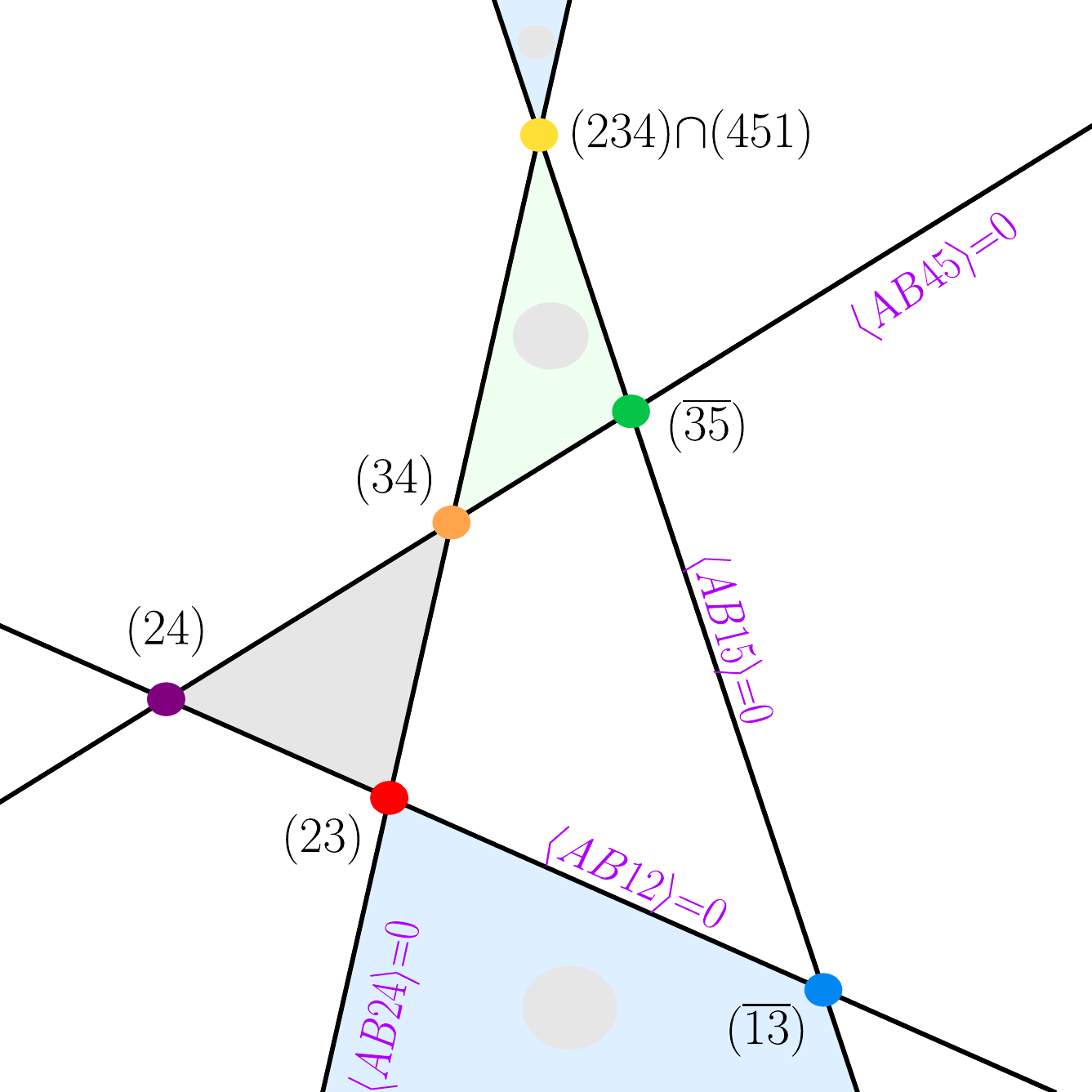}}    \underset{\text{dual to }}{\iff}
   \raisebox{-65pt}{\includegraphics[scale=0.45]{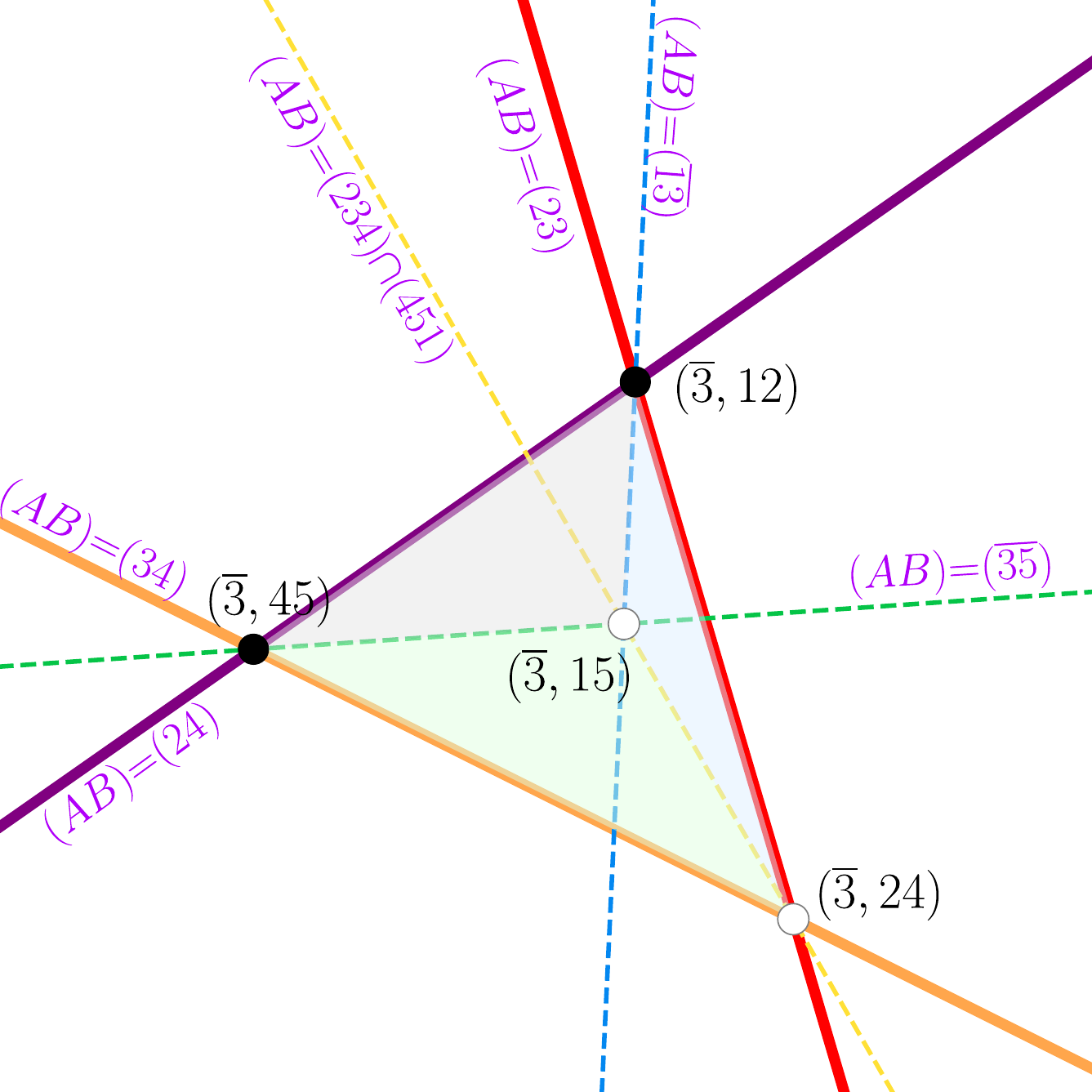}}
   \label{dual_A-prime_234}
\end{equation}
We have exhaustively verified for all remaining five and six point two-dimensional projections where $(AB)=(Ai)$ or $(AB)\subset(i{-}1ii{+}1)$ that the Amplituhedron-Prime internally triangulates the dual Amplituhedron up to contributions with zero-volume. In fact, in retrospect this conclusion would follow immediately from the existence of a spurious-boundary-free, zero-form space $\mathcal{B}$ which connects $\mathcal{A}$ and $\mathcal{A}'$ through $\mathcal{A}=\mathcal{A}'+\mathcal{B}$. We wrote down an explicit expression for $\mathcal{B}$ at four points in eq.~(\ref{eq:zero_form_space_4pt}) and, although we do not have an explicit formula for this space at $n$-points, we see no \emph{conceptual} obstruction which would preclude its existence. We leave an explicit construction of the zero-form space connecting the Amplituhedron and Amplituhedron-Prime to future work.

\subsection*{Rigidity of the dual space}
%

We have seen in the previous discussion that a single $d\log$ form gives rise to various positive geometries. We emphasized the importance of \emph{faithful geometries} in section \ref{subsec:dlog_form_to_geometry}, where all boundaries of the geometric space appear as poles in the $d\log$ form. Furthermore, we used these geometries to interpret the chiral pentagon expansion as external triangulation of Amplituhedron-Prime in eq.~(\ref{amp_prime_npt}). From our discussion it is clear, that all these spaces have different geometries, but one can naturally ask about the duals of the positive spaces which originate from the same $d\log$ form.

Here, we explicitly discuss different box spaces for the one-mass box $B_{45}$ summarized in eq.~(\ref{eq:b45_spaces}). We can dualize these spaces on the two-dimensional boundary where the line $(AB)$ passes through $Z_2$, c.f.~(\ref{2_proj_with_labels}) and (\ref{2_projection_5pt_dualization}). We repeat the same exercise from section \ref{subsec:dualizing_polygons} for the three alternative box spaces appearing in eq.~(\ref{2_proj_box45}) even though they are irrelevant for the Amplituhedron-Prime. Because these additional box spaces are all equivalent to eq.~(\ref{eq:dual_2_b45_ideal}) up to a zero-form region in the original two-dimensional projection, they all map to the same internal piece of the dual Amplituhedron \emph{up to lower-dimensional boundaries}. Indeed, the results of dualization for the alternative box spaces are, using the coloring convention of eq.~(\ref{2_proj_box45}):
 \begin{subequations}
 \begin{equation}
    \hspace{-.4cm}
    B_{45}^{(1)}\leftrightarrow
    \hspace{-.4cm}
     \raisebox{-75pt}{\includegraphics[scale=0.45]{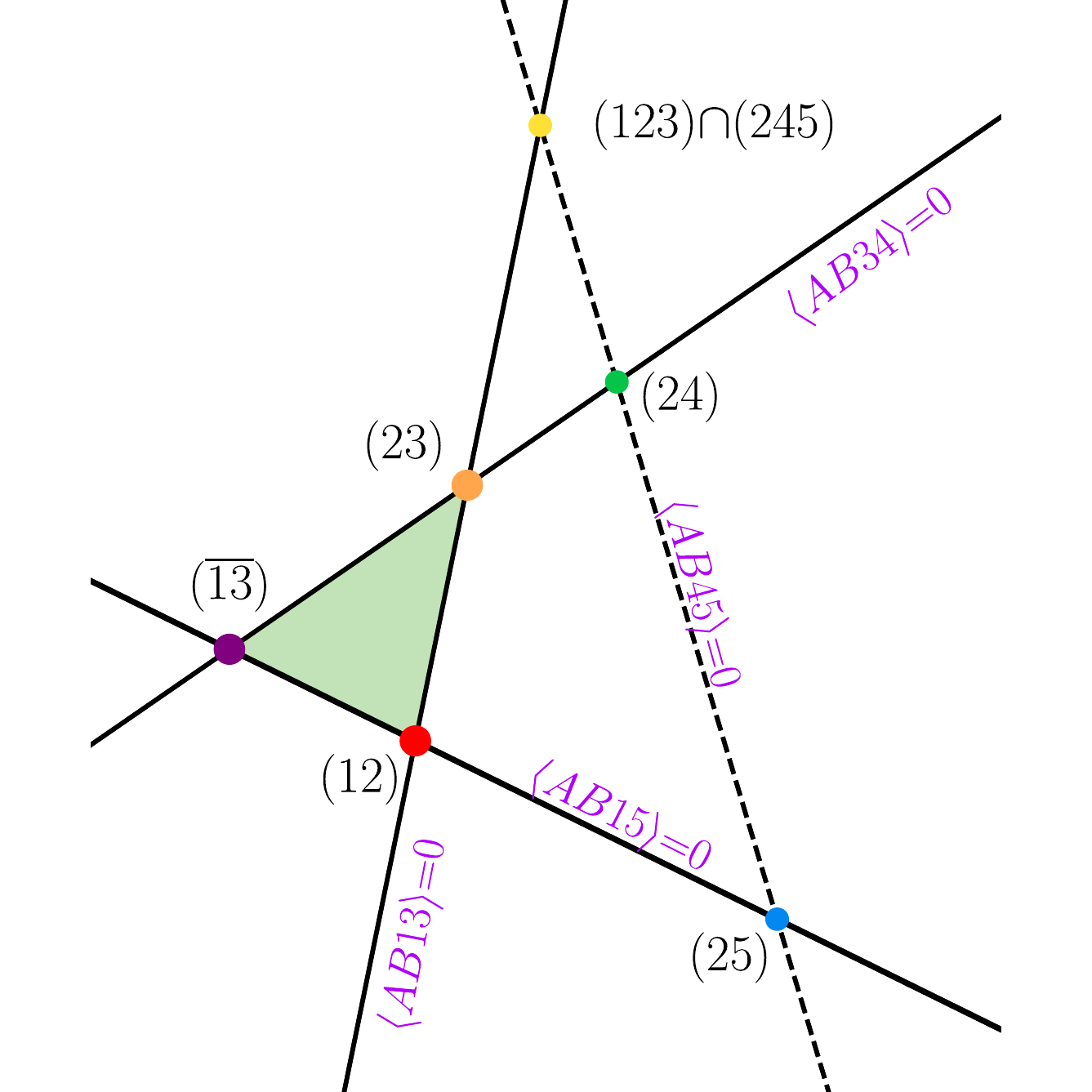}} 
     \hspace{-.4cm}
     \underset{\text{dual to}}{\iff}
   \raisebox{-75pt}{\includegraphics[scale=0.45]{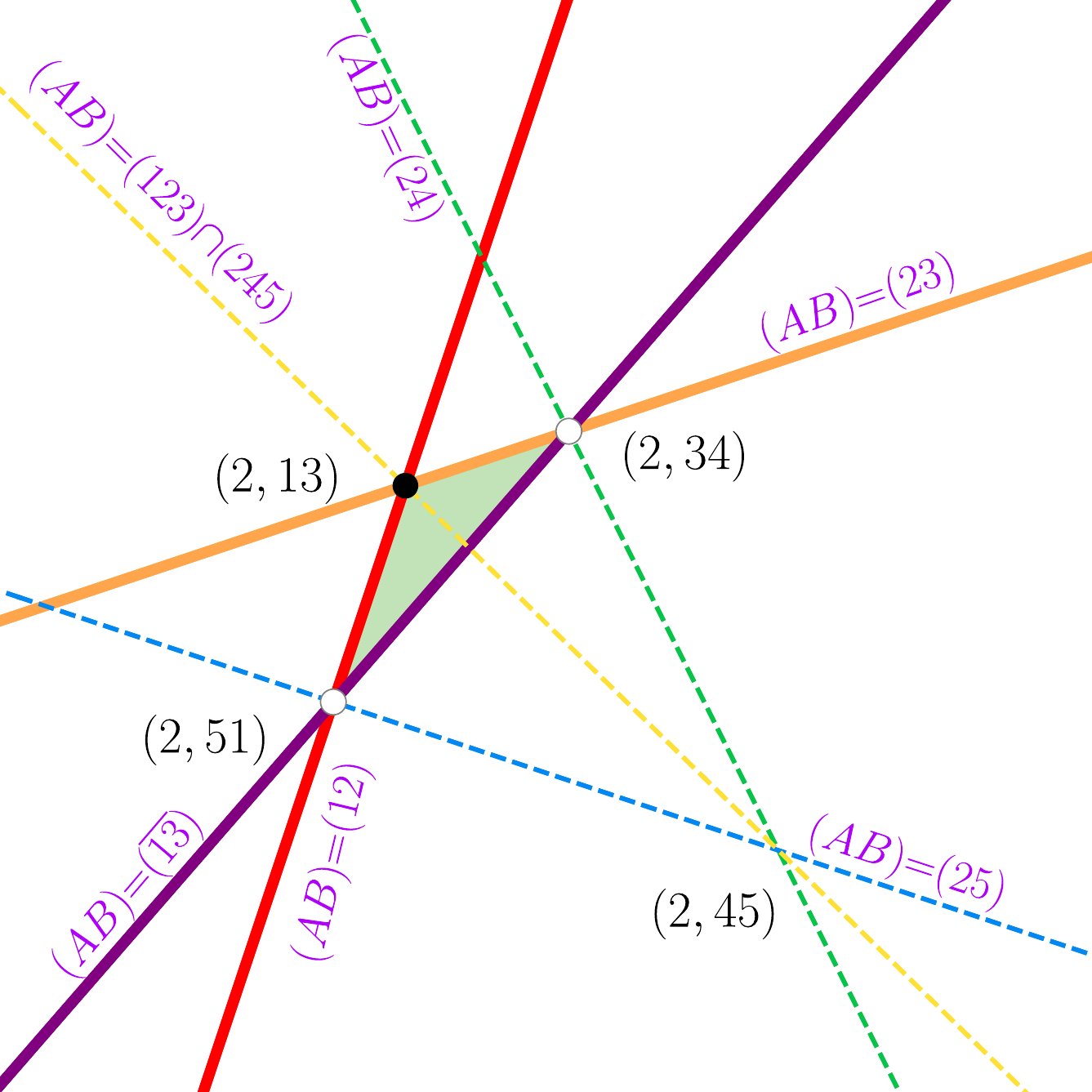}}
   \label{eq:box45_projection_A2_variant1}
\end{equation}
\begin{equation}
    B_{45}^{(2)}\leftrightarrow
    \hspace{-.4cm}
     \raisebox{-75pt}{\includegraphics[scale=0.45]{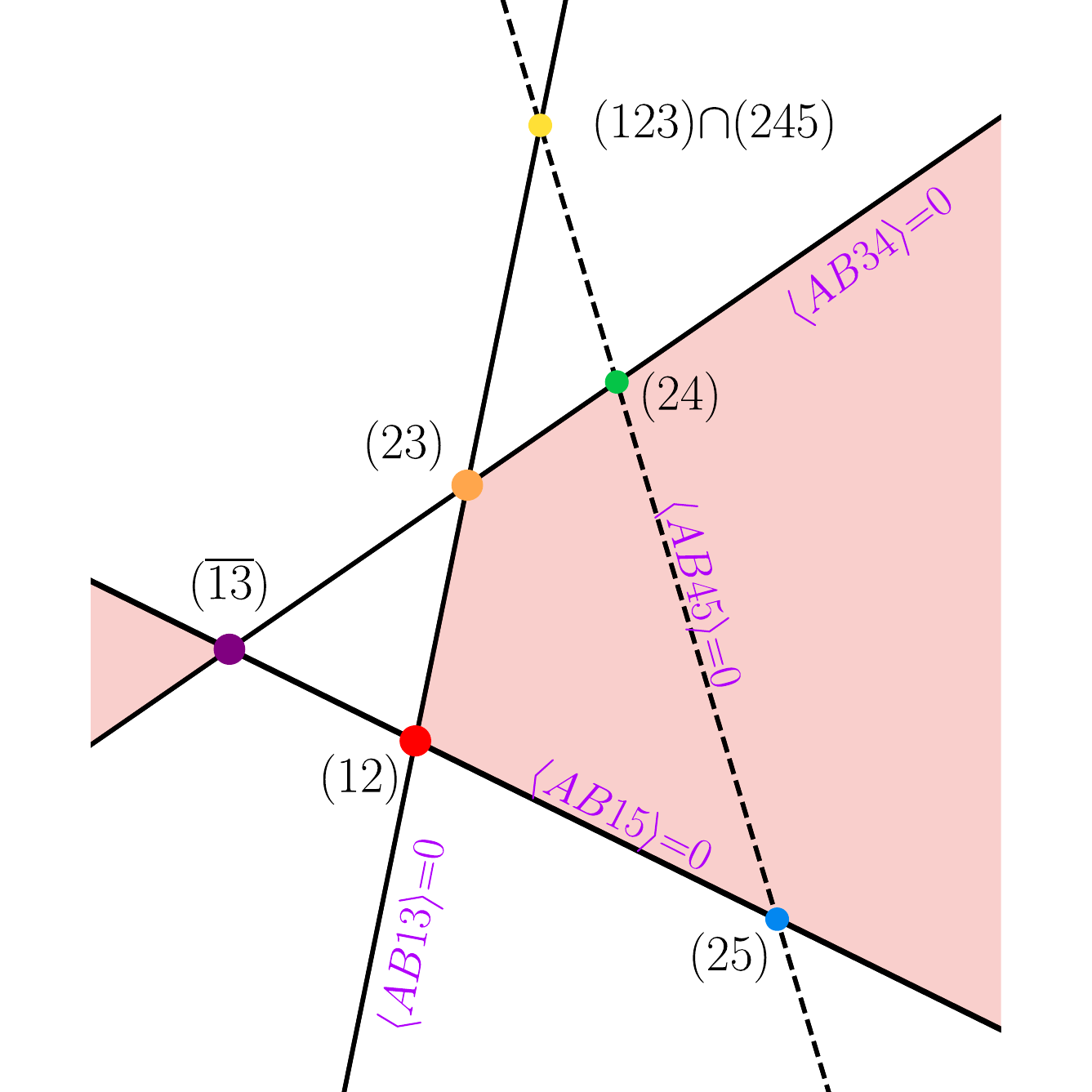}} 
     \hspace{-.4cm}
     \underset{\text{dual to}}{\iff}
   \raisebox{-75pt}{\includegraphics[scale=0.45]{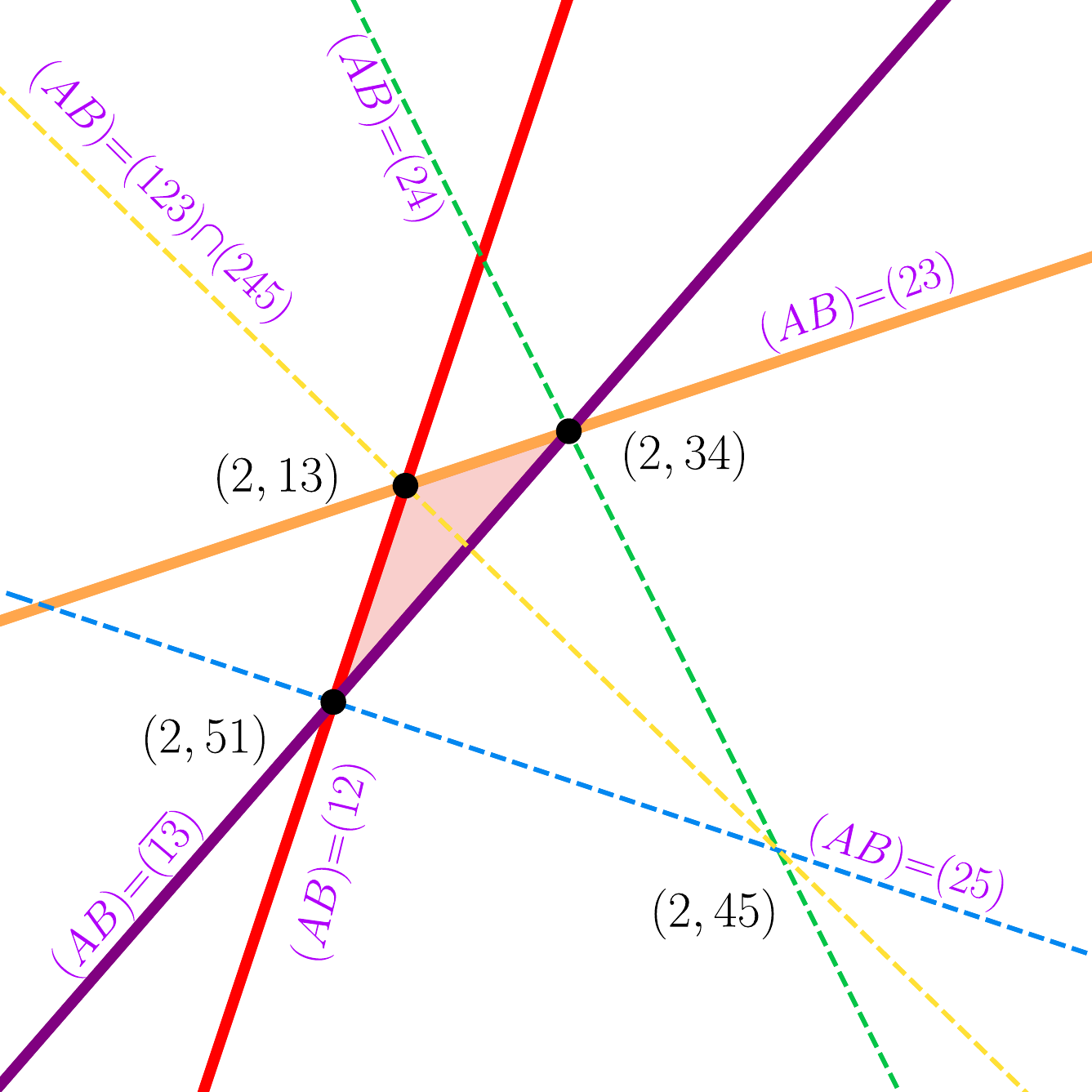}}
   \hspace{-.3cm}
   \label{eq:box45_projection_A2_variant3}
\end{equation}
\begin{equation}
\hspace{-.4cm}
    B_{45}^{(4)}\leftrightarrow
    \hspace{-.4cm}
     \raisebox{-75pt}{\includegraphics[scale=0.45]{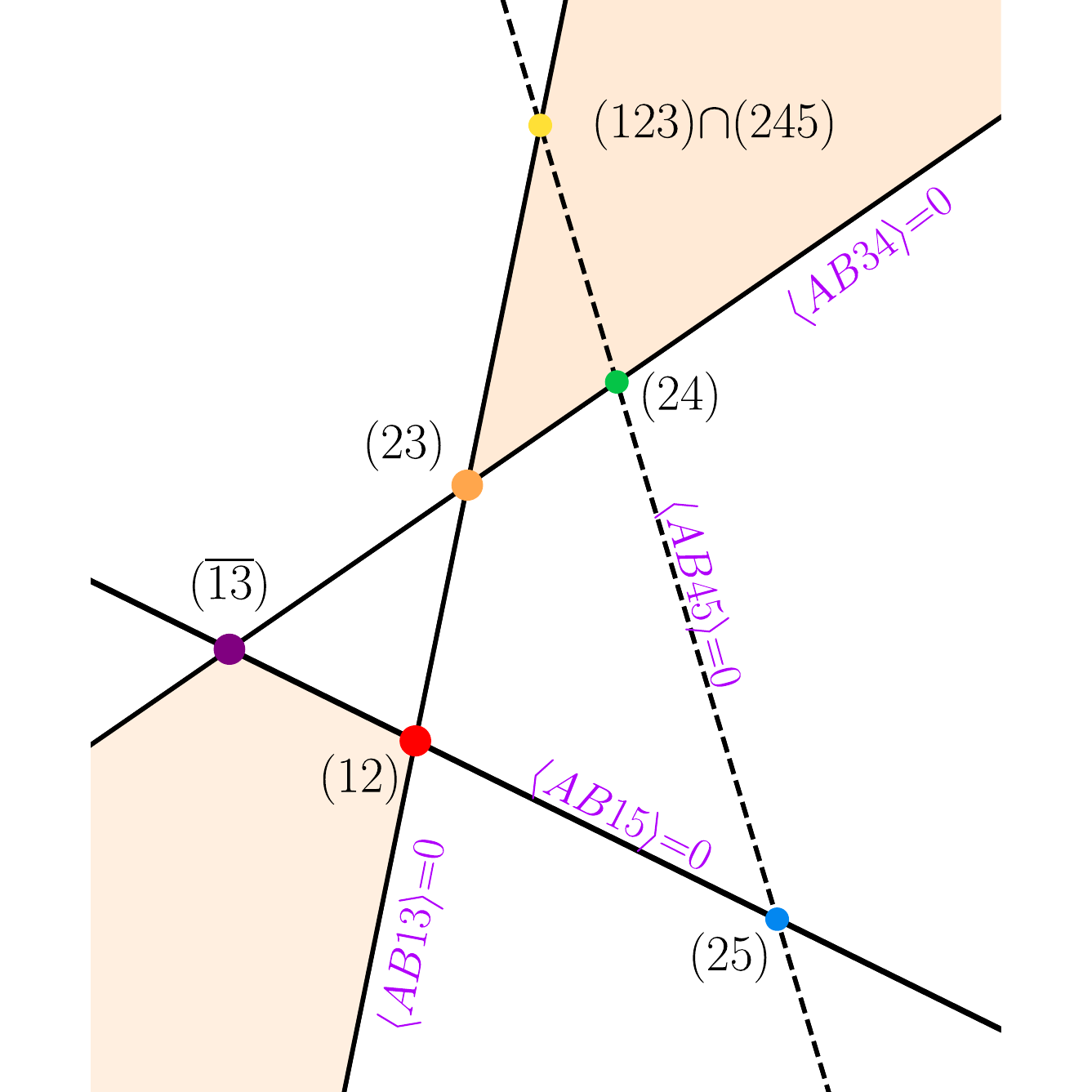}} 
     \hspace{-.4cm}
     \underset{\text{dual to}}{\iff}
   \raisebox{-75pt}{\includegraphics[scale=0.45]{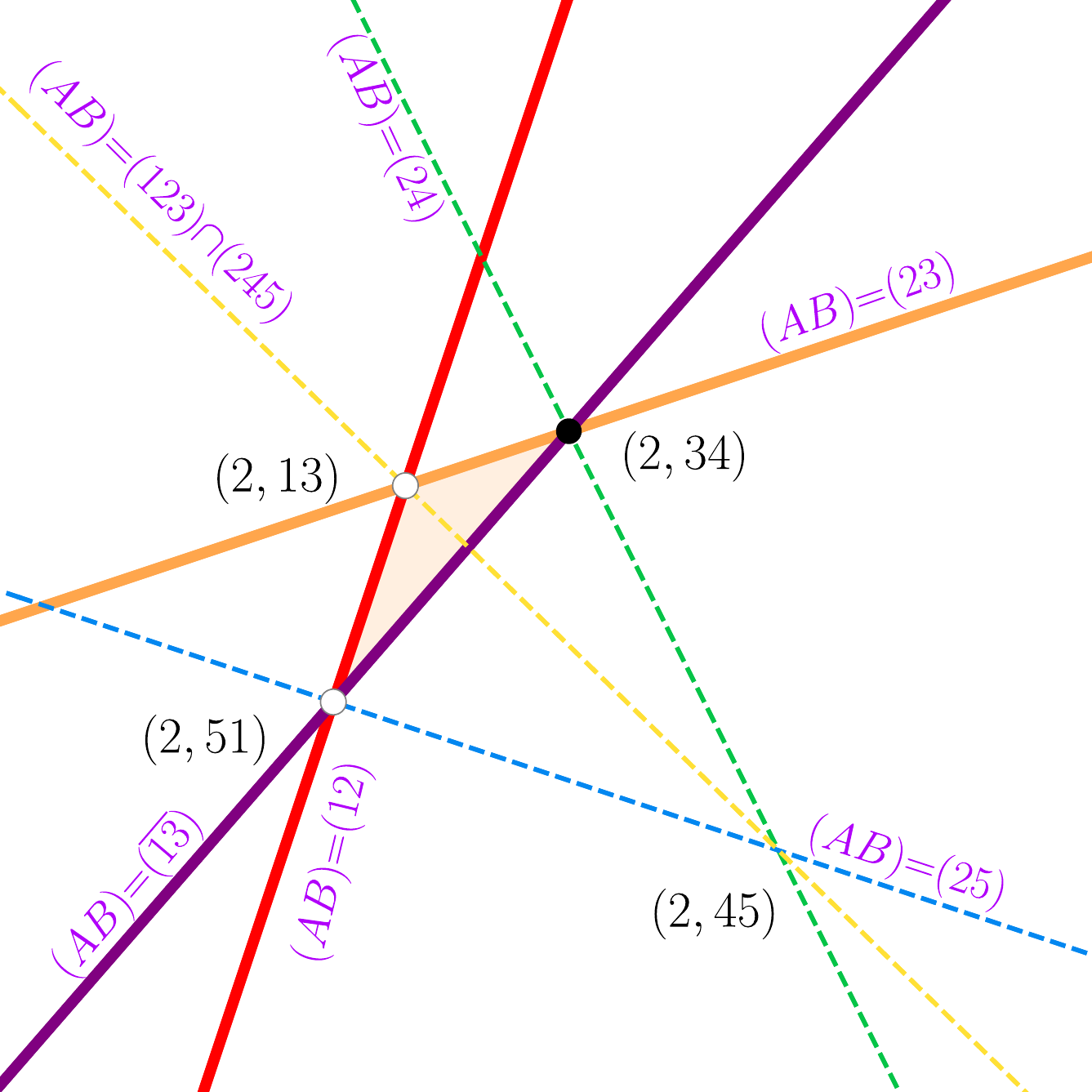}}
   \label{eq:box45_projection_A2_variant2}
\end{equation}
 \end{subequations}

This makes perfect sense because the positive geometries differ by spaces with vanishing form (wedges on the two-dimensional boundaries) which get mapped to points and lines in the dual two-dimensional geometry. Therefore, from the point of view of the dual Amplituhedron, it does not matter which positive geometry we use for a given $d\log$ form, it always represents the same dual geometry. This simply follows from the fact that the regions with vanishing $d\log$ form map to lower-dimensional objects in the dual which have zero volume.

Therefore, it is natural to expect that while the positive geometries ${\cal A}$ and ${\cal A}'$ are different, the putative dual Amplituhedron is unique, and the chiral pentagons triangulate it internally. We gave some evidence for this claim in this section.

\section{Conclusion}
\label{sec:conclusion}

In this paper we discussed various positive geometries in the context of the one-loop Amplituhedron and its variants. We have shown that for external data satisfying the MHV positivity conditions (\ref{eq:mhv_ext_positivity}), there are a number of further interesting positive geometries besides the original MHV and $\MHVbar$ Amplituhedra. In section \ref{sec:sign_flip_regions}, we classified all these spaces using topological sign-flip properties which are reminiscent of, but distinct from, the sign-flip definition of the Amplituhedron \cite{Arkani-Hamed:2017vfh}. Furthermore, we showed that these positive spaces can be used to give a geometric interpretation of the chiral pentagon expansion of the one-loop MHV amplitude of eq.~(\ref{pent}). In particular, the chiral pentagons externally triangulate a new Amplituhedron-Prime space (\ref{amp_prime_npt}) which is a non-overlapping twin of the original Amplituhedron with only physical boundaries and the same logarithmic form. Finally, in section \ref{sec:triangulation_dual_Amplituhedron}, we made more precise the statement that the chiral pentagon expansion can also be interpreted as the internal triangulation of the yet-to-be found dual Amplituhedron. We were able to demonstrate the internal triangulation of the dual on certain two-dimensional boundaries of the full space where the geometry reduces to that of polygons for which a dualization prescription exists.

Our work opens the door to various new research directions. The first question is how the stories about local geometries, as well as the study and classification of sign-flip spaces, extend to higher loops and to higher N$^k$MHV degree, where the positive external data will play a more important r\^{o}le. We already briefly touched upon this issue in \cite{Herrmann:2020oud} and found that sign-flip-six and higher spaces are allowed beyond MHV kinematics. Another interesting angle is to use this framework to generate IR-finite integrands at two-loops and beyond. The logarithmic forms for sign-flip-four spaces were given by chiral octagons, which form an IR-friendly dual conformal basis \cite{ArkaniHamed:2010gh} of integrands at one-loop---and separates naturally IR-divergent, IR-finite and parity-odd integrands. It would be very interesting to replicate this at two-loops and find a basis of IR-finite dual conformal invariant integrands. These integrands would be perfect targets for modern integration methods \cite{Bourjaily:2019jrk} and could play an important r\^{o}le in the expansion of IR-finite quantities, such as remainder and ratio functions. 

Finally, the main open direction is the exploration of the dual Amplituhedron. The link between chiral pentagons and the internal triangulations of this hypothetical space provides further evidence for its presumptive existence. However, we were able to establish this connection only on certain two-dimensional boundaries where the dual Amplituhedron geometry reduced to projective polygons. Exploring other two-dimensional boundaries of non-polygonal form, as well as three-dimensional boundaries and the r\^{o}le of internal triangulations, should bring us to the ultimate goal of the discovery of the dual Amplituhedron space.

\section*{Acknowledgements}

We thank Nima Arkani-Hamed, Akshay Yelleshpur Srikant, and Ryota Kojima for stimulating discussions. This research is supported in part by U.S. Department of Energy grant DE-SC0009999 and by the funds of University of California. E.H. is supported by the U.S. Department of Energy under contract DE-AC02-76SF00515. The research of C.L. is supported in part by an ERC Starting Grant (No. 757978) and a grant from the Villum Fonden (No. 15369).

\appendix

\newpage
\section{Configuration of lines in momentum twistor space}
\label{app:line_configs}

There is an intimate relation between configurations of (loop) lines in momentum twistor space and certain restricted kinematic configurations of loop momenta on unitarity cuts of loop integrands or local integrals. At one loop, we can depict the off-shell configuration of lines in twistor space corresponding to a generic loop integrand (either of the amplitude or of an integral) by a set of lines corresponding to external dual momenta, together with a line $(AB)$ in a generic configuration (parameterized via eq.~(\ref{eq:loop_line_par_gen})),
\begin{align}
    \raisebox{-52pt}{\includegraphics[scale=.5]{./figures/line_config_gen.pdf}}
    \qquad
    \leftrightarrow
    \qquad
    \raisebox{-40pt}{\includegraphics[scale=.5]{./figures/cut_config_gen.pdf}}
\end{align}
In this setup, the loop-line $(AB)$ does not intersect any of the lines associated to external kinematic points. In the next step, one could go to codimension-one configurations by imposing one condition, e.g.~$\ab{ABii{+}1} = 0$, so that the lines $(AB)$ and $(ii{+}1)$ intersect. 
\begin{align}
    \raisebox{-52pt}{\includegraphics[scale=.5]{./figures/line_config_codim_1.pdf}}
    \qquad
    \leftrightarrow
    \qquad
    \raisebox{-50pt}{\includegraphics[scale=.5]{./figures/cut_config_codim_1.pdf}}
\end{align}
At the level of cuts, this corresponds to setting a single propagator $\ab{ABii{+}1} = 0$ to zero. This codimension-one configuration for the line $(AB)$ can be parameterized by three degrees of freedom. The intersection implies that one of the defining points of the $(AB)$-loop lies on the line $(i i{+1})$. Taking into account the projectivity of the $Z$'s, one possible particular parametrization is 
\begin{align}
    Z_A = Z_i + \alpha_1 Z_{i+1}\,, \quad
    Z_B = Z_j + \alpha_2 Z_{k} + \alpha_3 Z_l\,.
\end{align}
In a second step, one can impose an additional constraint to end up on a codimension-two configuration of the line $(AB)$. Depending on the condition one imposes, there are three situations to consider
\begin{align}
\label{eq:codim_2_line_configs}
   &\raisebox{-60pt}{\includegraphics[scale=.5]{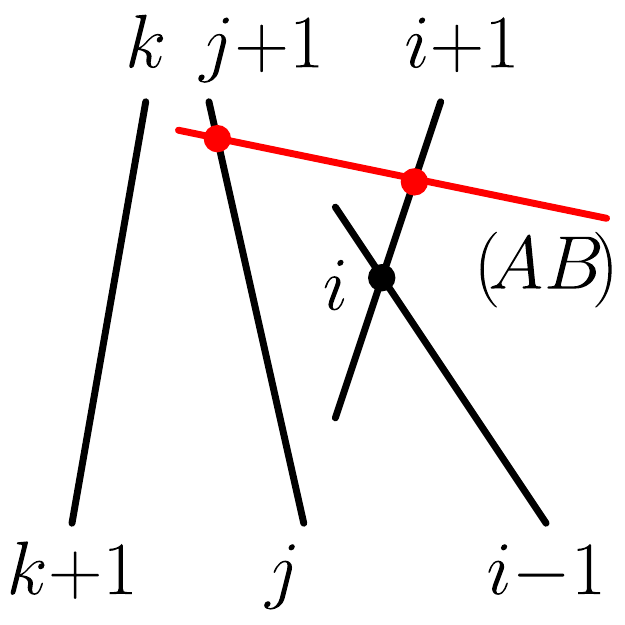}}
   && \raisebox{-60pt}{\includegraphics[scale=.5]{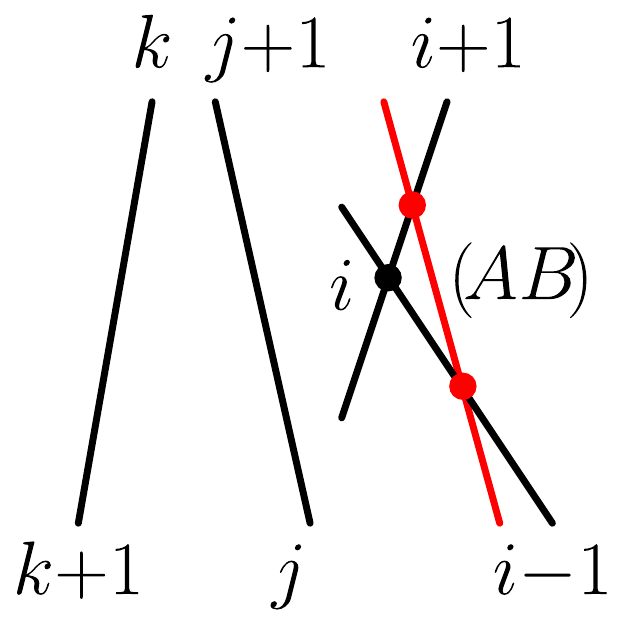}}
   &&& \raisebox{-60pt}{\includegraphics[scale=.5]{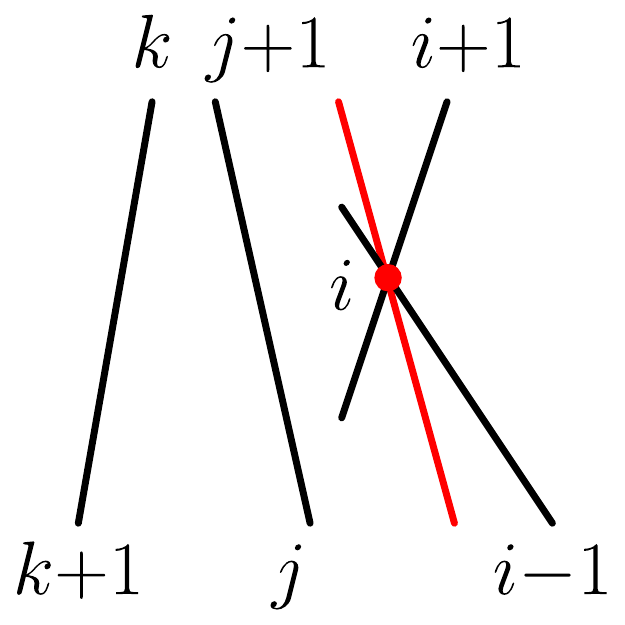}} 
    \\
    &\raisebox{-50pt}{\includegraphics[scale=.5]{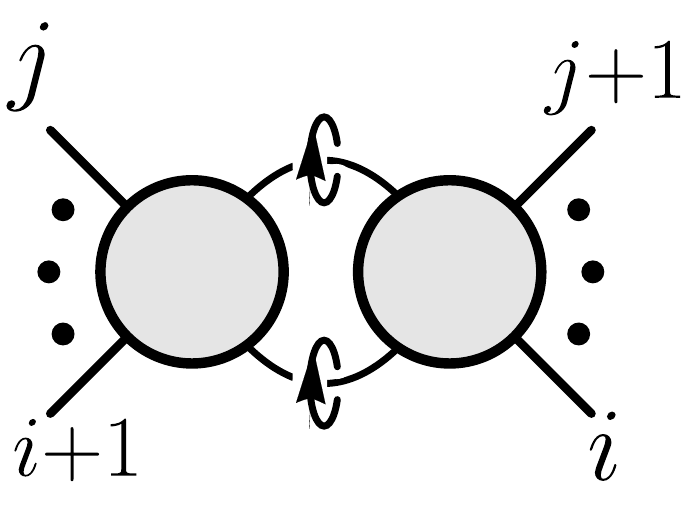}}
    &&\raisebox{-50pt}{\includegraphics[scale=.5]{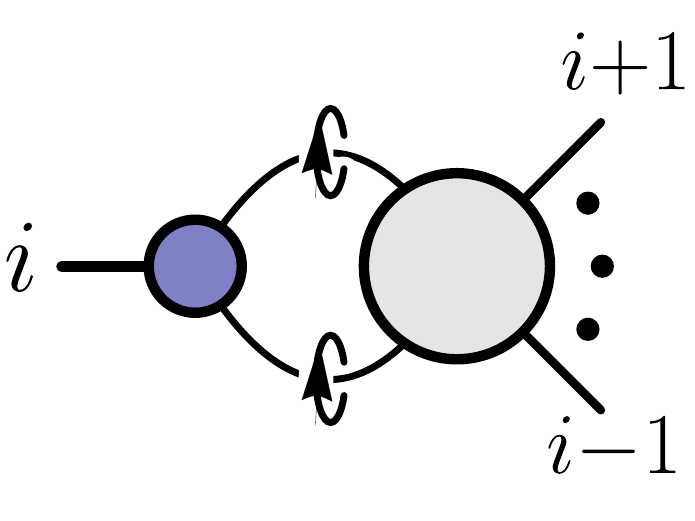}}
    &&&\raisebox{-50pt}{\includegraphics[scale=.5]{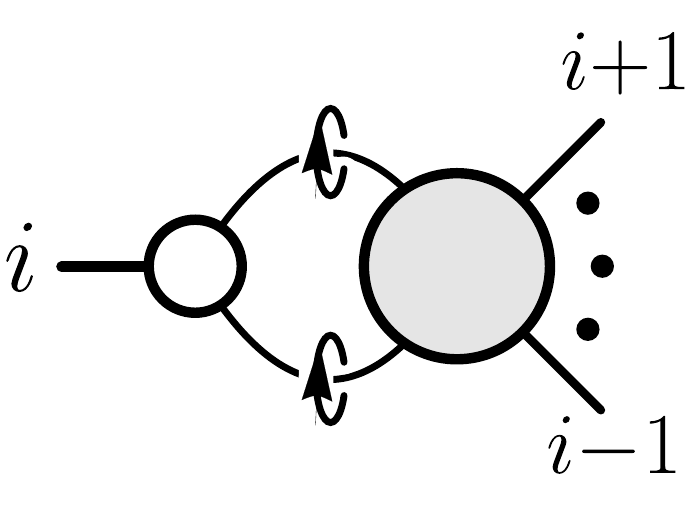}}
\end{align}
which have explicit two-dimensional parametrizations of the solution for the line $(AB)$ given by,
\begin{align}
\label{eq:codim_2_line_param}
\begin{split}
    &Z^{(1)}_A = Z_i + \gamma_1 Z_{i+1}\,, \quad  
    Z^{(2)}_A = Z_i + \gamma_1 Z_{i+1}\,,  \quad
    Z^{(3)}_A = Z_i \,, 
    \\
    &Z^{(1)}_B = Z_j + \gamma_2 Z_{j+1}\,, \quad  
    Z^{(2)}_B = Z_i + \gamma_2 Z_{i-1}\,, \quad
    Z^{(3)}_B = Z_j + \gamma_1 Z_{k} + \gamma_2 Z_{l} \,.
\end{split}
\end{align}
We can continue by imposing yet further constraints. At codimension-three, for the first time we encounter the situation associated with a \emph{composite} residue where we localize the loop-line into a collinear configuration, depicted in the second figure below.
\begin{align}
\label{eq:codim_3_line_configs}
\hspace{-1cm}
 &\raisebox{-55pt}{\includegraphics[scale=.5]{./figures/line_config_codim_3_gen.pdf}}
 &&\hspace{.4cm}\raisebox{-55pt}{\includegraphics[scale=.5]{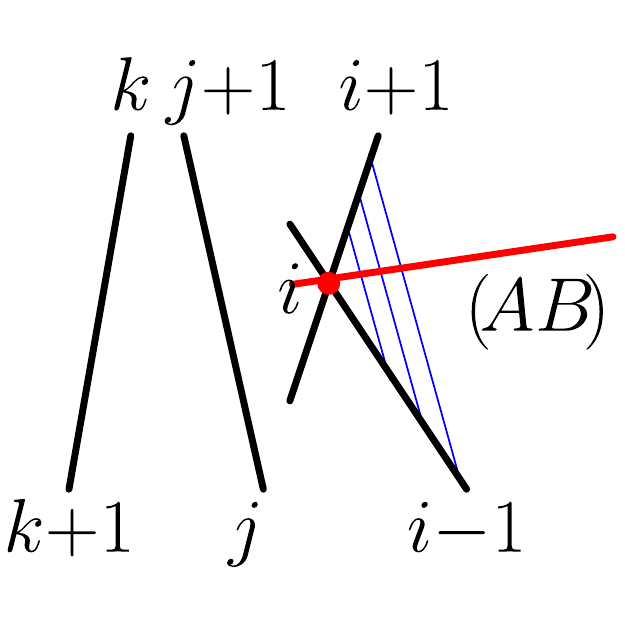}}
 &&& \raisebox{-55pt}{\includegraphics[scale=.5]{./figures/line_config_codim_3_plane.pdf}}
 &&&& \raisebox{-55pt}{\includegraphics[scale=.5]{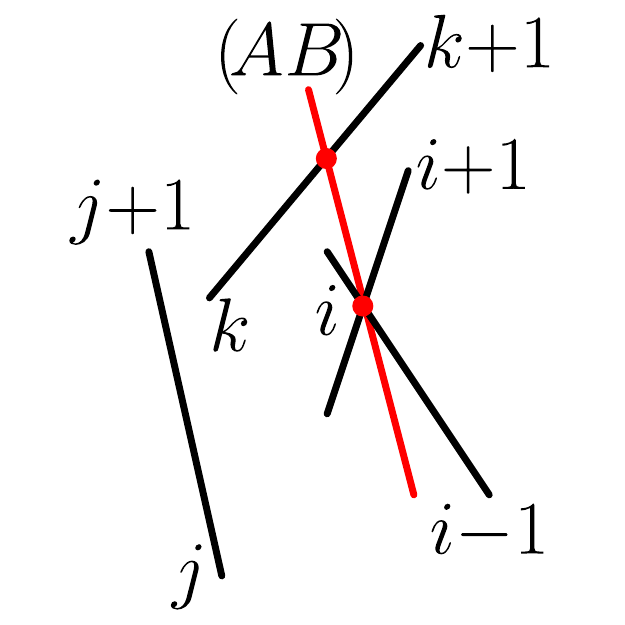}}
 \\
 &\raisebox{-45pt}{\includegraphics[scale=.45]{./figures/cut_config_codim_3_gen.pdf}}
 &&\raisebox{-30pt}{\includegraphics[scale=.45]{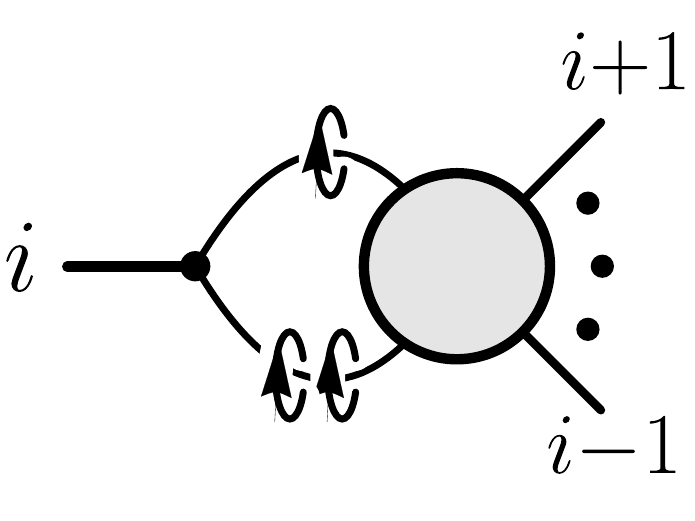}}
 &&&\raisebox{-45pt}{\includegraphics[scale=.45]{./figures/cut_config_codim_3_plane.pdf}}
 &&&&\raisebox{-45pt}{\includegraphics[scale=.45]{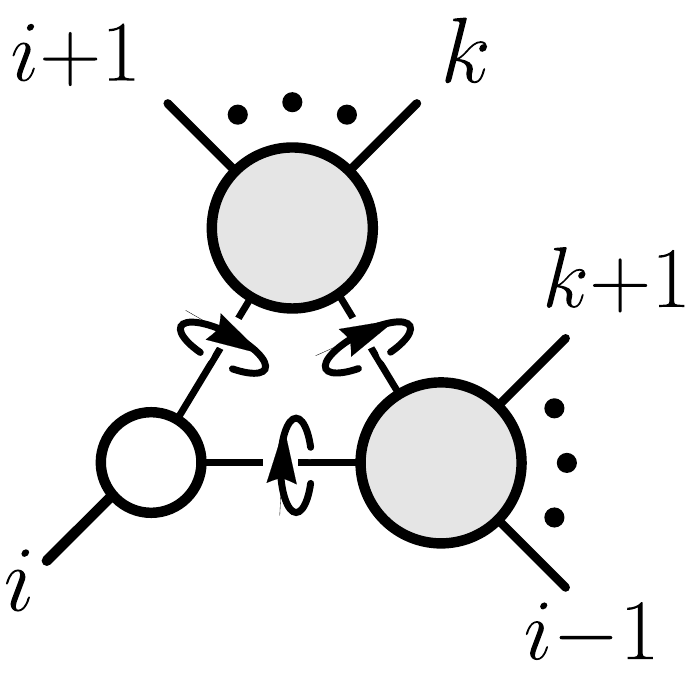}}
\end{align}
Here we have omitted two special configurations where the loop line $(AB)$ intersects three consecutive external lines $(i{-}2 i{-}1),\ (i{-}1 i)$ and $(i i{+}1)$. For the solutions depicted above, we can write one-parametric representations of the solution space for $(AB)$. The generic solution is a bit involved, so here we only give the parametrization for the simple configuration where $(AB)$ is in the plane $(i{-}1ii{+}1)$ and passes through the point $Z_i$,
\begin{align}
\begin{split}
    & Z^{(2)}_A = Z_i\,, \quad  Z^{(2)}_B = Z_{i-1} + \delta Z_{i+1}.
\end{split}
\end{align}
Finally, we can discuss codimension-four configurations of the line $(AB)$ where all degrees of freedom are completely fixed. Such configurations are related to leading singularities. Again, there are various cases to consider, some of which correspond to soft composite residues that are physical, as well as spurious residues where scattering amplitudes have no support (see the figure on the right below) 
\begin{align}
   &\raisebox{-55pt}{\includegraphics[scale=.5]{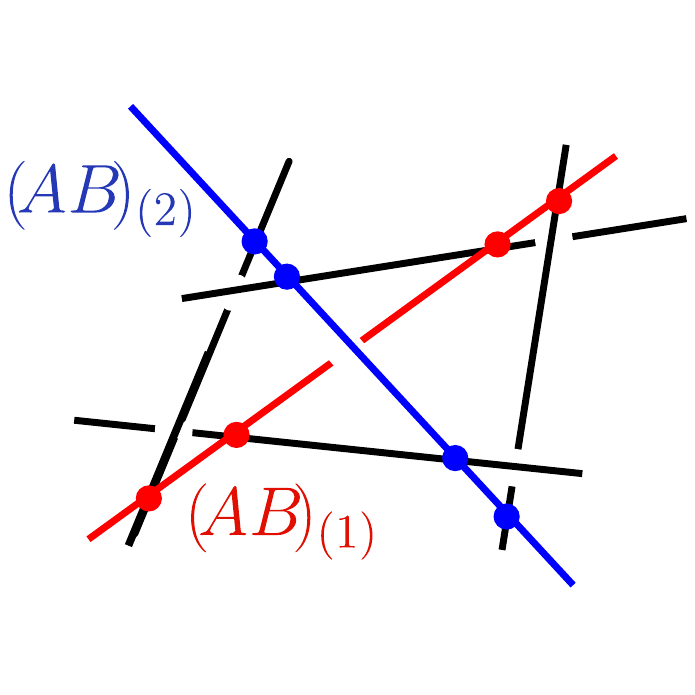}}
   && \raisebox{-55pt}{\includegraphics[scale=.5]{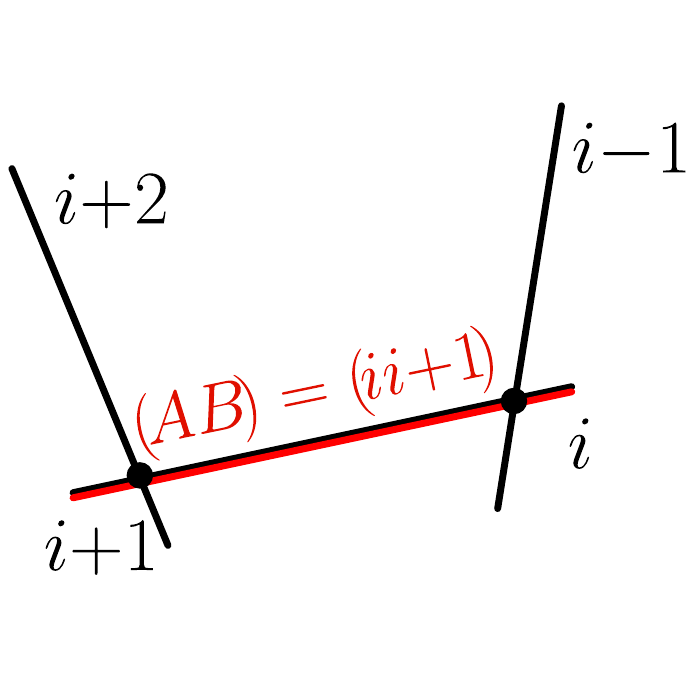}}
   &&& \raisebox{-55pt}{\includegraphics[scale=.5]{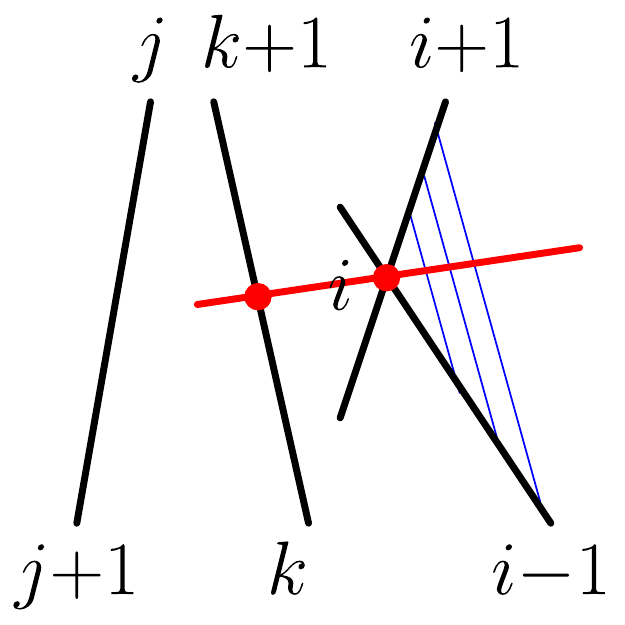}} 
    \\
    &\raisebox{-45pt}{\includegraphics[scale=0.6]{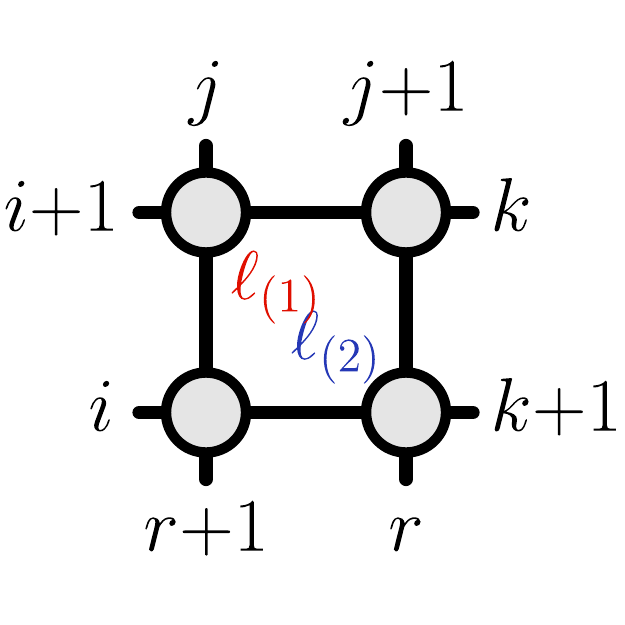}}
    &&\raisebox{-45pt}{\includegraphics[scale=.5]{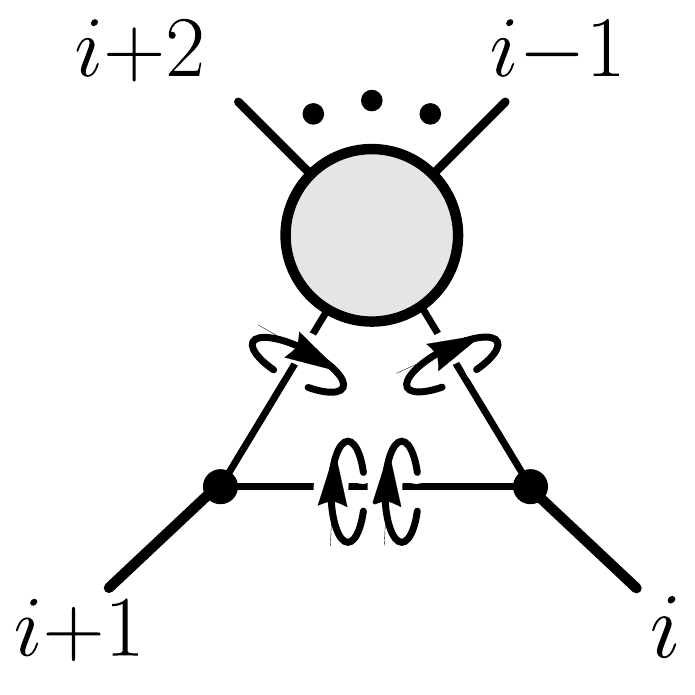}}
    &&&\raisebox{-45pt}{\includegraphics[scale=.5]{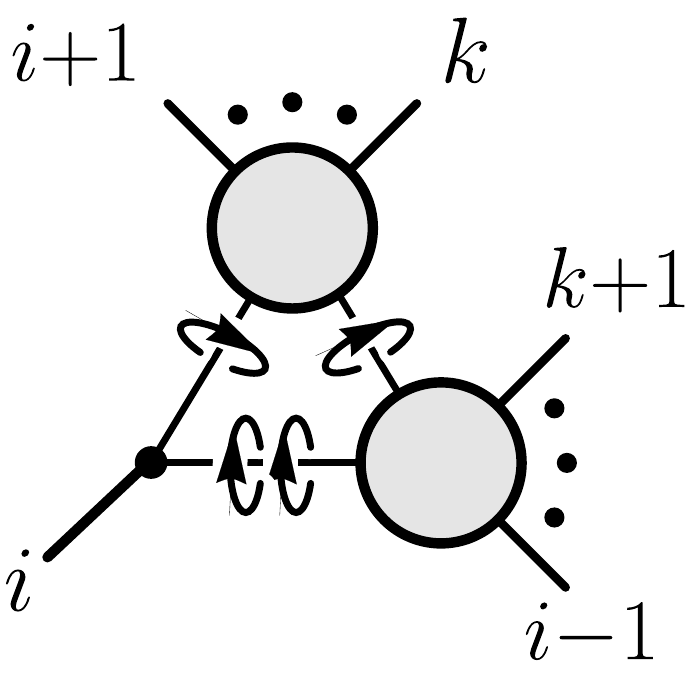}}
\end{align}
Besides the generic ``four-mass" configuration depicted in the left figure above, we can have special kinematic configurations where some of the external lines intersect (corresponding to massless corners in diagrams) and the Schubert problem simplifies. 
\begin{align}
    & 
    \raisebox{-40pt}{\includegraphics[scale=.5]{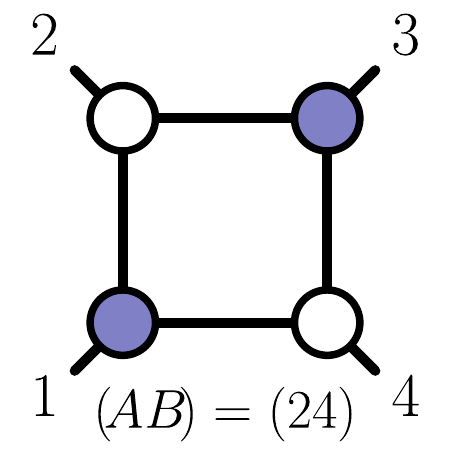}} 
    \hspace{.4cm}
    \quad \leftrightarrow \quad 
    \hspace{-.5cm}
    \raisebox{-25pt}{\includegraphics[scale=.3]{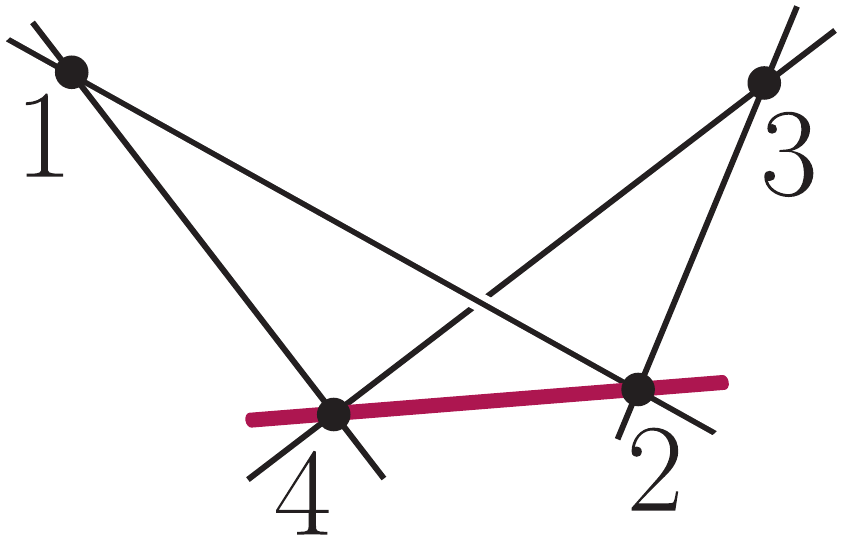}}
     &&
    \hspace{-.4cm}
    \raisebox{-25pt}{\includegraphics[scale=.3]{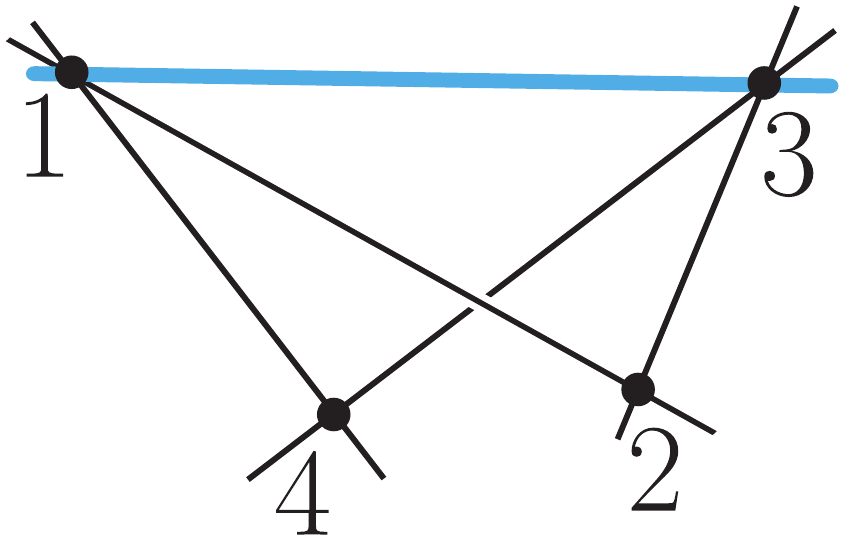} } 
    \hspace{-.4cm}
    \quad \leftrightarrow \quad 
    \raisebox{-40pt}{\includegraphics[scale=.5]{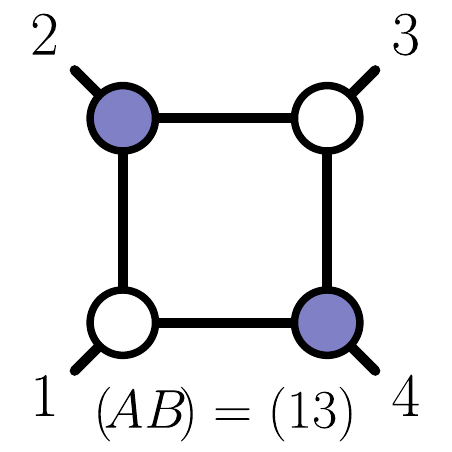}} 
   \nonumber \\
    &\raisebox{-40pt}{\includegraphics[scale=.6]{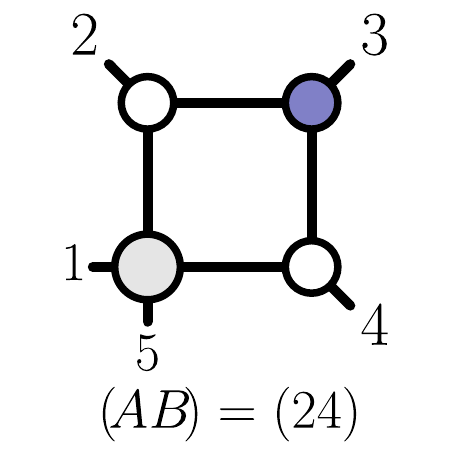}} 
    \quad \leftrightarrow \quad 
    \raisebox{-30pt}{\includegraphics[scale=.3]{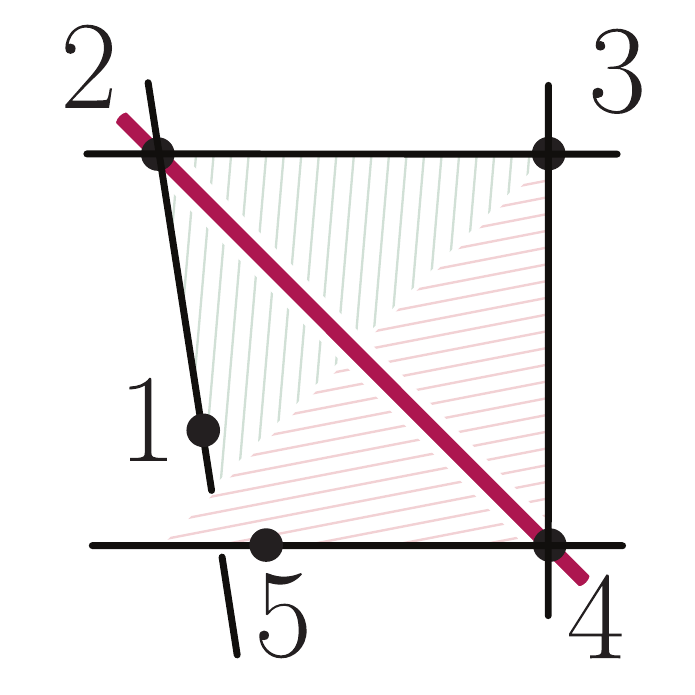}}
    \qquad &&
    \raisebox{-30pt}{\includegraphics[scale=.3]{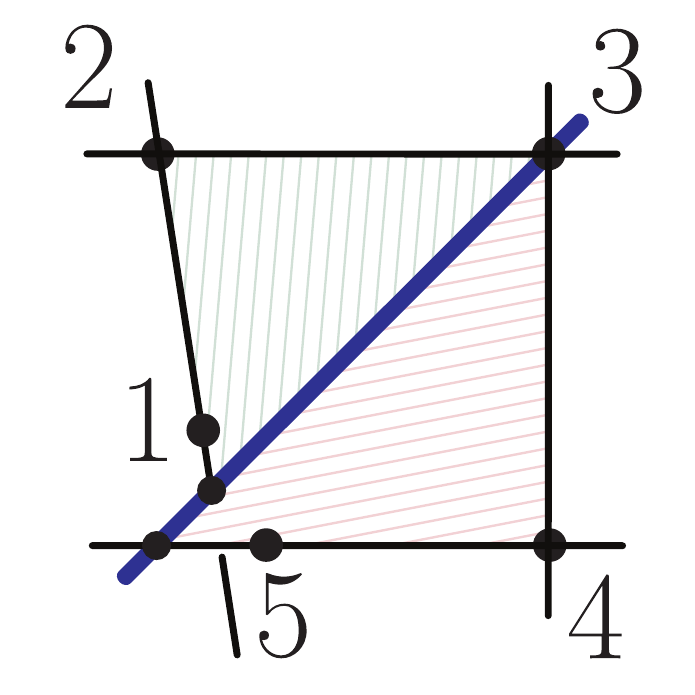}}  
    \quad \leftrightarrow \quad 
    \raisebox{-40pt}{\includegraphics[scale=.6]{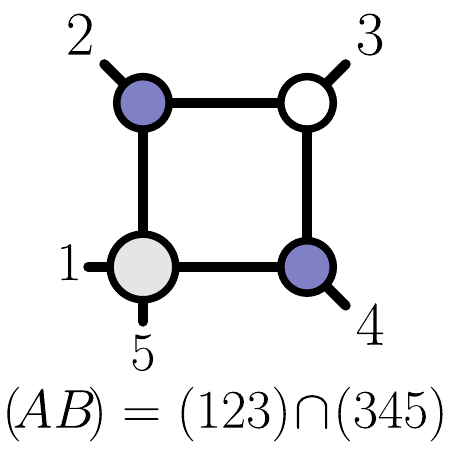}} 
  \nonumber  \\
    &\raisebox{-40pt}{\includegraphics[scale=.6]{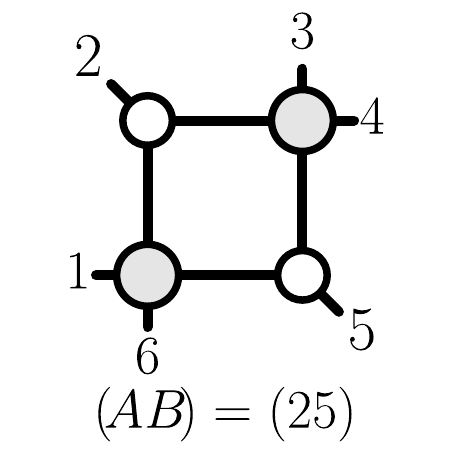}} 
    \quad \leftrightarrow \quad 
    \raisebox{-30pt}{\includegraphics[scale=.3]{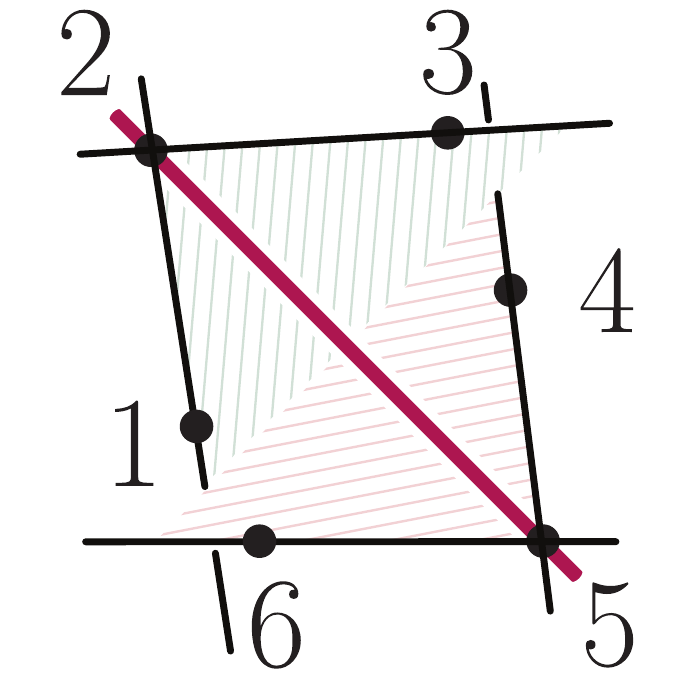}}
    \qquad &&
    \raisebox{-30pt}{\includegraphics[scale=.3]{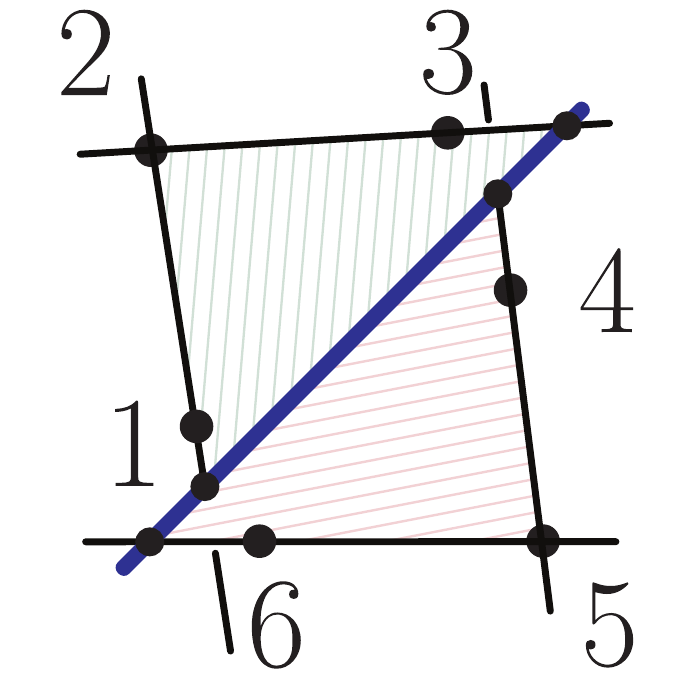} } 
    \quad \leftrightarrow \quad 
    \raisebox{-40pt}{\includegraphics[scale=.6]{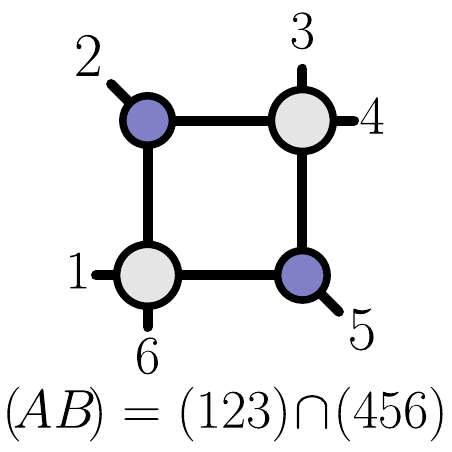}} 
    \\
    &\raisebox{-40pt}{\includegraphics[scale=.6]{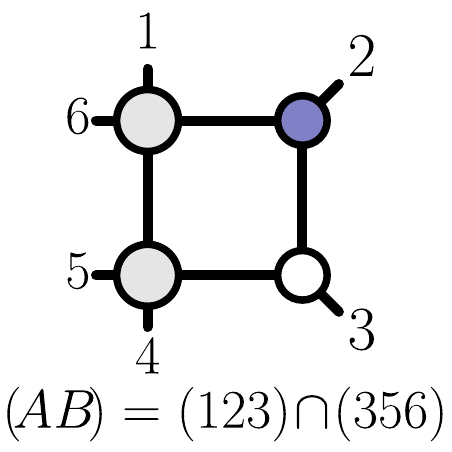}} 
    \quad \leftrightarrow \quad 
    \raisebox{-35pt}{\includegraphics[scale=.3]{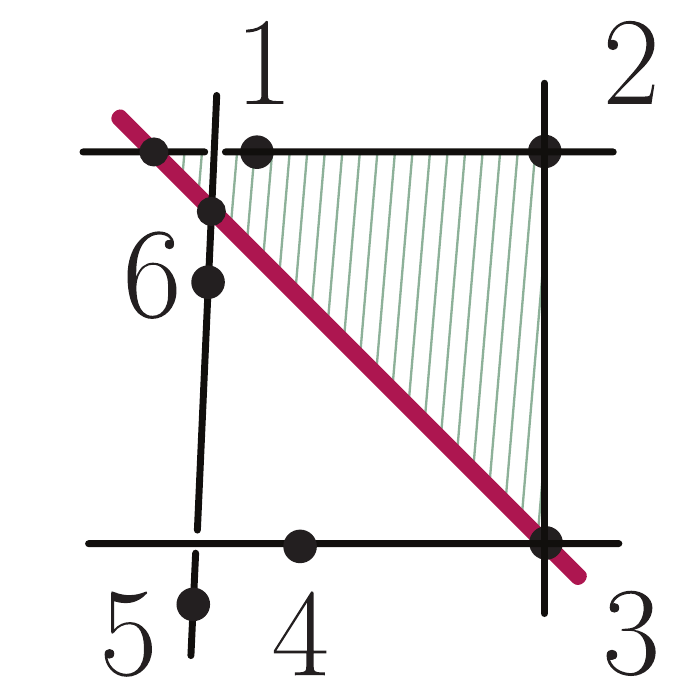}}
    \qquad  &&
    \raisebox{-35pt}{ \includegraphics[scale=.3]{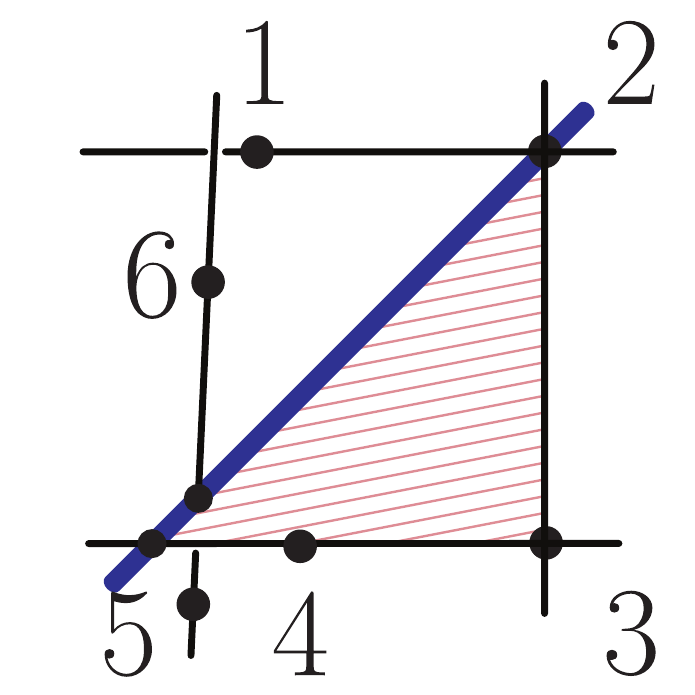}}  
    \quad \leftrightarrow \quad 
    \raisebox{-40pt}{\includegraphics[scale=.6]{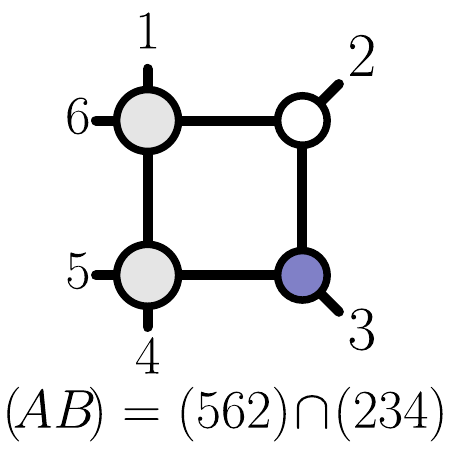}} 
    \nonumber\\
    &\raisebox{-40pt}{\includegraphics[scale=.6]{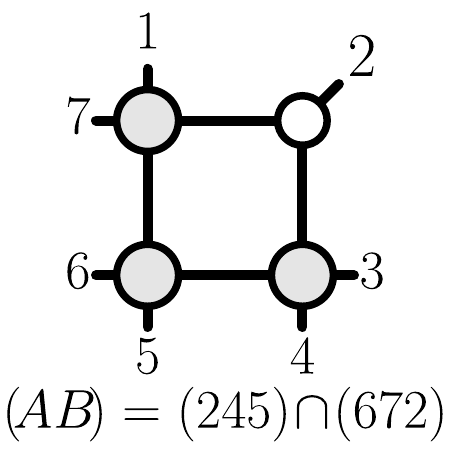}} 
    \quad \leftrightarrow \quad 
    \raisebox{-35pt}{\includegraphics[scale=.3]{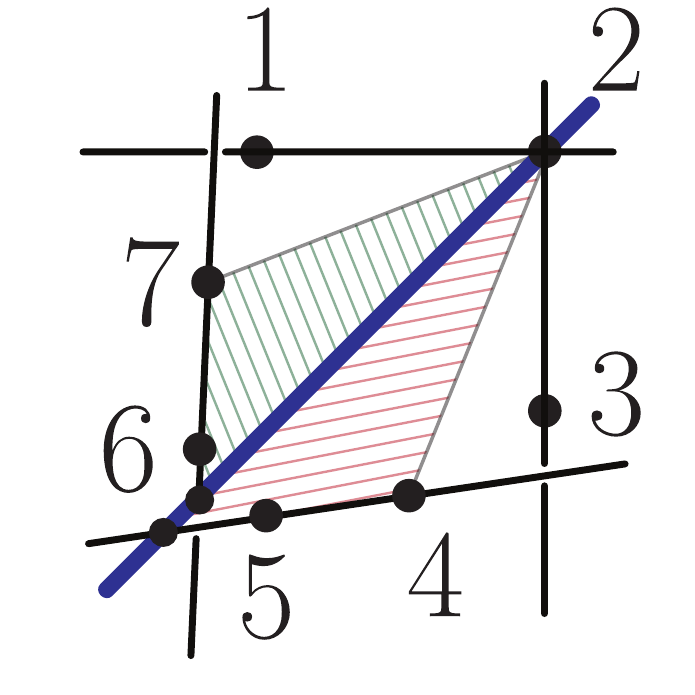}}
    \qquad &&
    \raisebox{-35pt}{\includegraphics[scale=.3]{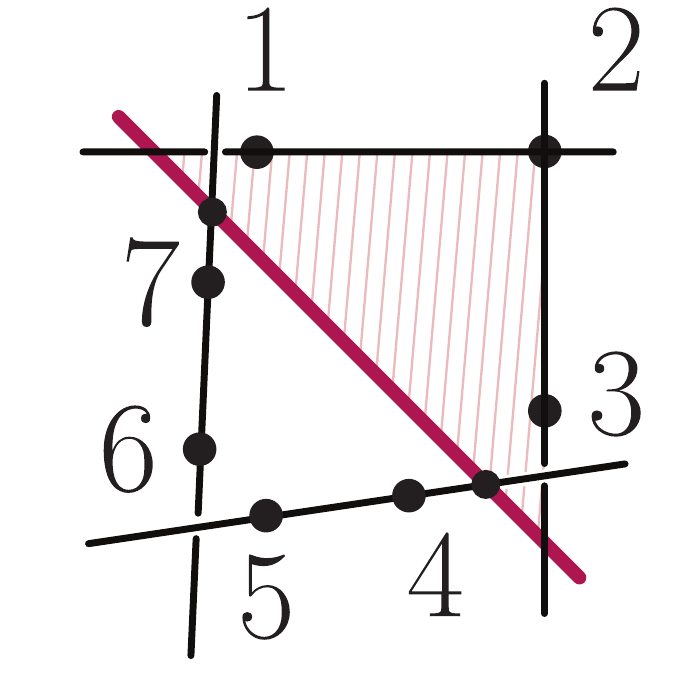}}  
    \quad \leftrightarrow \quad 
    \hspace{-.9cm}
    \raisebox{-40pt}{\includegraphics[scale=.6]{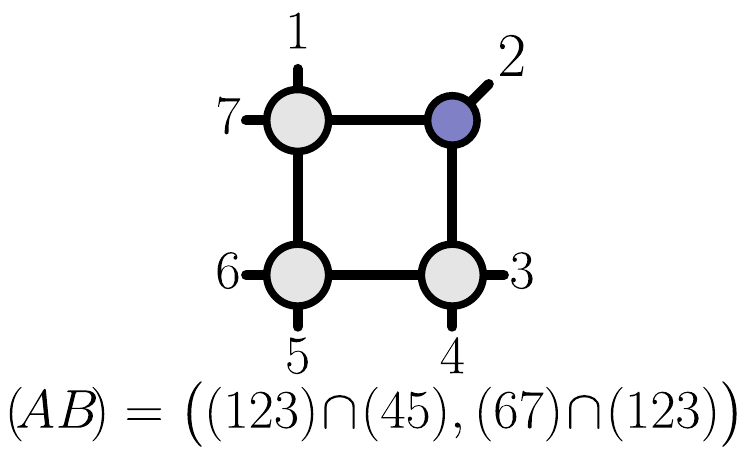}}\nonumber
\end{align}
For the maximal codimension configurations, we have explicitly written the final configuration of the loop-line $(AB)$ in terms of external twistors only. In these formulae, certain geometric quantities appear that we briefly discuss for completeness. In particular, $(abc) \cap (def)$ denotes the intersections of two planes, spanned by twistors $\{Z_a,Z_b,Z_c\}$ and $\{Z_d,Z_e,Z_f\}$ respectively. In $\mathbb{P}^3$, the intersection of two planes is a line, which can be represented as
\begin{align}
\begin{split}
    (abc)\cap (def) & = Z_a Z_b \ab{cdef} + Z_b Z_c \ab{adef} + Z_c Z_a \ab{bdef} \\ 
                    & = \ab{abcd}Z_eZ_f + \ab{abcf}Z_dZ_e+\ab{abce}Z_fZ_d
\end{split}                    
\end{align}
Additionally, there are points defined by the intersection of a line $(ab)$ and a plane $(cde)$, which can be represented as
\begin{align}
    (ab)\cap (cde) = Z_a \ab{bcde} +Z_b\ab{cdea} = - (Z_c \ab{deab} + Z_d \ab{eabc} + Z_e \ab{abcd})\,,
\end{align}
which naturally reflects the antisymmetry $(ab)\cap(cde) = -(cde)\cap (ab)$.

\section{External triangulations}
\label{sec:external_triangulations}
In this section we focus on \emph{external triangulations}, as a first application of the systematic classification of the sign-flip spaces introduced in section \ref{sec:sign_flip_regions}. As discussed briefly in section \ref{subsec:positivity_dual_Amplituhedron}, an external triangulation of a space introduces a ``larger'' space \emph{outside} of the original region using spurious vertices which violate the original positivity conditions. In the context of the sign-flip spaces above, this corresponds to flipping the signs of some brackets which defined the original space. As a warm-up, let us go back to the projective plane $\mathbb{P}^2$, where the analogue of a line $(AB)$ in $\mathbb{P}^3$ is a point $Y$ on the plane with two degrees of freedom. As heuristically described above, we can externally triangulate the quadrilateral with vertices $z_1,z_2,z_3,z_4$ with two triangles,
\begin{align}
\label{eq:plane_toy_ext_triang}
\hspace{-2cm}
\raisebox{-50pt}{\includegraphics[scale=.5]{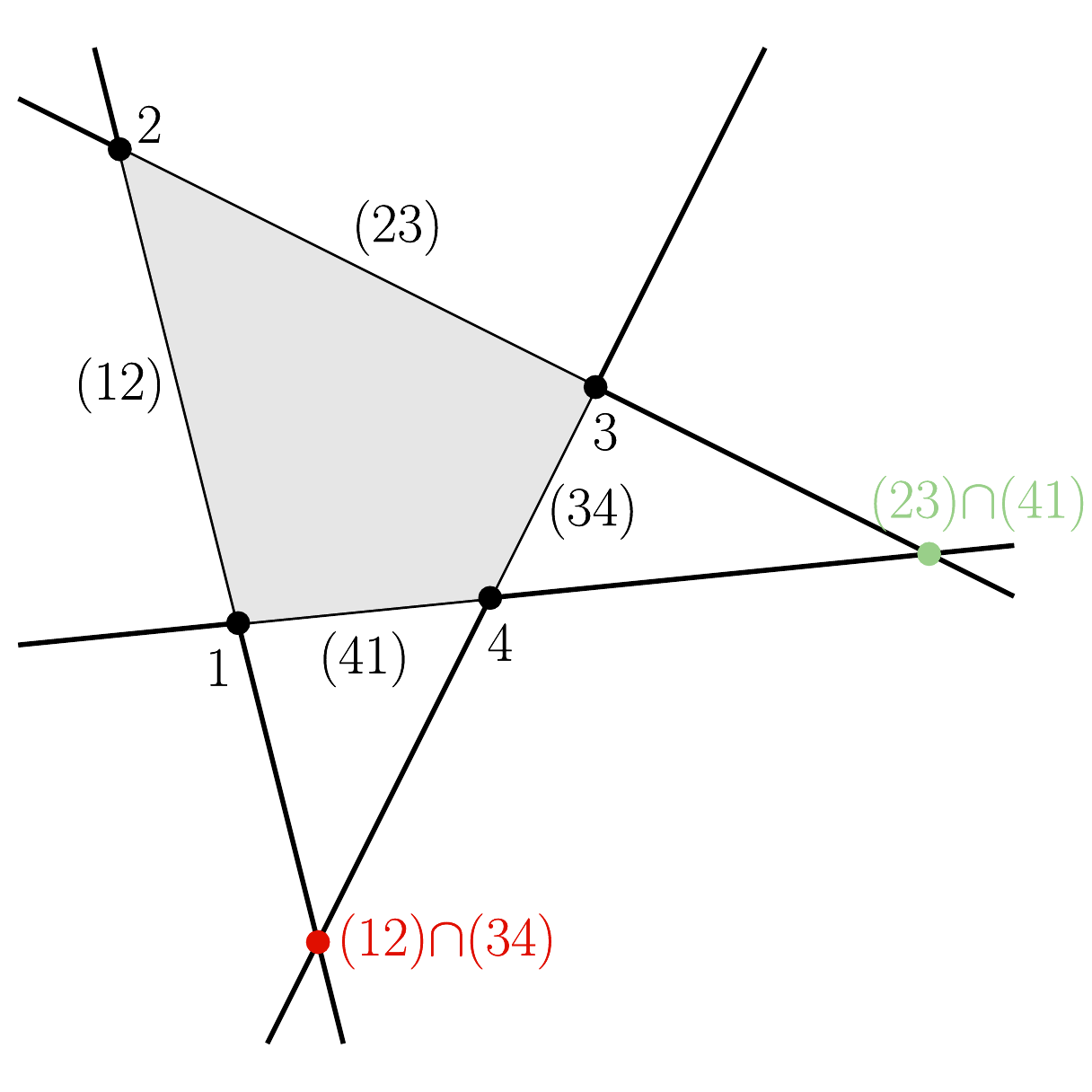}}
\hspace{-.5cm}
=
\hspace{-1cm}
\raisebox{-50pt}{\includegraphics[scale=.4]{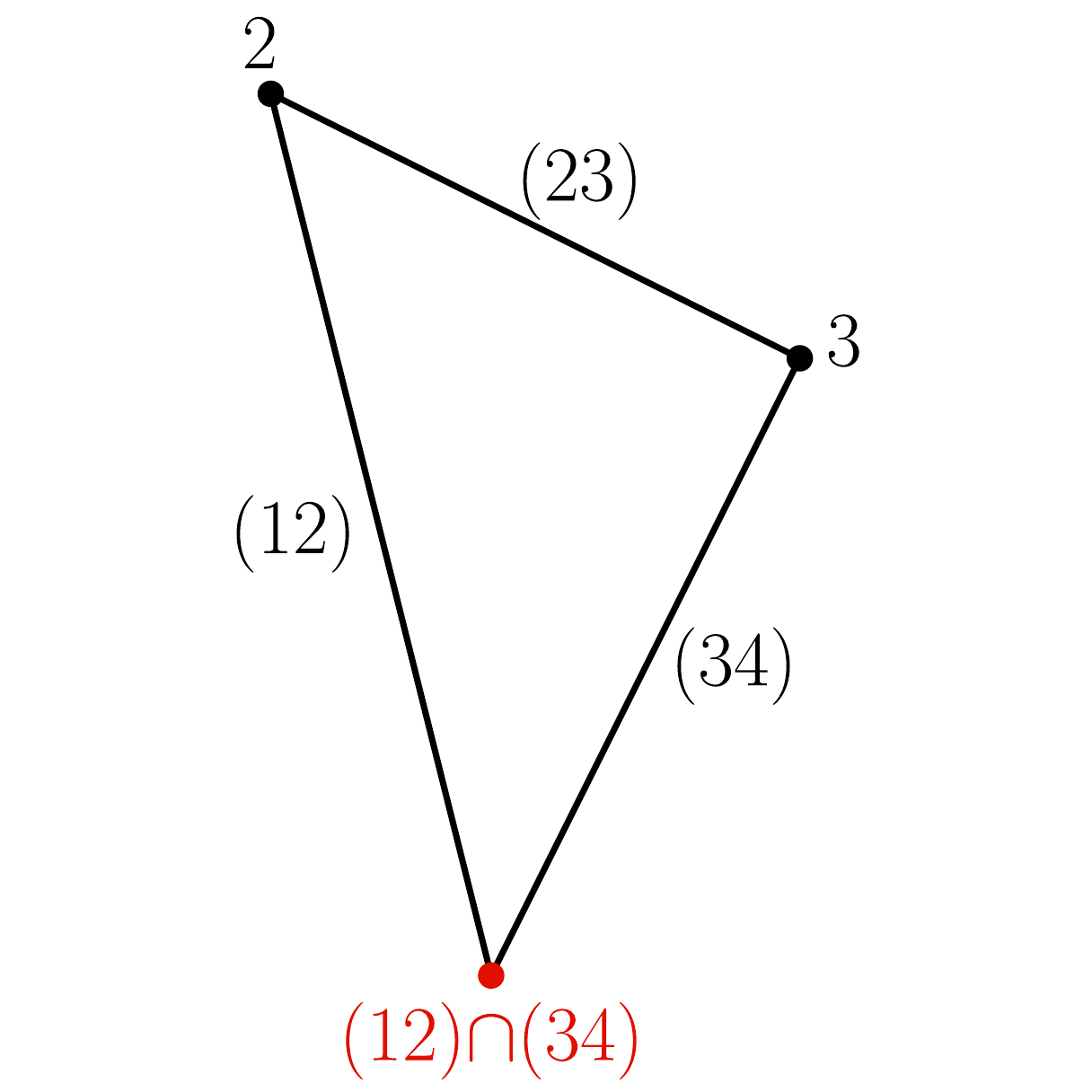}}
\hspace{-.8cm}
-
\hspace{-1cm}
\raisebox{-50pt}{\includegraphics[scale=.4]{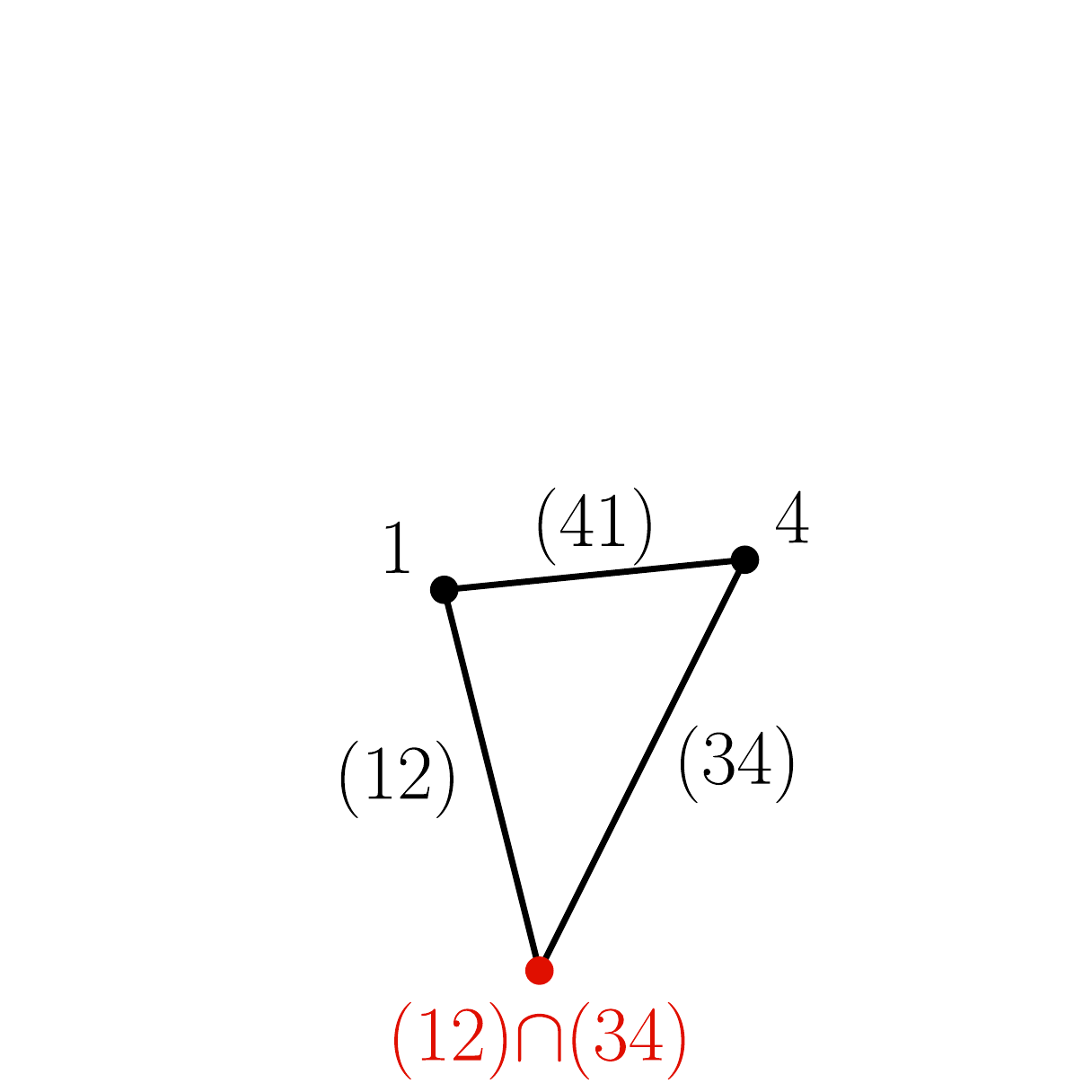}}
\hspace{-2cm}
\end{align}
Note that the codimension-one boundaries of each triangle (which are lines in $\mathbb{P}^2$) are subsets of boundaries of the original quadrilateral, but there is an additional spurious vertex $(12){\cap}(34):=z_1\ab{234}-z_2\ab{134}$ which is used as a triangulation point that cancels between triangles.\footnote{In this context, angle brackets denote contraction with a three-index Levi-Civita symbol, i.e., $\ab{abc}:=\epsilon_{IJK}z_a^I z_b^J z_c^K$.} We can also describe this triangulation in the language of the previous section as follows. The quadrilateral is defined by the following conditions:
\begin{equation}
\raisebox{-45pt}{\includegraphics[scale=.4]{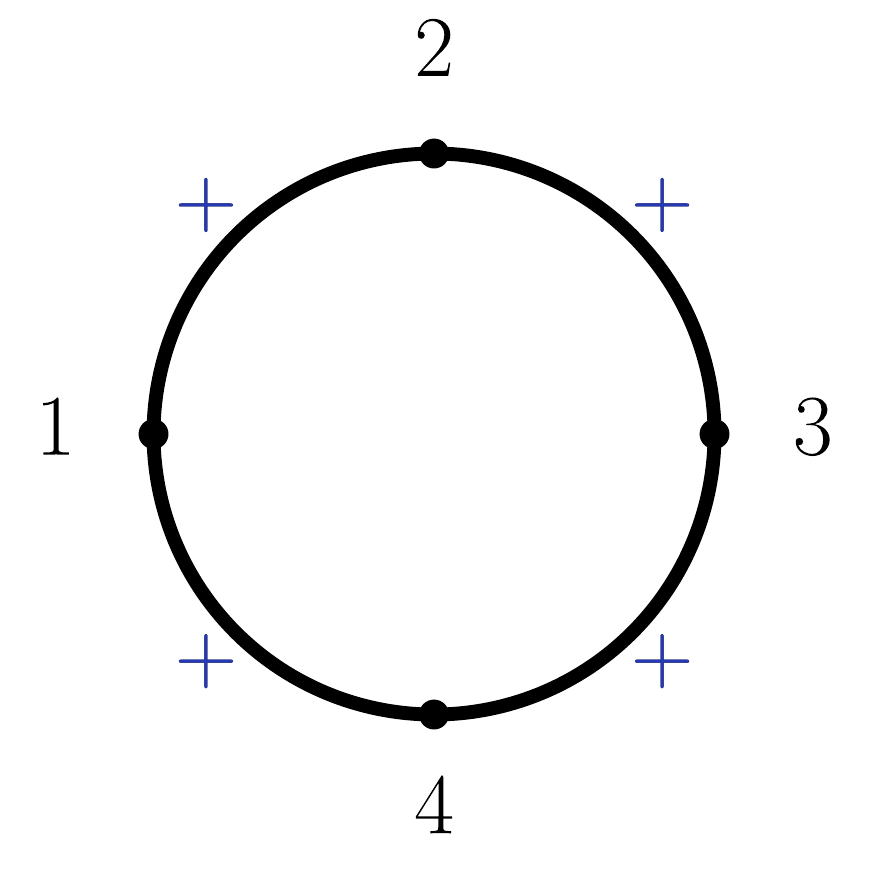}} 
\,\,\,\leftrightarrow \,\,\,
\ab{Y12}>0,\,\,\ab{Y23}>0,\,\,\ab{Y34}>0,\,\,\ab{Y14}>0,\\
\end{equation}
where we used the same circle to visualize the constraints, but now for $\la Yii{+}1\ra$. We can define the triangles in a similar way. Note that the first triangle with vertices $\{(12)\cap(34),2,3\}$ in eq.~(\ref{eq:plane_toy_ext_triang}) has an unfixed sign for $\la Y14\ra$ and therefore lacks this codimension-one boundary. Pictorially this will be denoted by $\ast$ for the relevant bracket in the circle-figures and can also be interpreted as marginalizing over both signs. Even though the second triangle with vertices $\{(12)\cap(34),1,4\}$ has $\la Y23\ra>0$ fixed, from the picture we see that this is not a boundary of the space. The two triangles are associated to the following circle diagrams:
\begin{align}
\hspace{-1cm}
\raisebox{-60pt}{\includegraphics[scale=.4]{./figures/2d_toy_triangulation_big_tri.pdf}}
\hspace{-1cm}
\,\,\,&\leftrightarrow \,\,\,
\raisebox{-45pt}{\includegraphics[scale=.4]{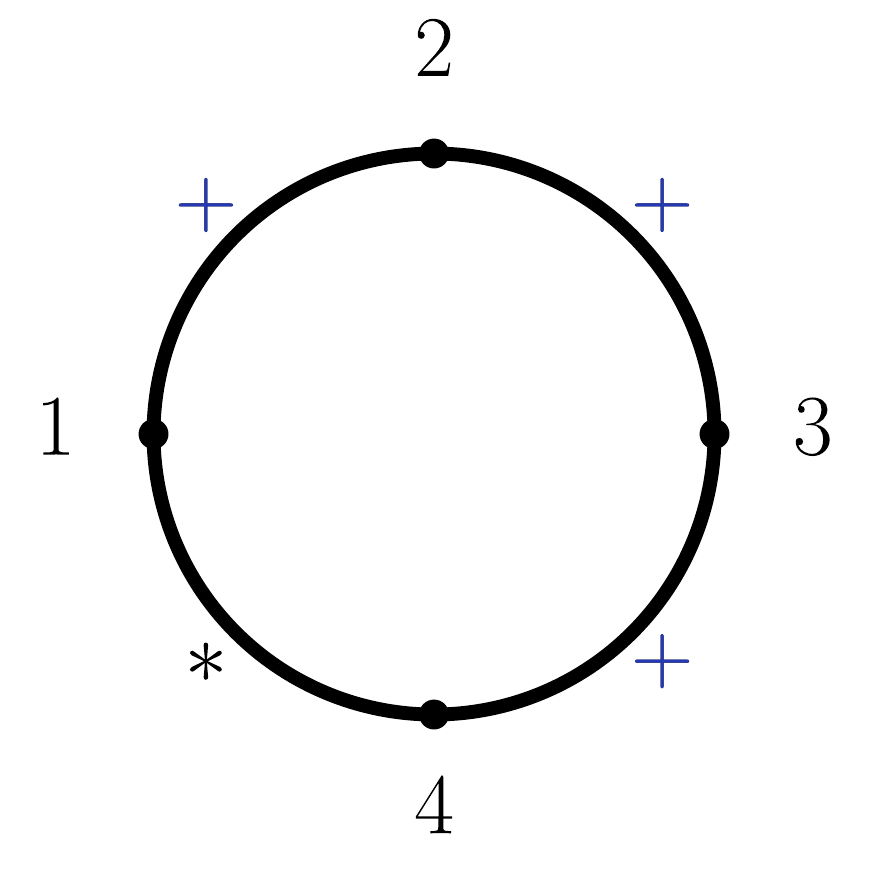}} ,
\\[-40pt]
\hspace{-1cm}
\raisebox{-42pt}{\includegraphics[scale=.4]{./figures/2d_toy_triangulation_little_tri.pdf}}
\hspace{-1cm}
\,\,\,&\leftrightarrow \,\,\,
\raisebox{-45pt}{\includegraphics[scale=.4]{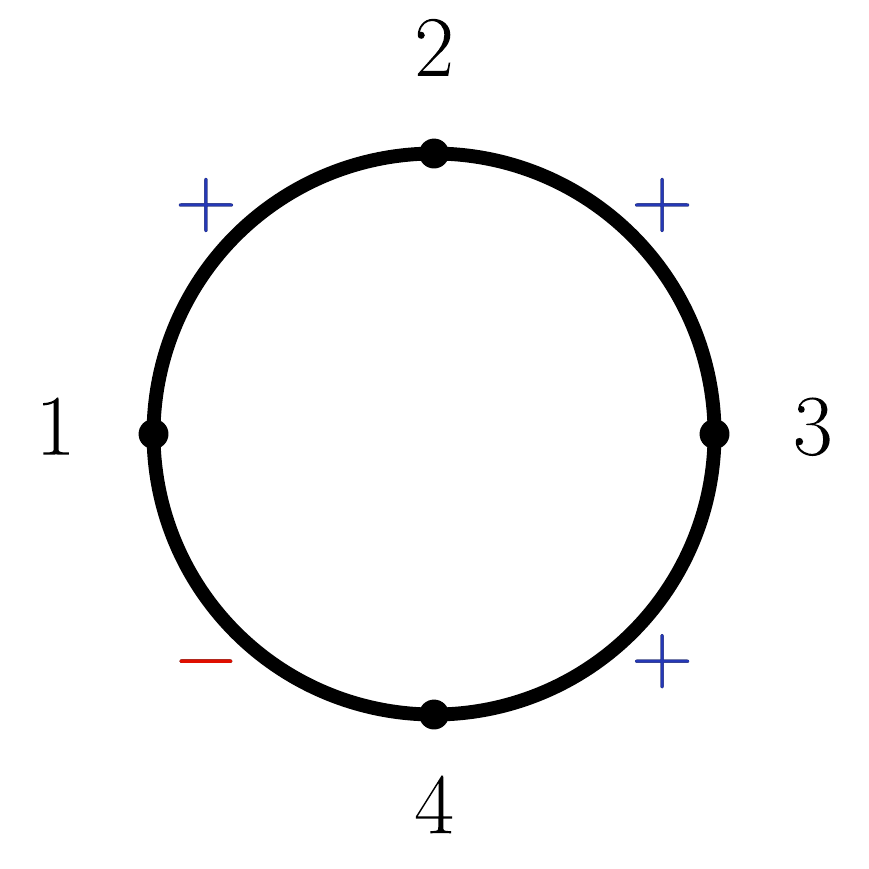}} .
\end{align}
In summary, we can re-interpret this triangulation as taking the space of the first triangle with unfixed sign of $\la Y14\ra$ and dividing it into two spaces: one where $\la Y14\ra>0$, which is the quadrilateral, and another where $\la Y14\ra<0$ which is the second triangle.
\begin{align}
\hspace{-2cm}
\raisebox{-60pt}{\includegraphics[scale=.4]{./figures/2d_toy_triangulation_big_tri.pdf}}
\hspace{-.5cm}
& \quad = \quad
\hspace{-.8cm}
\raisebox{-60pt}{\includegraphics[scale=.4]{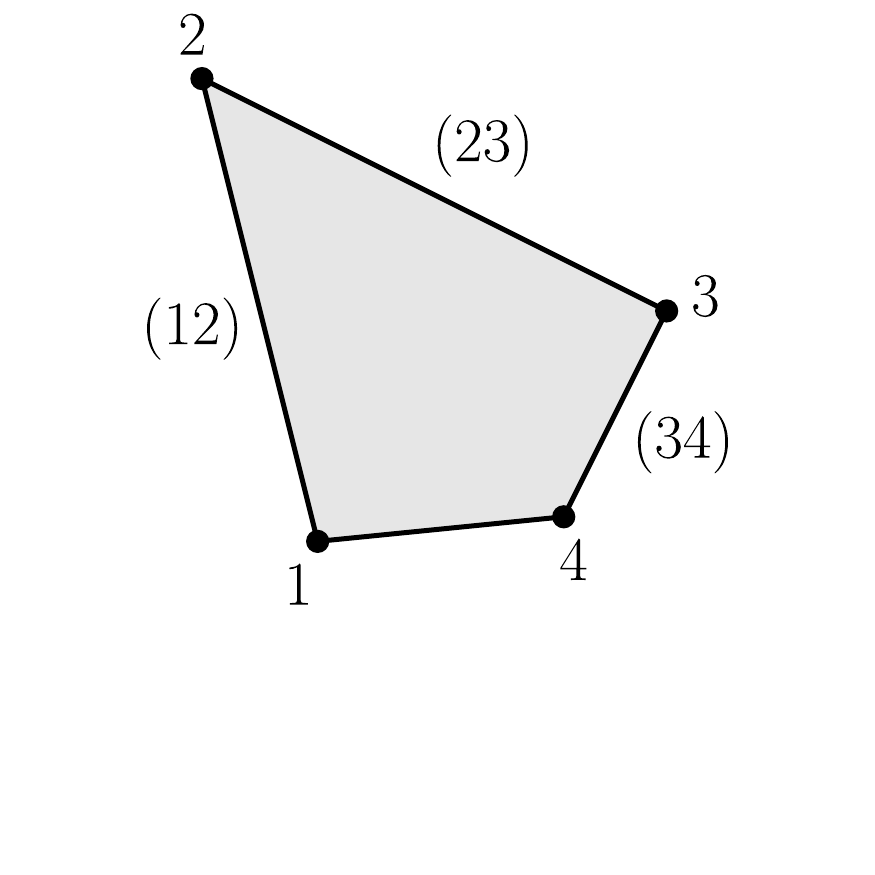}}
\hspace{-1cm}
&&\hspace{-1cm}+\qquad
\hspace{-1.1cm}
\raisebox{-60pt}{\includegraphics[scale=.4]{./figures/2d_toy_triangulation_little_tri.pdf}}
\hspace{-1cm}
\\
 \raisebox{-45pt}{\includegraphics[scale=.4]{./figures/external_triang_triangle_1_disc_representation_marginalized.pdf}} 
& \quad =
\raisebox{-45pt}{\includegraphics[scale=.4]{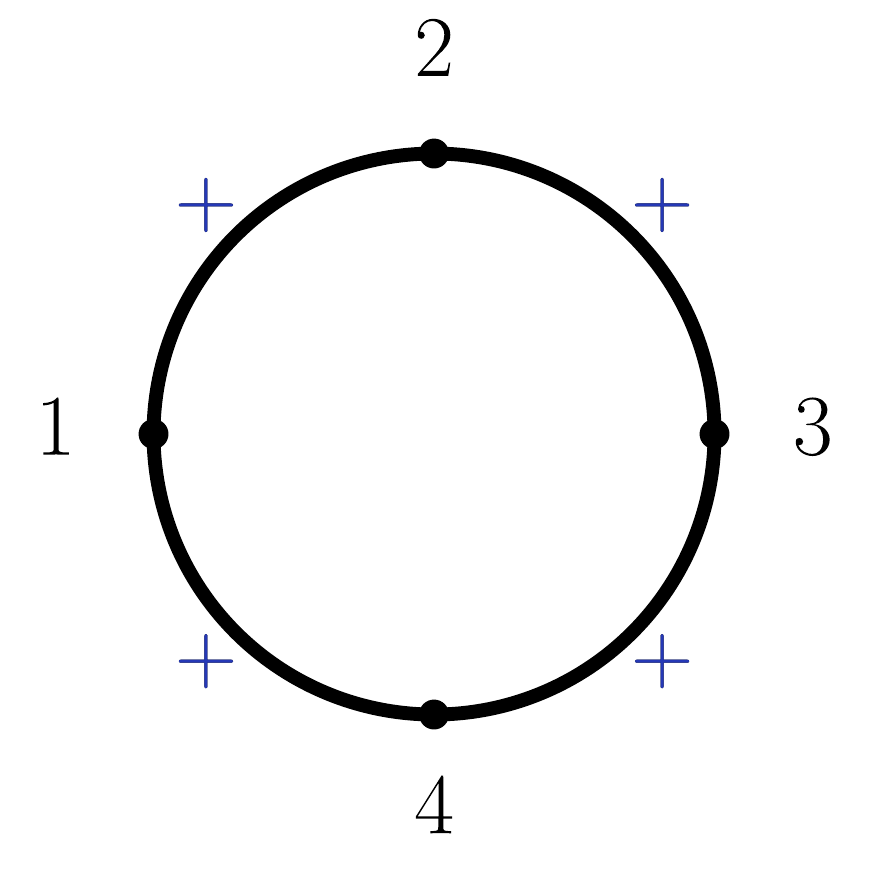}} 
&&\hspace{-1cm}+
\raisebox{-45pt}{\includegraphics[scale=.4]{./figures/external_triang_triangle_1_disc_representation_minus.pdf}} \nonumber
\end{align}
In $\mathbb{P}^2$ the triangle is the simplest geometric space with non-vanishing form. If we remove one more boundary by marginalizing over the corresponding $\la Yii{+}1\ra$, we get a ``wedge" defined by only two inequalities. In this case the canonical form vanishes, e.g.
\begin{equation}
\raisebox{-45pt}{\includegraphics[scale=.4]{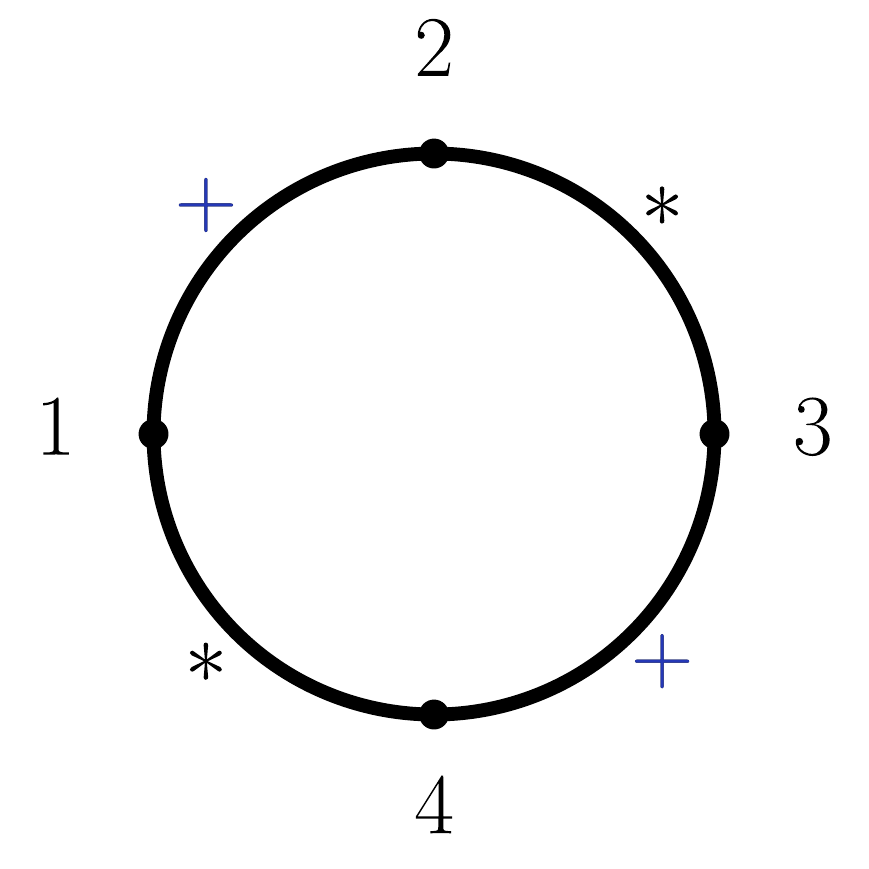}} 
\,\,\,\leftrightarrow \,\,\,
\Omega=0.\\
\end{equation}
This vanishing can also be understood from the $d\log$ form perspective as we need at least three brackets to form two independent projective ratios that enter the arguments of $d\log$'s.

\subsection*{Simplest sign-flip spaces} 

We can almost verbatim generalize the above discussion to the configuration space of lines $(AB)$ in $\mathbb{P}^3$ relevant to the MHV one-loop positive geometry. In $\mathbb{P}^3$, the simplest space has four boundaries, and the corresponding logarithmic form is given by the box integrand. However, in this case we need to include one additional inequality which does not correspond to a boundary but is nonetheless required to define the chiral sign-flip-four space. In the simplest four-point example, we define the space 
\begin{align}
\hspace{-1cm}
\raisebox{-45pt}{\includegraphics[scale=.4]{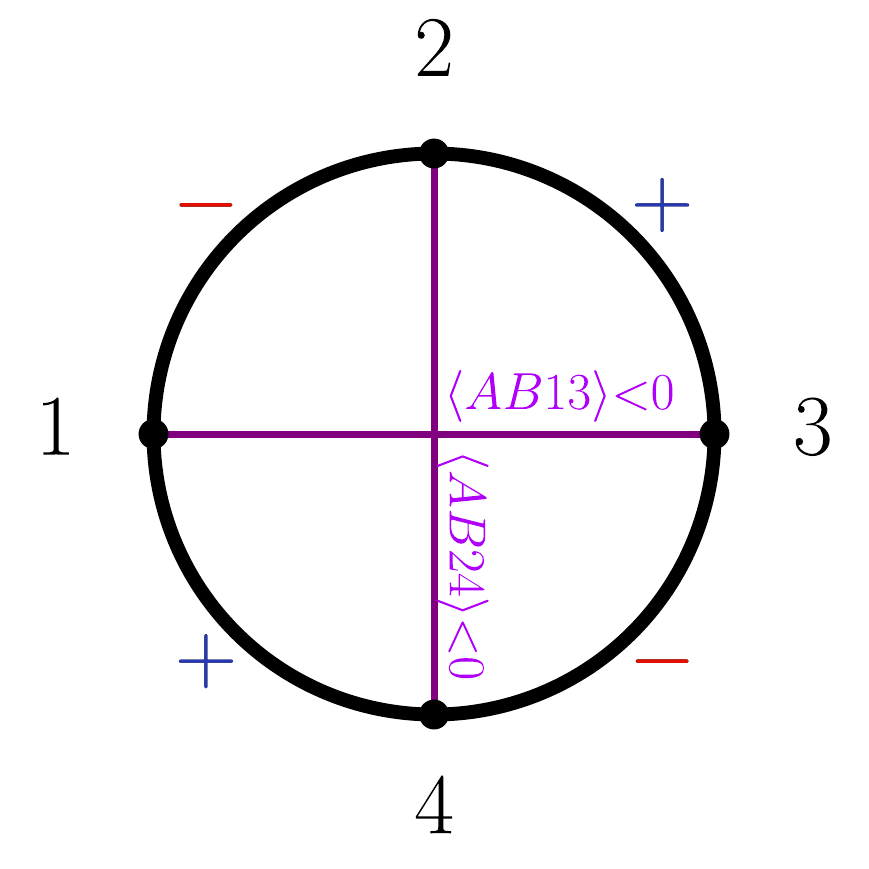}} 
\,\,\,\leftrightarrow \,\,\,
\omega^{(4),-}_{1234}=\frac{\ab{1234}^2}
                {\ab{AB12}\ab{AB23}\ab{AB34}\ab{AB14}}
\equiv
\hspace{-.2cm}
\raisebox{-32pt}{\includegraphics[scale=.7]{./figures/0mass_box}} 
\hspace{-.5cm}
\end{align}
with $\ab{AB13},\ab{AB24}<0$. The second chiral region has $\ab{AB13},\ab{AB24}>0$ and its form is the same up to a sign, $\Omega^{(4),+}_{1234}=-\Omega^{(4),-}_{1234}$. The union of these two regions is an achiral space which has vanishing form and unfixed signs for $\ab{AB13}$ and $\ab{AB24}$, 
\begin{equation}
\raisebox{-45pt}{\includegraphics[scale=.4]{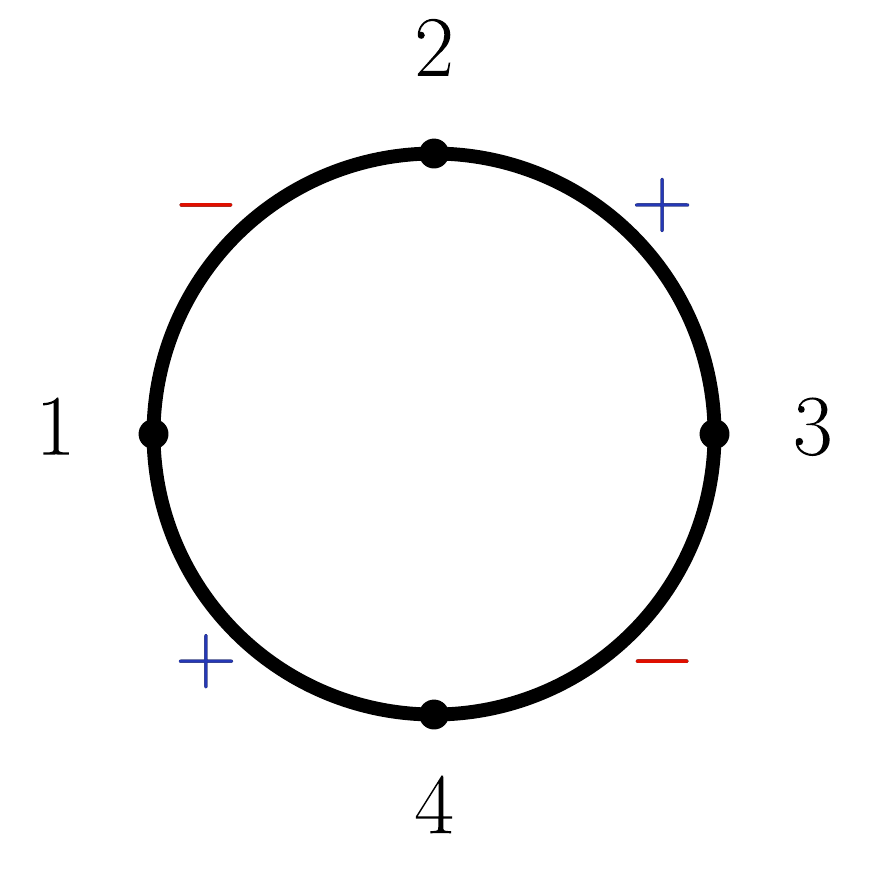}}
=
\raisebox{-45pt}{\includegraphics[scale=.4]{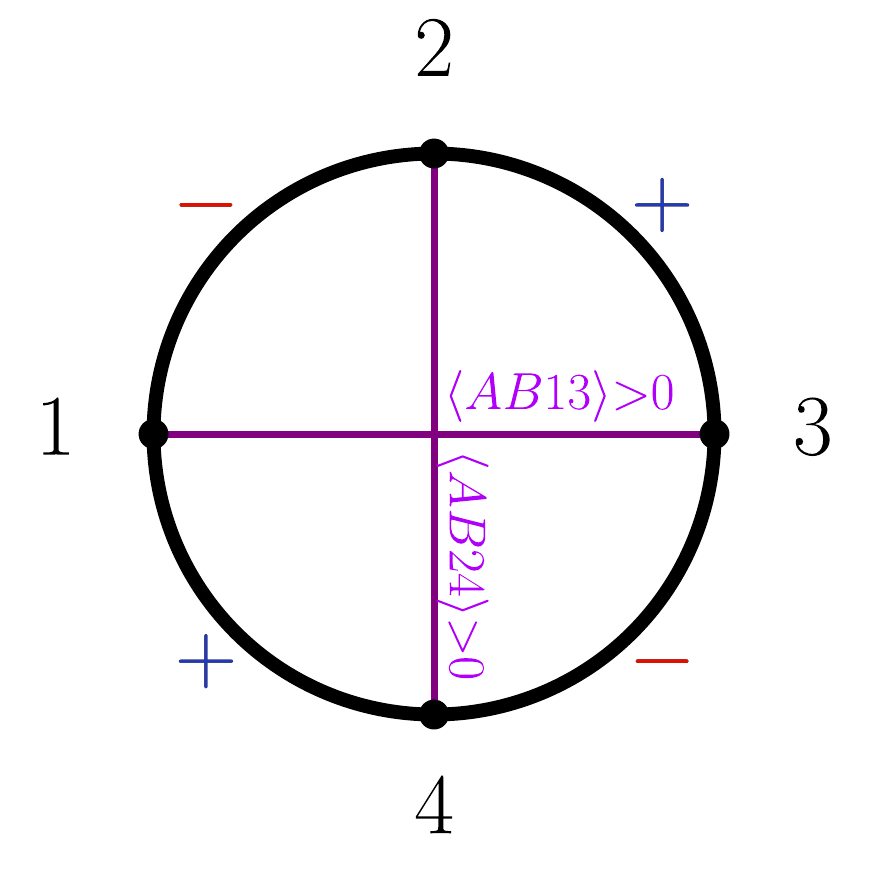}}
+
\raisebox{-45pt}{\includegraphics[scale=.4]{./figures/sf4_box_4pts_minus.pdf}}
\end{equation}
Directly at the level of $d\log$ forms, four brackets $\ab{AB12}$, $\ab{AB23}$, $\ab{AB34}$, $\ab{AB14}$ are insufficient to define four independent projective ratios that enter the arguments of the $d\log$s, and therefore the whole form must vanish. If we additionally impose $\ab{AB13}\lessgtr0$ to cut the space into chiral components, we have access to one further bracket to form four independent projective ratios, e.g., $\ab{ABii{+}1}/\ab{AB13}$. The same argument applies to any other space with only four boundaries $\ab{ABii{+}1}$. A less trivial example of that logic is the special sign-flip-two region we discussed in eq.~(\ref{eq:sf2_2minus_signs_npt_special_eg}). Even though there were many brackets with fixed signs, the space has only four boundaries -- so we get a zero-form space if we drop the chiralization condition $\ab{ABii{+}2}\gtrless0$. 

The simplest achiral space with non-vanishing form must therefore have five boundaries, and the integrand form is the general parity-odd pentagon (given by a suitable generalization of eq.~(\ref{eq:odd_pent_num}) where none of the external legs need to be massless). In fact, because the achiral space is defined by a set of inequalities which all correspond to physical codimension-one boundaries, any of the $2^5$ sign choices are allowed and lead to the same canonical form up to a sign. (This is distinct from the chiral components where only a subset of signs led to a consistent geometry.) For the five-point space we get 
\begin{equation}
\raisebox{-45pt}{\includegraphics[scale=.4]{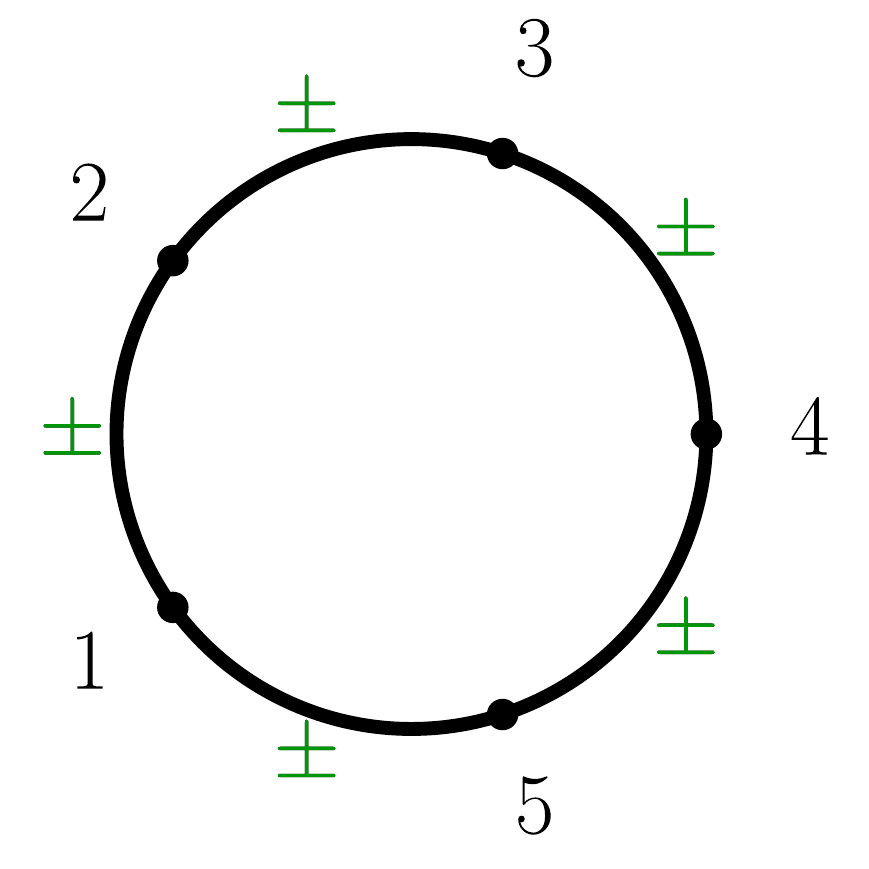}}
\,\,\,\leftrightarrow \,\,\,
\omega^{\text{odd}}_5 = \frac{ N_{\text{odd}}}
                {\ab{AB12}\ab{AB23}\ab{AB34}\ab{AB45}\ab{AB15}},\\
\end{equation}
where $\pm$ indicates an arbitrary (but fixed) sign for the corresponding bracket. Exactly the same is true for higher points: if we marginalize over the signs of the $n-5$ brackets $\ab{ABii{+}1}$ for $i\notin \{i_1,i_2,i_3,i_4,i_5\}$ and pick any of the $2^5$ possible sign choices for the remaining $\la ABi_r i_r{+}1\ra$, we find the general parity-odd pentagon
\begin{equation}
\label{eq:odd_pent_sf_form}
\raisebox{-70pt}{\includegraphics[scale=.5]{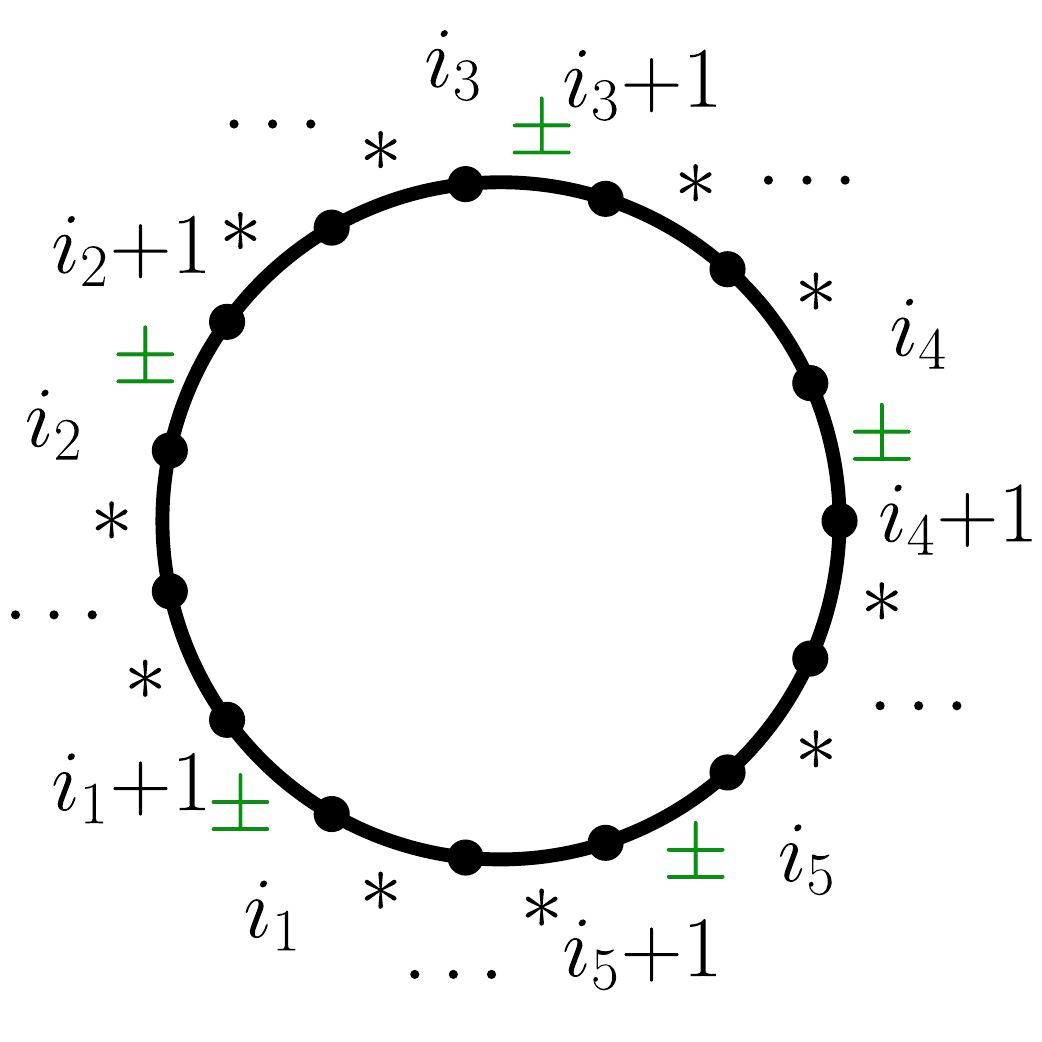}} 
\leftrightarrow
\omega^{\text{odd}}_n {=} \frac{N_{\text{odd}}}
{\begin{array}{c}\ab{ABi_1i_1{+}1}\ab{ABi_2i_2{+}1}\ab{ABi_3i_3{+}1}\\\times\ab{ABi_4i_4{+}1}\ab{ABi_5i_5{+}1}\end{array}}\,,\\
\end{equation}
where $\tbrgreen{\pm}$ indicates that either sign choice is acceptable and $\ast$ instructs us to marginalize over both sign choices.

In the rest of the section we will relate different sign-flip spaces via \emph{external triangulations}, which allows us to write the canonical forms for more complicated spaces with many boundaries. In our discussion, we use several important facts:
\begin{enumerate}
    \item Canonical forms of sign-flip-four spaces are chiral octagons, eq.~(\ref{eq:sf4_chiral_octagon_association}), and descendants. 
    \item The form for a chiral space with four boundaries is a box. 
    \item The form for an achiral space with five boundaries is a parity-odd pentagon, eq.~(\ref{eq:odd_pent_sf_form}). 
    \item Any chiral space with three or fewer boundaries has vanishing form. Any achiral space with four or fewer boundaries has vanishing form.
    \item Sign-flip-six and higher regions are empty; their form is identically zero.
\end{enumerate}

\subsection*{Triangulation of sign-flip-two regions}

We first consider the chiral sign-flip-two space, with the additional condition $\ab{ABij}{>}0$:
\begin{equation}
S^{(2),+}_{ij}=\raisebox{-45pt}{\includegraphics[scale=.4]{./figures/sf2_plus_components.pdf}}.
\end{equation}
As stated in section \ref{sec:sign_flip_regions}, the codimension-one boundaries of this space correspond to the four brackets adjacent to the sign flips, as well as all other `negative' brackets in the upper half of the circle, i.e., 
\begin{equation}
\text{boundaries:}\ \{\ab{ABi{-}1i},\ab{ABii{+}1},\ab{ABi{+}1i{+}2},\ldots,
                      \ab{ABj{-}1j},\ab{ABjj{+}1} \}. 
\end{equation}
(For the opposite chirality defined by $\ab{ABij}<0$, the boundaries correspond to the four brackets adjacent to the sign flips, as well as the \emph{positive} brackets in the lower half of the circle.) To externally triangulate this space, we use the fact that any space defined by four or fewer inequalities has a vanishing canonical form. Thus, if we marginalize over all signs in the sequence $\{\ab{ABi{+}1\,i{+}2},\ldots,\ab{ABj{-}2\,j{-}1}\}$ but leave all other signs unchanged, the corresponding space has four boundaries, and is therefore trivially a two-mass-easy box form:
\begin{equation}
\label{compl}
\raisebox{-45pt}{\includegraphics[scale=.4]{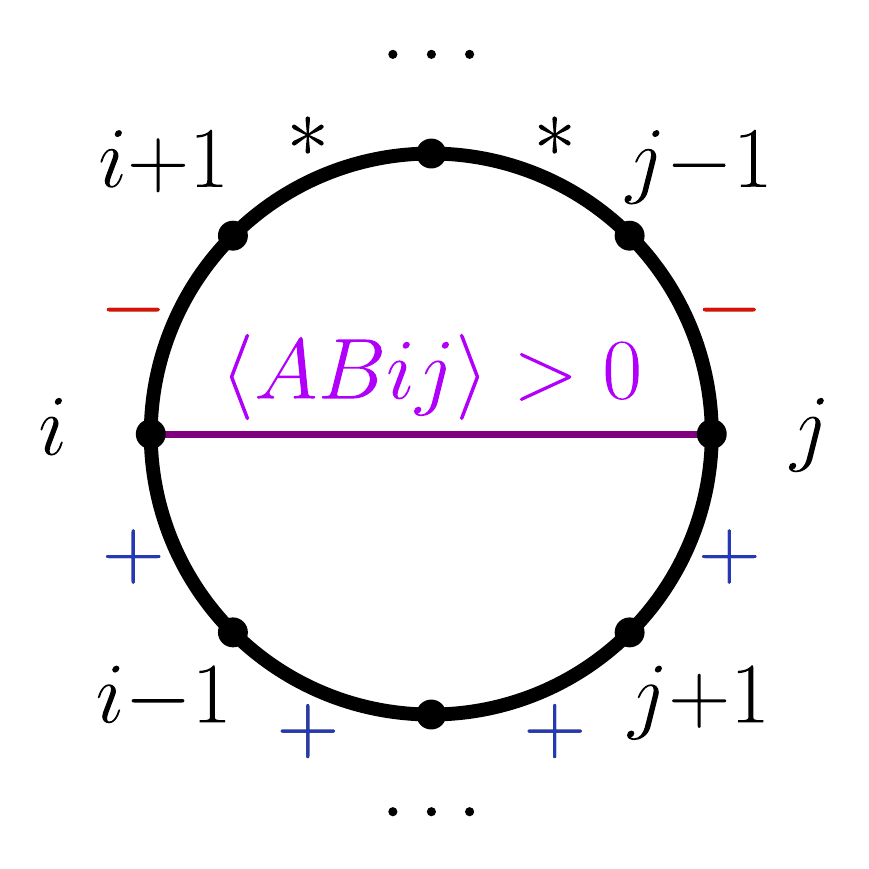}}
\,\,\,\leftrightarrow\,\,\,
\raisebox{-40pt}{\includegraphics[scale=.5]{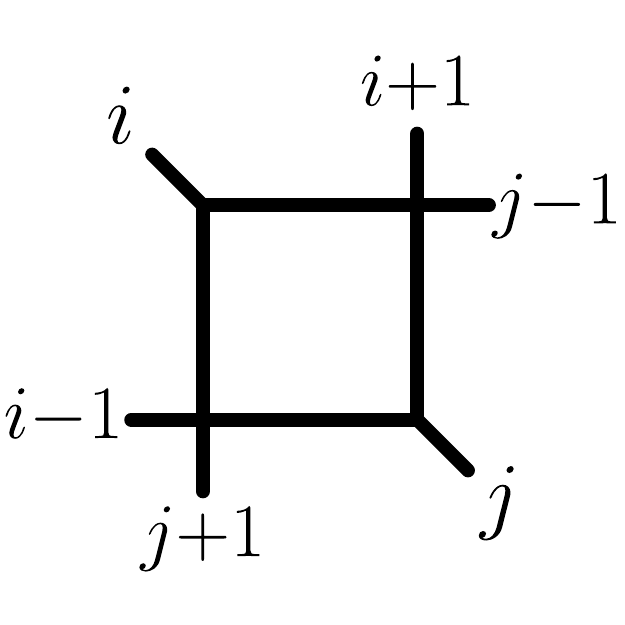}}. 
\end{equation}
This fact can now be used in a ``completeness relation'' to determine the canonical form for $S^{(2),+}_{ij}$. If we expand eq.~(\ref{compl}) in terms of regions with definite signs we encounter the sign-flip-two region whose form we want to determine, a collection of sign-flip-four regions, e.g.,
\begin{align}
\raisebox{-45pt}{\includegraphics[scale=.4]{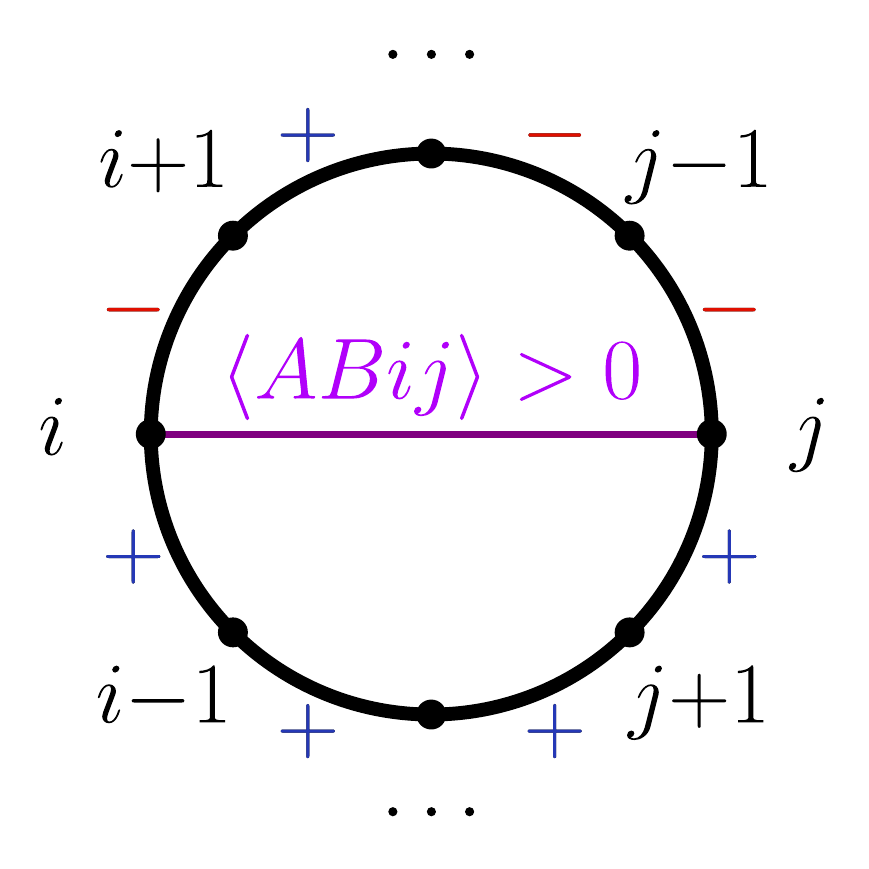}},
\end{align}
together with a number of sign-flip-six and higher regions which are empty. From this we can express the sign-flip-two region in terms of eq.~(\ref{compl}) and sign-flip-four regions\footnote{In the achiral sign-flip-four space, $\ab{ABij}>0$ is automatically satisfied once we impose the `external' inequalities. More details about these spaces and their fixed brackets are summarized in appendix \ref{app:fixed_signs_sf2_sf4_spaces}.}
\begin{equation}
\label{eq:chiral_sf2_ext_triangulation}
\hspace{-1cm}
\raisebox{-50pt}{\includegraphics[scale=.45]{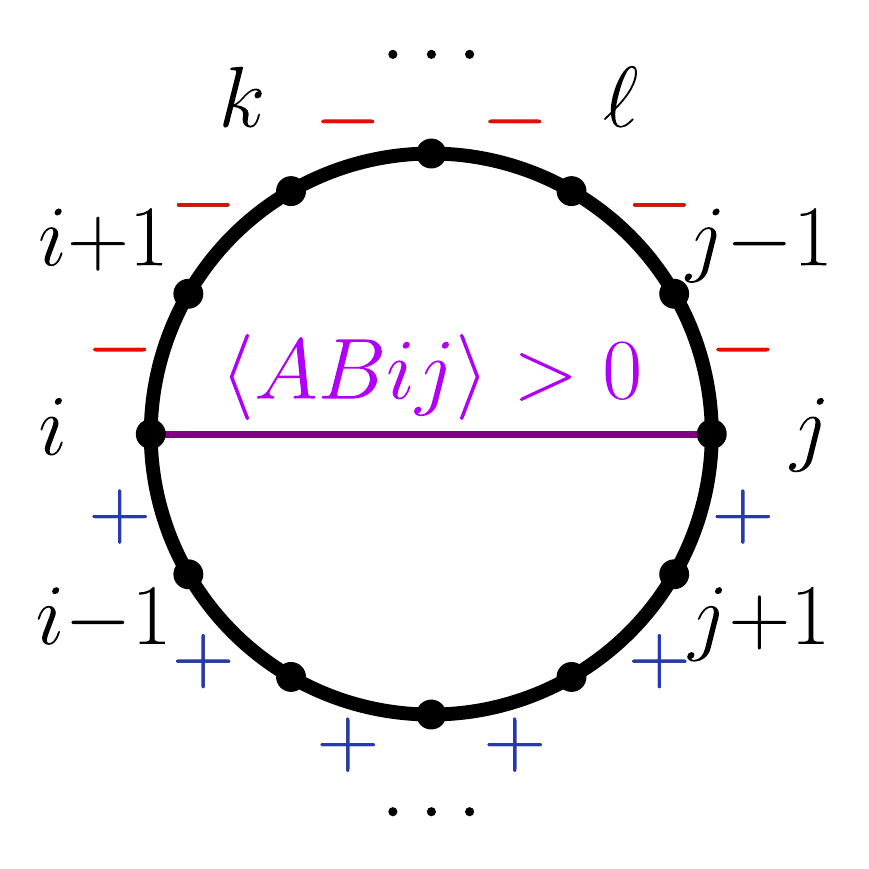}}
=
\raisebox{-50pt}{\includegraphics[scale=.45]{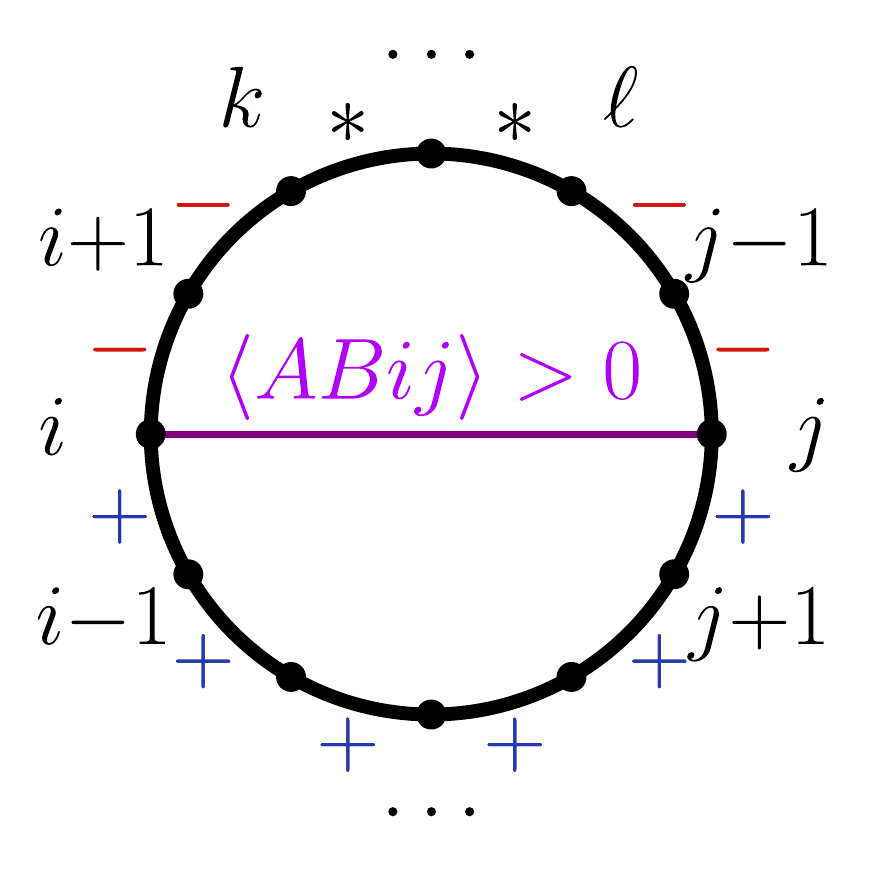}}
-
\hspace{-.5cm}
\raisebox{-50pt}{\includegraphics[scale=.45]{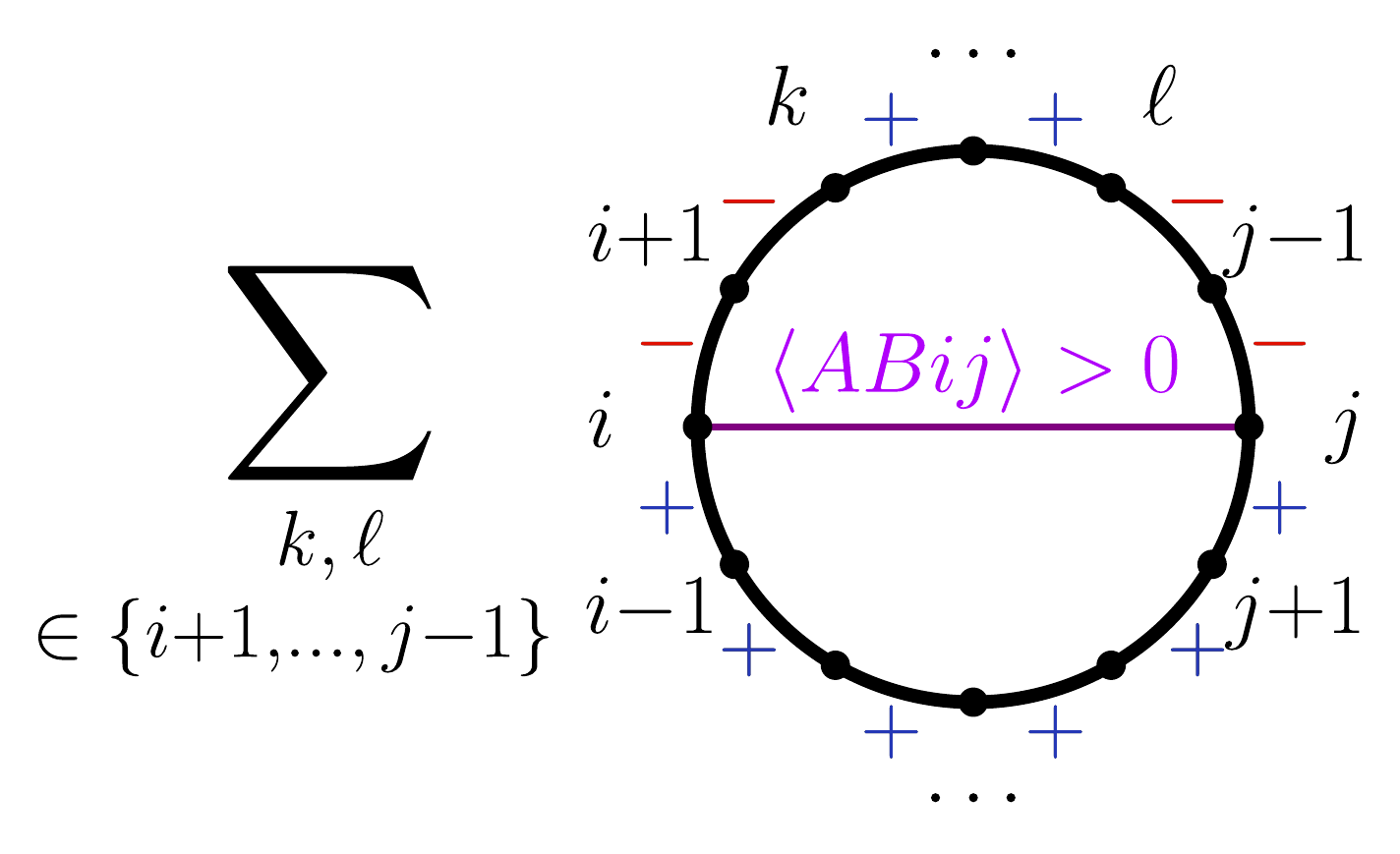}}
\end{equation}
Note that this is an external triangulation as also indicated by the minus sign between the two terms on the right-hand-side of eq.~(\ref{eq:chiral_sf2_ext_triangulation}). Geometrically we remove sign-flip-four regions from eq.~(\ref{compl}) leaving us with the chiral sign-flip-two region of interest. Since we already found all canonical forms associated to the regions on the right-hand-side of eq.~(\ref{eq:chiral_sf2_ext_triangulation}), we can immediately write down the canonical form of $\Omega^{(2),+}_{ij}$,
\begin{equation}
\label{eq:sf2_general_local_int_form_rep}
\hspace{-.5cm}
\Omega^{(2),+}_{ij}
=
\hspace{-.5cm}
\raisebox{-35pt}{\includegraphics[scale=.4]{./figures/2mass_easy_box.pdf}}
+
\hspace{-.7cm}
\raisebox{-50pt}{\includegraphics[scale=.55]{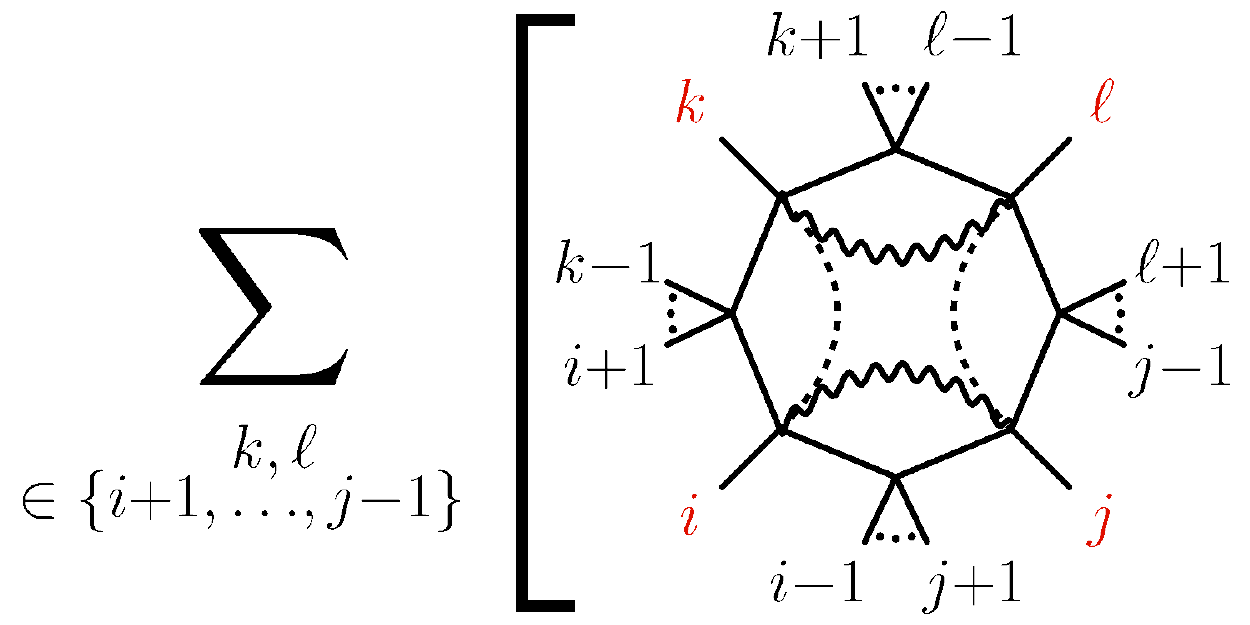}}
\ 
-
\ 
\raisebox{-50pt}{\includegraphics[scale=.55]{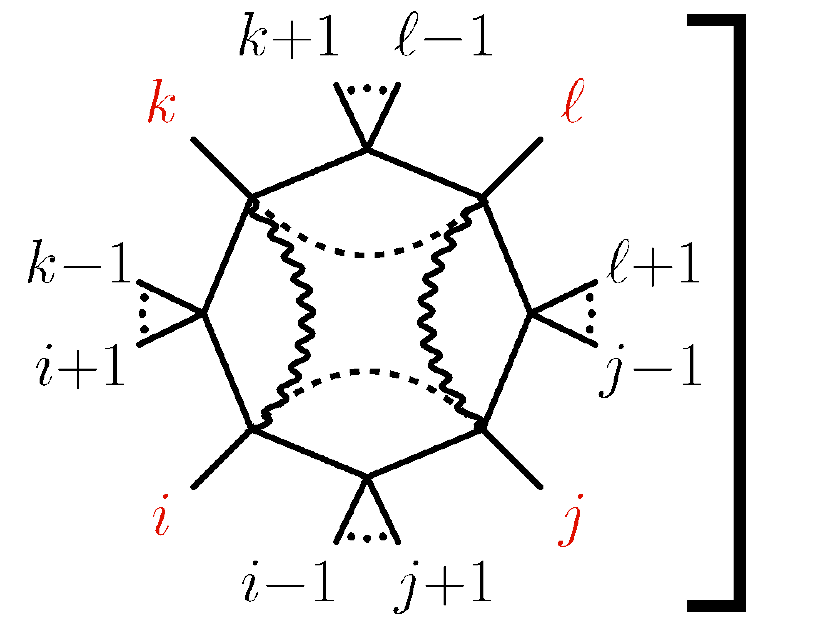}}
\end{equation}
where the sum over octagons includes all degenerations. Using this formula for the six-point example we find 
\begin{align}
\begin{split}
   & \raisebox{-45pt}{\includegraphics[scale=.4]{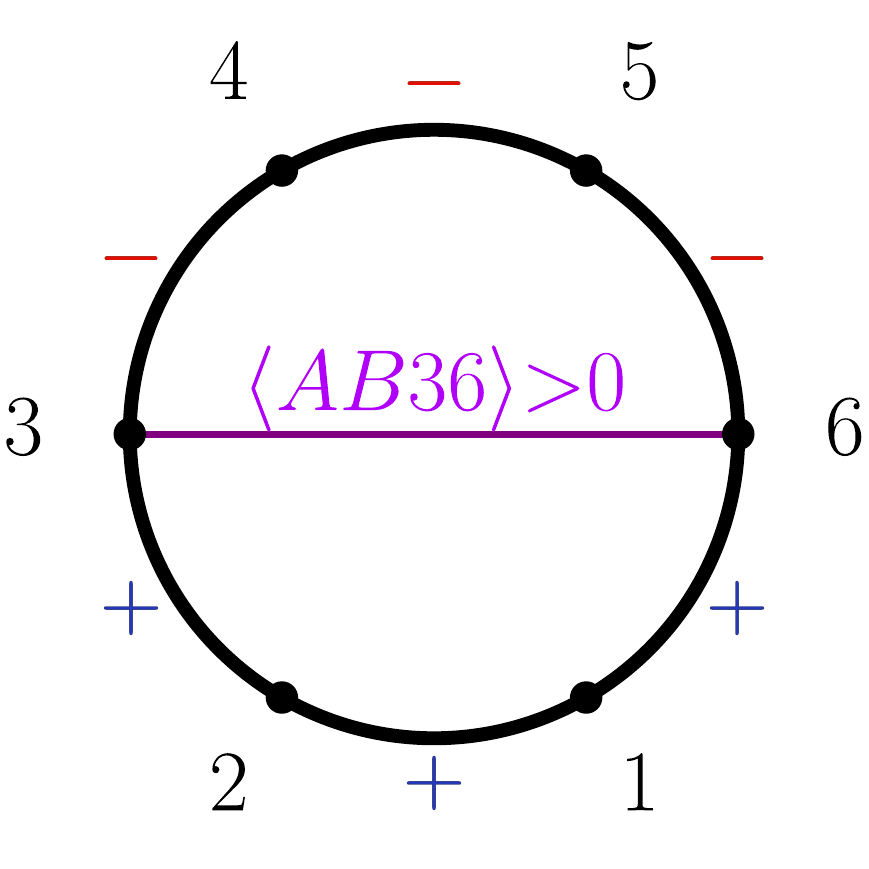}}
    =
    \raisebox{-45pt}{\includegraphics[scale=.4]{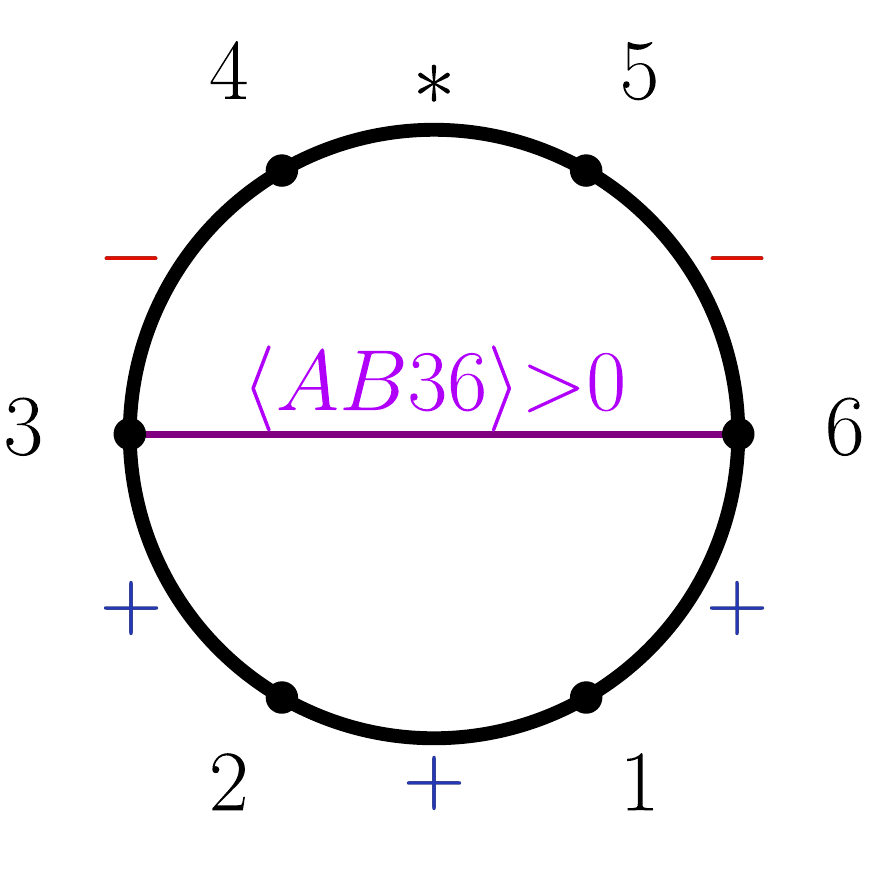}}
    -
    \raisebox{-45pt}{\includegraphics[scale=.4]{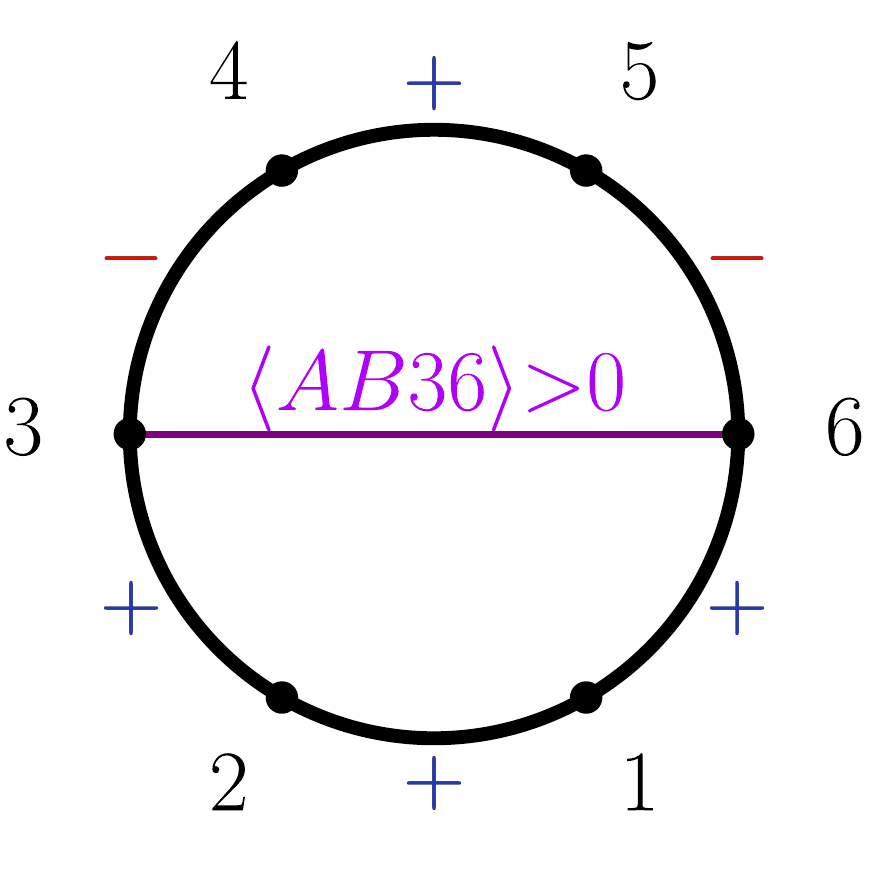}}\\
    &\Omega^{(2),+}_{36}=
     \raisebox{-38pt}{\includegraphics[scale=.4]{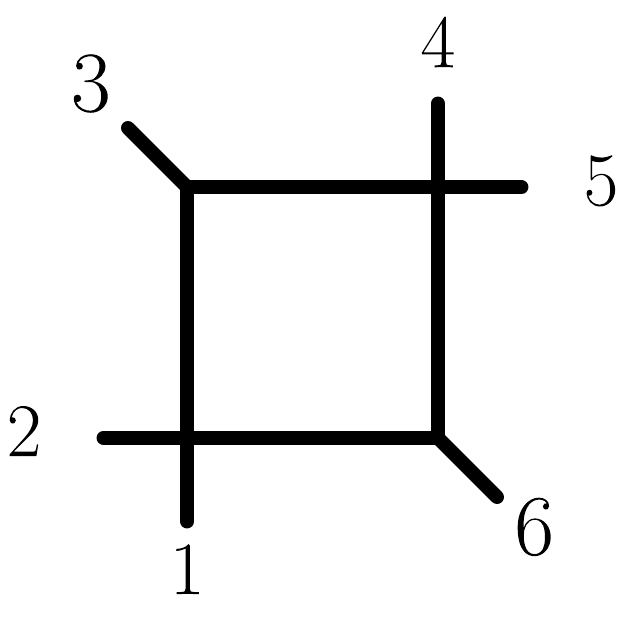}}
    +
     \raisebox{-45pt}{\includegraphics[scale=.4]{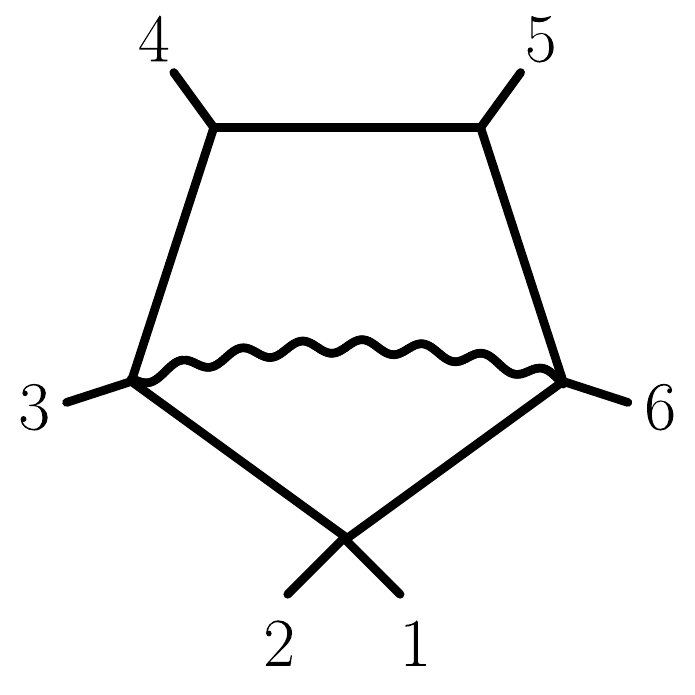}}
    -
    \raisebox{-45pt}{\includegraphics[scale=.4]{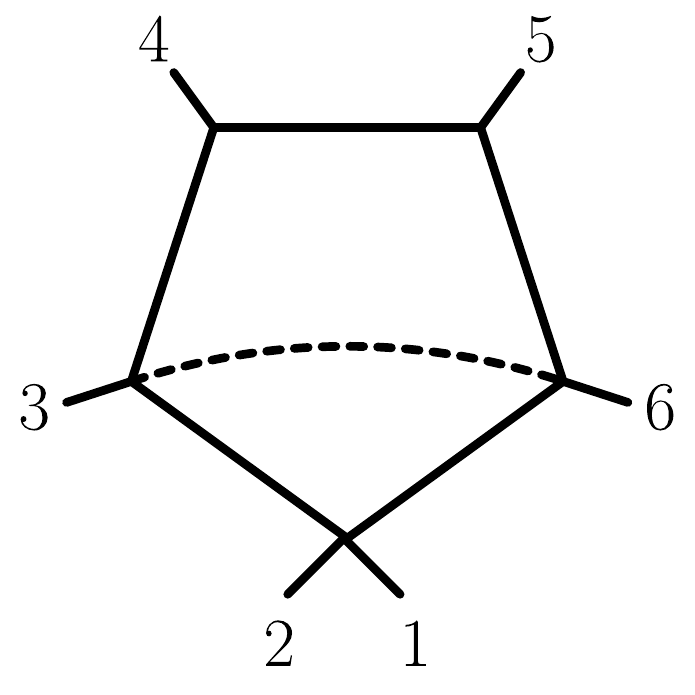}}
\end{split}
\end{align}
As we discussed earlier in section \ref{sec:sign_flip_regions}, there are a few special cases of sign-flip-two regions. The most special case is the chiral sign-flip-two region with only a single minus sign eq.~(\ref{eq:sf2_special_chiral_region_single_minus}) which is an empty space with vanishing canonical form.

In this degenerate case, for the opposite chirality $S_{ij}^{(2),-}$, we can triangulate the space using exactly the same procedure, only with different boundary data: the region where all $n-3$ `+' signs are replaced by `$\pm$' has vanishing form (as it only has three boundaries) and also the sign-flip-four regions are achiral rather than chiral. Therefore, the forms which appear on the right hand side are parity odd combinations of octagons (and descendants).

\subsection*{Triangulation of sign-flip-zero region}

As argued earlier in section \ref{subsec:no-go-local-triangulation}, there is no external triangulation of MHV or $\MHVbar$ Amplituhedra in terms of simple building blocks. However, this is not true for the achiral sign-flip-zero region eq.~(\ref{eq:sf0_achiral_form_abstract}) which is defined by $\ab{ABii{+}1}>0$ inequalities only. There are many ways to triangulate the $S^{(0)}$ region externally. The simplest (though certainly not the most efficient) is to fix four plus signs and marginalize over all other signs. In light of our earlier discussions, such a region has vanishing canonical form, but when we expand $* = + \oplus -$ we find the sign-flip-zero region of interest together with many sign-flip-two and four regions.
\begin{align}
\label{eq:sf0_achiral_triangulation}
\begin{split}
   \overbrace{
    \raisebox{-45pt}{\includegraphics[scale=.4]{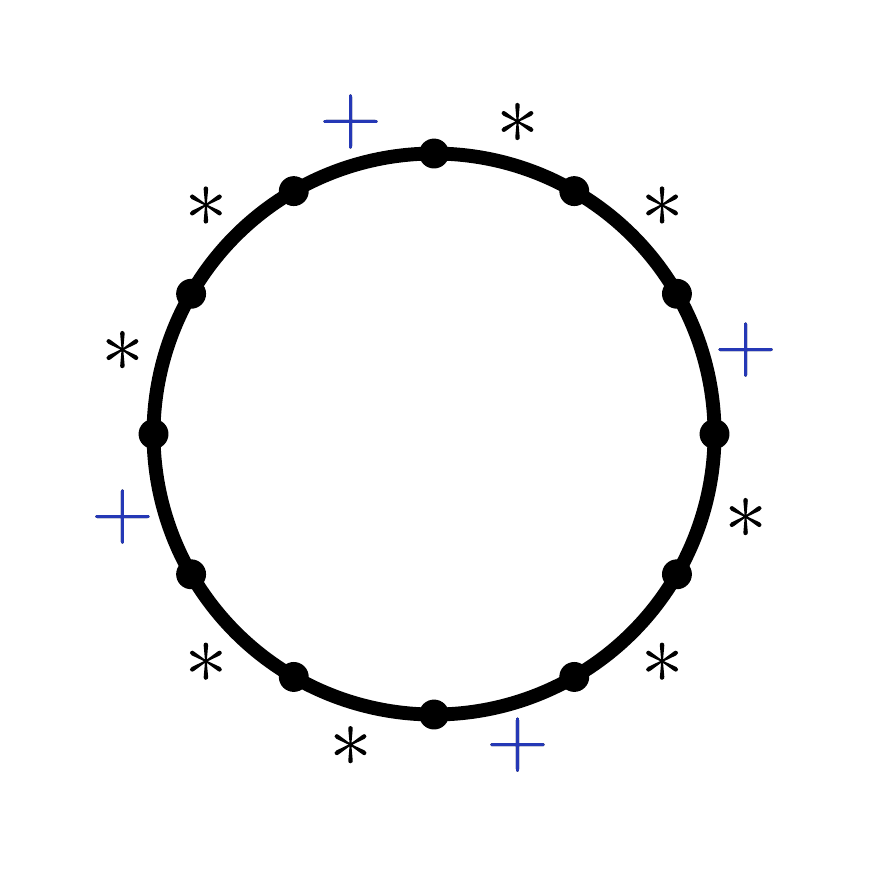}}}^{\text{zero-form space}}
    = & 
    \overbrace{
    \raisebox{-45pt}{\includegraphics[scale=.4]{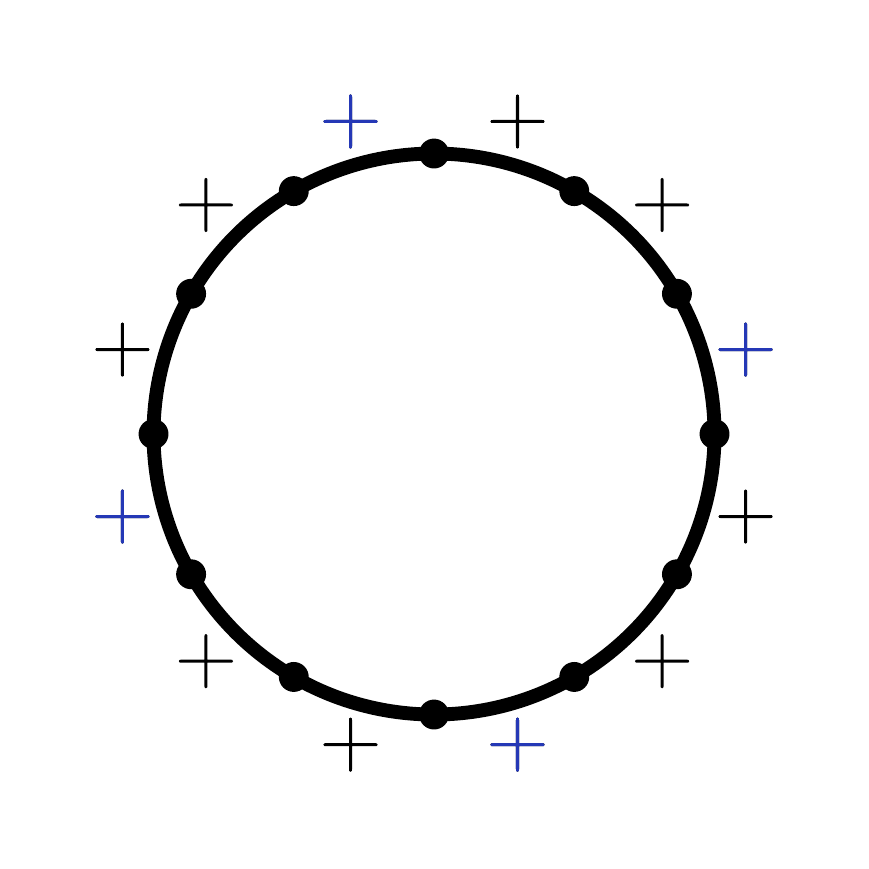}}}^{\text{achiral sf0 space}} \\[-15pt]
    + &
    \overbrace{
    \raisebox{-45pt}{\includegraphics[scale=.4]{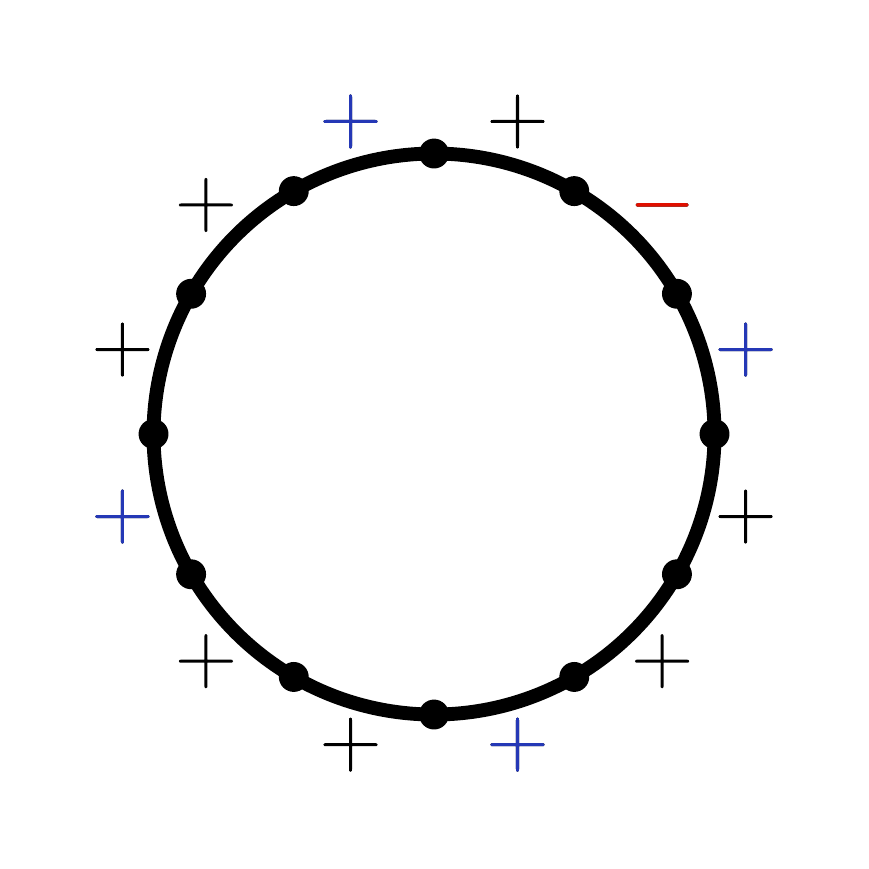}}
    \  + \  
    \raisebox{-45pt}{\includegraphics[scale=.4]{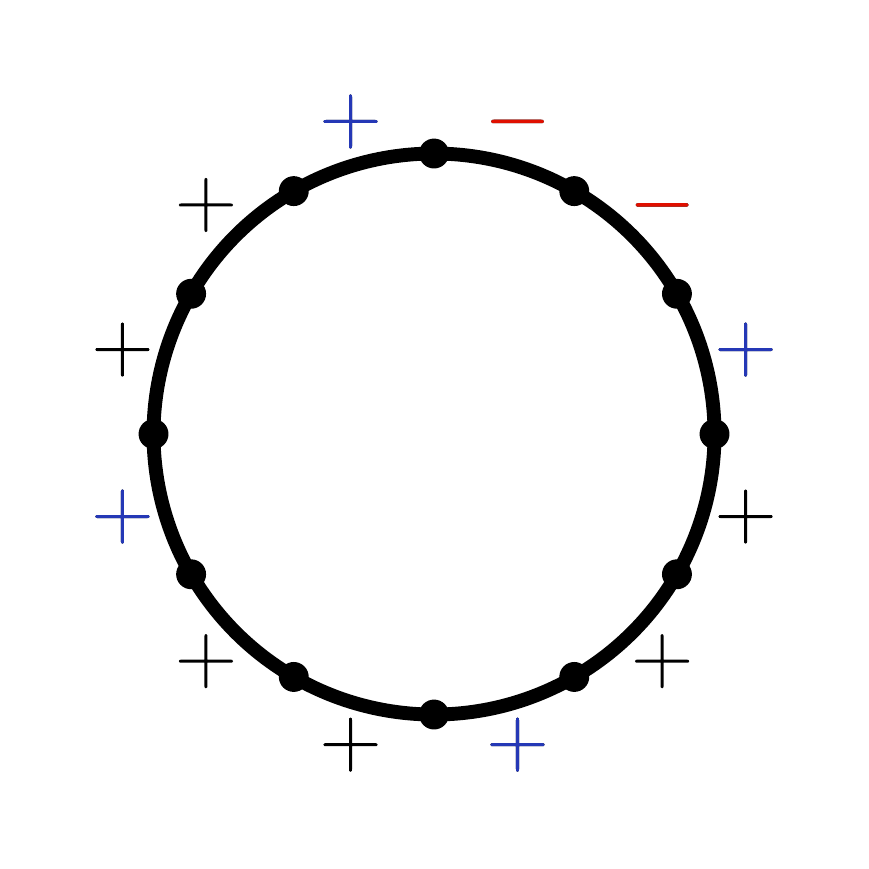}} \  + \ \cdots}^{\text{achiral sf2 spaces}}  \\[-15pt]
    + & 
    \overbrace{
    \raisebox{-45pt}{\includegraphics[scale=.4]{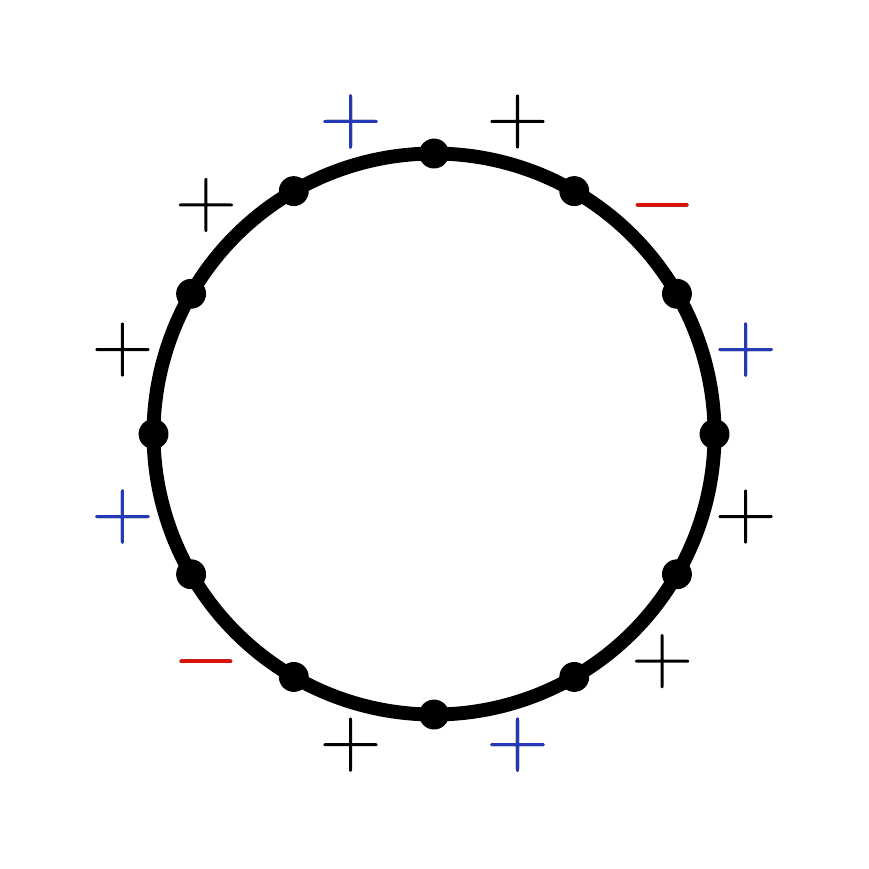}}
    \  + \  
    \raisebox{-45pt}{\includegraphics[scale=.4]{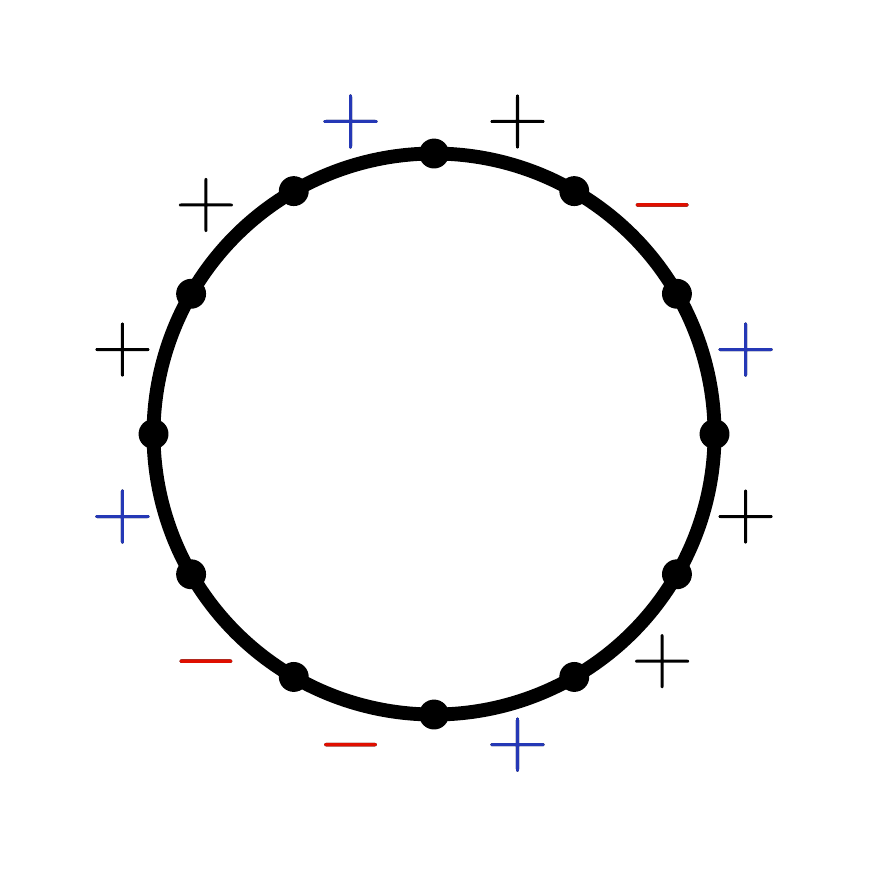}} \  + \ \cdots}^{\text{achiral sf4 spaces}}  \\[-15pt]
    + &
    \overbrace{
    \raisebox{-45pt}{\includegraphics[scale=.4]{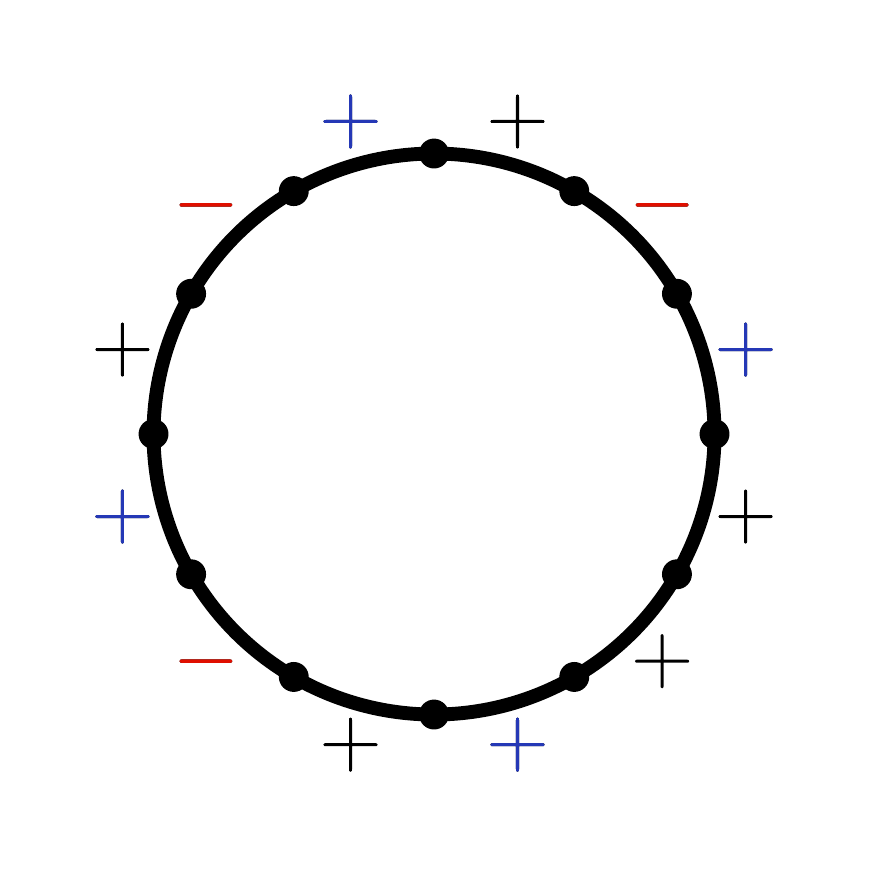}}
    \  + \  
    \text{other empty regions}}^{\text{empty sf6 and higher spaces with zero form}}.
\end{split}
\end{align}
We already calculated the necessary canonical forms for all regions appearing in eq.~(\ref{eq:sf0_achiral_triangulation}) before in eqs.~(\ref{eq:sf4_chiral_octagon_association}) and (\ref{eq:sf2_general_local_int_form_rep}), which allows us to write the form for the sign-flip-zero space $S^{(0)}$ eq.~(\ref{eq:sf0_achiral_form_abstract}) more explicitly in terms of known quantities. Note that all sign-flip-two and four regions are achiral, so we must use the relevant parity-odd forms associated to those spaces by combining both chiralities in eqs.~(\ref{eq:sf4_chiral_octagon_association}) and (\ref{eq:sf2_general_local_int_form_rep}).

Having discussed the canonical forms of all sign-flip spaces, we now revisit to the chiral pentagon expansion of eq.~(\ref{pent}), together with the parity-odd one-loop amplitude given as the difference of the MHV and $\MHVbar$ amplitudes. In the parity-odd case, the local expansion involves parity-odd pentagons, eq.~(\ref{odd_pent}), which can be associated to simple achiral spaces defined by only five inequalities (see our discussion earlier in this section) where the signs of these inequalities did not matter to get the correct parity-odd pentagon canonical form. The question is whether or not the parity-odd pentagon expansion eq.~(\ref{odd_pent}) can be understood geometrically as an external triangulation of some well-defined space. In other words: can we triangulate the sign-flip-zero space eq.~(\ref{eq:sf0_achiral_form_abstract}) not via eq.~(\ref{eq:sf0_achiral_triangulation}), but in terms of spaces with five boundaries only? The answer is yes, and in the following we give a straightforward description of such a triangulation:
\begin{enumerate}
    \item We start with the sign-flip-zero region $S^{(0)}$ and triangulate it externally via the space with the sign of $\ab{AB12}$ marginalized,
    \begin{align}
    \raisebox{-45pt}{\includegraphics[scale=.4]{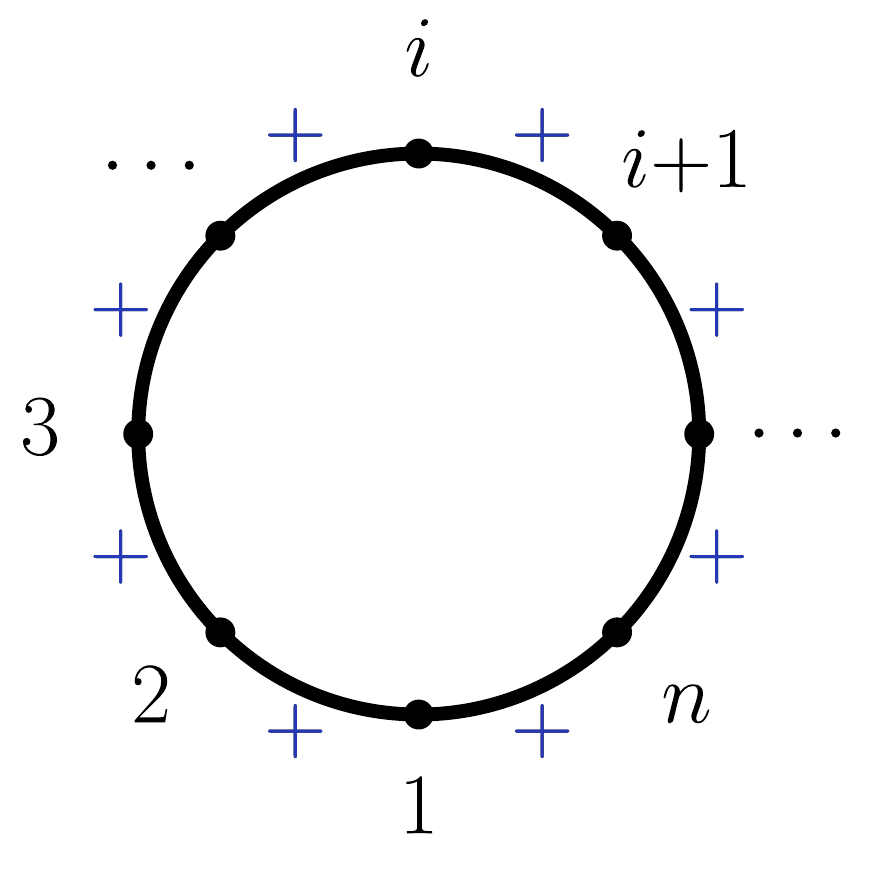}}
    =
    \raisebox{-45pt}{\includegraphics[scale=.4]{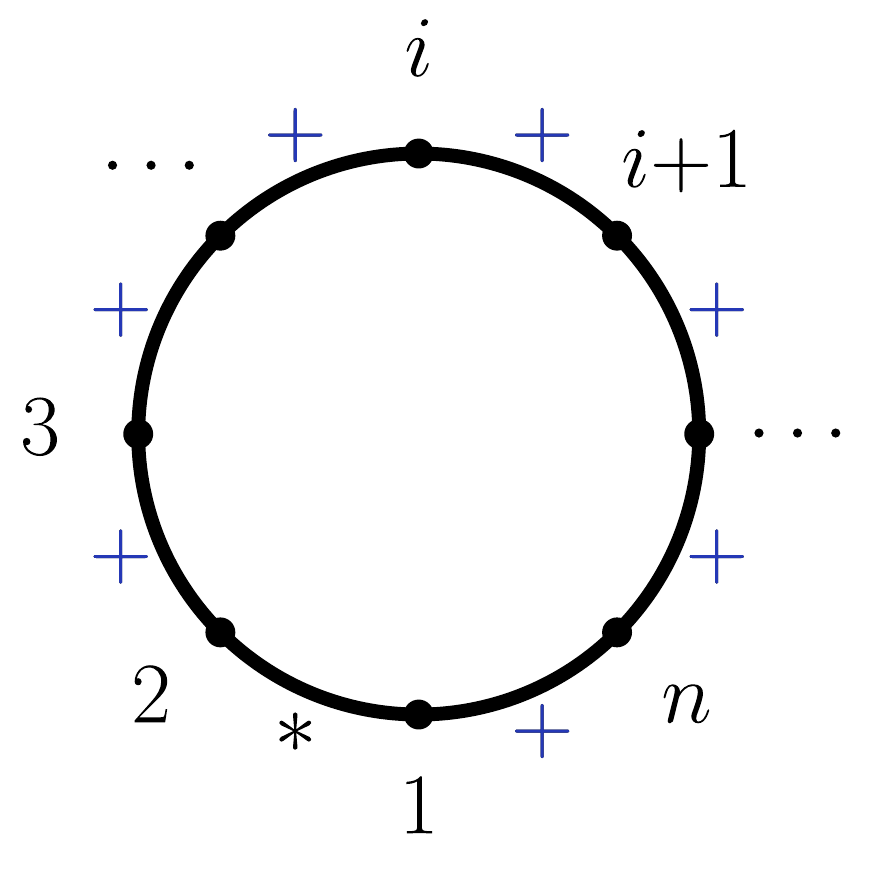}}
    -
    \raisebox{-45pt}{\includegraphics[scale=.4]{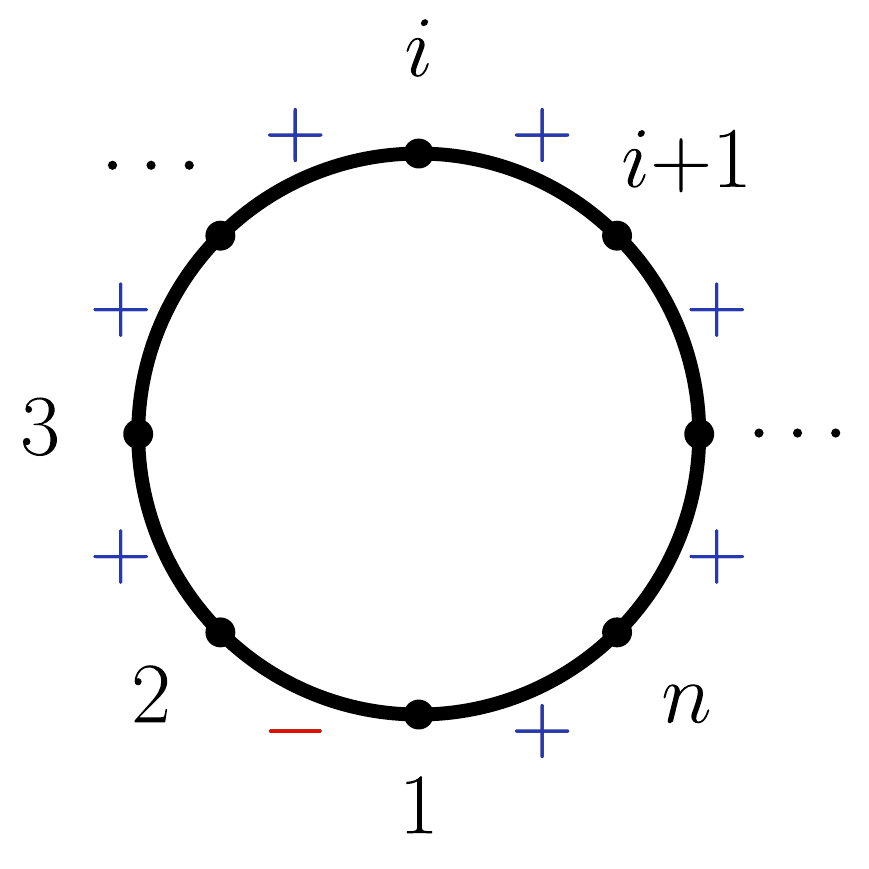}}
    \end{align}
    \item For the region with $\ab{AB12}=\ast$ (i.e., $\ab{AB12}{\gtrless}0$ does not have a fixed sign), we continue by marginalizing over the sign of $\ab{AB23}$ leaving us with two spaces, one where $\ab{AB23}=\ast$, and one where $\ab{AB23}<0$. 
    \begin{align}
        \raisebox{-45pt}{\includegraphics[scale=.4]{./figures/sf0_triangulation_step1_2.pdf}}
        =
        \raisebox{-45pt}{\includegraphics[scale=.4]{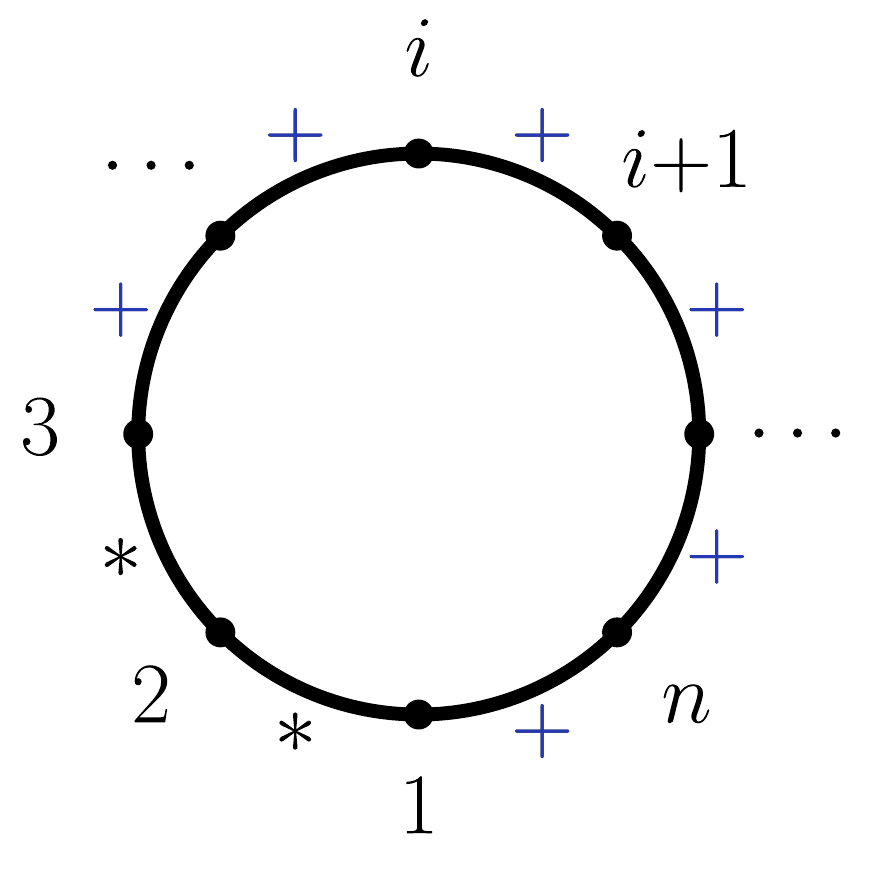}}
        -
        \raisebox{-45pt}{\includegraphics[scale=.4]{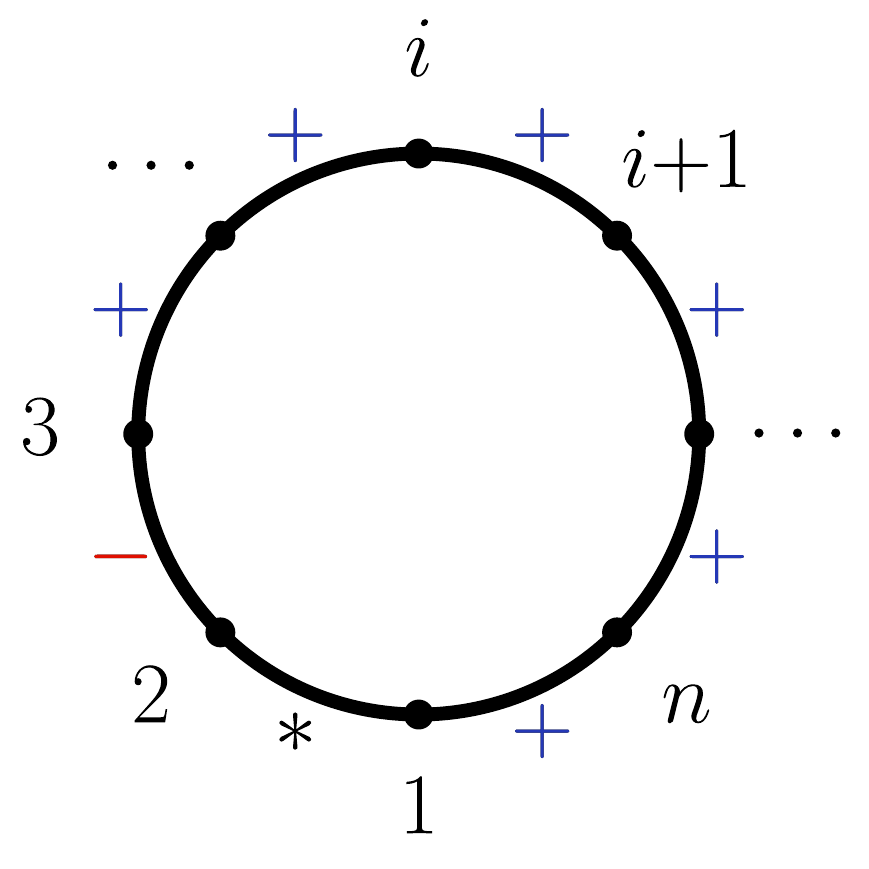}}
    \end{align}
    Whenever we encounter a space with a minus sign we stop, if we have $*$'s only we continue. This procedure results in a collection of spaces $S_i$ defined by  
    \begin{align}
    \begin{split}
        S_i :=
         \{&\ab{AB12}{\gtrless} 0, 
          {\ldots}, 
          \ab{ABi{-}2i{-}1}{\gtrless}0,
          \ab{ABi{-}1i}\tred{{<}0},\\
          &\ab{ABii{+}1}\tblue{{>}0},
          {\ldots},
          \ab{AB1n}{>}0
        \},\qquad i\in\{2,{\ldots},n{-}3\}\,,
    \end{split}
    \end{align}
    where, in the boundary case $i=2$, we start with $\ab{AB12}\tred{<0}$. Pictorially, $S_i$ is represented by the following circle diagram
    \begin{align}
     S_{i} = \raisebox{-45pt}{\includegraphics[scale=.4]{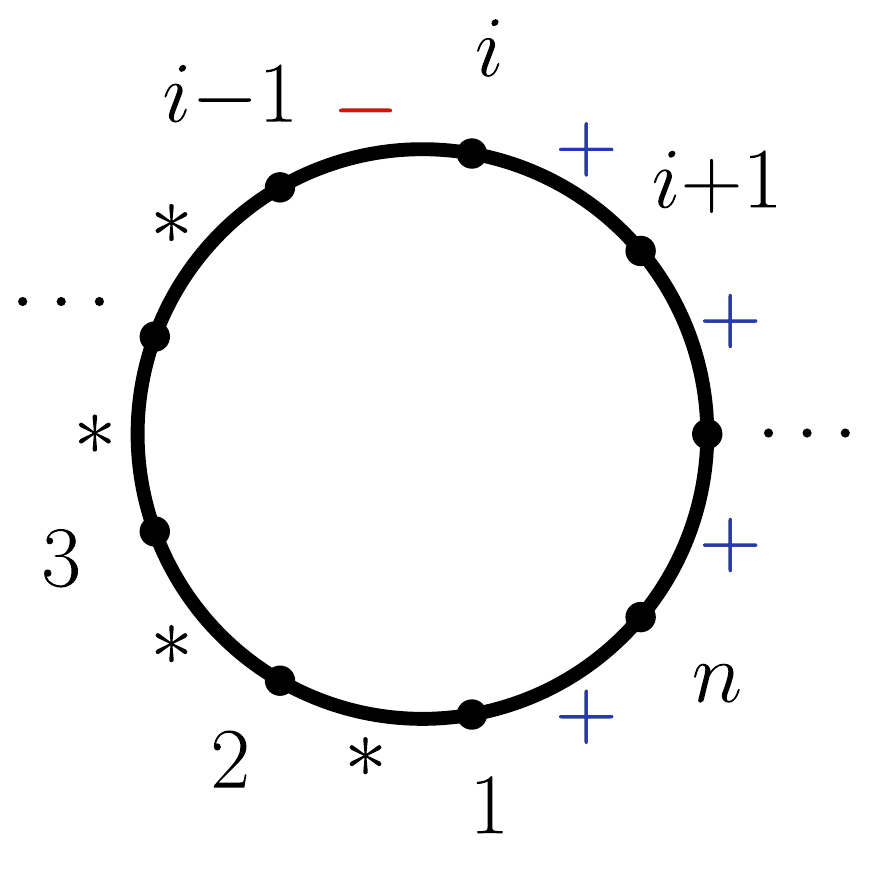}}\,.
    \end{align}
    This procedure stops at $i=n-3$ because we reach the end of the circle. Going beyond this point simply generates spaces with less than five boundaries that have vanishing form (and are therefore irrelevant for the purpose of obtaining the canonical form).
    \item In the third step, we continue the same procedure for each space $S_i$ but leave $\ab{ABii{+}1}>0$ untouched, marginalizing over $\ab{ABi{+}1i{+}2}$ 
     \begin{align}
      \raisebox{-45pt}{\includegraphics[scale=.4]{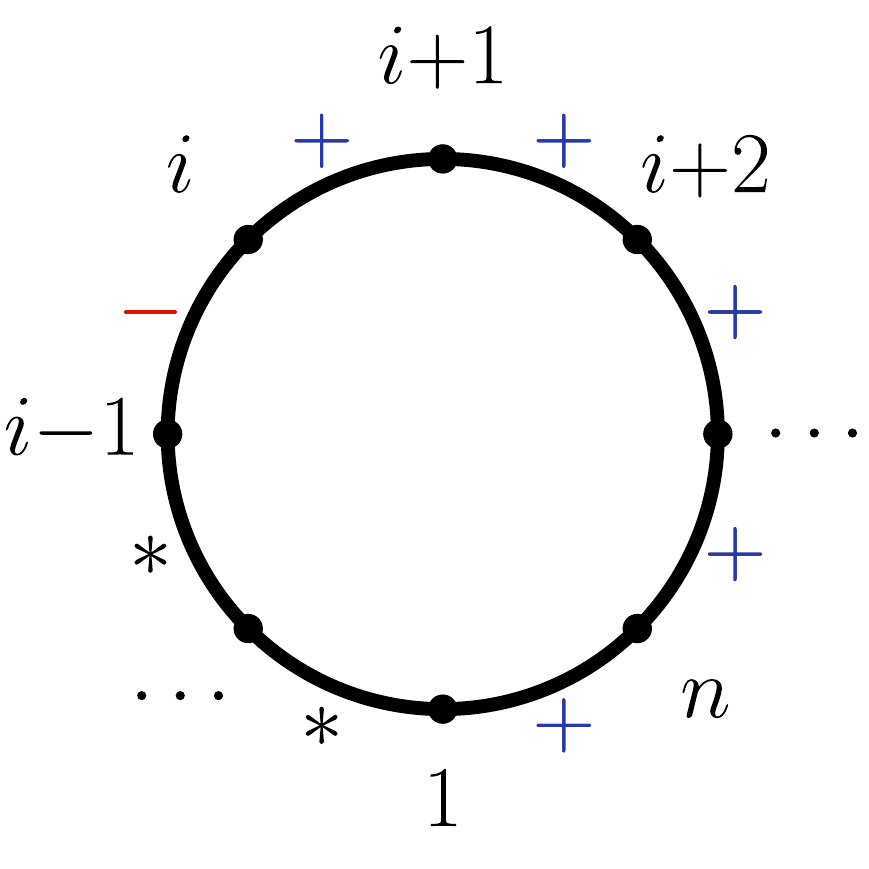}}
        =
      \raisebox{-45pt}{\includegraphics[scale=.4]{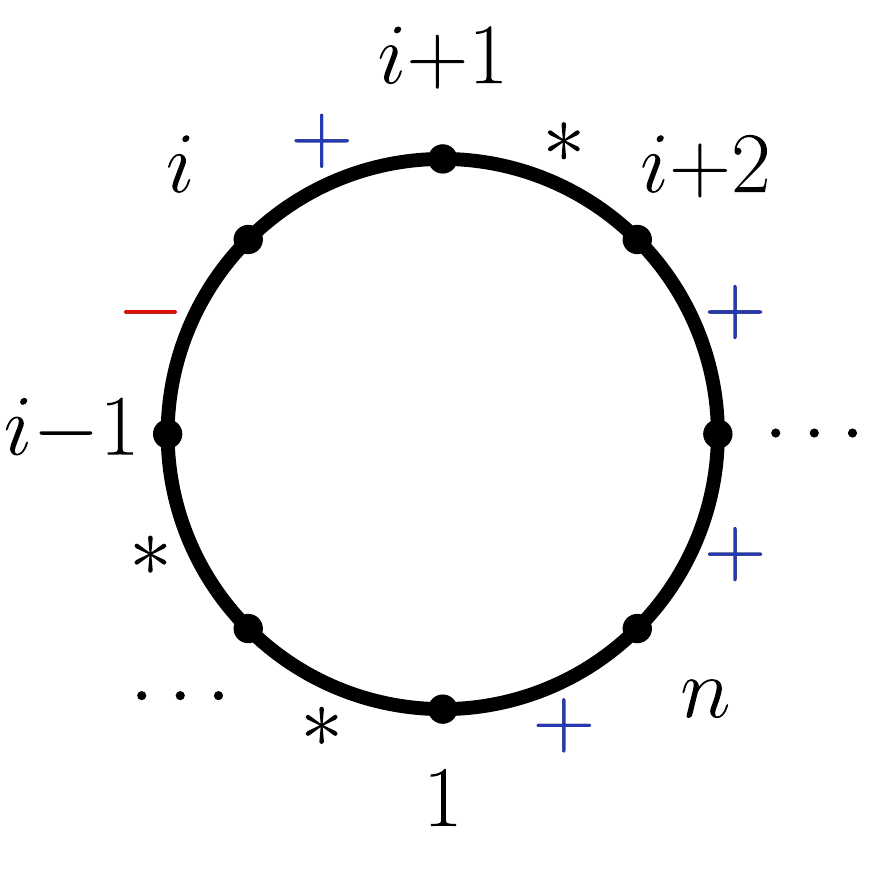}}
        -
      \raisebox{-45pt}{\includegraphics[scale=.4]{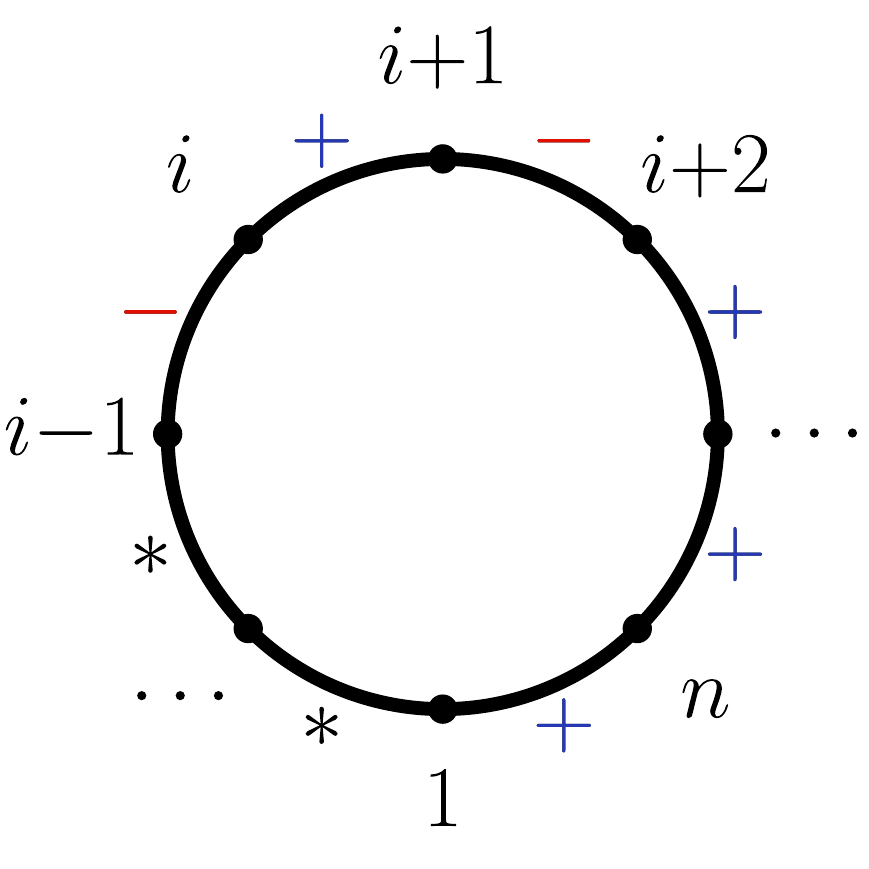}}\,.
    \end{align}
    We again keep the spaces with two minus signs and marginalize further over the spaces where we have $*$. As a result we generate a collection of spaces $S_{ij}$,
    \begin{align}
     S_{ij}= \raisebox{-45pt}{\includegraphics[scale=.5]{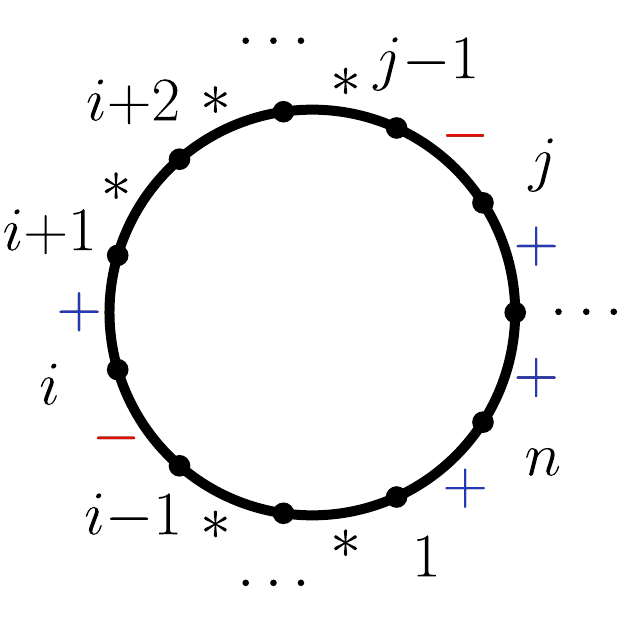}}
     \,,\quad
     i\in\{2,\ldots,n{-}3\},\,\,j\in\{i{+}2,\ldots,n{-}1\}.
    \end{align}
    For each region $S_{ij}$ the signs of $\{\ab{ABjj{+}1},\ldots \ab{AB1n}\}$ are all positive. However, we can freely replace all but the $\ab{ABjj{+1}}$ and $\ab{AB1n}$ positive signs by $+ \rightarrow \ast$. This is because $S_{ij}$ already has two non-adjacent minus signs; therefore, spaces where we introduce additional minus signs leads to empty sign-flip-six (or higher) regions, e.g.,
    \begin{align}
    \label{eq:sf0_triang_sij_empty_eg}
    \text{example of an empty region:}
    \qquad
    \raisebox{-45pt}{\includegraphics[scale=.5]{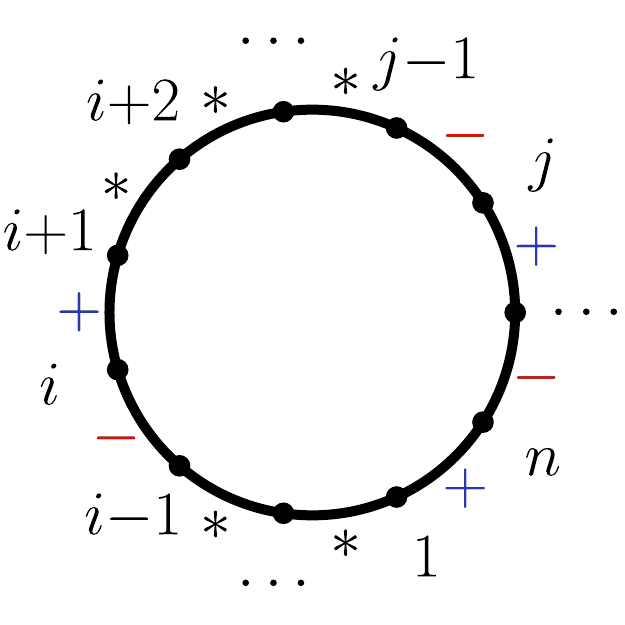}}.
    \end{align}
\end{enumerate}
As a result, we triangulate the sign-flip-zero region $S^{(0)}$ in eq.~(\ref{eq:sf0_achiral_form_abstract}) in terms of spaces with five boundaries only,
\begin{align}
\label{eq:sf0_odd_pentagon_triangulation_circles}
    \raisebox{-45pt}{\includegraphics[scale=.4]{./figures/sf0_triangulation_lhs}}
    =
    \raisebox{-45pt}{\includegraphics[scale=.4]{./figures/sf0_triangulation_rhs}},
\end{align}
which, at the level of canonical forms, corresponds to the parity-odd pentagon expansion eq.~(\ref{odd_pent}). Note that on the right-hand side of eq.~(\ref{eq:sf0_odd_pentagon_triangulation_circles}) we are suppressing spaces with zero-form which are necessary for the geometric triangulation, but unnecessary for the purposes of computing the canonical form. Geometrically, each space appearing in eq.~(\ref{eq:sf0_odd_pentagon_triangulation_circles}) is equivalent to that of a pentagon where we forget about the signs of the last two brackets in eq.~(\ref{pentsign2}) (see also eq.~(\ref{eq:odd_pent_sf_form})). The zero-form spaces implicit in eq.~(\ref{eq:sf0_odd_pentagon_triangulation_circles}) demonstrates that while we were successful in interpreting the pentagon expansion eq.~(\ref{odd_pent}) geometrically, spaces with vanishing form had to be added in order to construct a full triangulation. 

The natural question is why we can not use the same procedure to triangulate the MHV or $\MHVbar$ Amplituhedra. The problem arises due to additional inequalities (such as $\la AB\overline{ij}\ra>0$ in eq.~(\ref{eq:mhvBar_alt})) that are required in the definition of the MHV Amplituhedron. While these additional inequalities do not correspond to any boundaries in the original MHV space, if we attempt to externally triangulate the MHV space by marginalizing over the signs of $\ab{ABii{+}1}$ brackets as $+ \rightarrow*$, the resulting spaces now have $\la AB\overline{ij}\ra$ as boundaries. The corresponding canonical forms have spurious poles which can only be cancelled by adding additional spaces. Thus, this simple marginalization procedure does not extend straightforwardly to the chiral component spaces relevant for the MHV and $\MHVbar$ Amplituhedra.

\newpage
\section{Fixed signs in sign-flip-zero, two and four spaces}
\label{app:fixed_signs_sf2_sf4_spaces}

In this appendix, we discuss in more detail the sign-flip regions of section~\ref{sec:sign_flip_regions} and \ref{sec:external_triangulations}. As shown in the main text, each achiral sign-flip region is defined by imposing fixed signs for certain  $\ab{AB ii{+}1}$ brackets only. There is a nice hierarchy of these spaces based on the number of sign-flips in the $\{\ab{ABii{+}1}\}$ sequence. The only non-empty spaces are the ones with zero, two, or four sign flips. As we argued in (\ref{sec:sign_flip_regions}), the sign-flip-zero space is the most complicated, as measured by the number of boundaries and the complexity of the associated canonical form eqs.~(\ref{eq:sf0_achiral_form_abstract}) and (\ref{eq:sf0_achiral_triangulation}). The complexity of the sign-flip-two spaces is reduced (see section~\ref{sec:sign_flip_regions}), and even more so for the sign-flip-four regions of section~\ref{sec:sign_flip_regions} which have at most eight boundaries. The corresponding logarithmic form is linked to the chiral octagon integrals, eq.~(\ref{eq:sf4_chiral_octagon_association}), introduced in \cite{ArkaniHamed:2010gh}. Besides the achiral spaces alluded to above, each geometry can be cut into two chiral components by imposing further constraints on additional brackets $\ab{ABX} \gtrless 0$. 

Let us start our exposition with the achiral sign-flip-zero region $S^{(0)}$, eq.~(\ref{eq:sf0_achiral_form_abstract}),
\begin{equation}
\label{eq:sf0_def_app}
  S^{(0)}:\,\la ABii{+}1\ra>0,\quad i=1,\dots,n 
  \quad \leftrightarrow \quad
  \raisebox{-45pt}{\includegraphics[scale=.4]{./figures/sign_flip_zero.pdf}}
\end{equation}
No other signs of brackets are fixed inside $S^{(0)}$ besides the ones indicated in eq.~(\ref{eq:sf0_def_app}). We can cut this achiral region into two chiral components, which correspond to MHV and $\MHVbar$ one-loop amplitudes by imposing the inequalities of eq.~(\ref{eq:mhv_alt}) and eq.~(\ref{eq:mhvBar_alt}), respectively.  We denote the corresponding chiral spaces by $S^{(0)}_{\text{MHV}}$ and $S^{(0)}_{\MHVbar}$. Concretely, the $\MHVbar$ component $S^{(0)}_{\MHVbar}$ can be defined by imposing additional $n{-}3$ conditions eq.~(\ref{eq:mhvBar_alt}),
\begin{equation}
\label{eq:sf0_chiral_mhvbar}
  S^{(0)}_{\MHVbar}:\, 
  \left\{
  \begin{array}{c}
    \ab{ABi i{+}1}>0 \quad   i \in (1,\ldots, n{-}1) \\
    \ab{AB1i}>0      \quad   \ \ i \in (3,\ldots, n{-}1) \\
  \end{array}
  \right\}
  \quad \leftrightarrow \quad
  \raisebox{-45pt}{\includegraphics[scale=.4]{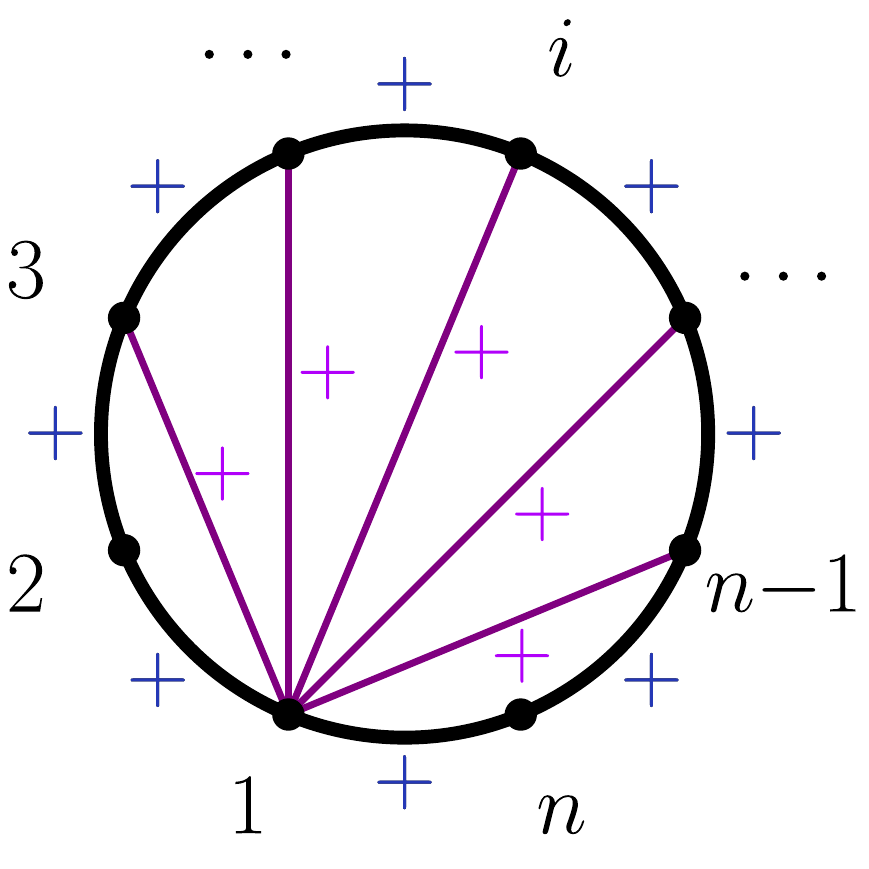}}
\end{equation}
Alternatively, imposing fixed signs for any sequence of brackets $\{\ab{AB2i}>0\}$ etc. leads to an equivalent definition of $S^{(0)}_{\MHVbar}$. While a given set of only $n-3$ fixed sings for the additional brackets (such as $\{\ab{AB2i}>0\}$ in eq.~(\ref{eq:sf0_chiral_mhvbar})) is sufficient to define the $\MHVbar$ region, in fact \emph{all} brackets $\la ABij\ra>0$ (for $i<j$) are fixed inside the $S^{(0)}_{\MHVbar}$ region. Let us note that no signs of any other brackets are fixed inside $S^{(0)}_{\MHVbar}$.

The MHV chiral component is traditionally defined by a certain sign-flip constraint on the sequence $\{\la AB1i\ra\}$, see the $k=0$ instance of eq.~(\ref{eq:abIneq}), but we can equivalently impose the $n{-}3$ conditions of eq.~(\ref{eq:mhv_alt}), $ \{\ab{AB\overline{1i}}>0\}\,, \  i \in (3,\ldots, n{-}1)$
\begin{equation}
\label{eq:sf0_chiral_mhv}
S^{(0)}_{\text{MHV}}:\, 
  \left\{
  \begin{array}{c}
    \ab{ABi i{+}1}>0 \quad   i \in (1,\ldots, n{-}1) \\
    \ab{AB\overline{1i}}>0      \quad   \ \ i \in (3,\ldots, n{-}1) \\
  \end{array}
  \right\}
  \quad \leftrightarrow \quad
    \raisebox{-45pt}{\includegraphics[scale=.4]{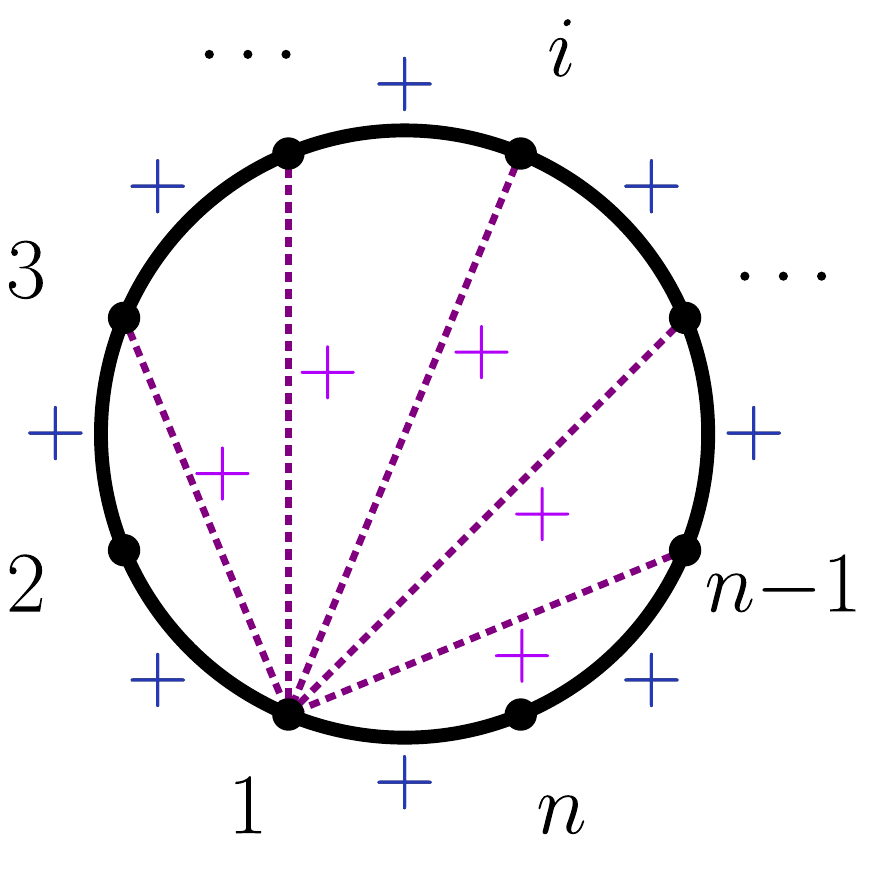}},
\end{equation}
where we used dotted lines in the circle figure eq.~(\ref{eq:sf0_chiral_mhv}) to denote the positivity of the respective $\ab{AB\overline{1i}}$. While this set of signs suffices to fix the MHV region, in fact all $\ab{AB\overline{ij}}{>}0$ (for $i{<}j$) are positive inside $S^{(0)}_{\text{MHV}}$, but no other signs of brackets are fixed. \\

The achiral sign-flip-two space $S^{(2)}_{ij}$ of eq.~(\ref{eq:sf2_achiral_abstract_def}) is defined by a set of inequalities on $\la ABii{+}1\ra$ brackets only that we graphically represent as
\begin{align}
\label{eq:sf2_achiral_app}
 S^{(2)}_{ij} \quad \leftrightarrow \quad 
 \raisebox{-45pt}{\includegraphics[scale=.4]{./figures/sf2_chiral_components_no_chord.pdf}}   
\end{align}
%
No other signs are fixed in $S^{(2)}_{ij}$. Compared to the sign-flip-zero space where we needed to impose $n{-}3$ additional signs to chiralize the space, the achiral sign-flip-two region can be cut into two chiral components by fixing a single sign of the bracket $\la ABij\ra$. This bracket is extremely natural and involves the two positions $i$ and $j$ where the two sign-flips occur,
\begin{align}
\begin{split}
& S^{(2),+}_{ij}:\, \{\text{eq.}~(\ref{eq:sf2_achiral_app}), \ab{ABij}>0\} \leftrightarrow
\raisebox{-45pt}{\includegraphics[scale=.4]{./figures/sf2_plus_components.pdf}} 
\,, \qquad i<j,
\\[-8pt]
& S^{(2),-}_{ij}:\, \{\text{eq.}~(\ref{eq:sf2_achiral_app}),\ab{ABij}<0\} \leftrightarrow
\raisebox{-45pt}{\includegraphics[scale=.4]{./figures/sf2_minus_components.pdf}} 
\,, \qquad i<j.
\end{split}
\end{align}
In the $S^{(2),+}_{ij}$ region we have a \emph{fully positive index space} for certain brackets involving $(j, j{+}1,\ldots i{-}1,i)$. By this we mean that all signs of the following brackets are fixed to be positive: 
\begin{equation}
\label{eq:sf2_plus_fixed_signs}
S^{(2),+}_{ij}:\, 
\{\ab{ABpq}>0, \ab{AB\overline{pq}}>0\} \,, 
\text{for }
p<q \in \{i,i{+1},\ldots,j{-}1,j\},
\end{equation}
where the inequalities on $p,q$ are to be understood in the cyclic sense. In contrast, for $S^{(2),+}_{ij}$, arbitrary non-adjacent brackets of the ``negative region'' do not have fixed signs.  

Similar to the discussion for $S^{(2),+}_{ij}$, for $S^{(2),-}_{ij}$ we have
\begin{equation}
\label{eq:sf2_minus_fixed_signs}
S^{(2),-}_{ij}:\, \{\ab{ABpq}<0, \ab{AB\overline{pq}}<0\}, \quad, 
\text{for } p<q \in \{i,i{+1},\ldots,j{-}1,j\},
\end{equation}
in the \emph{fully negative region} involving $(i,i{+}1,\ldots,j{-}1, j)$ the same set of brackets are fixed to be negative and the non-adjacent brackets in the ``positive region'' do not have a fixed sign. Note that the sign-inequalities eq.~(\ref{eq:sf2_plus_fixed_signs}) and eq.~(\ref{eq:sf2_minus_fixed_signs}) also apply for $q=p{+}1$ where both $\ab{ABpq}$ and $\ab{AB\overline{pq}}$ collapse to $\ab{ABpp{+}1}$ (up to a positive bracket of external twistors). \\ 

The sign-flip-four space $S^{(4)}_{ik\ell j}$ eq.~(\ref{eq:sf4}) is given by four patches of positive and negative $\ab{ABaa{+}1}$ brackets, graphically represented as
\begin{align}
\label{eq:sf4_achiral_app}
 S^{(4)}_{ik\ell j} \quad \leftrightarrow \quad 
 \raisebox{-52pt}{\includegraphics[scale=.5]{./figures/sf4_achiral.pdf}}  \,. 
\end{align}
In addition to these basic signs which define $S^{(4)}_{ik\ell j}$, there are many more signs which are fixed automatically just from the $\ab{ABaa{+}1}$ conditions alone even \emph{before} cutting the achiral space into its chiral sub-components. In fact, the four patches in index space are either fully positive or fully negative. This means that all signs of the following brackets are fixed
\begin{align}
\begin{split}
\label{fix4}
    & \{\ab{ABpq}>0, \ab{AB\overline{pq}}>0\} \,, \text{for }
    \left[
    \begin{array}{c}
      p<q \in \{j, j{+}1,\ldots, i{-}1,i\,\} \\
      p<q \in \{k, k{+}1,\ldots, \ell{-}1,\ell\}
    \end{array}  
    \right] 
    \\
    & \{\ab{ABpq}<0, \ab{AB\overline{pq}}<0\} \,, \text{for }
    \left[
    \begin{array}{c}
      p<q \in \{i, i{+}1,\ldots, k{-}1,k\} \\
      p<q \in \{\ell, \ell{+}1,\ldots, j{-}1,j\}
    \end{array}  
    \right]
\end{split}    
\end{align} 
Cutting $S^{(4)}_{ik\ell j}$ into two chiral components can be accomplished by specifying \emph{a single sign} of one of the diagonals, $\ab{ABi\ell}$, or $\ab{ABkj}$. The first chiral component,$S^{(4),+}_{ik\ell j}$, has both signs $\ab{ABi\ell},\ab{ABkj}>0$ positive, while the second component, $S^{(4),-}_{ik\ell j}$ has both signs $\ab{ABi\ell},\ab{ABkj}<0$ negative. But, in both cases fixing one sign implies the other. Hence we can represent the chiral components as:
\begin{align}
\label{eq:sf4_chiral_components_app}
\begin{split}
   & S^{(4),+}_{ik\ell j} \leftrightarrow
   \raisebox{-72pt}{\includegraphics[scale=.5]{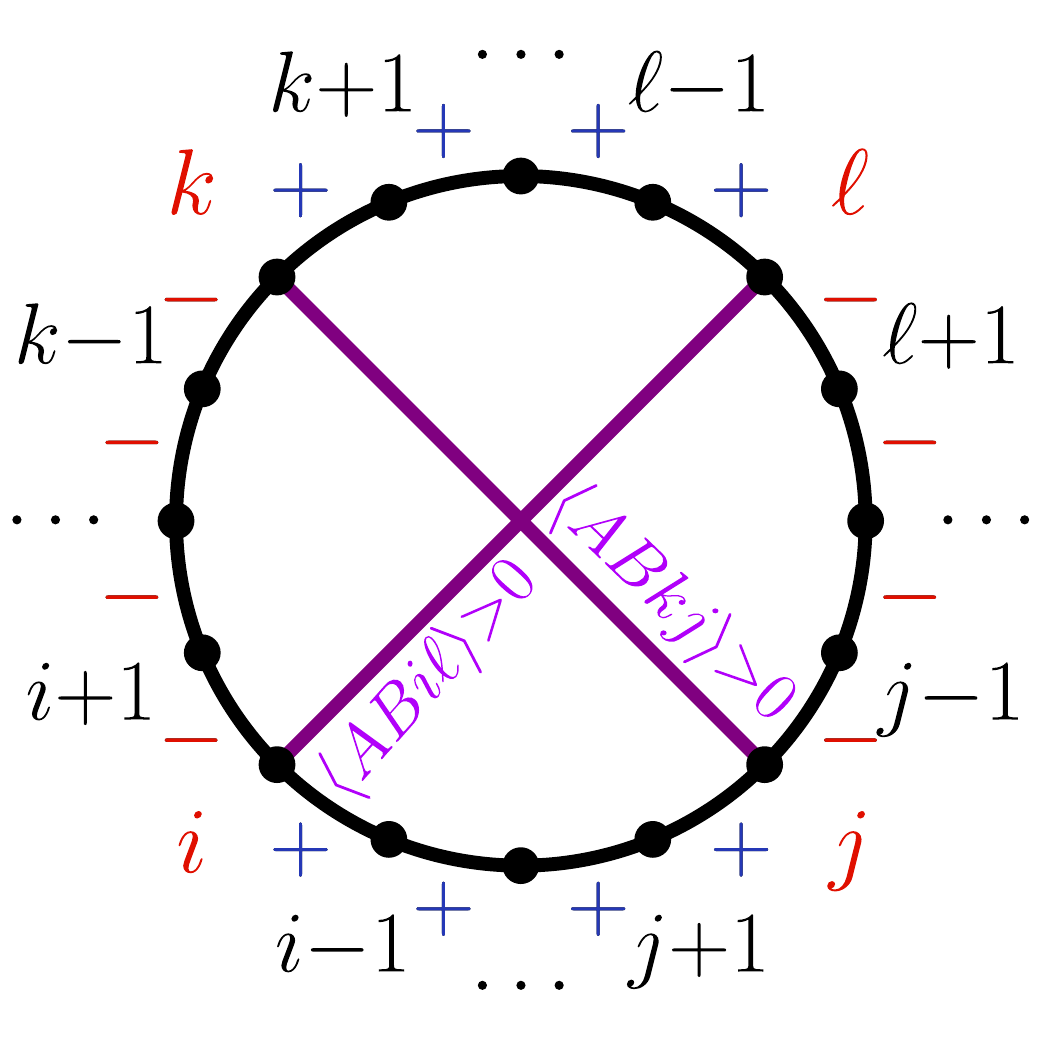}} \,, 
   \quad 
   S^{(4),-}_{ik\ell j} \leftrightarrow
   \raisebox{-72pt}{\includegraphics[scale=.5]{./figures/sf4_chiral_minus.pdf}}\,, \\[-25pt]
   & \hspace {6.5cm} i{<}k{<}\ell{<}j
\end{split}  
\end{align}
For each chiral component in eq.~(\ref{eq:sf4_chiral_components_app}) the signs of many other brackets are automatically fixed. In addition to the signs eq.~(\ref{fix4}) that were already fixed in the achiral space $S^{(4)}_{ik\ell j}$, we have 
\begin{align}
    S^{(4),+}_{ik\ell j}:\,
    \left[
    \begin{array}{c}
        \ab{ABpq}>0, \text{for } 
          p\in \{j,j{+}1,\ldots,i{-}1,i\},\,\, 
          q \in (k,k{+}1,\ldots,\ell{-}1,\ell) \\
        \ab{AB\overline{pq}}<0, \text{for }
          p\in (i,i{+}1,\ldots,k{-}1,k),\,\, 
          q\in (\ell,\ell{+}1,\ldots,j{-}1,j)
    \end{array}
    \right].
\end{align}
Note that these inequalities also cover the signs of the diagonals $\ab{ABi\ell}{>}0,\, \ab{ABkj}{>}0$, and also $\ab{ AB\overline{i\ell}}{<}0$, $\ab{AB\overline{kj}}{<}0$ so that in fact any single of these signs would be enough to specify the chiral subspace $S^{(4),+}_{ik\ell j}$.

The other chiral subspace $S^{(4),-}_{ik\ell j}$ has a corresponding set of fixed brackets
\begin{align}
    S^{(4),-}_{ik\ell j}:\,
    \left[
    \begin{array}{c}
        \ab{ABpq}<0, \text{for } 
          p\in \{j,j{+}1,\ldots,i{-}1,i\},\,\, 
          q \in (k,k{+}1,\ldots,\ell{-}1,\ell) \\
        \ab{AB\overline{pq}}>0, \text{for }
          p\in (i,i{+}1,\ldots,k{-}1,k),\,\, 
          q\in (\ell,\ell{+}1,\ldots,j{-}1,j)
    \end{array}
    \right],
\end{align}
which also covers the boundary cases such as $\ab{ABi\ell}<0$ which is enough to define $S^{(4),-}_{ik\ell j}$. Note that both set of signs for $S^{(4),+}_{ik\ell j}$ and $S^{(4),-}_{ik\ell j}$ are related by parity conjugation $\ab{ABpq}\leftrightarrow\ab{AB\overline{pq}}$; this is a consequence of the chiral nature of these spaces.

As usual, all the signs in this appendix are to be understood in the context of the usual twisted flips associated to the twisted cyclic symmetry. This means that $\ab{ABij}>0$ is valid for $1\leq i<j\leq n$ but we have to flip the sign if $j$ passes $n$ and now becomes smaller than $i$. Our sign-flip regions do not have an index space origin so that $n$ can in principle be anywhere. For the explicit sign-flip regions appearing in the main text, we always fixed $\ab{AB1n}>0$, which forced the points $n$ and 1 to be in a \emph{fully positive region}.

\newpage
\section{Gluing local geometries from two-dimensional projections}
\label{app:2d_gluing_details}
In this appendix, we summarize in detail how demanding a consistent geometry for the collection of local integral spaces selects a unique choice for the one-mass box, two-mass-hard box and chiral pentagons. This will be done by demanding that all spurious boundaries on various codimension-two projections cancel. The result of this exercise led us to the proposed spaces in section~\ref{subsec:gluing_regions}.

\subsection*{Five-point discussion}

First, let us look at the boundary when $(AB)$ passes through the point $Z_2$. On this cut surface, only the pentagon $P_{24}$ and the box $B_{45}$ in the expansion of the one-loop integrand in eq.~(\ref{eq:1loop_5pt_local_int_exp}) contribute. In eq.~(\ref{2_proj_with_labels}) we can identify the regions corresponding to the four options for the box $B_{45}$ eq.~(\ref{eq:b45_spaces}) as well as the two different options for the pentagon eqs.~(\ref{eq:pent_5_space_1})--(\ref{eq:pent_5_space_2}). The $B_{12}$ box does not contribute on this boundary. Using the labeling introduced in subsection~\ref{sec:five_point_choices}, the options for the $B_{45}$ spaces are:
\begin{align}
    \raisebox{-130pt}{\includegraphics[scale=0.6]{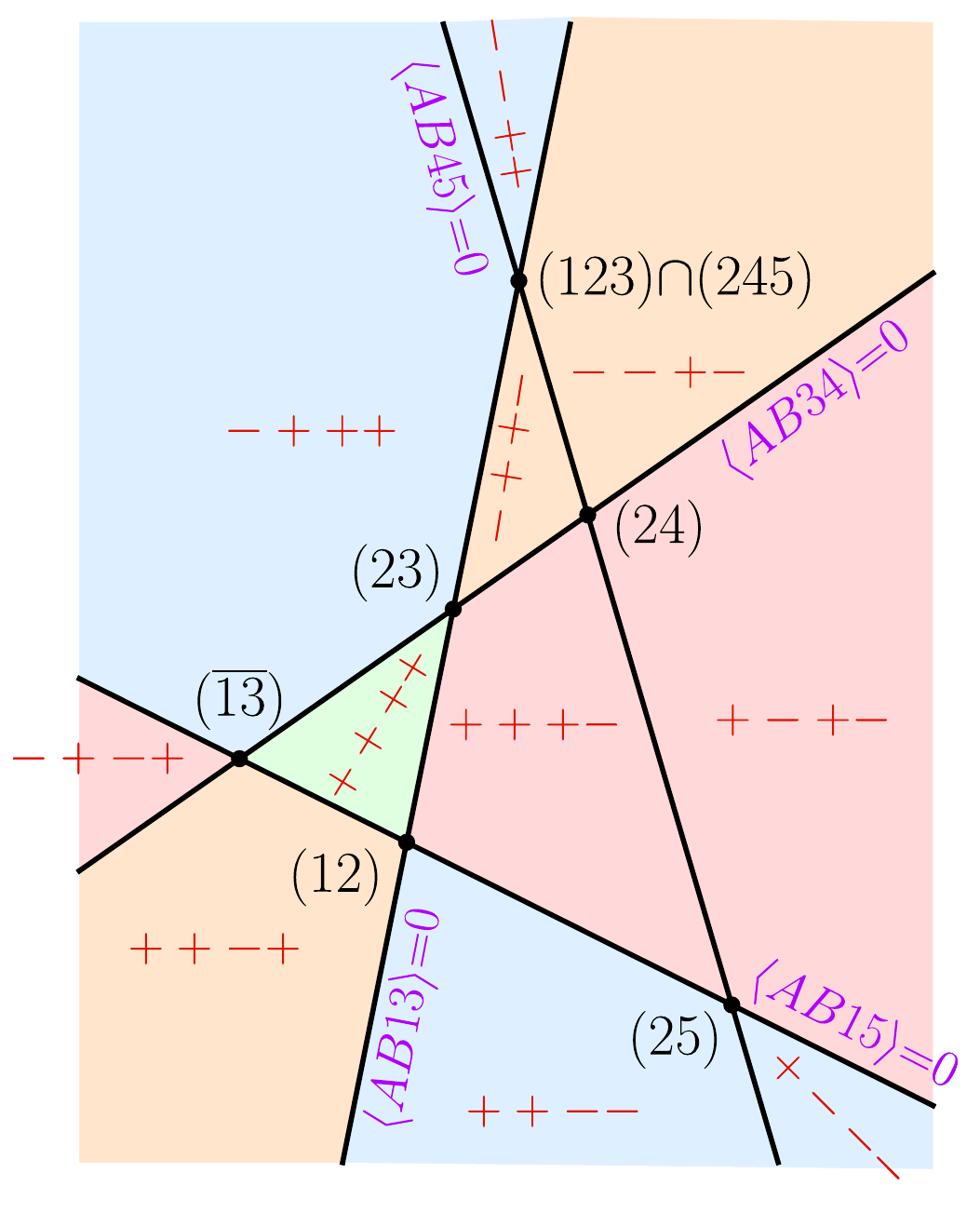}}
        \begin{array}{|c|c|c|c|c|}
        \hline
                     & \ab{AB34} & \ab{AB45}  & \ab{AB15} & \ab{AB13} \\
        \hline    
            \cellcolor{lightgreen} B^{(1)}_{45}  &  + & + & + & + \\
        \hline
            \cellcolor{lightred} B^{(2)}_{45}  & + & + & + & -\\
            \cellcolor{lightred}             & + & - & + & -\\
        \hline  
            \cellcolor{lightblue} B^{(3)}_{45}  & - & + & + & +\\
            \cellcolor{lightblue}             & - & - & + & +\\
        \hline
            \cellcolor{lightorange} B^{(4)}_{45}  & - & + & + & -\\
            \cellcolor{lightorange}            & - & - & + & -\\
        \hline              
        \end{array}
    \label{2_proj_box45}
\end{align}
Note that all four distinct regions $B^{(i)}_{45}$ have three vertices, $(12),(23)$ and $(\overline{13})$, which are the leading singularities of the box integral accessible from the codimension-two cut surface shown in eq.~(\ref{fig:on-shell-diags-boundary-structure}). The two choices for the pentagon spaces eqs.~(\ref{eq:pent_5_space_1})--(\ref{eq:pent_5_space_2}) correspond to the regions
\begin{align}
    \raisebox{-130pt}{\includegraphics[scale=0.6]{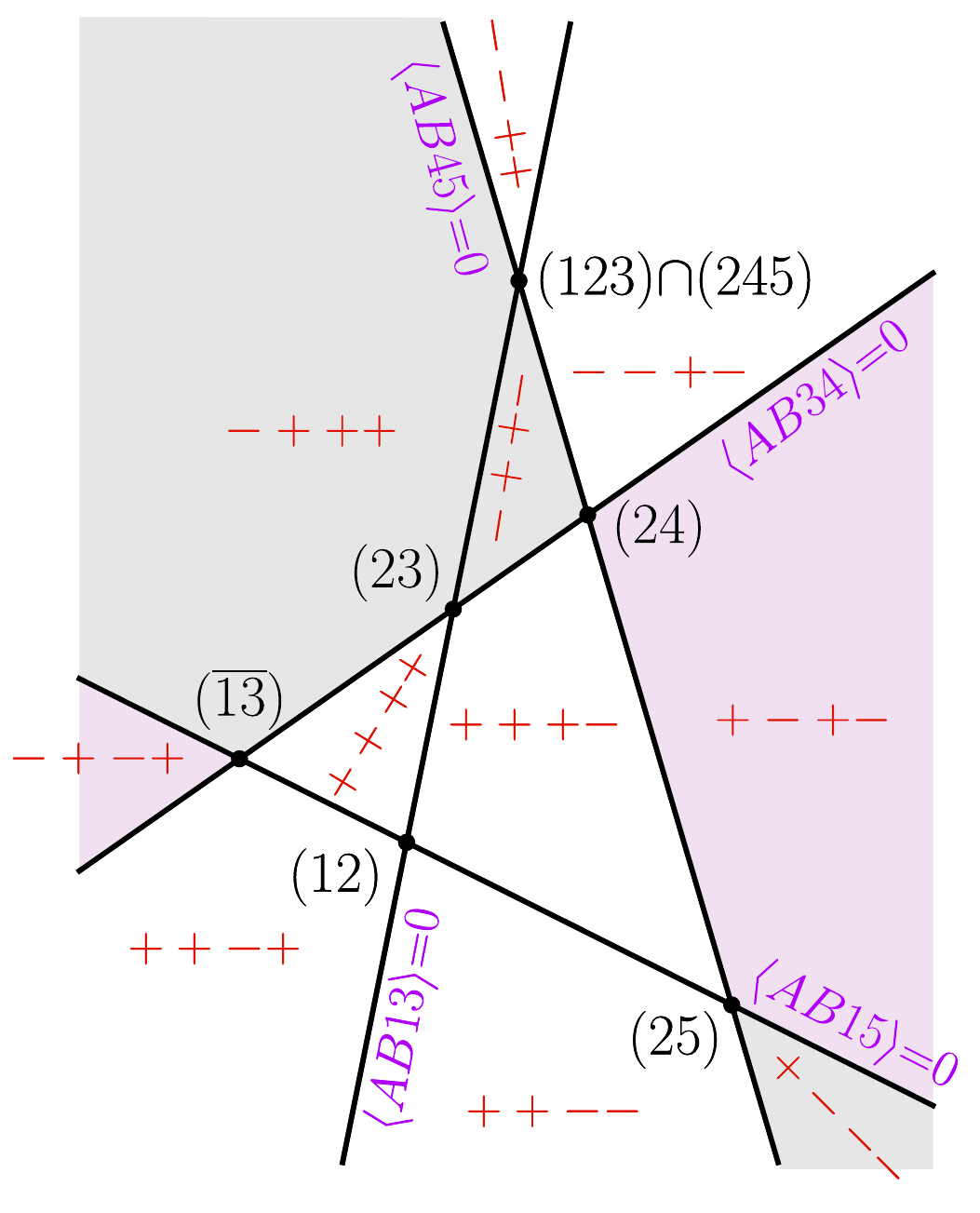}}
        \begin{array}{|c|c|c|c|c|}
        \hline
                     & \ab{AB34} & \ab{AB45}  & \ab{AB15} & \ab{AB13} \\
        \hline    
            \cellcolor{mygrey2} P^{(1)}_{24}  &  - & + & + & + \\
            \cellcolor{mygrey2}          &  - & + & + & - \\
        \hline
            \cellcolor{lightpurple} P^{(2)}_{24}  & + & - & + & -\\
        \hline  
        \end{array}
    \label{2_proj_pent}
\end{align}
In this case both pentagon spaces have three vertices, $(24),(25)$ and $(\overline{13})$. As discussed throughout  section~\ref{subsec:gluing_regions}, not all sign patterns which constitute a local integral necessarily contribute on a given cut surface; in this case, the space $P_{24}^{(1)}$ is composed of four sign patterns, only two of which have the boundary $(AB)=(A2)$.

For our purposes, we require that upon combining the two pictures in eqs.~(\ref{2_proj_box45}) and (\ref{2_proj_pent}) the spaces for $B_{45}$ and $P_{24}$ must be such that the spurious leading singularities $(\overline{13})$ and $(123){\cap}(245)$ cancel geometrically (however, since $(123){\cap}(245)$ was not present in individual integrals in the first place, we do not get any constraints from this spurious leading singularity). There are multiple ways of cancelling the spurious vertex $(\overline{13})$. The first way of cancelling this point is by simply covering it twice with an overlapping region; an example of this is given by combining $B_{45}^{(2)}$ and $P^{(2)}_{24}$ as shown in the left of eq.~(\ref{5pt_regions_spurious_pt_cancellation_egs}). Alternatively, we can cancel the spurious point by adding an additional region on one of the ``other sides'' of the vertex; an example of this is given by combining $B_{45}^{(1)}$ and $P^{(2)}_{24}$, which is illustrated in the right of eq.~(\ref{5pt_regions_spurious_pt_cancellation_egs}).
\begin{align}
\label{5pt_regions_spurious_pt_cancellation_egs}
    \raisebox{-100pt}{\includegraphics[scale=0.6]{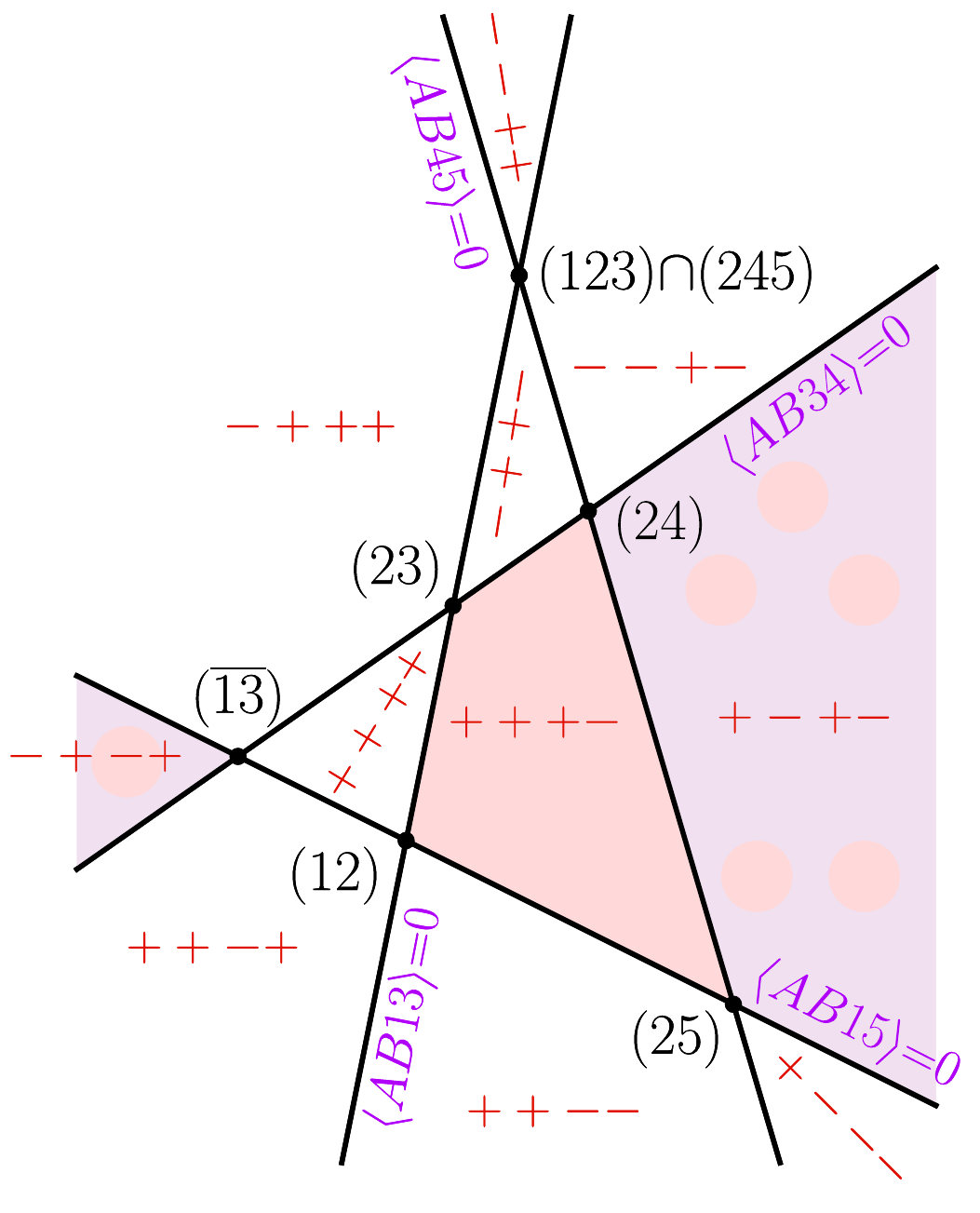}}
    \quad \text{and} \quad 
    \raisebox{-100pt}{\includegraphics[scale=0.6]{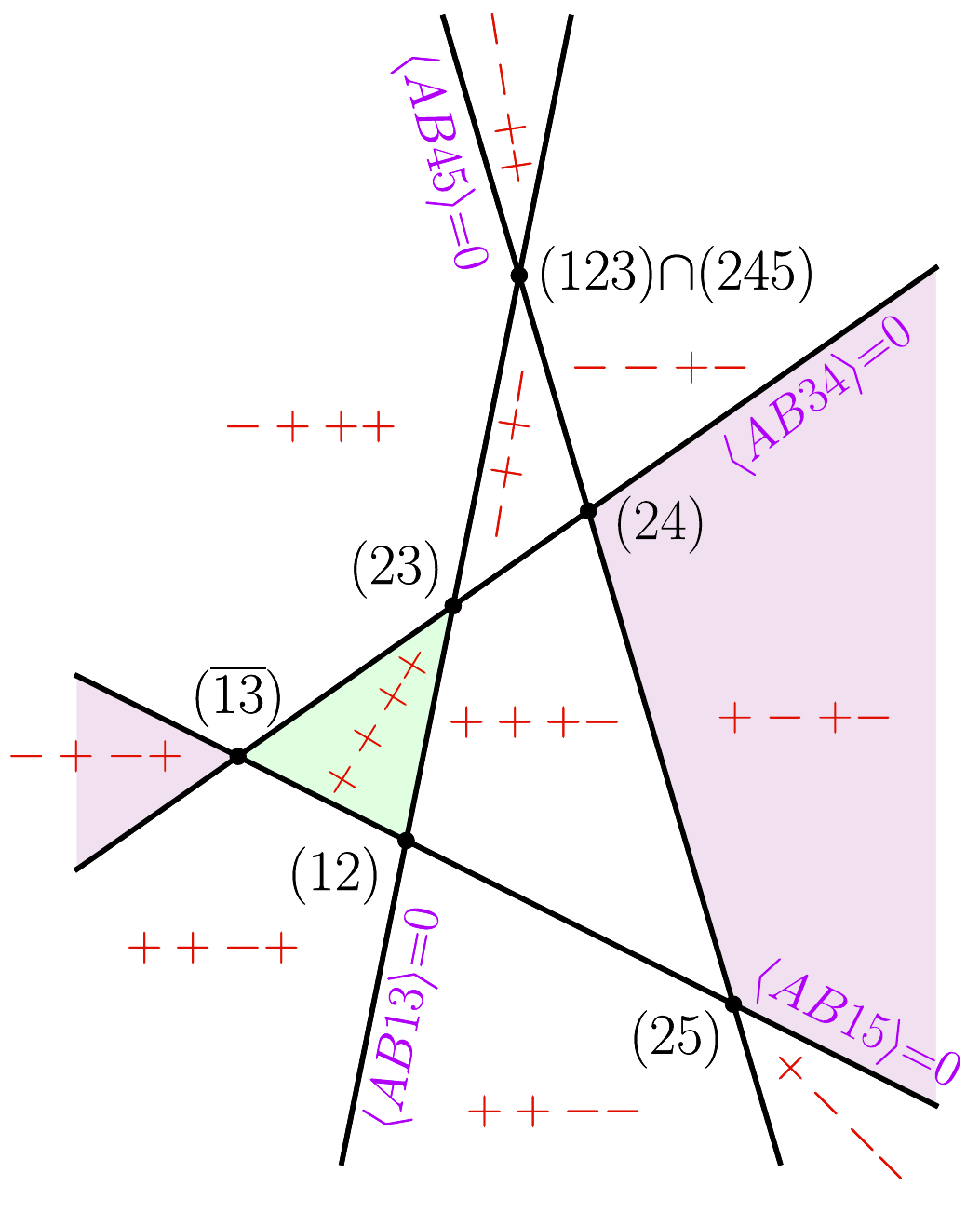}}
\end{align}
 In the first example, we see that after cancelling overlapping regions we are left with exactly the same region $\{+,+,+,-\}$ as the original MHV Amplituhedron, while in the latter we are left with a different region which has the same boundaries. In fact, it is easy to see that \emph{any} combination of the box and pentagon spaces will cancel this spurious boundary in a rather trivial fashion. The final result of any combination is equivalent to the MHV region, up to the addition of a region with zero form. The same argument holds for any other two-dimensional projection of the form $(AB)=(Ai)$, so these pictures do not yield any constraints. 

The parity conjugate configuration where $(AB)\subset(123)$ depicted in eq.~(\ref{eq:os_diag_5pt_AB_in_123}) also leads to a trivially correct space, no matter which combination of box and pentagon spaces we take. This is a consequence of the fact that only a single local integral, $B_{45}$, contributes on this cut. The chiral wavy-line numerator of the pentagon vanishes here. 

Let us now consider the configuration $(AB)\subset(234)$, which corresponds to the on-shell function of eq.~(\ref{eq:os_diag_5pt_234}), where we have the spurious boundary when $(AB)$ cuts the line $(15)$,
\begin{equation}
\raisebox{-35pt}{
\includegraphics[scale=.6]{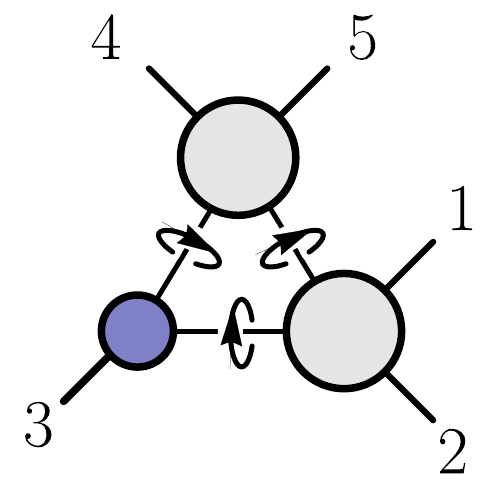}},
\end{equation}
which must be absent in the final space. For the boxes $B_{12},B_{45}$ and pentagon we have, respectively, the contributions
\begin{align}
    &
    \hspace{-1cm}
    \raisebox{-130pt}{\includegraphics[scale=0.5]{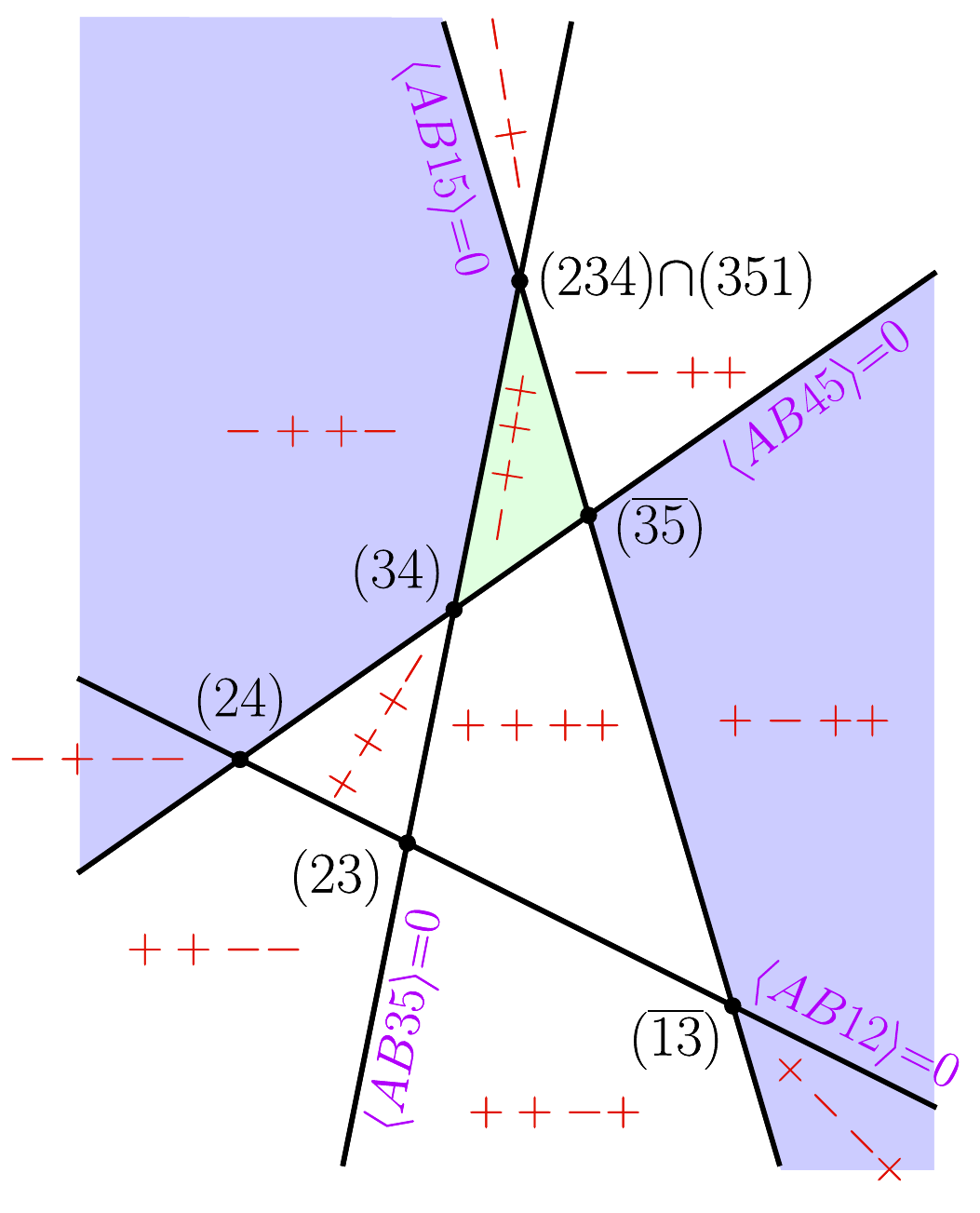}}
        \begin{array}{|c|c|c|c|c|}
        \hline
                     & \ab{AB45} & \ab{AB15}  & \ab{AB12} & \ab{AB35} \\
        \hline    
            \cellcolor{lightgreen} B^{(1,3)}_{12}  &  - & + & + & + \\
        \hline
            \cellcolor{lavenderblue} B^{(2,4)}_{12}  & - & + & + & -\\
            \cellcolor{lavenderblue}                 & - & + & - & -\\
        \hline  
        \end{array}
        \label{eq:box_b12_spaces_plane_234}
        \\[-10pt]
    &
    \hspace{-1cm}
    \raisebox{-130pt}{\includegraphics[scale=0.5]{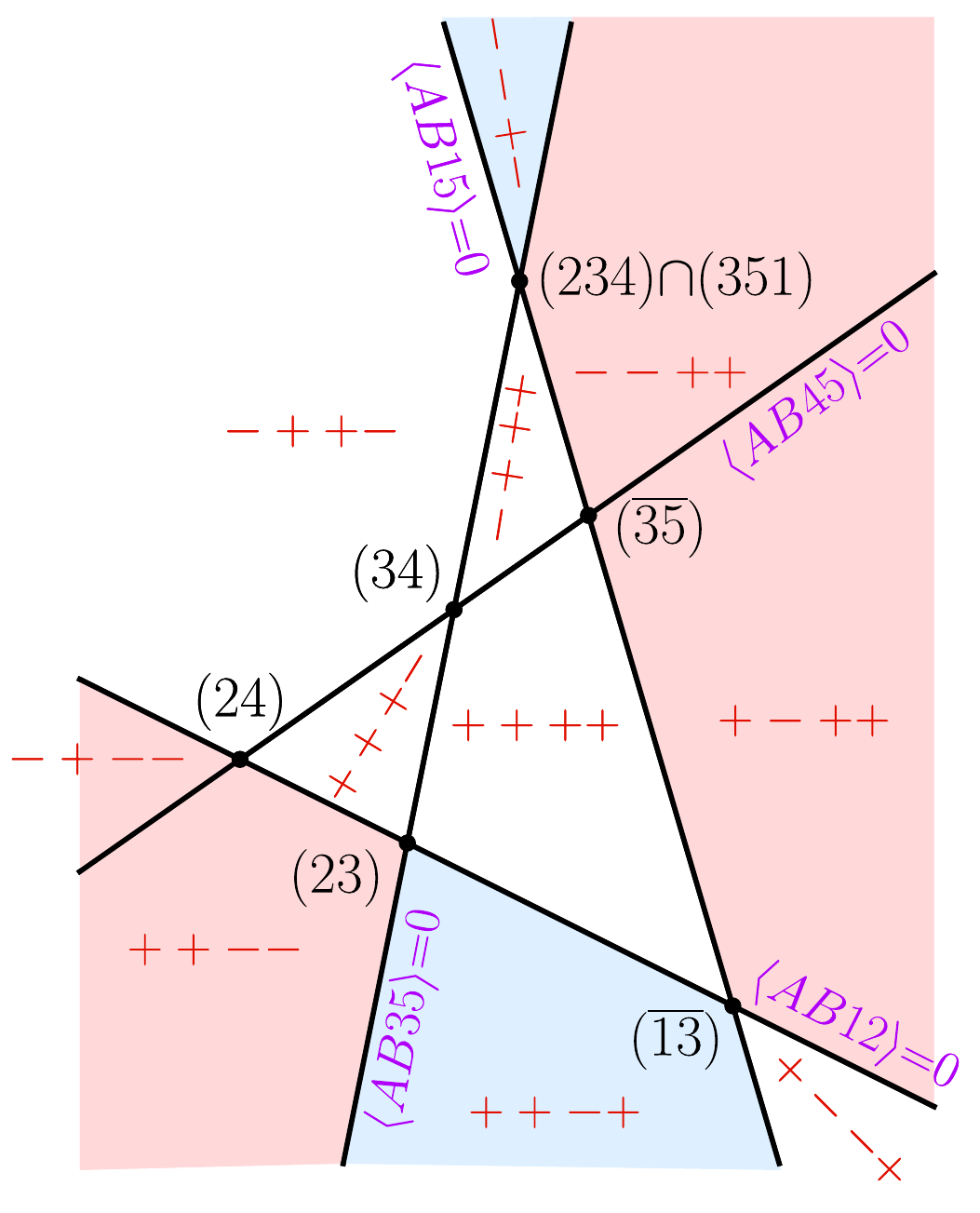}}
        \begin{array}{|c|c|c|c|c|}
        \hline
                     & \ab{AB45} & \ab{AB15}  & \ab{AB12} & \ab{AB35} \\
        \hline    
            \cellcolor{lightblue} B^{(1,3)}_{45}  &  + & + & - & + \\
        \hline
            \cellcolor{lightred} B^{(2,4)}_{45}  & + & + & - & -\\
            \cellcolor{lightred}                 & - & + & - & -\\
        \hline  
        \end{array} 
        \label{eq:box_b45_spaces_plane_234}
     \\[-10pt]
    &
    \hspace{-1cm}
    \raisebox{-130pt}{\includegraphics[scale=0.5]{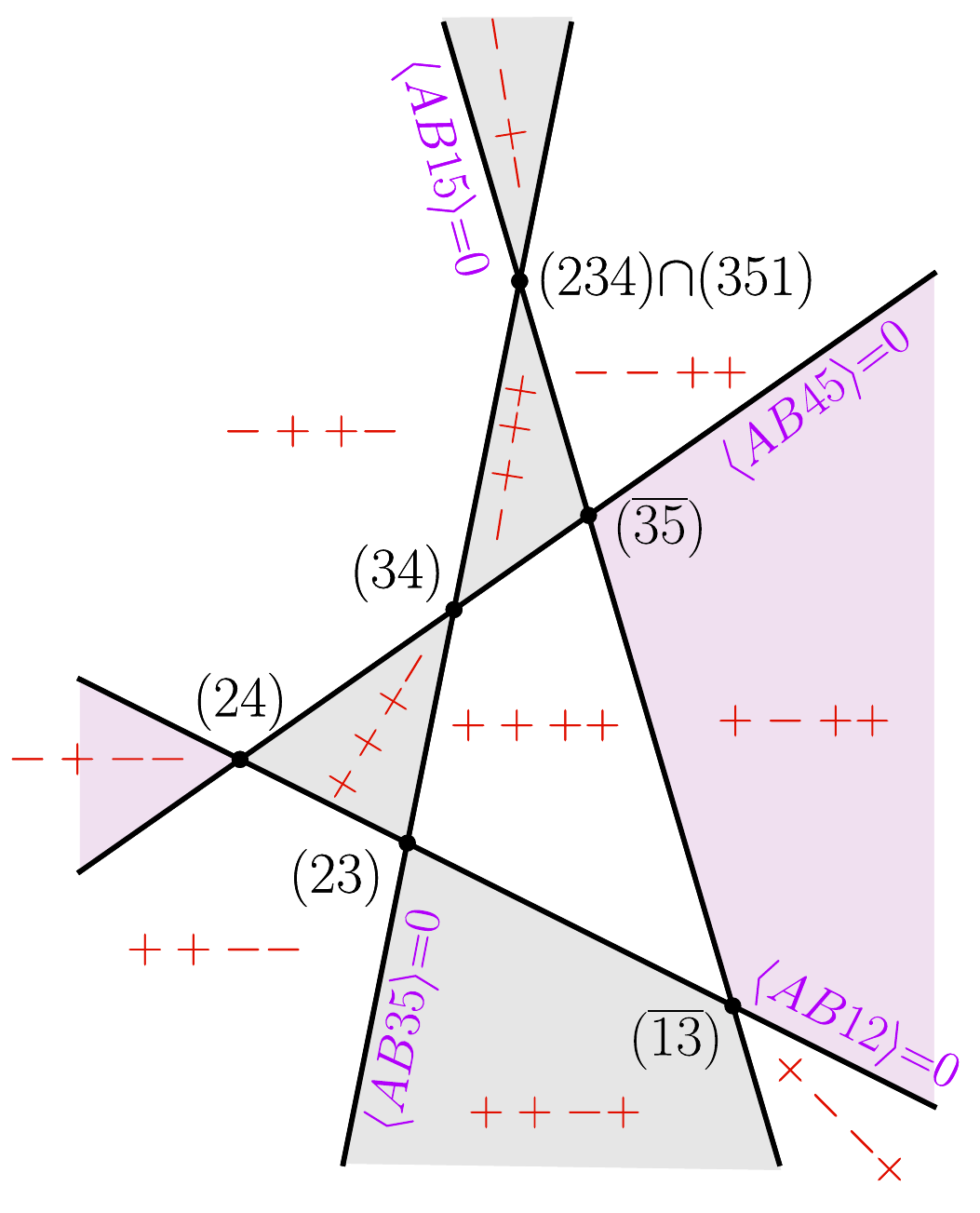}}
        \begin{array}{|c|c|c|c|c|}
        \hline
                     & \ab{AB45} & \ab{AB15}  & \ab{AB12} & \ab{AB35} \\
        \hline    
            \cellcolor{mygrey2} P^{(1)}_{24}  &  + & + & - & + \\
            \cellcolor{mygrey2}                 &  + & + & + & - \\
            \cellcolor{mygrey2}                 &  - & + & + & + \\
        \hline
            \cellcolor{lightpurple}     P^{(2)}_{24}  & - & + & - & -\\
        \hline  
        \end{array}   
        \label{eq:pent_spaces_plane_234}
\end{align}
In these pictures the entire line $\ab{AB15}=0$ is spurious i.e., it is not a boundary of the Amplituhedron; therefore, it cannot be a boundary of the Amplituhedron-Prime either. Note that both regions for the boxes $B_{12}$ and $B_{45}$, as well as the pentagon regions do have access to the $\ab{AB15}=0$ boundary. As we have seen throughout this paper, the geometric cancellation of this spurious boundary is a stronger constraint than what is na\"{i}vely observed at the level of adding canonical forms. There are two ways to combine these spaces to get the correct form: either we honestly cancel the boundary so it disappears from the full space, or we cover the \emph{entire} line. In the latter case the codimension-three line $\ab{AB15}=0$ is a geometric boundary of the space, although the codimension-four points $(234){\cap}(351)$, $(\overline{35})$ and $(\overline{13})$ on this line are not. This is unacceptable for our purposes here because geometrically it does not faithfully represent the correct boundary structure. In fact, such a combination can be seen explicitly by considering the following spaces: $B_{12}^{(1,3)}$, $B_{45}^{(1,3)}$ and $P^{(2)}_{24}$. The union of these spaces has the same vertices $(23),(24),(34)$ as the Amplituhedron, but also has the entire line $\ab{AB15}=0$:
\begin{align}
    &
    \hspace{-1cm}
    \raisebox{-130pt}{\includegraphics[scale=0.6]{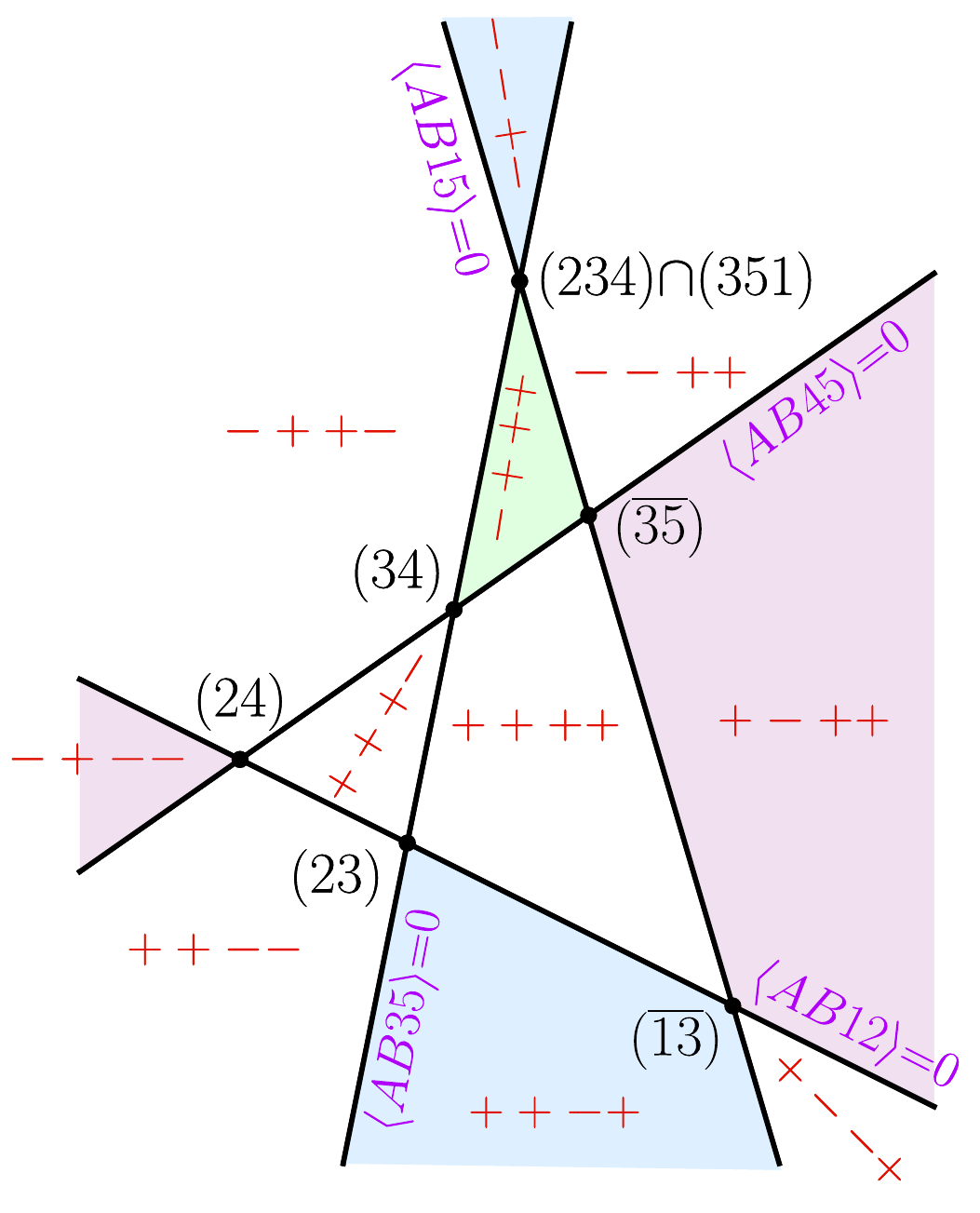}}
        \begin{array}{|c|c|c|c|c|}
        \hline
                     & \ab{AB45} & \ab{AB15}  & \ab{AB12} & \ab{AB35} \\
        \hline    
            \cellcolor{lightgreen} B^{(1,3)}_{12}  &  - & + & + & + \\
       \hline    
            \cellcolor{lightblue} B^{(1,3)}_{45}  &  + & + & - & + \\
            \hline
            \cellcolor{lightpurple}     P^{(2)}_{24}  & - & + & - & -\\
        \hline  
        \end{array}
\end{align}
If instead we use the space $P^{(1)}_{24}$, we cancel the spurious boundary:
\begin{align}
    &
    \hspace{-1cm}
    \raisebox{-130pt}{\includegraphics[scale=0.6]{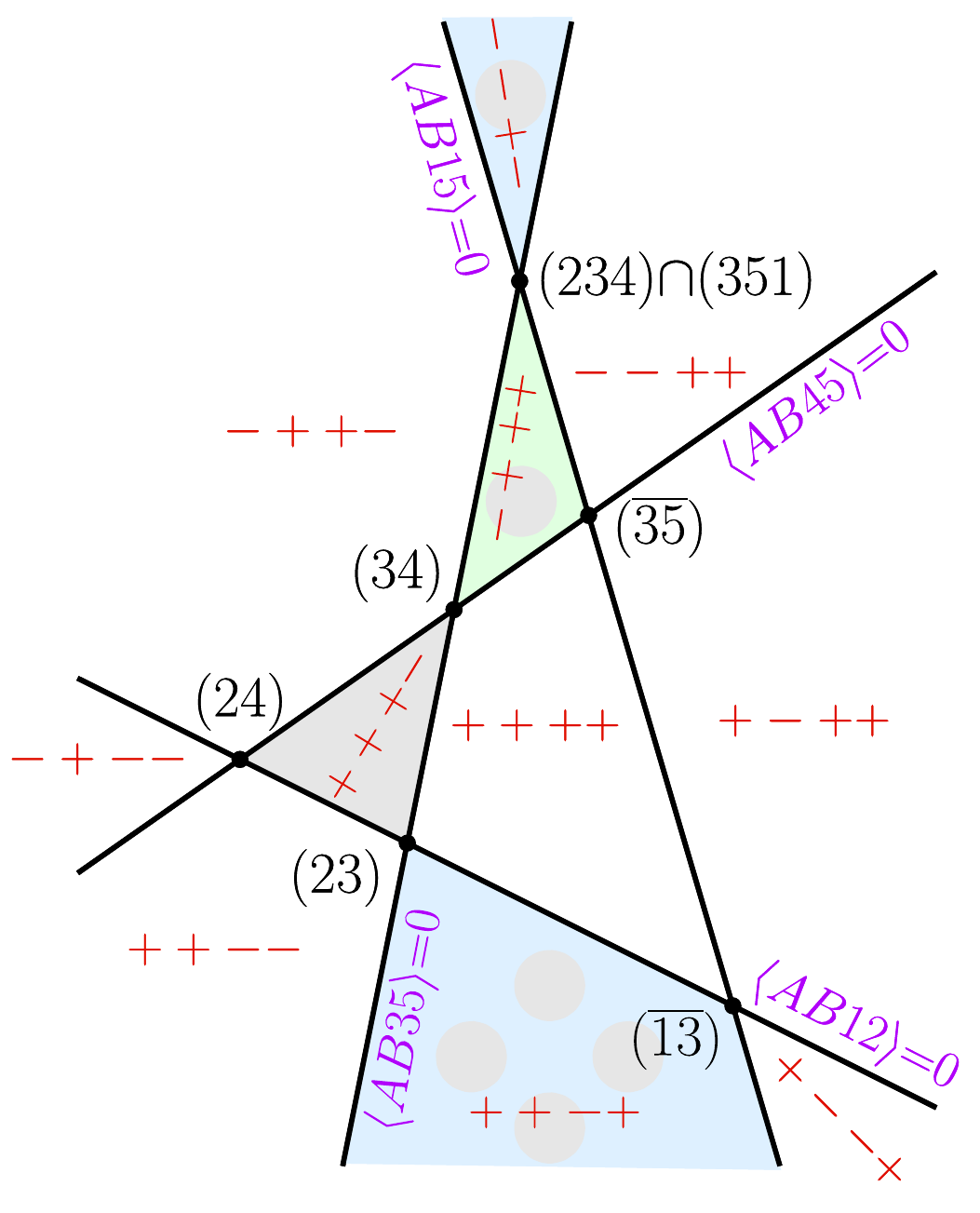}}
        \begin{array}{|c|c|c|c|c|}
        \hline
                     & \ab{AB45} & \ab{AB15}  & \ab{AB12} & \ab{AB35} \\
        \hline    
            \cellcolor{lightgreen} B^{(1,3)}_{12}  &  - & + & + & + \\
       \hline    
            \cellcolor{lightblue} B^{(1,3)}_{45}  &  + & + & - & + \\
            \hline
            \cellcolor{mygrey2} P^{(1)}_{24}  &  + & + & - & + \\
            \cellcolor{mygrey2}          &  + & + & + & - \\
            \cellcolor{mygrey2}          &  - & + & + & + \\
        \hline  
        \end{array}
\end{align}
Going through the eight possible combinations of spaces in eqs.~(\ref{eq:box_b12_spaces_plane_234})--(\ref{eq:pent_spaces_plane_234}), we see that only the following combinations cancel the spurious line $\ab{AB15}=0$:
\begin{equation}
\label{intermed_cases}
   B_{12}^{(1,3)},B_{45}^{(1,3)},P^{(1)}_{24},\quad \text{and} \quad
   B_{12}^{(2,4)},B_{45}^{(2,4)},P^{(1)}_{24}\,.
\end{equation}
Remarkably, we see that there is no uniform choice for the box spaces which cancels the spurious contributions of $P_{24}^{(2)}$ on this cut!

There are three remaining configurations $(AB)\subset(345),(451),(512)$ to check. The $(345)$ projection is trivially matched by any space for $B_{12}$ as no other term contributes. For the $(451)$ projection, only the box $B_{12}$ and the pentagon contribute. In terms of the labeling above, on this cut surface the box choices $(1,4)$ and $(2,3)$ become indistinguishable, respectively. 
The box $B_{12}$ and pentagon correspond to the regions
\begin{align}
&
    \hspace{-1cm}
    \raisebox{-130pt}{\includegraphics[scale=0.6]{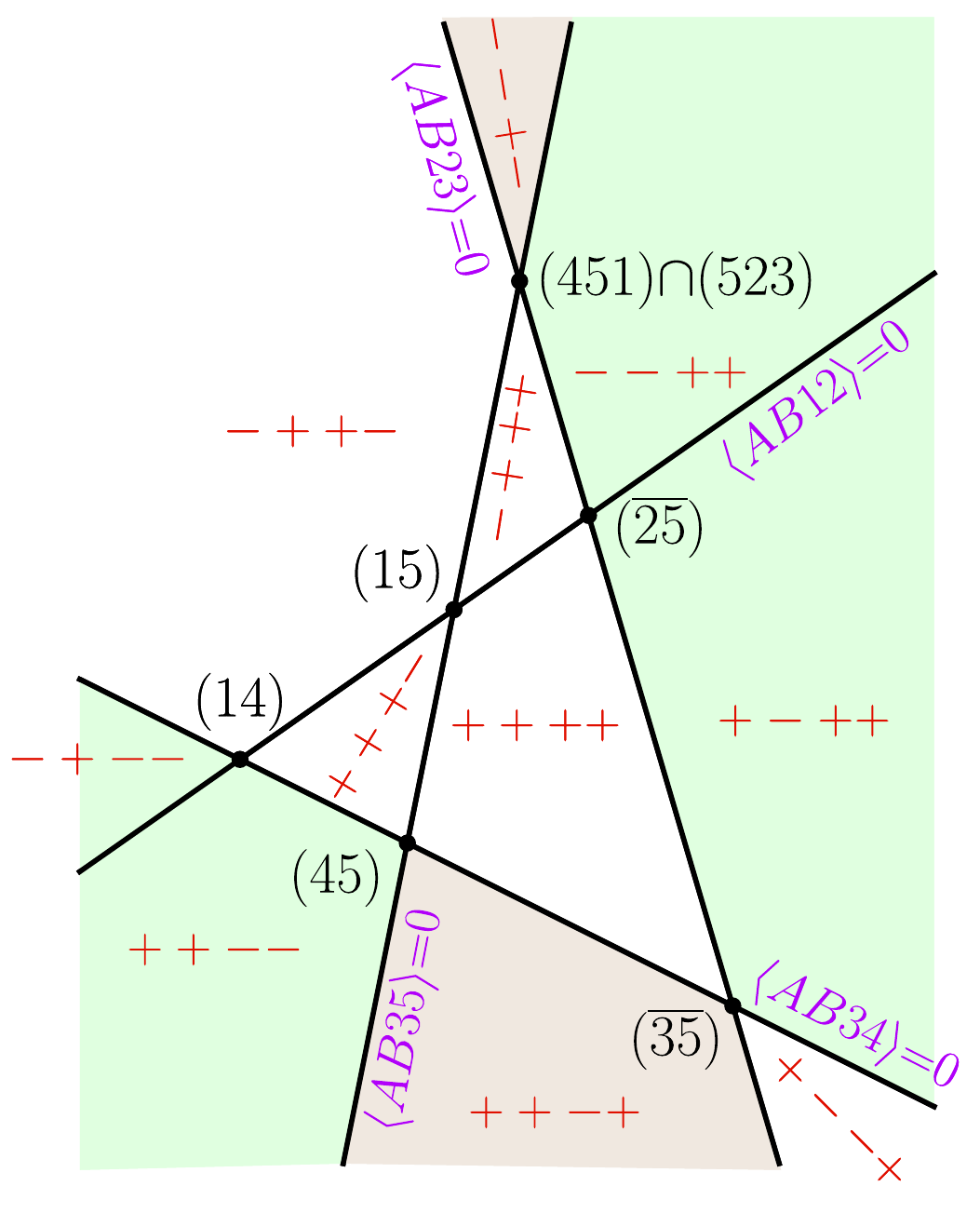}}
        \begin{array}{|c|c|c|c|c|}
        \hline
                     & \ab{AB12} & \ab{AB23}  & \ab{AB34} & \ab{AB35} \\
        \hline    
            \cellcolor{lightbrown}   B_{12}^{(1,4)}  & + & + & - & + \\
       \hline    
           \cellcolor{lightgreen}     B_{12}^{(2,3)}  & + & + & - & - \\
           \cellcolor{lightgreen}                     & - & + & - & - \\
        \hline  
        \end{array} \\
   &
    \hspace{-1cm}
    \raisebox{-130pt}{\includegraphics[scale=0.6]{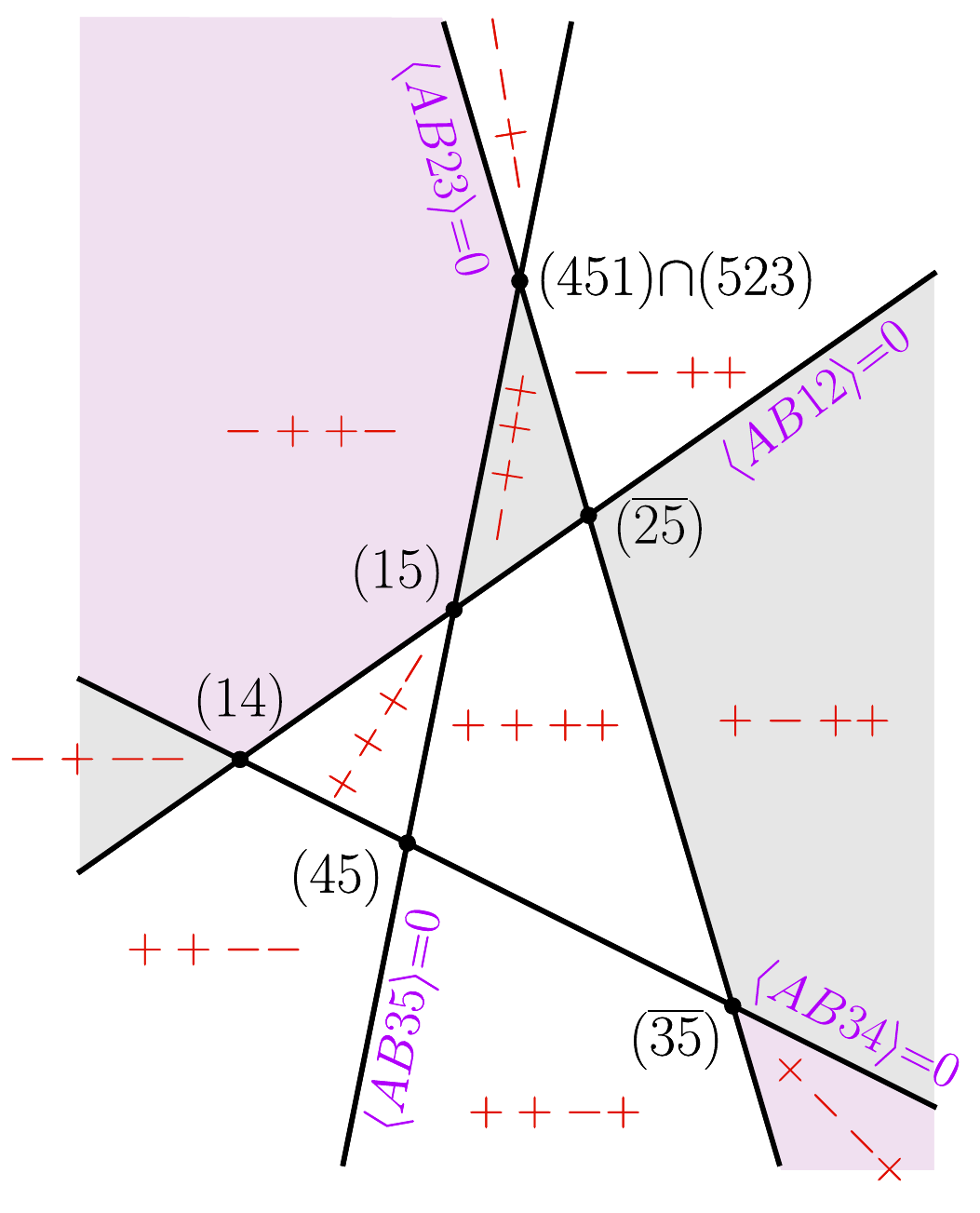}}
        \begin{array}{|c|c|c|c|c|}
        \hline
                     & \ab{AB12} & \ab{AB23}  & \ab{AB34} & \ab{AB35} \\
        \hline    
            \cellcolor{mygrey2} P^{(1)}_{24}  & - & + & + & + \\
            \cellcolor{mygrey2}          & + & - & + & + \\
       \hline    
           \cellcolor{lightpurple}     P^{(2)}_{24}  & - & + & + & -\\
        \hline  
        \end{array}
\end{align}
Demanding the cancellation of the boundary $\ab{AB23}=0$ we see that only two of the remaining four choices in eq.~(\ref{intermed_cases}) survive, 
\begin{equation}
\label{five_point_solns_app}
B_{12}^{(2)},B_{45}^{(2)},P^{(1)}_{24},\quad\text{and}\quad B_{12}^{(3)},B_{45}^{(3)},P^{(1)}_{24}.
\end{equation}
Following exactly the same procedure for the final configuration $(AB)\subset(512)$ we find that both choices once again cancel the spurious boundary $\ab{AB34}=0$.

Thus, at five points we are forced to choose the space, eq.~(\ref{eq:pent_5_space_1}), for the pentagon, and can cancel all spurious boundaries using two different choices for the boxes. Both choices are completely satisfactory at this multiplicity. However, only one of these solutions generalizes to higher points. This can be seen directly at six points, where an additional constraint arises: our five-point choice must be compatible with (at least) one of the two spaces in eqs.~(\ref{2mh_boxsign1})--(\ref{2mh_boxsign2}) for the two-mass hard box. 
\subsection*{Six-point discussion}

Let us examine the natural extensions of the five-point solutions in eq.~(\ref{five_point_solns}) relevant for the two-dimensional projection $(AB)\subset(234)$, where the box and pentagon spaces are 
\begin{align}
 B_{456}^{(2)}=&\raisebox{-45pt}{\includegraphics[scale=.37]{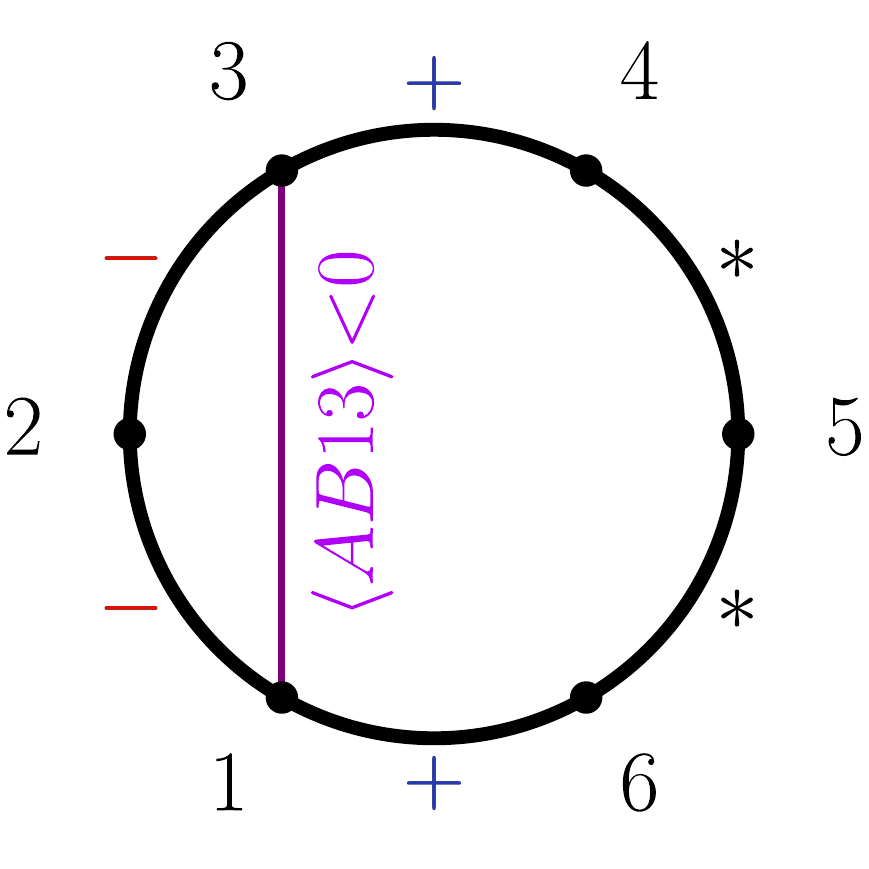}},\quad \text{and}\quad 
    B_{456}^{(3)}= \raisebox{-45pt}{\includegraphics[scale=.37]{./figures/choice3_b456.pdf}}.
\end{align}
\begin{align}
    \begin{split}
    P_{24}^{(1)}=&
    \raisebox{-45pt}{\includegraphics[scale=.37]{./figures/choice1_p24_1.pdf}}
    {+}\raisebox{-45pt}{\includegraphics[scale=.37]{./figures/choice1_p24_2.pdf}}
    {+}\raisebox{-45pt}{\includegraphics[scale=.37]{./figures/choice1_p24_3.pdf}}
   {+}\raisebox{-45pt}{\includegraphics[scale=.37]{./figures/choice1_p24_4.pdf}},
    \end{split}
\end{align}
For the two-mass-hard box $B_{12,56}$ we have two choices, see eqs.~(\ref{2mh_boxsign1})--(\ref{2mh_boxsign2}): 
\begin{equation}
    B_{12,56}^{(1)}=\raisebox{-48pt}{\includegraphics[scale=.37]{./figures/choice1_b12_56.pdf}},\quad \text{and}\quad
     B_{12,56}^{(2)}=\raisebox{-48pt}{\includegraphics[scale=.37]{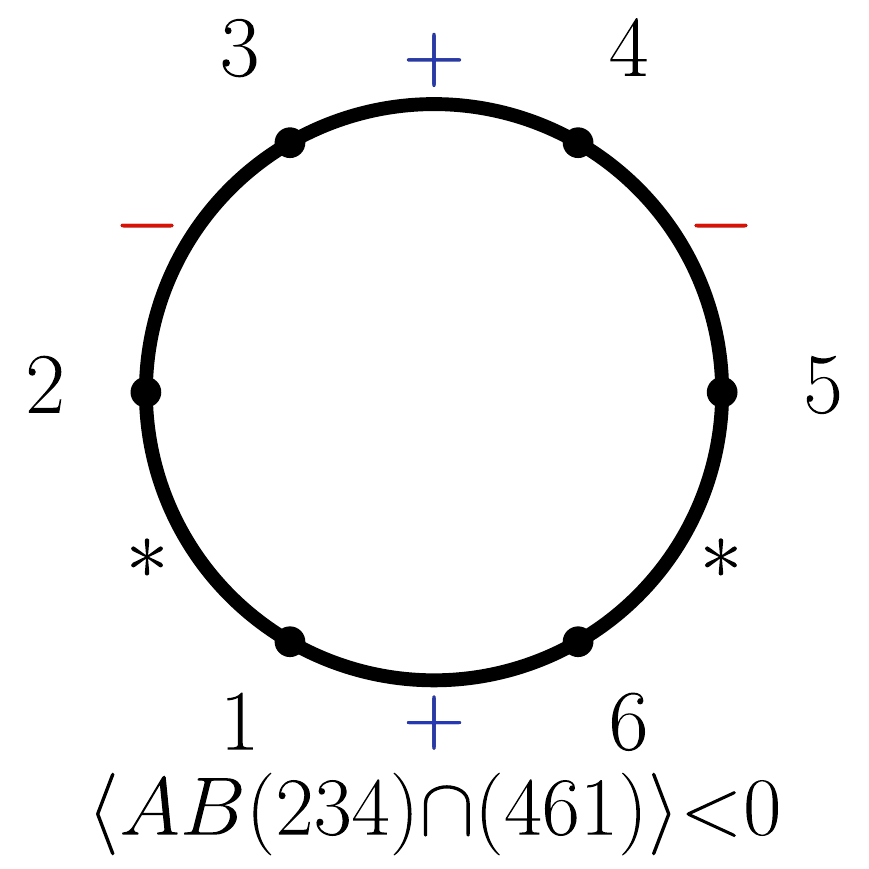}}.
\end{equation}
Filling in the corresponding regions in eq.~(\ref{2d_proj_6pt_with_labels}), the result for the one-mass box, pentagon and two-mass hard spaces are: 
\begin{align}
\begin{split}
&   \hspace{-1cm}
    \raisebox{-130pt}{\includegraphics[scale=.45]{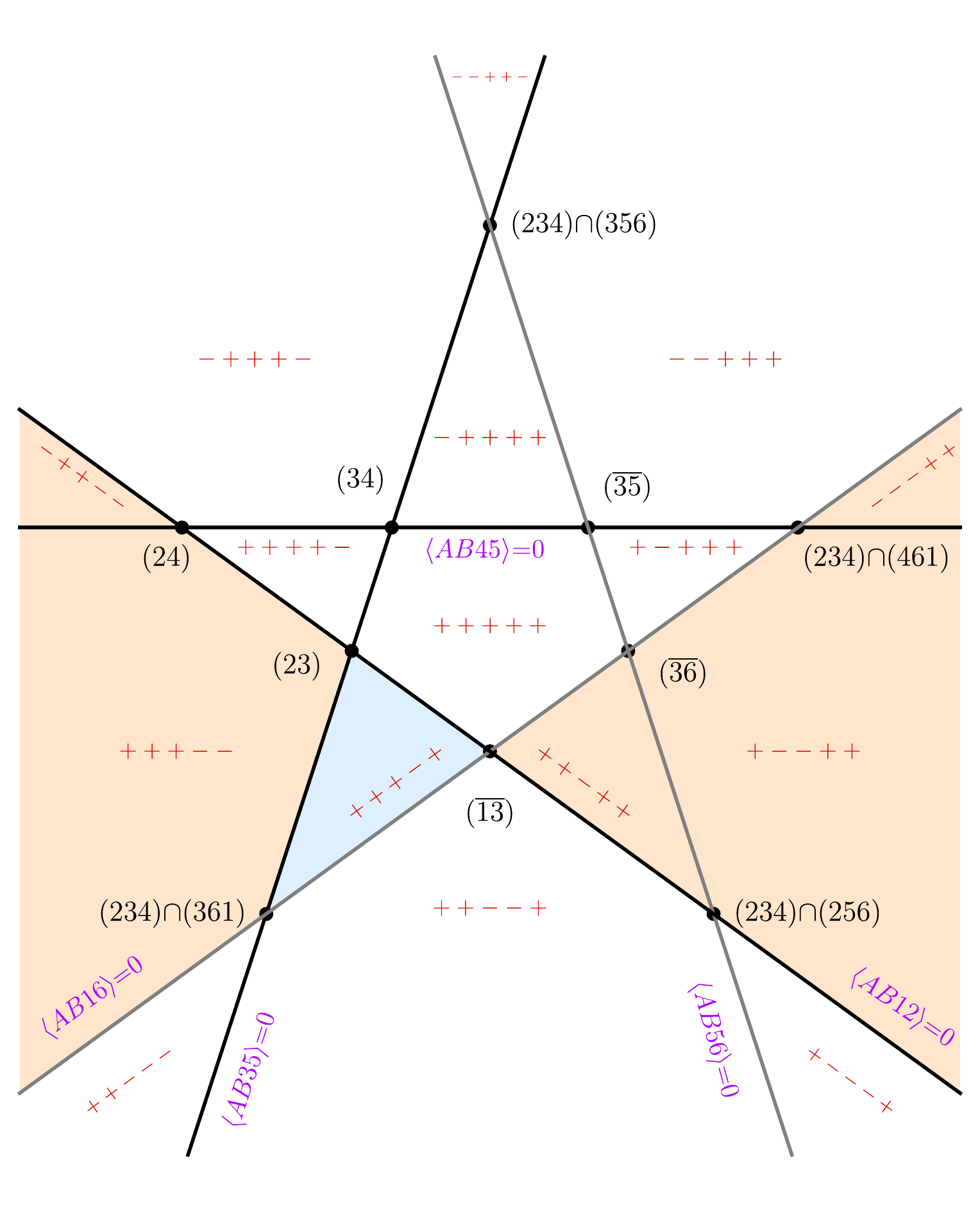}}
    \quad
    \raisebox{-130pt}{\includegraphics[scale=.45]{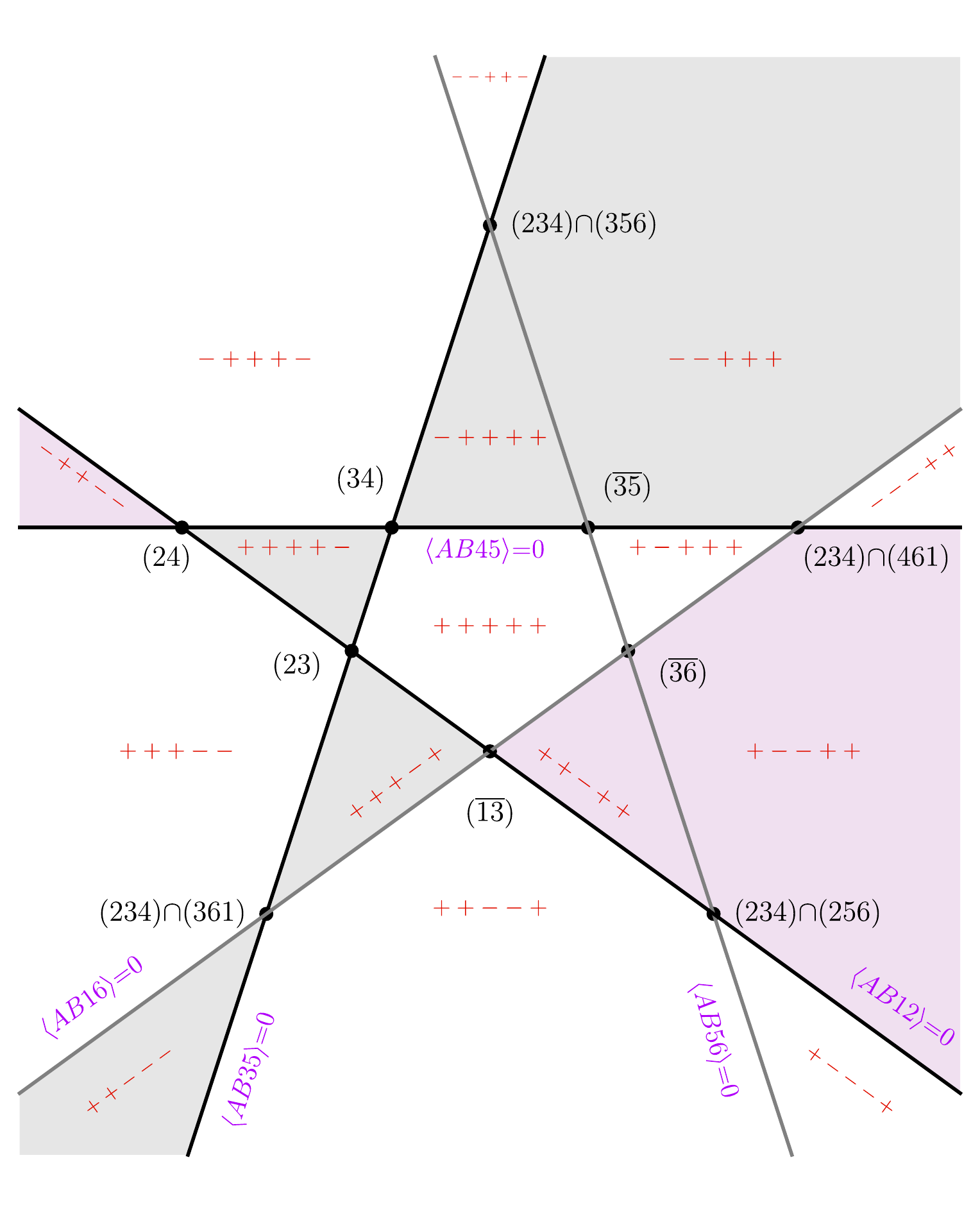}}
\\
&   \hspace{-1cm}
\raisebox{-130pt}{
    \includegraphics[scale=.45]{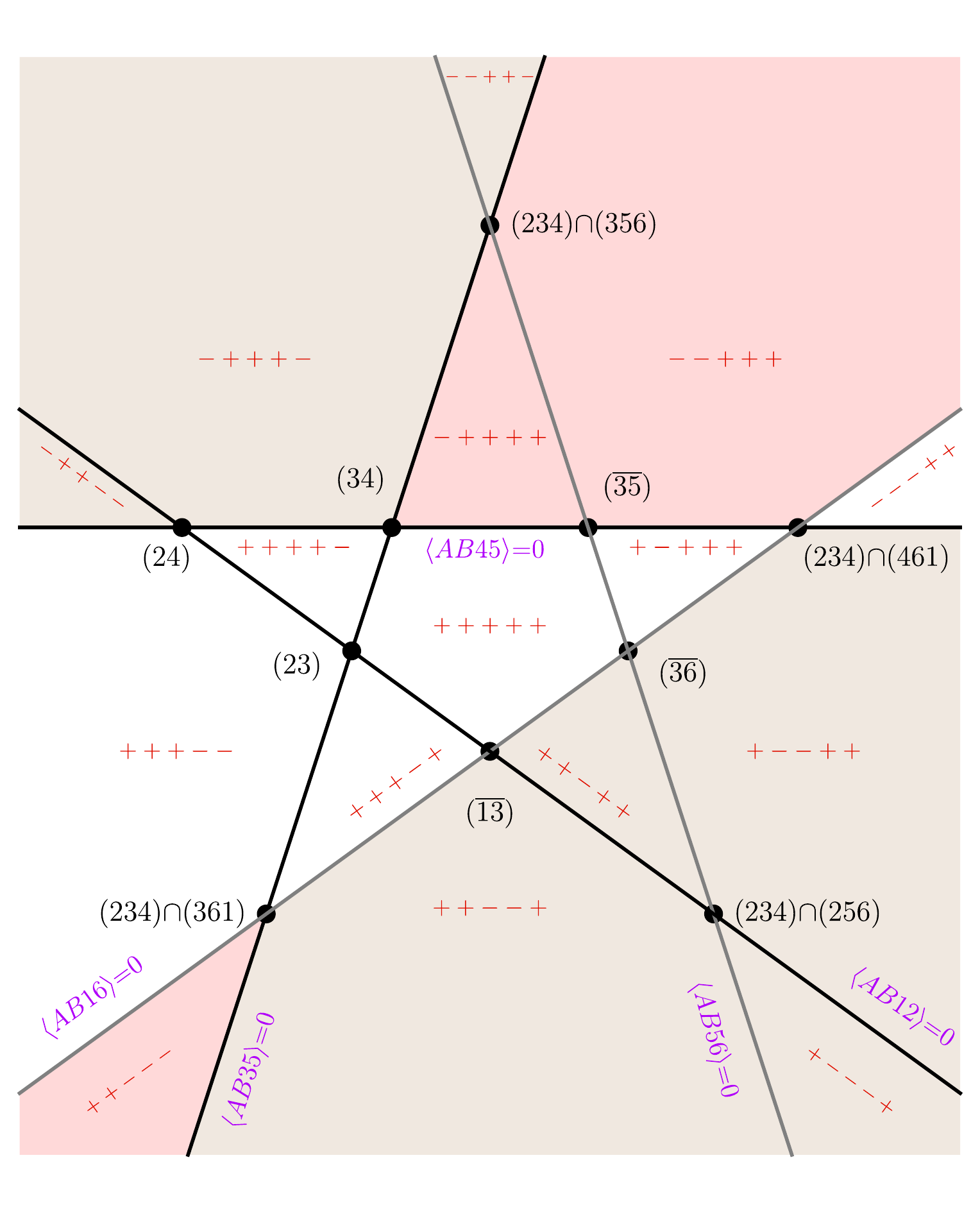}}\, 
    \begin{array}{|c|c|c|c|c|c|}
        \hline
                     & \ab{AB45} & \ab{AB56}  & \ab{AB16} & \ab{AB12} & \ab{AB35} \\
        \hline    
          \cellcolor{lightorange}     B_{456}^{(2)}  & + & + & + & - & +\\
           \cellcolor{lightorange}                    & - & + & + & - & - \\
           \cellcolor{lightorange}                    & - & - & + & - & - \\
       \hline
            \cellcolor{lightblue} B_{456}^{(3)}   & + & + & + & - & +\\
        \hline  
        \cellcolor{mygrey2}    P_{24}^{(1)}   & + & + & + & - & +\\[0pt]
            \cellcolor{mygrey2}                  & + & + & + & + & - \\[-5pt]   
            \cellcolor{mygrey2}                  & - & + & + & + & + \\[-5pt]   
            \cellcolor{mygrey2}                  & - & - & + & + & + \\               
       \hline    
           \cellcolor{lightpurple}     P_{24}^{(2)}   & + & - & - & + & +\\[0pt]
           \cellcolor{lightpurple}                    & + & + & - & + & + \\
           \hline
            \cellcolor{lightred}    B_{12,56}^{(1)}   & - & + & + & + & +\\
            \cellcolor{lightred}                      & - & - & + & + & + \\   
       \hline    
           \cellcolor{lightbrown}     B_{12,56}^{(2)}   & - & + & + & + & -\\
           \cellcolor{lightbrown}                       & - & - & + & + & - \\
           \cellcolor{lightbrown}                       & - & + & + & - & - \\   
           \cellcolor{lightbrown}                       & - & - & + & - & - \\       
        \hline  
    \end{array}
\end{split}    
\end{align}
Demanding the boundaries $\ab{AB56}=0$ and $\ab{AB16}=0$ cancel fixes the choice of the two-mass hard box for both solutions:
\begin{equation}
B_{456}^{(3)},P^{(1)}_{24},B_{12,56}^{(1)},\quad\text{and}\quad B_{456}^{(2)},P^{(1)}_{24},B_{12,56}^{(2)}.
\end{equation}
We claim that while the first option works on all two-dimensional projections, the second option is incompatible with the cut surface where $(AB)\subset(345)$. Indeed, repeating the above exercise, on this boundary the one-mass box $B_{123}$, pentagon $P_{35}$ and two-mass hard box contribute. Using the second option for the two-mass hard box $B_{12,56}^{(2)}$ we find:
\begin{align}
&
    \hspace{-1cm}
    \raisebox{-130pt}{\includegraphics[scale=.45]{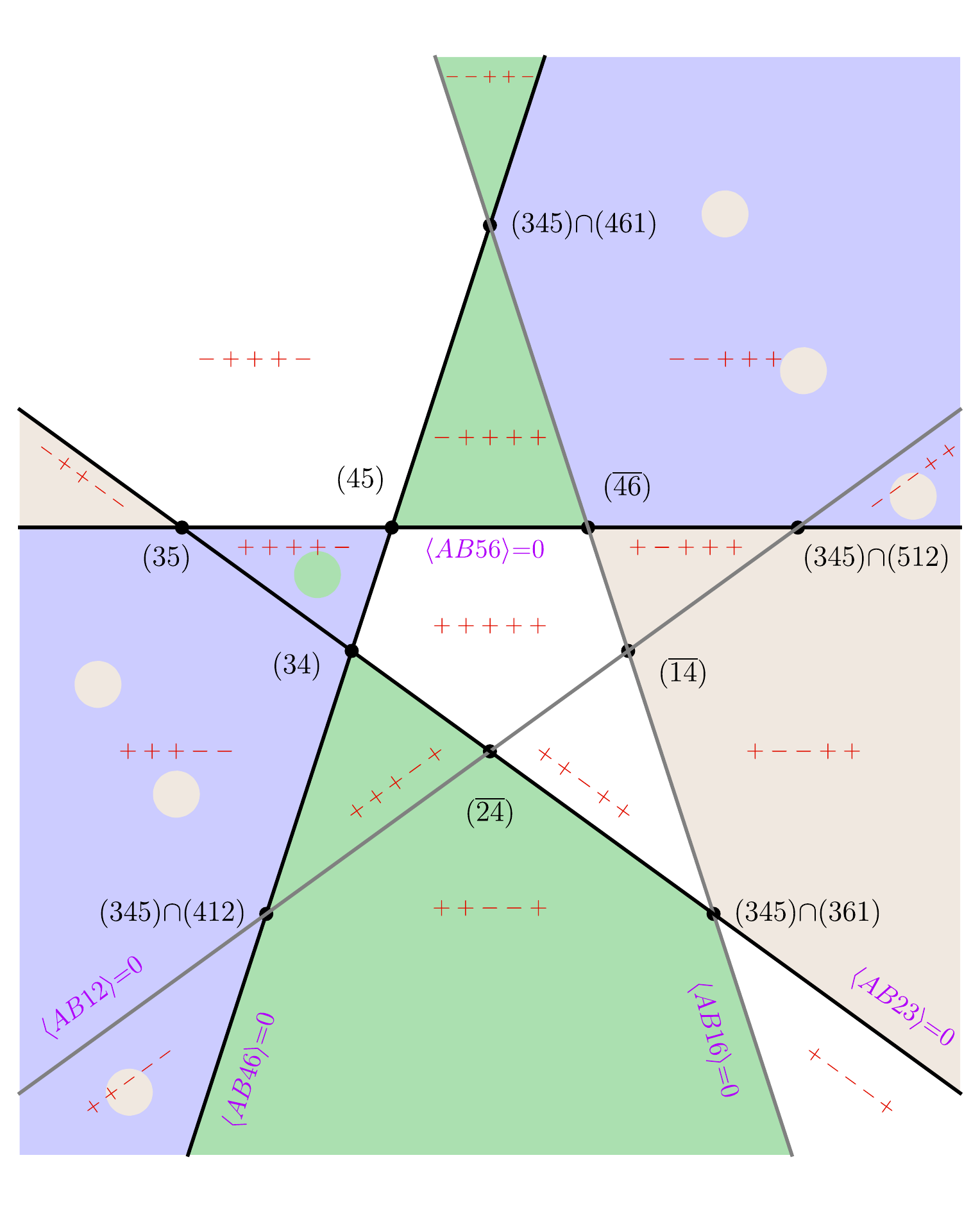}}
        \begin{array}{|c|c|c|c|c|c|}
        \hline
                     & \ab{AB45} & \ab{AB56}  & \ab{AB16} & \ab{AB12} & \ab{AB35} \\
        \hline    
            \cellcolor{lavenderblue}    B_{123}^{(2)}       & - & - & + & + & +\\
        \cellcolor{lavenderblue}   & - & - & - & + & +\\
         \cellcolor{lavenderblue}      & + & + & + & + & -\\
       \hline    
           \cellcolor{celadon}    P_{35}^{(1)}     & + & + & + & + & -\\
           \cellcolor{celadon}      & - & + & + & + & + \\         
            \cellcolor{celadon}      & + & + & + & - & + \\       
             \cellcolor{celadon}      & + & + & - & - & + \\         
       \hline    
           \cellcolor{lightbrown}     B_{12,56}^{(2)}   & + & - & + & + & +\\        
           \cellcolor{lightbrown}      & + & - & - & + & +\\    
           \cellcolor{lightbrown}      & - & - & + & + & +\\   
           \cellcolor{lightbrown}      & - & - & - & + & +\\    
        \hline  
        \end{array}
\end{align}
We see that although the boundary $\ab{AB12}{=}0$ is cancelled, the entire $\ab{AB16}{=}0$ boundary is present in the final space. This selects the union of $P^{(1)}_{35}$ $B_{456}^{(2)}$ and $B_{12,56}^{(1)}$ as the \emph{unique} (subject to the assumption that we make uniform choices for all boxes and pentagons, respectively) candidate space whose boundary structure is identical to the original Amplituhedron on this cut surface. At six points, we have verified that this combination (together with the other local integrals which did not contribute on the $(234),(345)$ boundaries) is free of all spurious boundaries. 


\newpage
\bibliographystyle{JHEP}
\phantomsection
\bibliography{amp_refs}
\clearpage

\end{document}